%% file: AM_thesis_November_2021.tex
\documentclass{report}
\usepackage{amsfonts}
\usepackage{makeidx}
\usepackage{latexsym,amsmath,amssymb,amscd}
\usepackage{makecell}
\usepackage{xcolor}
\usepackage[all]{xy}
\usepackage{float}
\usepackage{longtable}
\usepackage[pdftex]{graphicx}
\usepackage{rotating}
\usepackage{multirow}
\usepackage{framed}
\usepackage{enumitem}
\usepackage{mdframed}

\allowdisplaybreaks

\newtheorem{theorem}{Theorem}[section]

\newtheorem{example}{Example}[section]

\newtheorem{proposition}{Proposition}[section]
\newtheorem{remark}{Remark}[section]
\newenvironment{proof}[1][Proof]{\noindent\textbf{#1.} }{\ \rule{0.5em}{0.5em}}
\makeindex

\begin{document}

\begin{titlepage}
\begin{center}
{\Huge{Phd Thesis}}
\\[0.2cm]
\hrulefill
\\[0.8cm]
{\Huge{INTEGRABILITY OF DYNAMICAL}}
\\[0.3cm]
{\Huge{SYSTEMS: A GEOMETRICAL VIEWPOINT}}
\\[0.8cm]
\hrulefill
\\[1cm]
Author: Antonios Mitsopoulos \\ Email: antmits@phys.uoa.gr
\\[0.5cm]
Supervisor: Michael Tsamparlis, Professor NKUA
\\[0.5cm]
\begin{figure}[htb]
\begin{center}
\includegraphics[scale=0.6]{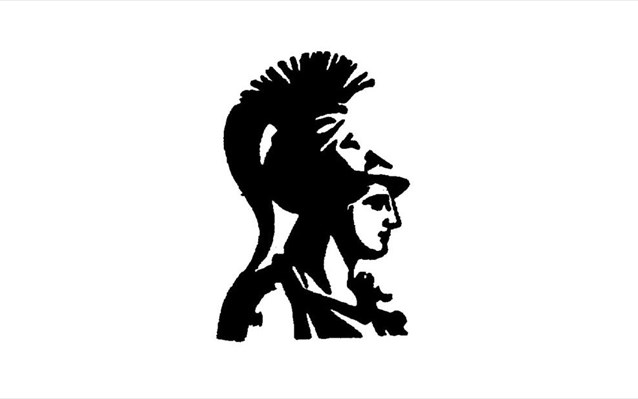}
\end{center}
\end{figure}
National and Kapodistrian University of Athens, \\
Department of Physics, Section of Astrophysics - Astronomy - Mechanics
\\[0.5cm]
Athens, 13 May 2022
\\[0.5cm]
\end{center}

Committee: \newline
1) Theocharis Apostolatos, Professor NKUA \newline
2) Nectarios Vlahakis, Professor NKUA \newline
3) Nikolaos Stergioulas, Professor AUTH \newline
4) Dimitrios Frantzeskakis, Professor NKUA \newline
5) Fotios Diakonos, Associate Professor NKUA \newline
6) George Pappas, Assistant Professor AUTH \newline
7) Kostas Kokkotas, Professor University of T\"{u}bingen

\end{titlepage}

\pagenumbering{roman}

\include{abstract}

\include{Introduction} 
\include{acknowledgements}
\include{notation}

\tableofcontents

\listoftables

\listoffigures

\newpage

\pagenumbering{arabic}

\part{Basic Mathematical Tools: Collineations, Integrability and Stability Theory}

\include{variation_calculus}

\include{collineations}

\include{integrability}

\include{stability}

\part{Symmetries in General Relativity}

\include{EMSF}

\include{CKVs_Bianchi_III_V} 

\part{Integrability of autonomous dynamical systems} \label{part3}

\include{QFIs_conservative_systems}

\include{2d_integrable_potentials}

\include{QFIs_damping}

\include{Ermakov}

\include{higher_order_FIs}

\part{Integrability of time-dependent dynamical systems}

\include{QFIs_timedependent}

\include{Brans_Dicke}

\include{td_central_pots}


\appendix

\include{stability_append2}

\include{stability_append3}

\include{proof_thm_QFIs}

\include{proof_thm_damping}

\include{proof_thm_higher_FIs}

\include{proof_thm_timedependent}

\printindex

\end{document}

%% file: abstract.tex
\chapter*{Abstract}

The physical phenomena are described by physical quantities related by specific physical laws. In the context of a Physical Theory, the physical quantities and the physical laws are described, respectively, by suitable geometrical objects and relations between these objects. These relations are expressed with systems of (mainly second order) differential equations. The solution of these equations is frequently a formidable task, either because the dynamical equations cannot be integrated by standard methods or because the defined dynamical system is non-integrable. Therefore, it is important that we have a systematic and reliable method to determine their integrability. This has led to the development of several (algebraic or geometric) methods, which determine if a dynamical system is integrable/superintegrable or not. Most of these methods concern the first integrals (FIs), that is, quantities that are constant along the evolution of the system. The FIs appear in the literature with many names such as constants of motion, conserved currents, and conservation laws. FIs are important, because they can be used to reduce the order of the system of the dynamical equations and, if there are `enough' of them, even to determine its solution by means of quadratures. In the latter case, the dynamical system is said to be Liouville integrable and it is associated with a canonical Lagrangian, whose kinetic energy defines a metric tensor known as kinetic metric. It is proved that there is a close relation between the geometric symmetries (collineations and Killing tensors) of this metric and the quantities defining the FIs. This correspondence makes it possible to use powerful results from Differential Geometry in the study of the integrability of dynamical systems. In this thesis, we study this correspondence and geometrize the determination of FIs by developing a new geometric method to compute them.

This thesis is divided in four parts, whose content is the following: \newline
a. The first part presents basic mathematical `tools', which are necessary for the methods developed in the next parts. In particular, notions from the symmetries of differential equations, the collineations of geometrical objects, the Liouville integrability of dynamical systems, and the stability theory are presented. Moreover, we discuss various methods for determining FIs (e.g. the Lie/Noether symmetry method, the Inverse Noether Theorem, the Hamilton-Jacobi method, and the direct method) by presenting their advantages and disadvantages. We point out that all these methods cannot replace the generality of Noether's theorem, but they act supplementary to this. Nevertheless, there are many cases where such methods are more convenient and can lead us to faster and safer conclusions about the integrability of the system. \newline
b. The second part is about symmetries in general relativity. \newline
First, we discuss a spacetime whose matter source is an electromagnetic string fluid (EMSF), that is, an isotropic charged string fluid interacting with a strong magnetic field. By considering the double congruence defined by the four-velocity $u^{a}$ of the fluid flow lines and the unit vector $n^{a}$ along the magnetic field lines, we determine the kinematic and the dynamic variables of the EMSF in both the $1+3$ and the $1+1+2$ decompositions. Then, we solve the resulting field equations by making simplifying geometric assumptions in the form of collineations. We consider the case of a conformal Killing vector (CKV) parallel to $u^{a}$ and of a CKV parallel to $n^{a}$. We apply the general results of the first case to the Friedman-Robertson-Walker (FRW) spacetime and of the second case to the Bianchi I spacetime. In the latter case, we find a new solution of the gravitational field equations. \newline
Next, we determine the CKVs of Bianchi III and Bianchi V spacetimes by using an algorithm which relates the CKVs of decomposable spacetimes with the collineations of the non-decomposable subspace. We find that there is only one Bianchi III spacetime and one Bianchi V which admit a single proper CKV. As an application in the spacetimes found, we study the kinematics of the comoving observers and the dynamics of the corresponding cosmological fluid. \newline
c. The third part concerns the integrability of autonomous dynamical systems. These are systems of second order ordinary differential equations (ODEs) of the general form $\ddot{q}^{a} = -\Gamma^{a}_{bc}(q)\dot{q}^{b}\dot{q}^{c}  +F^{a}(q,\dot{q})$, where $\Gamma^{a}_{bc}$ are the Riemannian connection coefficients defined by the kinetic metric $\gamma_{ab}(q)$ (kinetic energy) of the system and $F^{a}$ are the generalized forces. We consider three types of such systems: i) The conservative systems, where $F^{a}=-V^{,a}(q)$ and $V(q)$ denotes the potential of the conservative generalized forces. ii) The systems with a linear damping term, where $F^{a}= -Q^{a}(q) +A^{a}_{b}(q)\dot{q}^{b}$. iii) The systems without damping, where $F^{a}= -Q^{a}(q)$ stands for both the conservative and the non-conservative forces. For each of the above three types of autonomous dynamical systems, we determine FIs that are polynomials (of second or higher order) in the velocities $\dot{q}^{a}$ with coefficients that are totally symmetric tensors depending on the variables $t$ and $q^{a}$. The determination of the FIs is done by using the direct method. According to this method, one assumes a generic FI, say $I(t,q,\dot{q})$, which is of a polynomial form in terms of the velocities $\dot{q}^{a}$ with unknown coefficients and requires the condition $\frac{dI}{dt}=0$ along the dynamical equations. This condition leads to a system of partial differential equations (PDEs) involving the unknown coefficients (tensors) of $I$ together with the elements (quantities $F^{a}$) that characterize the dynamics of the system. We solve this system of PDEs in terms of the collineations and the Killing tensors (KTs) of the kinetic metric. The maximal order of the KTs is equal to the order of the considered FI. We collect our results in four theorems, which we apply in the following applications: 1) We compute the quadratic FIs (QFIs) of the autonomous generalized Kepler potential and we prove that it is superintegrable. 2) We determine the integrable and superintegrable two-dimensional (2d) Newtonian potentials $V(x,y)$, and we apply these results in order to prove in a direct way that the 2d conservative generalized Ermakov system is superintegrable. 3) We determine the QFIs of the autonomous linearly 2d damped harmonic oscillator and we find a plethora of new FIs. 4) We determine new integrable and superintegrable potentials that admit cubic FIs. 5) By using the inverse Noether theorem, we prove that all $m$th-order polynomial FIs are Noether FIs which can be associated with a gauged generalized weak Noether symmetry. It is also shown that there does not exist a one-to-one correspondence between Noether FIs and the type of Noether symmetry. \newline
d. The fourth part concerns the integrability of time-dependent dynamical systems. \newline
At first, we consider time-dependent dynamical systems of the form $\ddot{q}^{a}= -\Gamma^{a}_{bc}(q)\dot{q}^{b} \dot{q}^{c} -\omega(t)Q^{a}(q)$, where $\omega(t)$ is a non-zero arbitrary function and $\Gamma^{a}_{bc}(q)$ are the Riemannian connection coefficients defined by the kinetic metric $\gamma_{ab}(q)$ of the system. In order to determine the QFIs, we apply again the direct method for functional expressions of the general form $I= K_{ab}(t,q)\dot{q}^{a}\dot{q}^{b} +K_{a}(t,q)\dot{q}^{a} +K(t,q)$. This results in a system of PDEs involving the geometric quantities $K, K_{a}, K_{ab}$ and the dynamic quantities $\omega, Q^{a}$. We find that $K_{ab}$ is a second order KT of the kinetic metric. This is a result we use in two ways: i) We assume a general polynomial form in $t$ both for $K_{a}$ and $K_{ab}$ with coefficients, respectively, vectors and second order KTs depending on $q^{a}$. ii) We compute a basis for the vector space of the second order KTs of the kinetic metric and, then, we express the KT $K_{ab}$ in this basis by assuming coefficients that are functions of $t$. In both ways, we find a new system of PDEs, which we solve by specifying either the `frequencies' $\omega(t)$ or the quantities $Q^{a}(q)$. From the solutions found, we determine the corresponding QFIs in the following two special cases: 1) We assume that $\omega(t)$ is a general polynomial of $t$ and let the quantities $Q^{a}(q)$ to act as constraints. 2) We specify the quantities $Q^{a}(q)$ from the time-dependent generalized Kepler potential $V=-\frac{\omega (t)}{r^{\nu}}$ and determine the functions $\omega(t)$ for which QFIs are admitted. This potential for $\nu=-2,1,2$ includes, respectively, the 3d time-dependent harmonic oscillator, the time-dependent Kepler potential, and the Newton-Cotes potential. \newline
Next, using the well-known result that by a reparameterization $t=t(s)$ the linear damping term $\phi(t)\dot{q}^{a}$ of a dynamical equation is absorbed to a time-dependent force of the form $\bar{\omega}(s)Q^{a}(q)$, we also study the non-linear differential equation $\ddot{x}=-\omega(t)x^{\mu }+\phi (t)\dot{x}$ $(\mu \neq -1)$, and compute the relation between the coefficients $\omega(t)$ and $\phi(t)$ for which QFIs are admitted. It is found that a family of `frequencies' $\bar{\omega}(s)$ is admitted, which for $\mu =0, 1, 2$ is parameterized with functions, whereas for $\mu \neq -1,0,1,2$ is parameterized with constants. We apply these results in the following problems: i) To study the integrability of the well-known generalized Lane-Emden equation, and ii) To find new conservation laws (i.e. QFIs) for a modified Brans-Dicke cosmological model with an additional minimally coupled quintessence scalar field in a spatially flat FRW spacetime. Specifically, by assuming a power law potential function for the quintessence scalar field $\psi(t)$, we determine the QFIs associated with the equation of motion of $\psi(t)$. Then, we use these QFIs in order to find new exact solutions for the field equations. Our approach is more general and does not require the existence of a point-like Lagrangian, that is, of a minisuperspace description. Therefore, it can be applied in other gravitational models without minisuperspace (e.g. Class B Bianchi spacetimes). \newline
Finally, we determine the integrable time-dependent Newtonian central potentials which admit linear and quadratic FIs other than those constructed from the linear FIs of the angular momentum. It is shown explicitly that previous answers to this problem are incomplete. The results are collected in a theorem, which is applied in order to find: 1) The integrable time-dependent oscillators. 2) The integrable time-dependent generalized Kepler potentials. 3) A class of integrable binary systems with variable mass. 4) The integrable Yukawa and interatomic potentials with time-dependent parameters. 5) A solution of the Schr\"{o}dinger equation for a class of integrable central potentials which have been integrated.

%% file: Introduction.tex
\chapter*{Publications}

For the fulfillment of this Thesis, the following articles have been published:
\begin{enumerate}
  \item M. Tsamparlis, A. Mitsopoulos and A. Paliathanasis, \emph{`Symmetries of spacetimes embedded with an electromagnetic string fluid'}, Gen. Relativ. Gravit. \textbf{51}:6 (2019). \textbf{(Chapter \ref{ch.EMSF})}
  \item A. Mitsopoulos, M. Tsamparlis and A. Paliathanasis, \emph{`Constructing the CKVs of Bianchi III and V spacetimes'}, Mod. Phys. Lett. A \textbf{34}, 1950326 (2019). \textbf{(Chapter \ref{ch.CKVs.Bianchi.III.IV})}
  \item M. Tsamparlis and A. Mitsopoulos, \emph{`Quadratic first integrals of autonomous conservative dynamical systems'}, J. Math. Phys. \textbf{61}, 072703 (2020). \textbf{(Chapter \ref{ch1.QFIs.conservative})}
  \item A. Mitsopoulos, M. Tsamparlis and A. Paliathanasis, \emph{`Integrable and superintegrable potentials of 2d autonomous conservative dynamical systems'}, Symmetry \textbf{12}(10), 1655 (2020). \textbf{(chapter \ref{ch.2d.pots})}
  \item M. Tsamparlis and A. Mitsopoulos, \emph{`First integrals of holonomic systems without Noether symmetries'}, J. Math. Phys. \textbf{61}, 122701 (2020). \textbf{(Chapter \ref{ch.QFIs.damping})}
  \item A. Mitsopoulos and M. Tsamparlis, \emph{`The generalized Ermakov conservative system: A discussion'}, Eur. Phys. J. Plus \textbf{136}, 933 (2021). \textbf{(Chapter \ref{ch.Ermakov})}
  \item A. Mitsopoulos and M. Tsamparlis, \emph{`Higher order first integrals of autonomous dynamical systems'}, J. Geom. Phys. \textbf{170}, 104383 (2021). \textbf{(Chapter \ref{ch.Higher.order.FIs})}
  \item A. Mitsopoulos and M. Tsamparlis, \emph{`Quadratic first integrals of time-dependent dynamical systems of the form $\ddot{q}^{a}= -\Gamma^{a}_{bc}\dot{q}^{b} \dot{q}^{c} -\omega(t)Q^{a}(q)$'}, Mathematics \textbf{9}(13), 1503 (2021). \textbf{(Chapter \ref{ch.QFIs.timedependent})}
  \item A. Mitsopoulos, M. Tsamparlis, G. Leon, and A. Paliathanasis, \emph{`New conservation laws and exact cosmological solutions in Brans-Dicke cosmology with an extra scalar field'}, Symmetry \textbf{13}(8), 1364 (2021). \textbf{(Chapter \ref{ch.BransDicke})}
  \item A. Mitsopoulos and M. Tsamparlis, \emph{`Integrable time-dependent central potentials'}, Phys. Lett. A \textbf{423}, 127825 (2022). \textbf{(Chapter \ref{ch.td.central.pots})}
\end{enumerate}

%% file: acknowledgements.tex
\chapter*{Acknowledgements}

I would like to thanks my parents for supporting me in every step of my life; professor Michael Tsamparlis for his scientific guidance and the excellent collaboration; professors Theocharis Apostolatos, Nectarios Vlahakis, Nikolaos Stergioulas, and Theodosios Christodoulakis for their advices and the useful remarks; researcher Andronikos Paliathanasis for his help in the cosmological part of the Thesis; all the members of the Committee and the secretary of the Section of Astrophysics Sophia Zarbouti. 

%% file: notation.tex
\chapter*{Notation - Abbreviations}

\begin{itemize}

\item
AC $=$ affine collineation

\item
BD $=$ Brans-Dicke

\item
CFI $=$ cubic first integral

\item
CKV $=$ conformal Killing vector

\item
const. $=$ constant

\item
E-L $=$ Euler-Lagrange equations

\item
EM $=$ electromagnetic

\item
EMSF $=$ EM string fluid

\item
eq. $=$ equation

\item
FE $=$ fast eigendirection

\item
FI $=$ first integral

\item
FLRW $=$ Friedmann-Lema\^{i}tre-Robertson-Walker

\item
FRW $=$ Friedmann-Robertson-Walker

\item
GO $=$ geometrical object

\item
H-J $=$ Hamilton-Jacobi

\item
HV $=$ homothetic vector

\item
iff $=$ if and only if

\item
KT $=$ Killing tensor

\item
KV $=$ Killing vector

\item
LFI $=$ linear first integral

\item
LFX $=$ line of fixed points

\item
LRS $=$ Locally Rotational Symmetric

\item
MHD $=$ magnetohydrodynamics

\item
NBH $=$ Noether-Bessel-Hagen

\item
$N$d $=$ $N$-dimensional, e.g. 2d $=$ two-dimensional

\item
ODE $=$ ordinary differential equation

\item
PB $=$ Poisson bracket

\item
PC $=$ projective collineation

\item
PDE $=$ partial differential equation

\item
QFI $=$ quadratic first integral

\item
QUFI $=$ quartic first integral

\item
RMHD $=$ relativistic magnetohydrodynamics

\item
SCKV $=$ special CKV

\item
SE $=$ slow eigendirection

\item
sec. $=$ section

\item
SF $=$ string fluid

\item
SPC $=$ special projective collineation

\item
wrt $=$ with respect to

\end{itemize}

\bigskip

\bigskip

Mathematical conventions used frequently in the text:
\begin{itemize}

\item
Einstein summation convention is always used. When this is not the case, it is noted.

\item
The kinetic metric $\gamma_{ab}(q)$ of each dynamical system is used for lowering and raising the indices.

\item
Round (square) brackets indicate symmetrization (antisymmetrization) of the enclosed indices. We have
\[
T_{(i_1... i_k)} \equiv \frac{1}{k!} \sum_{\sigma} T_{i_{\sigma(1)}... i_{\sigma(k)}} \enskip \text{and} \enskip T_{[i_1... i_k]} \equiv \frac{1}{k!} \sum_{\sigma} (sign \sigma) T_{i_{\sigma(1)}... i_{\sigma(k)}}
\]
where $sign(\sigma)\equiv sign \sigma$ is the sign of the permutations $\sigma$ of the set $\{1,2,...,k\}$.

\item
Indices enclosed between vertical lines are overlooked by  antisymmetrization or symmetrization symbols.

\item
Curly brackets indicate cyclic permutation of the enclosed indices.

\item
A comma indicates partial derivative and a semicolon Riemannian covariant derivative.

\end{itemize}

%% file: variation_calculus.tex
\chapter{Calculus of variations and symmetries of differential equations}

\label{ch.VarCalSym}

In this chapter, we review some basic ideas from the Calculus of Variations (see e.g. \cite{VujanovicB}) in order to introduce the concept of symmetries of differential equations \cite{Bluman}; a concept of central importance in the chapters to follow.

\section{Variation of real-valued functions of several independent variables}

\label{con.mot.sec.4}

Let $F$ be a smooth real-valued function of $n$ independent variables $t^{i}$, $m$ dependent variables $q^{A}(t^{i})$ and up to $\kappa$-order partial derivatives $q^{A}{}_{,i_{1}...i_{\kappa}}$; that is,
\begin{eqnarray}
F &=& F \left( t^1, ..., t^n; q^1(t^j), ..., q^m(t^j); q^1{}_{,i}(t^j), ..., q^m{}_{,i}(t^j); ... \right) \notag \\
&\equiv& F \left( t^i, q^A(t^i), q^A{}_{,i_1}(t^i), q^A{}_{,i_1i_2}(t^i), ..., q^A{}_{,i_1..i_{\kappa}}(t^i) \right)
\label{con.mot.eq.gf1}
\end{eqnarray}
where $i, j, i_{\alpha} = 1, ..., n$, $\alpha = 1, ..., \kappa$, $A = 1, ..., m$ and $q^A{}_{,i} \equiv \frac{\partial q^A}{\partial t^i}$.

Then, an arbitrary infinitesimal (i.e. $0<\varepsilon\ll1$) transformation of variables:\index{Transformation! infinitesimal}
$\bar{t}^i \equiv t^i + \varepsilon \xi^i$ and $\bar{q}^A(t) \equiv q^A(t) + \varepsilon \chi^A$, where the generators\index{Generator} $\xi^{i}$ and $\chi^{A}$ are -in general- arbitrary smooth functions of the (independent and dependent) variables and their partial derivatives up to $\kappa$-order, induces to $F$ the following {\bf{(general) variation}}:\index{Variation! general}
\begin{eqnarray}
\delta F &\equiv& F \left( \bar{t}^i, \bar{q}^A(\bar{t}^i), \bar{q}^A{}_{,i_1}(\bar{t}^i), \bar{q}^A{}_{,i_1i_2}(\bar{t}^i), ..., \bar{q}^A{}_{,i_1..i_{\kappa}} (\bar{t}^i) \right) - \notag \\
&& -F \left( t^i, q^A(t^i), q^A{}_{,i_1}(t^i), q^A{}_{,i_1i_2}(t^i), ..., q^A{}_{,i_1..i_{\kappa}}(t^i) \right). \label{con.mot.eq.34}
\end{eqnarray}
If the independent variables do not vary (i.e. $\xi^{i}=0$), then variation $\delta$ becomes the \textbf{Lagrange (or simultaneous) variation}
\begin{eqnarray}
\delta_{0} F &\equiv& F \left( t^i, \bar{q}^A(t^i), \bar{q}^A{}_{,i_1}(t^i), \bar{q}^A{}_{,i_1i_2}(t^i), ..., \bar{q}^A{}_{,i_1..i_{\kappa}}(t^i) \right) - \notag \\
&& -F \left( t^i, q^A(t^i), q^A{}_{,i_1}(t^i), q^A{}_{,i_1 i_2}(t^i), ..., q^A{}_{,i_1..i_{\kappa}}(t^i) \right). \label{con.mot.eq.33}
\end{eqnarray}
From the above definitions $\delta t^i = \varepsilon \xi^i$ and $\delta_{0} q^A = \varepsilon \chi^A$.

Moreover, by using Taylor expansion up to second order terms (i.e $\varepsilon^{2} \to 0$), we find
\begin{equation}
\delta q^A = \delta_{0} q^A + q^A{}_{,i} \delta t^i = \varepsilon \left( \chi^A + q^A{}_{,i} \xi^i \right) = \varepsilon \eta^A
\label{con.mot.eq.35}
\end{equation}
where $\eta^A \equiv \chi^A + q^A{}_{,i} \xi^i$ is the induced generator\index{Generator! induced} of the general variation of $q^{A}$.

\begin{remark} \label{remark.new.5}
Consider a set of dynamical equations with generalized positions $q^{i}(t)$, where $i=1,2,...,n$ and $t$ the independent time variable. Then, the $\delta_{0}$-variation of the {\bf{solution (or trajectory or natural path)}}\index{Path! natural}\index{Trajectory} of the system measures the deviation of the perturbed path $\bar{q}^{i}(t)$ from the natural path $q^{i}(t)$ by comparing their points at the same time variable $t$. On the other hand, $\delta$-variation compares the points of the paths at different times $t$ and $t + \delta t$.
\end{remark}

Since the variation $\delta_{0} q^A$ is, in general, a function of $t^i, q^A(t^i), q^A{}_{,i_1}(t^i), ..., q^A{}_{,i_1..i_{\kappa}}(t^i)$, we should stress a subtlety concerning the computation of the variation $\delta_{0} q^A{}_{,i}$. Specifically, we have:
\begin{equation}
\delta_{0} q^A{}_{,i} = \bar{q}^A{}_{,i}(t^r) - q^A{}_{,i}(t^r) = \left( \bar{q}^A - q^A \right)_{,i} = \left( \delta_{0} q^A[t^{r}] \right)_{,i} = \frac{d}{dt^i}\left( \delta_{0} q^A \right) \implies\delta_{0} q^A{}_{,i} = \frac{d}{dt^i}\left( \delta_{0} q^A \right) \label{con.mot.eq.36}
\end{equation}
where $\frac{d}{dt^r} \equiv \frac{\partial}{\partial t^r} + q^A{}_{,r} \frac{\partial}{\partial q^A} + q^A{}_{,i_1r} \frac{\partial}{\partial q^A{}_{,i_1}} + ... \enskip.$

We note that equation (\ref{con.mot.eq.36}) is generalized as follows:
\begin{equation}
\delta_{0} q^A{}_{,i_1...i_{\kappa}} = \frac{d^{\kappa}}{dt^{i_1}...dt^{i_{\kappa}}} \left( \delta_{0} q^A \right). \label{con.mot.eq.38}
\end{equation}
If there is only one independent variable $t$ (e.g. time) and $n$ dependent variables $q^{i}(t)$, equation (\ref{con.mot.eq.38}) becomes $\delta_{0}(q^{i})^{(\kappa)} = \left( \delta_{0} q^{i} \right)^{(\kappa)}$ where $(\kappa)$ denotes the order of the total time derivative. For $\kappa=1$, we have $\delta_{0} \dot{q}^i = \left( \delta_{0} q^i \right)^{\cdot}$.

On the other hand, for the $\delta$-variation, it holds that
\begin{equation}
\delta q^A{}_{,i} = \frac{d}{dt^i} \left( \delta_{0} q^A \right) + q^A{}_{,ij} \delta t^j
\label{con.mot.eq.39}
\end{equation}
which is generalized as follows:
\begin{equation}
\delta q^A{}_{,i_1...i_{\kappa}} = \frac{d^{\kappa}}{dt^{i_1}...dt^{i_{\kappa}}} \left( \delta_{0} q^A \right) + q^A{}_{,i_1...i_{\kappa}j} \delta t^j = \varepsilon \eta^A_{i_1...i_{\kappa}}
\label{con.mot.eq.40}
\end{equation}
where
\begin{equation}
\eta^A_{i_1...i_{\kappa}} \equiv \frac{d^{\kappa}}{dt^{i_1}...dt^{i_{\kappa}}} \left( \eta^A - q^A{}_{,j} \xi^j \right) + q^A{}_{,i_1...i_{\kappa}j} \xi^j.
\label{con.mot.eq.41}
\end{equation}

Moreover, we have
\begin{equation}
\delta q^A{}_{,i} = \frac{d}{dt^i} \left( \delta q^A \right) - q^A{}_{,j} \frac{d}{dt^i} \left( \delta t^j \right) \label{con.mot.eq.42}
\end{equation}
and in general
\begin{equation}
\delta q^A{}_{,i_1...i_{\kappa}} = \frac{d}{dt^{i_{\kappa}}} \left( \delta q^A{}_{,i_1...i_{\kappa-1}} \right) - q^A{}_{,i_1...i_{\kappa-1}j} \frac{d}{dt^{i_{\kappa}}} \left( \delta t^j \right)
\label{con.mot.eq.43}
\end{equation}
which implies that
\begin{equation}
\eta^A_{i_1...i_{\kappa}} = \frac{d}{dt^{i_{\kappa}}} \left( \eta^A_{i_1...i_{\kappa-1}} \right) - q^A{}_{,i_1...i_{\kappa-1}j} \frac{d \xi^j}{dt^{i_{\kappa}}}.
\label{con.mot.eq.44}
\end{equation}

We compute:
\begin{eqnarray}
\delta_{0} F &=& \frac{\partial F}{\partial q^A} \delta_{0} q^A + \frac{\partial F}{\partial q^A{}_{,i_1}} \delta_{0} q^A{}_{,i_1} + ... + \frac{\partial F}{\partial q^A{}_{,i_1...i_{\kappa}}} \delta_{0} q^A{}_{,i_1...i_{\kappa}}
\label{con.mot.eq.45} \\
\delta F &=& \frac{\partial F}{\partial t^j} \delta t^j + \frac{\partial F}{\partial q^A} \delta q^A + \frac{\partial F}{\partial q^A{}_{,i_1}} \delta q^A{}_{,i_1} + ... + \frac{\partial F}{\partial q^A{}_{,i_1...i_{\kappa}}} \delta q^A{}_{,i_1...i_{\kappa}} = \varepsilon \mathbf{X}^{[\kappa]} F
\label{con.mot.eq.46}
\end{eqnarray}
where $\mathbf{X}^{[\kappa]}$ is the $\kappa$-prolongation\index{$\kappa$-prolongation} of the generator $\mathbf{X} = \xi^j \frac{\partial}{\partial t^j} + \eta^A \frac{\partial}{\partial q^A}$, that is,
\begin{equation}
\mathbf{X}^{[\kappa]} \equiv \xi^j \frac{\partial}{\partial t^j} + \eta^A \frac{\partial}{\partial q^A} + \eta^A_{i_1} \frac{\partial}{\partial q^A{}_{,i_1}} + ... + \eta^A_{i_1...i_{\kappa}} \frac{\partial}{\partial q^A{}_{,i_1...i_{\kappa}}} = \mathbf{X} + \eta^A_{i_1} \frac{\partial}{\partial q^A{}_{,i_1}} + ... + \eta^A_{i_1...i_{\kappa}} \frac{\partial}{\partial q^A{}_{,i_1...i_{\kappa}}}.
\label{con.mot.eq.47}
\end{equation}

An arbitrary function $F$ is said to be invariant\index{Invariant! function} under a given $\delta$-variation with generator $\mathbf{X}$ iff $\delta F = 0$ or, equivalently, $\mathbf{X}^{[\kappa]} F = 0$. The vector field $\mathbf{X}$ is called a {\bf{symmetry}} of $F$.\index{Symmetry}

Because $q^A{}_{,i_1...i_{\kappa}} = q^A{}_{,i_{\sigma(1)} ...i_{\sigma(\kappa)}}$ for all permutations $\sigma$ of the set $\{1,2,...,\kappa\}$, functions of the form (\ref{con.mot.eq.gf1}) are defined over manifolds $M$ whose dimension is:
\[
\begin{cases}
F = F \left(t^i, q^A\right) \implies \dim(M)=n+m \\
F = F \left(t^i, q^A, q^A_{,i_1}\right) \implies \dim(M)=n+m+mn \\
F = F \left(t^i, q^A, q^A_{,i_1}, q^A_{,i_1i_2}\right) \implies \dim(M)=n+m+mn+m\left( \frac{n^2-n}{2} + n \right) \\
....
\end{cases}
\]
Therefore, the Einstein summation in equations (\ref{con.mot.eq.45}), (\ref{con.mot.eq.46}) and (\ref{con.mot.eq.47}) is such that $i_1 \leq i_2 \leq ... \leq i_{\alpha}$ for all values of $\alpha$ from $1$ to $\kappa$.

\begin{proposition} \label{con.mot.pro.1}
Let $F = F \big(t, q(t), \dot{q}(t)\big)$ and $H = H \big(t, q(t), \dot{q}(t)\big)$ be arbitrary smooth real-valued functions of one independent variable $t$ and $n$ dependent variables $q^{i}(t)$. Then, the following properties hold: 1) $\delta \left( F + H \right) = \delta F + \delta H$, 2) $\delta \left( c F \right) = c \delta F$ where $c$ is an arbitrary constant, and 3) $\delta \left( F \cdot H \right) = \delta F \cdot H + F \cdot \delta H$ (Leibnitz rule).
\end{proposition}

\begin{remark} \label{remark.new.6}
For an arbitrary function $F = F \big(t, q(t), \dot{q}(t) \big)$ of one independent variable $t$ and $n$ dependent variables $q^{i}(t)$, we find:
\begin{eqnarray}
\delta_{0} F &=& \frac{\partial F}{\partial q^i} \delta_{0} q^i + \frac{\partial F}{\partial \dot{q}^i} \delta_{0} \dot{q}^i \label{con.mot.eq.9} \\
\delta F &=& \frac{\partial F}{\partial t} \delta t + \frac{\partial F}{\partial q^i} \delta q^i + \frac{\partial F}{\partial \dot{q}^i} \delta \dot{q}^i = \delta_{0} F + \dot{F} \delta t \label{con.mot.eq.10}
\end{eqnarray}
and the following mathematical identities:
\begin{eqnarray}
\delta( q^i )^{(\kappa)} &=& \delta_{0} ( q^i )^{(\kappa)} + (q^i)^{(\kappa + 1)} \delta t. \label{con.mot.eq.26} \\
\delta ( q^i )^{(\kappa)} &=& \left[ \delta ( q^i )^{(\kappa -1)} \right]^{\cdot} - ( q^i )^{(\kappa)} \left( \delta t \right)^{\cdot}, \enskip \kappa \geq 1 \label{con.mot.eq.27} \\
\delta_{0} \dot{F} &=& \left( \delta_{0} F \right)^{\cdot} \label{con.mot.eq.14} \\
\left( \delta F \right)^{\cdot} - \delta \dot{F} &=& \dot{F} \left( \delta t \right)^{\cdot} \label{con.mot.eq.17}
\end{eqnarray}
where $\kappa$ is an arbitrary natural number.

We note that $\left( \frac{d}{dt} \delta - \delta \frac{d}{dt} \right)q^i \neq 0$, i.e. $\delta \dot{q}^i = \dot{\bar{q}}^i(\bar{t}) - \dot{q}^i(t) \neq \left( \delta q^i \right)^{\cdot}$. We expected this result because the derivatives $\dot{\bar{q}}^i$ and $\dot{q}^i$ are determined in different times $\bar{t}$ and $t$, respectively. Furthermore, since $\dot{F} = \frac{\partial F}{\partial t} + \frac{\partial F}{\partial q^i} \dot{q}^i + \frac{\partial F}{\partial \dot{q}^i} \ddot{q}^i$, the term $\dot{F}$ may be thought of as a function of $t, q^i, \dot{q}^i$ and $\ddot{q}^i$.
\end{remark}

\section{Variation of functionals}

\label{con.mot.sec.2}

A {\bf{(real) functional}} \index{Functional} is a map that assigns functions to real numbers. For example, a definite integral of the form
\begin{equation}
I \big[ q(t) \big] = \int^{t_1}_{t_0} F \big( t, q(t), \dot{q}(t) \big) dt \label{con.mot.eq.19}
\end{equation}
where $t_0, t_1 \in \mathbb{R}$, is a functional from the set $\big[C^{\infty}(\mathbb{R})\big]^n$ to $\mathbb{R}$. $C^{\infty}(\mathbb{R})$ is the set of all smooth functions from $\mathbb{R}$ to itself. The functional (\ref{con.mot.eq.19}) is called an {\bf{action (or action-like functional)}}\index{Action}\index{Functional! action-like} of $F$. If $F$ is the Lagrangian $L$ of a dynamical system, then (\ref{con.mot.eq.19}) is the {\bf{(Hamilton's) action}} of the system and we write
\begin{equation}
S \big[ q(t) \big] = \int^{t_1}_{t_0} L \big( t, q(t), \dot{q}(t) \big) dt. \label{eq.Lev12}
\end{equation}

By using the results of sec. \ref{con.mot.sec.4}, the general variation of an action is defined as follows:
\begin{equation}
\delta I \equiv \delta \int^{t_1}_{t_0} F \big( t, q(t), \dot{q}(t) \big) dt \equiv \int^{\bar{t}_1}_{\bar{t}_0} F \big( \bar{t}, \bar{q}(\bar{t}), \dot{\bar{q}}(\bar{t}) \big) d\bar{t} - \int^{t_1}_{t_0} F \big( t, q(t), \dot{q}(t) \big) dt
\label{con.mot.eq.21}
\end{equation}
where $\bar{t}_0 = t_0 + \delta t (t_0)$ and $\bar{t}_1 = t_1 + \delta t (t_1)$. In the case that $\delta t=0$, the variation (\ref{con.mot.eq.21}) becomes
\begin{equation}
\delta_{0} I = \delta_{0} \int^{t_1}_{t_0} F \big( t, q(t), \dot{q}(t) \big) dt = \int^{t_1}_{t_0} F \big( t, \bar{q}(t), \dot{\bar{q}}(t) \big) dt - \int^{t_1}_{t_0} F \big( t, q(t), \dot{q}(t) \big) dt = \int^{t_1}_{t_0} \delta_{0} F dt. \label{con.mot.eq.20}
\end{equation}

Equation (\ref{con.mot.eq.21}) via equation (\ref{con.mot.eq.10}) implies that:
\begin{align*}
\delta I & = \int^{t_1}_{t_0} F \big( \bar{t}, \bar{q}(\bar{t}), \dot{\bar{q}}(\bar{t}) \big) \frac{d \bar{t}}{dt} dt - \int^{t_1}_{t_0} F \big( t, q(t), \dot{q}(t) \big) dt \\
& = \int^{t_1}_{t_0} F \big( \bar{t}, \bar{q}(\bar{t}), \dot{\bar{q}}(\bar{t}) \big) \left[ 1 + \left( \delta t \right)^{\cdot} \right] dt - \int^{t_1}_{t_0} F \big( t, q(t), \dot{q}(t) \big) dt \\
& = \int^{t_1}_{t_0} F \left( \delta t \right)^{\cdot} dt + \int^{t_1}_{t_0} \delta F dt + \int^{t_1}_{t_0} \underbrace{\delta F \left( \delta t \right)^{\cdot}}_{O(\varepsilon^2) \to 0} dt \\
&= \int^{t_1}_{t_0} \delta_{0} F dt + \int^{t_1}_{t_0} \dot{F} \delta t dt + \int^{t_1}_{t_0} F \left( \delta t \right)^{\cdot} dt = \delta_{0} I + \int^{t_1}_{t_0} \left( F \delta t \right)^{\cdot} dt \implies
\end{align*}
\begin{equation} \label{con.mot.eq.22}
\delta I = \delta_{0} I + \left[ F \delta t \right]^{t_1}_{t_0}.
\end{equation}

\begin{proposition} \label{con.mot.pro.1a}
Let $f: [t_0, t_1] \to \mathbb{R}$ be a continuous function such that $\int^{t_1}_{t_0} f(t) h(t) dt = 0$ for all continuous functions $h: [t_0, t_1] \to \mathbb{R}$ satisfying the condition $h(t_0) = h(t_1) = 0$. Then, $f = 0$.
\end{proposition}

\begin{proof}
Assume $f \neq 0$. Since $f$ is a continuous function, there exists a closed interval $[a,b]$ such that $t_0 < a < b < t_1$ where $f$ is either negative or positive. Without loss of generality, say $f > 0$ in $[a,b]$. If we choose the continuous function $h(t) =
\begin{cases}
(t - a) (b - t), \quad t \in [a,b] \\
0, \quad t \in [t_0,a) \cup (b,t_1]
\end{cases}$.
Then, $\int^{t_1}_{t_0} f(t) h(t) dt = \int^{b}_{a} f(t) h(t) dt > 0$ which is absurd. Therefore, $f = 0$.
\end{proof}

\begin{proposition}
[Hamilton's least action principle]\label{con.mot.pro.1b} \index{Hamilton's least action principle} If $\delta_{0} I = 0$ along a\index{Path} {\bf{path}}\footnote{
By this term we refer to a curve from $[t_0, t_1]$ to the configuration space $\{q^i\}$.
} $q^i(t)$ such that $\delta_{0} q^i(t_0) = \delta_{0} q^i(t_1) = 0$ for any $\delta_{0}$-variation, then $q^i(t)$ satisfies the second order system of ordinary differential equations (ODEs)
\begin{equation} \label{con.mot.eq.EL}
\frac{d}{dt} \left( \frac{\partial F}{\partial \dot{q}^i} \right) - \frac{\partial F}{\partial q^i} = 0.
\end{equation}
These ODEs are called {\bf{Euler-Lagrange (E-L) equations}}\index{Equations! Euler-Lagrange} and their solutions $q^i(t)$ are the {\bf{natural paths}}\index{Path! natural} of $F$ wrt the action $I$. In the case that $F$ is the Lagrangian of a system, ODEs (\ref{con.mot.eq.EL}) are the equations of motion of the system.
\end{proposition}

\begin{proof}
Along the path $q^i(t)$, we have:
\begin{align*}
0 & = \delta_{0} I = \int^{t_1}_{t_0} \delta_{0} F dt = \int^{t_1}_{t_0} \left( \frac{\partial F}{\partial q^i} \delta_{0} q^i + \frac{\partial F}{\partial \dot{q}^i} \delta_{0} \dot{q}^i \right) dt, \quad \delta_{0} \dot{q}^i = \left( \delta_{0} q^i \right)^{\cdot} \\
& = \int^{t_1}_{t_0} \left[ \frac{\partial F}{\partial q^i} - \frac{d}{dt} \left( \frac{\partial F}{\partial \dot{q}^i} \right) \right] \delta_{0} q^i dt + \underbrace{\left[ \frac{\partial F}{\partial \dot{q}^i} \delta_{0} q^i \right]^{t_1}_{t_0}}_{= 0}.
\end{align*}
Using proposition \ref{con.mot.pro.1a}, as $\delta_{0} q^i = \varepsilon \chi^i$, the E-L equations are derived.
\end{proof}

An action $I$ such that $\delta I = 0$ is called {\bf{absolutely invariant}}\index{Action! absolutely invariant} wrt the given infinitesimal transformation. We call $I$ a {\bf{gauge invariant}}\index{Action! gauge invariant} wrt the given variation if it satisfies the condition $\delta I = \varepsilon \int^{t_1}_{t_0} \dot{f} dt, ~\forall [t_0, t_1]$, where $f \big( t, q(t), \dot{q}(t) \big)$ is the {\bf{gauge function}} of the variation. We note that absolute invariance is a subcase of gauge invariance for $f = const$.

Recall that the variations $\delta_{0}$ and $\delta$ (in the following sections) are generated by the infinitesimal transformations:
\[
\begin{cases}
t \to \bar{t} = t + \varepsilon \xi \big( t, q(t), \dot{q}(t) \big) \\
q^i(t) \to \bar{q}^i(t) = q^i(t) + \varepsilon \chi^i \big( t, q(t), \dot{q}(t) \big) \\
\delta q^i \equiv \bar{q}^i(\bar{t}) - q^i(t) = \delta_{0} q^i + \dot{q}^i \delta t
\end{cases}
\implies
\begin{cases}
\delta t = \varepsilon \xi \\
\delta_{0} q^i = \varepsilon \chi^i \\
\delta q^i = \varepsilon \left( \chi^i + \dot{q}^i \xi \right) = \varepsilon\eta^i.
\end{cases}
\]
The quantities $\xi$ and $\eta^i$ are the time\index{Generator! time} and the spatial\index{Generator! space} \textbf{generators}\index{Generator} of the variation, respectively. The relation $\eta^i \equiv \chi^i + \dot{q}^i \xi$ can be used to determine the generator $\chi^i$ of the Lagrange variation.

\begin{proposition} \label{con.mot.pro.2}
If $I$ is gauge invariant wrt a given variation, then the generators of that variation satisfy the {\bf{Noether condition (or basic Noether identity)}} \index{Condition! Noether}
\begin{equation} \label{con.mot.eq.23}
\frac{\partial F}{\partial q^i} \eta^i + \frac{\partial F}{\partial \dot{q}^i} \dot{\eta}^i + \left( F - \frac{\partial F}{\partial \dot{q}^i} \dot{q}^i \right) \dot{\xi} + \frac{\partial F}{\partial t} \xi - \dot{f} = 0
\end{equation}
and vice versa.
\end{proposition}

\begin{proof}
We have (see results of sec. \ref{con.mot.sec.4}):
\begin{align*}
0 & = \delta I - \varepsilon \int^{t_1}_{t_0} \dot{f} dt =\int^{t_1}_{t_0} \left( \frac{\partial F}{\partial q^i} \delta_{0} q^i + \frac{\partial F}{\partial \dot{q}^i} \delta_{0} \dot{q}^i \right) dt + \left[ F \delta t \right]^{t_1}_{t_0} - \int^{t_1}_{t_0} \varepsilon \dot{f} dt \\
& = \int^{t_1}_{t_0} \left( \frac{\partial F}{\partial q^i} \delta q^i - \frac{\partial F}{\partial q^i} \dot{q}^i \delta t + \frac{\partial F}{\partial \dot{q}^i} \delta \dot{q}^i - \frac{\partial F}{\partial \dot{q}^i} \ddot{q}^i \delta t - \varepsilon \dot{f}\right) dt + \left[ F \delta t \right]^{t_1}_{t_0} \\
& = \int^{t_1}_{t_0} \left[ \frac{\partial F}{\partial q^i} \delta q^i + \frac{\partial F}{\partial \dot{q}^i} \left( \delta q^i \right)^{\cdot} - \left( F \delta t \right)^{\cdot} + F \left(\delta t\right)^{\cdot} + \frac{\partial F}{\partial t} \delta t - \frac{\partial F}{\partial \dot{q}^i} \dot{q}^i \left( \delta t \right)^{\cdot} - \varepsilon \dot{f} \right] dt + \left[ F \delta t \right]^{t_1}_{t_0} \\
& = \int^{t_1}_{t_0} \left[ \frac{\partial F}{\partial q^i} \delta q^i + \frac{\partial F}{\partial \dot{q}^i} \left( \delta q^i \right)^{\cdot} + \left( F - \frac{\partial F}{\partial \dot{q}^i} \dot{q}^i \right) \left( \delta t \right)^{\cdot} + \frac{\partial F}{\partial t} \delta t - \varepsilon \dot{f} \right] dt.
\end{align*}
Replacing with $\delta q^i = \varepsilon \eta^i$ and $\delta t = \varepsilon \xi$, we find the Noether condition because the resulting integral must hold $\forall$ $(t_0, t_1)$.
\end{proof}

\begin{theorem}[Noether's Theorem] \label{con.mot.pro.3}
Every set $\{ \xi, \eta^i; f \}$ that satisfies the Noether condition (\ref{con.mot.eq.23}) such that\index{Theorem! Noether's} $\delta_{0} q^i \neq 0$, i.e. $\eta^i - \dot{q}^i \xi \neq 0$, produces a constant of motion (or first integral (FI) or conservation law) \cite{Noether 1918, Flessas 1995}
\begin{equation}
\Lambda= \frac{\partial F}{\partial \dot{q}^i} \eta^i + \left( F - \frac{\partial F}{\partial \dot{q}^i} \dot{q}^i \right) \xi - f. \label{con.mot.eq.24}
\end{equation}
The function $\Lambda$ is constant along natural paths and is called a \textbf{Noether integral (or Noether FI)}.\index{First integral! Noether}
\end{theorem}

\begin{proof}

\underline{First method:} Since the set $\{ \eta^i, \chi; f \}$ satisfies the Noether condition, the action $I$ is gauge invariant. Therefore,
\begin{align*}
0 & = \delta I - \varepsilon \int^{t_1}_{t_0} \dot{f} dt = \int^{t_1}_{t_0} \left( \frac{\partial F}{\partial q^i} \delta_{0} q^i + \frac{\partial F}{\partial \dot{q}^i} \delta_{0} \dot{q}^i \right) dt + \left[ F \delta t - \varepsilon f \right]^{t_1}_{t_0} \\
& = \int^{t_1}_{t_0} \left[ \frac{\partial F}{\partial q^i} - \frac{d}{dt} \left( \frac{\partial F}{\partial \dot{q}^i} \right) \right] \delta_{0} q^i dt + \left[ \frac{\partial F}{\partial \dot{q}^i} \delta_{0} q^i + F \delta t - \varepsilon P \right]^{t_1}_{t_0}.
\end{align*}
Along natural paths, E-L equations are satisfied. Therefore, we find\footnote{We use that $\delta_{0} q^i = \varepsilon \left( \eta^i - \dot{q}^i \xi \right)$.}:
\[
\frac{\partial F}{\partial \dot{q}^i} \delta_{0} q^i + F \delta t - \varepsilon f = const \implies \frac{\partial F}{\partial \dot{q}^i} \eta^i + \left( F - \frac{\partial F}{\partial \dot{q}^i} \dot{q}^i \right) \xi - f = const.
\]

\underline{Second method:} By using the mathematical identity
\begin{equation} \label{con.mot.eq.25}
\frac{d}{dt} \left( F - \frac{\partial F}{\partial \dot{q}^i} \dot{q}^i \right) = \left[ \frac{\partial F}{\partial q^i} - \frac{d}{dt} \left( \frac{\partial F}{\partial \dot{q}^i} \right) \right] \dot{q}^i + \frac{\partial F}{\partial t}
\end{equation}
the Noether condition (\ref{con.mot.eq.23}) becomes
\[
\underbrace{\left[ \frac{\partial F}{\partial q^i} - \frac{d}{dt} \left( \frac{\partial F}{\partial \dot{q}^i} \right) \right]}_{= 0} \left( \eta^i - \dot{q}^i \xi \right) + \frac{d}{dt} \left[ \frac{\partial F}{\partial \dot{q}^i} \eta^i + \left( F - \frac{\partial F}{\partial \dot{q}^i} \dot{q}^i \right) \xi - f \right]=0
\]
which produces the Noether FI (\ref{con.mot.eq.24}).
\end{proof}

\begin{proposition} \label{con.mot.pro.gEL}
Consider the action
\begin{equation}
I \left[ q^A(t^i) \right] = \int_{\Omega} F \left( t^i, q^A(t^i), q^A{}_{,i_1}(t^i), q^A{}_{,i_1i_2}(t^i), ..., q^A{}_{,i_1..i_{\kappa}}(t^i) \right) d^nt
\label{con.mot.eq.48}
\end{equation}
where $d^nt \equiv dV \equiv dt^1 ... dt^n$ and $\Omega$ is a volume in the Euclidean space $E^n$ with coordinates $\{t^i\}$. If $\delta_{0}I = 0$ along a hypersurface\index{Hypersurface} $\{q^A(t^i)\}$ satisfying the boundary condition $\left. \delta_{0} q^A \right|_{\partial \Omega} = 0$ for all variations $\delta_{0}$, then $\{q^A(t^i)\}$ is called a {\bf{natural hypersurface}}\index{Hypersurface! natural} of $F$ and satisfies the {\bf{generalized E-L equations}}\index{Equations! generalized E-L}
\begin{equation}
\frac{\partial F}{\partial q^A} + \sum^{\kappa}_{\alpha=1} \sum_{i_1 \leq ... \leq i_{\alpha}} (-1)^{\alpha} \frac{d^{\alpha}}{dt^{i_1} ... dt^{i_{\alpha}}} \left( \frac{\partial F}{\partial q^A{}_{,i_1...i_{\alpha}}} \right) = 0. \label{con.mot.eq.gEL}
\end{equation}
\end{proposition}

\section{Symmetries of differential equations}

\label{con.mot.sec.3}

The calculus of the variations $\delta_{0}$ and $\delta$ is used in order to introduce the concept of symmetry for differential equations.

Replacing with $\delta t = \varepsilon \xi$, $\delta_{0} q^i = \varepsilon \chi^i$ and $\delta q^i = \varepsilon \left( \chi^i + \dot{q}^i \xi \right) \equiv \varepsilon \eta^i$, the identities (\ref{con.mot.eq.26}) and (\ref{con.mot.eq.27}) imply that
\begin{equation} \label{con.mot.eq.28}
\delta ( q^i )^{(\kappa)} = \varepsilon \eta^{i[\kappa]}
\end{equation}
where -in general- the generators $\xi, \chi^i, \eta^i$ are considered as functions of $t, q^i(t), \dot{q}^i(t), \ddot{q}^i(t), ...$, $\kappa$ is a natural number, and
\begin{align}
\eta^{i[\kappa]} & \equiv (\eta^i - \dot{q}^i \xi)^{(\kappa)} + ( q^i )^{(\kappa + 1)} \xi, \quad \text{for $\kappa \geq 0$} \label{con.mot.eq.28a} \\
\eta^{i[\kappa]} & = \dot{\eta}^{i[\kappa - 1]} - ( q^i )^{(\kappa)} \dot{\xi} , \quad \text{for $\kappa \geq 1$} \label{con.mot.eq.28b}.
\end{align}
If we replace (\ref{con.mot.eq.28a}) in (\ref{con.mot.eq.28b}), we find equation (\ref{con.mot.eq.28a}) for $\kappa \geq 1$. The quantity $\eta^{i[\kappa]}$ is the \textbf{$\boldsymbol{\kappa}$-prolongation} of the spatial infinitesimal generator $\eta^i$.\index{$\kappa$-prolongation}

According to the theory of the sec. \ref{con.mot.sec.4}, a real-valued function $F =$ $F \left(\right.$ $t, q^i(t),$ $\dot{q}^i(t), ...,$ $q^{i(\kappa)}(t)$ $\left.\right)$ is {\bf{invariant}}\index{Invariant! function} wrt a variation generated by the vector field
\begin{equation}
\mathbf{X} = \frac{1}{\varepsilon} \left( \delta t \frac{\partial}{\partial t} + \delta q^i \frac{\partial}{\partial q^i} \right) = \xi \frac{\partial}{\partial t} + \eta^i \frac{\partial}{\partial q^i} \label{eq.Lev14}
\end{equation}
iff it satisfies the condition
\begin{equation}
\delta F = 0 \iff \delta t \frac{\partial F}{\partial t} + \delta q^i \frac{\partial F}{\partial q^i} + \delta \dot{q}^i \frac{\partial F }{\partial \dot{q}^i} + ... + \delta q^{i(\kappa)} \frac{\partial F}{\partial q^{i(\kappa)}} = 0 \iff \mathbf{X}^{[\kappa]} F = 0 \label{con.mot.eq.29}
\end{equation}
where
\begin{equation}
\mathbf{X}^{[\kappa]} \equiv \mathbf{X} + \eta^{i[1]} \frac{\partial}{\partial \dot{q}^i} + ... + \eta^{i[\kappa]} \frac{\partial}{\partial q^{i(\kappa)}} \label{con.mot.eq.30}
\end{equation}
is the $\kappa$-prolongation\index{$\kappa$-prolongation} of the generator $\mathbf{X}$. The given $\delta$-variation is called a \textbf{symmetry} of the function $F$.\index{Symmetry}

Similarly, an arbitrary $\kappa$-order ODE $H \left(t, y(t), \dot{y}(t), ..., y^{(\kappa)}(t) \right) = 0$ is invariant\index{Invariant! ODE} wrt a variation generated by a vector $\mathbf{X}$ iff $\delta H = 0$ along the solutions $y=y(t)$ of $H = 0$. The generator $\mathbf{X} = \xi \frac{\partial}{\partial t} + \eta \frac{\partial}{\partial y}$ is a symmetry of the ODE $H=0$ and we have the \textbf{symmetry condition}\index{Symmetry! condition}
\begin{equation} \label{con.mot.eq.31}
\left. \delta H \right|_{y:H=0} = 0 \iff \left. \mathbf{X}^{[\kappa]}(H) \right|_{y:H=0} = 0 \iff \left. H \left(\bar{t}, \bar{y}(\bar{t}), \dot{\bar{y}}(\bar{t}), ..., \bar{y}^{(\kappa)}(\bar{t}) \right) \right|_{y:H=0} = 0
\end{equation}
where $\mathbf{X}^{[\kappa]} = \mathbf{X} + \eta^{[1]} \partial_{y^{(1)}} + ... + \eta^{[\kappa]} \partial_{y^{(\kappa)}}$ and $\eta^{[\kappa]} \equiv \left( \eta - \dot{y} \xi \right)^{(\kappa)} + y^{(\kappa+1)} \xi$.

Equation (\ref{con.mot.eq.31}) states that the variation $\delta H$ is not necessary to vanish identically, but only along solutions of the ODE. Therefore, a symmetry of an ODE maps a solution $y(t)$ into another solution $\bar{y}(\bar{t})$. For example, the vector field $\mathbf{X} = y \partial_y$ is a symmetry of the ODE $\ddot{y} + y = 0$ which maps a solution $y(t)$ to the solution $y(t) + \varepsilon y(t)$.

\section{Classification of symmetries of ODEs}

\label{sec.classym}

Consider the $\kappa$th-order system of ODEs
\begin{equation}
H^i \left(t, q^r(t), \dot{q}^r(t), ..., q^{r(\kappa)}(t) \right) = 0 \label{eq.Lev15}
\end{equation}
where $i, r = 1, 2, ..., n$. This system is said to be {\bf{(Lie) invariant}}\index{Invariant! ODE} wrt a $\delta$-variation generated by $\mathbf{X} = \xi \partial_t + \eta^i \partial_{q^i}$ iff
\begin{equation} \label{liesym.eq.1}
\left. \delta H^i \right|_{q^i : H^i = 0} = 0 \implies \left. \mathbf{X}^{[\kappa]}(H^i) \right|_{q^i : H^i = 0} = 0.
\end{equation}
The vector field $\mathbf{X}$ is called a {\bf{(Lie) symmetry}} of the system or, equivalently, we say that $\mathbf{X}$ is the generator of a Lie symmetry.\index{Symmetry! Lie}

We note that the study of symmetries of differential equations was initiated and systematized by Sophus Lie \cite{lie1, lie2, lie3}. Lie used the theory of continuous transformation groups in order to define the Lie symmetries. Using the Lie symmetries, one may construct appropriate sets of variables in which the differential equation is simplified and, in general, is brought into a solvable form.

The most widely studied types of Lie symmetries are the following: \newline
1) \underline{{\bf{Lie point symmetry}}:}\index{Symmetry! Lie point} $\xi = \xi(t, q^r)$ and $\eta^i = \eta^i(t, q^r)$.
\newline
2) \underline{{\bf{Lie-B\"{a}cklund symmetry (or dynamical Lie symmetry))}}:}\index{Symmetry! Lie-B\"{a}cklund} $\xi = \xi(t, q^r, \dot{q}^r, ..., q^{r(\kappa)})$ and \newline 
$\eta^i = \eta^i(t, q^r, \dot{q}^r, ..., q^{r(\kappa)})$. A special type of this symmetry is the {\bf{contact symmetry}}\index{Symmetry! contact}, where $\xi = \xi(t, q^r, \dot{q}^r)$ and $\eta^i = \eta^i(t, q^r, \dot{q}^r)$.\index{Symmetry! dynamical Lie}

In the case of the dynamical Lie symmetries, one has an extra degree of freedom \cite{StephaniB} which is removed, if one assumes an extra gauge condition. Then, one works with the so-called \textbf{gauged dynamical Lie symmetries}. One usually requires the gauge condition $\xi =0$ so that the generator is simplified to $\mathbf{X}=\eta ^{i}(t,q,\dot{q}, ...) \partial _{q^{i}}.$\index{Symmetry! gauged dynamical Lie}

\begin{example}
\label{exa.Lie.autonomous} The dynamical equations of a general holonomic dynamical system have the functional form
\begin{equation}
\ddot{q}^{a} = \omega^{a}(t,q,\dot{q})  \label{FL.0}
\end{equation}%
where $\omega^{a}= -\Gamma^{a}_{bc}(q) \dot{q}^{b} \dot{q}^{c} -V^{,a} -Q^{a}\left( t,q,\dot{q}\right)$, $-Q^{a}$ are the generalized (non-conservative) forces, $\Gamma_{bc}^{a}$ are the Riemannian connection coefficients determined from
the kinetic metric $\gamma_{ab}(q)$ (kinetic energy) and $-V^{,a}$ are the conservative forces.

Equation (\ref{FL.0}) defines in the jet space $J^{1}\left\{ t, q^{a}, \dot{q}^{a} \right\}$ the \textbf{Hamiltonian vector field}\index{Hamiltonian vector field}
\begin{equation}
\mathbf{\Gamma} =\frac{\partial }{\partial t} +\dot{q}^{a} \frac{\partial}{\partial q^{a}} +\omega^{a} \frac{\partial}{\partial \dot{q}^{a}}.
\label{FI.3}
\end{equation}

A Lie symmetry with generator $\mathbf{X} =\xi(t,q,\dot{q}) \partial_{t} +\eta^{a}(t,q,\dot{q})\partial_{q^{a}}$ is a point transformation in the jet space $J^{1}\{t,q^{a},\dot{q}^{a}\}$ which preserves the set of solutions of (\ref{FL.0}). By using analytical techniques (instead of $\delta$-variation formalism), it is proved that the vector field $\mathbf{X}$ is a Lie symmetry of (\ref{FL.0}) iff there exists a function $\lambda\left( t, q, \dot{q}\right)$ such that
\begin{equation}
\left[ \mathbf{X}^{\left[ 1\right] },\mathbf{\Gamma} \right] =\lambda (t,q,\dot{q}) \mathbf{\Gamma}.
\label{FL.0.1}
\end{equation}%
The first prolongation\footnote{This is the complete lift of $\mathbf{X}$ in the tangent bundle $TM$.} $\mathbf{X}^{[1]}=\xi (t,q,\dot{q})\partial_{t} +\eta^{a}(t,q,\dot{q})\partial_{q^{a}}+\left( \dot{\eta}^{a}-\dot{q}^{a}\dot{\xi}\right) \partial _{\dot{q}^{a}}$. It can be shown that condition (\ref{FL.0.1}) is equivalent to condition (\ref{liesym.eq.1}), which for the second order ODEs (\ref{FL.0}) reads
\begin{equation}
\left. \mathbf{X}^{[2]}(H^{a}) \right|_{q^a : H^a = 0}=0 \implies \eta^{a[2]} -\mathbf{X}^{[1]}\omega^{a}=0  \label{FL.0.2}
\end{equation}%
where $H^{a}\equiv \ddot{q}^{a}-\omega ^{a}$, $\eta ^{a[2]}=\ddot{\eta}^{a}-2%
\ddot{q}^{i}\dot{\xi}-\dot{q}^{i}\ddot{\xi}$ and $\mathbf{X}^{[2]}= \mathbf{X}^{[1]}+\eta
^{a[2]}\partial _{\ddot{q}^{a}}$ is the second prolongation of $\mathbf{X}$ in $J^{2}\left\{t, q^{a}, \dot{q}^{a}, \ddot{q}^{a}\right\}$. We note that in (\ref{FL.0.2}) the quantities $\ddot{q}^{a}$ -whenever they appear- must be replaced by the function $\omega^{a}$ or, equivalently, all total derivatives must be replaced by the Hamiltonian vector field $\mathbf{\Gamma}$.

In the gauge $\xi=0$, the Lie symmetry condition (\ref{FL.0.2}) becomes:
\begin{equation}
\mathbf{\Gamma} \left( \mathbf{\Gamma}(\eta^{a}) \right)= \eta^{b} \frac{\partial \omega^{a}}{\partial q^{b}} +\mathbf{\Gamma}(\eta^{b}) \frac{\partial \omega^{a}}{\partial \dot{q}^{b}}. \label{eq.gaugecon}
\end{equation}
\end{example}

Consider now a Lagrangian dynamical system $L = L(t, q^i, \dot{q}^i)$. This system is gauge invariant wrt a $\delta$-variation generated by $\mathbf{X} = \xi(t, q^r, \dot{q}^{r}) \partial_t + \eta^i(t, q^r, \dot{q}^{r}) \partial_{q^i}$ iff
\begin{equation}
\delta S \equiv \delta \int^{t_1}_{t_0} L dt = \varepsilon \int^{t_1}_{t_0} \dot{f} dt, \quad \forall \{q^i\}, \enskip \forall [t_0,t_1] \label{eq.Lev16}
\end{equation}
where $f = f(t, q^r, \dot{q}^{r})$ is the gauge (or Noether)\index{Function! Noether} function. The set $\left( \xi, \eta^i; f \right)$ is a {\bf{generalized (or dynamical) Noether symmetry}}\index{Symmetry! generalized} of the dynamical system. In general, a dynamical Noether symmetry has generators of the form $\xi(t, q,\dot{q},\ddot{q},...)$ and $\eta^{i}(t, q, \dot{q}, \ddot{q},...)$.

From proposition \ref{con.mot.pro.2}, the Noether condition is satisfied and takes the form:\index{Condition! Noether}
\begin{equation}
\mathbf{X}^{[1]} L + L \dot{\xi} = \dot{f}. \label{liesym.eq.2}
\end{equation}

Moreover, from proposition \ref{con.mot.pro.3}, every generalized Noether symmetry produces the Noether FI\index{First integral! Noether}
\begin{equation}
\Lambda \equiv f - L \xi - \frac{\partial L}{\partial \dot{q}^i} \left( \eta^i - \dot{q}^i \xi \right).
\label{liesym.eq.3}
\end{equation}

A Noether symmetry generated by the vector field $\mathbf{X}= \xi(t, q) \partial_t + \eta^i(t, q) \partial_{q^i}$ and the gauge $f=f(t,q)$ is called a {\bf{point Noether symmetry}}.\index{Symmetry! point Noether} For such symmetries, the corresponding FI is invariant\footnote{This is not the case for generalized Noether symmetries.}, i.e. $\mathbf{X}^{[1]}\Lambda =0$ (see proposition 2.2 in \cite{Sarlet Cantrijn 81}). Moreover, point Noether symmetries are a special class of Lie point symmetries because they leave E-L equations invariant. Indeed, we have $\delta S = \varepsilon \int^{t_1}_{t_0} \dot{f} dt \implies \bar{S} = S + \varepsilon \int^{t_1}_{t_0} \dot{f} dt \implies \delta_{0} \bar{S} = \delta_{0} S$, which leads to the same E-L equations.

We note that Noether point symmetries form a finite dimensional Lie algebra, whereas dynamical Noether symmetries form an infinite dimensional Lie algebra.

Concerning the geometric nature of Noether symmetries, it has been shown \cite{Tsamparlis 2012, Paliathanasis 2012} that the generators of Noether point symmetries of autonomous holonomic dynamical systems with a regular Lagrangian (i.e. $\det\frac{\partial ^{2}L}{\partial \dot{q}^{a} \partial \dot{q}^{b}}\neq 0$) of the form $L= \frac{1}{2}\gamma _{ab}\dot{q}^{a}\dot{q}^{b}-V(q)$, where $\gamma _{ab}=%
\frac{\partial ^{2}L}{\partial \dot{q}^{a} \partial\dot{q}^{b}}$ is the kinetic metric, are elements of the homothetic algebra (see sec. \ref{sub.dec.Lgij}) of $\gamma_{ab}$. A similar firm result does not exist for dynamical Noether symmetries.

\section{The variation $\delta_{\phi}$}

\label{con.mot.sec.gvar}

Consider a Lagrangian dynamical system $L = L \left(t, q^i(t), \dot{q}^i(t) \right)$. A more general variation $\delta_{\phi}$ can be defined as follows:
\begin{equation}
\delta_{\phi} L \equiv L \left( \bar{t}, \bar{q}(\bar{t}), \dot{\bar{q}}(\bar{t}) + \varepsilon \phi^i \right) - L \left(t, q(t), \dot{q}(t) \right)
\label{con.mot.eq.gvar1}
\end{equation}
where $\phi^i = \phi^i(t, q^r, \dot{q}^r)$ are given smooth functions and, as usual, $\delta t = \bar{t} - t = \varepsilon \xi(t, q, \dot{q}^r)$ and $\delta q^i = \bar{q}(\bar{t}) - q(t) = \varepsilon \eta^i(t,q,  \dot{q}^r)$.

Generalizing (\ref{con.mot.eq.gvar1}) to an arbitrary real-valued function $F = F(t,q,\dot{q})$, we find that $\delta_{\phi} t = \delta t$, $\delta_{\phi} q^i = \delta q^i$ and
\begin{equation}
\delta_{\phi} \dot{q}^i = \delta \dot{q}^i + \varepsilon \phi^i = \varepsilon \left( \eta^{i[1]} + \phi^i \right).
\label{con.mot.eq.gvar1b}
\end{equation}

We compute:
\begin{equation}
\delta_{\phi} L = \delta t \frac{\partial L}{\partial t} + \delta q^i \frac{\partial L}{\partial q^i} + \delta_{\phi} \dot{q}^i \frac{\partial L}{\partial \dot{q}^i} = \varepsilon \mathbf{X}^W L \implies \delta_{\phi} L = \varepsilon \mathbf{X}^W L \label{con.mot.eq.gvar2}
\end{equation}
where
\begin{equation}
\mathbf{X}^W \equiv \mathbf{X}^{[1]} + \phi^i \frac{\partial}{\partial \dot{q}^i} \label{con.mot.eq.gvar3}
\end{equation}
is the \textbf{weak first prolongation}.\index{First prolongation! weak} We note that
\begin{equation}
\delta_{\phi} L = \delta L + \varepsilon \phi^i \frac{\partial L}{\partial \dot{q}^i}. \label{con.mot.eq.gvar4}
\end{equation}

The $\delta_{\phi}$-variation generated by the set $\{ \xi, \eta^i; \phi^i \}$ is called a {\bf{symmetry}}\index{Symmetry} of a $\kappa$th-order system of ODEs $H^i(t, q^r, \dot{q}^r, ..., q^{r(\kappa)}) = 0$ iff $\left. \delta_{\phi} H^i \right|_{\{q^i\} : H^i=0} = 0$, which implies that $\mathbf{X}^W H^ i = 0$ along solutions of the system.

Following the procedure for deriving equations (\ref{con.mot.eq.23}) and (\ref{con.mot.eq.24}), Noether's theorem can be revisited wrt $\delta_{\phi}$ by taking a weaker form. Specifically, the Noether symmetry turns into a {\bf{weak Noether symmetry}} defined by the relation\index{Symmetry! weak Noether}
\begin{equation}
\delta_{\phi} I \equiv \delta_{\phi} \int^{t_1}_{t_0} L dt = \varepsilon \int^{t_1}_{t_0} \dot{f} dt, \quad \forall \{q^i\}, \enskip \forall [t_0,t_1]. \label{con.mot.eq.gvar5}
\end{equation}
The last equation thanks to (\ref{con.mot.eq.gvar4}) gives the {\bf{weak Noether condition}}\index{Condition! weak Noether}
\[
\delta_{\phi} \int^{t_1}_{t_0} L dt = \varepsilon \int^{t_1}_{t_0} \dot{f} dt \implies \underbrace{\delta \int^{t_1}_{t_0} L dt - \varepsilon \int^{t_1}_{t_0} \dot{f} dt}_{\text{see prop. \ref{con.mot.pro.2}}} \underbrace{+ \varepsilon \int^{t_1}_{t_0} \phi^i \frac{\partial L}{\partial \dot{q}^i} dt}_{\text{new term}} = 0 \implies
\]
\begin{equation}
\mathbf{X}^{[1]} L + L \dot{\xi} + \phi^i \frac{\partial L}{\partial \dot{q}^i} = \dot{f}. \label{con.mot.eq.gvar6}
\end{equation}
We note that
\begin{equation}
\delta_{\phi} I = \delta I + \varepsilon \int^{t_1}_{t_0} \phi^i \frac{\partial L}{\partial \dot{q}^i} dt. \label{con.mot.eq.gvar6b}
\end{equation}

We assume now that on the Lagrangian system $L$ act non-conservative generalized forces $F_i(t,q,\dot{q})$. Then, the E-L equations become $E_i L = F_i$ and Noether's theorem should be modified as follows (see proof of proposition \ref{con.mot.pro.3}):
\[
\underbrace{\left[ \frac{\partial L}{\partial q^i} - \frac{d}{dt} \left( \frac{\partial L}{\partial \dot{q}^i} \right) \right]}_{= - E_i L = - F_i} \left( \eta^i - \dot{q}^i \xi \right) + \frac{d}{dt} \underbrace{\left[ \frac{\partial L}{\partial \dot{q}^i} \eta^i + \left( L - \frac{\partial L}{\partial \dot{q}^i} \dot{q}^i \right) \xi - f \right]}_{\equiv - \Lambda} + \phi^i \frac{\partial L}{\partial \dot{q}^i} = 0 \implies
\]
\begin{equation}
\frac{d \Lambda}{dt} = - F_i \left( \eta^i - \dot{q}^i \xi \right) + \phi^i \frac{\partial L}{\partial \dot{q}^i} \label{con.mot.eq.gvar7}
\end{equation}
along natural paths $\{q^i(t)\}$. Therefore, the function $\Lambda$ is a FI of the system iff
\begin{equation}
F_i \left( \eta^i - \dot{q}^i \xi \right) = \phi^i \frac{\partial L}{\partial \dot{q}^i}. \label{con.mot.eq.gvar8}
\end{equation}

The additional requirement (\ref{con.mot.eq.gvar8}) is equivalent to Noether's condition because it implies that $\frac{d \Lambda}{dt} = 0$. It may be regarded also as an additional constraint from which the vectors $\phi^i$ can be computed, provided $F_i$ is known.

Replacing (\ref{con.mot.eq.gvar8}) in (\ref{con.mot.eq.gvar6}), we obtain the condition
\begin{equation}
\mathbf{X}^{[1]} L + L \dot{\xi} + F_i \left( \eta^i - \dot{q}^i \xi \right) = \dot{f}. \label{con.mot.eq.gvar9}
\end{equation}
From this condition, one computes directly for non-conservative dynamical systems all sets $\{\xi, \eta^{i}; f\}$ that produce Noether FIs $\Lambda$. The condition (\ref{con.mot.eq.gvar9}) is the well-known {\bf{Noether-Bessel-Hagen (NBH) equation}}.\index{Equation! NBH}

\section{Killing equation}

\label{con.mot.sec.killing}

Consider a free dynamical system $L = \frac{1}{2} g_{ij} \dot{q}^i \dot{q}^j$, where $g_{ij}$ is the kinetic metric. Then, the E-L equations are written as follows:
\begin{equation} \label{con.mot.eq.kil1}
\ddot{q}^i + \{^i_{jk}\} \dot{q}^j \dot{q}^k = 0.
\end{equation}
The quantities $\{^i_{jk}\}= \frac{1}{2}g^{ir} \left( g_{jr,k} +g_{kr,j} -g_{jk,r} \right)$ are the coefficients of the Riemannian connection defined by the metric $g_{ij}$. Equation (\ref{con.mot.eq.kil1}) is the geodesic equation in the configuration space of the system.

We assume: a) $\delta t = 0 \implies \left[ \xi = 0, \delta q^i = \delta_{0} q^i = \varepsilon \eta^i \right]$, b) $\delta$ is a variation such that $\delta I = 0$, i.e. $f = 0$ (absolutely invariant action), and c) $\eta^i = \eta^i(q)$.

Then, $\delta$ is a point Noether symmetry and the Noether condition (\ref{con.mot.eq.23}) implies that
\begin{equation} \label{con.mot.eq.kil2}
g_{ij,k} \eta^k + g_{kj} \eta^k_{,i} + g_{ik} \eta^k_{,j} = 0 \iff L_{\boldsymbol{\eta}} g_{ij} = 0 \iff \eta_{(i;j)} = 0
\end{equation}
where $L_{\boldsymbol{\eta}}$ is the Lie derivative along the generator $\eta^{i}$ and a semicolon denotes the Riemannian covariant derivative. Equation (\ref{con.mot.eq.kil2}) is the well-known {\bf{Killing equation}}.\index{Equation! Killing} In general, solutions of the Killing equation do not exist. However, when do exist, the associated Noether FI (\ref{liesym.eq.3}) is
\begin{equation}
\frac{\partial L}{\partial \dot{q}^k} \eta^k = const \implies g_{ij} \eta^i \dot{q}^j= \eta_{i}\dot{q}^{i} = const. \label{eq.Kil1}
\end{equation}

\section{The generalized Killing equations}

\label{con.mot.subsec.killing.2}

Consider a Lagrangian dynamical system that admits a generalized Noether symmetry with generators $\xi = \xi(t, q, \dot{q})$, $\eta^i = \eta^i(t, q, \dot{q})$ and Noether function $f = f(t, q, \dot{q})$. Then, the Noether condition (\ref{liesym.eq.2}) is written as follows:
\begin{align*}
0 &= \xi \frac{\partial L}{\partial t} + \eta^i \frac{\partial L}{\partial q^i} + \left( \frac{\partial \eta^i}{\partial t} + \dot{q}^j \frac{\partial \eta^i}{\partial q^j} - \dot{q}^i \frac{\partial \xi}{\partial t} - \dot{q}^i \dot{q}^j \frac{\partial \xi}{\partial q^j} \right) \frac{\partial L}{\partial \dot{q}^i} + L \left( \frac{\partial \xi}{\partial t} + \dot{q}^i \frac{\partial \xi}{\partial q^i} \right) - \frac{\partial f}{\partial t}- \\
& \quad - \dot{q}^i \frac{\partial f}{\partial q^i} +\ddot{q}^j \left[ \frac{\partial \xi}{\partial \dot{q}^j} L + \left( \frac{\partial \eta^i}{\partial \dot{q}^j} - \dot{q}^i \frac{\partial \xi}{\partial \dot{q}^j} \right) \frac{\partial L}{\partial \dot{q}^i} - \frac{\partial f}{\partial \dot{q}^j} \right].
\end{align*}
Since this condition must be satisfied identically for all $\ddot{q}^{i}$, it follows:
\begin{eqnarray}
\frac{\partial f}{\partial t} + \dot{q}^i \frac{\partial f}{\partial q^i} &=& \xi \frac{\partial L}{\partial t} + \eta^i \frac{\partial L}{\partial q^i} + \left( \frac{\partial \eta^i}{\partial t} + \dot{q}^j \frac{\partial \eta^i}{\partial q^j} - \dot{q}^i \frac{\partial \xi}{\partial t} - \dot{q}^i \dot{q}^j \frac{\partial \xi}{\partial q^j} \right) \frac{\partial L}{\partial \dot{q}^i} + \notag \\
&& +L \left( \frac{\partial \xi}{\partial t} + \dot{q}^i \frac{\partial \xi}{\partial q^i} \right) \label{con.mot.eq.kil3} \\
\frac{\partial f}{\partial \dot{q}^j} &=& \frac{\partial \xi}{\partial \dot{q}^j} L + \left( \frac{\partial \eta^i}{\partial \dot{q}^j} - \dot{q}^i \frac{\partial \xi}{\partial \dot{q}^j} \right) \frac{\partial L}{\partial \dot{q}^i}. \label{con.mot.eq.kil4}
\end{eqnarray}
These equations are the \textbf{generalized Killing equations}.\index{Equation! Generalized Killing} They coincide with conditions (A1) and (A2) in the appendix of \cite{Leach 1985}. The above solution of the Noether condition is called a solution in the space. Another method can be applied (see chapter \ref{ch.td.central.pots}) where $\ddot{q}^{a}$ can be replaced by the dynamical equations. Then, we have a solution along the trajectory and the split of the Noether condition is avoided.

We note that we have $n+2$ unknowns $\xi$, $\eta^{i}$ and $f$, but only $n+1$ equations. Therefore, there
is an extra degree of freedom which can be removed by a gauge condition (usually the condition $\xi =0$).

In the case of the weak Noether condition (\ref{con.mot.eq.gvar6}), the generalized Killing equations differ just by a term (due to the vector $\phi^i$). Indeed, we have:
\begin{eqnarray}
\frac{\partial f}{\partial t} + \dot{q}^i \frac{\partial f}{\partial q^i} &=& \xi \frac{\partial L}{\partial t} + \eta^i \frac{\partial L}{\partial q^i} + \left( \frac{\partial \eta^i}{\partial t} + \dot{q}^j \frac{\partial \eta^i}{\partial q^j} - \dot{q}^i \frac{\partial \xi}{\partial t} - \dot{q}^i \dot{q}^j \frac{\partial \xi}{\partial q^j} \right) \frac{\partial L}{\partial \dot{q}^i} + \notag \\
&& +L \left( \frac{\partial \xi}{\partial t} + \dot{q}^i \frac{\partial \xi}{\partial q^i} \right) \underbrace{+ \phi^i \frac{\partial L}{\partial \dot{q}^i}} \label{con.mot.eq.kil5} \\
\frac{\partial f}{\partial \dot{q}^j} &=& \frac{\partial \xi}{\partial \dot{q}^j} L + \left( \frac{\partial \eta^i}{\partial \dot{q}^j} - \dot{q}^i \frac{\partial \xi}{\partial \dot{q}^j} \right) \frac{\partial L}{\partial \dot{q}^i}. \label{con.mot.eq.kil6}
\end{eqnarray}
These are the {\bf{weak generalized Killing equations}}.\index{Equation! weak generalized Killing} In this case, we have again a system of $n+1$ equations but with $2n+2$ unknowns. Therefore, there are $n+1$ extra degrees of freedom which must be fixed (e.g. $\xi=0$ and $\phi^{i}$ is given).

%% file: collineations.tex
\chapter{Collineations and higher order symmetries}

\label{ch.collineations}

In sec. \ref{con.mot.sec.killing}, we derived the well-known Killing equations (\ref{con.mot.eq.kil2}). The vector field $\boldsymbol{\eta}= \eta^{i}\partial_{i}$ satisfying these equations is called a Killing vector (KV) of the metric $g_{ij}$ and it is said that such vectors determine the symmetries of the metric. For example, the condition $L_{\boldsymbol{\eta}}g_{ij} =0$ contains the concept of space symmetry in $E^{3}$, i.e. spherical symmetry, cylindrical symmetry, etc. . Therefore, there arises the following question: \emph{What is the geometrical meaning of equations of the form $L_{\mathbf{X}}A=B$? Is there any hidden, or higher order, symmetry?}

\section{Collineations of geometrical objects}

\label{sec.collineat.1}

A geometrical object (GO)\index{Geometrical object} is a set of components, defined on a smooth manifold $M$, which satisfies a specific transformation law. Tensors and connections are the most characteristic representatives of such objects.

The \textbf{collineation}\index{Collineation} of a GO $A$ is a vector field $\boldsymbol{\xi}$ such that\footnote{We recall that $L_{\boldsymbol{\xi}}A$ is the Lie derivative of $A$ along $\boldsymbol{\xi}$ and that the Lie derivative of a connection is a tensor field \cite{Yanob}.} $L_{\boldsymbol{\xi}}A =B$, where $B$ is a tensor field with the same number and the same symmetries of indices as $A$. If $A$ is a metrical GO (i.e. defined by the metric), the collineation $\boldsymbol{\xi}$ is called a \textbf{geometric collineation (or geometric symmetry)} \index{Collineation! geometric}\index{Symmetry! geometric} of $A$. In this case, the quantity $L_{\boldsymbol{\xi}}A$ can be expressed in terms of the fundamental quantity $L_{\boldsymbol{\xi}}g_{ij}$. Therefore, it is possible to characterize all geometric collineations in terms of the tensor $L_{\boldsymbol{\xi}}g_{ij}$. The collineations of the metric are called \textbf{generic collineations}.\index{Collineation! generic}

\section{The decomposition of $L_{\boldsymbol{\xi}}g_{ab}$: Generic collineations}

\label{sub.dec.Lgij}

Any (0,2)-tensor field $T_{ab}$ on a smooth $n$-dimensional Riemannian manifold $(M, g_{ab})$ is decomposed as follows:
\begin{equation} \label{dec.Tab}
T_{ab} = T_{(ab)} + T_{[ab]} = \psi g_{ab} + H_{ab} + F_{ab}
\end{equation}
where $\psi = \frac{1}{n} T_{(ab)} g^{ab}$, $H_{ab} = T_{(ab)} - \frac{1}{n} T_{(cd)} g^{cd} g_{ab}$ and $F_{ab} = T_{[ab]}$. We note that $H_{ab}$ is a symmetric traceless (i.e. $H^{a}{}_{a} =g^{ab}H_{ab}=0$) tensor of order two.

If $T_{ab} = \xi_{a;b}$, where $\boldsymbol{\xi}= \xi^{a}\partial_{a}$, the decomposition (\ref{dec.Tab}) reads
\begin{equation} \label{dec.Tab.1}
\xi_{a;b} = \psi(\boldsymbol{\xi}) g_{ab} + H_{ab}(\boldsymbol{\xi}) + F_{ab}(\boldsymbol{\xi})
\end{equation}
where $\psi(\boldsymbol{\xi}) = \frac{1}{n} \xi_{(a;b)} g^{ab} = \frac{1}{n} \xi^a{}_{;a}$, $H_{ab}(\boldsymbol{\xi}) = \xi_{(a;b)} - \frac{1}{n} \xi^c{}_{;c} g_{ab}$, and $F_{ab}(\boldsymbol{\xi}) = \xi_{[a;b]}$.

Using the well-known identity $L_{\boldsymbol{\xi}} g_{ab} = \xi_{a;b} + \xi_{b;a}= 2\xi_{(a;b)}$, we get the decomposition
\begin{equation}
L_{\boldsymbol{\xi}} g_{ab} = 2 \psi(\boldsymbol{\xi}) g_{ab} + 2 H_{ab}(\boldsymbol{\xi}). \label{eq.dec}
\end{equation}
The function $\psi(\boldsymbol{\xi})$ is the \textbf{conformal factor} \index{Conformal factor} of the metric $g_{ab}$ wrt the vector $\boldsymbol{\xi}$. It holds also that $\xi_{a;b} = \frac{1}{2} L_{\boldsymbol{\xi}} g_{ab} + F_{ab}$.

From the decomposition \eqref{eq.dec}, the generic collineation $\boldsymbol{\xi}$ is classified as follows: \newline
i) (Killing vector $=$ KV $=$ motion) $\iff$ $L_{\boldsymbol{\xi}} g_{ab} = 0$ $\iff$ $[$ $H_{ab} = 0$, $\psi = 0$ $]$.\index{Vector! Killing} \index{Motion} \newline \newline
ii) (Homothetic vector $=$ HV $=$ homothetic motion) $\iff$ $L_{\boldsymbol{\xi}} g_{ab} = 2 c g_{ab}$ $\iff$ $[$ $H_{ab} = 0$, $\psi = const \equiv c$ $]$.\index{Vector! homothetic} \index{Motion! homothetic} \newline
iii) (Conformal KV $=$ CKV $=$ conformal motion) $\iff$ $L_{\boldsymbol{\xi}} g_{ab} = 2 \psi g_{ab}$ $\iff$ $H_{ab} = 0$. \index{Vector! conformal Killing} \index{Motion! conformal} \newline
iv) (Special CKV $=$ SCKV $=$ special conformal motion) $\iff$ $[$ $H_{ab} = 0$, $\psi_{;ab} = 0$ $]$ $\iff$ $[$ $L_{\boldsymbol{\xi}} g_{ab} = 2 \psi g_{ab}$, $\psi_{;ab}=0$ $]$. \index{Vector! special conformal Killing} \index{Motion! special conformal}  \newline

A CKV that is not special (i.e. $\psi_{;ab}\neq0$) is called {\bf{proper CKV}}. \index{Vector! proper Conformal Killing} We note that CKVs $\supset$ SCKVs $\supset$ HVs $\supset$ KVs, and CKVs $=$ (proper CKVs) $+$ (SCKVs).

Why the computation of CKVs is important? \newline
a) In geometry: to construct coordinate systems in which the metric $g_{ab}$ takes a simplified form. \newline
b) In kinematics: to impose restrictions on the kinematic variables (see e.g. chapter \ref{ch.EMSF}). \newline
c) In dynamics: to obtain new solutions of the Einstein's field equations.

A CKV which is defined by a potential function $\phi(x)$, i.e. $\xi_a = \phi_{,a}$, is called a {\bf{gradient CKV}}.\index{Vector! gradient} It can be proved that $F_{ab}(\boldsymbol{\xi}) = 0$ $\iff$ $($ $\xi^a$ is gradient $)$. Furthermore, $\xi^a$ is a gradient KV iff it is covariantly constant (i.e. $\xi_{a;b} = 0$). We note that $\phi_{;ab}= \phi_{;ba}$.

Let $\boldsymbol{\xi}$ be a CKV. Then, we have the following identities:
\begin{eqnarray}
L_{\boldsymbol{\xi}} g^{ab} &=& -2 \psi g^{ab} \label{eq.CKV} \\
L_{\boldsymbol{\xi}} R^a{}_{bcd} &=& \delta^a_d \psi_{;bc} - \delta^a_c \psi_{;bd} + g^{ar} (\psi_{;rd} g_{bc} - \psi_{;rc} g_{bd}) \label{eq.CKV1} \\
L_{\boldsymbol{\xi}} R_{ab} &=& (2 - n) \psi_{;ab}  - g_{ab} \Box \psi \label{eq.CKV2} \\
L_{\boldsymbol{\xi}} R &=& - 2 \psi R + 2(1-n) \Box \psi \label{eq.CKV3} \\
F_{ab;c} &=& R_{abcd} \xi^d - 2 \psi_{;[a} g_{b]c} \label{eq.CKV4}
\end{eqnarray}
where $R^{a}{}_{bcd}$ is the Riemannian curvature tensor, $R_{ab}= R^{c}{}_{acb}$ is the Ricci tensor, $R= R^{a}{}_{a}= g^{ab}R_{ab}$ is the Ricci scalar, and $\Box\psi \equiv g^{ab}\psi_{;ab}$ is the D' Alembertian over $\psi$.

Let $\xi_a = \phi_{;a}$ be a gradient KV. Then, we have: $\phi_{;ab} = 0$, $\psi(\boldsymbol{\xi}) = 0$, $F_{ab}(\boldsymbol{\xi}) = 0$, $R_{abcd} \phi^{;d} = 0$, and $\phi_{;a} \phi^{;a} = const$.

\section{CKVs of a flat metric}

\label{sub.ckv.flat}

A generic {\bf{flat metric}}\index{Metric! flat}  is a second order non-degenerate tensor whose curvature tensor vanishes. For such a metric, there always exists a coordinate system $\{x^a\}$ such that\footnote{This is the pseudo-Euclidean (or reduced) form of a flat metric.} $\eta_{ab} = diag(-1, ..., -1, +1, ..,+1)$. A Riemannian manifold $(M, g)$ is said to be \textbf{flat}\index{Manifold! flat} iff around any point $p \in M$ there exists a chart $(U, \phi)$ such that $g_{ab}|_U = diag(-1,...,-1,+1,...,+1)$.

Let $\boldsymbol{\xi}$ be a CKV for a flat metric whose locally pseudo-Euclidean components are $\eta_{ab}$. Then, $R_{ab} = 0$ and $R = 0$ which when replaced into the identities \eqref{eq.CKV2} and \eqref{eq.CKV3} give ($n \geq 3$) $\psi_{;ab} = 0$. Therefore, a flat metric admits SCKVs alone, i.e. it does not admit proper CKVs. After some standard calculations, we find the generic SCKV
\begin{equation} \label{eq.flatCKV1}
\xi_a = \alpha_a + \alpha_{ab} x^b + \beta x_a + 2 (\beta_b x^b) x_a - \beta_a (x_b x^b)
\end{equation}
with conformal factor $\psi = \beta + 2 \beta_a x^a$, where $F_{ab} = \xi_{[a;b]} = \alpha_{ab} - 2 \beta_{[a} x_{b]}$, $\xi_{(a;b)} = \psi \eta_{ab}$, and $\alpha, \beta, \alpha^a, \beta^a$, $\alpha_{ab} = - \alpha_{ba}$ are integration constants. Equation (\ref{eq.flatCKV1}) is written equivalently as
\[
\boldsymbol{\xi} = \xi^a \partial_a = \alpha^a \mathbf{P}_a + \alpha^{BA} \mathbf{r}_{AB}  + \beta \mathbf{H} + 2 \beta^a \mathbf{K}_a
\]
where the summation over $A$, $B$ satisfies the inequality $1 \leq A < B \leq n$, $\mathbf{P}_a = \delta^b_a \partial_b$ ($n$ gradient KVs - translations), $\mathbf{r}_{ab} = 2 \delta^c_{[a} \delta^d_{b]} x_c \partial_d$ ($\frac{n(n-1)}{2}$ non-gradient KVs - rotations), $\mathbf{H} = x^a \partial_a$ (1 HV - dilatation), and $\mathbf{K}_a = \left( x_a x^b - \frac{1}{2} \delta^b_a x_c x^c \right) \partial_b = x_a \mathbf{H} - \frac{1}{2} (x_b x^b) \mathbf{P}_a$ ($n$ SCKVs). The $\frac{(n+1)(n+2)}{2}$ CKVs $(\mathbf{P}_a, \mathbf{r}_{AB}, \mathbf{H}, \mathbf{K}_a)$ with non-vanishing conformal factors $\psi(\mathbf{H}) = 1$ and $\psi(\mathbf{K}_a) = x_a$ span the \textbf{conformal algebra} of the flat metric $\eta_{ab}$.\index{Algebra! conformal}

We have the following commutation relations (Lie brackets):
\[
[ \mathbf{P}_a, \mathbf{P}_b ] = 0, \enskip [ \mathbf{P}_a, \mathbf{H} ] = \mathbf{P}_a, \enskip [ \mathbf{P}_a, \mathbf{K}_b ] = \eta_{ab} \mathbf{H} - \mathbf{r}_{ab}, \enskip [ \mathbf{P}_a, \mathbf{r}_{bc} ] = \eta_{bca}{}{^d} \mathbf{P}_d,
\]
\[
[ \mathbf{H}, \mathbf{K}_a ] = \mathbf{K}_a, \enskip [ \mathbf{H}, \mathbf{r}_{ab} ] = 0, \enskip [ \mathbf{K}_a, \mathbf{K}_b ] = 0,
\]
\[
[ \mathbf{r}_{ab}, \mathbf{K}_c ] = - \eta_{abr}{}^s \left[ \left( x_c x^r - \frac{1}{2} \delta^r_c x_d x^d \right) \mathbf{P}_s - x^r \eta_{sc} \mathbf{H} + x^r \mathbf{r}_{sc} \right]
\]
and
\[
[\mathbf{r}_{ab}, \mathbf{r}_{cd}] = \eta_{ab}{}^{ij} \eta_{cd}{}^{rs} \left( x_i \eta_{rj} \mathbf{P}_s - x_r \eta_{si} \mathbf{P}_j \right)
\]
where $\eta_{abcd} \equiv \eta_{ac} \eta_{bd} - \eta_{ad} \eta_{bc}$ and $\delta^{cd}_{ab} = 2 \delta^c_{[a} \delta^d_{b]} = \eta_{ab}{}^{cd}$.

\begin{example}
\label{exa.CKVs.Minkowski} The 15-dimensional conformal algebra of {\bf{Minkowski spacetime}} ($n=4$), i.e. Lorentz metric $\eta_{ij} =$ $diag(-1,+1,+1,+1)$, is the following: \index{Spacetime! Minkowski}

(10 KVs)
\[
\mathbf{P}_0 = \partial_0, \enskip \mathbf{P}_{\mu} = \partial_{\mu}
\]
\[
\mathbf{r}_{0 \mu} = x_0 \partial_{\mu} - x_{\mu} \partial_0 = - x^{\mu} \partial_0 - x^0 \partial_{\mu}, \enskip \mathbf{r}_{\mu \nu} = x_{\mu} \partial_{\nu} - x_{\nu} \partial_{\mu} = x^{\mu} \partial_{\nu} - x^{\nu} \partial_{\mu}
\]
(1 HV)
\[
\mathbf{H} = x^i \partial_i, \enskip \psi(\mathbf{H}) = 1
\]
and (4 SCKVs)
\[
\mathbf{K}_0 = - \frac{1}{2} \left[ (x^0)^2 + x_{\mu} x^{\mu} \right] \partial_0 - x^0 x^{\mu} \partial_{\mu}
\]
\[
\mathbf{K}_{\mu} = x^0 x^{\mu} \partial_0 + x^{\mu} x^{\nu} \partial_{\nu} - \frac{1}{2} (x_i x^i) \partial_{\mu}
\]
with conformal factors $\psi(\mathbf{K}_0) = - x^0$ and $\psi(\mathbf{K}_{\mu}) = x^{\mu}$.
\end{example}

\begin{example}
\label{exa.CKVs.E3} The 10-dimensional conformal algebra of the Euclidean space $E^3$ consists of:

(6 KVs)
\[
\mathbf{P}_1 = \partial_x, \enskip \mathbf{P}_2 = \partial_y, \enskip \mathbf{P}_3 = \partial_z
\]
\[
\mathbf{r}_{12} = x \partial_y - y \partial_x, \enskip \mathbf{r}_{13} = x \partial_z - z \partial_x, \enskip \mathbf{r}_{23} = y \partial_z - z \partial_y
\]
(1 HV)
\[
\mathbf{H} = x \partial_x + y \partial_y + z \partial_z, \enskip \psi(\mathbf{H}) = 1
\]
and (3 SCKVs)
\[
\mathbf{K}_1 = \frac{1}{2} (x^2 - y^2 - z^2) \partial_x + xy \partial_y + xz \partial_z
\]
\[
\mathbf{K}_2 = xy \partial_x + \frac{1}{2} (y^2 - x^2 - z^2) \partial_y + yz \partial_z
\]
\[
\mathbf{K}_3 = xz \partial_x + yz \partial_y + \frac{1}{2} (z^2 - x^2 - y^2) \partial_z
\]
with conformal factors $\psi(\mathbf{K}_1) = x$, $\psi(\mathbf{K}_2) = y$ and $\psi(\mathbf{K}_3) = z$.
\end{example}

\section{The Lie derivative of a connection}

\label{sub.LieC}

A \textbf{connection} $\Gamma^i_{jk}$ is a GO on an $n$-dimensional smooth manifold $M$ which satisfies the transformation law\index{Connection}
\begin{equation} \label{eq.LieC.1}
\Gamma^{i'}_{j'k'} = J^{i'}_iJ^j_{j'}J^k_{k'} \Gamma^i_{jk} + J^{i'}_{i} J^i_{j',k'}
\end{equation}
where $J^{i'}_i = \frac{\partial x^{i'}}{\partial x^i}$ and $\{x^i\}$, $\{x^{i'}\}$ are local coordinate systems of a chart on $M$.

If $M$ is equipped with a metric $g_{ab}$ (i.e. $M$ is a Riemannian manifold), then the resulting \textbf{Christoffel symbols} \index{Christoffel symbols}
\begin{equation} \label{eq.LieC.2}
\{^a_{bc}\} = \frac{1}{2} g^{ad} \left( g_{bd,c} + g_{cd,b} - g_{bc,d} \right)
\end{equation}
are components of a symmetric connection called \textbf{Riemannian connection}.\index{Connection! Riemannian}

The Lie derivative of a connection $\Gamma^i_{jk}$ wrt a vector field {\boldmath{$\xi$}}$=\xi^i\partial_i$ is a tensor field of type $(1,2)$ with local components
\begin{equation} \label{eq.LieC.3}
L_{\boldsymbol{\xi}} \Gamma^i{}_{jk} = \Gamma^i{}_{jk,c} \xi^c - \xi^i{}_{,c} \Gamma^c{}_{jk} + \xi^c{}_{,j} \Gamma^i{}_{ck} + \xi^c{}_{,k} \Gamma^i{}_{jc} + \xi^i{}_{,jk}.
\end{equation}
For a Riemannian connection, we find that\footnote{Recall that a semicolon denotes the Riemannian covariant derivative.}
\begin{equation} \label{eq.LieC.4}
L_{\boldsymbol{\xi}} \{^i_{jk}\} = \frac{1}{2} g^{ir} \left[ \left(L_{\boldsymbol{\xi}} g_{jr}\right)_{;k} + \left(L_{\boldsymbol{\xi}} g_{kr}\right)_{;j} - \left(L_{\boldsymbol{\xi}} g_{jk}\right)_{;r} \right]
\end{equation}
which implies that if $\xi^i$ is a HV, then $L_{\boldsymbol{\xi}} \{^i_{jk}\}$ vanishes (as $g_{ab;c}=0$).

A useful identity that relates the curvature tensor $R^i{}_{jkr}$ with the Lie derivative of a symmetric connection, i.e. $\Gamma^i_{[jk]}=0$, is the following \cite{Yanob}:
\begin{equation} \label{eq.LieC.5}
L_{\boldsymbol{\xi}} \Gamma^i{}_{jk} = \xi^i{}_{|jk} - R^i{}_{jkr} \xi^r
\end{equation}
where $|$ denotes the covariant derivative wrt $\Gamma^i_{jk}$ and
\begin{equation} \label{eq.LieC.6}
R^i{}_{jkr} = \Gamma^i_{jr,k} - \Gamma^i_{jk,r} + \Gamma^i_{sk} \Gamma^s_{jr} - \Gamma^i_{sr} \Gamma^s_{jk}.
\end{equation}
Moreover, we compute
\begin{equation} \label{eq.LieD4}
L_{\boldsymbol{\xi}} R^i{}_{jkr} = \left( L_{\boldsymbol{\xi}} \Gamma^i_{jr} \right)_{|k} - \left( L_{\boldsymbol{\xi}} \Gamma^i_{jk} \right)_{|r} + 2 \Gamma^s_{[rk]} L_{\boldsymbol{\xi}} \Gamma^i_{js}
\end{equation}

In the case of a Riemannian connection, equation (\ref{eq.LieC.5}) reads
\begin{equation} \label{eq.LieC.7}
L_{\boldsymbol{\xi}} \{^i_{jk}\} = \xi^i{}_{;jk} - B^i{}_{jkr} \xi^r
\end{equation}
where
\begin{equation} \label{eq.LieC.6}
B^i{}_{jkr} = \{^i_{jr}\}_{,k} - \{^i_{jk}\}_{,r} + \{^i_{sk}\} \{^s_{jr}\} - \{^i_{sr}\} \{^s_{jk}\}
\end{equation}
is the \textbf{Riemannian curvature tensor (or Riemann tensor)}. \index{Tensor! Riemann}

For a symmetric connection $\Gamma^{i}_{jk}$ and a general tensor field of type $(r,s)$, we have the following identities:
\begin{equation} \label{eq.LieD1}
L_{\boldsymbol{\xi}} T^{i_1...i_r}_{j_1...j_s} = T^{i_1...i_r}_{j_1...j_s|k} \xi^k - T^{k...i_r}_{j_1...j_s} \xi^{i_1}_{|k} - ... + T^{i_1...i_r}_{j_1...k} \xi^{k}_{|j_s}.
\end{equation}
and
\begin{eqnarray}
L_{\boldsymbol{\xi}} \nabla_k T^{i_1...i_r}_{j_1...j_s} - \nabla_k L_{\boldsymbol{\xi}} T^{i_1...i_r}_{j_1...j_s} &=& L_{\boldsymbol{\xi}} \left( \Gamma^{i_1}_{\ell k} \right) T^{\ell i_2...i_r}_{j_1 j_2...j_s} + ... + L_{\boldsymbol{\xi}} \left( \Gamma^{i_r}_{\ell k} \right) T^{i_1...i_{r-1} \ell}_{j_1...j_{s-1} j_s} - \nonumber \\
&& - L_{\boldsymbol{\xi}} \left( \Gamma^{\ell}_{j_1 k} \right) T^{i_1 i_2...i_r}_{\ell j_2...j_s} - ... - L_{\boldsymbol{\xi}} \left( \Gamma^{\ell}_{j_s k} \right) T^{i_1...i_{r-1} i_r}_{j_1...j_{s-1} \ell}. \label{eq.LieD2}
\end{eqnarray}

Using the irreducible decomposition \eqref{eq.dec}, we obtain the following mathematical identities:
\begin{eqnarray}
L_{\boldsymbol{\xi}} \{^i_{jk}\} &=& 2 \delta^i_{(j} \psi_{,k)} - \psi^{,i} g_{jk} + L^i{}_{jk} \label{eq.dec1} \\
\left( L_{\boldsymbol{\xi}} \{^i_{jk}\} \right)_{;i} &=& 2 \psi_{;jk} - g_{jk} \Box \psi + L^i{}_{jk;i} \label{eq.dec2} \\
\left( L_{\boldsymbol{\xi}} \{^i_{ji}\} \right)_{;r} &=& n \psi_{;jr} \label{eq.dec3} \\
\left( L_{\boldsymbol{\xi}} g_{ij} \right)_{;k} &=& g_{ir} L_{\boldsymbol{\xi}} \{^r_{jk}\} + g_{jr} L_{\boldsymbol{\xi}} \{^r_{ik}\} \label{eq.dec4} \\
F_{ij;k} &=& B_{ijkr} \xi^r + g_{r[i} L_{\boldsymbol{\xi}} \{^r_{j]k}\} \label{eq.dec5}
\end{eqnarray}
where $L^i{}_{jk} \equiv g^{ir} \left( H_{jr;k} + H_{kr;j} - H_{jk;r} \right)$ and $L^i{}_{ji} = 0$.

\section{Collineations of the connection}

\label{sub.ACPC}

An \textbf{affine collineation (AC) or affine motion}\index{Motion! affine} wrt a connection $\Gamma^{i}_{jk}$ is a vector field $\xi^a$ such that $L_{\boldsymbol{\xi}} \Gamma^i_{jk} = 0$.\index{Collineation! affine} Using the results of sec. \ref{sub.LieC}, it can be shown that a vector field $\xi^a$ is an AC of a Riemannian connection iff $\psi_{;i}=0$ and $H_{ij;k}= 0$.

A \textbf{projective collineation (PC)}\index{Collineation! projective} is a vector field $\xi^a$ which satisfies the condition $L_{\boldsymbol{\xi}}\Gamma^i_{jk} = 2 \delta^i_{(j} \phi_{,k)}$, where $\phi(x)$ is the \textbf{projection function} \index{Function! projection} of $\xi^a$. A PC is called \textbf{special PC (SPC)}\index{Collineation! special projective} iff $\phi_{;ij} = 0$, that is, $\phi_{,i}$ is a gradient KV. Using the results of sec. \ref{sub.LieC}, we obtain the following proposition.

\begin{proposition} \label{pro.ACPC.2}
A vector field $\xi^a$ is a PC of a Riemannian connection iff
\begin{eqnarray}
\psi &=& \frac{n+1}{n} \phi + c, \enskip c=const \label{eq.PC.0a} \\
H_{ab;c} &=& \frac{n}{2(n+1)} \left( g_{ac}\psi_{;b} + g_{bc} \psi_{;a} - \frac{2}{n} g_{ab} \psi_{;c} \right). \label{eq.PC.0b}
\end{eqnarray}
\end{proposition}

For a PC of a Riemannian connection $\Gamma^{i}_{jk}$, the identity \eqref{eq.LieC.5} becomes
\begin{equation} \label{eq.PC.1}
\xi^i{}_{;jk} -2\delta^i_{(j} \phi_{,k)} = R^i{}_{jkr} \xi^r.
\end{equation}
When $\phi=0$, the PC becomes an AC. The condition which defines an AC is
\begin{equation} \label{eq.PC.2}
\xi^i{}_{;jk} - R^i{}_{jkr} \xi^r = 0.
\end{equation}

In the following, we restrict our discussion to Riemannian connections alone.

From \eqref{eq.dec1}, we deduce that the HVs (include KVs) are ACs. An AC that is not a HV is called a \textbf{proper AC}. \index{Collineation! proper affine}

PCs can be defined by the gradient KVs and the HV as follows:

\begin{proposition} \label{pro.ACPC.3}
If in a space there exist $m$ gradient KVs $S_{I,a}$, where
$I=1,2,...,m$ and $S_{I}(x)$ are functions, and the gradient HV $H_{,a}$ with homothetic factor $\psi=const$, then the vectors $S_{I}H_{,a}$ are non-gradient SPCs with projection function $\psi S_{I}$.
\end{proposition}

ACs can be defined by gradient  KVs as follows:

\begin{proposition} \label{pro.ACPC.4}
If in a space there exist $m$ gradient KVs $S_{I,a}$, where $I=1,2,...,m$, then one construct $m^{2}$ non-gradient ACs by the formula $S_{I}S_{J,a}$
\end{proposition}

\section{Higher order symmetries: Killing tensors}

\label{sub.KT.1}

A Killing tensor (KT)\index{Tensor! Killing} of order $m$ in an $n$-dimensional Riemannian manifold $(V^{n}, g_{ab})$ is a totally symmetric tensor\footnote{The independent components of a totally symmetric tensor of rank $m$ in an $n$-dimensional manifold are $\frac{(n+m-1)!}{m!(n-1)!}$. For $n=2$ we have $m+1$, and for $n=3$ we have $\frac{(m+1)(m+2)}{2}$.} $K_{a_1...a_m}$ of type $(0,m)$ defined by the requirement
\begin{equation} \label{eq.KT.1}
K_{(a_1...a_m;b)}=0
\end{equation}
or equivalently (due to the totally symmetry)
\begin{equation} \label{eq.KT.1b}
K_{\{r_{1}...r_{m};k\}}=0.
\end{equation}
For $m=1$, $K_a$ is a KV.

KTs are important in the reduction of Lagrangian systems because they generate gauged generalized Noether symmetries (see e.g. chapter \ref{ch.Higher.order.FIs}). These symmetries produce polynomial in velocities FIs, which are of the same order with the associated KTs. Because the generators of the resulting Noether symmetries depend on the velocities $\dot{q}^{a}$, they are often referred to the literature as \textbf{hidden (or higher order) symmetries}. \index{Symmetry! hidden}

We have the following result \cite{Thomas1946, Collinson1965, Kalnins 1980, Thompson 1984, Thompson 1986, Horwood 2008}:

\begin{proposition} \label{pro.KT.1}
On a general $n$-dimensional (pseudo-Riemannian) smooth manifold $V^{n}$, the (vector) space $\mathcal{K}^{(m,n)}$ of $m$th-order KTs has dimension
\begin{equation*}
\dim\left(\mathcal{K}^{(m,n)}\right) \leq \frac{(n+m-1)!(n+m)!}{(n-1)!n!m!(m+1)!}.
\end{equation*}
The equality is attained iff $V^{n}$ is of constant curvature (or maximally symmetric). In the case of spaces of constant, all KTs can be expressed as a sum of symmetrized tensor products of KVs.
\end{proposition}

From proposition \ref{pro.KT.1}, we deduce that in an $n$-dimensional manifold $V^{n}$ the number $N_{m}$ of independent KTs of order $m=2, 3, 4$ is
\begin{equation*}
N_{2} \leq \frac{n(n+1)^{2}(n+2)}{12}, \enskip N_{3} \leq \frac{n(n+1)^{2}(n+2)^{2}(n+3)}{3!4!}, \enskip N_{4} \leq \frac{n(n+1)^{2}(n+2)^{2}(n+3)^{2}(n+4)}{4!5!}.
\end{equation*}
In the case of $E^{2}$, we have $N_{2}=6$, $N_{3}=10$, $N_{4}=15$; and for $E^{3}$, we have $N_{2}=20$, $N_{3}=50$, $N_{4}=105$.

If a Riemannian manifold admits $n_{0}$ KVs (gradient and non-gradient) $X_{Ia}$, where $I=1,2,...,n_{0}$, then one constructs the generic KT of order $m$
\begin{equation}
K_{i_{1}...i_{m}}= \alpha^{I_{1}...I_{m}} X_{I_{1}(i_{1}} X_{|I_{2}|i_{2}}... X_{|I_{m}|i_{m})} \label{eq.syKT2}
\end{equation}
where $\alpha^{I_{1}...I_{m}}$ are constants and the Einstein summation convention follows the inequality $1\leq I_{1} \leq I_{2} \leq ... \leq I_{m}\leq n_{0}$. According to the proposition \ref{pro.KT.1} (see also \cite{Thompson 1984, Thompson 1986, Horwood 2008, Takeuchi 1983, Eastwood, Nikiting}) in spaces of constant curvature, that admit $n_{0}$ KVs, all KTs of order $m$ are of the form (\ref{eq.syKT2}). We note that, in general, not all the symmetrized products in (\ref{eq.syKT2}) are linearly independent, that is, the parameters $\alpha^{I_{1}...I_{m}}$ are not all independent. Specifically, the number of these parameters is always larger or equal ($n=2$) to the dimension of the associated KT space computed in proposition \ref{pro.KT.1} (see also a useful Remark below eq. (2.12) in \cite{Horwood 2008}).

From the AC condition (\ref{eq.PC.2}) and the property $R_{abcd}=-R_{bacd}$, it follows that
\begin{equation*}
\xi_{(a;bc)}=0 \implies \xi_{(a;b);c} + \xi_{(b;c);a} + \xi_{(c;a);b} = 0 \implies \xi_{\{(a;b);c\}}=0.
\end{equation*}
Therefore, \emph{an AC $\xi_{a}$ defines a KT of order two of the form $\xi_{(a;b)}$.} This implies that: a) The $m^2$ ACs $S_{I}S_{J,a}$ found in proposition \ref{pro.ACPC.4} define the $m^{2}$ KTs\footnote{Recall that $S_{I;ab}=0$.} $\left( S_{I} S_{J,(a} \right)_{;b)} = S_{I,(a} S_{|J|,b)}$, b) a HV $H_a$ defines the trivial KT $g_{ab}$, and c) the KVs define the zero KTs.

If a Riemannian space admits a CKV $Y^{a}$ and a PC $X^{a}$ such that $\psi(\mathbf{Y})= -2\phi(\mathbf{X})$, we construct the reducible KT of order two $C_{ab}=L_{(a;b)}$ where $L_{a}=Y_{a}+X_{a}.$

Besides these KTs, new KTs of order two are constructed as follows:

\begin{proposition} \label{pro.KT.2}
Consider $m$ gradient KVs $S_{I,a}$, where $I=1,...,m$, and $r$ non-gradient KVs $M_{Aa}$, where $A=1,...,r$. Then: 1) the vectors $\eta_{IAa} = S_{I}M_{Aa}$ define the $mr$ KTs $\eta_{IA(a;b)} = S_{I,(a} M_{|A|b)}$, and 2) the $r^{2}$ quantities $M_{A(a} M_{|B|b)}$ are KTs.
\end{proposition}

From proposition \ref{pro.KT.2}, we infer the following theorem:

\begin{theorem} \label{thm.KT.1}
If an $n$-dimensional space admits $m$ gradient KVs $S_{I,a}$ and $r$ non-gradient KVs $M_{Aa}$, then we can construct $m^{2}+mr+ r^{2}$ $= (m+r)^{2} -mr$ KTs of order two. Therefore, such spaces admit KTs of the form:
\begin{equation} \label{eq.KT.3a}
C_{ab} = \alpha^{IJ} S_{I,(a} S_{|J|,b)} + \beta^{IA} S_{I,(a} M_{|A|b)} + \gamma^{AB} M_{A(a} M_{|B|b)}
\end{equation}
or, equivalently,
\begin{equation} \label{eq.KT.3b}
C_{ab} = \alpha^{(IJ)} S_{I,a} S_{J,b} + \beta^{IA} S_{I,(a} M_{|A|b)} + \gamma^{(AB)} M_{Aa} M_{Bb}
\end{equation}
where $\alpha^{IJ}, \beta^{IA}, \gamma^{AB}$ are arbitrary real coefficients. By introducing the vector\footnote{
In the vector $L_{a}$ given by (\ref{eq.KT.3c}), AC is only the first part $S_{I}S_{J,a}$, whereas the second part $S_{I}M_{Aa}$ is not an AC because it does not satisfy the AC condition (\ref{eq.PC.2}).}
\begin{equation} \label{eq.KT.3c}
L_{a}= \alpha^{IJ} S_I S_{J,a} + \beta^{IA} S_I M_{Aa}
\end{equation}
the KTs (\ref{eq.KT.3a}) are written as
\begin{equation} \label{eq.KT.3d}
C_{ab} = L_{(a;b)} + \gamma^{AB} M_{A(a} M_{|B|b)}.
\end{equation}
In manifolds of constant curvature \cite{Thompson 1986} any KT is of the form $C_{ab}$, while KTs of the reducible form $L_{(a;b)}$ are generated by vectors $L_{a}$ given by (\ref{eq.KT.3c}).
\end{theorem}

KTs of order $m$ generated by totally symmetric tensors of order $m-1$ are called \textbf{reducible KTs}. For example, the KT (\ref{eq.KT.3d}) for $\gamma^{AB}=0$ is a reducible KT. \index{Tensor! reducible Killing}

For recent works on KTs and conformal KTs see \cite{Barnes 2003, Garfinkle 2010} and references cited therein.

\section{PCs and KTs of order two in maximally symmetric spaces}

\label{sec.detour.KTs.3}

The special projective Lie algebra of a maximally symmetric space, or a space of constant curvature, consists of the vector fields of Table \ref{Table.collineations} $(I,J=1,2,...,n)$.

\begin{longtable}{|l|l|l|}
\hline
\multicolumn{1}{|l|}{Collineation} & Gradient & Non-gradient \\ \hline
\multicolumn{1}{|l|}{Killing vectors (KVs)} & $\mathbf{K}_{I}= \delta_{I}^{i} \partial_{i}$ & $\mathbf{X}_{IJ}= \delta_{[I}^{j} \delta_{J]}^{i} x_{j}\partial _{i}$ \\ \hline
\multicolumn{1}{|l|}{Homothetic vector (HV)} & $\mathbf{H}=x^{i}\partial
_{i}$ &  \\ \hline
\multicolumn{1}{|l|}{Affine Collineations (ACs)} & $\mathbf{A}%
_{II}=x_{I}\delta _{I}^{i}\partial _{i}$ & $\mathbf{A}_{IJ}=x_{J}\delta
_{I}^{i}\partial _{i}$, $I\neq J$ \\ \hline
\multicolumn{1}{|l|}{Special Projective collineations (SPCs)} &  & $\mathbf{P}_{I}=x_{I}\mathbf{H}$ \\ \hline
\caption{\label{Table.collineations} Collineations of the Euclidean space $E^{n}$.}
\end{longtable}

Therefore, a maximally symmetric space of dimension $n$ admits \newline
- $n$ gradient KVs and $\frac{n(n-1)}{2}$ non-gradient KVs, \newline
- one gradient HV, \newline
- $n^{2}$ non-proper ACs, and \newline
- $n$ PCs which are special (i.e. the partial derivative of the projective function is a gradient KV).

\begin{proposition}
\label{prop1} In a space $V^{n}$, the vector fields of the form
\begin{equation}
L_{a}=c^{1I}S_{I,a}+c^{2A}M_{Aa}+c^{3}HV_{a}+ c_{4}AC_{a}+c^{5IJ}S_{I}S_{J,a}+2c^{6IA}S_{I}M_{Aa}+ c^{7K}(PC_{Ka} + CKV_{Ka})
\label{FL.20}
\end{equation}%
where $S_{I,a}$ are the gradient KVs, $M_{Aa}$ are the non-gradient KVs, $HV_{a}$ is the HV, $AC_{a}$ are the proper ACs (not generated by KVs), $S_{I}S_{J,a}$ are ACs, $PC_{Ka}$ are the proper PCs with a projective factor $\phi_{K}$ and $CKV_{Ka}$ are the CKVs with conformal factor $-2\phi_{K}$, produce the KTs of order two of the form $C_{ab}=L_{(a;b)}$. In the case of maximally symmetric spaces \cite{Barnes 1993}, equation (\ref{FL.20}) takes the form
\begin{equation}
L_{a}=c^{1I}S_{I,a}+c^{2K}M_{Ka}+c_{3}HV_{a}+c^{5IJ}S_{I}S_{J,a}+ 2c^{6IK}S_{I}M_{Ka}.
\label{FL.21}
\end{equation}%
The general KT of order two is given by the formula (\ref{eq.KT.3a}) and the reducible KTs $C_{ab}=L_{(a;b)}$ are given by (\ref{eq.KT.3a}) for $\gamma^{AB}=0$, where the vector $L_{a}$ is given by\footnote{The vectors $L_{a}$ of the form (\ref{eq.KT.3c}) can be called master symmetries. They can be defined covariantly via the Schouten bracket as $[g,[g,L]]$.} (\ref{eq.KT.3c}). The KVs alone give the solution $C_{ab}=0$ and the HV generates the trivial KT $g_{ab}$.
\end{proposition}

Applying Theorem \ref{thm.KT.1}, Proposition \ref{prop1}, and the general formula (\ref{eq.syKT2}) in the case of the Euclidean plane $E^{2}$ and the Euclidean space $E^{3}$, we find the following results (see sections below).

\section{The geometric quantities of $E^{2}$}

\label{sec.KTE2}

$E^{2}$ admits two gradient KVs $\partial_{x}, \partial_{y}$ whose generating functions are $x, y$, respectively, and one
non-gradient KV (the rotation) $y\partial _{x}-x\partial _{y}$. These vectors are written collectively as
\begin{equation}
L_{a}=\left(
\begin{array}{c}
b_{1}+b_{3}y \\
b_{2}-b_{3}x%
\end{array}%
\right)  \label{FL.15}
\end{equation}%
where $b_{1}, b_{2}, b_{3}$ are arbitrary constants, possibly zero.

The symmetrized tensor products of the KVs produce the KTs of various orders in $E^{2}$.

\subsection{KTs of order two in $E^{2}$}

\label{sec.KTE2.1}

- The general KT of order two in $E^{2}$ is \cite{Chanu 2006, Adlam}
\begin{equation}
C_{ab}=\left(
\begin{array}{cc}
\gamma y^{2}+2\alpha y+A & -\gamma xy-\alpha x-\beta y+C \\
-\gamma xy-\alpha x-\beta y+C & \gamma x^{2}+2\beta x+B%
\end{array}%
\right)  \label{FL.14b}
\end{equation}
where $\alpha, \beta, \gamma, A, B, C$ are arbitrary constants.

- The vector $L_{a}$ generating KTs in $E^{2}$ of the form $%
C_{ab}=L_{(a;b)}$ is\footnote{%
Note that $L_{a}$ in (\ref{FL.14}) is the sum of the non-proper ACs of $%
E^{2} $ and not of its KVs which give $C_{ab}=0.$}
\begin{equation}
L_{a}=\left(
\begin{array}{c}
-2\beta y^{2}+2\alpha xy+Ax+(2C-a_{1})y+a_{2} \\
-2\alpha x^{2}+2\beta xy+a_{1}x+By+a_{3}%
\end{array}%
\right)  \label{FL.14}
\end{equation}
where $a_{1}, a_{2}, a_{3}$ are arbitrary constants.

- The KTs $C_{ab}=L_{(a;b)}$ in $E^{2}$ generated from the vector (\ref{FL.14}) are
\begin{equation}
C_{ab}=L_{(a;b)}=\left(
\begin{array}{cc}
L_{x,x} & \frac{1}{2}(L_{x,y}+L_{y,x}) \\
\frac{1}{2}(L_{x,y}+L_{y,x}) & L_{y,y}%
\end{array}%
\right) =\left(
\begin{array}{cc}
2\alpha y+A & -\alpha x-\beta y+C \\
-\alpha x-\beta y+C & 2\beta x+B%
\end{array}%
\right).  \label{FL.14.1}
\end{equation}
Observe that these KTs are special cases of the general KTs (\ref{FL.14b}) for $\gamma =0$.

We note that the vector $L_{a}$ given by (\ref{FL.14}) depends on eight parameters, while the generated KT $L_{(a;b)}$ depends on five of them the $\alpha, \beta, A, B, C$. This is because the remaining $8-5=3$ parameters $a_{1}, a_{2}, a_{3}$ of the vector $L_{a}$ generate the KVs in $E^{2}$, which generate the zero KTs.

\subsection{KTs of order three in $E^{2}$}

\label{sec.KTE2.2}

- The general KT $C_{abc}$ of order three in $E^{2}$ has independent components \cite{McLenaghan2004, HorwoodMc}
\begin{eqnarray}
C_{111}&=& a_{1}y^{3} +3a_{2}y^{2} + 3a_{3}y +a_{4}  \notag \\
C_{112}&=& -a_{1}xy^{2} -2a_{2}xy +a_{5}y^{2} -a_{3}x +a_{8}y +a_{9}  \notag
\\
C_{221}&=& a_{1}x^{2}y +a_{2}x^{2} -2a_{5}xy -a_{8}x -a_{6}y +a_{10}
\label{eq.KT1} \\
C_{222}&=&-a_{1}x^{3} +3a_{5}x^{2} +3a_{6}x +a_{7}  \notag
\end{eqnarray}
where $a_{K}$ with $K=1,2,...,10$ are arbitrary constants\footnote{These are the eqs. (3.3.11) - (3.3.14) found in \cite{Hietarinta 1987}. In the notation of \cite{Hietarinta 1987}, $A=C_{111}$, $B=3C_{112}$, $C=3C_{221}$ and $D=C_{222}$. The extra factor $3$ arises from the fact that the author in \cite{Hietarinta 1987} uses algebraic methods and not the techniques of differential geometry.}.

- The reducible KT $C_{abc}= L_{(ab;c)}$ of order three in $E^{2}$ is generated by the symmetric tensor
\begin{eqnarray}
L_{11}&=& 3b_{2}xy^{2} +3b_{5}y^{3} +3b_{3}xy +3(b_{10}+b_{8})y^{2} +b_{4}x
+3b_{15}y +b_{12}  \notag \\
L_{12}&=& -3b_{2}x^{2}y -3b_{5}xy^{2} -\frac{3}{2}b_{3}x^{2} -\frac{3}{2}%
(2b_{10}+b_{8})xy -\frac{3}{2}b_{6}y^2 +\frac{3}{2}(b_{9} -b_{15})x -\frac{3%
}{2}b_{11}y +b_{13}  \label{eq.KT2} \\
L_{22}&=& 3b_{2}x^{3} +3b_{5}x^{2}y +3b_{10}x^{2} +3b_{6}xy +3(b_{1}
+b_{11})x +b_{7}y +b_{14}  \notag
\end{eqnarray}
where $b_{1}, b_{2}, ..., b_{15}$ are arbitrary constants.

- The independent components of the generated KT $L_{(ab;c)}$ are
\begin{eqnarray}
L_{(11;1)}&=& 3b_{2}y^{2} +3b_{3}y +b_{4}  \notag \\
L_{(11;2)}&=& -2b_{2}xy +b_{5}y^{2} -b_{3}x +b_{8}y +b_{9}  \notag \\
L_{(22;1)}&=& b_{2}x^{2}-2b_{5}xy -b_{8}x -b_{6}y +b_{1}  \label{eq.KT3} \\
L_{(22;2)}&=& 3b_{5}x^{2} +3b_{6}x +b_{7}.  \notag
\end{eqnarray}
We note that the KT (\ref{eq.KT3}) is just a subcase of the general KT (\ref{eq.KT1}) for $a_{1}=0$.

Furthermore, we see that from the fifteen parameters of the symmetric tensor $L_{ab}$ given by (\ref{eq.KT2}), the nine first parameters $b_{1}, b_{2}, ..., b_{9}$ generate the reducible KT $L_{(ab;c)}$; while the remaining six parameters $b_{10}, b_{11}, ..., b_{15}$ generate all the second order KTs of the general form (\ref{FL.14b}), where $\alpha=\frac{3}{2}b_{15}$, $\beta= \frac{3}{2}b_{11}$, $\gamma=3b_{10}$, $A=b_{12}$, $B=b_{14}$ and $C=b_{13}$.

\subsection{KTs of order four in $E^{2}$}

\label{sec.KTE2.3}

- The general KT $C_{abcd}$ of order four in $E^{2}$ has independent components
\begin{eqnarray}
C_{1111} &=& a_{1}y^{4} + a_{2}y^{3} +a_{3}y^{2} +a_{4}y +a_{5} \notag \\
C_{1112} &=& -a_{1}xy^{3} -\frac{3}{4}a_{2}xy^{2} -\frac{a_{6}}{4}y^{3} -\frac{a_{3}}{2}xy +a_{10}y^{2} -\frac{a_{4}}{4}x +\frac{3}{2}a_{11}y +a_{12} \notag \\
C_{1122} &=& a_{1}x^{2}y^{2} +\frac{a_{2}}{2}x^{2}y +\frac{a_{6}}{2}xy^{2} +\frac{a_{3}}{6}x^{2} -\frac{4}{3}a_{10}xy +\frac{a_{7}}{6}y^{2} -a_{11}x -a_{13}y +a_{15} \label{eq.KT4} \\
C_{1222} &=& -a_{1}x^{3}y -\frac{a_{2}}{4}x^{3} -\frac{3}{4}a_{6}x^{2}y +a_{10}x^{2} -\frac{a_{7}}{2}xy +\frac{3}{2}a_{13}x -\frac{a_{8}}{4}y +a_{14} \notag \\
C_{2222} &=& a_{1}x^{4} +a_{6}x^{3} +a_{7}x^{2} +a_{8}x +a_{9} \notag
\end{eqnarray}
where $a_{1}, a_{2}, ..., a_{15}$ are arbitrary constants\footnote{The above equations coincide with the set of eqs. (3.4.5) of \cite{Hietarinta 1987}, where $f_{0}=C_{1111}$, $f_{1}=4C_{1112}$, $f_{2}= 6C_{1122}$, $f_{3}=4C_{1222}$ and $f_{4}=C_{2222}$. The extra factors 4 and 6 arise from the fact that the author in \cite{Hietarinta 1987} uses algebraic methods instead of geometric. We note also that in the expression for $f_{4}$ in eq. (3.4.5) there is an unnecessary 2 in the term of $x^{2}$, the correct eq. is $f_{4}= ax^{4} -kx^{3} +qx^{2} -tx +w$.}.

- The reducible KT $C_{abcd}= L_{(abc;d)}$ of order four in $E^{2}$ is generated by the totally symmetric tensor $L_{abc}$ with components
\begin{eqnarray}
L_{111}&=& b_{2}xy^{3} -b_{6}y^{4} +b_{3}xy^{2} +4b_{22}y^{3} +b_{4}xy +12b_{24}y^{2} +b_{5}x +24b_{15}y +b_{16} \notag \\
L_{112}&=& -b_{2}x^{2}y^{2} +b_{6}xy^{3} -\frac{2}{3}b_{3}x^{2}y +\frac{4}{3} (b_{10} -3b_{22})xy^{2} +\frac{b_{7}}{3}y^{3} -\frac{b_{4}}{3}x^{2} +2(b_{11}-4b_{24})xy + \notag \\
&& +2(2b_{20} -b_{13})y^{2} +4\left( \frac{b_{12}}{3} -2b_{15} \right)x +24b_{18}y +b_{19} \notag \\
L_{221}&=& b_{2}x^{3}y -b_{6}x^{2}y^{2} +\frac{b_{3}}{3}x^{3} -\frac{4}{3}(2b_{10}-3b_{22})x^{2}y -\frac{2}{3}b_{7}xy^{2} -2(b_{11}-2b_{24})x^{2} - \label{eq.KT5}\\
&& -2(4b_{20} -b_{13})xy -\frac{b_{8}}{3}y^{2} +2(b_{1}-12b_{18})x +4\left( \frac{b_{14}}{3} -2b_{21} \right)y +b_{23} \notag \\
L_{222}&=& -b_{2}x^{4} +b_{6}x^{3}y +4(b_{10}-b_{22})x^{3} +b_{7}x^{2}y +12b_{20}x^{2} +b_{8}xy +24b_{21}x +b_{9}y +b_{17} \notag
\end{eqnarray}
where $b_{1}, b_{2}, ..., b_{24}$ are arbitrary constants.

- The independent components of the generated KT $L_{(abc;d)}$ are
\begin{eqnarray}
L_{(111;1)}&=& b_{2}y^{3} +b_{3}y^{2} +b_{4}y +b_{5} \notag \\
L_{(111;2)}&=& -\frac{3}{4}b_{2}xy^{2} -\frac{b_{6}}{4}y^{3} -\frac{b_{3}}{2}xy +b_{10}y^{2} -\frac{b_{4}}{4}x +\frac{3}{2}b_{11}y +b_{12} \notag \\
L_{(112;2)}&=& \frac{b_{2}}{2}x^{2}y +\frac{b_{6}}{2}xy^{2} +\frac{b_{3}}{6}x^{2} -\frac{4}{3}b_{10}xy +\frac{b_{7}}{6}y^{2} -b_{11}x -b_{13}y +b_{1} \label{eq.KT6} \\
L_{(122;2)}&=& -\frac{b_{2}}{4}x^{3} -\frac{3}{4}b_{6}x^{2}y +b_{10}x^{2} -\frac{b_{7}}{2}xy +\frac{3}{2}b_{13}x -\frac{b_{8}}{4}y +b_{14} \notag \\
L_{(222;2)}&=& b_{6}x^{3} +b_{7}x^{2} +b_{8}x +b_{9}. \notag
\end{eqnarray}
This is a subcase of the general fourth order KT (\ref{eq.KT4}) for $a_{1}=0$.

We note that from the twenty four parameters defining the totally symmetric tensor $L_{abc}$ given by (\ref{eq.KT5}), only fourteen, the $b_{1}, b_{2}, ..., b_{14}$, generate the reducible KT $L_{(abc;d)}$. The remaining ten parameters $b_{15}, b_{16}, ..., b_{24}$ produce the general third-order KT (\ref{eq.KT1}).

\section{The geometric quantities of $E^{3}$}

\label{sec.KTE3}

- $E^{3}$ admits three gradient KVs $\partial_{x}, \partial_{y},\partial_{z}$ whose generating functions are $x, y, z$, respectively, and three
non-gradient KVs $y\partial _{x}-x\partial y$, $z\partial_{y}-y%
\partial_{z}$, $z\partial_{x} - x\partial _{z}$. These vectors are written collectively as
\begin{equation}
L_{a}=
\left(
  \begin{array}{c}
    b_{1} -b_{4}y +b_{5}z  \\
    b_{2} +b_{4}x -b_{6}z \\
    b_{3} -b_{5}x +b_{6}y \\
  \end{array}
\right) \label{eq.e3.2}
\end{equation}
where $b_{1}, b_{2}, ..., b_{6}$ are arbitrary constants.

- The general KT of order two in $E^{3}$ has independent components%
\begin{eqnarray}
C_{11} &=&\frac{a_{6}}{2}y^{2}+\frac{a_{1}}{2}%
z^{2}+a_{4}yz+a_{5}y+a_{2}z+a_{3}  \notag \\
C_{12} &=&\frac{a_{10}}{2}z^{2}-\frac{a_{6}}{2}xy-\frac{a_{4}}{2}xz-\frac{%
a_{14}}{2}yz-\frac{a_{5}}{2}x-\frac{a_{15}}{2}y+a_{16}z+a_{17}  \notag \\
C_{13} &=&\frac{a_{14}}{2}y^{2}-\frac{a_{4}}{2}xy-\frac{a_{1}}{2}xz -\frac{%
a_{10}}{2}yz-\frac{a_{2}}{2}x+a_{18}y-\frac{a_{11}}{2}z+a_{19}  \label{FL.E3}
\\
C_{22} &=&\frac{a_{6}}{2}x^{2}+\frac{a_{7}}{2}%
z^{2}+a_{14}xz+a_{15}x+a_{12}z+a_{13}  \notag \\
C_{23} &=&\frac{a_{4}}{2}x^{2}-\frac{a_{14}}{2}xy-\frac{a_{10}}{2}xz -\frac{%
a_{7}}{2}yz-(a_{16}+a_{18})x-\frac{a_{12}}{2}y-\frac{a_{8}}{2}z +a_{20}
\notag \\
C_{33} &=&\frac{a_{1}}{2}x^{2}+\frac{a_{7}}{2}%
y^{2}+a_{10}xy+a_{11}x+a_{8}y+a_{9}  \notag
\end{eqnarray}%
where $a_{I}$ with $I=1,2,...,20$ are arbitrary real constants.

- The vector $L_{a}$ generating the reducible KT $C_{ab}=L_{(a;b)}$ is
\begin{equation}
L_{a}=\left(
\begin{array}{c}
-a_{15}y^{2}-a_{11}z^{2}+a_{5}xy+a_{2}xz+2(a_{16}+a_{18})yz+a_{3}x +2a_{4}y+2a_{1}z+a_{6}
\\
-a_{5}x^{2}-a_{8}z^{2}+a_{15}xy-2a_{18}xz+a_{12}yz+ 2(a_{17}-a_{4})x+a_{13}y+2a_{7}z+a_{14}
\\
-a_{2}x^{2}-a_{12}y^{2}-2a_{16}xy+a_{11}xz+a_{8}yz+2(a_{19}- a_{1})x+2(a_{20}-a_{7})y+a_{9}z+a_{10}%
\end{array}%
\right) \label{eq.Kep.5}
\end{equation}
and the generated KT has independent components
\begin{eqnarray*}
C_{11}&=& a_{5}y+a_{2}z+a_{3} \notag \\
C_{12}&=& -\frac{a_{5}}{2}x-\frac{a_{15}}{2}y +a_{16}z+a_{17} \notag \\
C_{13}&=& -\frac{a_{2}}{2}x+a_{18}y-\frac{a_{11}}{2}z+a_{19} \label{eq.Kep.8} \\
C_{22}&=& a_{15}x+a_{12}z+a_{13} \notag \\
C_{23}&=& -(a_{16}+a_{18})x-\frac{a_{12}}{2}y -\frac{a_{8}}{2}z +a_{20} \notag \\
C_{33}&=& a_{11}x+a_{8}y+a_{9}. \notag \\
\end{eqnarray*}
The last KT is a subcase of the general KT (\ref{FL.E3}) for $a_{1}= a_{4} =a_{6} =a_{7} =a_{10} =a_{14}=0$.

Working in the same way, we can compute the KTs in a space of constant curvature of a larger dimension.

We note that the covariant expression of the most general KT $\Lambda _{ij}$ of order two in $E^{3}$ is \cite{Chanu 2006, Crampin 1984}
\begin{equation}
\Lambda _{ij}=(\varepsilon _{ikm}\varepsilon _{jln}+\varepsilon
_{jkm}\varepsilon _{iln})A^{mn}q^{k}q^{l}+(B_{(i}^{l}\varepsilon
_{j)kl}+\lambda _{(i}\delta _{j)k}-\delta _{ij}\lambda _{k})q^{k}+D_{ij}
\label{CRA.46}
\end{equation}%
where $A^{mn}, B_{i}^{l}, D_{ij}$ are constant symmetric tensors, $B_{i}^{l}$ is also traceless, $\lambda ^{k}$ is a constant vector, and $\varepsilon_{ijk}$ is the 3d Levi-Civita symbol.

Observe that $A^{mn}$ and $D_{ij}$ have each six independent components, $%
B_{i}^{l}$ has five independent components, and $\lambda ^{k}$ has three
independent components. Therefore, $\Lambda _{ij}$ depends on $6+6+5+3=20$
arbitrary real constants, a result  in accordance with the general KT (\ref{FL.E3}).

%% file: integrability.tex
\chapter{Integrability and first integrals}

\label{ch.integrability}

The precise meaning of the solution of a system of differential equations can be cast in several ways \cite{anleach}. We say that we have
determined a closed-form solution for a dynamical system, when we have
determined a set of explicit functions describing the variation of the
dependent variables in terms of the independent variable(s). On the other hand,
when we have proved the existence of a sufficient number of independent
explicit FIs and invariants for the dynamical system, we
say that we have found an analytic solution of the dynamical equations.
In addition, an algebraic solution is found when one has proved the
existence of a sufficient number of explicit transformations, which permit
the reduction of the system of differential equations to a system of algebraic equations. A feature, that is central to each of these three equivalent
prescriptions of integrability, is the existence of explicit functions which are FIs, or the coefficient functions of
the aforementioned transformations. In this chapter, we shall concentrate on FIs.

\section{About first integrals}

\label{sec.int.FI.1}

Consider a second order set of dynamical equations of the form
\begin{equation}
\ddot{q}^{a}= \omega^{a}(t,q,\dot{q}) \iff H^{a}\equiv  \ddot{q}^{a} -\omega^{a}= 0 \label{eq.intf.1}
\end{equation}
where $a=1,2,...,n$, $q^{a}(t)$ are the generalized coordinates and $n$ are the \textbf{degrees of freedom}\index{Degrees of freedom} of the system. A \textbf{first integral (FI) or constant of motion}\index{First integral} is a function $I(t,q,\dot{q})$ such that
\begin{equation}
\left. \frac{dI}{dt} \right|_{q^{a}: H^{a}=0} =0 \iff \mathbf{\Gamma}(I)=0 \label{eq.intf.2}
\end{equation}
where $\mathbf{\Gamma}$ is the associated Hamiltonian vector field (see eq. (\ref{FI.3}) ) defined by (\ref{eq.intf.1}). The FI condition (\ref{eq.intf.2}) implies that $I$ is constant along \textbf{solutions (or trajectories)}\index{Trajectory} $q^{a}(t)$ of (\ref{eq.intf.1}). We note that, in general, by the term \textbf{dynamical system}\index{Dynamical system} we refer to a set of differential equations.

FIs are important because they are used to reduce the order of the dynamical equations and, if there are `enough' of them, to determine the solution of the system by means of quadratures. In the latter case, the dynamical system is called \textbf{integrable};\index{Dynamical system! integrable} however, most of the actual problems (e.g. the $n$-body problem) cannot be integrated and they are called \textbf{non-integrable}.

\section{Hamiltonian systems}

\label{sec.int.FI.2}

Consider an arbitrary Lagrangian system $L(t,q,\dot{q})$. From the Lagrangian $L$, we introduce the \textbf{conjugate momenta} \index{Momentum! conjugate}
\begin{equation}
p_{a}(t,q,\dot{q})\equiv \frac{\partial L}{\partial \dot{q}^{a}}. \label{eq.intf.3}
\end{equation}
For example, in the case of autonomous conservative systems, we have the regular (i.e. $\det\left[ \frac{\partial^{2}L}{\partial \dot{q}^{a} \partial\dot{q}^{b}} \right] =\det[\gamma_{ab}] \neq0$) Lagrangian: \index{Lagrangian! regular}
\begin{equation}
L(q,\dot{q}) = \frac{1}{2} \gamma_{ab}(q) \dot{q}^{a} \dot{q}^{b} - V(q) \label{eq.Ham.1}
\end{equation}
which when replaced into (\ref{eq.intf.3}) gives
\begin{equation}
p_{a}= \gamma_{ab}\dot{q}^{b} \implies \dot{q}^{a} = \gamma^{ab} p_{b} = p^{a}. \label{eq.Ham.2}
\end{equation}
If the kinetic metric $\gamma_{ab}=\delta_{ab}$ (i.e. Euclidean metric), then $p_{a}= \dot{q}^{a}$ and we can work with generalized velocities instead of generalized momenta.

Using the Legendre transformation, we define the \textbf{Hamiltonian}\index{Hamiltonian} \index{Transformation! Legendre}
\begin{equation}
H(t,q,p) = \dot{q}^{a}(t,q,p) p_{a} -L. \label{eq.intf.4}
\end{equation}
From (\ref{eq.intf.4}), we see that the Hamiltonian of the system is a function of $t, q^{a}, p_{a}$, provided that the transformation (\ref{eq.intf.3}) is invertible (i.e. nonzero Jacobian $\det \left[ \frac{\partial p_{a}}{\partial \dot{q}^{b}} \right]\neq0$) wrt the generalized velocities as $\dot{q}^{a}= \dot{q}^{a}(t,q,p)$. The last inversion requires a regular Lagrangian.

Replacing the Lagrangian (\ref{eq.Ham.1}) and the transformation (\ref{eq.Ham.2}) into (\ref{eq.intf.4}), we find the Hamiltonian
\begin{equation}
H = \frac{1}{2} \gamma^{ab} p_{a} p_{b} + V(q). \label{eq.Ham.3}
\end{equation}

It can be proved that the equations of motion for a Hamiltonian system are the following:
\begin{equation}
\dot{q}^{a} = \frac{\partial H}{\partial p_{a}}, \enskip \dot{p}_{a} = -\frac{\partial H}{\partial q^{a}}. \label{eq.Ham.M}
\end{equation}
These equations are the well-known \textbf{Hamilton's equations} and they are equivalent to the E-L equations.\index{Hamilton's equations} If there exist generalized non-conservative forces $Q_{a}$, then Hamilton's equations become
\begin{equation}  \label{eq.Ham.Mb}
\dot{q}^{a} = \frac{\partial H}{\partial p_{a}}, \enskip \dot{p}_{a} = Q_{a} - \frac{\partial H}{\partial q^{a}}.
\end{equation}

Hamilton's equations (\ref{eq.Ham.M}) are a set of $2n$ first order ODEs with variables $t, q^{a}, p_{a}$. From the defining relation (\ref{eq.intf.2}), we deduce that a FI of the dynamical equations (\ref{eq.Ham.M}) is a function $I(t,q,p)$ such that
\begin{equation}
\mathbf{\Gamma}(I)=0 \iff \frac{\partial I}{\partial t} +\frac{\partial I}{\partial q^{a}}\frac{\partial H}{\partial p_{a}} - \frac{\partial I}{\partial p_{a}} \frac{\partial H}{\partial q^{a}}= \frac{\partial I}{\partial t} +\{ I, H \} =0 \iff \{H,I\}= \frac{\partial I}{\partial t} \label{eq.intf.5}
\end{equation}
where $\{.,.\}$ denotes the Poisson bracket (PB) and $\Gamma$ is the associated Hamiltonian vector field of the system.

If the hamiltonian $H$ of a dynamical system is independent of the $k$th-generalized position $q^{k}$, that is, $\frac{\partial H}{\partial q^{k}}=0$, then Hamilton's equations imply that $p_{k}=const$. Therefore, $p_{k}$ is a FI of the system. The coordinate $q^{k}$ is called a \textbf{cyclic (or an ignorable) coordinate}\index{Coordinate! cyclic} of the system.

Moreover, a coordinate transformation $(t,q,p) \rightleftarrows (t,Q,P)$ that preserves the form of Hamilton's equations is called \textbf{canonical transformation}.\index{Transformation! canonical} A canonical transformation that is time-independent, i.e. $(q,p) \rightleftarrows (Q,P)$, is called restricted. The ultimate goal of Classical Mechanics is to construct a canonical transformation such that the new generalized coordinates $Q^{a}$ are all cyclic coordinates of the system. Then, the solution of Hamilton's equation is $P_{a}=c_{a}$ and $Q^{a}= d^{a}t +Q^{a}(0)$, where $c_{a}$ and $d^{a}$ are arbitrary constants. These special type of canonical coordinates are known in the literature as \textbf{action-angle variables}. \index{Variables! action-angle}

\section{Liouville integrability}

\label{sec.int.FI.4}

It is well-known \cite{Arnold 1989} that in the special case of autonomous Hamiltonian systems $H(q,p)$ with $n$ degrees of freedom, the definition for integrable systems given in sec. \ref{sec.int.FI.1} becomes more specific. Indeed, such systems are called \textbf{(Liouville) integrable}\index{Dynamical system! Liouville integrable} if they admit $n$ (functionally) independent\footnote{A set of FIs is said to be functionally independent if their gradient vectors (or equivalently the 1-forms defined by the FIs) $\mathbf{\nabla} I$ over the phase space $q^{a},p_{a}$ are linearly independent.} autonomous FIs $I_{a}(q,p)$ which are in involution\footnote{We note that there can be at most $n$ independent FIs in involution.} (i.e. $\{I_{a}, I_{b}\}=0$ for all indices).\index{First integral! autonomous}

The last definition is carried over \cite{Kozlov 1983, Vozmishcheva 2005} for non-autonomous Hamiltonian systems $H(t,q,p)$ and general time-dependent FIs $I(t,q,p)$. This means that time-dependent FIs can be used to establish the integrability of a dynamical system.\index{First integral! time-dependent} In the case of non-autonomous Hamiltonian systems, the Hamiltonian is not a FI.

If there exist $2n-1$ independent FIs, an integrable Hamiltonian system $H(q,p,t)$ is called \index{Dynamical system! superintegrable} \textbf{(maximally) superintegrable}. If there are $k$ independent FIs such that $n < k < 2n -1$, the system is called \textbf{minimally superintegrable}. The maximum number of independent FIs is $2n-1$ only when the considered FIs are autonomous. If time-dependent FIs are used, this maximum limit can be exceeded.

A general first order autonomous system $\dot{x}_{i} =F_{i}(x)$, where $i=1,...,n$ and $F_{i}$ are arbitrary smooth functions of the variables $x_{i}$, is always integrable if there exist $n-1$ independent FIs \cite{Yoshida II}. However, the existence of fewer FIs may also be sufficient since in the case of Hamiltonian systems, where $n=2m$, $m$ independent FIs in involution are enough for establishing (Liouville) integrability.

The motion of an integrable autonomous Hamiltonian system takes place on the $n$-dimensional surface $I_{a}(q, p)= c_{a}$ defined by the FIs $I_{a}$, where $c_{a}$ are arbitrary constants. Using Lie's theory, it is proved \cite{Arnold 1989} that this surface is diffeomorphic to an $n$-torus $T^{n}= \underbrace{S^{1} \times ... \times S^{1}}_{n}$, where $S^{1}$ denotes a circle. Then, one can define a canonical transformation $(q^{a}, p_{a}) \rightleftarrows (\phi^{a}, J_{a})$ such that the \textbf{angles} $0 \leq \phi^{a} \leq 2\pi$ are coordinates on $T^{n}$ and the \textbf{actions}\footnote{If we set $J_{a}=I_{a}$, then $(q^{a}, p_{a}) \rightleftarrows (\phi^{a}, I_{a})$ is not a canonical transformation in general.} $J_{a}= J_{a}(I_{1}, ..., I_{n})$ are FIs\footnote{Note that if $I_{1},I_{2},...,I_{k}$ are FIs of a given dynamical system, then any function $f(I_{1},...,I_{k})$ is also a FI of the system.}. These are the \textbf{action-angle coordinates} of the system. In these coordinates, Hamilton's equations (\ref{eq.Ham.M}) are written as follows:\index{Coordinates! action-angle}
\begin{eqnarray}
\dot{J}_{a}&=& 0 \implies J_{a}=const= m_{a}, \enskip \frac{\partial H}{\partial \phi^{a}}=0 \label{eq.holt13a} \\
\dot{\phi}^{a}&=& \frac{\partial H}{\partial J_{a}} \equiv \omega^{a}(J)=const \implies \phi^{a}(t)= \omega^{a}t +\phi^{a}(0). \label{eq.holt13b}
\end{eqnarray}
Equation (\ref{eq.holt13a}) implies that the Hamiltonian $H= H(J) \equiv H(J_{1}, ..., J_{n})$; therefore, the angles $\phi^{a}$ are cyclic coordinates. The solution $\phi^{a}(t)= \omega^{a}t +\phi^{a}(0)$ describes a \textbf{conditionally-periodic} motion on $T^{n}$ with frequencies the constants $\omega^{a}$.

\section{First integrals and weak Noether symmetries}

\label{sec.int.FI.3}

Consider holonomic dynamical systems with a regular Lagrangian $L= L(t,q,\dot{q})$ and generalized non-conservative forces $F^{a}(t,q,\dot{q})$. The equations of motion for such systems are given by the E-L equations
\begin{equation}
E_{a}(L)\equiv \frac{d}{dt}\left( \frac{\partial L}{\partial \dot{q}^{a}}\right) -\frac{\partial L}{\partial q^{a}} = F_{a}. \label{eq.inv1}
\end{equation}
Because the Lagrangian is considered to be \textbf{regular},\index{Lagrangian! regular} it holds that $\det\left[\gamma_{ab}\right]\neq0$ where $\gamma_{ab}\equiv \frac{\partial^{2}L}{\partial \dot{q}^{a} \partial \dot{q}^{b}}$ is the kinetic metric of the system. This implies that the kinetic metric is invertible (i.e.  $\gamma^{ac}\gamma_{cb}= \delta^{a}_{b}$) and, therefore, eq. (\ref{eq.inv1}) takes the form
\begin{equation}
\ddot{q}^{a} = \gamma^{ab} \left( F_{b} + \frac{\partial L}{\partial q^{b}} - \frac{\partial^{2} L}{\partial \dot{q}^{b}\partial t} - \frac{\partial^{2} L}{\partial \dot{q}^{b} \partial q^{c}} \dot{q}^{c} \right). \label{eq.inv2}
\end{equation}

For such dynamical systems, a vector field
\begin{equation}
\mathbf{X}= \xi(t,q,\dot{q}) \frac{\partial}{\partial t} +\eta^{a}(t,q,\dot{q}) \frac{\partial}{\partial q^{a}} \label{eq.inv3}
\end{equation}
is a weak Noether symmetry\index{Symmetry! weak Noether} iff it satisfies\footnote{Not along solutions of E-L equations!} the weak Noether condition\index{Condition! weak Noether}
\begin{equation}
\mathbf{X}^{[1]}(L) +\phi^{a} \frac{\partial L}{\partial \dot{q}^{a}} +L\dot{\xi} =\dot{f} \label{eq.inv4}
\end{equation}%
where $f(t,q,\dot{q})$ is the Noether function and $\phi^{a}(t,q,\dot{q})$ is an additional vector generator.

As we have shown in sec. \ref{con.mot.sec.gvar}, the weak Noether condition (\ref{eq.inv4}) is written equivalently as
\begin{equation}
E_{a}(L) \left( \eta^{a} -\dot{q}^{a}\xi \right) - \phi^{a} \frac{\partial L}{\partial \dot{q}^{a}} + \frac{d}{dt} \left[ f -L\xi -\frac{\partial L}{\partial \dot{q}^{a}} \left( \eta^{a} -\dot{q}^{a}\xi \right) \right]=0. \label{eq.inv6}
\end{equation}
This condition along E-L equations (\ref{eq.inv1}) leads to the Noether FI
\begin{equation}
I=f-L\xi -\frac{\partial L}{\partial \dot{q}^{a}}\left( \eta^{a}- \dot{q}^{a}\xi \right) \label{eq.inv7}
\end{equation}%
provided that the functions $\phi^{a}$ are defined by the
condition
\begin{equation}
F_{a}\left( \eta^{a} - \dot{q}^{a}\xi \right) = \phi^{a}\frac{\partial L}{\partial \dot{q}^{a}}. \label{eq.inv8}
\end{equation}

Someone now may ask if the above result can be reversed, that is, is it always possible an arbitrary FI of the system to be associated with a weak Noether symmetry? The answer to this question is affirmative and is given in the following Theorem \cite{Djukic 1975}:

\begin{theorem}
\label{Inverse Noether Theorem} \textbf{(Inverse Noether Theorem).} Suppose $\Lambda$ is a FI of a holonomic dynamical system with regular Lagrangian $L(t,q,\dot{q})$ and generalized non-conservative forces $F^{a}(t,q,\dot{q})$. Then, the vector $\mathbf{X}=\xi(t,q,\dot{q})\frac{\partial}{\partial t} + \eta^{a}(t,q,\dot{q})\frac{\partial}{\partial q^{a}}$ with a weak first prolongation \index{Theorem! Inverse Noether}
\begin{equation}
\mathbf{X}^{W}= \mathbf{X}^{[1]} + \phi^{a}(t,q,\dot{q})\partial _{\dot{q}^{a}} = \xi\partial _{t}+\eta^{a}\partial_{q^{a}} +\left( \dot{\eta}^{a}-\dot{q}^{a}\dot{\xi} + \phi^{a}\right)
\partial_{\dot{q}^{a}}   \label{eq.inv9}
\end{equation}%
is the generator of a weak Noether symmetry with gauge function $f(t,q,\dot{q})$ provided that
\begin{eqnarray}
\eta^{a} &=& -\gamma^{ab} \frac{\partial \Lambda}{\partial \dot{q}^{b}} + \xi \dot{q}^{a} \label{eq.inv10a} \\
\phi^{a} \frac{\partial L}{\partial \dot{q}^{a}} &=& -F^{a} \frac{\partial \Lambda}{\partial \dot{q}^{a}} \label{eq.inv10b} \\
\xi &=& \frac{1}{L} \left( f - \Lambda + \gamma^{ab} \frac{\partial L}{\partial \dot{q}^{a}} \frac{\partial \Lambda}{\partial \dot{q}^{b}} \right). \label{eq.inv10c}
\end{eqnarray}
This weak Noether symmetry produces the given FI $\Lambda$. Therefore, any FI for such systems can be associated to a weak Noether symmetry.
\end{theorem}

\begin{proof}

First, we have the identity
\begin{equation}
E_{a}(L)\equiv \frac{d}{dt}\left( \frac{\partial L}{\partial \dot{q}^{a}}\right)  -\frac{\partial L}{\partial q^{a}}\iff \ddot{q}^{a} = \gamma^{ab} \left[ P_{b}(L,F) +F_{b} + \frac{\partial L}{\partial q^{b}} - \frac{\partial^{2} L}{\partial \dot{q}^{b}\partial t} - \frac{\partial^{2} L}{\partial \dot{q}^{b} \partial q^{c}} \dot{q}^{c} \right] \label{eq.inv11}
\end{equation}
where $P_{b}(L,F)\equiv E_{b}(L) -F_{b}$.

Using eq. (\ref{eq.inv11}), the total time derivative of an arbitrary function $N(t,q,\dot{q})$ is
\[
\dot{N}= \frac{\partial N}{\partial t} +\frac{\partial N}{\partial q^{a}}\dot{q}^{a} +\frac{\partial N}{\partial \dot{q}^{a}}\gamma^{ab} \left[ P_{b}(L,F) +F_{b} + \frac{\partial L}{\partial q^{b}} - \frac{\partial^{2} L}{\partial \dot{q}^{b}\partial t} - \frac{\partial^{2} L}{\partial \dot{q}^{b} \partial q^{c}} \dot{q}^{c} \right] \implies
\]
\begin{equation}
\dot{N}= M(N) +\gamma^{ab}\frac{\partial N}{\partial \dot{q}^{b}} P_{a}(L,F) \label{eq.inv12}
\end{equation}
where the function
\begin{equation}
M(N)\equiv \frac{\partial N}{\partial t} +\frac{\partial N}{\partial q^{a}}\dot{q}^{a} +\frac{\partial N}{\partial \dot{q}^{a}}\gamma^{ab} \left( F_{b} +\frac{\partial L}{\partial q^{b}} - \frac{\partial^{2} L}{\partial \dot{q}^{b}\partial t} - \frac{\partial^{2} L}{\partial \dot{q}^{b} \partial q^{c}} \dot{q}^{c} \right). \label{eq.inv13}
\end{equation}
We note that by replacing $\ddot{q}^{a}$ from the identity (\ref{eq.inv11}), the resulting function $M(N)$ depends only on $t, q^{a},$ and $\dot{q}^{a}$.

If $N=\Lambda$ where $\Lambda$ is an arbitrary FI of the system, eq. (\ref{eq.inv12}) becomes
\begin{equation}
\dot{\Lambda}= \gamma^{ab}\frac{\partial \Lambda}{\partial \dot{q}^{b}} P_{a}(L,F) \label{eq.inv14}
\end{equation}
because the function $M(\Lambda)$ should vanish identically for all $t, q^{a}, \dot{q}^{a}$ in order to have $\dot{\Lambda}=0$ along solutions of the E-L equations $P_{a}(L,F)=0$.

Taking into account the above results, it is sufficient to show that the weak Noether condition in the form (\ref{eq.inv6}) is satisfied identically for the set $(\xi, \eta^{a}, \phi^{a}, f)$ defined by the conditions (\ref{eq.inv10a}) - (\ref{eq.inv10c}).

Substituting equations (\ref{eq.inv10a}) - (\ref{eq.inv10c}) and (\ref{eq.inv14}) in (\ref{eq.inv6}), we find that
\[
-\gamma^{ab} \frac{\partial \Lambda}{\partial \dot{q}^{b}} P_{a}(L,F) + \dot{\Lambda}=0 \implies -\gamma^{ab} \frac{\partial \Lambda}{\partial \dot{q}^{b}} P_{a}(L,F) +\gamma^{ab}\frac{\partial \Lambda}{\partial \dot{q}^{b}} P_{a}(L,F) = 0 \implies 0=0.
\]
Therefore, the weak Noether condition is satisfied identically by the generators defined by the conditions (\ref{eq.inv10a}) - (\ref{eq.inv10c}). Moreover, the condition (\ref{eq.inv8}) is satisfied as well; therefore, substituting in (\ref{eq.inv7}), we produce the Noether FI $I=\Lambda$ which completes the proof.
\end{proof}

In the case of the gauge $\xi=0$, the conditions defining a gauged weak (generalized) Noether symmetry are reduced as follows:
\begin{eqnarray}
\eta^{a} &=& -\gamma^{ab} \frac{\partial \Lambda}{\partial \dot{q}^{b}} \label{eq.inv14a} \\
\phi^{a} \frac{\partial L}{\partial \dot{q}^{a}} &=& -F^{a} \frac{\partial \Lambda}{\partial \dot{q}^{a}} \label{eq.inv14b} \\
f&=& \Lambda -\gamma^{ab} \frac{\partial L}{\partial \dot{q}^{a}} \frac{\partial \Lambda}{\partial \dot{q}^{b}}. \label{eq.inv14c}
\end{eqnarray}

\section{Methods for determining FIs}

\label{sec.methods.determine.FIs}

As we have seen in the previous sections, it is important to have a systematic (i.e. algorithmic) method for determining FIs. In the course of time, there have been developed various such methods either algebraic or geometric. A brief review of the major such methods has as follows.

\subsection{The Lie symmetry method}

A Lie symmetry (see sec. \ref{sec.classym})\index{Symmetry! Lie} of a differential equation is a point transformation in the solution space of the equation which preserves the set of solutions of the equation. Therefore, it is possible such symmetries to produce FIs (see e.g. \cite{Katzin 1973}); however, in general, it is not, and one has to stick with Noether symmetries. These are a special class of Lie symmetries which satisfy the additional requirement of the Noether condition. The method of Noether symmetries is the most widely used tool for the determination of FIs (see e.g. \cite{Katzin 1974, Katzin 1976, Ranada 1997, Damianou 2004, Mei et all, Hadler Paliathanasis Leach 2018}). According to Noether's Theorem \ref{con.mot.pro.3}, every Noether symmetry leads to a Noether FI.

\subsection{The Inverse Noether theorem}

If $I(t,q,\dot{q})$ is a FI of a second order dynamical system whose Lagrangian $L(t,q,\dot{q})$ is regular, then by means of the Inverse Noether theorem \ref{Inverse Noether Theorem} one may associate to $I$ a gauged generalized Noether symmetry and finally compute the FIs. This is done as follows.

From the Inverse Noether theorem, the FI $I$ is associated to the generalized Noether symmetry (see e.g. \cite{Sarlet Cantrijn 81, Djukic 1975}):
\begin{eqnarray}
\eta^{a} &=& -\gamma^{ab} \frac{\partial I}{\partial \dot{q}^{b}} +\xi\dot{q}^{a}  \label{Cartan.1} \\
\xi &=& \frac{1}{L} \left( f - I + \gamma^{ab} \frac{\partial L}{\partial \dot{q}^{a}} \frac{\partial I}{\partial \dot{q}^{b}} \right) \label{Cartan.1.1}
\end{eqnarray}%
where $f(t,q,\dot{q})$ is the Noether function and the kinetic metric $\gamma_{ab}$ is used for lowering and raising the indices. Equation (\ref{Cartan.1}) is the well-known \textbf{Cartan condition}.\index{Condition! Cartan} In the gauge $\xi =0$, conditions (\ref{Cartan.1}) and (\ref{Cartan.1.1}) become:
\begin{eqnarray}
\eta^{a} &=& -\gamma^{ab} \frac{\partial I}{\partial \dot{q}^{b}} \label{Cartan.1.2} \\
f &=& I -\gamma^{ab} \frac{\partial L}{\partial \dot{q}^{a}} \frac{\partial I}{\partial \dot{q}^{b}}. \label{Cartan.1.3}
\end{eqnarray}
If one looks for QFIs of the form $I=K_{ab}(t,q)\dot{q}^{a}\dot{q}^{b}+K_{a}(t,q)\dot{q}^{a}+K(t,q)$ where $K_{ab}(t,q), K_{a}(t,q), K(t,q)$ are symmetric tensor quantities, then from conditions (\ref{Cartan.1.2}) - (\ref{Cartan.1.3}) it follows that the generator $\eta_{a}= -2K_{ab}\dot{q}^{b}-K_{a}$ and the Noether function $f= -K_{ab}\dot{q}^{a}\dot{q}^{b} +K$. Replacing these results into the Noether condition, one obtains a set of PDEs whose solution provides the corresponding Noether integrals \cite{Leach 1985}.

\subsection{The Lax pair method}

In this method (see e.g. \cite{Lax 1968, Calogero, Olshanetsky, Babelon 1990, ArutyunovB}), one brings the dynamical equations into a special matrix form called a \textbf{Lax representation}.\index{Representation! Lax} Then, the existence of an extended set of FIs is guaranteed. Specifically, Hamilton's equations have to be written in the form
\begin{equation}
\dot{A}= [B,A]= BA -AB \label{lax1}
\end{equation}
where $A$ and $B$ are two square matrices whose entries are functions on the phase space $q, p$ of the system. If this is possible, then it is said that the system admits a Lax representation with $A$ being the corresponding Lax matrix. The pair of matrices $A, B$ is called a \textbf{Lax pair}.\index{Pair! Lax}

If one finds a Lax representation, then the functions
\begin{equation}
I_{k}= tr(A^{k}) \label{lax2}
\end{equation}
where $tr$ denotes the trace and $k$ is a positive integer, are FIs. Indeed, we have
\begin{eqnarray*}
\dot{I}_{k}&=& k ~tr\left( A^{k-1} \dot{A} \right) = k ~tr\left( A^{k-1} [B,A] \right)= k ~tr\left( A^{k-1}BA \right) -k ~tr\left( A^{k} B\right) \\
&=& k ~tr\left( A^{k}B \right) -k ~tr\left( A^{k} B\right) = 0
\end{eqnarray*}
because the trace is invariant under cyclic permutations and $tr(A+B)= tr(A) + tr(B)$. In fact, the matrix equation (\ref{lax1}) has the general solution
\begin{equation}
A(t)= F(t) A(0) F(t)^{-1} \label{lax3}
\end{equation}
where the invertible matrix $F(t)$ is such that $B= \dot{F} F^{-1}$.

A Hamiltonian system may admit more than one Lax pair. These pairs may be: 1) represented by square matrices of different size, and 2) related by transformations of the type $A' = G A G^{-1}$ and $B'= GBG^{-1} +\dot{G}G^{-1}$, where $G$ is an arbitrary invertible matrix.

\subsection{The Hamilton-Jacobi (H-J) method}

This is also a widely applied method which --as a rule-- concerns autonomous conservative dynamical systems and FIs with small degrees of freedom. In this method, one
considers in the phase space (cotangent bundle) the Hamiltonian $H= \frac{1}{2}\gamma^{ab}(q)p_{a}p_{b} +V(q)$, where $V(q)$ denotes the potential and $q^{a}, p_{a}$ are the canonical coordinates. The coordinates $q^{a}$ and the Hamiltonian are called \textbf{separable} if the corresponding Hamilton-Jacobi (H-J) equation\index{Equation! Hamilton-Jacobi}
\begin{equation*}
\frac{1}{2}\gamma^{ab}W_{,a}W_{,b} +V =h
\end{equation*}%
has a complete solution of the form $W(q;c)=W_{1}(q^{1};c) + ... +W_{n}(q^{n};c)$, where $W, W_{1}, ..., W_{n}$ are smooth functions of $q^{a}$, $W_{,a}= \partial_{q^{a}}W$, $h$ is an arbitrary constant and $c=(c_{1}, ..., c_{n})$ are integration constants. Separable Hamiltonian systems form a large class of integrable systems. We note that the additive separation of the H-J equation is related to the multiplicative separation of the corresponding Helmholtz (or Schr\"{o}dinger) equation. The separation of variables in the H-J equation, corresponding to a natural Hamiltonian $H= \frac{1}{2}\gamma^{ab}(q)p_{a}p_{b} +V(q)$ with a kinetic metric of any signature, is intrinsically characterized by geometrical objects on the Riemannian configuration manifold, i.e. KVs, KTs, and Killing webs. The intrinsic characterization in terms of Riemannian geometry of the additive separation of variables in the H-J equation is discussed, e.g., in \cite{Benenti 1997, Tsiganov 2000, Benenti 2002} and references cited therein. The H-J theory in the context of the moving frames formalism of E. Cartan is discussed in \cite{Adlam}.

One application of the H-J theory, which is relevant to the present work, is the determination of the autonomous conservative dynamical systems with two degrees of freedom which
are superintegrable with one cubic FI (CFI) and either one linear FI (LFI) or a QFI. It is found in \cite{Marquette Winternitz 2008} that the case of LFIs gives the
well-known cases of the harmonic oscillator and the Kepler potential, while
the case of QFIs gives five irreducible potentials whose finite trajectories
are all closed. In another relevant work \cite{McLenaghan2004}, concerning the classification of autonomous CFIs
of autonomous Hamiltonians with two degrees of freedom, the authors classify
the non-trivial third order KTs using the group invariants
of KTs defined on pseudo-Riemannian spaces of constant curvature
under the action of the isometry group. Higher order FIs are also discussed in \cite{Tsiganov 2008}. In all cases mentioned above, the studies concern autonomous Hamiltonians and autonomous FIs (see e.g. \cite{Kalnins 1980, Fokas 1979, Fokas 1980, Evans 1990, McLenaghan 2002}).

\subsection{The direct method}

\label{par.direct}

The direct method\index{Method! direct} applies to second order holonomic dynamical systems which are not necessarily conservative. In this method, instead of Lie/Noether symmetries, one assumes a generic FI, say $I(t,q,\dot{q})$, which is polynomial in the velocities $\dot{q}^{a}$ with unknown coefficients and requires the condition  $\frac{dI}{dt}=0$ along the dynamical equations. This condition leads to a system of PDEs involving the unknown coefficients (tensors) of $I$ together with the elements which characterize the dynamical system, that is, the potential $V$
and the non-conservative generalized forces $F^{a}$. The solution of this system provides the class of FIs defined by $I$. It appears that the direct method has been introduced for the first time by Bertrand \cite{Bertrand 1852} in the study of integrable surfaces and, later, used by Whittaker \cite{Whittaker} in the determination of the integrable autonomous
conservative Newtonian systems with two degrees of freedom. In the course of time,
this method has been used and extended by various authors (see e.g. \cite{Thompson 1984, Katzin 1973, Marquette Winternitz 2008, Fris, Katzin 1981, Hall 1983, Sen 1985, Sen 1987B, Gravel 2002, Horwood 2007, Post 2015}, and chapters \ref{ch1.QFIs.conservative} and \ref{ch.QFIs.damping} of this Thesis).

\begin{figure}[H]
\begin{center}
\includegraphics[scale=0.6]{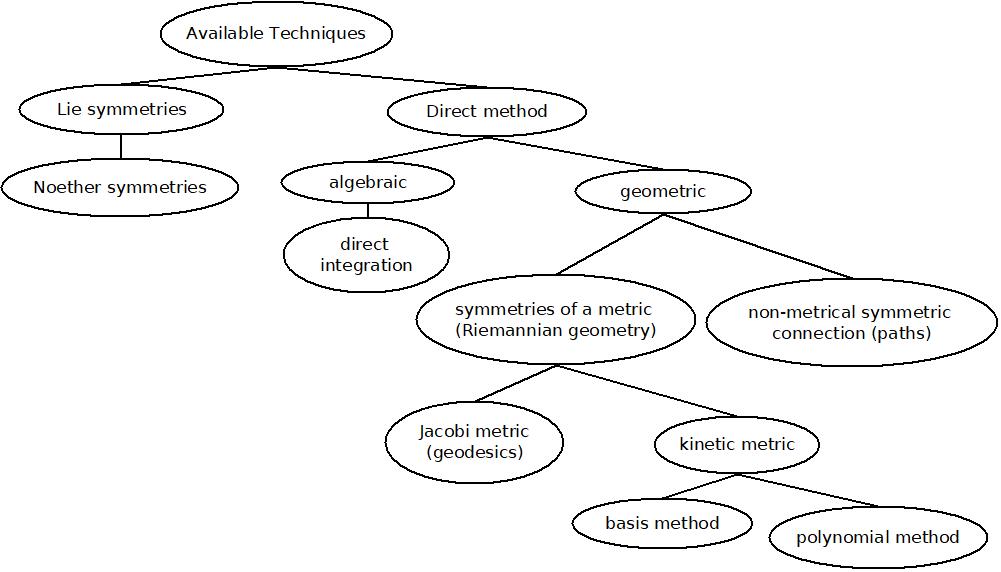}
\caption{\label{fig.tree} A tree diagram of the available techniques in the determination of FIs.}
\end{center}
\end{figure}

Concerning the solution of the system of PDEs resulting from the requirement $\frac{dI}{dt}=0$, there have been employed two methods (see Figure \ref{fig.tree}): a) the algebraic method and b) the geometric method.

In the algebraic method,\index{Method! algebraic} the system of PDEs is solved using the standard approach, i.e direct integration and/or change of variables (see e.g. \cite{Hietarinta 1987, Ranada 1997, Holt 1982, Daskalogiannis 2006}). As expected, the algebraic method becomes impossible even for small degrees of freedom or for FIs of higher order than QFIs even in the Euclidean space.

On the other hand, in the geometric method,\index{Method! geometric} one uses either the results of Riemannian geometry concerning the collineations of the metric, or the collineations of the non-metrical symmetric connection defined by the dynamical equations in the case of a non-Riemannian configuration space. Because these results are covariant, they make possible the systematic computation of FIs, autonomous or time-dependent, of any order and in a curved space.

If the dynamical equations are defined on a Riemannian configuration space, there have been used two different approaches depending on the Riemannian metric considered.

\subsubsection*{The Jacobi metric}

It is well-known that the geodesics of the Jacobi metric\index{Metric! Jacobi} which is defined by the dynamical equations (see e.g. \cite{Pin 1975, Uggla 1990, Rosquist 1995}) coincide with the trajectories/solutions of the dynamical system. Because the FIs of any order of the geodesic equations of a Riemannian space are computed in terms of the KTs of the metric (see e.g. \cite{Katzin 1981}), in order to compute the FIs of any order of a dynamical system it is enough to compute the KTs of the Jacobi metric.

With this approach many new CFIs and quartic FIs (QUFIs) have been found (see e.g. \cite{Rosquist 1995, Karlovini 2000, Karlovini 2002}). However, certain drawbacks exist concerning the Jacobi metric: a. It is not a metric of constant curvature, where we know how to compute the KTs, (see chapter \ref{ch.collineations}); and b. It has one more dimension than the degrees of freedom of the system. Both these facts make the computation of the KTs of the Jacobi metric, hence the computation of the FIs of higher order, a difficult task.

\subsubsection*{The kinetic metric}

The kinetic metric\index{Metric! kinetic} is defined by the kinetic energy of the dynamical equations and one `solves' the system of PDEs resulting from the requirement $\frac{dI}{dt}=0$ in terms of the symmetries (collineations) of this metric. This approach has been used extensively in the works of Katzin (see e.g. \cite{Katzin 1973, Katzin 1974, Katzin 1981, Katzin 1982}).

The difficulty in this approach is again the computation of the KTs of higher order. However, the situation is much easier than the case of the Jacobi metric. For example, for all holonomic Newtonian systems whose dynamical equations can be written in the form $\ddot{q}^{a} =F^{a}(t,q)$, the kinetic metric is the flat Euclidean metric whose KTs are known (they follow directly from the KVs of the metric). This approach can also be used directly in special relativistic problems without any change but the change of signature of the metric. For example, the kinetic metric has been used in scalar field cosmology in order to determine the FIs (Noether symmetries) of the mini superspace metric defined by the flat Friedmann-Robertson-Walker (FRW) metric (see e.g. \cite{Tsamparlis 2018}).

\begin{remark} \label{remark.new.7}
It is to be noted that the direct method cannot replace the generality of the Noether approach, which uses the computational tools of the Lie symmetries method, but it acts supplementary to Noether's theorem. However, there are cases (e.g. configuration spaces of constant curvature or decomposable spaces or spaces of low dimension) where the use of the direct method is more convenient due to the use of powerful results from differential geometry, which concern mainly the collineations of GOs.
\end{remark}

\subsection{Hojman approach}

An additional different approach has been developed by Hojman \cite{Hojman 1992} who showed that under certain conditions, a Lie symmetry leads to a FI called Hojman integral.\index{First integral! Hojman} However, these FIs are coordinate-dependent; therefore, they
are not useful\footnote{At least in Physics where the Covariance Principle requires
that the physical quantities must be covariant wrt the fundamental group of
the theory.}. It has been shown also that under certain conditions a\index{Symmetry! form-invariance} form-invariance symmetry\footnote{%
In the case of a holonomic dynamical system $E_{a}(L)=F_{a}$, where $F_{a}$
are the non-conservative generalized forces and $E_{a}=\frac{d}{dt}\frac{%
\partial }{\partial \dot{q}^{a}}-\frac{\partial }{\partial q^{a}}$ is the
Euler-Lagrange vector field, the form-invariance symmetry satisfies the
condition
\begin{equation*}
E_{a}\left( \delta L(t,q,\dot{q})\right) =\delta F_{a}(t,q,\dot{q})\implies
E_{a}\left( \mathbf{X}^{[1]}(L)\right) =\mathbf{X}^{[1]}\left( F_{a}\right).
\end{equation*}
} is possible to give a FI \cite{Mei et all}.

\begin{remark} \label{remark.4.Hojman}
The Hojman FIs \cite{Hojman 1992} and the ones defined by the form-invariance symmetry \cite{Mei et all} are often referred to as \textbf{non-Noetherian FIs}\index{First integral! non-Noetherian} because the generators of the corresponding
point transformations do not satisfy the weak Noether condition. However, there is an alternative approach \cite{Flessas 1995} to look at the non-Noetherian and the Noetherian FIs. Indeed, according to the Inverse Noether theorem \ref{Inverse Noether Theorem}, to every FI one may associate (in general) a velocity-dependent gauged Noether symmetry whose generator is not necessarily the same with the one generating the non-Noetherian FI. Therefore, in a sense, all FIs are or can be Noether integrals.
\end{remark}

%% file: stability.tex
\chapter{Stability of dynamical systems}

\label{ch.stability}

\section{Fixed points}

\label{sec.fixed.points}

Consider the autonomous system of $n$ first order ODEs
\begin{equation}
\dot{q}^{a}= f^{a}(q) \label{eq.fx1}
\end{equation}
where $q^{a}(t)$ are generalized coordinates with $a=1,2,...,n$ and $f^{a}(q)$ are arbitrary smooth functions. A point $q_{0}= (q^{1}_{0}, ..., q^{n}_{0}) \in \mathbb{R}^{n}$ is called a \textbf{fixed (or stationary or equilibrium) point}\index{Fixed point}\index{Equilibrium point} iff $f^{a}(q_{0})=0$ for all indices $a$. This means that for initial conditions $q^{a}(t_{0})=q^{a}_{0}$, the system will remain at $q_{0}$ for all times $t>t_{0}$. In other words, the constant function $q^{a}(t)= q^{a}_{0}$ is a solution of (\ref{eq.fx1}). In real experiments, it is very difficult to achieve initial conditions that lead to a fixed point; therefore, it is more realistic to study the behaviour of the system near $q_{0}$. In the case that $f^{a}= A^{a}_{b}q^{b}$, where $A^{a}_{b}\in \mathbb{R}$, the system (\ref{eq.fx1}) is called a \textbf{linear system};\index{System! linear} otherwise, it is a \textbf{non-linear system}.\index{System! non-linear}

\subsection{Autonomous conservative dynamical systems}

\label{sec.stability.1}

Consider an $n$-dimensional autonomous conservative dynamical system with regular (i.e. $\det[\gamma_{ab}] \neq0$) Lagrangian $L= \frac{1}{2} \gamma_{ab}(q) \dot{q}^{a} \dot{q}^{b} -V(q)$, where $\gamma_{ab}$ is the kinetic metric and $V(q)$ denotes the potential. Then, the E-L equations lead to the following the system of $n$ second order ODEs:
\begin{equation}
\ddot{q}^{a}= -\Gamma^{a}_{bc}\dot{q}^{b}\dot{q}^{c} -V^{,a} \label{eq.fx6}
\end{equation}
where $\Gamma^{a}_{bc}$ is the Riemannian connection defined by the kinetic metric.

The system (\ref{eq.fx6}) can be brought into the form (\ref{eq.fx1}) as follows:
\begin{eqnarray}
\dot{q}^{a}&=& u^{a} \label{eq.fx7.1} \\
\dot{u}^{a}&=& -\Gamma^{a}_{bc}u^{b}u^{c} -V^{,a}(q). \label{eq.fx7.2}
\end{eqnarray}
Therefore, the system of the $n$ second order ODEs (\ref{eq.fx6}) is equivalent to the system of the $2n$ first order ODEs (\ref{eq.fx7.1}) - (\ref{eq.fx7.2}), which can be written more compactly as
\begin{equation}
\dot{X}^{\mu} = F^{\mu}(X) \label{eq.fx8}
\end{equation}
where $X^{\mu}= (q^{1}, ..., q^{n}; u^{1}, ..., u^{n})\equiv (q^{a}; u^{b})$ is the state vector of the system in $\mathbb{R}^{2n}$. \index{Vector! state}

Equations (\ref{eq.fx7.1}) and (\ref{eq.fx7.2}) imply that $X^{\mu}_{0}= (q^{a}_{0}; 0, ..., 0)$ is a fixed point of the system iff $\left. V_{,a} \right|_{q_{0}}= 0$ for all indices $a$ (i.e. $q_{0}$ is a critical point of the potential). Observe that $u^{a}_{0}=0$. \emph{If the critical point $q_{0}$ is a strict local minimum (maximum) of $V$, then $q_{0}$ is stable (unstable).} In sec. 22B of \cite{Arnold 1989}, it is noted that in an analytic system with $n$ degrees of freedom a fixed point, which is not a minimum of $V$, is likely to be unstable; however, this has never proved for $n>2$.

\section{Linearization}

\label{sec.linearization}

Consider a solution of (\ref{eq.fx1})
\begin{equation}
q^{a}(t)= q^{a}_{0} + \varepsilon^{a}(t) \label{eq.fx9}
\end{equation}
close to a fixed point $q^{a}_{0}$, that is, $|\varepsilon^{a}|<<1$; therefore, second order terms $\varepsilon^{a}\varepsilon^{b} \to 0$.

Replacing (\ref{eq.fx9}) in (\ref{eq.fx1}), we obtain the linearized system \index{System! linearized}
\begin{equation}
\dot{\varepsilon}^{a}= \underbrace{f^{a}(q_{0})}_{=0} + \left. \frac{\partial f^{a}}{\partial q^{b}} \right|_{q_{0}} \varepsilon^{b} + O(\varepsilon^{2}) \implies \dot{\varepsilon}^{a}= A^{a}_{b}(q_{0}) \varepsilon^{b} \label{eq.fx10}
\end{equation}
where $A^{a}_{b}(q_{0}) \equiv \left. \frac{\partial f^{a}}{\partial q^{b}} \right|_{q_{0}} \in \mathbb{R}^{n \times n}$ is the \textbf{linearization (or Jacobian) matrix}\index{Lnearization matrix} at the fixed point $q_{0}$. The advantage of the linearized system (\ref{eq.fx10}) is that it admits the general solution
\begin{equation}
\varepsilon(t)= e^{A(q_{0})(t-t_{0})} \varepsilon(t_{0}). \label{eq.fx11}
\end{equation}
We note that equation (\ref{eq.fx11}) is in matrix form. The vector $\varepsilon(t) \in \mathbb{R}^{n\times 1}$ and the real $n \times n$ exponential matrix $e^{At} = I +At +A^{2}\frac{t^{2}}{2!} +A^{3}\frac{t^{3}}{3!} +\ldots \enskip.$

\subsection{Linearization of the equivalent Lagrangian system (\ref{eq.fx8}) around the fixed point $X_{0}= (q_{0}; 0)$}

\label{sec.stability.2}

We consider solutions $X^{\mu}(t)= X^{\mu}_{0} +\varepsilon^{\mu}(t)$.

The linearization matrix at $X_{0}$ is (in block form)
\begin{equation}
\left. \frac{\partial F^{\mu}}{\partial X^{\nu}} \right|_{X_{0}}= \left(
  \begin{array}{cc}
    0_{n \times n} & I_{n} \\
    -\gamma^{ab}(q_{0}) V_{,bc}(q_{0}) & 0_{n \times n} \\
  \end{array}
\right) \label{eq.fx13}
\end{equation}
where we used that $\left. V_{,a} \right|_{q_{0}}=0$. Therefore, the linearized system is written as:
\begin{eqnarray}
\dot{\varepsilon}^{a}&=& \varepsilon^{n+a} \label{eq.fx14.1} \\
\dot{\varepsilon}^{n+a}&=& -B^{a}_{c}(q_{0}) \varepsilon^{c} \label{eq.fx14.2}
\end{eqnarray}
where $B^{a}_{c}(q_{0}) \equiv \gamma^{ab}(q_{0}) V_{,bc}(q_{0})$.

The system of the $2n$ first order ODEs (\ref{eq.fx14.1}) - (\ref{eq.fx14.2}) is equivalent to the $n$ second order ODEs
\begin{equation}
\ddot{\varepsilon}^{a}= -B^{a}_{c}(q_{0}) \varepsilon^{c}. \label{eq.fx14.3}
\end{equation}
These equations are the E-L equations of the linearized Lagrangian\index{Lagrangian! linearized}
\begin{equation}
\tilde{L}= \frac{1}{2}\gamma_{ab}(q_{0}) \dot{\varepsilon}^{a} \dot{\varepsilon}^{b} -\frac{1}{2} B_{ab}(q_{0}) \varepsilon^{a} \varepsilon^{b} \label{eq.fx15}
\end{equation}
where $B_{ab}(q_{0})= V_{,ab}(q_{0})$.

\section{Lyapunov's stability test}

\label{sec.Lyapunov}

Consider the autonomous non-linear first order system of ODEs (\ref{eq.fx1}) and let $q_{0}$ be a fixed point of the system. Concerning its stability, $q_{0}$ can be either a (Lyapunov) stable or an unstable fixed point. We have the following classification scheme for the fixed point $q_{0}$: \newline
a. It is said to be \textbf{(Lyapunov) stable}\index{Fixed point! stable} if $\forall$ $\epsilon >0$, there exists a $\delta >0$ such that solutions $q(t)$ for which $\|q(t_{0}) -q_{0}\| < \delta$ satisfy the further inequality $\| q(t) -q_{0}\| < \epsilon$ for all $t \geq t_{0}$. \emph{This means that solutions starting `close enough' to $q_{0}$ remain `close enough' forever.} \newline
b. It is said to be \textbf{asymptotically stable}\index{Fixed point! asymptotically stable} if it is stable and for solutions $q(t)$ `close enough' to $q_{0}$ it holds that $\lim_{t\to\infty} q(t) =q_{0}$. \emph{This means that solutions starting `close enough' to $q_{0}$ not only remain `close enough' forever, but also eventually converge to $q_{0}$.} \newline
c. It is said to be \textbf{exponentially stable}\index{Fixed point! exponentially stable} if it is asymptotically stable and for solutions $q(t)$ `close enough' to $q_{0}$ there exist $\epsilon_{1}, \epsilon_{2} >0$ such that $\| q(t) -q_{0}\| \leq \alpha \| q(t_{0}) -q_{0}\| e^{-\beta(t-t_{0})}$ for all $t \geq t_{0}$. \emph{This means that solutions not only converge to $q_{0}$ but, in fact, converge faster than or at least as fast as a particular known rate $\alpha \| q(t_{0}) -q_{0}\| e^{-\beta(t-t_{0})}$.}

\begin{theorem}
[Liapunov's theorem] \label{thm.Liapunov} Let $\lambda_{i}$, $i=1,2,...,n$, be the eigenvalues (possibly complex) of the linearization matrix at the fixed point $q_{0}$ of the system (\ref{eq.fx1}). Then, $q_{0}$ is asymptotically stable (in fact exponentially stable) iff $Re(\lambda_{i}) <0$ for all values of $i$; and unstable iff $Re(\lambda_{i}) >0$ for some $i$. When all the eigenvalues are on the imaginary axis, further investigation is needed.
\end{theorem}

\section{Two-dimensional phase portraits}

\label{sec.2dnonlin}

In this case, the system (\ref{eq.fx1}) becomes $\dot{q}^{1}= f^{1}(q^{1}, q^{2})$ and $\dot{q}^{2}= f^{2}(q^{1}, q^{2})$. The plane with coordinates $(q^{1}, q^{2})$ is the \textbf{phase plane}\index{Phase plane} of the system. The entire phase plane is filled with trajectories since each point can play the role of an initial condition.

For non-linear systems, analytic solutions are very difficult (or impossible) to be found. Even when such solutions are known, they are often too complicate to provide much insight. Therefore, it is better to do a qualitative analysis and try to find the system's phase portrait directly from the properties of the vector field $f(q)$. For this purpose, we should: a) find all the fixed points, b) compute the associated linearization matrices, and c) study the equivalent linearized system around each fixed point.

Different trajectories never intersect (i.e. each initial conditions give a unique trajectory). Consider now that in a 2d phase space there is a closed orbit $C$. Then, any trajectory starting inside $C$ will be trapped in this region forever. If there is a fixed point inside $C$, then the trapped trajectory may converge to this; but if there is not, according to \textbf{Poincar\'{e}-Bendixson theorem},\index{Theorem! Poincar\'{e}-Bendixson} the trajectory must eventually approach $C$.

\section{The linear system $\dot{q}= Aq$}

\label{sec.stability.3}

We consider the 2d linear first order system
\begin{equation}
\dot{q}= Aq \iff
\begin{cases}
\dot{q}^{1}= A^{1}_{1}q^{1} +A^{1}_{2}q^{2} \\
\dot{q}^{2}= A^{2}_{1}q^{1} +A^{2}_{2}q^{2}
\end{cases} \label{eq.fx16}
\end{equation}
where the square matrix $A= \left[ A^{i}_{j} \right] \in \mathbb{R}^{2\times2}$. The $q_{1}, q_{2}$ plane is the \textbf{phase plane}\index{Phase plane} and all the solutions/trajectories $q(t)$ of (\ref{eq.fx16}) consist the \textbf{phase portrait}.\index{Phase portrait} It is well-known that (\ref{eq.fx16}) admits the general solution
\begin{equation}
q(t)= e^{At} q(0). \label{eq.fx17}
\end{equation}
\emph{The solution $(0,0)$ is always a fixed point of (\ref{eq.fx16}). This is the only fixed point of the system when $\det(A)\neq0$. If $\det(A)=0$, equation $Aq=0$ gives more fixed points.} Specifically, we have the following proposition.

\begin{proposition} \label{pro.fx1}
When $\det(A)=0$, the linear system (\ref{eq.fx16}) has two different families of fixed points: 1) For $A\neq0$, it admits a line of fixed points passing through $(0,0)$; and 2) For $A=0$, the whole phase plane is filled with fixed points.
\end{proposition}

Next, we solve the eigenvalue problem $Av =\lambda v$ where, in general, $\lambda \in \mathbb{C}$ and $v \in \mathbb{C}^{2\times1}$. Non-zero eigenvectors occur when $\det(A -\lambda I_{2})=0$. This is the characteristic equation of $A$ which gives\index{Equation! characteristic}
\begin{equation}
\lambda^{2} -tr(A) \lambda +\det(A) =0. \label{eq.fx18}
\end{equation}
Solving (\ref{eq.fx18}), we find the two eigenvalues
\begin{equation}
\lambda_{\pm} = \frac{tr(A) \pm \sqrt{\Delta}}{2} \label{eq.fx19}
\end{equation}
where the discriminant $\Delta =tr(A)^{2} -4\det(A)$. We study several cases concerning the sign of $\Delta$.

\subsection{Case $\Delta >0$ - real distinct eignevalues $\lambda_{+} \neq \lambda_{-}$}

In this case, the corresponding eigenvectors $v_{\pm}$ constitute a basis of the phase plane. Hence, any initial condition $q_{0}= c_{+}v_{+} +c_{-}v_{-}$ where $c_{\pm}$ are arbitrary real constants. The solution (\ref{eq.fx17}) is written as
\begin{equation}
q(t)= c_{+}e^{\lambda_{+}t}v_{+} +c_{-}e^{\lambda_{-}t}v_{-} \label{eq.fx20}
\end{equation}
because\footnote{We use that $Av=\lambda v$.} $e^{At}v = \left( I + At + \frac{A^{2}t^{2}}{2!} + ... \right) v = v +\lambda v t + \lambda^{2}v \frac{t^{2}}{2!} + ... = e^{\lambda t} v.$ We consider the following subcases.
\bigskip

\underline{Subcase $\det(A)=0$:}

In this subcase, we have a line of fixed points passing through $(0,0)$ and the discriminant $\Delta = tr(A)^{2} >0$. We have two subcases concerning the sign of $tr(A)$.

1) For $tr(A)>0$.

From (\ref{eq.fx19}), we find the real eigenvalues $\lambda_{+}= tr(A)>0$ and $\lambda_{-}= 0$. The solution (\ref{eq.fx20}) becomes $q(t)= c_{+} e^{tr(A) t} v_{+} +c_{-}v_{-}$. Therefore, \textbf{all the points on the line $c_{-}v_{-}$ are unstable fixed points}.

2) For $tr(A)<0$.

From (\ref{eq.fx19}), we find the real eigenvalues $\lambda_{+}=0$ and $\lambda_{-}= tr(A)<0$. The solution (\ref{eq.fx20}) becomes $q(t)= c_{+} v_{+} +c_{-}e^{tr(A)t}v_{-}$. Therefore, \textbf{all the points on the line $c_{+}v_{+}$ are stable fixed points}.
\bigskip

\underline{Subcase $\det(A)<0$:}

In this case, $\Delta > tr(A)^{2}$; therefore, the eigenvalues $\lambda_{\pm}$ given by (\ref{eq.fx19}) have opposite signs, i.e. $\lambda_{-} < 0 < \lambda_{+}$. In the special case where $tr(A)=0$, we have $\lambda_{\pm}= \pm \sqrt{-\det(A)}$.

The fixed point $(0,0)$ is a \textbf{saddle point}\index{Fixed point! saddle} (i.e. an unstable fixed point) because as $t \to \infty$ the contribution of the eigenvector $v_{-}$ decays, whereas the contribution of $v_{+}$ grows exponentially.
\bigskip

\underline{Subcase $\det(A)>0$:}

Since $\det(A)>0$ and $\Delta >0$, we have $tr(A)^{2} - 4\det(A) >0 \implies tr(A)^{2}> 4\det(A) >0 \implies tr(A)\neq 0$ and the origin $(0,0)$ is the only fixed point. We study two subcases.

1) For $tr(A)>0$.

We find that $\lambda_{+} > \lambda_{-} >0$. Therefore, the origin $(0,0)$ is a \textbf{source (or an unstable node)}\index{Fixed point! source} because as $t \to \infty$ the contribution of both eigenvectors $v_{\pm}$ grows exponentially. It is said that the eigenvector $v_{+}$ defines the `fast eigendirection' (FE),\index{Eigendirection! fast} whereas $v_{-}$ the `slow eigendirection' (SE).\index{Eigendirection! slow} The trajectories become parallel to the FE. In backwards time ($t \to -\infty$), the trajectories approach the fixed point tangentially to the SE.

2) For $tr(A)<0$.

We find that $\lambda_{-} < \lambda_{+} <0$. As $t \to \infty$ along both eigendirections $v_{\pm}$, the solution decays exponentially; therefore, the origin $(0,0)$ is a \textbf{sink (or a stable node)}.\index{Fixed point! sink} This is an asymptotically stable fixed point. Since $\lambda_{-} < \lambda_{+} <0$, the trajectories approach the origin tangentially to the eigendirection $v_{+}$. In backwards time ($t \to -\infty$) the trajectories become parallel to $v_{-}$.

\subsection{Case $\Delta=0$ - one real eigenvalue with multiplicity two}

In this case,
\begin{equation}
\det(A)= \frac{1}{4} tr(A)^{2} \label{eq.fx21.1}
\end{equation}
is a parabola on the plane $\left( tr(A), \det(A) \right)$. We consider two subcases.
\bigskip

\underline{Subcase $tr(A)=0$:}

For $tr(A)=0$, we find that $\det(A)=0$. There are two subcases due to proposition \ref{pro.fx1}.

1) Subcase $A=0$. All the points of the phase plane are fixed points.

2) Subcase $A\neq0$. We have a line of fixed point passing through $(0,0)$.

The eigenvalues $\lambda_{\pm}=0$ and the corresponding eigendirections $v_{\pm}$ are solutions of the linear homogeneous system $Av=0$. The solution of the latter system computes also the fixed points of (\ref{eq.fx16}). However, due to proposition \ref{pro.fx1}, there must exist only one eigenvector $v$ (for the double eigenvalue $\lambda=0$) along which lies the line of the fixed points. If $u$ is another vector linearly independent with $v$, then any initial condition $q(0)= c_{1}v + c_{2}u$ where $c_{1}$ and $c_{2}$ are arbitrary real constants. Therefore, the solution (\ref{eq.fx17}) of the system (\ref{eq.fx16}) becomes
\begin{equation}
q(t)= e^{At} \left( c_{1}v + c_{2}u \right) \implies q(t)= \left(c_{1} +c_{2}t \right)v +c_{2}u, \enskip \text{where $Au=v$}. \label{eq.fx22}
\end{equation}
The last result arises from the following fact: \emph{When $tr(A)=\det(A)=0$, the matrix $A^{2}=0$.}

If $u$ is an arbitrary vector in the phase plane, then $A^{2}u= 0 \implies A(Au)=0$. This implies that $Au$ is an eigenvector of $A$. The vector $Au$ exists and is non-zero because $u, v$ are linearly independent. If $Au=0$, then $u,v$ would be linearly dependent.

\begin{remark}
\label{remark.new.1} Solving the linear system $Av=0$, we find an eigenvector $v$ which --in fact-- is a family of vectors defining a straight line passing through the origin. The resulting vector $Au$ for each arbitrary vector $u$ belongs to the family of the eigenvectors of $v$. In other words, each vector $u$ chooses an eigenvector $Au=v$. To write down the solution (\ref{eq.fx22}), we choose from the line family of eigenvectors the vector $v=Au$. For more details on this subject see appendix \ref{append2}.
\end{remark}

The solution (\ref{eq.fx22}) implies that the line $c_{1}v$ along the unique eigendirection $v$ of the matrix $A$ is a line of fixed points (LFX) including the origin. Any other solution is a line parallel to the LFX. We have \textbf{uniform motion} with \textbf{unstable fixed points}. The LFX separates the phase plane in two half-planes whose trajectories are lines with opposite directions.
\bigskip

\underline{Subcase $tr(A)\neq0$:}

In this case, we have one double non-zero eigenvalue $\lambda= \lambda_{\pm}= \frac{tr(A)}{2}$ and $\det(A)\neq0$ (i.e. the origin is the only fixed point). According to the proposition \ref{pro.app2} (see appendix \ref{append2}), there are three different types (i, ii, iii) of matrices $A$. The types i and iii have only one eigenvector $v$, whereas the type ii has two distinct eigenvectors which form a basis of the phase space. The solution (\ref{eq.fx17}) in each of these cases is written as follows:

1) \underline{$A$ has two distinct eigenvectors.}

From proposition \ref{pro.app2}, the matrix $A=
\left(
  \begin{array}{cc}
    a & 0 \\
    0 & a \\
  \end{array}
\right)$, $\lambda=a$ and the eigenvector $v=
\left(
  \begin{array}{c}
    v_{1} \\
    v_{2} \\
  \end{array}
\right)$ is an arbitrary vector of the phase space. The solution (\ref{eq.fx17}) becomes $q(t) = e^{at} q(0)$.

- If $tr(A)>0$, then $a>0$ and as $t\to \infty$ the solutions are straight lines growing exponentially along the initial direction $q(0)$. The origin is a \textbf{star source} (unstable fixed point).\index{Fixed point! star}

- If $tr(A)<0$, then $a<0$ and the origin is a \textbf{star sink} (stable fixed point).

2) \underline{$A$ has one eigenvector $v$.}

From proposition \ref{pro.app2}, the matrix $A$ is either of the type i or of the type iii. Using proposition \ref{pro.app3}, the solution (\ref{eq.fx17}) becomes $q(t)= (c_{1} +c_{2}t)e^{\lambda t} v +c_{2}e^{\lambda t}u$ where $u, v$ are linearly independent such that $Au= \lambda u +v$. The initial condition $q(0) =c_{1}v +c_{2}u$ where $c_{1}, c_{2}$ are arbitrary constants. By fixing the eigenvector $v$, we find the vector $u$ (see appendix \ref{append2}). We note that
\[
e^{At}u= \sum_{k=0}^{\infty} \frac{t^{k}}{k!} A^{k}u = \sum_{k=0}^{\infty} \frac{t^{k}}{k!} \left( \lambda^{k}u +k\lambda^{k-1}v \right) = \sum_{k=0}^{\infty} \frac{(\lambda t)^{k}}{k!} u + t \sum_{k=1}^{\infty} \frac{(\lambda t)^{k-1}}{(k-1)!} v = e^{\lambda t} u +t e^{\lambda t} v
\]
where we used equation (\ref{app2.4}).

- If $tr(A)>0$, then $\lambda>0$ and the origin is a \textbf{degenerate source} (unstable fixed point). As $t\to \infty$ the trajectory becomes parallel to $v$.\index{Fixed point! degenerate source}

- If $tr(A)<0$, then $\lambda<0$ and the origin is a \textbf{degenerate sink}\index{Fixed point! degenerate sink} (stable fixed point). As $t\to -\infty$, the trajectory becomes parallel to $v$.

\begin{remark}
\label{remark.new.2} We observe that it is possible a point $(tr(A), \det(A))$ of the parabola (\ref{eq.fx21.1}) to correspond to two different matrices $A$. For example, this the case for the type ii and iii matrices of proposition \ref{pro.app2}. However, the stability of the origin is different. In the type ii the origin is a star source (or sink), whereas in the type iii is a degenerate source (or sink).
\end{remark}

\subsection{Case $\Delta <0$ - complex distinct eigenvalues}

In this case, the matrix $A$ admits two distinct (non-zero) complex conjugate eigenvalues $\lambda_{\pm}= \frac{tr(A) \pm i\sqrt{-\Delta}}{2}$. We note that $\lambda_{-}= \bar{\lambda}_{+}$, where $\bar{\lambda}_{+}$ is the complex conjugate of $\lambda_{+}$. Therefore, we can set $\lambda_{+}\equiv\lambda$. Then, $\lambda_{-}= \bar{\lambda}$ and $\lambda= \sigma +i\omega$, where $\sigma\equiv \frac{tr(A)}{2}$ and $\omega\equiv \frac{\sqrt{-\Delta}}{2}$. The corresponding eigenvectors are $v_{\pm}= w \pm iz$, where $w,z \in \mathbb{R}^{n\times1}$..

It has been shown in proposition \ref{pro.app5} of appendix \ref{append3} that the linear system (\ref{eq.fx16}) admits the general solution
\begin{equation}
q(t)= Re^{\sigma t} \left[ \cos(\omega t +\theta) w -\sin(\omega t +\theta)z \right] \label{eq.fx27}
\end{equation}
where $R= \sqrt{c_{1}^{2} +c_{2}^{2}}$, $\cos(-\theta)= \frac{c_{1}}{R}$ and $\sin(-\theta)= \frac{c_{2}}{R}$.

Moreover, $\Delta <0 \implies \det(A) > \frac{tr(A)^{2}}{4} \geq 0 \implies \det(A)>0$. Therefore, the linear system has only one fixed point the origin.

We consider the following subcases.
\bigskip

\underline{Subcase $tr(A)=0$:}

In this subcase, $\Delta= -4\det(A)<0$, $\sigma=0$ and $\omega= \sqrt{\det(A)}$. The general solution (\ref{eq.fx27}) becomes $q(t)= R \left[ \cos(\sqrt{\det(A)} t +\theta) w -\sin(\sqrt{\det(A)} t +\theta)z \right]$. This solution describes \emph{closed and periodic trajectories in the phase plane with period $T= \frac{2\pi}{\sqrt{\det(A)}}$}. The origin is said to be a \textbf{center} (neutrally stable).\index{Fixed point! center}

\underline{Subcase $tr(A)>0$:}

In this subcase, the real part of the eigenvalue is $\sigma >0$ and the solutions (\ref{eq.fx27}) as $t\to +\infty$ grow exponentially. The origin is a \textbf{spiral source} (unstable fixed point).\index{Fixed point! spiral}

\underline{Subcase $tr(A)<0$:}

In this subcase, the real part of the eigenvalue is $\sigma <0$ and the solutions (\ref{eq.fx27}) as $t\to +\infty$ come closer to the origin. The origin is a \textbf{spiral sink} (stable fixed point).

\subsection{Summary: Stability theory of 2d first order linear systems}

\label{sec.summary}

\begin{figure}[H]
\begin{center}
\includegraphics[scale=1]{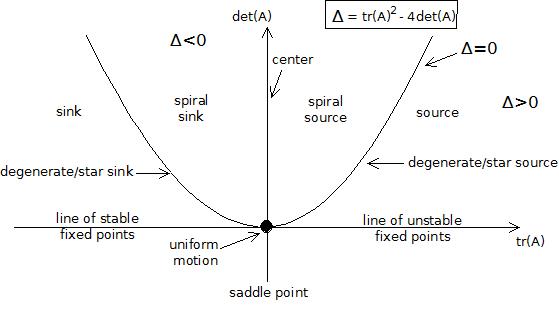}
\caption{The plane $(tr(A), \det(A))$.} \label{fig.st15}
\end{center}
\end{figure}

We have the following cases of fixed points (when $\det(A)\neq0$ the origin is the only fixed point): \index{Fixed point}

- For $\Delta>0$ (below the parabola).\newline
1) $\det(A)=0$, $tr(A)>0$ ($\lambda_{+}>0$, $\lambda_{-}=0$): line of unstable fixed points. \newline
2) $\det(A)=0$, $tr(A)<0$ ($\lambda_{+}=0$, $\lambda_{-}<0$): line of stable fixed points. \newline
3) $\det(A)<0$ ($\lambda_{-} < 0 <\lambda_{+}$): saddle point. \newline
4) $\det(A)>0$, $tr(A)>0$ ($\lambda_{+} > \lambda_{-} >0$): source (unstable node). \newline
5) $\det(A)>0$, $tr(A)<0$ ($\lambda_{-} < \lambda_{+} <0$): sink (stable node).

- For $\Delta=0$ (on the parabola). \newline
1) $tr(A)=0$ ($\lambda_{\pm} =0$): a. $A=0$, the phase space is filled with fixed points; and b. $A\neq0$, uniform motion (line of unstable fixed points). \newline
2) $tr(A)>0$ ($\lambda_{+}=\lambda_{-}>0$): a. two distinct eigenvectors, star source; and b. one eigenvector, degenerate source. \newline
3) $tr(A)<0$ ($\lambda_{+}=\lambda_{-}<0$): a. two distinct eigenvectors, star sink; and b. one eigenvector, degenerate sink.

- For $\Delta<0$ (above the parabola). \newline
1) $tr(A)=0$ ($\lambda_{\pm}= \pm i\omega$): center (neutrally stable, closed trajectories). \newline
2) $tr(A)>0$ ($\lambda_{\pm}= \sigma \pm i\omega$, $\sigma>0$): spiral source. \newline
3) $tr(A)<0$ ($\lambda_{\pm}= \sigma \pm i\omega$, $\sigma<0$): spiral sink.

A fixed point is said to be \textbf{isolated}\index{Fixed point! isolated} if there exists a neighborhood around it such that no other fixed points lie within. Fixed points that are not isolated are called \textbf{non-isolated} (i.e. line or plane of fixed points).

In Figure \ref{fig.st15}, we see that saddle points, nodes (sinks or sources) and spirals are the most common types of fixed points, because they occur in large open regions of the plane $(tr(A), \det(A))$. On the other hand, centers, degenerate/star nodes and non-isolated fixed points occur along the line $(0,\det(A)>0)$, the parabola $\Delta=0$ and $tr(A)$-axis of the plane $(tr(A), \det(A))$, respectively. The later types of fixed points are called \textbf{borderline cases}.\index{Fixed point! borderline}

As we noted in section \ref{sec.2dnonlin}, when we deal with first order non-linear systems, we compute the fixed points and we study the linearized system around each fixed point. \emph{If the linearization method predicts non-borderline cases, it can be trusted and gives the correct phase portrait around the fixed points.} But when borderline cases occur, it is not safe to neglect small quadratic or higher order non-linear terms. In other words, if the linearized system predicts a saddle, a node, or a spiral, then the fixed point of the non-linear system is really of these types \cite{Strogatz}.

Applying Theorem \ref{thm.Liapunov}, we also see that even though the type of an isolated borderline case fixed point can be altered by small non-linearities, its stability (stable or unstable) is preserved in the special cases of stars and degenerate nodes. However, this does not hold for centers. Since all the trajectories around a center are closed and periodic, small non-linear terms can convert them into spirals. \emph{All the above can be perceived directly from the diagram \ref{fig.st15}. We observe that stars and degenerate nodes lie completely within the stable or unstable region, whereas centers lie on the border between stability and instability.}

Concerning the stability of the fixed points we may classify them as follows \cite{Strogatz}: \newline
a. \textbf{Repellers:}\index{Fixed point! repeller} both eigenvalues have positive real part (i.e. spiral sources, degenerate/star sources, unstable node); unstable fixed points. \newline
b. \textbf{Attractors:}\index{Fixed point! attractor} both eigenvalues have negative real part (i.e. spiral sink, degenerate/star sink, stable node); asymptotically stable fixed points. \newline
c. \textbf{Saddles:} one eigenvalue is positive and the other is negative; unstable fixed points. \newline
d. \textbf{Centers:} both eigenvalues are pure imaginary; they meet the criterion of stable fixed points (not asymptotically). \newline
e. \textbf{Non-isolated fixed points:} at least one eigenvalue is zero.\index{Fixed point! non-isolated}

Repellers, attractors and saddles are also known as \textbf{robust cases (or hyperbolic fixed points)},\index{Fixed point! hyperbolic} whereas centers and non-isolated fixed points are called \textbf{marginal cases (or non-hyperbolic fixed points)}.\index{Fixed point! non-hyperbolic} We note that in the marginal cases at least one eigenvalue has zero real part. Moreover, the stability of the hyperbolic fixed points is unaffected by small non-linear terms; the converse holds for non-hyperbolic fixed points. Finally, there are issues with the application of the linearization method when the fixed points are non-isolated.

%% file: EMSF.tex
\chapter{Symmetries of spacetimes embedded with an electromagnetic string fluid (EMSF)}

\label{ch.EMSF}

\section{Introduction}

Relativistic magnetohydrodynamics (RMHD)\index{Magnetohydrodynamics! relativistic} is the main theory which describes various phenomena in modern astrophysics \cite{Gourgoulhon 2006, Grozdanov 2017, Hernandez 2017, Armas 2019}. When a charged plasma\index{Plasma} enters a strong magnetic field, it is possible the pressures along the magnetic field and perpendicular to the magnetic field to be unequal. This results in a physical system, which we call an anisotropic \textbf{electromagnetic string fluid (EMSF)}.\index{Fluid! electromagnetic string}

In this chapter, we study the dynamics of an \textbf{isolated anisotropic gravitating fluid}\index{Fluid! isolated anisotropic gravitating} which for the observers $u^{a}$ ($u_{a}u^{a}=-1)$ has an \textbf{energy momentum tensor}\index{Tensor! energy momentum} of the form
\begin{equation}
T_{ab}=\mu u_{a}u_{b}+p_{\parallel }n_{a}n_{b}+p_{\perp }p_{ab}
\label{Eqn2.1}
\end{equation}%
where \newline
- $n^{a}$ is a unit spacelike vector which is characteristic of the fluid. \newline
- $\mu$ is the matter density of the fluid observed by the observers $u^{a}$. \newline
- $p_{ab}$ is the tensor projecting normal to both the vectors $u^{a}$ and $n^{a}$ defined by the relation
\begin{equation}
p_{ab}= h_{ab} -n_{a}n_{b}  \label{NST.2}
\end{equation}
where
\begin{equation}
h_{ab}= g_{ab} +u_{a}u_{b}  \label{NST.3}
\end{equation}
is the tensor projecting normal to the vector $u^{a}$.\index{Tensor! projection}

This type of fluid is a special case of a string fluid. A general \textbf{string fluid (SF)}\index{Fluid! string} is defined \cite{Letelier 1979, Smalley 1996} as a fluid consisting of the mixture of a general fluid with energy momentum tensor $T_{ab}$ and a second fluid, which is characterized by an antisymmetric tensor field $F_{ab}$ with energy momentum tensor $_{F}T_{ab}$ of the form
\begin{equation}
_{F}T_{ab}=\sigma n_{a}n_{b}  \label{NST.3a}
\end{equation}%
where $\sigma$ is a function and the spacelike vector $n^{a}=F^{ab}u_{b}$. For certain types of $T_{ab}$ it is possible that the energy momentum tensor $_{F}T_{ab}+T_{ab}$ of the string fluid has the form (\ref{Eqn2.1}).

There have been considered many types of sting fluids in the literature involving different combinations of the two fluids (see e.g. \cite{Grozdanov 2017, Hernandez 2017, Armas 2019, Letelier 1980, Letelier 1981, Letelier 1983}). The recent works on the topic study the string fluid mainly from the point of view of thermodynamics by considering conserved currents and the corresponding chemical potentials including --in most of the studies-- the entropy and the temperature. They seem to make no extensive use of the gravitational field equations and concentrate rather on the catastatic equations in order to complete the set of the field equations. The problem
with this approach is that the geometry of spacetime does not enter explicitly into the study; therefore, one cannot use the results for the various important spacetimes considered in General Relativity and, particularly, on the various cosmological models.

The approach presented in this chapter follows the line of older research on string fluids, where emphasis was given in the macroscopic study of a gravitating string fluid using geometric methods and, especially, collineations (see chapter \ref{ch.collineations}) to simplify the dynamical equations. Specifically, the quantity $L_{\mathbf{X}}R_{ab}$, which enters the gravitational field equations, allows one to express these equations in terms of the quantity $L_{\mathbf{X}}g_{ab}$ (or the tensors $\psi$ and $H_{ab}$) so that for each particular collineation $\mathbf{X}$, one simply replaces the appropriate
expression of $L_{\mathbf{X}}g_{ab}$ and obtains the gravitational equations in a form that already incorporates the collineation.

Concerning the study of a gravitating string fluid using collineations, this must be based on a scenario which shall be: a) independent of the particular type of the collineation; and b)systematic, in the sense that it will describe the steps one has to follow in order to get to the required answers in the easiest and safest way. A first attempt to develop such a scenario was done in \cite{Tsamparlis M GRG 2006}. In this chapter, we shall discuss again this scenario in a more systematic way and we shall extend it in a more general framework.

The study of a string fluid requires two main vector fields: the
four-velocity $u^{a}$ and the vector field $n^{a}$ ($u^{a}n_{a}=0$) describing a dynamical variable of the fluid. These vector fields define a set of variables which are classified in the following two sets: \newline
a. \textbf{Kinematic variables},\index{Variables! kinematic} which are due to the vector field $u^{a}$; and \newline
b. \textbf{Dynamical variables},\index{Variables! dynamical} which are defined by the vector field $n^{a}$.

The kinematic and the dynamical variables are not independent because they are constrained by the following conditions: \newline
i. Certain geometric identities which the vector fields $u^{a}$ and $n^{a}$ must satisfy (e.g. Ricci identity). \newline
ii. The gravitational field equations $G_{ab}=T_{ab}+_{F}T_{ab}$ from which results the conservation law $\left( T^{ab}+_{F}T^{ab}\right) _{;b}=0$. \newline
iii. The field equations of the tensor field $F_{ab}$ together with the constraint $F_{;b}^{ab}=0.$

The standard way to define the dynamical variables of all fluids is the
covariant decomposition of the tensor fields defining the fluid, by means of
the $1+3$ decomposition defined by the \textbf{projection tensor}\index{Tensor! projection} $h_{ab}$ in (\ref{NST.3}) and the $1+1+2$ decomposition defined by the \textbf{screen projection tensor}\index{Tensor! screen projection} $p_{ab}$ in (\ref{NST.2}).

Subsequently, all the above conditions i.-iii. must be $1+3$ and $1+1+2$ decomposed in order to define the physical variables and,
eventually, to obtain a set of equations which incorporate fully the
dynamics of the string fluid in all spacetimes and for all possible
collineations. As expected, the set of the final equations is not enough for
the determination of all physical variables and one has to introduce new
assumptions, which are the equations of state and, possibly, other additional physical assumptions.

We demonstrate the above scenario to a particular string fluid, which has been considered in various forms in the literature (see \cite{Letelier 1979, Smalley 1996} and references therein). This string
fluid consists of a mixture of a charged perfect fluid interacting with an
electromagnetic field which is described in the RMHD approximation (i.e. vanishing electric field and infinite conductivity). This particular string
fluid, in the following, we shall call the \textbf{electromagnetic string fluid (EMSF)}.

We note that in the following sections we shall use pieces of results from  \cite{Tsamparlis Mason 1990}, which have required
long and tedious calculations, which there is no point to be repeated.

\section{Mathematical preliminaries}

\label{section2.0}

\subsection{Properties of spacetimes and relative tensors}

By the term {\bf{spacetime}},\index{Spacetime} we refer to a Minkowski ($sign(g) = -1$) four-dimensional manifold $M^4$ with a non-degenerate metric $g_{ij}$ in a coordinate system $\{x^i\}$, where $i,j = 1,2,3,4$.

We introduce the notation
\[
\dot{S}^{a_1...}_{b_1...} = S^{a_1...}_{b_1...;c} u^c, \enskip \overset{\ast}{S}{}^{a_1...}_{b_1...} = S^{a_1...}_{b_1...;c} n^c.
\]

We work with {\bf{orientable}}\index{Spacetime! orientable} spacetimes, i.e. $J > 0$. In general, $J \neq 0$ and thus either $J < 0$ (negative orientation) or $J > 0$ (positive orientation). We recall that the Jacobian $J^{i}_{i'}\equiv \frac{\partial x^{i}}{\partial x^{i'}}$ and the determinant $J \equiv \det[J^i_{i'}]$. The orientation of spacetime in general relativity is still an open problem.

From the theory of determinants, the adjoint matrix $adj(J^i_{i'}) = J \cdot [J^{i'}_i]$ and $J_{,a} = J J^{b'}_b J^b_{b',a}$, which implies that $J_{,a'} = J_{,a} J^a_{a'} = J J^a_{a',a}$.

The sign function: $sign(g) \equiv \epsilon(g) = \pm 1$, i.e. $\epsilon^2 = 1$ and $\epsilon = \frac{1}{\epsilon}$.

We recall that $g\equiv \det[g_{ij}] \neq 0$, $J \equiv \det[J^i_{i'}] \neq 0$ and $g' = J^2 g$. Therefore, $\epsilon(g) = \epsilon(g')$, $|J| = \epsilon(J) J = \sqrt{\frac{g'}{g}} = \sqrt{\frac{|g'|}{|g|}}$ and $J = \epsilon(J) \sqrt{\frac{|g'|}{|g|}}$.

The covariant derivative along a vector field $\mathbf{X}$ wrt a general connection $\Gamma^{i}_{jk}$ of a relative $(r,s)$-tensor of weight $w$ is a relative tensor\index{Tensor! relative} of the same kind\footnote{
We recall that such GOs satisfy the transformation law
\[
T^{i_1'...i_r'}_{j_1'...j_s'} = J^w J^{i_1'}_{i_1} ... ~J^{i_r'}_{i_r} J^{j_1}_{j_1'} ... ~J^{j_s}_{j_s'} T^{i_1...i_r}_{j_1...j_s}
\]
where $w \in \mathbb{Z}$.
} with components
\begin{align*}
\nabla_{\mathbf{X}} T^{i_1...i_r}_{j_1...j_s} & =  T^{i_1...i_r}_{j_1...j_s|c} X^c \\
& = T^{i_1...i_r}_{j_1...j_s,c} X^c + \Gamma^{i_1}_{kc} X^c T^{ki_2...i_r}_{j_1j_2...j_s} + ... + \Gamma^{i_r}_{kc} X^c T^{i_1...i_{r-1}k}_{j_1...j_{s-1}j_s} - \\
& \quad - \Gamma^k_{j_1c} X^c T^{i_1i_2...i_r}_{kj_2...j_s} - ... - \Gamma^k_{j_sc} X^c T^{i_1...i_{r-1}i_r}_{j_1...j_{s-1}k} - w \Gamma^k_{kc} X^c T^{i_1...i_r}_{j_1...j_s}.
\end{align*}

The quantity $\sqrt{|g|}^w$ is an oriented relative tensor of weight $w$ because $\sqrt{|g'|}^w = \left[ \epsilon(J) \right]^w J^w \sqrt{|g|}^w$. The Riemannian covariant derivative gives $(\sqrt{|g|}^w)_{;a} = 0$, but in general $(\sqrt{|g|}^w)_{|a} = (\sqrt{|g|}^w)_{,a} - w \Gamma^b_{ba} \sqrt{|g|}^w$.

\subsection{Levi-Civita pseudotensor}

\label{subsec.levicivita}

Assume an $n$-dimensional smooth manifold. The {\bf{generalized Kronecker delta}}\index{Kronecker delta! generalized} is a constant totally antisymmetric $(r,r)$-tensor field with components ($n \geq r$)
\[
\delta^{i_1...i_r}_{j_1...j_r} = \sum_{\sigma} (sign \sigma) \delta^{i_1}_{j_{\sigma(1)}} ... ~\delta^{i_r}_{j_{\sigma(r)}} = \sum_{\sigma} (sign \sigma) \delta_{j_1}^{i_{\sigma(1)}} ... ~\delta_{j_r}^{i_{\sigma(r)}}.
\]
We note that $\delta^{i_1...i_r}_{j_1...j_r|c} = 0$, $\delta^{i_1...i_r}_{j_1...j_r} = r! \delta^{i_1}_{[j_1} ... ~\delta^{i_r}_{j_r]} = r! \delta^{[i_1}_{j_1} ... ~\delta^{i_r]}_{j_r}$ and $\delta^{i_1...i_r}_{j_1...j_r} T_{i_1...i_r} = r! T_{[j_1...j_r]}$.

The quantities
\[
\varepsilon^{i_1...i_n} \equiv \delta^{i_1...i_n}_{1...n} \enskip \text{and} \enskip \varepsilon_{i_1...i_n} \equiv \delta_{i_1...i_n}^{1...n}
\]
are tensor densities called {\bf{Levi-Civita symbols}}.\index{Levi-Civita symbols} They satisfy the transformation laws: $\varepsilon^{i_1'...i_n'} = J J^{i_1'}_{i_1} ... ~J^{i_n'}_{i_n} \varepsilon^{i_1...i_n}$ $=$ $\epsilon(J) \sqrt{\frac{|g'|}{|g|}} J^{i_1'}_{i_1} ... ~J^{i_n'}_{i_n} \varepsilon^{i_1...i_n}$ (relative $(n,0)$-tensor of weight $+1$, i.e. a tensor density\index{Tensor! density} of type $(n,0)$) and $\varepsilon_{i_1'...i_n'} = J^{-1} J_{i_1'}^{i_1} ... ~J_{i_n'}^{i_n} \varepsilon_{i_1...i_n}$ $=$ $\epsilon(J) \sqrt{\frac{|g|}{|g'|}} J^{i_1'}_{i_1} ... ~J^{i_n'}_{i_n} \varepsilon^{i_1...i_n}$ (relative $(0,n)$-tensor of weight $-1$). The only relation which involves all the totally antisymmetric symbols is  $\varepsilon^{i_1...i_n} \varepsilon_{j_1...j_n} = \delta^{i_1...i_n}_{j_1...j_n}$.

Some useful properties are the following: \newline
a) $\varepsilon^{i_1...i_si_{s+1}...i_n} \varepsilon_{j_1...j_si_{s+1}...i_n} = (n-s)! \delta^{i_1...i_s}_{j_1...j_s}$. \newline
b) $\varepsilon^{i_1...i_ti_{t+1}...i_n}$ $\varepsilon_{i_1...i_tj_{t+1}...j_n}$ $=$ $t! \delta^{i_{t+1}...i_n}_{j_{t+1}...j_n}$. \newline
c) (for $t=n$ or $s=0$) $\varepsilon^{i_1...i_n} \varepsilon_{i_1...i_n} = n!$. \newline
d) $\det(A) = \varepsilon^{i_1...i_n} A^1_{i_1} ... A^n_{i_n} = \varepsilon_{i_1...i_n} A_1^{i_1} ... A_n^{i_n}$ $=$
$\frac{1}{n!}$ $\delta^{j_1...j_n}_{i_1...i_n}$ $A^{i_1}_{j_1}...A^{i_n}_{j_n}$. \newline
e) $\det[a_{ij}] \varepsilon_{i_1...i_n}$ $=$ $\varepsilon^{j_1...j_n} a_{i_1j_1} ... a_{i_nj_n}$ and $\det[a^{ij}] \varepsilon^{i_1...i_n}$ $=$ $\varepsilon_{j_1...j_n} a^{i_1j_1} ... a^{i_nj_n}$.

By the term\index{Tensor! oriented} {\bf{pseudotensor (or oriented tensor)}},\index{Pseudotensor} we refer to a GO satisfying the transformation law
\[
T^{i_1'...i_r'}_{j_1'...j_s'} = \epsilon(J) J^{i_1'}_{i_1} ... ~J^{j_s}_{j_s'} T^{i_1...i_r}_{j_1...j_s}.
\]
The quantities
\[
\eta^{i_1...i_n} \equiv \frac{\epsilon(g)}{\sqrt{|g|}} \varepsilon^{i_1...i_n} \enskip \text{and} \enskip \eta_{i_1...i_n} \equiv \sqrt{|g|} \varepsilon_{i_1...i_n}
\]
are examples of pseudotensors called\index{Pseudotensor! Levi-Civita} {\bf{Levi-Civita pseudotensors}}. We note that (see paragraphs 7.210, 7.211 and 7.100 in \cite{Synge1978}) the raising and lowering of indices wrt the metric $g_{ij}$ do not hold for Levi-Civita symbols. Indeed, we have
\[
\varepsilon_{i_1...i_n} = \frac{1}{g} g_{i_1j_1} ... g_{i_nj_n} \varepsilon^{j_1...j_n} \enskip \text{and} \enskip \varepsilon^{i_1...i_n} = g g^{i_1j_1} ... g^{i_nj_n} \varepsilon_{j_1...j_n}.
\]
However,
\[
\eta_{i_1...i_n} = g_{i_1j_1} ... g_{i_nj_n} \eta^{j_1...j_n} \enskip \text{and} \enskip \eta^{i_1...i_n} = g^{i_1j_1} ... g^{i_nj_n} \eta_{j_1...j_n}.
\]
Moreover, it holds that $\varepsilon^{i_1...i_n}{}_{|c} = \varepsilon_{i_1...i_n|c} = 0$ and $\eta^{i_1...i_n}{}_{;c} = \eta_{i_1...i_n;c} = 0$.

For the spacetime $M^4$, we have the following:
\[
\eta_{abcd} = \sqrt{-g} \varepsilon_{abcd}, \enskip \eta^{abcd} =  - \frac{1}{\sqrt{-g}} \varepsilon^{abcd}, \enskip \eta^{abcd} \eta_{arst} = - \delta^{bcd}_{rst} = - 3! \delta^b_{[r} \delta^c_s \delta^d_{t]},
\]
\[
\eta^{abcd} \eta_{abrs} = - 2! \delta^{cd}_{rs} = - 4 \delta^c_{[r} \delta^d_{s]}, \enskip \eta^{abcd} \eta_{abcr} = - 3! \delta^d_r, \enskip \eta^{abcd} \eta_{abcd} = - 4!, \enskip \eta^{abcd} = g^{ar} g^{bs} g^{ct} g^{dq} \eta_{rstq}.
\]
Since $M^4$ is assumed to be orientable ($J > 0$), the quantities $\eta^{abcd}$ and $\eta_{abcd}$ are tensors called {\bf{Levi-Civita tensors}}.\index{Tensor! Levi-Civita}

\subsection{Projection tensors}

\label{subsec.projection.tensor}

Consider\index{Tensor! projection} a unit timelike four-vector $u^a$, i.e. $u^a u_a = -1$, and a unit spacelike four-vector $n^a$, i.e. $n^a n_a = 1$, such that $u^a n_a = 0$. When we study a fluid, $u^a$ describes the family of the observers (four-velocity) and $n^a$ is a vector field characteristic of the internal geometry of the fluid.

We define the tensors: $h_{ab} = g_{ab} + u_a u_b$ projecting normal to $u^a$, and $p_{ab} = h_{ab} - n_a n_b$ projecting normal to both $u^a$ and $n^a$. It is proved that: $h^a_b \equiv h^a{}_b = h_b{}^a$, $h^a_b = \delta^a_b + u^a u_b$, $h^a_a = 3$, $h_{ab} u^b = 0$, $h^a_b n^b = n^a$, $h^{ac} h_{bc} = h^a_b$, $p_{ab} u^b = p_{ab} n^b = 0$, $p^a_a = 2$, $p^a_c p^c_b = p^a_b$ and $h^a_c p^c_b = p^a_b$. Moreover, $\dot{u}^a u_a = \overset{\ast}{u}{}^a u_a = 0$, $u^a{}_{;b} u_a = 0$ and $u^a{}_{;b} n_a = - n^a{}_{;b} u_a$.

\section{The definition of the physical variables}

\label{section2}

The definition of the physical variables of a relativistic fluid is done with the use of the 1+3 and the 1+1+2 decomposition of the characteristic fields $u^{a}$ and $n^{a}$ of the fluid. The 1+3 decomposition generated by $u^{a}$ can be found, among others, in the early (and excellent) paper of Ellis \cite{Ellis 1998}; whereas, the 1+1+2 in \cite{Tsamparlis Mason 1990, Saridakis Tsamparlis 1991}. In the following, we review briefly the application of these decompositions for the case of a string fluid.

\subsection{The 1+3 decomposition}

\label{section2.1}

Consider a spacetime $M$ with metric $g_{ab}$ and a fluid of observers with four-velocity $u^{a}$ ($u^{a}u_{a}=-1$). The four-vector $u^{a}$ defines the projection operator $h_{ab}=g_{ab}+u_{a}u_{b}$ wrt which all GOs defined on $M$ can be $1+3$ decomposed.

The kinematical variables are defined by the the 1+3 decomposition of
\begin{equation*}
u_{a;b}=-\dot{u}_{a}u_{b}+\omega _{ab}+\underbrace{\sigma _{ab}+\frac{1}{3}\theta h_{ab}}_{=\theta _{ab}}
\end{equation*}
where $\omega _{ab}=h_{a}^{c}h_{b}^{d}u_{[c;d]}$ is called the \textbf{vorticity tensor}\index{Tensor! vorticity}, $\theta_{ab} =h_{a}^{c}h_{b}^{d}u_{(c;d)}$, $\theta =\theta
_{a}^{a}=h^{ab}u_{a;b}=u^{a}{}_{;a}$ is called the \textbf{expansion (isotropic strain)},\index{Expansion} and\index{Isotropic strain} $\sigma _{ab}=\theta _{ab}-\frac{1}{3}\theta h_{ab}=\left[h_{(a}^{c}h_{b)}^{d} -\frac{1}{3}h^{cd}h_{ab}\right] u_{c;d}$ is called the \textbf{shear stress tensor}\index{Tensor! shear stress} or simply the \textbf{shear}.\index{Shear}

From the vorticity tensor, one defines the \textbf{vorticity vector}\index{Vector! vorticity} $\omega ^{a}=\frac{1}{2}\eta ^{abcd}u_{b;c}u_{d}$. We note that $\omega_{ab} = \eta_{abcd} \omega^c u^d$, $\omega_{ab} u^b = 0$, $\omega^a u_a = 0$, $\omega \equiv \omega^a_a = 0$, $\omega_{ab} = h^r_a h^s_b \omega_{rs}$, $\omega_{ab} = h^r_a \omega_{rb}$, $\sigma_{ab} u^b = 0$, $\sigma \equiv \sigma^a_a = 0$, $\sigma_{ab} = h^r_a h^s_b \sigma_{rs}$, and $\sigma_{ab} = h^r_a \sigma_{rb}$.

At the level of dynamics, the 1+3 decomposition concerns the energy momentum tensor\footnote{We recall that $T_{ab}=T_{ba}$ and  $T^{ab}{}_{;b} = 0$.} $T_{ab}$ of a fluid in $M$ and defines the dynamical variables of the fluid, as observed by the observers $u^{a}$, as follows:
\begin{equation}
T_{ab}=\mu u_{a}u_{b}+ph_{ab}+2q_{(a}u_{b)}+\pi_{ab}.  \label{EqnSS.0}
\end{equation}%
The variables $\mu ,p,q^{a},\pi _{ab}$ have the following physical interpretation: \newline
a. The scalars $\mu = T_{ab} u^a u^b$ and $p = \frac{1}{3} h^{ab} T_{ab}$ correspond, respectively, to the \textbf{energy (mass) density}\index{Energy (mass) density} and the \textbf{isotropic pressure}\index{Pressure! isotropic} of the fluid. \newline
b. The spacelike vector\footnote{It holds that $q^{a}q_{a}>0$ and $q^{a}u_{a}=0$.} $q_a = - h^d_a T_{dc} u^c$ is the \textbf{energy (heat) flux}\index{Vector! energy (heat) flux} in the three-space defined by the projection tensor $h_{ab}$. \newline
c. $\pi_{ab} = \left( h^c_a h^d_b - \frac{1}{3} h^{cd} h_{ab} \right) T_{cd}$ is the traceless (i.e. $\pi _{a}^{a}=0$) \textbf{stress tensor}\index{Tensor! stress} (measures the anisotropy) with  $h^{ab} \pi_{ab} = 0$ and $\pi_{ab} u^b = 0$.

The fluids are classified, according to the dynamical variables, in: \textbf{dust}\index{Dust} $(p=q_{a}=\pi _{ab}=0),$ \textbf{perfect fluid}\index{Fluid! perfect} $(q_{a}=\pi _{ab}=0),$ \textbf{heat conducting fluid}\index{Fluid! heat conducting} $(q_{a}\neq 0)$, and \textbf{anisotropic fluid}\index{Fluid! anisotropic} $(\pi _{ab}\neq 0).$

We compute the trace $T \equiv T^a_a = - \mu + 3p$.

As an example, we see that the 1+3 decomposition of the energy momentum tensor (\ref{NST.3a}) describes an unusual fluid with $\mu = 0$, $p = \frac{\sigma}{3} n^a n_a$, $q^a = 0$ and $\pi_{ab} = \sigma \left( n_a n_b - \frac{n^c n_c}{3} h_{ab} \right)$.

At the same level, one considers the 1+3 decomposition of the conservation equation $T^{ab}{}_{;b}=0$ which leads to the following two equations (see e.g. \cite{Saridakis Tsamparlis 1991}):
\begin{eqnarray}
\overset{.}{\mu }+(\mu +p)\theta +\pi ^{ab}\sigma _{ab}+q_{;a}^{a}+q^{a}%
\overset{.}{u}_{a} &=&0  \label{CE.1} \\
(\mu +p)\overset{.}{u}_{a} + h^c_a (p_{,c} + \pi^b_{c;b} + \overset{.}{q}%
_{c}) + \left( \omega_{ac} + \sigma_{ac} + \frac{4}{3} \theta h_{ac} \right)
q^c &=& 0.  \label{CE.2}
\end{eqnarray}

At this point, we should point out that the conservation law $T^{ab}{}_{;b} = 0$ holds for the total energy momentum tensor of the matter (see e.g. \cite{Ellis book}) and not for the energy momentum tensors of each matter component separately. This is the case due to the interaction of the different matter components. In particular, for a mixture of $m$ fluids $T^{ab}_{(I)}$ with $I = 1, ..., m$, we have
\[
T^{ab}_{(tot)} = \sum_I T^{ab}_{(I)}, \enskip T^{ab}_{(tot);b} = 0, \enskip T^{ab}_{(I);b} = Q^a_{(I)}
\]
where $Q^a_{(I)}$ is the rate of the energy and momentum density of the $I$-component. Therefore, $\sum_I Q^a_{(I)} = 0$. For example, the Minkowski energy momentum tensor ${}_{EM}T^{ab}$ of the electromagnetic field has divergence ${}_{EM}T^{ab}{}_{;b} = -\lambda F^{ab} J_b$. If $F^{ab} J_b \neq 0$, then the total conservation of the energy momentum implies the existence of an extra fluid component $T^{ab}$ such that $T^{ab}{}_{;b} = \lambda F^{ab} J_b$.

Finally, the 1+3 decomposition of the Ricci identity\index{Identity! Ricci} $u_{a;bc}-u_{a;cb}= R^{d}{}_{abc} u_{d}$ leads to two more sets of equations, which are called \textbf{the propagation and the constraint equations}\index{Equations! propagation} \cite{Ellis 1998}.\index{Equations! constraint}

\subsection{The 1+1+2 decomposition}

\label{section2.2}

The pair of vectors $u^{a}, n^{a}$ ($n^{a}n_{a}=1, u^{a}n_{a}=0$) constitutes a double congruence and defines the plane projection operator\index{Tensor! plane projection}
\begin{equation*}
p_{ab}=h_{ab}-n_{a}n_{b}
\end{equation*}%
which projects normal to the plane defined by $u^{a}$ and $n^{a}$. This 2d space is called the \textbf{screen space} of the congruence $u^{a}, n^{a}$.\index{Space! screen}

The 1+1+2 decomposition defines the kinematic variables of the spacelike vector field $n^{a}$ (see \cite{Saridakis Tsamparlis 1991, Mason Tsamparlis 1985} and references cited therein) as follows:
\begin{eqnarray}
n_{a;b} &=& A_{ab} + \overset{\ast}{n}_a n_b - \dot{n}_a u_b + u_a \left[ n^c u_{c;b} + (n^c \dot{u}_c) u_b - (n^c \overset{\ast}{u}_c) n_b \right] \notag \\
&=& A_{ab} + \overset{\ast}{n}_a n_b - \dot{n}_a u_b + u_a \left( - N_b + 2 n^c \omega_{cb} + p^c_b \dot{n}_c \right) \notag \\
&=& A_{ab} + \overset{\ast}{n}_a n_b - \dot{n}_a u_b + u_a p^c_b \left( \dot{n}_c + 2 n^d \omega_{dc} - N_c \right) \label{N.17}
\end{eqnarray}
where\footnote{It holds that $A_{ab} u^b = A_{ab} n^b = 0$ and $p^c_a A_{cb} = A_{ab}$.} $A_{ab} = p^c_a p^d_b n_{c;d}$ is the {\bf{screen tensor}}\index{Tensor! screen} and $N_b = p_{bc} ( \dot{n}^c - \overset{\ast}{u}{}^c)$ is the {\bf{Greenberg vector}}.\index{Vector! Greenberg} We compute:
\begin{equation}
N_b = p_{bc} (n^c{}_{;d} u^d - u^c{}_{;d} n^d) = p_{bc} L_{\mathbf{u}} n^c = p_{bc} [\mathbf{u}, \mathbf{n}]^c, \label{N.22}
\end{equation}
$p^b_a N_b = N_a$ and $N_b u^b = N_b n^b = 0$.

The Greenberg vector is important because it vanishes
iff the vector fields $u^{a}$ and $n^{a}$ are surface forming (i.e. iff $L_{\mathbf{u}}n^{b}=Au^{b}+Bn^{b}$ where $A, B$ are arbitrary constants). From the kinematics point of view, the
vector $N_{a}$ vanishes iff the vector field $n^{a}$ is `frozen' along the observers $u^{a}$.

The screen tensor $A_{ab}$ is decomposed further into its irreducible parts (the kinematic variables
of $n^{a}$) as follows:
\begin{equation}
A_{ab}= \mathcal{S}_{ab} +\mathcal{R}_{ab} +\frac{1}{2}\mathcal{E}p_{ab}
\label{N.17a}
\end{equation}%
where $\mathcal{S}_{ab}=\mathcal{S}_{ba}$ ($\mathcal{S}_{\text{\ }b}^{b}=0$) is the traceless part (\textbf{screen shear}),\index{Tensor! screen shear} $\mathcal{R}_{ab}= -\mathcal{R}_{ba}$ is the antisymmetric part (\textbf{screen rotation})\index{Tensor! screen rotation} and $\mathcal{E}$ is the trace (\textbf{screen expansion}).\index{Expansion! screen} We have the defining relations:
\begin{align}
\mathcal{S}_{ab}& =A_{(ab)}  =\left(p_{a}^{c}p_{b}^{d} -\frac{1}{2}p^{cd} p_{ab} \right)n_{(c;d)}  \label{N.18} \\
\mathcal{R}_{ab}& =A_{[ab]} -\frac{1}{2}\mathcal{E}p_{ab} =p_{a}^{c}p_{b}^{d}n_{[c;d]}  \label{N.19} \\
\mathcal{E}& =p^{cd}n_{c;d}=n^{c}{}_{;c} +\dot{n}^{c}u_{c}.
\label{N.20}
\end{align}
One can define also the screen rotation vector\index{Vector! screen rotation} $\mathcal{R}^{a} =\frac{1}{2}\eta^{abcd} \mathcal{R}_{bc}u_{d}$. It holds that $\mathcal{R}^a u_a = 0$.

We consider now a vector field $X^a$ such that $X^a u_a = 0$. The $1+1+2$ decomposition gives
\begin{equation}
X^a = \nu n^a + K^a \label{eq.112decXa}
\end{equation}
where $\nu = X^b n_b$ and $K^a = p^a_b X^b$.

Concerning the dynamical variables, the energy momentum tensor is 1+1+2 decomposed as follows:
\begin{align}
T_{ab} & = \mu u_a u_b + 2 \kappa u_{(a}n_{b)} + \gamma n_a n_b + \frac{1}{2} \alpha p_{ab} + 2 Q_{(a}u_{b)} + 2 P_{(a}n_{b)} + D_{ab} \nonumber \\
& = \mu u_a u_b + p h_{ab} + 2 \kappa u_{(a} n_{b)} + \bar{\gamma} \left( n_a n_b - \frac{1}{2} p_{ab} \right) + 2 Q_{(a} u_{b)} + 2 P_{(a} n_{b)} + D_{ab} \nonumber \\
& = \mu u_a u_b + p h_{ab} + \underbrace{2 \kappa n_{(a} u_{b)} + 2 Q_{(a} u_{b)}}_{= 2 q_{(a} u_{b)}} + \underbrace{\bar{\gamma} \left( n_a n_b - \frac{1}{2} p_{ab} \right) + 2 P_{(a} n_{b)} + D_{ab}}_{= \pi_{ab}} \label{eq.dec112Tab}
\end{align}
where we have introduced the new dynamical variables $\kappa = - T_{ab} u^a n^b$, $\gamma = T_{ab} n^a n^b$, $\alpha = p^{ab} T_{ab}$ $=$ $3p - \gamma$, $\bar{\gamma} = \pi_{ab} n^a n^b$ $=$ $\gamma - p$, $Q^a = - p^{ab} u^c T_{bc}$, $P^a = p^{ab} n^c T_{bc}$ and $D_{ab}$ $=$ $\left( p^c_a p^d_b - \frac{1}{2} p^{cd} p_{ab} \right) T_{cd}$ $=$ $\left( p^c_a p^d_b - \frac{1}{2} p^{cd} p_{ab} \right) \pi_{cd}$ as observed by the observers $u^{a}$. The physical meaning of each of the new dynamical variables is the following: \newline
a. The scalar $\kappa$ and the screen vector $Q^a$ are related to the heat conduction of the fluid. \newline
b. The scalar $\gamma$, the screen vector $P^a$ and the traceless screen tensor $D_{ab}$ have to do with the anisotropy of the fluid.

From \eqref{eq.dec112Tab}, the 1+1+2 dynamical variables are related to the 1+3 dynamical variables as follows:
\begin{align}
q^a & = \kappa n^a + Q^a  \label{eq.dec112.1} \\
\pi_{ab} & = \bar{\gamma} \left( n_a n_b - \frac{1}{2} p_{ab} \right) + 2 P_{(a} n_{b)} + D_{ab}.  \label{eq.dec112.2}
\end{align}
We note that equation \eqref{eq.dec112.1} is the $1+1+2$ decomposition (see eq. \eqref{eq.112decXa}) of $q^a$ because $q^a u_a = 0$.

The above decompositions are general and hold for all fluids.

\subsection{The string fluid defined by the electromagnetic field}

\label{section2.3}

The energy momentum tensor (\ref{Eqn2.1}) of an isolated anisotropic gravitating fluid can be 1+3 decomposed wrt $u^{a}$ as follows:
\begin{equation}
T_{ab}=\mu u_{a}u_{b}+\frac{1}{3}(p_{\parallel }+2p_{\perp
})h_{ab}+(p_{\perp }-p_{\parallel })\left( \frac{1}{3}h_{ab}-n_{a}n_{b}%
\right).  \label{EqnSS.2}
\end{equation}%
Therefore,
\begin{eqnarray}
p &=&\frac{1}{3}(p_{\parallel }+2p_{\perp })  \label{EqnSS.3} \\
\pi_{ab} &=&\left( p_{\perp }-p_{\parallel} \right) \left( \frac{1}{3}h_{ab}-n_{a}n_{b} \right)  \label{EqnSS.4} \\
q^{a} &=&0.  \label{EqnSS.4a}
\end{eqnarray}

It follows, from (\ref{EqnSS.2}), that the energy momentum tensor (\ref{Eqn2.1}) corresponds to an anisotropic fluid with vanishing heat flux. Furthermore, we note that
\[
\pi_{ab} n^b = - \frac{2}{3} \left( p_{\perp} - p_{\parallel} \right) n_a.
\]
This implies that $n^{a}$ is an eigenvector of the anisotropic stress tensor $\pi_{ab}$ with eigenvalue $-\frac{2}{3} (p_{\perp }-p_{\parallel}).$ We assume $p_{\perp }-p_{\parallel}\neq 0$; otherwise, the string fluid reduces to a perfect fluid with energy momentum tensor $T_{ab}= \mu u_{a}u_{b} +p_{\perp}h_{ab}$ which under the unphysical\footnote{
Because in this case (since $p_{\parallel}= p_{\perp}$) the pressure $p$ of the fluid becomes negative ($p = -\mu < 0$). We note that, in general, one of the two components of the pressure may be negative ($p_{\parallel} = - \mu < 0$) but the observed pressure $p$ is not! To have the physical condition $p > 0$, it must hold that $p_{\perp} > -\frac{p_{\parallel}}{2}$.
} equation of state $\mu +p_{\perp}=0$ reduces to $T_{ab}=p_{\perp} g_{ab}$.

A {\bf{perfect string fluid}}\index{Fluid! perfect string} (the case we study in this work) has energy momentum tensor
\begin{equation} \label{eq.perSF}
T_{ab} = \mu (u_a u_b - n_a n_b) + q p_{ab}.
\end{equation}
This is an isolated anisotropic gravitating fluid (\ref{Eqn2.1}) for $p_{\parallel} = - \mu$ and $p_{\perp} = q$. Then, $p = \frac{1}{3} (2q - \mu)$, $q^a = 0$ and $\pi_{ab} = (q + \mu) \left( \frac{1}{3} h_{ab} - n_a n_b \right)$.

Furthermore, in the 1+1+2 decomposition the tensor $\pi_{ab}$ is
written as
\begin{equation}
\pi_{ab} =-\frac{2}{3}(p_{\perp }-p_{\parallel })\left( n_{a}n_{b} -\frac{1}{2}p_{ab}\right).  \label{EqnSS.5}
\end{equation}%
By replacing (\ref{EqnSS.5}) in (\ref{eq.dec112.2}), we find that the only non-vanishing irreducible part of $\pi_{ab}$ is
\begin{equation}
\bar{\gamma}=-\frac{2}{3}(p_{\perp }-p_{\parallel }).
\end{equation}%
Hence, we conclude that the string fluid defined by (\ref{EqnSS.2}) is the `simplest' anisotropic fluid.

An important example of a string fluid is the electromagnetic field in the \textbf{RMHD approximation}\index{Approximation! RMHD} with \textbf{infinite conductivity and vanishing electric field} (see e.g. \cite{Grozdanov 2017, Hernandez 2017, Armas 2019}).

Indeed, in this approximation, the electromagnetic tensor $F_{ab}$ is given by the expression\index{Tensor! electromagnetic}
\begin{equation}
F_{ab}=\eta_{abcd}H^{c}u^{d}  \label{EqnSS.5b}
\end{equation}%
where $H^{a}$ is the magnetic field and the vector $n^{a}=\frac{H^{a}}{H}$ is the unit vector in the direction of the magnetic field.

The \textbf{Minkowski energy momentum tensor}\index{Tensor! Minkowski energy momentum} of the electromagnetic field $_{EM}T^{ab}$ is given by
\begin{equation}
_{EM}T^{ab}=\lambda \left( F^{ac}F^{b}{}_{c} -\frac{1}{4}g^{ab}F_{cd}F^{cd} \right)  \label{ME.14}
\end{equation}%
where $\lambda $ is a constant. Using Maxwell equations, one shows that
\begin{equation}
_{EM}T^{ab}{}_{;b}=- \lambda F^{ab}J_{b}.  \label{ME.15}
\end{equation}%
Replacing $F_{ab}$ from (\ref{EqnSS.5b}) in (\ref{ME.14}), we find
\begin{equation}
_{EM}T_{ab}=\frac{1}{2}\lambda H^{2}u_{a}u_{b} +\frac{1}{6}\lambda H^{2}h_{ab} +\lambda H^{2}\left( \frac{1}{3}h_{ab} -n_{a}n_{b}\right).
\label{EqnSS.6}
\end{equation}

Replacing $h_{ab}=p_{ab}+n_{a}n_{b}$, we end up with the expression
\begin{equation}
_{EM}T_{ab}=\frac{1}{2}\lambda H^{2}(u_{a}u_{b}-n_{a}n_{b})+\frac{1}{2}%
\lambda H^{2}p_{ab}  \label{EqnSS.7}
\end{equation}
which defines a perfect string fluid with irreducible parts:
\begin{eqnarray}
\mu &=&\frac{1}{2}\lambda H^{2}  \label{EqnSS.10} \\
p &=&\frac{1}{6}\lambda H^{2}  \label{EqnSS.11} \\
\pi_{ab} &=&\lambda H^{2}\left(\frac{1}{3}h_{ab} -n_{a}n_{b}\right) \label{EqnSS.12} \\
q^{a} &=&0.  \label{EqnSS.13}
\end{eqnarray}
Therefore, the equation of state is always $p = \frac{\mu}{3}$. Since $\mu \geq 0$, then necessarily $p \geq 0$ and the fluid of the electromagnetic field satisfies the strong energy condition.

\section{The electromagnetic string fluid (EMSF)}

\label{section3}

We consider the dynamical system consisting of a charged perfect fluid with
isotropic pressure $p$ and energy density $\rho$, which interacts with the
electromagnetic field in the RMHD approximation. Physically, this situation is considered to be
the case in various plasmas \cite{Duarte 2014}.

Due to the interaction of the fluid with the electromagnetic field, it is possible that the magnetic field produces a different fluid pressure perpendicular and parallel to the magnetic field; therefore, the perfect fluid becomes an anisotropic fluid with pressure distribution $p_{\parallel
}n_{a}n_{b}+p_{\perp }p_{ab}$. Then, the energy momentum tensor of the interacting fluid is\footnote{We note that the total energy momentum conservation $T^{ab}{}_{;b} = 0$ implies that $_{SF}T^{ab}{}_{;b} = \lambda F^{ab} J_b$.}
\begin{eqnarray}
T_{ab}&=& _{SF}T_{ab} + _{EM}T_{ab} \notag \\
&=& \left( \rho u_a u_b +p_{\parallel} n_a n_b + p_{\perp} p_{ab} \right) + \left[ \frac{1}{2} \lambda H^2 (u_a u_b - n_a n_b) + \frac{1}{2} \lambda H^2 p_{ab} \right] \notag \\
&=& \left( \rho + \frac{1}{2} \lambda H^2 \right) u_a u_b + \left( p_{\parallel } -\frac{1}{2} \lambda H^2 \right) n_a n_b + \left(p_{\perp} + \frac{1}{2} \lambda H^2\right) p_{ab}. \label{EqnSS.8}
\end{eqnarray}

The interacting fluid is not a perfect string fluid. For this to be the case, the following condition must be satisfied:
\begin{equation}
\rho +\frac{1}{2}\lambda H^{2}= -\left( p_{\parallel} -\frac{1}{2}\lambda H^{2}\right) \implies p_{\parallel}= -\rho.  \label{EqnSS.15}
\end{equation}
With this condition assumed, \textbf{the energy momentum tensor of the EMSF} is \index{Tensor! EMSF energy momentum}
\begin{equation}
T_{ab}= \left( \rho +\frac{1}{2}\lambda H^{2}\right) \left( u_{a}u_{b} -n_{a}n_{b}\right) +\underbrace{\left( p_{\perp} +\frac{1}{2} \lambda H^{2} \right)}_{\equiv q} p_{ab}. \label{eq.EMSF}
\end{equation}
Then, the 1+3 decomposition gives
\begin{eqnarray}
\mu &=&\rho +\frac{1}{2}\lambda H^{2}  \label{EqnSS.17} \\
p &=& \frac{1}{3}\left( 2p_{\perp } -\rho +\frac{1}{2}\lambda
H^{2}\right)  \label{EqnSS.18} \\
q^{a} &=&0  \label{EqnSS.19} \\
\overline{\pi}_{ab} &=&\left( \rho +p_{\perp } +\lambda H^{2} \right) \left( \frac{1}{3}h_{ab}-n_{a}n_{b} \right). \label{EqnSS.20}
\end{eqnarray}
Concerning the 1+1+2 dynamical variables, we find that
\begin{equation*}
\mu = \rho +\frac{1}{2}\lambda H^{2}, \enskip \kappa =0, \enskip \alpha =2p_{\perp} +\lambda H^{2}, \enskip \gamma= -\rho -\frac{1}{2}\lambda H^{2}, \enskip Q_{a}=P_{a}=0,\enskip D_{ab}=0
\end{equation*}
and
\begin{equation*}
\bar{\gamma}= \gamma -p \implies \bar{\gamma}=-\frac{2}{3}\left( \rho +p_{\perp} +\lambda H^{2}\right).
\end{equation*}

Finally, we note the relations:
\begin{equation*}
\mu + q = \rho + p_{\perp }+ \lambda H^{2} \enskip \text{and} \enskip T= 2(p_{\perp} - \rho) =2(q - \mu)
\end{equation*}
which are useful in the calculations. Here $q = p_{\perp} +\frac{1}{2} \lambda H^{2}$ and, thus, for the EMSF the energy momentum tensor is written $T_{ab} = \mu (u_a u_b - n_a n_b) + q p_{ab}$ which is the general form (\ref{eq.perSF}) for any perfect string fluid.

One direction in which the string fluids have been studied is the simplification of the field equations for various types of collineations of spacetime \cite{Tsamparlis M GRG 2006, Yavuz Yilmaz 1997, Yilmaz 2001, Baysal Camci et all 2002, Baysal Yilmaz 2002, U Camci 2002, Sharif M Sheikh U 2006}. In the next sections, we extend these studies to the case of the EMSF.

\subsection{The Ricci tensor of the EMSF}

\label{section3.1}

{\bf{Einstein field equations}}\index{Equations! Einstein field} have the general form
\begin{equation} \label{eq.einstein1}
G_{ab} + \Lambda g_{ab} = T_{ab}
\end{equation}
where $G_{ab} = R_{ab} - \frac{1}{2} R g_{ab}$ is the \textbf{Einstein tensor}\index{Tensor! Einstein} ($G^{ab}{}_{;b} = 0$), $R_{ab} = R^c{}_{acb}$ is the \textbf{Ricci tensor}\index{Tensor! Ricci} and $\Lambda$ is the Einstein cosmological constant. Taking the trace of \eqref{eq.einstein1}, we find $R = 4 \Lambda - T$ which when replaced into \eqref{eq.einstein1} gives the equivalent \textbf{trace-reversed form} of field equations
\begin{equation} \label{eq.einstein2}
R_{ab} = T_{ab} + \left( \Lambda - \frac{1}{2} T \right) g_{ab}.
\end{equation}
In what follows, we use this form.

For a perfect string fluid (\ref{eq.perSF}), equation (\ref{eq.einstein2}) gives
\begin{equation}
R_{ab}= \left(q - \Lambda\right) (u_a u_b - n_a n_b) + \left( \mu + \Lambda \right) p_{ab}. \label{eq.einstein3}
\end{equation}
Therefore, for the EMSF, we have
\begin{equation}
R_{ab}= \left( p_{\perp }+\frac{1}{2}\lambda H^{2}-\Lambda \right) (u_{a}u_{b}-n_{a}n_{b}) +\left( \rho +\frac{1}{2}\lambda H^{2}+\Lambda \right)p_{ab}.  \label{EqnSS.25}
\end{equation}
We note that $R_{ab}$ is found immediately from $T_{ab}$ given by (\ref{eq.perSF}) if we interchange $\mu \leftrightarrow p_{\perp} +\frac{1}{2}\lambda H^{2}-\Lambda$, $p_{\perp} \leftrightarrow \rho +\frac{1}{2}\lambda H^{2}+\Lambda$ and vice versa. This is a useful observation because it
allows us to compute various results for $R_{ab}/T_{ab}$ and write down the
answer for the corresponding quantities for $T_{ab}/R_{ab}$ by interchanging
the string variables as indicated above. For example, the 1+3 decomposition of $R_{ab}$ is written directly from (\ref{EqnSS.2}) for $p_{\parallel}=-\mu$ as follows:
\begin{eqnarray}
R_{ab}&=& \left( p_{\perp }+\frac{1}{2}\lambda H^{2}-\Lambda \right) u_{a}u_{b} +\frac{1}{3}\left( 2\rho -p_{\perp} +\frac{1}{2}\lambda H^{2} +3\Lambda \right) h_{ab}+ \notag \\
&& +(\rho +p_{\perp} +\lambda H^{2})\left( \frac{1}{3}h_{ab}-n_{a}n_{b} \right).  \label{EqnSS.26}
\end{eqnarray}

\subsection{The conservation equations for the EMSF in the 1+1+2
decomposition}

\label{section3.2}

In the case of the EMSF, the conservation equations (\ref{CE.1}) and (\ref{CE.2}) are simplified as follows:
\begin{eqnarray}
\overset{.}{\mu} +(\mu +q)\left( \frac{2}{3}\theta -\sigma
_{ab}n^{a}n^{b}\right) &=&0  \label{CE.3} \\
(\mu +q)\left[ \overset{.}{u}_{a}-(\mathcal{E}-\overset{.}{n}%
_{b}u^{b})n_{a}-h_{a}^{b}\overset{\ast }{n}_{b}\right] +p_{a}^{b}q_{,b}-%
\overset{\ast }{\mu }n_{a} &=&0.  \label{CE.4}
\end{eqnarray}

Furthermore, by projecting (\ref{CE.4}) along $n^{a}$ and using the tensor $p_{b}^{a}$, we get the two equations:
\begin{eqnarray}
\overset{\ast }{\mu }+(\mu +q)\mathcal{E} &=&0 \\
p_{a}^{b}\left[ q_{,b}+(\mu +q)(\overset{.}{u}_{b}-\overset{\ast }{n}_{b})%
\right] &=&0.
\end{eqnarray}%
Replacing the energy density $\mu$ and the heat coefficient $q$ from equation (\ref{eq.EMSF}), we find that the 1+1+2
decomposition of the conservation equations for an EMSF are:
\begin{eqnarray}
\overset{.}{\rho }+\lambda H\overset{.}{H}+(\rho +p_{\perp }+\lambda H^{2})
\left( \frac{2}{3}\theta -\sigma _{ab}n^{a}n^{b} \right) &=& 0
\label{EqnSS.37} \\
\overset{\ast }{\rho }+\lambda H\overset{\ast }{H}+(\rho +p_{\perp }+\lambda H^{2})\mathcal{E} &=&0  \label{EqnSS.38} \\
p_{a}^{b}\left[ p_{\perp }{}_{,b}+\lambda HH_{,b}+(\rho +p_{\perp }+\lambda H^{2})(\overset{.}{u}_{b}-\overset{\ast }{n}_{b})\right] &=&0.
\label{EqnSS.39}
\end{eqnarray}%
\emph{These equations are independent of any other assumptions which one might do concerning the fluid, including the symmetries.}

We continue our analysis with the 1+1+2 decomposition of Maxwell equations.

\subsection{Maxwell equations in the 1+3 and the 1+1+2 formalisms}

\label{section3.3}

All the information about the electromagnetic field is squeezed into an antisymmetric tensor field $F_{ab}$ which is called {\bf{electromagnetic tensor}}.\index{Tensor! electromagnetic} This tensor satisfies the well-known {\bf{Maxwell equations}}\index{Equations! Maxwell} (see e.g. \cite{Tsamparlis Book})
\begin{equation}
F_{[ab;c]}=0, \enskip F^{ab}{}_{;b}= J^{a}  \label{ME.1}
\end{equation}%
where $J^{a}$ is the \textbf{four-current}.\index{Four-current} The four-current and the electromagnetic field tensor in the 1+3
decomposition are decomposed as follows:
\begin{eqnarray}
J^{a} &=& eu^{a} +j^{a}  \label{ME.2} \\
F^{ab} &=& u^{a}E^{b} -u^{b}E^{a} +\eta^{abcd}H_{c}u_{d}.  \label{ME.3}
\end{eqnarray}%
The various physical quantities introduced are (as measured by the observer $u^{a}$): (a) $e$ the \textbf{charge density},\index{Charge density} (b) $j^{a}$ the \textbf{conduction current},\index{Current! conduction} (c) $E^{a}$ the \textbf{electric field},\index{Electric field} and (d) $H^{a}$ the \textbf{magnetic field}.\index{Magnetic field} Inverting (\ref{ME.2}) and (\ref{ME.3}), we find:
\begin{eqnarray}
e &=& -u^{a}J_{a}, \enskip j^{a}=h_{b}^{a}J^{b}  \label{ME.4} \\
E^{a} &=&F^{ab}u_{b}, \enskip H^{a}=\frac{1}{2} \eta^{abcd} F_{bc}u_{d}.  \label{ME.5}
\end{eqnarray}
In general, the magnetic field is a pseudovector (see sec. \ref{subsec.levicivita}); however, for the spacetime we study is just a vector field. We note that $H^a u_a = E^a u_a = 0$ (i.e. $H^a$ and $E^a$ are spacelike) and $f^{ab}\equiv h^a_c h^b_d F^{cd}$ $=$ $\eta^{abcd} H_c u_d$. Therefore, the chosen direction of the magnetic field lines $n^a = \frac{H^a}{H}$ satisfies the relations $n^a u_a = 0$ and $n^a n_a = 1$. \emph{That's why the direction of the magnetic field wrt the observers $u^{a}$ is always taken orthogonal.} We also recall the \textbf{continuity equation} $J^a{}_{;a} = 0$.\index{Equation! continuity}

Taking into account the 1+3 kinematic variables, Maxwell equations are 1+3 decomposed\footnote{
To apply such a decomposition in $F_{[ab;c]} = 0$, we recall that $F^{*ab}{}_{;b} = 0$ iff $F_{[ab;c]} = 0$, where $F^{*ab} = \frac{1}{2} \eta^{abcd} F_{cd}$. In particular, we have
\[
F^{*ab}{}_{;b} = 0 \iff \eta^{abcd} F_{bc;d} = 0 \iff \eta_{arst} \eta^{abcd} F_{bc;d} = 0 \iff \delta^{bcd}_{rst} F_{bc;d} = 0 \iff F_{[ab;c]} = 0.
\]
} wrt the observers $u^{a}$ into the following equations (see \cite{Ellis 1998} and sec. 13.10.6 in \cite{Tsamparlis Book}):
\begin{eqnarray}
h_{b}^{a}H^{b}{}_{;a} &=& 2\omega^{a}E_{a}  \label{ME.6} \\
h_{b}^{a}E^{b}{}_{;a} &=& e -2\omega^{a}H_{a}  \label{ME.7} \\
h_{b}^{a}\overset{.}{H}^{b} &=& u^{a}{}_{;b} H^{b} -\theta H^{a} -I^{a}(E) \label{ME.8} \\
h_{b}^{a}\overset{.}{E}^{b} &=& u^{a}{}_{;b}E^{b} -\theta E^{a} +I^{a}(H) -j^{a} \label{ME.9}
\end{eqnarray}
where we introduced the `currents':
\begin{eqnarray}
I^{a}(E) &=&\eta ^{abcd}u_{b}(\overset{.}{u}_{c}E_{d}-E_{c;d})  \label{ME.10}
\\
I^{a}(H) &=&\eta ^{abcd}u_{b}(\overset{.}{u}_{c}H_{d}-H_{c;d}).  \label{ME.11}
\end{eqnarray}%
Equations (\ref{ME.6}) - (\ref{ME.7}) are the \textbf{constraint equations}\index{Equations! constraint} and (\ref{ME.8}) - (\ref{ME.9}) are the \textbf{propagation equations}.\index{Equations! propagation} Also, we recall that $\omega ^{a}=\frac{1}{2}\eta ^{abcd}u_{b;c}u_{d}$ and $\theta =u^{a}{}_{;a}$ are the the vorticity vector and the expansion of the fluid, respectively, as measured by the observers $u^{a}$.

If we operate on (\ref{ME.10}) and (\ref{ME.11}) with $\eta^{abcd}u_{d}$, then a direct calculation yields the following two mathematical identities:
\begin{eqnarray}
E_{[r;s]} &=& u_{[r}\overset{.}{E}_{s]} +\overset{.}{u}_{[r}E_{s]} +u^{t}E_{t;[r}u_{s]} +\frac{1}{2}\eta_{rstm}u^{t}I^{m}(E) \label{ME.12} \\
H_{[r;s]} &=& u_{[r}\overset{.}{H}_{s]} +\overset{.}{u}_{[r}H_{s]} +u^{t}H_{t;[r}u_{s]} +\frac{1}{2}\eta_{rstm}u^{t}I^{m}(H). \label{ME.13}
\end{eqnarray}

From the identity (\ref{ME.13}), the screen rotation tensor of the magnetic field lines is
\begin{equation}
\mathcal{R}_{ab}= p_{a}^{c}p_{b}^{d}n_{[c;d]} = \frac{1}{H}%
p_{a}^{c}p_{b}^{d}H_{[c;d]}= -\frac{1}{2H}p_{a}^{c}p_{b}^{d} \eta_{cdrs}I^{r}(H)u^{s}.  \label{ME.23}
\end{equation}

\begin{proposition}
\label{Spacelike MHD Rotation Vector} The screen rotation vector of the magnetic field lines is proportional to the magnetic field, that is,
\begin{equation}
\mathcal{R}^{a}= -\frac{H_{b}I^{b}(H)}{2H^{3}}H^{a}  \label{ME.24}
\end{equation}
\end{proposition}

\begin{proof}
Expanding $p_{a}^{c}$ and $p_{b}^{d}$ in (\ref{ME.23}), we get
\begin{equation*}
\mathcal{R}_{ab}=-\frac{1}{2H}\eta _{abrs}I^{r}(H)u^{s}-\frac{1}{H^{3}}%
H_{[a}\eta _{b]crs}H^{c}I^{r}(H)u^{s}.
\end{equation*}

We operate with $\eta^{abpq}$ on both sides and find:%
\begin{eqnarray*}
-\frac{1}{2H}\eta ^{abpq}\eta _{abrs}I^{r}(H)u^{s} &=&\frac{1}{H}%
[I^{p}(H)u^{q}-I^{q}(H)u^{p}]  \\
-\frac{1}{H^{3}}\eta ^{abpq}H_{[a}\eta _{b]crs}H^{c}I^{r}(H)u^{s} &=&\frac{1%
}{H}\left[ -I^{p}(H)u^{q}+I^{q}(H)u^{p}\right] + \frac{H_{c}I^{c}(H)}{H^{3}}%
\left[ H^{p}u^{q}-H^{q}u^{p}\right] .
\end{eqnarray*}

Therefore,
\begin{equation}
\eta ^{abpq}\mathcal{R}_{ab}=\frac{H_{c}I^{c}(H)}{H^{3}}\left[
H^{p}u^{q}-H^{q}u^{p}\right] .  \label{ME.25}
\end{equation}

In terms of the screen rotation vector $\mathcal{R}^{a}=\frac{1}{2}\eta
^{abcd}u_{b}\mathcal{R}_{cd}$, equation (\ref{ME.25}) is written as
\begin{equation*}
\mathcal{R}^{a}=-\frac{H_{b}I^{b}(H)}{2H^{3}}H^{a}
\end{equation*}
which completes the proof.
\end{proof}

From Proposition \ref{Spacelike MHD Rotation Vector}, we infer that the screen rotation of the magnetic field congruence vanishes iff $H_{a}I^{a}(H)=0$.

\subsection{Maxwell equations in the RMHD approximation}

\label{section3.4}

In the RMHD approximation with infinite electric conductivity and vanishing electric field, the electromagnetic tensor (\ref{ME.3}) is $F^{ab}= \eta^{abcd} H_{c}u_{d}$ and Maxwell equations (\ref{ME.6}) - (\ref{ME.9}) become:
\begin{eqnarray}
h_{b}^{a}H^{b}{}_{;a} &=&0  \label{ME.16} \\
e&=&2\omega ^{a}H_{a}  \label{ME.17} \\
h_{b}^{a}\overset{.}{H}^{b} &=&u_{;b}^{a}H^{b}-\theta H^{a}  \label{ME.18} \\
I^{a}(H) &=&j^{a}.  \label{ME.19}
\end{eqnarray}

Let $n^{a}=H^{a}/H$ be the unit vector in the direction of the magnetic
field. Geometrically, $n^{a}$ is the unit tangent to the spacelike magnetic
field lines. The pair ($u^{a},n^{a})$ forms a double congruence. Maxwell
equations in terms of the irreducible parts defined by this double
congruence take a geometric form. The constraint (\ref{ME.16}) for the magnetic field gives $Hh_{b}^{a}n^{b}{}_{;a} +H_{,a}n^{a}=0$. But $h_{b}^{a}n^{b}{}_{;a}= p_{b}^{a}n^{b}{}_{;a}=\mathcal{E}$, where $\mathcal{E}$ is the screen expansion of the magnetic field lines. Therefore,
\begin{equation}
\mathcal{E}= -\left( \ln H \right)^{\ast}  \label{ME.20}
\end{equation}%
From this equation, we infer that the stronger the magnetic field, the denser the magnetic field lines on the screen space, that is, the greater is the magnetic flux through the screen space (as expected).

We examine now the propagation equation (\ref{ME.18}) of the magnetic field. We have
\begin{equation*}
\frac{\dot{H}}{H} n^{a} + h_{b}^{a} \overset{.}{n}^{b} =
u^{a}{}_{;b}n^{b}-\theta n^{a}.
\end{equation*}%
Contracting with $n^{a}$ and projecting with $p_{b}^{a}$, we get the pair of equations:%
\begin{eqnarray}
(\ln H)^{\cdot } &=&\sigma _{ab}n^{a}n^{b}-\frac{2}{3}\theta  \label{ME.21} \\
N^{a} &\equiv & p_{b}^{a}L_{\mathbf{u}}n^{b}=0.  \label{ME.22}
\end{eqnarray}

Equation (\ref{ME.21}) involves the change of the strength of the magnetic field
along the flow lines $u^{a}$ of the fluid. A kinematic
interpretation is that the vector $n^{a}$ is an eigenvector of the shear with eigenvalue $(\ln H)^{\cdot }+\frac{2}{3}\theta$.

Equation (\ref{ME.22}) is the geometric condition that the magnetic field lines are material lines in the fluid. This corresponds to the statement that the magnetic field is `frozen' \index{Magnetic field! frozen} along the fluid. \emph{Physically, this means that each particle of the fluid moves always on the same magnetic field line.}

Relation (\ref{ME.19}) due to (\ref{ME.24}) gives
\begin{equation}
\mathcal{R}^{a}=-\frac{H_{c}j^{c}}{2H^{3}}H^{a}.  \label{ME.26}
\end{equation}%
Therefore, in the RMHD approximation the screen rotation of the magnetic
field lines vanishes iff the conduction current $j^{a}$ is normal to the
magnetic field.

\textbf{Ohm's Law}\index{Law! Ohm} in its generalized form, which includes the Hall current, \index{Current! Hall} is written \cite{dunn}
\begin{equation}
J^{a}=e u^{a} +\underbrace{\frac{1}{\left( 1+\lambda ^{2}H^{c}H_{c} \right)}\left[ kE^{a}+\lambda k\eta^{abcd}E_{b}u_{c}H_{d} +\lambda^{2}k(E^{c}H_{c})H^{a}\right]}_{=j^{a}} \label{Ohm.9}
\end{equation}
where $k$ is the electric conductivity and $\lambda$ is the transverse conductivity (Hall effect).\index{Effect! Hall}

In the RMHD approximation, we have that the spatial part $j^{a}$ of the four-current $J^{a}$ vanishes; therefore, $\mathcal{R}^{a}=0$. Hence, in a \textbf{perfectly conducting fluid}, for which generalized Ohm's law applies, the
magnetic field lines have zero rotation as measured by $u^{a}$. Because (\ref{Ohm.9}) is not the most general form of Ohm's Law, we shall assume in the following $\mathcal{R}^{a}$ to be given by (\ref{ME.26}), that is, there exist additional terms in (\ref{Ohm.9}) such that $j^{a}\neq0$.

Summarizing, we have that in the RMHD approximation Maxwell equations are:
\begin{eqnarray}
\mathcal{E} &=&-(\ln H)^{\ast }  \label{ME.27} \\
\mathcal{R}^{a} &=&-\frac{H_{c}j^{c}}{2H^{3}}H^{a}  \label{ME.28} \\
e &=&2\omega ^{a}H_{a}  \label{ME.29} \\
(\ln H)^{\cdot } &=&\sigma _{ab}n^{a}n^{b}-\frac{2}{3}\theta  \label{ME.30}
\\
N^{a} &\equiv &p_{b}^{a}\mathcal{L}_{u}n^{b}=0.  \label{ME.31}
\end{eqnarray}
We note that equation (\ref{ME.31}) can be written in the equivalent form
\begin{equation}
p_{b}^{a}\overset{.}{n}^{b}=\left( p^{a}{}_{c}\sigma_{b}^{c} +\omega^{a}{}_{b}\right) n^{b}.  \label{ME.32}
\end{equation}

Equations (\ref{ME.27}) - (\ref{ME.31}) are general and independent of further simplifying assumptions (e.g. symmetry assumptions) we might do.

\section{The field equations for the EMSF}

\label{section4}

The EMSF must satisfy three sets of equations: a) Maxwell equations, b) Conservation laws, and c) Einstein field equations. We have already given in sec. \ref{section3} Maxwell equations and the conservation equations.

Concerning Einstein field equations, we shall consider their Lie derivative along some characteristic direction of the EMSF. The reason for this is that we want to employ symmetry assumptions, that is, equations of the form $L_{\boldsymbol{\xi}}M_{ab} = A_{ab}$, where $M_{ab}$ is a metrical tensor and $A_{ab}$ is an arbitrary tensor having the same symmetries as the $M_{ab}$. Due to the form of
Einstein field equations, we compute $L_{\boldsymbol{\xi}}R_{ab}$ in terms of
$L_{\boldsymbol{\xi}}g_{ab}$ using various identities of Riemannian Geometry. Then, we impose the symmetry assumption by choosing a specific form for $A_{ab}$. For example, if $\boldsymbol{\xi}$ is a CKV, then $M_{ab}=g_{ab}$ and $A_{ab}=2\psi g_{ab}$ where $\psi(x^{a})$ is the conformal factor. Next, we replace $L_{\boldsymbol{\xi}}R_{ab}$ in the Lie derivative of the field equations and we find the field equations in a form that incorporates already the imposed geometric symmetry assumption.

In a previous work on string fluids \cite{Tsamparlis M GRG 2006}, the author has computed Einstein equations for a perfect string fluid and many types of symmetries in the cases that the symmetry vector is either $\xi^{a}=\xi u^{a}$ or $\xi ^{a}=\xi n^{a}$. Therefore, we could write straight away the field equations in the case of an EMSF by simply specifying $n^{a}=\frac{H^{a}}{H}$. Of course in this case the resulting equations will be supplemented by Maxwell equations. In the following, we recall briefly some important intermediate steps in order to make the present work more readable and self-contained. Details can be found in \cite{Tsamparlis M GRG 2006}.

The Lie derivative of the Ricci tensor wrt a general time-like vector $\xi^{a}=\xi u^{a}$ has been computed (see eq. (3.9) in \cite{Saridakis
Tsamparlis 1991}) in terms of the standard dynamic variables $\mu
,p,q_{a},\pi _{ab}.$ By using this general expression, $L_{\boldsymbol{\xi}}R_{ab}$ can
be written in terms of the perfect string fluid parameters $\mu ,q$. In
a similar way, the Lie derivative of the Ricci tensor along the spacelike
vector $\xi ^{a}=\xi n^{a}$ can be expressed in terms of the 1+1+2 dynamic
quantities.

Using Maxwell equations (\ref{ME.27})-(\ref{ME.31}) in RMHD approximation, we show easily that the conservation equations (\ref{EqnSS.37}) - (\ref{EqnSS.39}) (which must also be
satisfied in all cases) are simplified as follows:
\begin{eqnarray}
\overset{.}{\rho }-(\rho +p_{\perp })(\ln H)^{\cdot } &=&0  \label{MC.9} \\
\overset{\ast }{\rho }-(\rho +p_{\perp })(\ln H)^{\ast } &=&0  \label{MC.10} \\
p_{a}^{b}\left[ p_{\perp ,b}+\lambda HH_{,b}+(\rho +p_{\perp} +\lambda H^{2})(\overset{.}{u}_{b}-\overset{\ast }{n}_{b})\right] &=&0.  \label{MC.11}
\end{eqnarray}

Concerning the Einstein filed equations, the Lie derivative of (\ref{eq.einstein3}) is \cite{Tsamparlis M GRG 2006, Saridakis Tsamparlis 1991} for a timelike vector $\xi^{a}=\xi u^{a}$
\begin{eqnarray}
\frac{1}{\xi }L_{\boldsymbol{\xi}}R_{ab} &=&\left[ \overset{.}{q}+2(q-\Lambda )(\ln \xi
)^{\cdot }\right] u_{a}u_{b}+2(q-\Lambda )\left[ \overset{.}{u}_{c}-(\ln \xi
)_{,c}\right] u_{(a}h_{b)}^{c} +  \notag \\
&&+\frac{1}{3}\left[ 2\overset{.}{\mu }-\overset{.}{q}+\frac{2}{3}(2\mu
-q+3\Lambda )\theta \right] h_{ab} +  \notag \\
&&+\left[ \overset{.}{\mu }+\overset{.}{q}+\frac{2}{3}(\mu +q)\theta %
\right] \left( \frac{1}{3}h_{ab}-n_{a}n_{b} \right) + \frac{2}{3}(2\mu
-q+3\Lambda )\sigma_{ab} +  \notag \\
&& + 2 (\mu +q) \left(\frac{1}{3} h_{cd} - n_{c} n_{d} \right)
\delta_{(a}^{d}(\omega_{.b)}^{c}+\sigma _{.b)}^{c}) -  \notag \\
&&-2(\mu +q)\overset{.}{n}_{d}h_{(a}^{d}n_{b)}  \label{N.33a}
\end{eqnarray}
and for a spacelike vector $\xi^{a}=\xi n^{a}$
\begin{eqnarray}
\frac{1}{\xi }L_{\boldsymbol{\xi}}R_{ab} &=&\left[ \overset{\ast }{q}+2(q-\Lambda )%
\overset{.}{u}_{c}n^{c}\right] u_{a}u_{b}-2(q-\Lambda )\left[ \overset{\ast }%
{u}_{c}n_{c}-(\ln \xi )^{\cdot }\right] u_{(a}n_{b)} -  \notag \\
&&-\left[ \overset{\ast }{q}+2(q-\Lambda )(\ln \xi )^{\ast }\right]
n_{a}n_{b} -2\left[ (\mu +\Lambda )N_{c}+2(q-\Lambda )\omega _{dc}n^{d}\right]
u_{(a}p_{b)}^{c} -  \notag \\
&&-2(q-\Lambda )p_{c}^{d}\left[ \overset{\ast }{n}_{d}+(\ln \xi )_{,d}\right]
n_{(a}p_{b)}^{c}+\left[ \overset{\ast }{\mu }+(\mu +\Lambda )\mathcal{E}%
\right] p_{ab} +2(\mu +\Lambda )\mathcal{S}_{ab}.  \notag \\
&& \label{N.34a}
\end{eqnarray}

\emph{Expressions (\ref{N.33a}) and (\ref{N.34a}) are general and hold for all collineations and all perfect string fluids.} The left-hand sides of the expressions (\ref{N.33a}) and (\ref{N.34a}) are specified by the type of the chosen collineations, whereas the right-hand sides by the type of the considered perfect string fluid. Equating the two parts, one finds immediately Einstein
field equations for the specific string fluid considered and the specific
symmetry assumed.

In the case of the EMSF, we have $\mu = \rho + \frac{1}{2}\lambda H^2$ and $q = p_{\perp} + \frac{1}{2} \lambda H^2$. This result applies to all collineations concerning the EMSF.

For easy reference we collect below the results of the calculations for an EMSF.
\bigskip

\textbf{Maxwell equations:}
\begin{eqnarray*}
N^{a} &=&0 \iff p_{b}^{a}\overset{.}{n}^{b} = \left( p_{.c}^{a}\sigma_{b}^{c}+\omega _{.b}^{a}\right) n^{b} \\
\mathcal{E} &\mathcal{=}&-(\ln H)^{\ast } \\
\sigma _{ab}n^{a}n^{b}-\frac{2}{3}\theta &=& \left( \ln H\right)^{\cdot} \\
e &=& 2 \omega^a H_a, \quad \mathcal{R}^a = - \frac{H_c j^c}{2H^3} H^a, \quad I^{a}(H)=j^{a}.
\end{eqnarray*}

\textbf{Conservation equations:}
\begin{eqnarray*}
\overset{.}{\rho }-(\rho +p_{\perp })(\ln H)^{\cdot } &=&0 \\
\overset{\ast }{\rho }-(\rho +p_{\perp })(\ln H)^{\ast } &=&0 \\
p_{a}^{b}\left[ p_{\perp }{}_{,b}+\lambda HH_{,b}+(\rho +p_{\perp }+\lambda H^{2})(\overset{.}{u}_{b}-\overset{\ast}{n}_{b})\right] &=&0.
\end{eqnarray*}

\textbf{Einstein field equations:}
\begin{eqnarray}
\frac{1}{\xi }L_{\boldsymbol{\xi}}R_{ab} &=&\left[ \overset{.}{p_{\perp }}+\lambda H%
\overset{.}{H}+2\left( p_{\perp }+\frac{1}{2}\lambda H^{2}-\Lambda \right)
(\ln \xi )^{\cdot }\right] u_{a}u_{b} + \notag \\
&& +2\left( p_{\perp }+\frac{1}{2}\lambda
H^{2}-\Lambda \right) \left[ \overset{.}{u}_{c}-(\ln \xi )_{,c}\right]
u_{(a}h_{b)}^{c}+  \notag \\
&&+\frac{1}{3}\left[ 2\overset{.}{\rho }-\overset{.}{p}_{\perp }+\lambda H%
\overset{.}{H}+\frac{2}{3}\left( 2\rho -p_{\perp }+\frac{1}{2}\lambda
H^{2}+3\Lambda \right) \theta \right] h_{ab}+  \notag \\
&&+\left[ \overset{.}{\rho }+\overset{.}{p}_{\perp }+2\lambda H\overset{.}{H} +\frac{2}{3}\left( \rho +p_{\perp }+\lambda H^{2} \right) \theta \right] \left( \frac{1}{3}%
h_{ab}-n_{a}n_{b}\right) + \notag \\
&& +2 \left( \rho +p_{\perp }+\lambda H^{2} \right) \left( \frac{1}{3}h_{cd}-n_{c}n_{d}\right) \delta _{(a}^{d}(\omega _{.b)}^{c}+\sigma _{.b)}^{c})+  \notag \\
&&+\frac{2}{3}\left( 2\rho -p_{\perp }+\frac{1}{2}\lambda H^{2}+3\Lambda \right) \sigma _{ab} -\notag \\
&& -2\left( \rho +p_{\perp }+\lambda H^{2} \right) \overset{.}{n}_{d}h_{(a}^{d}n_{b)}, \quad \xi^{a}=\xi u^{a} \label{N.35a}
\end{eqnarray}
and
\begin{eqnarray}
\frac{1}{\xi }L_{\boldsymbol{\xi}}R_{ab} &=&\left[ \overset{\ast }{p_{\perp }}+\lambda H%
\overset{\ast }{H}+2\left( p_{\perp }+\frac{1}{2}\lambda H^{2}-\Lambda
\right) \overset{.}{u}_{c}n^{c}\right] u_{a}u_{b} -\notag \\
&& -2\left( p_{\perp }+\frac{1}{2}%
\lambda H^{2}-\Lambda \right) \left[ \overset{\ast }{u}_{c}n^{c}-(\ln \xi )^{\cdot }%
\right] u_{(a}n_{b)}-  \notag \\
&&-\left[ \overset{\ast }{p_{\perp }}+\lambda H\overset{\ast }{H}+2\left(
p_{\perp }+\frac{1}{2}\lambda H^{2}-\Lambda \right) (\ln \xi )^{\ast }\right]
n_{a}n_{b}-  \notag \\
&& -2\left( p_{\perp }+\frac{1}{2}\lambda H^{2}-\Lambda \right) p_{c}^{d}%
\left[ \overset{\ast }{n}_{d}+(\ln \xi )_{,d}\right] n_{(a}p_{b)}^{c}- \notag \\
&& -4\left( p_{\perp }+\frac{1}{2}\lambda H^{2}-\Lambda \right) \omega
_{dc}n^{d}u_{(a}p_{b)}^{c} +2\left( \rho +\frac{1}{2}\lambda H^{2}+\Lambda \right) \mathcal{S}_{ab} + \notag \\
&& +\left[
\overset{\ast }{\rho }+\lambda H\overset{\ast }{H} +\left( \rho +\frac{1}{2}%
\lambda H^{2}+\Lambda \right) \mathcal{E}\right] p_{ab}, \quad \xi^{a}=\xi n^{a}.
\label{N.36a}
\end{eqnarray}

For a double congruence pair $(u^a,n^a)$, we have: a) For the timelike congruence $u^a$, the kinematic quantities $\theta$, $\dot{u}_a$, $\sigma_{ab}$, $\omega_{ab}$; and b) For the spacelike congruence $n^a$. the kinematic quantities $\mathcal{E}$, $\dot{n}_a$, $\overset{\ast}{n}_a$, $\overset{\ast}{u}_a$, $\mathcal{S}_{ab}$, $\mathcal{R}_{ab}$.

\subsection{Decomposition of $L_{\boldsymbol{\xi}} g_{ab}$ along $u^a$ and $n^a$}

\label{section4.1}

Recall the general decomposition (see sec. \ref{sub.dec.Lgij})
\begin{equation} \label{eq.Lgab1}
L_{\boldsymbol{\xi}} g_{ab} = 2 \psi g_{ab} + 2 H_{ab}
\end{equation}
where $\psi = \frac{1}{4} g^{ab} \xi_{(a;b)} = \frac{1}{4} \xi^a{}_{;a}$ and $H_{ab} = \xi_{(a;b)} - \frac{1}{4} \xi^c{}_{;c} g_{ab}$. We consider the following cases:
\bigskip

\underline{i) $\xi^a = \xi u^a$, $\xi > 0$:}
\begin{equation} \label{eq.Lgab2}
\psi = \frac{1}{4} \xi \left[ (\ln\xi)^{\cdot} + \theta \right]
\end{equation}
and
\begin{equation} \label{eq.Lgab3}
H_{ab} = \xi \left\{ \sigma_{ab} + \frac{1}{3} \theta h_{ab} - \dot{u}_{(a} u_{b)} + (\ln\xi)_{,(a} u_{b)} - \frac{1}{4} \left[ (\ln\xi)^{\cdot} + \theta \right] g_{ab} \right\}.
\end{equation}

\underline{ii) $\xi^a = \xi n^a$, $\xi > 0$:}
\begin{equation} \label{eq.Lgab4}
\psi = \frac{1}{4} \xi \left[ (\ln\xi)^{\ast} - \dot{n}^a u_a + \mathcal{E} \right]
\end{equation}
and
\begin{align}
H_{ab} & = \xi \left\{ \mathcal{S}_{ab} + \frac{1}{2} \mathcal{E} p_{ab} + \overset{\ast}{n}_{(a} n_{b)} - \dot{n}_{(a} u_{b)} + u_{(a} p^c_{b)} \left( \dot{n}_c + 2 \omega_{dc} n^d - N_c \right) + \right. \nonumber \\
& \quad \left. + (\ln\xi)_{,(a} n_{b)} - \frac{1}{4} \left[ \mathcal{E} + (\ln\xi)^{\ast} - \dot{n}^c u_c \right] g_{ab} \right\}. \label{eq.Lgab5}
\end{align}

Moreover, recall that for an arbitrary vector field $\xi^{a}$, we can always select $n^{a}$ such that
\[
\xi^{a}= \xi(u) u^{a} +\xi(n) n^{a}.
\]
Hence, the Lie derivative $L_{\boldsymbol{\xi}}M_{ab}$ can be always expressed as
\[
L_{\boldsymbol{\xi}}M_{ab}= L_{\xi(u)u^{a}}M_{ab} +L_{\xi(n)n^{a}}M_{ab}
\]
which means that we have two components. We continue by studying the special cases $\xi(n)=0$ (i.e. $\xi^{a}$ is timelike), and $\xi(u)=0$ (i.e. $\xi^{a}$ is spacelike).

\section{The role of the symmetry assumption}

\label{section5}

As it has been mentioned, there are two types of equations constraining the evolution of a gravitational system which admits a symmetry. These are the kinematic conditions and the dynamic equations.

The kinematic conditions are equations among the kinematic variables of the
gravitational system which result from geometric identities and additional
geometric assumptions (such as symmetries). However, this is not the case with the dynamic equations which do not
necessarily inherit the kinematic symmetries of the system. In the following,
we consider two types of symmetries (a) CKVs defined by timelike vectors $%
\xi ^{a}=\xi u^{a}$ $(\xi > 0)$, and (b) CKVs defined by the spacelike
vectors $\xi ^{a}=\xi n^{a}$ $(\xi > 0).$ Since in both cases $\xi^{a}$ is a CKV, it holds that $L_{\boldsymbol{\xi}}g_{ab}= 2\psi g_{ab}$.

\section{The EMSF in spacetimes admitting a timelike CKV $\protect\xi^{a} = \protect\xi u^{a}
\enskip (\protect\xi > 0)$}

\label{section6}

We look first on the kinematic implications of the assumed symmetry and, then, on the dynamical ones.

\subsection{The kinematic implications}

\label{section6.1}

From previous works \cite{Norris 1977}, we have the following kinematic conditions for a CKV $\xi^{a}=\xi u^{a}$.

\begin{proposition}
\label{Timelike CKV Kinematics.1} A fluid spacetime $(u^{a}, g_{ab})$ admits a CKV $\xi ^{a}=\xi u^{a}$ iff
\begin{enumerate}
\item
$\sigma _{ab}=0$

\item
$\overset{.}{u}_{a}=(\ln \xi )_{,a}+\frac{1}{3}\theta u_{a}$, where $\sigma _{ab},\theta$ and $\overset{.}{u}^{a}$ are, respectively, the shear, expansion and four-acceleration of the timelike congruence generated by $u^{a}$. The conformal factor $\psi =\frac{1}{3}\xi \theta = \dot{\xi}$.
\end{enumerate}
\end{proposition}

The conditions imposed by Proposition \ref{Timelike CKV Kinematics.1} supplement Maxwell equations and simplify the conservation equations. Because $\sigma_{ab}=0$, the `energy' conservation equation (\ref{MC.9}) gives
\begin{equation}
\overset{.}{\rho }+\frac{2}{3}(\rho +p_{\perp })\theta =0.  \label{EqnSS.40}
\end{equation}

Equation (\ref{MC.10}) remains the same and equation (\ref{MC.11}) becomes:
\begin{equation}
p_{a}^{b}\left\{ p_{\perp }{}_{,b}+\lambda HH_{,b}+(\rho +p_{\perp }+\lambda H^{2})\left[(\ln \xi )_{,b}-\overset{\ast }{n}_{b}\right]\right\} =0.  \label{EqnSS.41}
\end{equation}

Eventually, the conservation equations (\ref{MC.9}) - (\ref{MC.11}) reduce to equations (\ref{MC.10}), (\ref{EqnSS.40}) and (\ref{EqnSS.41}).

\subsection{The dynamic implications}

\label{section6.2}

For a CKV $\xi^{a}=\xi u^{a}$, we have the identity
\begin{equation*}
L_{\boldsymbol{\xi}}R_{ab}= -2\psi _{;ab}-g_{ab}\square \psi.
\end{equation*}
The 1+3 decomposition of $\psi_{;ab}$ wrt $u^{a}$ is (note that $\psi_{;ab}= \psi_{;ba}$)
\begin{equation}
\psi _{;ab}=\lambda _{\psi }u_{a}u_{b}+p_{\psi }h_{ab}+2q_{\psi
(a}u_{b)}+\pi _{\psi ab}  \label{N.5}
\end{equation}%
where
\begin{equation}
\lambda _{\psi }=\psi _{;ab}u^{a}u^{b},\;p_{\psi }=\frac{1}{3}h^{ab}\psi_{;ab},\;q_{\psi a}=-h_{a}^{b}\psi _{;bc}u^{c},\;\pi_{\psi ab}=
\left( h_{a}^{c}h_{b}^{d}-\frac{1}{3}h_{ab}h^{cd} \right) \psi_{;cd}.
\label{N.6}
\end{equation}
We also compute
\begin{equation}
\square \psi =g^{ab}\psi _{;ab}=-\lambda _{\psi }+3p_{\psi}.  \label{N6.a}
\end{equation}

Therefore, for a CKV $\xi ^{a}=\xi u^{a}$, we have that
\begin{equation}
L_{\boldsymbol{\xi}}R_{ab}=-\left[ 3(\lambda _{\psi }-p_{\psi })u_{a}u_{b}-(\lambda
_{\psi }-5p_{\psi })h_{ab}+4q_{\psi (a}u_{b)}+2\pi _{\psi ab}\right].
\label{N6.b}
\end{equation}

Using the kinematic conditions and the conservation equations, the right-hand side of equation (\ref{N.35a}) simplifies as follows\footnote{%
It is easy to show (use Maxwell equations in RMHD approximation) that $\overset{.}{n}_{d}h_{(a}^{d}n_{b)} +n_{c}n_{d} \delta_{(a}^{d} \omega_{.b)}^{c}=0$.}:
\begin{eqnarray}
\frac{1}{\xi }L_{\boldsymbol{\xi}}R_{ab} &=&\left[ \overset{.}{p_{\perp }}+\frac{1}{2}%
\lambda H\overset{.}{H}+2(p_{\perp }-\Lambda )\frac{1}{3}\theta \right]
u_{a}u_{b}-\frac{1}{3}\left[ \overset{.}{p}_{\perp }+2(p_{\perp }-\Lambda
)\theta -\frac{1}{2}\lambda H\overset{.}{H}\right] h_{ab}+  \notag \\
&& +\left[ \overset{.}{p}_{\perp }+\lambda H\overset{.}{H}\right] \left( \frac{%
1}{3} h_{ab}-n_{a}n_{b} \right).  \label{N6.c}
\end{eqnarray}

From (\ref{N6.b}) and (\ref{N6.c}), we find that the field equations for an EMSF admitting the CKV $\xi ^{a}=\xi u^{a}$ are
\begin{eqnarray*}
&& \left[ \overset{.}{p_{\perp }}+\frac{1}{2}\lambda H\overset{.}{H}%
+2(p_{\perp }-\Lambda )\frac{1}{3}\theta \right] u_{a}u_{b}-\frac{1}{3}\left[
\overset{.}{p}_{\perp }+2(p_{\perp }-\Lambda )\theta -\frac{1}{2}\lambda H%
\overset{.}{H}\right] h_{ab}+ \\
&&+\left[ \overset{.}{p}_{\perp }+\lambda H\overset{.}{H}\right] \left(
\frac{1}{3} h_{ab}-n_{a}n_{b} \right)= \\
&=&-\frac{1}{\xi }\left[ 3(\lambda _{\psi }-p_{\psi
})u_{a}u_{b}-(\lambda_{\psi } - 5 p_{\psi }) h_{ab} + 4 q_{\psi(a} u_{b)} +
2 \pi_{\psi ab}\right].
\end{eqnarray*}
This relation implies the field equations\footnote{%
The same equations are found from \cite{Tsamparlis M GRG 2006} where the
field equations were:
\par
{}%
\begin{eqnarray*}
\overset{.}{q} &=& - \frac{3}{\xi} (p_{\psi} + \lambda_{\psi }) \\
(q - \Lambda) \theta &=& \frac{9}{\xi} p_{\psi} \\
3(p_{\psi }+\lambda _{\psi })\left( \frac{1}{3}h_{ab}-n_{a}n_{b} \right)
&=&2\pi _{\psi ab}. \\
q_{\psi }^{a} &=&0
\end{eqnarray*}%
}:
\begin{eqnarray}
\overset{.}{p_{\perp }}+\lambda H\overset{.}{H} &=& - \frac{3}{\xi} (p_{\psi
}+\lambda_{\psi })  \label{EqnSS.45} \\
\left( p_{\perp }+\frac{1}{2}\lambda H^{2}-\Lambda \right) \theta &=& \frac{9%
}{\xi} p_{\psi }  \label{EqnSS.46} \\
3(p_{\psi }+\lambda _{\psi }) \left( \frac{1}{3}h_{ab}-n_{a}n_{b} \right)
&=& 2 \pi_{\psi ab}  \label{EqnSS.47} \\
q_{\psi }^{a} &=&0.  \label{EqnSS.48}
\end{eqnarray}

Equation (\ref{EqnSS.46}), using that\footnote{
By multiplying the kinematic condition $\dot{u}_{a}= \left(\ln\xi\right)_{,a} +\frac{1}{3}\theta u_{a}$ with $u^{a}$ and using the result
\[
u^{a}u_{a}=-1 \implies u^{a}{}_{;b}u^{b} u_{a} +u^{a}(g_{ac}u^{c})_{;b}u^{b}=0 \implies \dot{u}^{a}u_{a}=0, \enskip (g_{ab;c}=0)
\]
we find that $(ln\xi )^{\cdot }=\frac{\theta}{3}$.
} $(ln\xi )^{\cdot }=\frac{1}{3}%
\theta $, gives
\begin{equation}
p_{\psi }=\frac{1}{3} \left( p_{\perp }+\frac{1}{2}\lambda H^{2}-\Lambda
\right) \dot{\xi}  \label{EqnSS.50}
\end{equation}
which when substituted into (\ref{EqnSS.45}) yields
\begin{eqnarray}
\left( p_{\perp }+\frac{1}{2}\lambda H^{2}-\Lambda \right)^{\cdot } &=& -
\left( p_{\perp }+\frac{1}{2}\lambda H^{2}-\Lambda \right)(ln\xi )^{\cdot
}-3\lambda _{\psi } \implies  \notag \\
\lambda_{\psi } &=&-\frac{1}{3}\left[ \left( p_{\perp }+\frac{1}{2}\lambda
H^{2}-\Lambda \right) \xi \right] ^{\cdot }.  \label{EqnSS.49}
\end{eqnarray}

The final set of equations which results from the assumption that the EMSF admits the CKV $\xi ^{a}=\xi u^{a}$ is the following:
\bigskip

Geometric implications:
\begin{eqnarray*}
\sigma _{ab} &=&0 \\
\overset{.}{u}_{a} &=&(\ln \xi )_{,a}+\frac{1}{3}\theta u_{a}.
\end{eqnarray*}

Maxwell equations:
\begin{eqnarray}
N^{a} &=&0  \iff p_{b}^{a}\dot{n}^{b} =\omega^{a}{}_{b} n^{b}  \label{EqnSS.51} \\
\mathcal{E} &=&-(\ln H)^{\ast }  \label{EqnSS.52} \\
-\frac{2}{3}\theta &=&(\ln H)^{\cdot }  \label{EqnSS.54} \\
e &=& 2 \omega^a H_a, \enskip \mathcal{R}^a= - \frac{H_b j^b}{2 H^3} H^a, \enskip I^a(H) = j^a. \label{EqnSS.53.new}
\end{eqnarray}

Conservation equations:
\begin{eqnarray}
\overset{.}{\rho }+\frac{2}{3}(\rho +p_{\perp })\theta &=&0  \label{EqnSS.55}
\\
\overset{\ast }{\rho }-(\rho +p_{\perp })(\ln H)^{\ast } &=&0
\label{EqnSS.56} \\
p_{a}^{b}\left\{ p_{\perp }{}_{,b}+\lambda HH_{,b}+(\rho +p_{\perp }+\lambda H^{2})\left[(\ln \xi )_{,b}-\overset{\ast}{n}_{b} \right] \right\} &=&0.  \label{EqnSS.57}
\end{eqnarray}

Gravitational field equations:
\begin{eqnarray}
\lambda _{\psi } &=&-\frac{1}{3}\left[ \left( p_{\perp }+\frac{1}{2}\lambda
H^{2}-\Lambda \right) \xi \right] ^{\cdot }  \label{EqnSS.58} \\
p_{\psi } &=&\frac{1}{3} \left( p_{\perp }+\frac{1}{2}\lambda H^{2}-\Lambda
\right) \dot{\xi}  \label{EqnSS.59} \\
2\pi_{\psi ab} &=& - \xi \left( p_{\perp }+\frac{1}{2}\lambda H^{2}-\Lambda
\right)^{\cdot } \left( \frac{1}{3}h_{ab}-n_{a}n_{b} \right)
\label{EqnSS.60} \\
q_{\psi }^{a} &=&0.  \label{EqnSS.61}
\end{eqnarray}

\subsection{The case of an EMSF admitting a timelike CKV $\protect\xi ^{a}=%
\protect\xi u^{a}$ in the FRW spacetime}

\label{section6.3}

We apply the results of the last section in the case of the FRW spacetime.\index{Spacetime! FRW}
The FRW spacetime has metric (in conformal coordinates):%
\begin{equation}
ds^{2}=R^{2}(\tau )\left[ -d\tau ^{2}+U^{2}(x^{\mu }) d\sigma_{E}^{2}\right]
\label{FRW.1}
\end{equation}
where $d\sigma_{E}^{2}$ is the Euclidean 3d metric and the function $U^{2}(x^{\mu})=\left( 1+\frac{k}{4}\mathbf{x}\cdot
\mathbf{x}\right) ^{-1}$ with $k=0 ,\pm 1.$ This metric admits the gradient
CKV $\partial _{\tau }$ whose conformal factor is $\psi = \frac{dR}{d\tau}$.
We define the timelike unit vector $u^{a}=\frac{1}{R}\partial _{\tau }.$ If
we define the new coordinate $t$ by the requirement
\begin{equation}
d\tau =\frac{1}{R(t)}dt  \label{FRW.2}
\end{equation}%
then the metric is written as
\begin{equation}
ds^{2}=-dt^{2}+R^{2}(t)U^{2}(x^{\mu })d\sigma _{E}^{2}  \label{FRW.3}
\end{equation}%
and the unit vector $u^{a}=\partial _{t}.$ The conformal factor becomes
\begin{equation}
\psi =\frac{dR}{dt}\equiv \overset{.}{R}(t). \label{FRW.4}
\end{equation}
Then\footnote{
In the coordinates $(t,x,y,z)$, we have $g_{ab}= diag \left(-1, R^{2}U^{2}, R^{2}U^{2}, R^{2}U^{2} \right)$, that is, $g^{ab}= diag \left(-1, R^{-2}U^{-2}, R^{-2}U^{-2}, R^{-2}U^{-2} \right)$. The vector $u^{a}=(1,0,0,0)$, $u_{a}=(-1,0,0,0)$, and the timelike CKV $\xi^{a}=Ru^{a}$ with $\xi=R$.
},
\[
\psi_{;ab}= \psi_{,ab} -\psi_{,c}\{^{c}_{ab}\}= \overset{...}{R} \delta^{0}_{a} \delta^{0}_{b} -\psi_{,0}\{^{0}_{ab}\}= \overset{...}{R}u_{a}u_{b} -\frac{1}{2}\ddot{R}g_{ab,0}
\]
where
\[
\{^{0}_{ab}\}= \frac{1}{2}g^{0c} \left( g_{ac,b} +g_{bc,a} -g_{ab,c} \right)= \frac{1}{2}g^{00} \left( g_{a0,b} +g_{b0,a} -g_{ab,0} \right)= \frac{1}{2}g_{ab,0}.
\]
For $\mu=1,2,3$, we find $\{^{0}_{\mu\mu}\}= \frac{1}{2}g_{\mu\mu,0}= U^{2}R\dot{R}$. From 1+3 decomposition over $\psi_{;ab}$, we obtain
\[
\lambda_{\psi}= \overset{...}{R}, \enskip p_{\psi}= -\frac{\dot{R}\ddot{R}}{R}, \enskip q^{a}_{\psi}=0, \enskip \pi_{\psi ab}= -\frac{1}{2}\ddot{R}g_{ab,0} +\frac{\dot{R}\ddot{R}}{R} h_{ab}.
\]
Equation (\ref{EqnSS.59}) gives ($\dot{R}\neq0$)
\[
\ddot{R}=-\frac{1}{3} \left( p_{\perp} +\frac{1}{2}\lambda H^{2} -\Lambda \right)R.
\]
Then, $p_{\perp} +\frac{1}{2}\lambda H^{2} -\Lambda= 0$ is a possible assumption which implies that $\ddot{R}=0$; thus, $R(t)= c_{1}t +c_{2}$ and $\psi=\dot{R}=c_{1}$ where $c_{1}, c_{2}$ are arbitrary real constants.

For a different 1+3 decomposition over $\psi_{;ab}$, we have
\begin{equation}
\lambda _{\psi }=\overset{...}{R}, \enskip p_{\psi }=0, \enskip q_{\psi a}=0, \enskip \pi_{\psi ab}= -\frac{1}{2}\ddot{R} g_{ab,0}.  \label{FRW.5}
\end{equation}
Then, field equation (\ref{EqnSS.59}) gives ( $\dot{R}\neq 0$)
\begin{equation*}
p_{\perp }=-\frac{1}{2}\lambda H^{2}+\Lambda
\end{equation*}%
and equation (\ref{EqnSS.58}) implies that $\lambda_{\psi}= \overset{...}{R}$ $=0.$ We deduce the following: \newline
1) $\partial _{\tau }=R(t)u^{a}$ is a SCKV, or one of its
specializations (If $\psi =0 \implies R(t)=const$ the spacetime reduces to an Einstein space). \newline
2) The conformal factor is $\psi =c_{1}t +c_{2}$. \newline
3) The spacetime admits the gradient SCKV $\psi_{,a}= c_{1}\delta_{a}^{t} $. \newline
4) The scale factor\index{Scale factor} $R(t)=\frac{1}{2}c_{1}t^{2}+c_{2}t+c_{3}$ where $c_{1}, c_{2}$ and $c_{3}$ are arbitrary real constants.

We work now with the conservation equations (\ref{EqnSS.55}) - (\ref{EqnSS.57}). Equation (\ref{EqnSS.55}), using (\ref{EqnSS.54}) and $p_{\perp }=-\frac{1}{2}\lambda H^{2} +\Lambda $, gives
\begin{eqnarray}
\overset{.}{\rho }-\left( \rho -\frac{1}{2}\lambda H^{2}+\Lambda \right)
(\ln H)^{\cdot } &=&0\implies \overset{.}{\rho }-(\rho +\Lambda )(\ln
H)^{\cdot }+\frac{1}{2}\lambda H^{2}(\ln H)^{\cdot }=0\implies  \notag \\
\frac{\overset{.}{\rho }}{\rho +\Lambda }-\frac{\overset{.}{H}}{H}+\frac{1}{2%
}\lambda \frac{H}{\rho +\Lambda }\overset{.}{H} &=&0\implies \left( \ln
\frac{\rho +\Lambda }{H}\right) ^{\cdot }+\frac{1}{2}\lambda \frac{H}{\rho
+\Lambda }\overset{.}{H}=0\implies  \notag \\
\left( \frac{\rho +\Lambda }{H}\right) ^{\cdot }+\frac{1}{2}\lambda \overset{%
.}{H} &=&0\implies  \notag \\
\left( \rho +\frac{1}{2}\lambda H^{2}\right) ^{\cdot } &=&0.  \label{FRW.6}
\end{eqnarray}
This equation means that the energy of the EMSF is constant
along the flow lines of the observers.

Working similarly, we show that equation (\ref{EqnSS.56}) becomes
\begin{equation}
\left( \rho +\frac{1}{2}\lambda H^{2}\right) ^{\ast }=0  \label{FRW.7}
\end{equation}%
which implies that the total energy of the EMSF is also conserved along the
magnetic field lines.

Both these results are compatible with: \newline
a. The fact that the \emph{magnetic field lines are frozen along the flow lines of the fluid} (there is no relative motion of the two sets of lines) due to the condition $N^{a}=0$. \newline
b. The dynamic equation $q_{\psi}^{a}=0$, i.e. there is no heat flux wrt the observers $u^{a}.$

There remains equation (\ref{EqnSS.57}). Taking into account the fact that $n^{a}u_{a}=0 \implies n^{0}=0$ and $\xi= R(t)$ (hence $p_{b}^{a}\xi _{,a}=0)$, we find that this equation becomes
\begin{equation*}
\left( \rho +\Lambda +\frac{1}{2}\lambda H^{2}\right) p_{a}^{b}\overset{\ast}{n^{a}}=0.
\end{equation*}

Because the total energy of the fluid (including the cosmological constant) is considered to be positive, the above equation gives the condition $p_{a}^{b}\overset{\ast}{n^{a}}=0$. This is a dynamical equation which involves the magnetic field only. However, we also have the kinematical identity ($n^{a}u_{a}=0$)
\[
\theta_{ab}= \sigma_{ab} +\frac{\theta}{3}h_{ab} \implies h^{c}_{a}h^{d}_{b}u_{(c;d)}n^{a}n^{b}= \sigma_{ab}n^{a}n^{b} +\frac{\theta}{3}h_{ab}n^{a}n^{b} \implies \overset{\ast}{n}^{a} u_{a} = -\sigma_{ab} n^{a} n^{b} - \frac{\theta}{3}.
\]
But since $\sigma_{ab} = 0$, we find that $\overset{\ast}{n}^{a} u_{a} = - \frac{\theta}{3}$ which from $p_{a}^{b}\overset{\ast}{n}^{a}=0$ gives
\begin{equation}
\left( \delta^{b}_{a}  +u^{b}u_{a} -n^{b}n_{a} \right)\overset{\ast}{n}^{a}=0 \implies \overset{\ast}{n}^{a} = \frac{\theta}{3}u^a = (\ln R)^{\cdot} u^a. \label{frw0}
\end{equation}

Eventually, we have that the magnetic field lines are carried along with the fluid so that the total energy density (i.e. the fluid energy and the magnetic field energy) remains constant. Furthermore, the fluid does not heat.

\emph{The magnetic field lines are coplanar with the fluid lines, but they are not Lie transported along these lines except in the case of Minkowski spacetime.} Indeed, from the condition $N^{a}=0$ we have $L_{u}n^{a}=f_{1}u^{a}+f_{2}n^{a}$, where
$f_{1}$ and $f_{2}$ are quantities which have to be computed. From the definition of the Lie derivative, we have $L_{u}n^{a}= \overset{.}{n}^{a}-\overset{\ast}{u}{}^{a}$. Therefore,
\begin{equation*}
\overset{.}{n}^{a}-\overset{\ast }{u}^{a} =f_{1}u^{a} +f_{2}n^{a}.
\end{equation*}
Contracting in turn with $u^{a}$ and $n^{a}$, we have the following (recall proposition \ref{Timelike CKV Kinematics.1} and equation (\ref{frw0}) ):
\begin{eqnarray*}
f_{1}&=& -\dot{n}^{a}u_{a} = \dot{u}_{a}n^{a} =(\ln\xi)^{\ast} =(\ln R)^{\ast} \\
f_{2}&=& -\overset{\ast }{u}{}^{a}n_{a} =\overset{\ast}{n}_{a}u^{a} = -\frac{\theta}{3}= -(\ln R)^{\cdot}.
\end{eqnarray*}
Therefore,
\begin{equation}
L_{u}n^{a}=(\ln \xi)^{\ast}u^{a} -\frac{\theta}{3} n^{a} \label{frw1}
\end{equation}
which proves our assertion. From (\ref{frw1}), it follows that $p_{b}^{a}L_{u}n^{a}=0$, that is, $N^{a}=0$.

Concerning the magnetic field, we have (use equations (\ref{frw1}) and  (\ref{EqnSS.54} )
\begin{eqnarray}
L_{u}H^{a}&=& L_{u}(Hn^{a})= \dot{H}n^{a} +H L_{u}n^{a} \notag \\
&=& \frac{\dot{H}}{H}H^{a} +(\ln \xi)^{\ast}Hu^{a} -\frac{\theta}{3} H^{a} \notag \\
&=& (\ln H)^{\cdot}H^{a} +(\ln \xi)_{,b} H^b u^{a} -\frac{\theta}{3} H^{a} \notag \\
&=& (\ln \xi)_{,b} H^b u^{a} -\theta H^{a}. \label{frw2}
\end{eqnarray}

\section{The EMSF in spacetimes admitting a spacelike CKV $\ \protect\xi^{a}=\protect\xi n^{a}$}

\label{section7}

We derive again the kinematic and the dynamic equations as we did in section \label{section6} for the case of $\xi ^{a}=\xi u^{a}.$

\subsection{The kinematic implications}

\label{section7.1}

For a double congruence $u^{a}$ and $n^{a}$, one has the kinematic quantities $\theta$, $\dot{u}_{a}$, $\sigma_{ab}$, $\omega_{ab}$ for the timelike congruence $u^{a}$, and the kinematic quantities $\mathcal{E}$, $\overset{.}{n}_{a}$, $\overset{\ast}{n}_{a}$, $\overset{\ast}{u}_{a}$, $\mathcal{S}_{ab}$, $\mathcal{R}_{ab}$ for the spacelike congruence $n^{a}$. Therefore, the kinematic restrictions, in this case, involve in general all these quantities plus the parameters $\psi$
and $H_{ab}$, and their derivatives. To find the kinematic conditions
resulting from a collineation relative to a double congruence, we need the
1+1+2 decomposition of $H_{ab}$. To do that, we consider the symmetry defining equation and contract it to get (see sec. \ref{section4.1})
\begin{equation*}
\psi= \frac{\xi}{4}\left[ \mathcal{E} +(\ln \xi)^{\ast} -\overset{.}{n}^{c}u_{c}\right].
\end{equation*}

For the case of a CKV\ $\xi ^{a}=\xi n^{a}$, we find the following kinematic conditions \cite{Tsamparlis Mason 1990}.

\begin{proposition}
\label{Spacelike CKV Kinematic} A fluid spacetime $(u^{a}, g_{ab})$ with a spacelike congruence $n^{a}$ ($u^{a}n_{a}=0$) admits the spacelike CKV\footnote{$\xi$ is not necessarily equal to $H$!} $\xi^{a}=\xi n^{a}$ ($\xi >0)$ iff
\begin{eqnarray}
\mathcal{S}_{ab} &=& 0  \label{KC.18} \\
\dot{n}_{a}u^{a} &=& -\frac{\mathcal{E}}{2} \label{KC.19} \\
\overset{\ast}{n}^{a} &=& (\ln \xi)^{\cdot}u^{a} -p^{ab}(\ln \xi)_{,b} \label{KC.20} \\
N_{a} &=& -2\omega_{ab}n^{b}.  \label{KC.21}
\end{eqnarray}
\end{proposition}

The conformal factor is
\begin{equation}
\psi= \frac{1}{2}\xi\mathcal{E}= \overset{\ast}{\xi}  \label{KC.22}
\end{equation}
which implies that $(\ln \xi)^{\ast}= \frac{\mathcal{E}}{2}$. Then, from (\ref{KC.20}), we find that
\begin{equation}
\overset{\ast}{n}_{a}= \frac{\mathcal{E}}{2}n_{a} -(\ln\xi)_{,a}, \enskip \overset{\ast}{n}^{a}u_{a}= -(\ln\xi)^{\cdot}. \label{KC.22.1}
\end{equation}

Replacing the defining relations $N_{a}= p_{ab} \left(\dot{n}^{b} -\overset{\ast}{n}^{b}\right)$ and $\omega_{ab}= h^{c}_{a} h^{d}_{b} u_{[c;d]}$ into (\ref{KC.21}), we find the identity $p^{ab} (\dot{n}_b + u^c n_{c;b}) = 0$. It also holds that
\[
\omega^a = (\omega^b n_b) n^a + \frac{1}{2} \eta^{abcd} (h^r_b \dot{n}_r - \overset{\ast}{u}_b) u_c n_d \iff
- 2 \omega_{ab} n^b = N_a = h^b_a \dot{n}_b - \overset{\ast}{u}_a - (u^b \overset{\ast}{n}_b) n_a.
\]

Furthermore, we can show that the Lie derivatives \cite{Maartens 1986}
\begin{eqnarray}
L_{\boldsymbol{\xi}}n^{a} &=&-\psi n^{a}, \enskip L_{\boldsymbol{\xi}} n_a = \psi n_a  \label{KC.23} \\
L_{\boldsymbol{\xi}}u^{a} &=&=-\psi u^{a} -\xi N^{a}, \enskip L_{\boldsymbol{\xi}} u_a = \psi u_a - \xi N_a.  \label{KC.24}
\end{eqnarray}
We note that
\begin{equation}
\mathcal{E}= (\ln \xi ^{2})^{\ast }.  \label{KC.24a}
\end{equation}

The quantity $\overset{\ast}{n^{a}}$ is the principal normal to the magnetic field
lines. We note that, in general, these lines are not straight lines. The
main results on the kinematics of a CKV $\xi ^{a}=\xi n^{a}$ are given in
the following proposition (see Theorem 4.1. of \cite{Saridakis Tsamparlis 1991}).

\begin{proposition}
\label{Theorem 1} Let $\xi ^{a}=\xi n^{a}$ be a spacelike CKV orthogonal to $u^{a}$. Then, $L_{\boldsymbol{\xi}}n_{a}=\psi n_{a}$. Furthermore, the following statements are equivalent: 1) $N^{a}=0$. 2) $\omega ^{a}\parallel \xi ^{a}$ or $\omega ^{a}=0$. 3) $L_{\xi }u_{a}=\psi u_{a}$. 4) $L_{\boldsymbol{\xi}}\omega _{ab}=\psi \omega _{ab}$. 5) $L_{\boldsymbol{\xi}}\sigma _{ab}=\psi \sigma _{ab}$. 6) $L_{\boldsymbol{\xi}} \overset{.}{u}_{a} = \psi_{,a} +\overset{.}{\psi} u_{a}$. 7) $L_{\boldsymbol{\xi}}\theta =-\psi \theta +3\overset{.}{\psi}$.
\end{proposition}

We have the obvious identity $\overset{.}{n}^{a} =-(\overset{.}{n}_{b}u^{b})u^{a} +p_{b}^{a}\overset{.}{n}^{b}$ and
\begin{eqnarray}
N^{a} &=& p_{b}^{a}(L_{u}n^{b}) =p_{b}^{a}(\dot{n}^{b}-%
\overset{\ast}{u}{}^{b}) =p_{b}^{a}\overset{.}{n}^{b}-p_{b}^{a}\overset{\ast }{u}{}^{b}  \notag \\
&=&p_{b}^{a}\overset{.}{n}^{b}-p_{b}^{a}\left( \sigma _{c}^{b}+\omega
_{c}^{b}\right) n^{c}=p_{b}^{a}\overset{.}{n}^{b}-p_{b}^{a}\sigma
_{c}^{b}n^{c}-\omega _{\;\;c}^{a}n^{c}  \notag \\
&=&p_{b}^{a}\overset{.}{n}^{b}-p_{b}^{a}\sigma _{c}^{b}n^{c}+\frac{1}{2}%
N^{a} \implies  \notag \\
p_{b}^{a}\overset{.}{n}^{b} &=&p_{b}^{a}\sigma _{c}^{b}n^{c}+\frac{1}{2}%
N^{a}.  \label{KC.24b}
\end{eqnarray}
Using the symmetry equation we find
\begin{equation}
\overset{.}{n}{}^{a}=\frac{1}{2}\mathcal{E}u^{a} +p_{b}^{a}\sigma^{b}{}_{c}n^{c}+\frac{1}{2}N^{a}.  \label{KC.24c}
\end{equation}

Moreover, a spacelike CKV $\xi^a = \xi n^a$ satisfies the following relations:
\begin{eqnarray}
L_{\boldsymbol{\xi}} \{^a_{bc}\} &=& \delta^a_b \psi_{,c} + \delta^a_c \psi_{,b} - g_{bc} \psi^{,a} \label{eq.ckvn1} \\
L_{\boldsymbol{\xi}} p_{ab} &=& 2 \psi p_{ab} - 2 \xi u_{(a} N_{b)} \label{eq.ckvn2} \\
L_{\boldsymbol{\xi}} h_{ab} &=& 2 \psi h_{ab} - 2 \xi u_{(a} N_{b)} \label{eq.ckvn3} \\
L_{\boldsymbol{\xi}} u_{a;b} &=& \psi u_{a;b} - \xi_{,b} N_a - \xi N_{a;b} - u_b \psi_{,a} + \dot{\psi} g_{ab} \label{eq.ckvn4} \\
L_{\boldsymbol{\xi}} \dot{u}_a &=& - \dot{\xi} N_a - \xi \dot{N}_a - \xi u_{a;b} N^b + \psi_{,a} + \dot{\psi} u_a \label{eq.ckvn5} \\
L_{\boldsymbol{\xi}} \theta &=& - \psi \theta - \xi_{,a} N^a - \xi N^a{}_{;a} + 3 \dot{\psi} \label{eq.ckvn6} \\
L_{\boldsymbol{\xi}} h^{ab} &=& - 2 \psi h^{ab} - 2 \xi u^{(a} N^{b)} \label{eq.ckvn7} \\
L_{\boldsymbol{\xi}} h^a_b &=& - \xi u^a N_b - \xi N^a u_b. \label{eq.ckvn8}
\end{eqnarray}

\subsection{The dynamic implications}

\label{section7.2}

We have to consider three sets of equations, i.e. Maxwell equations, the conservation equations and the gravitational field equations.

\subsubsection{Maxwell equations}

\label{section7.2.1}

The above results hold for any spacelike CKV and any string fluid. For the special case of the EMSF, we have to supplement these equations with Maxwell equations which become:
\begin{eqnarray}
N^{a} &=&0 \iff p_{b}^{a}\overset{.}{n}{}^{b} =p^{a}{}_{c}\sigma_{b}^{c}n^{b} \label{KC.25} \\
\mathcal{E} &=&-(\ln H)^{\ast }  \label{KC.26} \\
\sigma _{ab}n^{a}n^{b}-\frac{2}{3}\theta &=&\left( \ln H\right)^{\cdot } \label{KC.27} \\
e &=&2\omega ^{a}H_{a}, \enskip \mathcal{R}^{a}= -\frac{j^{b}H_{b}}{2H^{3}} H^{a}. \label{KC.28}
\end{eqnarray}

Using (\ref{KC.26}) and (\ref{KC.24a}), we find that $(\xi ^{2}H)^{\ast }=0$, i.e. the quantity $\xi ^{2}H$ is constant along the magnetic field lines.

We also conclude (from $N^{a}=0$) that $\omega ^{a}\parallel H^{a}$ (or $\omega^{a}=0$), that is, the magnetic field congruence coincides with the vorticity congruence.

\textbf{Using Maxwell equations}, we show the following important proposition.

\begin{proposition} \label{pro.CKV.screen.metric}
The vector $\xi ^{a}$ is a CKV both of the screen metric $p_{ab}$ and the projection metric $h_{ab}$ with conformal factor $\psi =\frac{1}{2}\xi \mathcal{E}$ (the same for both metrics).
\end{proposition}

\begin{proof}
In \cite{Tsamparlis M GRG 2006}, it has been shown (see eqs. (26) and (27) of \cite{Tsamparlis M GRG 2006}) that the following general relations/identities hold for the Lie derivatives of the projection tensors $h_{ab}$ and $p_{ab}:$
\begin{eqnarray}
\frac{1}{\xi }L_{\boldsymbol{\xi}}p_{ab} &=&2\left( \mathcal{S}_{ab}+\frac{1}{2}\mathcal{%
E}p_{ab}\right) -2u_{(a}N_{b)}  \label{KC.28a} \\
\frac{1}{\xi }L_{\boldsymbol{\xi}}h_{ab} &=&2\left( \mathcal{S}_{ab}+\frac{1}{2}\mathcal{%
E}p_{ab}\right) -2u_{(a}N_{b)}+2(\ln \xi )_{,(a}n_{b)}+2\overset{\ast }{n}%
_{(a}n_{b)}.  \label{KC.28b}
\end{eqnarray}

From (\ref{KC.25}) and the kinematic condition (\ref{KC.18}), equations (\ref{KC.28a}) and (\ref{KC.28b}) reduce as follows:
\begin{eqnarray}
\frac{1}{\xi }L_{\boldsymbol{\xi}}p_{ab} &=&\mathcal{E}p_{ab}  \label{KC.28c} \\
\frac{1}{\xi }L_{\boldsymbol{\xi}}h_{ab} &=&\mathcal{E}p_{ab}+2(\ln \xi )_{,(a}n_{b)}+2%
\overset{\ast }{n}_{(a}n_{b)}.  \label{KC.28d}
\end{eqnarray}%
From  (\ref{KC.28c}), it follows that $\xi ^{a}$ is a CKV of the screen metric $p_{ab}$ in the screen space with conformal factor $\frac{1}{2}\xi \mathcal{E}.$

To show that $\xi ^{a}$ is a CKV for the projection metric $h_{ab}$, we 1+1+2
decompose $(\ln \xi )_{,a}$ in terms of the vectors $u^{a},n^{a}$ and find
\begin{equation*}
(\ln \xi )_{,a}=-(\ln \xi )^{\cdot }u_{a}+(\ln \xi )^{\ast }n_{a}+p^c_a(\ln
\xi )_{,c}.
\end{equation*}

From (\ref{KC.28d}) and (\ref{KC.20}), we get:
\begin{eqnarray*}
\frac{1}{\xi }L_{\boldsymbol{\xi}}h_{ab} &=&\mathcal{E}p_{ab}+2(\ln \xi )_{,(a}n_{b)}+2%
\overset{\ast }{n}_{(a}n_{b)} \\
&=& \mathcal{E}p_{ab}+2\left[ \overset{\ast }{n}_{d}-(\ln \xi )^{\cdot
}u_{d}+(\ln \xi )^{\ast }n_{d}+ p^c_d (\ln \xi )_{,c}\right] \delta^d_{(a}
n_{b)} \\
&=& \mathcal{E}p_{ab}+2(\ln \xi )^{\ast }n_{a}n_{b}.
\end{eqnarray*}

But equation (\ref{KC.22}) implies that $\mathcal{E=}2(\ln \xi )^{\ast }$. Therefore,
\begin{equation*}
\frac{1}{\xi }L_{\boldsymbol{\xi}}h_{ab} =\mathcal{E(}p_{ab}+n_{a}n_{b})=\mathcal{E}h_{ab}
\end{equation*}
from which it follows that $\xi ^{a}$ is a CKV of $h_{ab}$
with conformal factor $\frac{1}{2}\xi \mathcal{E}.$
\end{proof}

From (\ref{KC.24c}), we also have\footnote{%
One could possibly expect to get information on $\sigma _{bc}n^{b}n^{c}$ from this equation, but this is not so. Indeed, by expanding $p_{b}^{a}$, we find $\dot{n}^{a}=-\frac{1}{2}(\ln H)^{\ast }u^{a} +\sigma_{c}^{a}n^{c} -(\sigma _{bc}n^{b}n^{c})n^{a}$ from which we get no information on $\sigma _{bc}n^{b}n^{c}.$} $\overset{.}{n}{}^{a}= -\frac{1}{2}(\ln H)^{\ast }u^{a}+p_{b}^{a}\sigma_{c}^{b}n^{c}$.

\subsubsection{Conservation equations}

\label{section7.2.2}

These equations are the same as before, that is, we have:
\begin{eqnarray}
\overset{.}{\rho }-(\rho +p_{\perp })(\ln H)^{\cdot } &=&0  \label{KC.29} \\
\overset{\ast }{\rho }-(\rho +p_{\perp })(\ln H)^{\ast } &=&0  \label{KC.30}
\\
p_{a}^{b}\left[ p_{\perp }{}_{,b}+\lambda HH_{,b}+(\rho +p_{\perp }+\lambda
H^{2})(\overset{.}{u}_{b}-\overset{\ast }{n}_{b})\right] &=&0.  \label{KC.31}
\end{eqnarray}

\subsubsection{Gravitational field equations}

\label{section7.2.3}

We use (\ref{N.36a}) to compute these equations. Of course, we can also take them directly from \cite{Tsamparlis M GRG 2006}, but we prefer to derive them here in order to make clear the methods we follow.

First, we compute the $L_{\boldsymbol{\xi}}R_{ab}$.

We note that
\begin{equation}
p_{\psi }=\frac{1}{3}\psi _{;ab}h^{ab}=\frac{1}{3}\psi
_{;ab}(p^{ab}+n^{a}n^{b})=\frac{1}{3}(\gamma _{\psi }+\alpha_{\psi }).
\label{KC.32}
\end{equation}

We have
\begin{eqnarray}
L_{\boldsymbol{\xi}}R_{ab} &=&-2\psi _{;ab}-g_{ab}\square \psi  \notag \\
&=&-2\left[ \lambda _{\psi }u_{a}u_{b}+2k_{\psi }u_{(a}n_{b)}+2\mathcal{S}
_{\psi (a}u_{b)}+\gamma _{\psi }n_{a}n_{b}+2P_{\psi (a}n_{b)}+\frac{1}{2}%
\alpha _{\psi }p_{ab}+D_{\psi ab}\right] -  \notag \\
&&-(3p_{\psi }-\lambda _{\psi })(p_{ab}+n_{a}n_{b}-u_{a}u_{b})  \notag \\
&=&3(p_{\psi }-\lambda _{\psi })u_{a}u_{b}+(\lambda _{\psi }-3p_{\psi
}-2\gamma _{\psi })n_{a}n_{b}+(\lambda _{\psi }-3p_{\psi }-\alpha_{\psi
})p_{ab}+rest.  \label{KC.33}
\end{eqnarray}

From (\ref{N.36a}), we get the following field equations (including equations $k_{\psi }=0$, $S_{\psi a}=0$, $P_{\psi a}=0$ and $D_{\psi ab}=0$ which result directly from the kinematic conditions over (\ref{N.36a}) ):
\begin{eqnarray*}
\overset{\ast }{p_{\perp }}+\lambda H\overset{\ast }{H}+2\left( p_{\perp }+%
\frac{1}{2}\lambda H^{2}-\Lambda \right) \frac{1}{2}\mathcal{E} &=&\frac{1}{%
\xi }3(p_{\psi }-\lambda _{\psi }) \\
\overset{\ast }{p_{\perp }}+\lambda H\overset{\ast }{H+}2\left( p_{\perp }+%
\frac{1}{2}\lambda H^{2}-\Lambda \right) \frac{1}{2}\mathcal{E} &=&-\frac{1}{%
\xi }(\lambda _{\psi }-3p_{\psi }-2\gamma _{\psi }) \\
\overset{\ast }{\rho }+\lambda H\overset{\ast }{H}+\left( \rho +\frac{1}{2}%
\lambda H^{2}+\Lambda \right) \mathcal{E} &\mathcal{=}&\frac{1}{\xi }%
(\lambda _{\psi }-3p_{\psi }-\alpha_{\psi }).
\end{eqnarray*}

Using (\ref{KC.26}) to replace $\mathcal{E}$ in terms of $(\ln
H)^{\ast }$ and observing that the first two equations have identical left-hand sides, we end up with the following equations:
\begin{eqnarray}
\overset{\ast }{p_{\perp }}- \left( p_{\perp }-\frac{1}{2}\lambda
H^{2}-\Lambda \right)(\ln H)^{\ast } &=&\frac{1}{\xi }3(p_{\psi }-\lambda_{\psi }) \label{eq.new1} \\
\lambda _{\psi } &=&-\gamma _{\psi }. \label{eq.new2}
\end{eqnarray}
The last equation is written as $\overset{\ast }{\rho }-\left( \rho -\frac{1}{2}\lambda H^{2}+\Lambda \right) (\ln H)^{\ast }\mathcal{=}\frac{1}{\xi }(\lambda _{\psi }-3p_{\psi }-\alpha_{\psi})$. Using (\ref{KC.30}) and \eqref{KC.32}, we find $\left( p_{\perp }+\frac{1}{2}\lambda H^{2}-\Lambda \right) (\ln H)^{\ast }=-\frac{2}{\xi }(\gamma _{\psi } +\alpha_{\psi})$.

Finally, we have that the field equations in the case of a spacelike vector $\xi ^{a}=\xi n^{a}$ are\footnote{
To find equation (\ref{KC.34}), we replace (\ref{eq.new2}) into (\ref{eq.new1}) and, then, we add with equation (\ref{KC.35}).
}:
\begin{eqnarray}
\overset{\ast }{p_{\perp }}+\lambda H\overset{\ast }{H} &=&\frac{1}{\xi }%
(2\gamma _{\psi }-\alpha_{\psi })  \label{KC.34} \\
\left( p_{\perp }+\frac{1}{2}\lambda H^{2}-\Lambda \right) (\ln H)^{\ast }&=&-\frac{2}{\xi }(\gamma _{\psi }+\alpha_{\psi }) \label{KC.35}
\end{eqnarray}
where
\begin{equation}
\psi _{;ab}=-\gamma _{\psi }(u_{a}u_{b}-n_{a}n_{b})+\frac{1}{2}\alpha _{\psi
}p_{ab}.  \label{KC.36}
\end{equation}

We see that $\psi _{;ab}$ is the energy momentum tensor or, equivalently, the
Ricci tensor of a perfect string fluid of the type (\ref{eq.perSF}) with $\mu =-\gamma _{\psi }$ and $q=\frac{1}{2}\alpha_{\psi}$.

The result we found coincides with the one found in \cite{Tsamparlis M GRG 2006} for string fluids.

From (\ref{KC.34}) and (\ref{KC.35}), ones shows easily that
\begin{equation}
\left[ \left(p_{\perp }+\frac{1}{2}\lambda H^{2}-\Lambda \right) H\right]^{\ast } =-\frac{3H}{\xi}\alpha_{\psi }. \label{KC.37}
\end{equation}

This equation shows that if $\alpha _{\psi }=p^{ab}\psi _{;ab}=0$, then the quantity $\left( p_{\perp }+\frac{1}{2}\lambda H^{2}-\Lambda \right)H$ is constant
along the magnetic field lines.

The constraint equation\footnote{%
This equation follows form the identity $(R^{ab}\xi _{b})_{;a}=-3\square
\psi $ which holds for all CKVs.} for a general anisotropic fluid of the
form we consider is
\begin{equation}
(\bar{\rho} -2\bar{p}_{\perp }+p_{\parallel }+2\Lambda )\psi =2(\lambda _{\psi}-a_{\psi }).  \label{KC.38}
\end{equation}
Setting $\bar{\rho} =-p_{\parallel }$ and $\bar{p}_{\perp }=p_{\perp }+\frac{1}{2}\lambda H^{2}$, we obtain the EMSF. In this case, equation (\ref{KC.38}) becomes
\begin{equation}
(p_{\perp }+\frac{1}{2}\lambda H^{2}-\Lambda )\psi =a_{\psi }+\gamma _{\psi}.  \label{KC.39}
\end{equation}%
But $\psi =\frac{1}{2}\xi \mathcal{E}=-\frac{1}{2}\xi (\ln H)^{\ast }$; therefore, we obtain the same result.

We collect the above results in the following proposition.

\begin{proposition}
\label{Spacelike CKV} An EMSF spacetime admits a CKV of the form $\xi^{a}=\xi n^{a}$ where $n^{a}=\frac{H^{a}}{H}$ iff the following system of equations is satisfied:
\begin{eqnarray}
0&=&\overset{.}{\rho }-(\rho +p_{\perp })(\ln H)^{\cdot } \label{KC.40} \\
0&=& \overset{\ast }{\rho }-(\rho +p_{\perp })(\ln H)^{\ast } \label{KC.41}
\\
0&=& p_{a}^{b}\left[ \left( p_{\perp }+\frac{1}{2}\lambda H^{2}-\Lambda \right)_{,b}+(\rho +p_{\perp }+\lambda H^{2})(\overset{.}{u}_{b}-\overset{\ast }{n}_{b})\right] \label{KC.42} \\
\psi _{;ab} &=&-\gamma_{\psi}(u_{a}u_{b}-n_{a}n_{b}) +\frac{1}{2}\alpha_{\psi }p_{ab}  \label{KC.43} \\
0&=& \left[ \left( p_{\perp }+\frac{1}{2}\lambda H^{2}-\Lambda \right) H\right]^{\ast } +\frac{3H}{\xi }\alpha _{\psi }  \label{KC.44} \\
0&=& \left( p_{\perp }+\frac{1}{2}\lambda H^{2}-\Lambda \right) (\ln H)^{\ast } +\frac{2}{\xi }(\gamma _{\psi }+a_{\psi })  \label{KC.45} \\
\mathcal{S}_{ab} &=&0, \enskip \mathcal{E}=-(\ln H)^{\ast }  \label{KC.46}
\\
\overset{.}{n}^{a} &=&-\frac{1}{2}(\ln H)^{\ast }u^{a} +p_{b}^{a}\sigma_{c}^{b}n^{c}  \label{KC.47} \\
\overset{\ast }{n}^{a} &=&(\ln \xi )^{\ast }n^{a}-(\ln \xi )^{,a} \label{KC.48} \\
N_{a} &=&0  \label{KC.49} \\
\sigma _{ab}n^{a}n^{b}-\frac{2}{3}\theta &=&\left( \ln H\right) ^{\cdot } \label{KC.49a} \\
e &=&2\omega ^{a}H_{a}, \enskip \mathcal{R}^{a}=-\frac{j^{b}H_{b}}{2H^{3}}H^{a}. \label{KC49.b}
\end{eqnarray}
Furthermore, the rotation $\omega^{a}$ is either parallel to $H^{a}$, or vanishes, and the conformal factor $\psi =\frac{1}{2}\xi \mathcal{E} =\overset{\ast}{\xi}$.
\end{proposition}

One important result is that if the vorticity vanishes, then the same must be
true for the charge density and conversely. This is a restriction of physical nature resulting from the considered geometrical symmetry assumption.

Coley and Tupper \cite{Coley Tupper 1989} have shown that if an anisotropic fluid space-time admits a proper SCKV $\xi ^{a}=\xi n^{a}$, then (assuming $\Lambda =0)$ $\mu =-p_{\parallel }=\frac{R}{2}$ and $p_{\perp }=0$ where $R$ is the Ricci scalar. From Einstein field equations, it follows that for this case the energy momentum tensor is of the form
\begin{equation}
T_{ab}=\frac{R}{2}(u_{a}u_{b}-n_{a}n_{b}).  \label{KC.51}
\end{equation}

For the case of a perfect string fluid this result gives $\mu =\frac{1}{2}R$ and $q=0.$ Obviously $R\neq 0$, otherwise we do not have a fluid at all.
Let us check if our results are compatible with this general result.

The SCKV condition $\psi_{;ab} = 0$ implies identically from the $1+1+2$ decomposition of $\psi_{;ab}$ that $\gamma_{\psi} = \alpha_{\psi} = 0$. Then, from (\ref{KC.45}), assuming $\overset{\ast }{H}\neq 0$, we have
\begin{equation}
p_{\perp }+\frac{1}{2}\lambda H^{2}-\Lambda =0  \label{KC.52}
\end{equation}%
which when replaced into (\ref{EqnSS.25}) gives
\begin{equation}
R_{ab}= \left( \rho +\frac{1}{2}\lambda H^{2}+\Lambda \right) p_{ab}. \label{KC.52.0}
\end{equation}
The last equation implies that
\begin{equation}
R=2\left( \rho +\frac{1}{2}\lambda H^{2}+\Lambda \right) .  \label{KC.52a}
\end{equation}%
Replacing (\ref{KC.52a}) into (\ref{KC.52.0}), we find $R_{ab}=\frac{R}{2}p_{ab}$.

Assuming $\Lambda=0$, the energy
momentum tensor for an EMSF is
\begin{equation*}
T_{ab}= R_{ab} - \frac{R}{2}g_{ab} \implies T_{ab}=\frac{R}{2}(u_{a}u_{b}-n_{a}n_{b})
\end{equation*}%
which is in agreement with the quoted result.

\begin{proposition}
\label{Prop SCKV} Let $\xi ^{a}=\xi n^{a}$ be a proper SCKV in an EMSF space-time and let the total energy of the EMSF $\rho +\frac{1}{2}\lambda H^{2}+\Lambda \neq 0.$ Then, for $\overset{\ast }{H}\neq 0$ we have
the following: \newline
(a) The Ricci tensor satisfies the property\footnote{%
It is not necessarily an Einstein space!}  $R_{ab}=\frac{R}{2}p_{ab}$.\newline
(b) The quantity $\frac{R}{H}$ is constant along the magnetic field lines and along the fluid flow lines. \newline
(c) The following equations hold:
\begin{eqnarray*}
\overset{\ast }{\rho }-\left( \rho -\frac{1}{2}\lambda H^{2}+\Lambda \right)(\ln H)^{\ast } &=&0 \\
\overset{.}{\rho }-(\rho +p_{\perp })(\ln H)^{\cdot } &=&0
\end{eqnarray*}
together with equations (\ref{KC.46}) - (\ref{KC49.b}).
\end{proposition}

\begin{proof}
The first part (a) has been shown above.

Concerning (b), we note that using (\ref{KC.52}), equation (\ref{KC.41}) gives
\[
\overset{\ast }{\rho }-\left( \rho -\frac{1}{2}\lambda H^{2}+\Lambda \right)
(\ln H)^{\ast } =0\implies
\overset{\ast }{\rho }-\left( \frac{R}{2}-\lambda H^{2}\right) (\ln H)^{\ast} =0\implies
\]
\[
\overset{\ast }{\rho }+\lambda HH^{\ast }-\frac{R}{2}(\ln H)^{\ast }=0\implies  (\rho +\frac{1}{2}\lambda H^{2}+\Lambda )^{\ast }-\frac{R}{2}(\ln H)^{\ast }=0\implies
\]
\[
(\ln R)^{\ast }-(\ln H)^{\ast } =0\implies \left( \ln \frac{R}{H}\right) ^{\ast } =0.
\]
Therefore, the quantity $\frac{R}{H}$ is constant along the magnetic field lines.

Working in exactly the same way for the other conservation equation (\ref{KC.40}), it is shown that $\left(\ln \frac{R}{H} \right)^{\cdot }=0$, which implies that the quantity $\frac{R}{H}$ is constant along the fluid flow lines.
\end{proof}

Concerning the case of KVs, we have the following result.

\begin{proposition}
\label{Spacelike KV} An EMSF spacetime admits a KV\footnote{In this case, $\psi=0 \implies \mathcal{E} = 0$.} of the form $\xi ^{a}=\xi n^{a}$ where $n^{a}=\frac{H^{a}}{H}$ iff \newline
(a) $\overset{\ast }{H}=\overset{\ast }{\xi }=0$. \newline
(b) The following equations hold:
\begin{eqnarray}
\overset{\ast }{\rho } &=&\overset{\ast }{p}_{\perp}=0  \label{KC.65a} \\
\overset{.}{\rho }-(\rho +p_{\perp })(\ln H)^{\cdot } &=&0  \label{KC.61} \\
\mathcal{S}_{ab} &=&0, \enskip \mathcal{R}^{a}=-\frac{j^{b}H_{b}}{2H^{3}}H^{a}
\label{KC.62} \\
\overset{\ast }{n}^{a} &=&-(\ln \xi )^{,a}  \label{KC.63} \\
\overset{.}{n}{}^{a} &=&p_{b}^{a}\sigma _{c}^{b}n^{c}  \label{KC.64} \\
\sigma _{ab}n^{a}n^{b}-\frac{2}{3}\theta &=&\left( \ln H\right) ^{\cdot }
\label{KC.65} \\
N_{a} &=&0, \enskip e=2\omega ^{a}H_{a}. \label{KC.66}
\end{eqnarray}
\end{proposition}

We conclude that when an EMSF admits the KV $\xi ^{a}=\xi n^{a}$, the following results hold: \newline
i) Because $n^{a}=\frac{\omega^{a}}{\omega} =\frac{H^{a}}{H}$ ($\omega \neq 0$), the string lies over the 2d timelike surface spanned by $u^{a}$ and the vorticity $\omega^{a}$ (Nambu geometric string), or $\omega ^{a}=0$. \newline
ii) From equations (\ref{KC.28a}) and (\ref{KC.28b}), it follows that $L_{\boldsymbol{\xi}}p_{ab}=0$ and $L_{\boldsymbol{\xi}} h_{ab}=0$, that is, $\xi ^{a}$ is also a KV of the metric $h_{ab}$ of the 3-space normal to $u^{a}$, and a KV of the screen space metric $p_{ab}$. \newline
iii) From Proposition \ref{Theorem 1}, we have that the Killing symmetry is inherited by the geometric and the dynamic variables. This means that $L_{\boldsymbol{\xi}}u_{a} =L_{\boldsymbol{\xi}} n_{a}=L_{\boldsymbol{\xi} } \overset{.}{u}_{a}=0$, $L_{\boldsymbol{\xi} }\sigma_{ab}=L_{\boldsymbol{\xi}}\omega _{ab}=0$, and $L_{\boldsymbol{\xi}}\theta=0$.
\newline
iv) If $\omega^{a}\neq 0$, then the vectors $u^{a}$ and $n^{a}=\frac{\omega^{a}}{\omega}$ must commute.

Obviously, these restrictions are severe and allow only few special choices for the string fluids in given spacetimes.

\subsection{Application: The EMSF in the Bianchi I spacetime}

\label{section7.3}

The Bianchi I spacetime with metric defined by the square line element
\begin{equation}
ds^{2}=-dt^{2}+A_{1}^{2}(t)dx^{2}+A_{2}^{2}(t)dy^{2} +A_{3}^{2}(t)dz^{2} \label{KC.57}
\end{equation}%
has been a platform for studying anisotropy and, more specifically, string fluids and electromagnetic fields. For example, Letelier \cite{Letelier 1980}
studied string dust in Bianchi I spacetimes, whereas the electromagnetic field
in the RMHD has been studied (among many others) in \cite{Jacobs Bianchi 1969}. Following this line of research, we shall use the
results obtained in sections \ref{section7.1} and \ref{section7.2} to compute all possible Bianchi I
spacetimes (if any), which carry a magnetic field satisfying the RMHD approximation and admit a spacelike CKV or a spacelike KV.

In order to get comparable results with the literature, we consider the comoving observers $u^{a}=(1,0,0,0)$. This choice has a double effect. Firstly, it implies that the vorticity $\omega ^{a}=0$; therefore, the Maxwell equation $e=2\omega^a H_a$ implies that the charge density $e=0$. This excludes all analytical solutions found in \cite{Jacobs Bianchi 1969}.
Secondly, it must satisfy the geometric condition $N^{a}=0$, which restricts heavily the possible symmetry vectors $\xi ^{a}=\xi n^{a}.$ All the CKVs of the Bianchi I spacetime have been found in \cite{TsamPalKarp 2015}.

We have checked that for this choice of $u^{a}$ none of these vectors
satisfies the condition $N^{a}=0.$ Therefore, the only remaining choice is the KVs so that the system of equations, we have to solve, consists of equations (\ref{KC.65a}) - (\ref{KC.66}).

Consider now the KV $\xi ^{a}=\xi (t)n^{a}$, where $n^{a}=\partial
_{z}=\left( 0, 0, 0, \frac{1}{A_{3}(t)} \right)$. Equation (\ref{KC.65a}) implies that $\rho(t)$ and $p_{\perp}(t)$. We prove easily that equation (\ref{KC.63}) gives $\xi =1$. Therefore, the KV is the $\partial _{z}$. Equation (\ref{KC.64}) is satisfied identically, while equation (\ref{KC.65}) gives $H(t)=\left[A_{1}(t)A_{2}(t)\right]^{-1}$. Therefore, the
magnetic field is given by%
\begin{equation*}
H^{a}=\left[ A_{1}(t)A_{2}(t) \right]^{-1}\partial _{z}.
\end{equation*}%
The remaining equation (\ref{KC.61}) is written as
\begin{equation}
\overset{.}{\rho } +(\rho +p_{\perp }) \left\{ \ln \left[ A_{1}(t)A_{2}(t)\right]\right\}^{\cdot }=0.  \label{KC.69}
\end{equation}

For each equation of state, we determine a Bianchi I spacetime which admits a string fluid. For example, let us consider the equation of state $p_{\perp}= \rho\neq0$. Then, from (\ref{eq.EMSF}), we have that the EMSF has the energy momentum tensor
\begin{equation}
T_{ab}= \left( \rho +\frac{1}{2}\lambda H^{2} \right) \left( u_{a}u_{b} -n_{a}n_{b} +p_{ab} \right) \label{eq.emsf.biaI}
\end{equation}
and equation (\ref{KC.69}) gives
\begin{equation}
\rho= \frac{c}{A_{1}^{2}(t) A_{2}^{2}(t)}. \label{eq.emsf.biaI.2}
\end{equation}
Therefore, in the Bianchi I spacetime, we know the string fluid as well as the magnetic field.

\section{Conclusions}

\label{section8}

In this chapter, we have applied the 1+1+2 decomposition to the case of the EMSF in the RMHD approximation. We have shown that a geometric assumption in the form of a symmetry effects both the kinematics and the dynamics of the resulting EMSF. We have approached the problem in two steps: a. In full generality independently of a particular symmetry; and b. In the case of a CKV which is either of the form $\xi ^{a}=\xi u^{a}$ or of the form $\xi ^{a}=\xi n^{a}$ with $n^{a}=\frac{H^{a}}{H}$, where $u^{a}$ is the four-velocity of the fluid and $H^{a}$ is the magnetic field. We applied the results of the case $\xi ^{a}=\xi u^{a}$ in the FRW spacetime, and the results of the case $\xi ^{a}=\xi n^{a}$ in the Bianchi I spacetime. In the latter case, we found new solutions for the gravitational field.

It is apparent that the results stated in this chapter due to their generality can be used in many different situations involving the electromagnetic field and various types of symmetries. However, one may ask, if all the initial conditions are viable for the solutions which follow from the existence of symmetries. In particular, the existence of a symmetry in a solution is a strong argument which when it is violated leads to other kinds of solutions. On the other hand, from the theory of similarity solutions of differential equations \cite{Hydon 2005, Cherniha 2012}, we know that for a given differential equation a similarity solution satisfies the initial value problem/boundary conditions iff the later are also invariant under the action of the symmetries which provide the similarity transformations. This property can be applied in order to define initial conditions for which a nonsymmetric solution can be related with a symmetric one. For instance, to relate the inner and outer solutions in a compact body.

%% file: CKVs_Bianchi_III_V.tex
\chapter{Constructing the CKVs of Bianchi III and V spacetimes}

\label{ch.CKVs.Bianchi.III.IV}

\section{Introduction}

The knowledge of the proper CKVs (see sec. \ref{sub.dec.Lgij}) of a given spacetime is important because they act as geometric constraints which can be used in the study of the kinematics and the dynamics of a given spacetime. For example, a CKV can be used to reduce the number of unknowns of a gravitational (or cosmological) model and, also, to increase the possibility of finding new solutions of Einstein's field equations (see e.g. \cite{Maartens 1986, Herrera 1985A, Herrera 1985B, Colley 1990A, Colley 1990B, Coley1992, Coley1994, Maartens 1995} and chapter \ref{ch.EMSF}). Furthermore, the conformal algebra can be used to classify spaces (e.g. Finsler manifolds and pseudo-Euclidean manifolds) \cite{Frances, Johar}. For example, one may use the CKVs of a space in order to determine the classes of manifolds which are conformally related to the given space; or in order to study the locally conformal flatness of a space around a singularity (i.e. a point $x_{0}$ where the CKV vanishes). The special class of the CKVs, the KVs, have been used in numerous applications. For example, Geroch in \cite{Geroch1} and \cite{Geroch2} has used the KVs in order to derive new solutions of the gravitational field equations. The gradient KVs can be used also to decompose the spacetime metric and to simplify the field equations. Moreover, the KVs are related with the conservation laws for the geodesic equations. Indeed, the KVs form a subalgebra on the Noether symmetries for the geodesic Lagrangian \cite{Tsamparlis 2010B}.

Apart from the above applications, another important area where the CKVs and
the more general PCs have been used is the geometric
study of Lie symmetries of differential equations. For example, early studies of the geodesic equations \cite{Katzin 1981, Katzin 1968, Aminova 1995, Aminova 2000} have shown a unique connection of the Lie point symmetries of the
geodesic equations in a Riemannian space with the elements of the projective algebra of this space. Furthermore, in a conservative dynamical system one may consider the kinetic energy as a metric (i.e. the kinetic metric) and study the Lie and the Noether point symmetries of the dynamical equations using the collineations of this metric. In \cite{Tsamparlis 2010A}, it has been shown that in such systems, the Lie point symmetries are generated by the special projective algebra and the Noether point symmetries by the Homothetic algebra of the kinetic metric. Similar results have been found for some PDEs of special interest in curved spacetimes, as the wave and the heat equation (see e.g. \cite{Bozhkov 2010, Paliathanasis 2016, Paliathanasis 2018} and references therein), where it has been shown that the Lie point symmetries involve the CKVs.

In this chapter, we apply the propositions and the methodology developed in \cite{TsamPalKarp 2015, Tsamp-Nikol-Apost, Tsamp-Apost} in order to determine all Bianchi III and V spacetimes that admit proper CKVs. The Bianchi I spacetimes which admit proper CKVs have been determined in\footnote{ For completeness, we restate these results in sec. \ref{subsec.bia.results} below.} \cite{TsamPalKarp 2015}. Bianchi III and V spacetimes are of special interest and have many applications in the study of anisotropic cosmologies. Bianchi spacetimes can provide a different cosmological behavior in the early universe \cite{Karagiorgos 2018}, while also they can been seen as the homogeneous limits of exact inhomogeneous cosmological models \cite{Krasinski 1997}.

\textbf{Bianchi spacetimes}\index{Spacetime! Bianchi} are spatially homogeneous spacetimes of the general form
\begin{equation}
ds^{2}=-dt^{2} +A^{2}(t)(\omega _{1})^{2}+B^{2}(t)(\omega
_{2})^{2}+C^{2}(t)(\omega _{3})^{2}  \label{ds.bianchi}
\end{equation}%
where $\omega _{i}$, $i=1,2,3$, are basis 1-forms and $A(t)$, $B(t)$, $C(t)$ are functions of the time coordinate (see e.g. \cite{Stephani, Wald, Ryan}). For instance:
\begin{eqnarray*}
\text{Bianchi I} &:&\omega _{1}=dx,~\omega _{2}=dy,~\omega _{3}=dz \\
\text{Bianchi III} &\text{:}&\omega _{1}=dx,~\omega _{2}=dy,~\omega
_{3}=e^{-x}dz \\
\text{Bianchi V} &\text{:}&\omega _{1}=dx,~\omega _{2}=e^{x}dy,~\omega
_{3}=e^{x}dz.
\end{eqnarray*}
In the case of $B^{2}(t)=C^{2}(t)$, the Bianchi spacetimes contain a fourth isometry (the rotation of the $yz$-plane) and reduce to the important subclass of Locally Rotational Symmetric (LRS) spacetimes (see e.g. \cite{Tsamparlis2001} and citations therein).\index{Spacetime! LRS}

\section{Preliminaries: Decomposable spacetimes}

\label{sec.Bianchi.1n}

We recall that a Riemannian manifold is \textbf{decomposable}\index{Spacetime! decomposable} along the coordinate $t$ iff the metric $g_{ab}$ admits the non-null gradient KV $u^{a}=\partial_t$, where $u^{a}u_{a}=\varepsilon u^{2}$ and $\varepsilon =\pm 1$. In this case, one defines the projection operator
\begin{equation}
h_{ab}=g_{ab}-\frac{\varepsilon }{u^{2}}u_{a}u_{b}  \label{f1}
\end{equation}%
and decomposes the tensor algebra along $u^{a}$ and normal to $u^{a}.$ For an $n$-dimensional decomposable Riemannian manifolds $M^{n}$ with $n\geq 3$ an
algorithm has been developed \cite{Tsamp-Nikol-Apost} which determines
the proper CKVs in terms of the (gradient) proper CKVs of the $\left(n-1\right)$-dimensional non-decomposable space.

In particular, it has been shown that \emph{an $n$-dimensional decomposable space $M^{n}$ admits proper CKVs iff the $\left( n-1\right)$-space $M^{n-1}$ admits a gradient proper CKV whose conformal factor is the gradient factor which constructs the (gradient) CKV.} In addition, any gradient proper
CKV of the $M^{n-1}$ provides two proper CKVs for the $M^{n}$. For a four-dimensional manifold, the following result is shown in \cite{Tsamp-Nikol-Apost}.

If $M^{n}$, where $n=4$, is a decomposable Riemannian manifold with line element ($\mu ,\nu =1,2,3)$%
\begin{equation}
ds^{2}=\varepsilon dt^{2}+h_{\mu \nu }\left( x^{\sigma }\right) dx^{\mu}dx^{\nu }  \label{sx2.1}
\end{equation}%
where $\varepsilon =\pm 1$, then the vector field
\begin{equation}
X^{a}\partial _{a}=-\frac{\varepsilon }{p}\dot{\lambda}\left( t\right) \psi
\left( x^{\sigma }\right) \partial _{t}+\frac{1}{p}\lambda \left( t\right)
\xi ^{\mu }\left( x^{\sigma }\right) \partial _{\mu }+L^{\mu }\partial _{\mu
}  \label{sx2.0a}
\end{equation}
is a proper CKV of (\ref{sx2.1}), where: \newline
a. $L^{\mu }$ is a non-gradient KV or HV of $M^{n-1}$. \newline
b. $\xi ^{\mu }\left( x^{\sigma }\right) $ is a gradient proper CKV of $M^{n-1}$ with conformal factor $\psi \left( x^{\sigma }\right)$, i.e. the Lie derivative $\mathit{L}_{\boldsymbol{\xi}} h_{\mu \nu }\left( x^{\sigma }\right) =2\psi \left(
x^{\sigma }\right) h_{\mu \nu }\left( x^{\sigma }\right) $. \newline
c. The function $\lambda \left( t\right) $ is given by
\begin{equation}
\lambda \left( t\right) =\lambda _{1}e^{i\sqrt{\varepsilon p}t}+\lambda
_{2}e^{-i\sqrt{\varepsilon p}t},\enskip \text{for $\varepsilon p>0$}
\label{sx2.2}
\end{equation}%
or
\begin{equation}
\lambda \left( t\right) =\lambda _{1}e^{\sqrt{-\varepsilon p}t}+\lambda
_{2}e^{-\sqrt{-\varepsilon p}t},\enskip \text{for $\varepsilon p<0$}
\label{sx2.2b}
\end{equation}
where $p$ is a non-vanishing constant and $\lambda _{1}$, $\lambda _{2}$ are
independent constants, provided the function $\psi \left( x^{\sigma }\right)$ satisfies the
condition
\begin{equation}
\psi _{;\mu \nu }=p\psi h_{\mu \nu }.  \label{sx2.1a}
\end{equation}

Concerning the HV, it has been shown in \cite{Tsamp-Nikol-Apost} that when the $M^{n-1}$ space admits a HV $H^{\mu
}\left( x^{\sigma }\right)$ with conformal factor $C$, the $M^{n}$ admits
the HV
\begin{equation}
H^{a}\partial _{a}=Ct\partial _{t}+H^{\mu }\partial _{\mu }. \label{sx2.3}
\end{equation}

Finally, concerning the KV fields, it has been shown that the
KV fields of $M^{n}$ are
\begin{eqnarray}
K^{a}\partial_{a}&=&k_{0}\partial _{t} + \left[ k_{1I}h^{\mu \nu }\left( x^{\sigma }\right) K_{\nu
}^{I}\left( x^{\sigma }\right) +k_{2I}h^{\mu \nu }\left( x^{\sigma }\right)
S_{,\nu }^{I}\left( x^{\sigma }\right) \right] \partial_{\mu}+ \notag \\
&& +k_{3I} \left[ -\varepsilon
S^{I}\left( x^{\sigma }\right) \partial _{t}+h^{\mu \nu }\left( x^{\sigma
}\right) S_{,\nu }^{I}\left( x^{\sigma }\right)\partial_{\mu} \right] \label{sx2.4}
\end{eqnarray}
where $K_{\nu }^{I}\left( x^{\sigma }\right) $ are the non-gradient KVs of $%
M^{n-1}$ and $S_{,\nu }^{I}\left( x^{\sigma }\right) $ are the gradient KVs
of $M^{n-1}$. The parameters $k_{0}$, $k_{1I}$, $k_{2I}$ and $k_{3I}$ are independent constants.

However, another possibility that the space (\ref{sx2.1}) admits
proper CKVs is when it is conformally flat. That case was found to be
important in the classification of Bianchi I spacetimes in \cite{TsamPalKarp 2015}
according to the admitted CKVs, but it does not provide any result in the
case of Bianchi III and Bianchi V spacetimes. Thus, we omit it from the
present discussion.

The concept of \textbf{conformally related metrics}\index{Metric! conformally related} plays a crucial role in the
computation of the CKVs; therefore, we review the basic definitions
concerning these metrics. Two metrics $\widehat{g}_{ab}$ and $g_{ab}$ are said to be conformally related iff there is a function $%
N^{2}(x^{r})$ such that $\widehat{g}_{ab}=N^{2}(x^{r})g_{ab}$. The
conformally related metrics share the same conformal algebra but with
different conformal factors. For a given vector field $\mathbf{X}$, we have
the following decompositions/identities (see sec. \ref{sub.dec.Lgij}):
\begin{equation*}
L_{\mathbf{X}}\widehat{g}_{ab}=2\widehat{\psi }(\mathbf{X})\widehat{g}_{ab}+2\widehat{H}_{ab}(\mathbf{X}) \enskip \text{and} \enskip L_{\mathbf{X}}g_{ab}=2\psi (\mathbf{X})g_{ab} +2H_{ab}(\mathbf{X})
\end{equation*}
where $H_{ab}(\mathbf{X})$ and $\widehat{H}_{ab}(\mathbf{X})$ are symmetric traceless tensors. It can be shown that
\begin{equation*}
\widehat{\psi }(\mathbf{X})=\mathbf{X}(\ln N)+\psi (\mathbf{X}), \enskip \widehat{H}_{ab}(\mathbf{X})=N^{2}H_{ab}(\mathbf{X}) \enskip \text{and} \enskip
\widehat{F}_{ab}(\mathbf{X})=N^{2}F_{ab}(\mathbf{X}) -2NN_{,[a}X_{b]}
\end{equation*}
where $\widehat{F}_{ab}(\mathbf{X})= \widehat{X}_{[a\hat{;}b]}= \widehat{X}_{[a;b]}$ and $F_{ab}(\mathbf{X})=X_{[a;b]}$. Moreover,
\[
\widehat{X}_{a\hat{;}b}= \frac{1}{2}L_{\mathbf{X}}\widehat{g}_{ab}+%
\widehat{F}_{ab}(\mathbf{X}) \enskip \text{and} \enskip
X_{a;b} =\frac{1}{2}L_{\mathbf{X}}g_{ab}+F_{ab}(\mathbf{X}).
\]

A metric $g_{ab}$ is called \textbf{conformally flat}\index{Metric! conformally flat} iff it is conformally related to the flat metric\footnote{For more details on the conformal algebra of the flat metric, see sec. \ref{sub.ckv.flat}.} $\eta_{ab}$. A metric conformally related to a conformally flat metric is also conformally flat. It is well-known that all the 2d spacetimes are conformally flat and admit an infinite number of CKVs, while only the flat 2d metrics admit SCKVs.

\section{CKVs of Bianchi III spacetimes}

\label{sec.Bianchi.2n} 

Consider the 3d decomposable spacetime of Lorentzian signature%
\begin{equation}
ds_{(1+2)}^{2}=\Gamma ^{2}\left( \tau \right) \left( -d\tau
^{2}+dx^{2}\right) +dy^{2}.  \label{b3.01}
\end{equation}%
The line element (\ref{b3.01}) for arbitrary $\Gamma \left( \tau \right)$ admits a 2d conformal Killing algebra consisting by the KVs $\partial _{x}$ and $\partial _{y}$.

For the conformal spacetime
\begin{equation}
d\bar{s}_{\left( 1+2\right) }^{2}=B^2\left( \tau \right)
e^{2x}ds_{(1+2)}^{2} \label{b3.02}
\end{equation}%
the vector field $\partial _{y}$ remains a KV but $\partial _{x}$ now becomes a proper HV.

Consider now the 4d decomposable spacetime
\begin{equation}
ds_{\left( 1+3\right) }^{2}=d\bar{s}_{\left( 1+2\right)}^{2} +dz^{2} \label{b3.03}
\end{equation}%
which admits a 3d conformal algebra consisting of the KVs $\partial_{y}, \partial _{z}$ and the proper HV $\partial_{x} +z\partial_{z}$. Then, the conformally related spacetime $ds_{\left( III\right)
}^{2}=A^{2}\left( \tau \right) e^{-2x}ds_{\left( 1+3\right) }^{2}$ which can
be written equivalently\footnote{%
Here $\alpha ^{2}\left( t\right) =A^{2}\left( t\right) B^{2}\left( t\right)
\Gamma ^{2}\left( t\right) $, $\beta ^{2}\left( t\right) =A^{2}\left(
t\right) B^{2}\left( t\right) $ and $\gamma ^{2}\left( t\right) =A^{2}\left(
t\right), $ while $t=\int a\left( \tau \right) d\tau $.}%
\begin{equation}
ds_{\left( III\right) }^{2}=-dt^{2}+\alpha ^{2}\left( t\right) dx^{2}+\beta
^{2}\left( t\right) dy^{2}+\gamma ^{2}\left( t\right) e^{-2x}dz^{2}
\label{b3.04}
\end{equation}%
is a Bianchi III spacetime and the vector fields $\partial _{y},~\partial_{z},~\partial _{x}+z\partial _{z}$ form the Killing algebra of (\ref{b3.04}). Therefore, in order the Bianchi III spacetime (\ref{b3.04}) to admit a greater
conformal algebra, the functions $\alpha \left( t\right) ,~\beta \left(
t\right) $ and $\gamma \left( t\right) $ must be specified. Recall that when
$\alpha \left( t\right) =\gamma \left( t\right)$, the spacetime (\ref{b3.04}) is
locally rotational and admits as extra KV the rotation in the 2d space $ds^{2}=dx^{2}+e^{-2x}dz^{2}$.

The 3d spacetime (\ref{b3.01}) admits a greater conformal algebra
for specific functions $\Gamma \left( \tau \right) $. From the discussion of sec. \ref{sec.Bianchi.1n}, it follows that $\Gamma \left( \tau \right) $ must be
such that the 2d space
\begin{equation}
ds_{\left( 2\right) }^{2}=\Gamma ^{2}\left( \tau \right) \left( -d\tau
^{2}+dx^{2}\right) \label{b3.05}
\end{equation}%
admits proper gradient CKVs or a greater Killing algebra. For
2d spaces, it is well-known that the admitted KVs can be zero,
one or three and, in the latter case, the space is maximally symmetric. Since (%
\ref{b3.05}) admits always the KV $\partial _{x}$, the $\Gamma \left( \tau
\right) $ must be specified so that (\ref{b3.05}) is maximally symmetric.
Without loss of generality, we can select either $\Gamma ^{2}\left( \tau \right)
=e^{m\tau }$ in which case (\ref{b3.05}) is the flat space with Ricci Scalar
$R_{\left( 2\right) }=0$, or $\Gamma ^{2}\left( \tau \right) =\kappa
^{-2}\cos ^{-2}\left( \tau \right) $ in which case $R_{\left( 2\right)
}=2\kappa ^{2}$.

Furthermore, because all the 2d spaces admit infinitely many CKVs,
the requirement that at least one of the proper CKVs is to be gradient
specifies the spacetime to be of nonzero constant curvature (i.e. maximally symmetric space and admits five gradient proper CKVs).

\subsection{Case $\Gamma^2 \left( \protect\tau \right) =e^{m \tau}$}

\label{sec.Bianchi.2n.1} 

In the case of $\Gamma ^{2}\left( \tau \right) =e^{m \tau}$, the 3d space
\begin{equation}
ds_{(1+2)}^{2}=e^{m\tau}\left( -d\tau ^{2}+dx^{2}\right) +dy^{2}.  \label{b3.06}
\end{equation}%
is flat (see sec. \ref{sub.ckv.flat}) and admits a ten-dimensional conformal algebra. This algebra
consists of the six KVs:
\[
\mathbf{Y}_{1} =\frac{2}{m}e^{-\frac{m}{2}(\tau -x)}\partial _{\tau }-\frac{2}{m}e^{-\frac{m}{2}(\tau -x)}\partial _{x}, \enskip \mathbf{Y}_{2}= -\frac{2}{m}e^{-\frac{m}{2}(\tau +x)}\partial _{\tau }-\frac{2}{m}e^{-\frac{m}{2}(\tau +x)}\partial _{x},
\]
\[
\mathbf{Y}_{3}= \partial _{x}, \enskip \mathbf{Y}_{4}=\partial _{y}, \enskip \mathbf{Y}_{5} = ye^{-\frac{m}{2}(\tau +x)}\partial _{\tau }+ye^{-\frac{m}{2}(\tau +x)}\partial _{x}+\frac{2}{m}e^{\frac{m}{2}(\tau -x)}\partial _{y},
\]
\[
\mathbf{Y}_{6}= -ye^{-\frac{m}{2}(\tau -x)}\partial _{\tau }+ye^{-\frac{m}{2}%
(\tau -x)}\partial _{x}-\frac{2}{m}e^{\frac{m}{2}(\tau +x)}\partial _{y},
\]
the HV $\mathbf{Y}_{7}=\frac{2}{m}\partial _{\tau }+y\partial _{y}$ with homothetic factor $\psi_{(1+2)}(\mathbf{Y}_{7})=1$
and the three SCKVs:
\begin{eqnarray*}
\mathbf{Y}_{8} &=& \left[ \frac{2}{m^{2}}e^{\frac{m}{2}(\tau -x)}+\frac{y^{2}}{2}%
e^{-\frac{m}{2}(\tau +x)}\right] \partial _{\tau }+\left[ -\frac{2}{m^{2}}e^{%
\frac{m}{2}(\tau -x)}+\frac{y^{2}}{2}e^{-\frac{m}{2}(\tau +x)}\right]
\partial _{x}+\frac{2y}{m}e^{\frac{m}{2}(\tau -x)} \\
\mathbf{Y}_{9} &=& -\left[ \frac{2}{m^{2}}e^{\frac{m}{2}(\tau +x)}+\frac{y^{2}}{2%
}e^{-\frac{m}{2}(\tau -x)}\right] \partial _{\tau }+\left[ -\frac{2}{m^{2}}%
e^{\frac{m}{2}(\tau +x)}+\frac{y^{2}}{2}e^{-\frac{m}{2}(\tau -x)}\right]
\partial _{x}-\frac{2y}{m}e^{\frac{m}{2}(\tau +x)} \\
\mathbf{Y}_{10} &=& my\partial _{\tau }+\left[ \frac{m^{2}y^{2}}{4}+e^{m\tau }%
\right] \partial _{y}
\end{eqnarray*}
with conformal factors $\psi _{(1+2)}(\mathbf{Y}_{8})=\frac{2}{m}e^{\frac{m}{%
2}(\tau -x)}$, $\psi _{(1+2)}(\mathbf{Y}_{9})=-\frac{2}{m}e^{\frac{m}{2}%
(\tau +x)}$ and $\psi _{(1+2)}(\mathbf{Y}_{10})=\frac{m^{2}y}{2}$, respectively.

The conformally flat space%
\begin{equation}
d\bar{s}_{\left( 1+2\right) }^{2}=B^{2}\left( \tau \right) e^{2x}\left[
e^{m\tau}\left( -d\tau ^{2}+dx^{2}\right) +dy^{2}\right]  \label{b3.07}
\end{equation}%
admits the same elements of conformal algebra with (\ref{b3.06}), but
with different conformal factors 
\begin{equation}
\bar{\psi}_{(1+2)}(\mathbf{Y}_{A})=\mathbf{Y}_{A}\left[ \ln (Be^{x})\right]
+\psi _{(1+2)}(\mathbf{Y}_{A}).  \label{b3.08}
\end{equation}
When we impose the condition (\ref{sx2.1a}), we find that there does not exist a function $B\left( \tau \right)$ such that the factors $\bar{\psi}_{(1+2)}(\mathbf{Y}_{A})$ to satisfy (\ref{sx2.1a}). On the other hand, we observe that for
\begin{equation}
B\left( \tau \right) =e^{\mu \tau },\enskip\mu =\frac{m(\lambda -1)}{2} \label{b3.09}
\end{equation}%
it follows $\bar{\psi}_{(1+2)}(\mathbf{Y}_{7})=\lambda= const$, which means that $Y_{7}$ is reduced to a HV for (\ref{b3.07}). At this point, it is important
to mention that $\bar{\psi}_{1+2}\left( \mathbf{Y}_{3}\right) =1$; however, as expected,
there is only one proper HV and not two. We assume $\mathbf{Y}%
_{7}$ to be the proper HV and $\mathbf{Y}_{3}-\frac{1}{\lambda }\mathbf{Y}%
_{7}$ to be a KV.

For the 4d decomposable spacetime
\begin{equation}
ds_{(1+3)}^{2}=e^{2x}e^{2\mu \tau }\left[ e^{m\tau}\left( -d\tau
^{2}+dx^{2}\right) +dy^{2}\right] +dz^{2}  \label{b3.10}
\end{equation}%
from $\mathbf{Y}_{7}$, we find the proper HV%
\begin{equation}
\mathbf{L}_{1}\equiv \mathbf{Y}_{7}+\lambda z\partial _{z}=\frac{2}{m}%
\partial _{\tau }+y\partial _{y}+\lambda z\partial _{z}.  \label{b3.11}
\end{equation}

We conclude that the Bianchi III spacetime%
\begin{equation}
ds_{(III)}^{2}=e^{m\lambda \tau }A^{2}\left( \tau \right) \left( -d\tau
^{2}+dx^{2}+e^{-m\tau }dy^{2}+e^{-m\lambda \tau }e^{-2x}dz^{2}\right)
\label{b3.12}
\end{equation}%
admits the proper CKV $\mathbf{L}_{1}$ with conformal factor $\psi _{(III)}(%
\mathbf{L}_{1})=\frac{2}{m}\frac{A_{,\tau }}{A}+\lambda$, which reduces to a
HV when $A\left( \tau \right) $ is an exponential. In the last case, the line element is
\begin{equation}
ds_{(III)}^{2}=-e^{m\kappa \tau }d\tau ^{2}+e^{m\kappa \tau
}dx^{2}+e^{m(\kappa -1)\tau }dy^{2}+e^{m(\kappa -\lambda )\tau
}e^{-2x}dz^{2}  \label{b3.14}
\end{equation}%
or, equivalently,
\begin{equation}
ds_{(III)}^{2}=-dt^{2}+\frac{m^{2}\kappa ^{2}t^{2}}{4}dx^{2}+\left( \frac{%
m^{2}\kappa ^{2}t^{2}}{4}\right) ^{\frac{\kappa -1}{\kappa }}dy^{2}+\left(
\frac{m^{2}\kappa ^{2}t^{2}}{4}\right) ^{\frac{\kappa -\lambda }{\kappa }%
}e^{-2x}dz^{2}  \label{b3.15}
\end{equation}%
and we write $\mathbf{L}_{1}=\kappa t\partial _{t}+y\partial_{y}+\lambda z\partial _{z}$ with $\psi_{(III)}(\mathbf{L}_{1})=const\equiv \kappa \neq 0$. Recall that $dt=e^{\frac{m\kappa}{2}\tau}d\tau$.

Performing the same analysis for the second case of $\Gamma ^{2}\left( \tau
\right) =\kappa ^{-2}\cos ^{-2}\left( \tau \right)$, we find that the
resulting Bianchi III spacetime does not admit any proper CKV or a proper
HV; hence, we omit the presentation of this analysis.

We summarize our results in the following Proposition.

\begin{proposition} \label{pro.BianchiIII}
The only Bianchi III spacetime\index{Spacetime! Bianchi III} which admits a proper CKV is
\begin{equation}
ds^{2}=A^{2}\left( \tau \right) \left[ e^{m\lambda \tau } \left( -d\tau ^{2}+dx^{2}\right) + e^{m\left( \lambda -1\right) \tau}
dy^{2} + e^{-2x} dz^{2} \right].  \label{Bianchi3.1}
\end{equation}%
The CKV is $\mathbf{L}_{1}=\frac{2}{m}\partial _{\tau }+y\partial
_{y}+\lambda z\partial _{z}$ and it has conformal factor $\psi _{(III)}(%
\mathbf{L}_{1})=\frac{2}{m}\frac{A_{,\tau }}{A}+\lambda$, where $A\left(\tau \right)$ is an arbitrary function.
\end{proposition}

\section{Bianchi V spacetimes which admit a CKV}

\label{sec.Bianchi.3n} 

For the computation of the CKVs for the Bianchi V spacetime, we apply the
same procedure with sec. \ref{sec.Bianchi.2n}, but for this case, we start from the 2d spacetime%
\begin{equation}
ds_{\left( 2\right) }^{2}=\Gamma^2 \left( \tau \right) e^{-2x}\left(
-d\tau^{2}+dx^{2}\right).  \label{b4.01}
\end{equation}%
The latter space is maximally symmetric only for $\Gamma^2 \left( \tau \right) =e^{\gamma \tau }$, where the Ricci Scalar is calculated to be $R_{\left( 2\right) }=0$. It is important to mention that there does not exist a function $\Gamma \left( \tau \right) $ such that the space (\ref{b4.01}) is of constant curvature.

We omit the intermediary calculations and we summarize the results in the following Proposition.

\begin{proposition} \label{pro.BianchiV} 
The Bianchi V spacetime\index{Spacetime! Bianchi V}
\begin{equation}
ds^{2}=A^{2}\left( \tau \right) \left[ \Gamma ^{2}\left( \tau \right) \left(
-d\tau ^{2}+dx^{2}\right) +e^{2x}\left( B^{2}\left( \tau \right)
dy^{2}+dz^{2}\right) \right]  \label{Bianchi5}
\end{equation}%
admits the unique proper CKV $\mathbf{L}_{1}=\frac{2}{m}\partial _{\tau
}+y\partial _{y}+\lambda z\partial _{z}$ with $\psi _{(V)}(\mathbf{L}_{1})=%
\frac{2}{m}\frac{A_{,\tau }}{A}+\lambda $ only when $\Gamma ^{2}\left( \tau
\right) =e^{m\lambda \tau }$ and $B^{2}\left( \tau \right) =e^{m(\lambda -1)\tau
}$. For $A^{2}\left( \tau \right) =e^{m(\kappa -\lambda )\tau }$, the CKV reduces to a HV with conformal factor $\psi_{(V)}(\mathbf{L}_{1}) =const =\kappa \neq 0$.
\end{proposition}

\section{Applications}

\label{sec.Bianchi.4n} 

\subsection{Bianchi III cosmological fluid}

\label{sec.Bianchi.4n.1} 

In this section, we study some of the kinematic and the dynamic  properties of spacetime given by equation (\ref{b3.12}) for the comoving observers $u^{a}=\frac{e^{-\frac{m\lambda }{2} \tau }}{A\left( \tau \right) }\delta _{\tau }^{a}$, where $u^{a}u_{a}=-1$. As it is well-known (see e.g. \cite{Ellis 1998} and chapter \ref{ch.EMSF}), the four-velocity of a class of observers introduces the $1+3$ decomposition of tensor fields in spacetime. The decomposition of $u_{a;b}$ gives the kinematic quantities $\theta,~\sigma ^{2},~\omega ^{2}$ and $\alpha^a$ defined by the identity
\begin{equation}
u_{a;b}=-\alpha _{a}u_{b}+\omega _{ab}+\sigma _{ab}+\frac{1}{3}\theta h_{ab}
\label{eq.1plus3u}
\end{equation}
where $\alpha^{a}=\dot{u}^{a}=u^{a}{}_{;b}u^{b}$, $\omega
_{ab}=h_{a}^{c}h_{b}^{d}u_{[c;d]}$, $\sigma _{ab}=\left( h_{a}^{c}h_{b}^{d}-%
\frac{1}{3}h^{cd}h_{ab}\right) u_{(c;d)}$, $\theta
=h^{ab}u_{a;b}=u^{a}{}_{;a}$, $\sigma ^{2}\equiv \frac{1}{2}\sigma
_{ab}\sigma ^{ab}$ and $\omega ^{2}\equiv \frac{1}{2}\omega _{ab}\omega ^{ab}.$
In this decomposition, $\alpha^{a}$ is the four-acceleration of the observers $u^{a}$,
and the quantities $\omega _{ab},\sigma _{ab},\theta $ concern the
variation of the projected ($\bot\delta x^{a}u_a=0$) connecting vector $\bot\delta x^{a}$ along the congruence (i.e.
the integral lines) defined by the vector field $u^{a}$. The antisymmetric tensor $\omega _{ab}$ measures the relative rotation; the tensor $\sigma_{ab}$, the anisotropic expansion; and the scalar $\theta $, the isotropic
expansion of $\bot \delta x^{a}$. The dynamic variables of the spacetime are
defined by the 1+3 decomposition of the Einstein tensor $G_{ab}$ as follows \cite{Ellis 1998}:
\begin{equation}
G_{ab}=\rho u_{a}u_{b}+2q_{(a}u_{b)}+ph_{ab}+\pi _{ab}  \label{eq.1plus3G}
\end{equation}%
where $\rho =G_{ab}u^{a}u^{b}$ is the energy-mass density of the fluid, $p=\frac{1}{3}h^{ab}G_{ab}$ is the isotropic pressure, $q^{a}=-h^{ac}G_{cd}u^{d}$ is the heat flux tensor and $\pi_{ab} = \left( h_{a}^{c}h_{b}^{d} - \frac{1}{3} h^{cd} h_{ab} \right) G_{cd}$ is the traceless anisotropic tensor (measures the anisotropy of the fluid).

Applying the above for the comoving observers in Bianchi III spacetime (\ref{b3.12}), we compute that the kinematic quantities are $\omega ^{2}=0$, $\alpha^{a}=0$,
\begin{equation}
\theta =\frac{e^{-\frac{m\lambda }{2}\tau }}{A}\left[ 3\frac{d(\ln A)}{d\tau
}+\frac{m(2\lambda -1)}{2}\right]  \label{b3.17}
\end{equation}%
and
\begin{equation}
\sigma ^{2}=\frac{m^{2}(\lambda ^{2}-\lambda +1)}{12}\frac{e^{-m\lambda \tau
}}{A^{2}}.  \label{b3.18}
\end{equation}%
Similarly for the dynamic quantities, we find that the (nonzero) components
for the cosmological fluid defined by the Bianchi III spacetime (\ref{b3.12}) are
\begin{equation}
\rho =\frac{e^{-m\lambda \tau }}{4A^{2}}\left[ 4\frac{d\left( \ln A\right) }{%
d\tau }\left( 3\frac{d\left( \ln A\right) }{d\tau }+m\left( 2\lambda
-1\right) \right) +m^{2}\lambda \left( \lambda -1\right) -4\right]
\label{b3.19}
\end{equation}%
\begin{equation}
p=\frac{e^{-m\lambda \tau }}{A^{2}}\left[ -\frac{2}{A}\frac{d^{2}A}{d\tau
^{2}}+\frac{d\left( \ln A\right) }{d\tau }\left( \frac{d\left( \ln A\right)
}{d\tau }+\frac{m}{3}(2-\lambda )\right) -\frac{m^{2}}{12}(\lambda
-1)(\lambda -2)+\frac{1}{3}\right]   \label{b3.20}
\end{equation}%
\begin{equation}
q^{a}=\left( 0,\frac{m\lambda }{2}\frac{e^{-\frac{3m\lambda }{2}\tau }}{A^{3}%
},0,0\right)  \label{b3.21}
\end{equation}%
\begin{equation}
\pi _{xx}=\frac{m(\lambda +1)}{3}\frac{d\left( \ln A\right) }{d\tau }+\frac{%
m^{2}(\lambda ^{2}-1)}{12}-\frac{1}{3}  \label{b3.22}
\end{equation}%
\begin{equation}
\pi _{yy}=e^{-m\tau }\left[ \frac{m(\lambda -2)}{3}\frac{d\left( \ln
A\right) }{d\tau }+\frac{m^{2}}{12}(\lambda -1)(\lambda -2)+\frac{2}{3}%
\right]  \label{b3.23}
\end{equation}%
and
\begin{equation}
\pi _{zz}=-e^{-m\lambda \tau -2x}\left[ \frac{m(2\lambda -1)}{3}\frac{%
d\left( \ln A\right) }{d\tau }+\frac{m^{2}}{12}(\lambda -1)(2\lambda -1)+%
\frac{1}{3}\right] .  \label{b3.24}
\end{equation}

In the case of $A^{2}(\tau )=e^{m(\kappa -\lambda )\tau }$, where the CKV $%
\mathbf{L}_{1}$ becomes a HV, the above nonzero quantities are simplified
as follows:
\begin{equation}
\theta =\frac{m(3\kappa -\lambda -1)}{2}e^{-\frac{m\kappa }{2}\tau }
\label{b3.25}
\end{equation}%
\begin{equation}
\sigma ^{2}=\frac{m^{2}(\lambda ^{2}-\lambda +1)}{12}e^{-m\kappa \tau }
\label{b3.26}
\end{equation}%
\begin{equation}
\rho =\rho _{0}\left( m,\kappa ,\lambda \right) e^{-m\kappa \tau
}~,~p=p_{0}\left( m,\kappa ,\lambda \right) e^{-m\kappa \tau }
\label{b3.27}
\end{equation}%
\begin{equation}
q^{a}=\left( 0,\frac{m\lambda }{2}e^{-\frac{3m\kappa }{2}\tau },0,0\right)
\label{b3.28}
\end{equation}
and
\begin{equation}
\pi _{xx}=\pi _{xx0}\left( m,\kappa ,\lambda \right) ~,~\pi _{yy}=\pi
_{yy0}\left( m,\kappa ,\lambda \right) e^{-m\tau }~,~\pi _{zz}=\pi
_{zz0}\left( m,\kappa ,\lambda \right) e^{-m\lambda \tau -2x}.  \label{b3.29}
\end{equation}

From the latter expressions, we infer that for large $\tau$ and $m \kappa >0$, all the kinematical quantities, the mass density, the isotropic pressure and
the heat flux vector vanish. If, in addition, $\pi_{xx0}\left( m,\kappa
,\lambda \right)=0$, $m>0$ and $\lambda >0$, then, for large $\tau$, the
fluid source vanishes and the solution describes an isotropic empty
spacetime.

\subsection{Bianchi V cosmological fluid}

\label{sec.Bianchi.4n.2} 

We consider the extended Bianchi V spacetime of Proposition \ref{pro.BianchiV}
\begin{equation}
ds_{(V)}^{2}=-A^{2}(\tau )e^{m\lambda \tau }d\tau ^{2}+A^{2}(\tau
)e^{m\lambda \tau }dx^{2}+A^{2}(\tau )e^{m(\lambda -1)\tau
}e^{2x}dy^{2}+A^{2}(\tau )e^{2x}dz^{2}  \label{eq.biaV}
\end{equation}%
and repeat the calculations of the previous section for the comoving observers. We find that the
kinematic quantities are exactly the same with those of the Bianchi III
spacetime, while the (nonzero) dynamic variables of the cosmological
fluid are
\begin{equation}
\rho =\frac{e^{-m\lambda \tau }}{4A^{2}}\left[ 4\frac{d\left( \ln A\right) }{%
d\tau }\left( 3\frac{d\left( \ln A\right) }{d\tau }+m\left( 2\lambda
-1\right) \right) +m^{2}\lambda \left( \lambda -1\right) -12\right]
\label{b4.19}
\end{equation}%
\begin{equation}
p=\frac{e^{-m\lambda \tau }}{A^{2}}\left[ -\frac{2}{A}\frac{d^{2}A}{d\tau
^{2}}+\frac{d\left( \ln A\right) }{d\tau }\left( \frac{d\left( \ln A\right)
}{d\tau }+\frac{m}{3}(2-\lambda )\right) -\frac{m^{2}}{12}(\lambda
-1)(\lambda -2)+1\right]   \label{b4.20}
\end{equation}%
\begin{equation}
q^{a}=\left( 0,-\frac{m(\lambda +1)}{2}\frac{e^{-\frac{3m\lambda }{2}\tau }}{%
A^{3}},0,0\right)  \label{b4.21}
\end{equation}%
\begin{equation}
\pi _{xx}=\frac{m(\lambda +1)}{3}\frac{d\left( \ln A\right) }{d\tau }+\frac{%
m^{2}(\lambda ^{2}-1)}{12}  \label{b4.22}
\end{equation}%
\begin{equation}
\pi _{yy}=e^{-m\tau +2x}\left[ \frac{m(\lambda -2)}{3}\frac{d\left( \ln
A\right) }{d\tau }+\frac{m^{2}}{12}(\lambda -1)(\lambda -2)\right]
\label{b4.23}
\end{equation}%
and
\begin{equation}
\pi _{zz}=-e^{-m\lambda \tau +2x}\left[ \frac{m(2\lambda -1)}{3}\frac{%
d\left( \ln A\right) }{d\tau }+\frac{m^{2}}{12}(\lambda -1)(2\lambda -1)%
\right].  \label{b4.24}
\end{equation}%
In the case $\mathbf{L}_{1}$ is a HV, we deduce the same conclusions with the Bianchi III case of sec. \ref{sec.Bianchi.4n.1}.

\subsection{Lie point symmetries of the wave equation}

\label{sec.Bianchi.4n.3} 

Collineations of spacetimes can be used to construct symmetries and
conservation laws for some differential equations defined in curved
spacetimes. In \cite{Tsamparlis 2011}, it has been shown that there exists a unique
connection between the point Noether symmetries for the geodesic Lagrangian of a
given Riemannian space and the elements of the admitted homothetic algebra.
Similar results have been proved for other PDEs of
special interest \cite{Paliathanasis 2012, Paliathanasis 2014}.

In this section, we consider the \textbf{wave equation}\index{Equation! wave}
\begin{equation}
\frac{1}{\sqrt{-g}}\frac{\partial }{\partial x^{\mu }}\left( \sqrt{-g}g^{\mu
\nu }\frac{\partial }{\partial x^{\nu }}\right) u\left( x^{\lambda }\right)
=0  \label{wave1}
\end{equation}%
in the Bianchi III spacetime (\ref{Bianchi3.1}) and in the Bianchi V spacetime (%
\ref{Bianchi5}), and determine its Lie point symmetries.\index{Symmetry! Lie point} By following the generic
results of \cite{Paliathanasis 2014}, we find that the admitted Lie point symmetries of the wave equation in the Bianchi III spacetime (\ref{Bianchi3.1}) are the three KVs, the vector field $Y_{u}=u\partial _{u}$, and the infinitely many vectors $Y_{\infty }=b\left( x^{\mu }\right) \partial _{u}$, where $b\left( x^{\mu
}\right) $ is a solution of the original equation (\ref{wave1}). The latter
symmetry vector fields exist because equation (\ref{wave1}) is a linear PDE.

For a higher-dimensional conformal algebra, equation (\ref{wave1})
admits extra Lie point symmetries. Indeed, from our analysis and for the
case where the Bianchi III and Bianchi V spacetimes admit a proper HV, the
wave equation becomes, respectively,
\begin{equation}
\left( -u_{tt}+u_{xx}+u_{yy}+e^{m\lambda t+2x}u_{zz}\right) +\frac{m}{2}%
\left( \lambda -2\kappa +1\right) u_{t}-u_{x}=0  \label{wave2}
\end{equation}%
and
\begin{equation}
\left( -u_{tt}+u_{xx}+e^{mt-2x}u_{yy}+e^{m\lambda t-2x}u_{zz}\right) +\frac{m%
}{2}\left( \lambda +2\kappa -1\right) u_{t}+4u_{x}=0.  \label{wave3}
\end{equation}

Then, we find that equation (\ref{wave2}) admits the generic Lie point symmetry vector
\begin{equation}
Y_{III}=\left( a_{1}\frac{2}{m}\right) \partial _{t}+a_{2}\partial
_{x}+\left( a_{1}y+a_{3}\right) \partial _{y}+\left( a_{1}\lambda
z+a_{2}z+a_{4}\right) \partial _{z}+\left[ a_{u}u+a_{\infty }b\left(
t,x,y,z\right) \right] \partial _{u} \label{wave4}
\end{equation}
while equation (\ref{wave3}) is invariant under the one-parameter point transformation with generator%
\begin{equation}
Y_{V}=\left( a_{1}\frac{2}{m}\right) \partial _{t}+a_{2}\partial _{x}+\left(
a_{1}y-a_{1}y+a_{3}\right) \partial _{y}+\left( a_{1}\lambda
z-a_{2}z+a_{4}\right) \partial _{z}+\left[ a_{u}u+a_{\infty }b\left(t,x,y,z\right) \right] \partial _{u}.  \label{wave5}
\end{equation}

The latter symmetry vectors can be applied to construct conservation laws or similarity solutions for the wave equation. However, such an analysis is beyond the scope of this work.

\section{Summary of results}

\label{subsec.bia.results}

\textbf{The Bianchi I spacetimes:}
\begin{enumerate}
\item
The spacetime
\[
ds^2_{(I)} = - C^2 e^{m \lambda \tau}  d\tau^2 + C^2 e^{m \lambda \tau} dx^2 + C^2 e^{m(\lambda - 1) \tau} dy^2 + C^2 dz^2
\]
admits the proper CKV  $\mathbf{K}_2 = \frac{2}{m} \partial_{\tau} + y \partial_y + \lambda z \partial_z$ with $\psi_{(I)}(\mathbf{K}_2) = \frac{2}{m} \frac{C_{,\tau}}{C} + \lambda$. When $\mathbf{K}_2$ is a HV, then
\begin{align*}
ds^2_{(I)} & = - e^{m \kappa \tau}  d\tau^2 + e^{m \kappa \tau} dx^2 + e^{m(\kappa - 1) \tau} dy^2 + e^{m(\kappa - \lambda)\tau} dz^2 \\
& = - dt^2 + \frac{m^2 \kappa^2 t^2}{4} dx^2 + \left( \frac{m^2 \kappa^2 t^2}{4} \right)^{\frac{\kappa -1}{\kappa}} dy^2 + \left( \frac{m^2 \kappa^2 t^2}{4} \right)^{\frac{\kappa - \lambda}{\kappa}} dz^2
\end{align*}
and $\mathbf{K}_2 = \kappa t \partial_t + y \partial_y + \lambda z \partial_z$ with $\psi_{(I)}(\mathbf{K}_2) = const \equiv \kappa \neq 0$.

\item
The spacetime
\begin{align*}
ds^2_{(I)} & = - C^2 b^2 \tau^{-2} d\tau^2 + C^2 b^2 \tau^{-2} dx^2 + C^2 b^2 \tau^{-2c} dy^2 + C^2 dz^2 \\
& = - C^2 dt^2 + C^2 b^2 e^{-2t/b} dx^2 + C^2 b^2 e^{-2ct/b} dy^2 + C^2 dz^2
\end{align*}
admits the proper CKV $\mathbf{K} = \tau \partial_{\tau} + x \partial_x + c y \partial_y$ $=$ $b \partial_t + x \partial_x + c y \partial_y$ with $\psi_{(I)}(\mathbf{K}) = \tau \frac{C_{,\tau}}{C} = b \frac{\dot{C}}{C}$. When $\mathbf{K}$ is a HV, then
\begin{align*}
ds^2_{(I)} & = - e^{2 \psi_0 t} dt^2 + e^{2 \psi_0 t} b^2 e^{-2t/b} dx^2 + e^{2 \psi_0 t} b^2 e^{-2ct/b} dy^2 + e^{2 \psi_0 t} dz^2 \\
& = - d\bar{t}^2 + b^2 (\psi_0 \bar{t})^{\frac{2(\psi_0 b -1)}{\psi_0 b}} dx^2 + b^2 (\psi_0 \bar{t})^{\frac{2(\psi_0 b -c)}{\psi_0 b}} dy^2 + (\psi_0 \bar{t})^2 dz^2.
\end{align*}

\item
The spacetime
\begin{align*}
ds^2_{(I)} & = - C^2 b_2^2 \tau^{2(b_1-1)} d\tau^2 + C^2 b_2^2 \tau^{2(b_1-1)} dx^2 + C^2 b_2^2 \tau^{2(b_1-c)} dy^2 + C^2 dz^2 \\
& = - C^2 dt^2 + C^2 b_1^2 \left( \frac{b_1}{b_2} \right)^{-\frac{2}{b_1}} t^{\frac{2(b_1-1)}{b_1}} dx^2 + C^2 b_1^2 \left( \frac{b_1}{b_2} \right)^{-\frac{2c}{b_1}} t^{\frac{2(b_1-c)}{b_1}} dy^2 + C^2 dz^2
\end{align*}
admits the proper CKV $\mathbf{K}_1 = \tau \partial_{\tau} + x \partial_x + c y \partial_y + b_1 z \partial_z = b_1 t \partial_t + x \partial_x + c y \partial_y + b_1 z \partial_z$ with $\psi_{(I)}(\mathbf{K}_1) = \tau \frac{C_{,\tau}}{C} + b_1 = b_1 t \frac{\dot{C}}{C} + b_1$. When $\mathbf{K}_1$ is a HV, then $C = t^{\frac{\psi_0 - b_1}{b_1}}$.

\item
The spacetimes
\[
ds^2_{(RT)} = - dt^2 + \sin^2 \left( \frac{t}{a} \right) dx^2 + \cos^2 \left( \frac{t}{a} \right) dy^2 + dz^2
\]
and
\[
ds^2_{(ART)} = - dt^2 + \sinh^2 \left( \frac{t}{a} \right) dx^2 + \cosh^2 \left( \frac{t}{a} \right) dy^2 + dz^2
\]
admit a 15-dimensional conformal algebra with a 7-dimensional Killing subalgebra (see appendix 2 in \cite{TsamPalKarp 2015}); its 3-dimensional space is of constant curvature.
\end{enumerate}

\textbf{The Bianchi III spacetimes:}
\begin{enumerate}
\item
The spacetime
\[
ds^2_{(III)} = - C^2(\tau) e^{m \lambda \tau}  d\tau^2 + C^2(\tau) e^{m \lambda \tau} dx^2 + C^2(\tau) e^{m(\lambda - 1) \tau} dy^2 + C^2(\tau) e^{-2x} dz^2
\]
admits the proper CKV $\mathbf{L}_1 = \frac{2}{m} \partial_{\tau} + y \partial_y + \lambda z \partial_z$ with  $\psi_{(III)}(\mathbf{L}_1) = \frac{2}{m} \frac{C_{,\tau}}{C} + \lambda$. When $\mathbf{L}_1$ is a HV, then
\begin{align*}
ds^2_{(III)} & = - e^{m \kappa \tau}  d\tau^2 + e^{m \kappa \tau} dx^2 + e^{m(\kappa - 1) \tau} dy^2 + e^{m(\kappa - \lambda)\tau} e^{-2x} dz^2 \\
& = - dt^2 + \frac{m^2 \kappa^2 t^2}{4} dx^2 + \left( \frac{m^2 \kappa^2 t^2}{4} \right)^{\frac{\kappa -1}{\kappa}} dy^2 + \left( \frac{m^2 \kappa^2 t^2}{4} \right)^{\frac{\kappa - \lambda}{\kappa}} e^{-2x} dz^2
\end{align*}
and $\mathbf{L}_1 = \kappa t \partial_t + y \partial_y + \lambda z \partial_z$ with $\psi_{(III)}(\mathbf{L}_1) = const \equiv \kappa \neq 0$.

\item
The spacetimes
\[
ds^2_{(III)} = - B^2 e^{m \tau} d\tau^2 + B^2 e^{m \tau} dx^2 + B^2 dy^2 + C^2 e^{-2x} dz^2
\]
and
\[
ds^2_{(III)} = - \frac{16 \alpha^2 C^2 e^{\tau}}{(e^{\tau} - 8c)^2} d\tau^2 + \frac{16 \alpha^2 C^2 e^{\tau}}{(e^{\tau} - 8c)^2} dx^2 + \alpha^2 C^2 dy^2 + C^2 e^{-2x} dz^2
\]
admit the KV $\mathbf{L} = \lambda \partial_x + \lambda z \partial_z$.
\end{enumerate}

\textbf{The Bianchi V spacetimes:}
\begin{enumerate}
\item
The spacetime
\[
ds^2_{(V)} = - C^2 e^{\lambda \tau} d\tau^2 +  C^2 e^{\lambda \tau} dx^2 + C^2 e^{(\lambda - 1)\tau} e^{2x} dy^2 + C^2 e^{2x} dz^2
\]
admits the proper CKV $\mathbf{L}_2 = 2 \partial_{\tau} + y \partial_y + \lambda z \partial_z$ with $\psi_{(V)}(\mathbf{L}_2) = 2 \frac{C_{\tau}}{C} + \lambda$. When $\mathbf{L}_2$ is a HV, then 
\begin{align*}
ds^2_{(V)} & = - e^{\kappa \tau} d\tau^2 + e^{\kappa \tau} dx^2 + e^{(\kappa - 1)\tau} e^{2x} dy^2 + e^{(\kappa - \lambda)\tau} e^{2x} dz^2 \\
& = - dt^2 + \frac{\kappa^2 t^2}{4} dx^2 + \left( \frac{ \kappa^2 t^2}{4} \right)^{\frac{\kappa -1}{\kappa}} e^{2x} dy^2 + \left( \frac{\kappa^2 t^2}{4} \right)^{\frac{\kappa - \lambda}{\kappa}} e^{2x} dz^2
\end{align*}
and $\mathbf{L}_2 = \kappa t \partial_t + y \partial_y + \lambda z \partial_z$ with $\psi_{(V)}(\mathbf{L}_2) = const \equiv \kappa \neq 0$.

\item
Any Bianchi V spacetime
\[
ds^2_{(V)} = - dt^2 + A^2(t) dx^2 + B^2(t) e^{2x} dy^2 + C^2(t) e^{2x} dz^2
\]
admits the KV $\mathbf{M}_2 = \partial_x - y \partial_y - z \partial_z$.

\end{enumerate}

\section{Conclusions}

\label{sec7}

In this chapter, we have shown that there is only one type of Bianchi III and
Bianchi V spacetime given, respectively, in (\ref{b3.12}) and (\ref{Bianchi5}) which admit a single proper CKV. Furthermore, two more spacetimes are found which admit a HV. In order to arrive at this result, we applied an
algorithm which relates the CKVs of decomposable spacetimes with the
collineations of their non-decomposable subspaces. The kinematics of the fluid of the comoving observers in all these four spacetimes is not accelerating
and rotating and has only expansion and shear; a result compatible with the
anisotropy of the Bianchi spacetimes. Concerning the dynamics, it has been
shown that the fluid of these observers is heat conducting and anisotropic,
that is, it is a general fluid. Finally, we have used the CKVs we
found in each case in order to determine the generators of the Lie point
symmetries of the wave equation in the Bianchi III spacetime (\ref{b3.12})
and in the Bianchi V spacetime (\ref{Bianchi5}).

%% file: QFIs_conservative_systems.tex
\chapter{Quadratic first integrals of autonomous conservative dynamical systems}

\label{ch1.QFIs.conservative}

\section{Introduction}

As we have seen in chapter \ref{ch.integrability}, FIs are used to reduce the order of the dynamical equations. The standard method to determine the FIs of a Lagrangian dynamical system is to use a special class of Lie symmetries, the Noether symmetries. We recall that a Noether symmetry is a Lie symmetry which in addition satisfies the Noether condition \cite{Sarlet Cantrijn 81, Lutzky, Sarlet, kalotas} (see also sec. \ref{sec.classym})
\begin{equation}
\mathbf{X}^{[1]}L+\frac{d\xi }{dt}L=\frac{df}{dt}.  \label{FI.1}
\end{equation}
According to Noether's theorem \ref{con.mot.pro.3}, to every Noether symmetry corresponds the Noether FI
\begin{equation}
I=\xi \left( \dot{q}^{a}\frac{\partial L}{\partial \dot{q}^{a}}-L\right) -\eta ^{a}\frac{\partial L}{\partial \dot{q}^{a}}+f  \label{FI.4}
\end{equation}%
which can be easily determined if one knows the generator $\mathbf{X}= \xi(t,q,\dot{q},...) \partial_{t} +\eta^{a}(t,q,\dot{q},....) \partial_{q^{a}}$ of the Noether symmetry.

In this chapter, we restrict to autonomous conservative dynamical systems and we determine their quadratic FIs (QFIs) by following a different approach (see e.g. \cite{StephaniB, Katzin 1973, kalotas, Kaplan}). In this latter approach, one assumes the QFI to be of the general form\footnote{
The case of linear FIs (LFIs) also included for $K_{ab}=0$.
}:\index{First integral! quadratic}
\begin{equation}
I=K_{ab}(t,q)\dot{q}^{a}\dot{q}^{b}+K_{a}(t,q)\dot{q}^{a}+K(t,q)  \label{FL.5}
\end{equation}%
where the coefficients $K_{ab}, K_{a}, K$ are symmetric tensors depending on the coordinates $t, q^{a}$ and imposes directly the condition $\frac{dI}{dt}=0$ along the dynamical equations. This condition leads to a system of PDEs involving the unknown quantities $K_{ab}, K_{a}, K$ whose solution provides the QFIs (\ref{FL.5}). In all occasions considered so far, the system of these conditions has been solved for specific cases only. The aim of the present chapter\footnote{A recent preliminary work along this line is presented in \cite{Leonidas 2018}.} is double: a. To give the general solution of the system of PDEs resulting from the condition $\frac{dI}{dt}=0$; and b. To geometrize the answer to the maximum possible degree.

\section{The conditions for a QFI}

\label{sec.ch1.conditions.QFIs}

\subsection{The case of a general dynamical system}

\label{sec.ch1.conditions.QFIs.1}

In this section, we consider an $n$-dimensional dynamical system defined by the equations of motion
\begin{equation}
\ddot{q}^{a}=-\Gamma^{a}_{bc}(q) \dot{q}^{b}\dot{q}^{c} -V^{,a}(q) +Q^{a}\left( t,q,\dot{q}\right) \label{FL.0.3}
\end{equation}%
where $Q^{a}$ are the non-conservative generalized forces, $\Gamma^{a}_{bc}$ are the Riemannian connection coefficients defined from the kinetic metric $\gamma_{ab}(q)$ (kinetic energy) of the system and $-V^{,a}$ are the conservative generalized forces.

We consider, next, a function $I(t,q^{a},\dot{q}^{a})$, which is linear and quadratic in the velocities, with coefficients which depend only on the coordinates $t, q^{a}$, that is, $I$ is of the form:
\begin{equation}
I=K_{ab}(t,q)\dot{q}^{a}\dot{q}^{b}+K_{a}(t,q)\dot{q}^{a}+K(t,q)
\label{FI.5}
\end{equation}%
where $K_{ab}$ is a symmetric tensor, $K_{a}$ is a vector and $K$ is an invariant.

We demand that $I$ is a FI of (\ref{FL.0.3}). This requirement leads to the condition $\frac{dI}{dt}=0$ which gives a system of PDEs for the coefficients $K_{ab},$ $K_{a}$ and $K$. Using the dynamical equations (\ref{FL.0.3}) to replace $\ddot{q}^{a}$
whenever it appears, we find
\begin{align}
\frac{dI}{dt}& =K_{(ab;c)}\dot{q}^{a}\dot{q}^{b}\dot{q}^{c}+\left(
K_{ab,t}+K_{a;b}\right) \dot{q}^{a}\dot{q}^{b}+2K_{ab}\dot{q}%
^{(b}(Q^{a)}-V^{,a)})+\left( K_{a,t}+K_{,a}\right) \dot{q}^{a}+  \notag \\
& \quad +K_{a}(Q^{a}-V^{,a})+K_{,t}.  \label{eq.veldep3}
\end{align}%
In order to get a working environment, we restrict our considerations to linear generalized forces, that is, we consider the case $Q^{a}=A_{b}^{a}(q)\dot{q}^{b}$. Then, the general result (\ref{eq.veldep3}) becomes
\begin{align*}
0& =K_{(ab;c)}\dot{q}^{a}\dot{q}^{b}\dot{q}^{c}+\left(
K_{ab,t}+K_{a;b}+2K_{c(b}A_{a)}^{c}\right) \dot{q}^{a}\dot{q}^{b}+\left(
K_{a,t}+K_{,a}-2K_{ab}V^{,b}+\right. \\
& \quad \left. +K_{b}A_{a}^{b}\right) \dot{q}^{a}+K_{,t}-K_{a}V^{,a}
\end{align*}%
from which follows the system of PDEs\footnote{A subcase of these equations, for $K_{a}=0$ and $A^{a}_{b}=0$, has been found before by e.g. Kalotas (see eqs. (12a) - (12d) in \cite{kalotas}) who considered their solution in certain special cases.}:
\begin{eqnarray}
K_{(ab;c)} &=&0  \label{eq.veldep4.1N} \\
K_{ab,t}+K_{(a;b)}+2K_{c(b}A_{a)}^{c} &=&0  \label{eq.veldep4.2N} \\
-2K_{ab}V^{,b}+K_{a,t}+K_{,a}+K_{b}A_{a}^{b} &=&0  \label{eq.veldep4.3N} \\
K_{,t}-K_{a}V^{,a} &=&0.  \label{eq.veldep4.4N}
\end{eqnarray}
Condition $K_{(ab;c)}=0$ implies that $K_{ab}$ is a KT of order two (possibly zero) of the kinetic metric $\gamma _{ab}$.

Because $\gamma_{ab}$ is autonomous, the condition $K_{(ab;c)}=0$ is satisfied if the KT $K_{ab}$ is of the form
\begin{equation}
K_{ab}(t,q)=g(t)C_{ab}(q) \label{choice1}
\end{equation}%
where $g(t)$ is an arbitrary analytic function and $C_{ab}(q)$ is a KT of order two of the metric $\gamma _{ab}$. This choice of $K_{ab}$ and equation (\ref{eq.veldep4.2N}) indicate that we set
\begin{equation}
K_{a}(t,q)=f(t)L_{a}(q)+B_{a}(q) \label{choice2}
\end{equation}
where $f(t)$ is an arbitrary analytic function and $L_{a}(q), B_{a}(q)$ are
arbitrary vectors.

Replacing the choices (\ref{choice1}) and (\ref{choice2}) in the system of equations (\ref{eq.veldep4.1N}) - (\ref{eq.veldep4.4N}), we find the following system of PDEs\footnote{Equation (\ref{eq.veldep4.1N}) is satisfied identically, because the quantities $C_{ab}(q)$ are assumed to second order KTs.}:
\begin{eqnarray}
g_{,t}C_{ab}+f(t)L_{(a;b)}+B_{(a;b)}+2g(t)C_{c(b}A_{a)}^{c} &=&0
\label{eq.veldep6N} \\
-2g(t)C_{ab}V^{,b}+f_{,t}L_{a}+K_{,a}+(fL_{b}+B_{b})A_{a}^{b} &=&0
\label{eq.veldep7N} \\
K_{,t}-(fL_{a}+B_{a})V^{,a} &=&0.  \label{eq.veldep8N}
\end{eqnarray}

Conditions (\ref{eq.veldep6N}) - (\ref{eq.veldep8N}) must be supplemented with the integrability conditions $K_{,at}=K_{,ta}$ and $K_{,[ab]}=0$ for the scalar function $K$. The integrability condition $K_{,at}=K_{,ta}$ gives --if we make use of (\ref{eq.veldep7N}) and (\ref{eq.veldep8N})-- the equation
\begin{equation}
f_{,tt}L_{a}+f_{,t}L_{b}A_{a}^{b}+f\left( L_{b}V^{;b}\right) _{;a}+\left(
B_{b}V^{;b}\right) _{;a}-2g_{,t}C_{ab}V^{,b}=0.  \label{eq.veldep9N}
\end{equation}
Condition $K_{,[ab]}=0$ gives the equation
\begin{equation}
2g\left( C_{[a\left\vert c\right\vert }V^{,c}\right) _{;b]}-f_{,t}L_{\left[
a;b\right] }-(fL_{c;[b}+B_{c;[b})A_{a]}^{c}-(fL_{c}+B_{c})A_{[a;b]}^{c}=0
\label{eq.veldep10N}
\end{equation}
which is known as the \textbf{second order Bertrand-Darboux equation}.\index{Equation! Bertrand-Darboux}

Finally, the system of equations which we have to solve consists of equations (\ref{eq.veldep6N}) - (\ref{eq.veldep10N}), where $C_{ab}(q)$ is a KT.

\subsection{The case of autonomous conservative dynamical systems}

\label{sec.tables.theorem}

We restrict further our considerations to the case of autonomous
conservative dynamical systems so that $V=V(q)$ and $Q^{a}=0$. In this case,
the system of equations (\ref{eq.veldep6N}) - (\ref{eq.veldep10N}) reduces as
follows:
\begin{eqnarray}
g_{,t}C_{ab}+fL_{(a;b)}+B_{(a;b)} &=&0  \label{FL.1.a} \\
-2gC_{ab}V^{,b}+f_{,t}L_{a}+K_{,a} &=&0  \label{FL.1.b} \\
K_{,t}-fL_{a}V^{,a}-B_{a}V^{,a} &=&0  \label{FL.1.c} \\
f_{,tt}L_{a}+f(L_{b}V^{,b})_{;a}+(B_{b}V^{,b})_{;a}-2g_{,t}C_{ab} V^{,b} &=&0 \label{FL.1.d} \\
2g\left( C_{[a\left\vert c\right\vert }V^{,c}\right) _{;b]}-f_{,t}L_{[a;b]}
&=&0.  \label{FL.1.e}
\end{eqnarray}

Obviously, the solution of this system of PDEs is quite involved and requires the consideration of many cases and subcases. The general solution of the system is stated in the following theorem (the proof is given in appendix \ref{app2.proof.QFIs}).

\begin{theorem}
\label{The first integrals of an autonomous holonomic dynamical system} Assume that the functions $g(t)$ and $f(t)$ are analytic so that they may be represented by polynomial functions as follows:
\begin{equation}  \label{eq.thm1}
g(t) = \sum^n_{k=0} c_k t^k = c_0 + c_1 t + ... + c_n t^n
\end{equation}
\begin{equation}  \label{eq.thm2}
f(t) = \sum^m_{k=0} d_k t^k = d_0 + d_1 t + ... + d_m t^m
\end{equation}
where $n, m \in \mathbb{N}$, or may be infinite, and $c_k, d_k \in \mathbb{R}$. Then, the independent QFIs  of an autonomous conservative dynamical system are the following:
\bigskip

\textbf{Integral 1.}
\begin{equation*}
I_{1} = -\frac{t^{2}}{2} L_{(a;b)}\dot{q}^{a}\dot{q}^{b} + C_{ab}\dot{q}^{a} \dot{q}^{b} + t L_{a} \dot{q}^{a} + \frac{t^{2}}{2} L_{a}V^{,a} + G(q)
\end{equation*}
where $C_{ab}$ and $L_{(a;b)}$ are KTs, $\left(L_{b}V^{,b}\right)_{,a} =
-2L_{(a;b)} V^{,b}$ and $G_{,a}= 2C_{ab}V^{,b} - L_{a}$.

\textbf{Integral 2.}
\begin{equation*}
I_{2} = -\frac{t^{3}}{3} L_{(a;b)}\dot{q}^{a}\dot{q}^{b} + t^{2} L_{a} \dot{q%
}^{a} + \frac{t^{3}}{3} L_{a}V^{,a} - t B_{(a;b)} \dot{q}^{a}\dot{q}^{b} +
B_{a}\dot{q}^{a} + tB_{a}V^{,a}
\end{equation*}
where $L_{a}$ and $B_{a}$ are such that $L_{(a;b)}$ and $B_{(a;b)}$ are KTs, $\left(L_{b}V^{,b}\right)_{,a} = -2L_{(a;b)} V^{,b}$ and $\left(B_{b}V^{,b}\right)_{,a} = -2B_{(a;b)} V^{,b} - 2L_{a}$.

\textbf{Integral 3.}
\begin{equation*}
I_{3} = -e^{\lambda t} L_{(a;b)}\dot{q}^{a}\dot{q}^{b} + \lambda e^{\lambda
t} L_{a} \dot{q}^{a} + e^{\lambda t} L_{a} V^{,a}
\end{equation*}
where $\lambda \neq 0$, $L_{a}$ is such that $L_{(a;b)}$ is a KT and $\left(L_{b}V^{,b}\right)_{,a} = -2L_{(a;b)} V^{,b} - \lambda^{2} L_{a}$.
\end{theorem}

\begin{remark} \label{remark.new.3}
Concerning the Lie bracket between an arbitrary vector field $B^{a}$ and a gradient vector field $V^{,a}$, we have the following:\index{Lie bracket}
\[
\left[\mathbf{B},\mathbf{\nabla}V\right]^{a} =B^{b}V^{,a}{}_{,b} -V^{,b}B^{a}{}_{,b}= B^{b}V^{,a}{}_{;b} -V^{,b}B^{a}{}_{;b} \implies
\]
\[
\left[ \mathbf{B},\mathbf{\nabla}V \right]_{a} =B^{b}V_{;ba}-B_{a;b}V^{,b}= \left(B_{b}V^{,b}\right)_{;a} -2B_{(a;b)}V^{,b}.
\]
Therefore, $\left[ \mathbf{B},\mathbf{\nabla }V \right]_{a}=0\iff \left(B_{b}V^{,b}\right)_{;a} =2B_{(a;b)}V^{,b}$. If $B^{a}$ is a KV such that $B_{a}V^{,a}=const$, then the vectors $B^{a}$ and $V^{,a}$ commute (i.e. $\left[ \mathbf{B}, \mathbf{\nabla }V \right]^{a}=0$).
\end{remark}

For easier reference, we collect the QFIs and the LFIs  of Theorem \ref{The first integrals of an autonomous holonomic dynamical system} in Tables \ref{Table.QFIs.thm1} and \ref{Table.LFIs.thm1}, respectively.

\begin{longtable}{|l|l|}
\hline
QFI & Conditions \\ \hline
\makecell[l]{$I_{1} = -\frac{t^{2}}{2} L_{(a;b)}\dot{q}^{a}\dot{q}^{b} + C_{ab}\dot{q}^{a} \dot{q}^{b} + t L_{a} \dot{q}^{a} +$ \\ \qquad \enskip $+ \frac{t^{2}}{2} L_{a}V^{,a} + G(q)$} & \makecell[l]{$C_{ab}, L_{(a;b)}$ are KTs, $\left(L_{b}V^{,b}\right)_{,a} = -2L_{(a;b)} V^{,b}$, \\ $G_{,a}=2C_{ab}V^{,b} - L_{a}$} \\ \hline
\makecell[l]{$I_{2} = -\frac{t^{3}}{3} L_{(a;b)}\dot{q}^{a}\dot{q}^{b} +
t^{2} L_{a} \dot{q}^{a} + \frac{t^{3}}{3} L_{a}V^{,a} -$ \\ \qquad \enskip
$- t B_{(a;b)} \dot{q}^{a}\dot{q}^{b} + B_{a}\dot{q}^{a} + tB_{a}V^{,a}$} & %
\makecell[l]{$L_{(a;b)}, B_{(a;b)}$ are KTs, $\left(L_{b}V^{,b}\right)_{,a}
= -2L_{(a;b)} V^{,b}$, \\ $\left(B_{b}V^{,b}\right)_{,a} = -2B_{(a;b)}
V^{,b} - 2L_{a}$} \\ \hline
$I_{3} = e^{\lambda t} \left( -L_{(a;b)}\dot{q}^{a}\dot{q}^{b} + \lambda
L_{a} \dot{q}^{a} + L_{a}V^{,a} \right)$ & $L_{(a;b)} = KT$, $%
\left(L_{b}V^{,b}\right)_{,a} = -2L_{(a;b)} V^{,b} - \lambda^{2} L_{a}$ \\
\hline
\caption{\label{Table.QFIs.thm1} The QFIs of Theorem \ref{The first integrals of an autonomous holonomic dynamical system}.}
\end{longtable}

\begin{longtable}{|l|l|}
\hline
LFI & Conditions \\ \hline
$I_{1}=-tG_{,a}\dot{q}^{a}-\frac{s}{2}t^{2}+G(q)$ & $G_{,a}=KV$, $G_{,a}V^{,a}=s$ \\
$I_{2}=(t^{2}L_{a}+B_{a})\dot{q}^{a}+\frac{s}{3}t^{3}+tB_{a}V^{,a}$ & $%
L_{a},B_{a}$ are KVs, $L_{a}V^{,a}=s$, $\left( B_{b}V^{,b}\right)
_{,a}=-2L_{a}$ \\
$I_{3}=e^{\lambda t}\left( \lambda L_{a}\dot{q}^{a}+L_{a}V^{,a}\right) $ & $%
L_{a}=KV$, $\left( L_{b}V^{,b}\right) _{,a}=-\lambda ^{2}L_{a}$ \\ \hline
\caption{\label{Table.LFIs.thm1} The LFIs of Theorem \ref{The first integrals of an autonomous holonomic dynamical system}.}
\end{longtable}

We note that all the QFIs reduce to LFIs when the KT $K_{ab}$ vanishes. Moreover, it can be checked that the FIs listed in Theorem \ref{The first integrals of an autonomous holonomic dynamical system} produce all the potentials, which admit a LFI or a QFI, given in \cite{TsaPal 2011} and are due to point Noether symmetries. Since, as it is shown in Theorem \ref{Inverse Noether Theorem}, these FIs also follow from a gauged velocity-dependent Noether symmetry, we conclude that \emph{there does not exist a one-to-one correspondence between Noether FIs and the type of Noether symmetry.} To illustrate this important statement, we give some examples.

The QFI of the total energy (Hamiltonian) $E= \frac{1}{2} \gamma_{ab}\dot{q}^{a}\dot{q}^{b} +V(q)$ (case \textbf{Integral 1} for $L_{a}=0$ and $C_{ab}=\frac{\gamma_{ab}}{2}$) is generated by the point Noether symmetry $\Big(\xi=1, \eta_{a}=0; f=0\Big)$ and, also, by the gauged generalized Noether symmetry
\[
\Big( \xi=0, \enskip \eta_{a}= -\gamma_{ab}\dot{q}^{b}; \enskip f= -\frac{1}{2}\gamma_{ab}\dot{q}^{a}\dot{q}^{b} +V(q) \Big).
\]

The QFI $-I_{2}(L_{a}=0)$ for $B_{a}$ be a HV with homothetic factor $\psi=const$ is generated by the point Noether symmetry $\Big( \xi=2\psi t, \eta_{a}=B_{a}; f=ct \Big)$ such that $B_{a}V^{,a} +2\psi V +c =0$, where $c$ is an arbitrary constant, and, also, by the gauged generalized Noether symmetry \[
\Big( \xi=0, \enskip \eta_{a}= -2t\psi\gamma_{ab}\dot{q}^{b} +B_{a}; \enskip f= -t\psi\gamma_{ab} \dot{q}^{a} \dot{q}^{b} -tB_{a}V^{,a} \Big).
\]

As a final example, we consider the QFI $-\frac{I_{3}}{\lambda}$ for the gradient HV $L_{a}=\Phi(q)_{,a}$ where  $\Phi_{;ab}= \psi\gamma_{ab}$ with $\psi=const$. This QFI is generated by the point Noether symmetry
\[
\Big( \xi= \frac{2\psi}{\lambda}e^{\lambda t}, \eta_{a}= e^{\lambda t} \Phi(q)_{,a}; f= \lambda e^{\lambda t} \Phi(q) -\frac{c}{\lambda} e^{\lambda t} \Big)
\]
where $\lambda, c$ are non-zero constants and $\Phi_{,a}V^{,a}= -2\psi V -\lambda^{2}\Phi +c$, and, also, by the gauged generalized Noether symmetry
\[
\Big( \xi=0, \eta_{a} = -\frac{2\psi}{\lambda} e^{\lambda t} \gamma_{ab}\dot{q}^{b} + e^{\lambda t} \Phi_{,a}; f= -\frac{\psi}{\lambda} e^{\lambda t} \gamma_{ab}\dot{q}^{a} \dot{q}^{b} -\frac{e^{\lambda t}}{\lambda} \Phi_{,a} V^{,a} \Big).
\]

As a first application of Theorem \ref{The first integrals of an autonomous holonomic dynamical system}, we determine in the next section the QFIs of geodesic equations.

\section{The QFIs of geodesic equations of an $n$-dimensional Riemannian space}

\label{sec.QFIs.geodesics}

Concerning the FIs of geodesic equations, we have the following well-known result (see sec. 39 of \cite{Eisenhart}):

\begin{proposition}
\label{prop.Eisen} The geodesic equations in an $n$-dimensional Riemannian manifold $(M, g_{ab})$ admit $m$th-order FIs
of the form
\begin{equation}
A_{r_{1}...r_{m}}\lambda ^{r_{1}}...\lambda ^{r_{m}}=const  \label{FOIG.1}
\end{equation}%
where $\lambda ^{a}\equiv \dot{q}^{a}$ and $A_{r_{1}...r_{m}}$ is an $m$th-order KT of the metric $g_{ab}$.
\end{proposition}

In order to determine the QFIs of the geodesic
equations\index{Equations! geodesic} in an $n$-dimensional Riemannian space with metric $\gamma_{ab}$, we apply Theorem \ref{The first integrals of an autonomous holonomic dynamical system} with $V=0$. For each case of Theorem \ref{The first integrals of an autonomous holonomic dynamical system}, we have the following:
\bigskip

\textbf{Integral 1.}
In this case $L_{a}=-G_{,a}$ and the QFI is written
\begin{equation*}
I_{1} = \frac{t^{2}}{2} G_{;ab} \dot{q}^{a}\dot{q}^{b} + C_{ab}\dot{q}^{a} \dot{q}^{b} - t G_{,a} \dot{q}^{a}+ G(q)
\end{equation*}
where $C_{ab}$ and $G_{;ab}$ are KTs.

The QFI $I_{1}$ consists of two independent QFIs, which are the following:
\[
I_{1a} = C_{ab}\dot{q}^{a} \dot{q}^{b} \enskip \text{and} \enskip I_{1b} = \frac{t^{2}}{2} G_{;ab} \dot{q}^{a}\dot{q}^{b} - t G_{,a} \dot{q}^{a}+ G(q).
\]

\textbf{Integral 2.}
Since $V=0$, the condition $\left(B_{b}V^{,b}\right)_{,a} = -2B_{(a;b)} V^{,b} - 2L_{a}$ implies that $L_{a}=0$. Therefore, the QFI is written
\begin{equation*}
I_{2} = - t B_{(a;b)} \dot{q}^{a}\dot{q}^{b} +
B_{a}\dot{q}^{a}
\end{equation*}
where $B_{a}$ is such that $B_{(a;b)}$ is a KT.

\textbf{Integral 3.}
Since $V=0$ and $\lambda\neq0$, the condition $\left(L_{b}V^{,b}\right)_{,a} = -2L_{(a;b)} V^{,b} - \lambda^{2} L_{a}$ implies that $L_{a}=0$. Therefore, the QFI $I_{3}=0$.
\bigskip

We collect the above results in Table \ref{Table.QFIs.geodesics}.

\begin{longtable}{|l|l|}
\hline
QFI & Condition \\ \hline
$I_{1a} = C_{ab}\dot{q}^{a}\dot{q}^{b}$ & $C_{ab}=$ KT \\
$I_{1b} = \frac{t^{2}}{2} G_{;ab} \dot{q}^{a}\dot{q}^{b} - tG_{,a} \dot{q}^{a} + G(q)$ & $G_{;ab}=$ KT \\
$I_{2} = -tB_{(a;b)}\dot{q}^{a}\dot{q}^{b} + B_{a}\dot{q}^{a}$ & $B_{(a;b)}= $ KT \\ \hline
\caption{\label{Table.QFIs.geodesics} The QFIs of geodesic equations.}
\end{longtable}

\section{The general Kepler problem $V= -\frac{k}{r^{\ell}}$}

\label{sec.GKepler}

The general Kepler problem\index{Potential! generalized Kepler} is a 3d Euclidean dynamical system with kinetic metric $\delta _{ij}=diag(1,1,1)$ and potential $V=-\frac{k}{r^{\ell }}$, where $k,\ell$ are non-zero real constants and $r=(x^{2}+y^{2}+z^{2})^{\frac{1}{2}}$. This dynamical system reduces to the 3d harmonic oscillator\index{Oscillator! 3d harmonic} for $k<0$ and $\ell =-2$ (which is the probe dynamical system for checking the validity of arguments and calculations); and to the classical Kepler problem\index{Potential! Kepler} considered earlier by Kalotas \cite{kalotas} for $\ell =1$. The Lagrangian of the system is
\begin{equation}
L=\frac{1}{2}(\dot{x}^{2}+\dot{y}^{2}+\dot{z}^{2}) +\frac{k}{r^{\ell }} \label{eq.GKep.1}
\end{equation}%
with equations of motion:
\begin{equation}
\ddot{x}=-\frac{\ell k}{r^{\ell +2}}x,\enskip\ddot{y}=-\frac{\ell k }{r^{\ell +2}}y,\enskip\ddot{z}=-\frac{\ell k}{r^{\ell +2}}z.
\label{eq.GKep.1a}
\end{equation}

To determine the QFIs of the above dynamical system, we apply Theorem \ref{The first integrals of an autonomous holonomic dynamical system} using the geometric quantities of $E^{3}$ (see sec. \ref{sec.KTE3}).
\bigskip

\textbf{Integral 1.}
\begin{equation*}
I_{1} = -\frac{t^{2}}{2} L_{(a;b)}\dot{q}^{a}\dot{q}^{b} + C_{ab}\dot{q}^{a} \dot{q}^{b} + t L_{a} \dot{q}^{a} + \frac{t^{2}}{2} L_{a}V^{,a} + G(q)
\end{equation*}
where $C_{ab}$, $L_{(a;b)}$ are KTs, $\left(L_{b}V^{,b}\right)_{,a} =
-2L_{(a;b)} V^{,b}$ and $G_{,a}= 2C_{ab}V^{,b} - L_{a}$.
\bigskip

Since $C_{ab}$ and $L_{(a;b)}$ are KTs, the results of sec. \ref{sec.KTE3} imply that:
\begin{eqnarray*}
C_{11} &=&\frac{a_{6}}{2}y^{2}+\frac{a_{1}}{2}%
z^{2}+a_{4}yz+a_{5}y+a_{2}z+a_{3}  \\
C_{12} &=&\frac{a_{10}}{2}z^{2}-\frac{a_{6}}{2}xy-\frac{a_{4}}{2}xz-\frac{%
a_{14}}{2}yz-\frac{a_{5}}{2}x-\frac{a_{15}}{2}y+a_{16}z+a_{17} \\
C_{13} &=&\frac{a_{14}}{2}y^{2}-\frac{a_{4}}{2}xy-\frac{a_{1}}{2}xz-\frac{%
a_{10}}{2}yz-\frac{a_{2}}{2}x+a_{18}y-\frac{a_{11}}{2}z+a_{19}
\\
C_{22} &=&\frac{a_{6}}{2}x^{2}+\frac{a_{7}}{2}%
z^{2}+a_{14}xz+a_{15}x+a_{12}z+a_{13}  \\
C_{23} &=&\frac{a_{4}}{2}x^{2}-\frac{a_{14}}{2}xy-\frac{a_{10}}{2}xz -\frac{a_{7}}{2}yz-(a_{16}+a_{18})x-\frac{a_{12}}{2}y -\frac{a_{8}}{2}z +a_{20}
 \\
C_{33} &=&\frac{a_{1}}{2}x^{2}+\frac{a_{7}}{2}%
y^{2}+a_{10}xy+a_{11}x+a_{8}y+a_{9}
\end{eqnarray*}
\begin{equation*}
L_{a}=\left(
\begin{array}{c}
-b_{15}y^{2}-b_{11}z^{2}+b_{5}xy+b_{2}xz +2(b_{16}+b_{18})yz+b_{3}x+2b_{4}y+2b_{1}z+b_{6}
\\
-b_{5}x^{2}-b_{8}z^{2}+b_{15}xy-2b_{18}xz+b_{12}yz+ 2(b_{17}-b_{4})x+b_{13}y+2b_{7}z+b_{14}
\\
-b_{2}x^{2}-b_{12}y^{2}-2b_{16}xy+b_{11}xz+b_{8}yz+2(b_{19}- b_{1})x+2(b_{20}-b_{7})y+b_{9}z+b_{10}
\end{array}%
\right)
\end{equation*}
and
\[
L_{(1;1)}= b_{5}y+b_{2}z+b_{3}, \enskip L_{(1;2)}= -\frac{b_{5}}{2}x -\frac{b_{15}}{2}y +b_{16}z+b_{17}, \enskip L_{(1;3)}= -\frac{b_{2}}{2}x+b_{18}y-\frac{b_{11}}{2}z+b_{19},
\]
\[
L_{(2;2)}= b_{15}x+b_{12}z+b_{13}, \enskip L_{(2;3)}= -(b_{16}+b_{18})x-\frac{b_{12}}{2}y-\frac{b_{8}}{2}z+b_{20}, \enskip L_{(3;3)}= b_{11}x+b_{8}y+b_{9}.
\]

Substituting the above quantities in $\left(L_{b}V^{,b} \right)_{,a} =-2L_{(a;b)} V^{,b}$ and taking the integrability conditions $G_{,[ab]}=0$ of the constraint $G_{,a}= 2C_{ab}V^{,b} - L_{a}$, we find:
\[
a_{16}=a_{18}=0,\enskip(\ell+2)a_{17}=0, \enskip(\ell+2)a_{19}=0,\enskip (\ell+2)a_{20}=0,
\]
\[
(\ell-1)a_{2}=0,\enskip (\ell-1)a_{5}=0, \enskip (\ell-1)a_{11}=0, \enskip a_{2}=a_{12}, \enskip a_{5}=a_{8}, \enskip a_{11}=a_{15},
\]
\[
(\ell+2)(a_{3}-a_{13})=0,\enskip(\ell+2)(a_{3}-a_{9})=0, \enskip b_{3}=b_{9}=b_{13}, \enskip (\ell-2)b_{3}=0
\]
and $b_{1}=b_{2}=b_{4}=b_{5}=b_{6}=b_{7}=b_{8}=b_{10}= b_{11}=b_{12}=b_{14}=b_{15}=b_{16}=b_{17}=b_{18}=b_{19}=b_{20} =0$.

The above conditions lead to the following four cases: a) $\ell=-2$ (3d harmonic oscillator); b) $\ell=1$ (the Kepler problem); c) $\ell=2$ (Newton-C\^{o}tes potential); and d) $\ell \neq -2,1,2$.
\bigskip

a) Case $\ell=-2$.

We have $L_{a}=0$ and $a_{2}=a_{5} =a_{8}=a_{11}= a_{12}=a_{15}= a_{16}=a_{18}=0$.

Then, the independent components of the KT $C_{ab}$ are
\begin{eqnarray*}
C_{11} &=&\frac{a_{6}}{2}y^{2}+\frac{a_{1}}{2}z^{2}+a_{4}yz+a_{3}  \\
C_{12} &=&\frac{a_{10}}{2}z^{2}-\frac{a_{6}}{2}xy-\frac{a_{4}}{2}xz-\frac{%
a_{14}}{2}yz+a_{17} \\
C_{13}&=&\frac{a_{14}}{2}y^{2}-\frac{a_{4}}{2}xy -\frac{a_{1}}{2}xz-\frac{a_{10}}{2}yz+a_{19} \\
C_{22} &=&\frac{a_{6}}{2}x^{2}+\frac{a_{7}}{2}z^{2}+a_{14}xz+a_{13} \\
C_{23} &=&\frac{a_{4}}{2}x^{2}-\frac{a_{14}}{2}xy-\frac{a_{10}}{2}xz -\frac{a_{7}}{2}yz+a_{20} \\
C_{33} &=& \frac{a_{1}}{2}x^{2}+\frac{a_{7}}{2}y^{2}+a_{10}xy+a_{9}.
\end{eqnarray*}

Substituting these components in $G_{,a}=2C_{ab}V^{,b}$ and integrating wrt each coordinate, we find
\begin{equation*}
G(x,y,z)=-2k\left( a_{3}x^{2}+a_{13}y^{2}+a_{9}z^{2}+2a_{17}xy+
2a_{19}xz+2a_{20}yz \right).
\end{equation*}

The QFI is
\begin{eqnarray*}
I_{1} &=& C_{ab}\dot{q}^{a}\dot{q}^{b} + G(x,y,z) \\ &=&\frac{a_{1}}{2}(z\dot{x}-x\dot{z})^{2}+\frac{a_{6}}{2} (y%
\dot{x}-x\dot{y})^{2}+\frac{a_{7}}{2}(z\dot{y}-y\dot{z})^{2} +a_{3}\left( \dot{x}^{2}-2kx^{2}\right) +a_{9}(\dot{z}^{2}-2kz^{2}) + \\
&&+a_{13}(\dot{y}^{2}- 2ky^{2}) + a_{4}(y\dot{x} - x\dot{y}) (z\dot{x} - x\dot{z}) + a_{10}(z\dot{x} - x\dot{z}) (z\dot{y} - y\dot{z})- \\
&& -a_{14} (z\dot{y} - y\dot{z}) (y\dot{x} - x\dot{y}) + 2a_{17}(\dot{x}\dot{y}-2kxy)+ 2a_{19}(\dot{x}\dot{z}-2kxz) +2a_{20}(\dot{y}\dot{z}-2kyz)
\end{eqnarray*}
which consists of the following independent FIs:
\[
M_{1} = y\dot{z} - z\dot{y}, \enskip M_{2}= z\dot{x} - x\dot{z}, \enskip M_{3}= x\dot{y} - y\dot{x}, \enskip B_{ij} = \dot{q}_{i} \dot{q}_{j} - 2k q_{i}q_{j}
\]
where $q_{i}=(x,y,z)$, $M_{i}$ are the components of the angular momentum,\index{Momentum! angular} and $B_{ij}$ are the components of a symmetric tensor. From these nine autonomous FIs, the maximum number of functional independent FIs is $2n-1=5$ since $n=3$ is the dimension of the configuration system.

The total energy of the system is written
\begin{equation*}
H \equiv E = \frac{1}{2} (B_{11} + B_{22} + B_{33}) = \frac{1}{2}(\dot{x}^{2}+\dot{y}^{2} +\dot{z}^{2}) -kr^{2}.
\end{equation*}

For $k=-\frac{1}{2}$, the tensor $B_{ij}= \dot{q}_{i}\dot{q}_{j}+q_{i}q_{j}$ is the
\textbf{Jauch-Hill-Fradkin tensor} \cite{Leach 1988}.\index{Tensor! Jauch-Hill-Fradkin}

The Poisson brackets (PBs) for the components of the angular momentum give the well-known relation
\begin{equation}
\{M_{a}, M_{b}\}= \varepsilon_{abc} M^{c} \label{eq.angm.1}
\end{equation}
where $\varepsilon_{abc}$ is the totally-antisymmetric Levi-Civita symbol. This means that the LFIs $M_{1}, M_{2}, M_{3}$ are not in involution and, hence, they cannot be used for Liouville integrability.

However, the 3d harmonic oscillator is integrable because the triplet $H, M_{a}, B_{aa}$ (the index $a$ is fixed) is functionally independent and in involution, i.e. $\{H,B_{aa}\}=$ $\{H,M_{a}\}=$ $\{B_{aa}, M_{a}\}=0$. The same properties also hold for the triplet $B_{11}, B_{22}, B_{33}$.

We compute: $\{M_{1},B_{22}\}=\{B_{33},M_{1}\}=2B_{23}$, $\{M_{2},B_{33}\}=\{B_{11},M_{3}\}=2B_{13}$ and \newline $\{M_{3},B_{11}\}=\{B_{22},M_{3}\}=2B_{12}$.

Finally, the system of the 3d harmonic oscillator is also superintegrable because it is integrable and the five FIs $H,L_{1},L_{2},L_{3},B_{aa}$ are functionally independent.
\bigskip

b) Case $\ell=1$.

We have $L_{a}=0$, $a_{16}=a_{17}=a_{18}=a_{19}=a_{20}=0$, $a_{2}=a_{12}$, $a_{3}=a_{9}=a_{13}$, $a_{5}=a_{8}$, and $a_{11}=a_{15}$.

Then, the independent components of the KT $C_{ab}$ are
\begin{eqnarray*}
C_{11} &=& \frac{a_{6}}{2}y^{2}+ \frac{a_{1}}{2}%
z^{2}+a_{4}yz+a_{5}y+a_{2}z+a_{3} \\
C_{12} &=&\frac{a_{10}}{2}z^{2}-\frac{a_{6}}{2}xy-\frac{a_{4}}{2}xz-\frac{%
a_{14}}{2}yz-\frac{a_{5}}{2}x-\frac{a_{11}}{2}y \\
C_{13} &=&\frac{a_{14}}{2}y^{2}-\frac{a_{4}}{2}xy-\frac{a_{1}}{2}xz-\frac{%
a_{10}}{2}yz-\frac{a_{2}}{2}x-\frac{a_{11}}{2}z \\
C_{22} &=&\frac{a_{6}}{2}x^{2}+\frac{a_{7}}{2}%
z^{2}+a_{14}xz+a_{11}x+a_{2}z+a_{3} \\
C_{23} &=&\frac{a_{4}}{2}x^{2}-\frac{a_{14}}{2}xy-\frac{a_{10}}{2}xz-\frac{%
a_{7}}{2}yz-\frac{a_{2}}{2}y-\frac{a_{5}}{2}z \\
C_{33} &=&\frac{a_{1}}{2}x^{2}+\frac{a_{7}}{2}%
y^{2}+a_{10}xy+a_{11}x+a_{5}y+a_{3}.
\end{eqnarray*}

Substituting the above quantities in $G_{,a}=2c_{0}C_{ab}V^{,b}$ and integrating wrt each coordinate, we find the function $G(x,y,z)=-\frac{k}{r}(a_{11}x+a_{5}y+a_{2}z+2a_{3})$.

The QFI is
\begin{eqnarray*}
I_{1} &=&\frac{a_{1}}{2}(z\dot{x}-x\dot{z})^{2} +\frac{a_{6}}{2} (y\dot{x}-x\dot{y})^{2} + \frac{a_{7}}{2} (z\dot{y}-y\dot{z})^{2} + a_{4} (y\dot{x}-x\dot{y}) (z\dot{x}-x\dot{z}) + \\
&& +a_{10} (z\dot{x}-x\dot{z}) (z\dot{y}-y\dot{z}) -a_{14} (z\dot{y}-y\dot{z}) (y\dot{x}-x\dot{y}) +2a_{3}\left[
\frac{1}{2}(\dot{x}^{2}+\dot{y}^{2}+\dot{z}^{2}) -\frac{k}{r}\right] + \\
&& +a_{2}\left[ z(\dot{x}^{2}+\dot{y}^{2}) -\dot{z}(x\dot{x}+y\dot{y}) -\frac{k}{r}z\right] +a_{5}\left[ y(\dot{x}^{2}+\dot{z}^{2}) -\dot{y}(x\dot{x}+z\dot{z}) -\frac{k}{r}y\right] + \\
&& +a_{11}\left[ x(\dot{y}^{2}+\dot{z}^{2}) -\dot{x}(y\dot{y}+z\dot{z}) -\frac{k}{r}x\right].
\end{eqnarray*}

The QFI $I_{1}$ contains the following irreducible FIs: \newline
i. The three components of the angular momentum $M_{1}=y\dot{z}-z\dot{y}$, $M_{2}=z\dot{x}-x\dot{z}$ and $M_{3}= x\dot{y}-y\dot{x}$. . \newline
ii. The total energy (Hamiltonian) of the system $E \equiv \frac{1}{2}(\dot{x}^{2}+\dot{y}^{2}+\dot{z}^{2}) -\frac{k}{r}$.
\newline
iii. The three components of the Runge-Lenz vector\index{Vector! Runge-Lenz}
\begin{eqnarray*}
R_{1} &=& x(\dot{y}^{2}+\dot{z}^{2})-\dot{x}(y\dot{y}+z\dot{z})- \frac{k}{r}x \\
R_{2} &=& y(\dot{x}^{2}+\dot{z}^{2})- \dot{y}(x\dot{x}+z\dot{z}) -\frac{k}{r}y \\
R_{3} &=& z(\dot{x}^{2}+\dot{y}^{2})-\dot{z}(x\dot{x}+y\dot{y}) -\frac{k}{r}z
\end{eqnarray*}
which can be written in the compact form
\begin{equation}
R_{i}=(v^{j}v_{j})x_{i}-(x^{j}v_{j})v_{i}-\frac{k}{r}x_{i} \label{eq.Kep.6}
\end{equation}
where $x_{i}=(x,y,z)$ and $v_{i}=\dot{x}_{i}= (\dot{x},\dot{y},\dot{z})$. The linear combination $\mu^{i}R_{i}$, where $\mu^{i}$ are arbitrary constants, is the Noether invariant found in \cite{kalotas}.

Using the vector identity $\mathbf{A}\times \left( \mathbf{B}\times \mathbf{C}\right) =\left( \mathbf{A}\cdot \mathbf{C}\right) \mathbf{B}-\left( \mathbf{A}\cdot \mathbf{B}\right)\mathbf{C}$, equation (\ref{eq.Kep.6}) takes the well-known vector form
\begin{equation}
\mathbf{R}=\mathbf{v}\times (\mathbf{x}\times \mathbf{v})-\frac{k}{r}\mathbf{x}.  \label{eq.Kep.7}
\end{equation}

We should point out that the above seven FIs are not all independent because they are related via the relations $\mathbf{R} \cdot \mathbf{M} =0$ and $\mathbf{R}^{2} = k^{2} + 2E\mathbf{M}^{2}$. From these relations, we deduce that there exist only five independent FIs: the total energy $E$, the three components of the angular momentum $\mathbf{M}$, and the direction of the Runge-Lenz vector $\mathbf{R}$.

The Kepler potential\index{Potential! Kepler} is Liouville integrable because the three FIs $E,M_{a},R_{a}$ are functionally independent and in involution, i.e. $\{M_{a},E\}=0$, $\{R_{a},E\}=0$ and $\{M_{a},R_{a}\}=0$. It is also superintegrable because it has dimension $n=3$ and admits $2n-1=5$ independent FIs.

For the components of the Runge-Lenz vector, we find that $\{R_{a}, M_{b}\} = \varepsilon_{abc} R^{c}$ and $\{R_{a},R_{b}\}$ $= -2\varepsilon_{abc}M^{c}E$.
\bigskip

c) Case $\ell=2$.

We have: $L_{a}= b_{3}
\left(
  \begin{array}{c}
    x \\
    y \\
    z \\
  \end{array}
\right)$, $L_{(a;b)}= b_{3}\delta_{ab}$ and
\[
C_{11}=\frac{a_{6}}{2}y^{2}+\frac{a_{1}}{2}z^{2}+a_{4}yz+a_{3}, \enskip C_{12}= \frac{a_{10}}{2}z^{2}-\frac{a_{6}}{2}xy -\frac{a_{4}}{2}xz-\frac{a_{14}}{2}yz,
\]
\[
C_{13}= \frac{a_{14}}{2}y^{2}-\frac{a_{4}}{2}xy -\frac{a_{1}}{2}xz -\frac{a_{10}}{2}yz, \enskip C_{22}= \frac{a_{6}}{2}x^{2}+\frac{a_{7}}{2}z^{2}+a_{14}xz+a_{3}
\]
\[
C_{23}=\frac{a_{4}}{2}x^{2}-\frac{a_{14}}{2}xy-\frac{a_{10}}{2}xz -\frac{a_{7}}{2}yz, \enskip C_{33}= \frac{a_{1}}{2}x^{2} +\frac{a_{7}}{2}y^{2}+a_{10}xy+a_{3}.
\]

Substituting in $G_{,a}=2c_{0}C_{ab}V^{,b}$ and integrating wrt each coordinate, we find the function $G(x,y,z)= -\frac{2ka_{3}}{r^2} - \frac{b_{3}}{2}r^{2}$.

The QFI is
\begin{eqnarray*}
I_{1} &=& - b_{3} t^{2} \left[ \frac{1}{2}(\dot{x}^{2} + \dot{y}^{2} + \dot{z}^{2}) - \frac{k}{r^{2}}  \right] + \frac{a_{1}}{2}(z\dot{x}-x\dot{z})^{2}+ \frac{a_{6}}{2}(y\dot{x}-x\dot{y})^{2} +\frac{a_{7}}{2}(z\dot{y} -y\dot{z})^{2} + \\
&& + 2a_{3}\left[ \frac{1}{2}(\dot{x}^{2}+ \dot{y}^{2}+\dot{z}^{2})- \frac{k}{r^{2}}\right] + a_{4} (y\dot{x}-x\dot{y}) (z\dot{x}-x\dot{z}) + a_{10} (z\dot{x}-x\dot{z}) (z\dot{y}-y\dot{z}) - \\
&& - a_{14} (z\dot{y}-y\dot{z}) (y\dot{x}-x\dot{y}) + b_{3} t (x\dot{x} + y\dot{y} + z\dot{z}) - \frac{b_{3}}{2} r^{2}.
\end{eqnarray*}

This QFI contains the three components of the angular momentum, the total energy $E =\frac{1}{2}(\dot{x}^{2} + \dot{y}^{2} + \dot{z}^{2}) - \frac{k}{r^{2}}$ of the resulting system, and also the time-dependent FI
\[
I_{1a}(\ell=2)= - E t^{2} + t(x\dot{x} + y\dot{y} + z\dot{z}) - \frac{r^{2}}{2}.
\]
\bigskip

d) Case $\ell \neq -2,1,2$.

In this case, the KT $C_{ab}$ is that of the case $\ell=2$, the vector $L_{a}=0$, and $G(x,y,z)=-\frac{2ka_{3}}{r^{\ell}}$.

The QFI is
\begin{eqnarray*}
I_{1} &=& \frac{a_{1}}{2}(z\dot{x}-x\dot{z})^{2}+ \frac{a_{6}}{2}(y\dot{x}-x\dot{y})^{2} +\frac{a_{7}}{2}(z\dot{y} -y\dot{z})^{2} + 2a_{3}\left[ \frac{1}{2}(\dot{x}^{2}+ \dot{y}^{2}+\dot{z}^{2})- \frac{k}{r^{\ell}}\right]+ \\
&& + a_{4} (y\dot{x}-x\dot{y}) (z\dot{x}-x\dot{z}) + a_{10} (z\dot{x}-x\dot{z}) (z\dot{y}-y\dot{z}) - a_{14} (z\dot{y}-y\dot{z}) (y\dot{x}-x\dot{y})
\end{eqnarray*}
which consists of the three LFIs of the angular momentum and the total energy $E = \frac{1}{2}(\dot{x}^{2}+ \dot{y}^{2}+\dot{z}^{2})- \frac{k}{r^{\ell}}$ of the system.
\bigskip

\textbf{Integral 2.}
\begin{equation*}
I_{2} = -\frac{t^{3}}{3} L_{(a;b)}\dot{q}^{a}\dot{q}^{b} + t^{2} L_{a} \dot{q%
}^{a} + \frac{t^{3}}{3} L_{a}V^{,a} - t B_{(a;b)} \dot{q}^{a}\dot{q}^{b} +
B_{a}\dot{q}^{a} + tB_{a}V^{,a}
\end{equation*}
where $L_{a}$ and $B_{a}$ are such that $L_{(a;b)}$ and $B_{(a;b)}$ are KTs, $\left(L_{b}V^{,b}\right)_{,a} = -2L_{(a;b)} V^{,b}$, and $\left(B_{b}V^{,b}\right)_{,a} = -2B_{(a;b)} V^{,b} - 2L_{a}$.
\bigskip

Since $L_{(a;b)}$ and $B_{(a;b)}$ are KTs, we have the following (see sec. \ref{sec.KTE3}):
\begin{equation*}
L_{a}=\left(
\begin{array}{c}
-a_{15}y^{2}-a_{11}z^{2}+a_{5}xy+a_{2}xz+2(a_{16}+a_{18})yz +a_{3}x +2a_{4}y+2a_{1}z+a_{6}
\\
-a_{5}x^{2}-a_{8}z^{2}+a_{15}xy-2a_{18}xz+a_{12}yz+ 2(a_{17}-a_{4})x+a_{13}y+2a_{7}z+a_{14}
\\
-a_{2}x^{2}-a_{12}y^{2}-2a_{16}xy+a_{11}xz+a_{8}yz+2(a_{19}- a_{1})x+2(a_{20}-a_{7})y+a_{9}z+a_{10}%
\end{array}%
\right)
\end{equation*}
\begin{equation*}
B_{a}=\left(
\begin{array}{c}
-b_{15}y^{2}-b_{11}z^{2}+b_{5}xy+b_{2}xz +2(b_{16}+b_{18})yz+b_{3}x+2b_{4}y+2b_{1}z+b_{6}
\\
-b_{5}x^{2}-b_{8}z^{2}+b_{15}xy-2b_{18}xz+b_{12}yz+ 2(b_{17}-b_{4})x+b_{13}y+2b_{7}z+b_{14}
\\
-b_{2}x^{2}-b_{12}y^{2}-2b_{16}xy+b_{11}xz+b_{8}yz+2(b_{19}- b_{1})x+2(b_{20}-b_{7})y+b_{9}z+b_{10}%
\end{array}%
\right)
\end{equation*}
\[
L_{(1;1)}= a_{5}y+a_{2}z+a_{3}, \enskip L_{(1;2)}= -\frac{a_{5}}{2}x -\frac{a_{15}}{2}y +a_{16}z+a_{17}, \enskip L_{(1;3)}= -\frac{a_{2}}{2}x+a_{18}y-\frac{a_{11}}{2}z+a_{19},
\]
\[
L_{(2;2)}= a_{15}x+a_{12}z+a_{13}, \enskip L_{(2;3)}= -(a_{16}+a_{18})x -\frac{a_{12}}{2}y -\frac{a_{8}}{2}z+a_{20} , \enskip L_{(3;3)}= a_{11}x+a_{8}y+a_{9}
\]
and
\[
B_{(1;1)}= b_{5}y+b_{2}z+b_{3}, \enskip B_{(1;2)}= -\frac{b_{5}}{2}x -\frac{b_{15}}{2}y+b_{16}z+b_{17}, \enskip B_{(1;3)}=-\frac{b_{2}}{2}x+b_{18}y-\frac{b_{11}}{2}z+b_{19},
\]
\[
B_{(2;2)}= b_{15}x+b_{12}z+b_{13}, \enskip B_{(2;3)}= -(b_{16}+b_{18})x -\frac{b_{12}}{2}y-\frac{b_{8}}{2}z+b_{20} , \enskip B_{(3;3)}=b_{11}x+b_{8}y+b_{9}.
\]

From the constraint $\left(L_{b}V^{,b}\right)_{,a} = -2L_{(a;b)} V^{,b}$, we find: $(\ell-2)a_{3}=0$, $L_{a}=$ $a_{3}
\left(
  \begin{array}{c}
    x \\
    y \\
    z \\
  \end{array}
\right)$, and $L_{(a;b)}= a_{3}\delta_{ab}$.

Replacing in the remaining constraint $\left(B_{b}V^{,b} \right)_{,a} = -2B_{(a;b)} V^{,b} - 2L_{a}$, we obtain $a_{3}=0 \implies$ $L_{a}=0$,  $(\ell-2)b_{3}=0$, $B_{a}= b_{3}
\left(
  \begin{array}{c}
    x \\
    y \\
    z \\
  \end{array}
\right)$, and $B_{(a;b)}= b_{3}\delta_{ab}$.

For $\ell\neq 2$, we have $B_{a}=0$ and the QFI $I_{2}=0$. On the other hand, for $\ell=2$, we get the non-trivial time-dependent LFI
\[
I_{2}(\ell=2) = -Et + \frac{1}{2} (x\dot{x} + y\dot{y} + z\dot{z}).
\]
\bigskip

\textbf{Integral 3.}
\begin{equation*}
I_{3} = -e^{\lambda t} L_{(a;b)}\dot{q}^{a}\dot{q}^{b} + \lambda e^{\lambda
t} L_{a} \dot{q}^{a} + e^{\lambda t} L_{a} V^{,a}
\end{equation*}
where $\lambda \neq 0$, $L_{a}$ is such that $L_{(a;b)}$ is a reducible KT, and $\left(L_{b}V^{,b}\right)_{,a} = -2L_{(a;b)} V^{,b} - \lambda^{2} L_{a}$.
\bigskip

Since $L_{(a;b)}$ is a KT, we have the following:
\begin{equation*}
L_{a}=\left(
\begin{array}{c}
-a_{15}y^{2}-a_{11}z^{2}+a_{5}xy+a_{2}xz+2(a_{16}+a_{18})yz +a_{3}x +2a_{4}y+2a_{1}z+a_{6}
\\
-a_{5}x^{2}-a_{8}z^{2}+a_{15}xy-2a_{18}xz+a_{12}yz+ 2(a_{17}-a_{4})x+a_{13}y+2a_{7}z+a_{14}
\\
-a_{2}x^{2}-a_{12}y^{2}-2a_{16}xy+a_{11}xz+a_{8}yz+2(a_{19}- a_{1})x+2(a_{20}-a_{7})y+a_{9}z+a_{10}%
\end{array}%
\right)
\end{equation*}
and
\[
L_{(1;1)}= a_{5}y+a_{2}z+a_{3}, \enskip L_{(1;2)}= -\frac{a_{5}}{2}x -\frac{a_{15}}{2}y+a_{16}z+a_{17}, \enskip L_{(1;3)}=-\frac{a_{2}}{2}x+a_{18}y-\frac{a_{11}}{2}z+a_{19},
\]
\[
L_{(2;2)}= a_{15}x+a_{12}z+a_{13}, \enskip L_{(2;3)}= -(a_{16}+a_{18})x-\frac{a_{12}}{2}y-\frac{a_{8}}{2}z+a_{20} , \enskip L_{(3;3)}=a_{11}x+a_{8}y+a_{9}.
\]

Substituting in $\left(L_{b}V^{,b}\right)_{,a} = -2L_{(a;b)} V^{,b} - \lambda^{2} L_{a}$, we find that the vector $L_{a}$ survives only for $\ell=-2$. Therefore, in what it follows, we consider only that case.

We compute
\begin{equation*}
\left( L_{b}V^{,b}\right) _{,a}=-2k\left(
\begin{array}{c}
2a_{3}x+2a_{17}y+2a_{19}z+a_{6} \\
2a_{13}y+2a_{17}x+2a_{20}z+a_{14} \\
2a_{9}z+2a_{19}x+2a_{20}y+a_{10}%
\end{array}%
\right)
\end{equation*}
and
\begin{equation*}
L_{(a;b)}V^{,b}=-k\left(
\begin{array}{c}
-a_{15}y^{2}-a_{11}z^{2}+a_{5}xy+a_{2}xz+2(a_{16}+a_{18})yz +2a_{3}x+2a_{17}y+2a_{19}z
\\
-a_{5}x^{2}-a_{8}z^{2}+a_{15}xy-2a_{18}xz+a_{12}yz+2a_{17}x +2a_{13}y+2a_{20}z
\\
-a_{2}x^{2}-a_{12}y^{2}-2a_{16}xy+a_{11}xz+a_{8}yz+2a_{19}x +2a_{20}y+2a_{9}z%
\end{array}%
\right).
\end{equation*}
Therefore, the condition $\lambda ^{2}L_{a}+\left( L_{b}V^{,b}\right)_{,a} +2L_{(a;b)}V^{,b}=0$ gives the following set of algebraic equations:
\begin{eqnarray*}
0 &=&-a_{15}(\lambda ^{2}-2k)y^{2}-a_{11}(\lambda
^{2}-2k)z^{2}+a_{5}(\lambda ^{2}-2k)xy+a_{2}(\lambda
^{2}-2k)xz+ \\
&& +2(a_{16}+a_{18})(\lambda ^{2}-2k)yz +a_{3}(\lambda ^{2}-8k)x+2(\lambda ^{2}a_{4}-4ka_{17})y + \\
&& +2(\lambda^{2}a_{1}-4ka_{19})z+(\lambda ^{2}-2k)a_{6} \\
0 &=& - a_5 (\lambda^2 - 2k)x^2 - a_8 (\lambda^2 - 2k)z^2 + a_{15}
(\lambda^2 - 2k) xy - 2a_{18} (\lambda^2 - 2k) xz + \\
&& + 2a_{12} (\lambda^2 - 2k)
yz +2 \left[ a_{17} (\lambda^2 -4k) - \lambda^2a_4\right] x +
a_{13}(\lambda^2 - 8k)y + \\
&& +2(\lambda^2 a_7 - 4ka_{20}) z +(\lambda^2-2k)a_{14} \\
0 &=& - a_2 (\lambda^2 - 2k)x^2 - a_{12} (\lambda^2 - 2k)y^2 - 2a_{16}
(\lambda^2 - 2k) xy + a_{11} (\lambda^2 - 2k) xz + \\
&& + a_8 (\lambda^2 - 2k) yz + 2 \left[ a_{19} (\lambda^2 -4k) - \lambda^2a_1\right] x + 2 \left[
a_{20} (\lambda^2 -4k) - \lambda^2a_7\right] y + \\
&& +a_9(\lambda^2 - 8k)z +(\lambda^2-2k)a_{10}.
\end{eqnarray*}

We consider the following cases:
\bigskip

a) For $\lambda ^{2}=2k$, we have $a_{1}=a_{3}=a_{4}=a_{7} =a_{9}=a_{13}=a_{17}=a_{19}=a_{20}=0$.

Then,
\begin{equation*}
L_{a}=\left(
\begin{array}{c}
-a_{15}y^{2}-a_{11}z^{2}+a_{5}xy+a_{2}xz+2(a_{16}+a_{18})yz+a_{6} \\
-a_{5}x^{2}-a_{8}z^{2}+a_{15}xy-2a_{18}xz+a_{12}yz+a_{14} \\
-a_{2}x^{2}-a_{12}y^{2}-2a_{16}xy+a_{11}xz+a_{8}yz+a_{10}%
\end{array}%
\right)
\end{equation*}%
and
\begin{equation*}
L_{(a;b)}=\left(
\begin{array}{ccc}
a_{5}y+a_{2}z & -\frac{a_{5}}{2}x-\frac{a_{15}}{2}y+a_{16}z & -\frac{a_{2}}{2%
}x+a_{18}y-\frac{a_{11}}{2}z \\
-\frac{a_{5}}{2}x-\frac{a_{15}}{2}y+a_{16}z & a_{15}x+a_{12}z &
-(a_{16}+a_{18})x-\frac{a_{12}}{2}y-\frac{a_{8}}{2}z \\
-\frac{a_{2}}{2}x+a_{18}y-\frac{a_{11}}{2}z & -(a_{16}+a_{18})x-\frac{a_{12}%
}{2}y-\frac{a_{8}}{2}z & a_{11}x+a_{8}y%
\end{array}%
\right).
\end{equation*}

The QFI is
\begin{eqnarray*}
I_{3a}(\ell=-2) &=&\left( a_{6} -\frac{a_{2}}{\lambda} M_{2} +\frac{a_{5}}{\lambda} M_{3} \right) e^{\lambda t} (\dot{x}-\lambda x) + \\
&& +\left( a_{14}+ \frac{a_{12}}{\lambda} M_{1} -\frac{2a_{16}}{\lambda} M_{2} -\frac{a_{15}}{\lambda} M_{3} \right) e^{\lambda t} (\dot{y}-\lambda y)+ \\
&& +\left( a_{10} -\frac{a_{8}}{\lambda} M_{1} +\frac{a_{11}}{\lambda} M_{2} +\frac{2a_{18}}{\lambda} M_{3} \right) e^{\lambda t} (\dot{z}-\lambda z).
\end{eqnarray*}
From this expression, we have the following irreducible time-dependent LFIs:
\[
I_{3a1}= e^{\lambda t}(\dot{x}-\lambda x), \enskip I_{3a2}=e^{\lambda t}(\dot{y}-\lambda y), \enskip I_{3a3}= e^{\lambda t}(\dot{z}-\lambda z).
\]

If $k>0$, then $\lambda= \pm \sqrt{2k}$; and if $k<0$, then $\lambda= \pm i \sqrt{-2k}$. Therefore, for all values of the non-zero parameter $k$ of the system, there exist two constants $\lambda_{\pm}$ each generating three independent LFIs of the system. We have\footnote{The calculations are the same for either $k>0$ or $k<0$. We continue for $k<0$, which is the case of the 3d harmonic oscillator.}:
\[
I_{3a1\pm}= e^{\pm i \sqrt{-2k} t}(\dot{x} \mp i \sqrt{-2k} x), \enskip I_{3a2\pm}=e^{\pm i \sqrt{-2k} t}(\dot{y} \mp i \sqrt{-2k} y), \enskip I_{3a3\pm}= e^{\pm i \sqrt{-2k} t}(\dot{z} \mp i \sqrt{-2k} z).
\]
Using the above six LFIs, we can derive all the FIs found in the case \textbf{Integral 1} for $\ell=-2$. Indeed, we have
\[
I_{3a1+}I_{3a1-}=B_{11}, \enskip I_{3a2+}I_{3a2-}=B_{22}, \enskip I_{3a3+}I_{3a3-}=B_{33},
\]
\[
I_{3a1\pm}I_{3a2\mp}=B_{12} \mp i\sqrt{-2k}M_{3}, \enskip I_{3a1\pm}I_{3a3\mp}=B_{13} \pm i\sqrt{-2k}M_{2}, \enskip I_{3a2\pm}I_{3a3\mp}=B_{23} \mp i\sqrt{-2k}M_{1}.
\]
Therefore, \emph{all the components of the Jauch-Hill-Fradkin tensor $B_{ij}$ can be constructed by the three components of the angular momentum and the six time-dependent LFIs $I_{3a1\pm}$, $I_{3a2\pm}$ and $I_{3a3\pm}$}.
\bigskip

b) For $\lambda^2=4k$, we get $L_a=0$. Therefore, the QFI vanishes. \bigskip

c) For $\lambda^2 = 8k$, we have $a_2=a_5=a_6=a_8=a_{10} =a_{11}=a_{12}=a_{14}=a_{15}=a_{16}= a_{18} = 0$, $a_{17}=2a_4$, $a_{19}=2a_1$ and $a_{20}= 2a_7$.

Then,  $L_a = \left(
\begin{array}{c}
a_3x + a_{17}y + a_{19}z \\
a_{17}x + a_{13}y + a_{20}z \\
a_{19}x + a_{20}y + a_9z%
\end{array}
\right)$ and $L_{(a;b)} = \left(
\begin{array}{ccc}
a_3 & a_{17} & a_{19} \\
a_{17} & a_{13} & a_{20} \\
a_{19} & a_{20} & a_9%
\end{array}
\right)$.

The QFI is
\begin{align*}
I_{3c}(\ell=-2) &=-\frac{a_{3}}{\lambda }e^{\lambda t}\left( \dot{x}-\frac{\lambda }{2}x\right)^{2}-\frac{a_{9}}{\lambda }e^{\lambda t}\left(
\dot{z}-\frac{\lambda }{2}z\right) ^{2}-\frac{a_{13}}{\lambda }e^{\lambda
t}\left( \dot{y}-\frac{\lambda }{2}y\right) ^{2}- \\
&-\frac{a_{17}}{\lambda }e^{\lambda t}\left[ 2\dot{x}\dot{y}+\frac{\lambda
^{2}}{2}xy-\lambda (y\dot{x}+x\dot{y})\right] -\frac{a_{19}}{\lambda }%
e^{\lambda t}\left[ 2\dot{x}\dot{z}+\frac{\lambda ^{2}}{2}xz-\lambda (z\dot{x%
}+x\dot{z})\right] - \\
&-\frac{a_{20}}{\lambda }e^{\lambda t}\left[ 2\dot{y}\dot{z}+\frac{\lambda
^{2}}{2}yz-\lambda (y\dot{z}+z\dot{y})\right].
\end{align*}
This expression consists of the time-dependent FIs: $I_{3b1}= e^{\lambda t}\left( \dot{x}-\frac{\lambda }{2}x\right)^{2}$, $I_{3b2}= e^{\lambda t}\left( \dot{y}-\frac{\lambda }{2}y\right)^{2}$, $I_{3b3}= e^{\lambda t}\left( \dot{z}-\frac{\lambda }{2}z\right)^{2}$, $I_{3b4}= e^{\lambda t}\left[ \dot{x}\dot{y}+ \frac{\lambda^{2}}{4} xy- \frac{\lambda}{2} (y\dot{x}+x\dot{y})\right], \enskip I_{3b5} = e^{\lambda t}\left[ \dot{x}\dot{z}+ \frac{\lambda^{2}}{4} xz - \frac{\lambda}{2} (z\dot{x}+x\dot{z})\right]$ and \newline $I_{3b6}= e^{\lambda t}\left[ \dot{y}\dot{z}+ \frac{\lambda^{2}}{4} yz - \frac{\lambda}{2} (y\dot{z}+z\dot{y})\right]$.

If $k>0$, $\lambda= \pm 2\sqrt{2k}$; and if $k<0$, $\lambda =\pm 2i\sqrt{-2k}$. Similarly to the calculations of the case a), we find that (we continue for $k<0$ and adopt the notation of the case a) for the FIs)
\[
I_{3b1\pm}=(I_{3a1\pm})^{2}, \enskip I_{3b2\pm}=(I_{3a2\pm})^{2}, \enskip I_{3b3\pm}=(I_{3a3\pm})^{2}, \enskip I_{3b4\pm}= I_{3a1\pm} I_{3a2\pm},
\]
\[
I_{3b5\pm} = I_{3a1\pm} I_{3a3\pm}, \enskip I_{3b6\pm}= I_{3a2\pm}I_{3a3\pm}.
\]
Therefore, this case gives again the six time-dependent LFIs $I_{3a1\pm}$, $I_{3a2\pm}$ and $I_{3a3\pm}$ of the case a).
\bigskip

We collect the results of this section in Table \ref{Table.QFIs.Kepler.gen}, where $q^{i}=(x,y,z)$.

\begin{longtable}{|l|l|}
\hline
$V=-\frac{k}{r^{\ell}}$ & LFIs and QFIs \\ \hline
$\forall$ $\ell$ & $M_{1} = y\dot{z} - z\dot{y}$, $M_{2}= z\dot{x} - x\dot{z}$, $M_{3}= x\dot{y} - y\dot{x}$, $H= \frac{1}{2}(\dot{x}^{2} + \dot{y}^{2} + \dot{z}^{2}) - \frac{k}{r^{\ell}}$ \\
$\ell=-2$ & $B_{ij} = \dot{q}_{i} \dot{q}_{j} - 2k q_{i}q_{j}$ \\
$\ell=-2$, $k>0$ & $I_{3a\pm}= e^{\pm \sqrt{2k} t}(\dot{q}_{a} \mp \sqrt{2k} q_{a})$ \\
$\ell=-2$, $k<0$ & $I_{3a\pm}= e^{\pm i \sqrt{-2k} t}(\dot{q}_{a} \mp i \sqrt{-2k} q_{a})$ \\
$\ell=1$ & $R_{i}= (\dot{q}^{j} \dot{q}_{j}) q_{i} - (\dot{q}^{j}q_{j})\dot{q}_{i}- \frac{k}{r}q_{i}$ \\
$\ell=2$ & $I_{1}= - Ht^{2} + t(\dot{q}^{i}q_{i}) - \frac{r^{2}}{2}$, $I_{2}= - Ht + \frac{1}{2} (\dot{q}^{i}q_{i})$
 \\ \hline
\caption{\label{Table.QFIs.Kepler.gen} The LFIs/QFIs of the general Kepler problem.}
\end{longtable}

\section{The time-dependent FIs}

\label{subsec.Kep.3}

As it has been shown, Theorem \ref{The first integrals of an autonomous holonomic dynamical system} produces all LFIs/QFIs of the autonomous conservative dynamical equations, i.e. the autonomous and the time-dependent FIs, the latter being equally important as the former. Furthermore, this is achieved in a way that is independent of the dimension, the signature, and the curvature of the kinetic metric defined by the kinetic energy/Lagrangian of the specific dynamical system. On the contrary, the standard methods determine mainly the autonomous FIs, usually for low degrees of freedom, and consider principally the `usual' dynamical systems.

The time-dependent FIs can be used to test the integrability of a dynamical system and, of course, they can be used to obtain the solution of the dynamical equations in terms of quadratures. As we have seen in sec. \ref{sec.int.FI.4}, the Liouville integrability (see p. 271, sec. 49 in \cite{Arnold 1989}) requires $n$ functionally independent FIs in involution of the form $I(q,p)$. However, it has been pointed out\footnote{See Theorem 1, p.17, chap. II, para. 2 in \cite{Kozlov 1983}, and Theorem 3.4 in \cite{Vozmishcheva 2005}.} that we can also use time-dependent FIs of the form $I(q,p,t)$ for the same purpose. It is to be noted that both Theorems in \cite{Kozlov 1983} and \cite{Vozmishcheva 2005} refer to non-autonomous Hamiltonians $H(q,p,t)$. Moreover, the usefulness of the time-dependent FIs can be seen from the examples I and II of sec. VII in \cite{KatzinLev1985}.

In order to show the use of the time-dependent FIs in the solution of the dynamical equations, we consider two cases of the general Kepler equations (\ref{eq.GKep.1a}) considered in  sec. \ref{sec.GKepler}.
\bigskip

\begin{example} \label{exa.td.1}
In the case of the potential $V=-kr^{2}$ ($\ell=-2$, $k>0$), we found the six time-dependent LFIs $I_{3a\pm}= e^{\pm \sqrt{2k} t}(\dot{q}_{a} \mp \sqrt{2k} q_{a})$. We use these LFIs to obtain the solution of the corresponding dynamical equations. We have the following:
\[
\begin{cases}
e^{\sqrt{2k} t}(\dot{x} - \sqrt{2k}x) = A_{+} \\
e^{-\sqrt{2k} t}(\dot{x} + \sqrt{2k}x) = A_{-}
\end{cases}
\implies
\begin{cases}
\dot{x} - \sqrt{2k}x = A_{+}e^{-\sqrt{2k} t} \\
\dot{x} + \sqrt{2k}x = A_{-}e^{\sqrt{2k} t}
\end{cases} \implies
\]
\[
\dot{x} = \frac{1}{2} \left( A_{+}e^{-\sqrt{2k} t} + A_{-} e^{\sqrt{2k} t}\right)
\implies
x(t)= \frac{1}{2} \left( -\frac{A_{+}}{\sqrt{2k}} e^{-\sqrt{2k} t} + \frac{A_{-}}{\sqrt{2k}} e^{\sqrt{2k} t}\right)
\]
where $A_{\pm}$ are arbitrary constants. Similarly to the other LFIs, we find:
\[
y(t)= \frac{1}{2} \left( -\frac{B_{+}}{\sqrt{2k}} e^{-\sqrt{2k} t} + \frac{B_{-}}{\sqrt{2k}} e^{\sqrt{2k} t}\right), \enskip z(t)= \frac{1}{2} \left( -\frac{C_{+}}{\sqrt{2k}} e^{-\sqrt{2k} t} + \frac{C_{-}}{\sqrt{2k}} e^{\sqrt{2k} t}\right)
\]
where $B_{\pm}$ and $C_{\pm}$ are arbitrary constants.
\end{example}

\begin{example} \label{exa.td.2}
For the case of the 3d harmonic oscillator (i.e. $\ell=-2$, $k<0$), using the time-dependent LFIs $I_{3a\pm}= e^{\pm i \sqrt{-2k} t}(\dot{q}_{a} \mp i \sqrt{-2k} q_{a})$, we find (by working as in the Example \ref{exa.td.1}):
\[
x(t)= \frac{1}{2} \left( \frac{iD_{+}}{\sqrt{-2k}} e^{-i \sqrt{-2k}t} - \frac{iD_{-}}{\sqrt{-2k}} e^{i \sqrt{-2k}t} \right), \enskip y(t)= \frac{1}{2} \left( \frac{iE_{+}}{\sqrt{-2k}} e^{-i \sqrt{-2k}t} - \frac{iE_{-}}{\sqrt{-2k}} e^{i \sqrt{-2k}t} \right),
\]
\[
z(t)= \frac{1}{2} \left( \frac{iF_{+}}{\sqrt{-2k}} e^{-i \sqrt{-2k}t} - \frac{iF_{-}}{\sqrt{-2k}} e^{i \sqrt{-2k}t} \right)
\]
where $D_{\pm}$, $E_{\pm}$ and $F_{\pm}$ are arbitrary constants.
\end{example}

\section{Conclusions}

\label{sec.qfi.concl}

As we have seen in previous sections, the FIs play a crucial role in the solution of the dynamical equations. Therefore, it is important that one has a systematic method to compute them for a given dynamical system. In Theorem  \ref{The first integrals of an autonomous holonomic dynamical system}, we
have developed such a method for the case of autonomous conservative dynamical systems. It has been shown that these FIs are closely related to the KTs and the symmetries of the kinetic metric, which is defined by the kinetic energy or
the Lagrangian of the particular dynamical system.

Finally, from Theorem \ref{The first integrals of an autonomous holonomic dynamical system}, it follows that the determination of a LFI/QFI of an autonomous conservative dynamical system consists of two parts. One part, which is entirely characteristic of the kinetic metric and it is common to
all dynamical systems sharing the same metric; and a second part, which consists of constraints which involve in addition the potential defining the specific dynamical system.

%% file: 2d_integrable_potentials.tex
\chapter{Second order integrable and superintegrable potentials of 2d autonomous conservative dynamical systems}

\label{ch.2d.pots}

In this chapter, we determine the integrable and superintegrable potentials $V(x,y)$ of Newtonian autonomous conservative dynamical systems in the Euclidean plane $E^{2}$ by applying Theorem \ref{The first integrals of an autonomous holonomic dynamical system}. These potentials are widely known as \textbf{second order integrable potentials} because only FIs up to second order (i.e. LFIs and QFIs) are used for establishing their integrability.\index{Potential! second order integrable}

\section{Introduction}

\label{sec.2d.intro}

The determination of integrable and superintegrable systems is a topic that is in continuous investigation. Obviously, a universal method, which computes the FIs for all types of dynamical equations independently of their complexity and degrees of freedom, is not available. For this reason, the
existing studies restrict their considerations to flat spaces or spaces of constant curvature of low dimension (see e.g. \cite{Thompson 1984, Ranada 1997, Whittaker, TsaPal 2011, Darboux, Dorizzi Grammatikos Ramani 1983, Sen, Kalnins 2001} and references therein). The prevailing cases involve the autonomous conservative dynamical systems with two degrees of freedom and the classification of the potential functions in integrable and superintegrable.

The problem of finding integrable and superintegrable potentials in $E^{2}$ is not new. It was raised for the first time by Darboux \cite{Darboux} and Whittaker (see ch. XII of \cite{Whittaker}) who considered the Newtonian autonomous holonomic systems with two degrees of freedom and determined most potentials $V(x,y)$ for which the system has an autonomous QFI other than the Hamiltonian (energy). Additional potentials found much later by G. Thompson \cite{Thompson 1984, Thompson 1984 II}. Furthermore, a comprehensive review of the known integrable and superintegrable 2d autonomous potentials is given in \cite{Hietarinta 1987}.

In most of the studies mentioned above, Noether's theorem and the direct method (see sec. \ref{sec.methods.determine.FIs}) were used. However, other approaches have also appeared. For example, Koenigs in \cite{Koenigs} used coordinate transformations in order to
solve the system of equations resulting from the condition $\{H,I\}=0$. The
solution of that system of equations gives the general functional form of
the QFIs and the superintegrable free Hamiltonians, that is, the ones which
possess two more QFIs --in addition to the Hamiltonian-- which are
functionally independent. Koenigs's method has been generalized in several
works (see e.g. \cite{Daskalogiannis 2006} and references cited therein) for 2d autonomous conservative systems.

In the following sections, Theorem \ref{The first integrals of an autonomous holonomic dynamical system}, which is a `product' of the direct method, is applied to the case of 2d autonomous conservative dynamical systems in order to determine the integrable and superintegrable potentials that admit LFIs/QFIs. It is found that the integrable potentials are classified in \textbf{Class I} and \textbf{Class II}, and that superintegrable potentials exist in both classes. All potentials together with their QFIs are listed in tables for easy reference. Moreover, all the results listed in the review paper of \cite{Hietarinta 1987} as well as in more recent works (see e.g. \cite{Ranada 1997, Kalnins 2001}) are recovered, while some new ones are found which admit time-dependent LFIs/QFIs.

\section{The determination of the QFIs}

\label{The determination of the QFIs}

From theorem \ref{The first integrals of an autonomous holonomic dynamical
system}, it follows that for the determination of the QFIs the following
problems have to be solved: \newline
a. Determine the KTs of order two of the kinetic metric $\gamma_{ab}$. \newline
b. Determine the special subspace of KTs of order two of the form $C_{ab}=L_{(a;b)}$ where $L^{a}$ is a vector. \newline
c. Determine the KTs satisfying the constraint $G_{,a}= 2C_{ab} V^{,b}$. \newline
d. Find all KVs $L_{a}$ of the kinetic metric which satisfy the constraint $L_{a}V^{,a}=s$ where $s$ is a constant, possibly zero.

We note that constraints a. and b. depend only on the kinetic metric.
Because the kinetic energy is a positive definite non-singular quadratic
2-form, we can always choose coordinates in which this form reduces either to
$\delta _{ab}$ or to $A(q)\delta _{ab}.$ Since we know the KTs and all the
collineations of a conformally flat metric (of Euclidean or Lorentzian
character) \cite{Barnes 2003}, we already have the results for all
autonomous (Newtonian or special relativistic) conservative dynamical systems.

The involvement of the potential function is only in the constraints c.
and d. which also depend on the geometric characteristics of the kinetic
metric. There are two different ways to proceed.

\subsection{The potential $V\left(q\right)$ is known}

\label{subsec.pot.given}

In this case, the following procedure is used:\newline
a) Substitute $V$ in the constraints $L_{a}V^{,a}=s$ and $%
G_{,a}=2C_{ab}V^{,b}$ and find conditions for the defining parameters of the vector $L_{a}$ and the KT $C_{ab}$. \newline
b) From these conditions determine $L_{a}$ and $C_{ab}$. \newline
c) Substitute $C_{ab}$ in the constraint $G_{,a}=2C_{ab}V^{,b}$ and find the function $G(q)$. \newline
d) Using the above results, write the LFI/QFI $I$ in each case and determine directly from Theorem  \ref{Inverse Noether Theorem} the gauged generalized Noether symmetry. \newline
e) Examine if $I$ can be reduced to simpler independent FIs or if it is new. \newline

\subsection{The potential $V\left(q\right)$ is unknown}

\label{subsec.pot.not.given}

In this case, the following algorithm is used:\newline
a) Compute the KTs and the KVs of the kinetic metric. \newline
b) Solve the PDE $L_{a}V^{,a}=s$ or the\footnote{
The integrability conditions for the scalar $G$ are very general PDEs
from which one can find only special solutions by making additional
simplifying assumptions (e.g. symmetries) involving $L_{a}$, $C_{ab}$ and $V(q)$ itself. Therefore, one does not find the most general solution. For example, in \cite{Markakis 2014}, it is required that the QFI $I$ is
axisymmetric, that is, $\phi^{[1]}I=0$, where $\phi
^{i[1]}=-y\partial_{x}+x\partial_{y} -\dot{y}\partial_{\dot{x}} +\dot{x}\partial_{\dot{y}}$ is the first prolongation of the rotation $\phi^{i}=-y\partial x+x\partial y$. It is proved easily that in this case we have also the constraints $L_{\phi}K_{a}=0$ and $L_{\phi}K_{ab}=0$.} integrability conditions $G_{,[ab]}=0$ and find the possible potentials $V(q)$. \newline
c) Substitute the potentials and the KTs found in the constraint $G_{,a}=2C_{ab}V^{,b}$ and compute (if it exists) the function $G(q)$. \newline
d) Write the LFI/QFI $I$ for each potential and determine the gauged generalized Noether symmetry. \newline
e) Examine if $I$ can be reduced further to simpler independent FIs or if it is a new FI. \newline

In the following sections, we assume the potential is not given and apply the second procedure. For that we need the geometric quantities of the 2d Euclidean plane $E^2$, which have been already determined in sec. \ref{sec.KTE2}. According to Theorem \ref{The first integrals of an autonomous holonomic
dynamical system}, these quantities in $E^{2}$ are common to all 2d Newtonian systems and what changes in each particular case are the constraints $G_{,a}=2C_{ab}V^{,b}$
and $L_{a}V^{,a}=s$, which determine the potential $V(q)$.

\section{Computing the potentials and the FIs}

\label{sec.find.Pots}

The application of Theorem \ref{The first integrals of an autonomous
holonomic dynamical system} in the case of $E^{2}$ indicates that there are
three different ways to find potentials that admit QFIs (other than the
Hamiltonian):
\bigskip

1) The constraint $L_{a}V^{,a}=s$, which leads to the PDE
\begin{equation}
(b_{1}+b_{3}y)V_{,x}+(b_{2}-b_{3}x)V_{,y}-s=0.  \label{eq.PDE1}
\end{equation}

2) The constraint $G_{,a}=2C_{ab}V^{,b}$, which leads to the second order Bertrand-Darboux PDE ($G_{,xy}=G_{,yx}$)\index{Equation! Bertrand-Darboux}
\begin{eqnarray}
0 &=&(\gamma xy+ \alpha x+\beta y-C)(V_{,xx}-V_{,yy})+\left[ \gamma(y^{2}-x^{2})-2\beta x+2\alpha y+A-B\right] V_{,xy}-  \notag \\
&&-3(\gamma x+\beta )V_{,y}+3(\gamma y+\alpha)V_{,x}.  \label{eq.PDE2}
\end{eqnarray}

3) The constraint $\left( L_{b}V^{,b}\right) _{,a} =-2L_{(a;b)}V^{,b}-\lambda^{2}L_{a}$ with\footnote{For $\lambda=0$ this constraint is a subcase of the constraint $G_{,a}=2C_{ab}V^{,b}$. Hence, only the case $\lambda \neq 0$ must be considered.} $\lambda \neq 0$ and the integrability
condition $\left( L_{b}V^{,b}\right) _{,xy}=\left( L_{b}V^{,b}\right) _{,yx}$,
which lead to the PDEs:
\begin{eqnarray}
0 &=& \left[ -2\beta y^{2}+2\alpha xy+ Ax+ (2C-a_{1})y +a_{2} \right]V_{,xx}+ \left( -2\alpha x^{2}+2\beta
xy+ a_{1}x+ \right. \notag \\
&& \left. +By+a_{3} \right) V_{,xy}+\left[-6\alpha x+2a_{1}+(2C-a_{1}) \right]V_{,y} +3(2\alpha y+A)V_{,x}+  \notag \\
&&+\lambda ^{2} \left[ -2\beta y^{2}+2\alpha xy+Ax +(2C-a_{1})y +a_{2} \right]  \label{eq.PDE3.1}
\end{eqnarray}
\begin{eqnarray}
0 &=&(-2\alpha x^{2}+2\beta xy + a_{1}x +By +a_{3})V_{,yy} +\left[-2\beta y^{2}+2\alpha xy+Ax + \right. \notag \\
&& \left. +(2C-a_{1})y +a_{2} \right]V_{,xy}+3(2\beta x+B)V_{,y}+ \left[ -6\beta y+ 2(2C-a_{1}) +a_{1} \right] V_{,x}+  \notag \\
&& +\lambda^{2}(-2\alpha x^{2} +2\beta xy +a_{1}x+By+a_{3})  \label{eq.PDE3.2}
\end{eqnarray}%
\begin{eqnarray}
0 &=&(\alpha x+\beta y-C)(V_{,xx}-V_{,yy})+\left( -2\beta x +2\alpha y+A-B\right)V_{,xy} -3\beta V_{,y} +3\alpha V_{,x}+  \notag \\
&&+\lambda^{2} (3\alpha x- 3\beta y +C -a_{1}).
\label{eq.PDE3.3}
\end{eqnarray}

For $\alpha=\beta =0$ and $a_{1}=C$, the PDE (\ref{eq.PDE3.3}) reduces to PDE (\ref{eq.PDE2}). Therefore, in order to find new potentials, one of these
conditions must be relaxed. This case of finding potentials is the most
difficult because the problem is overdetermined, i.e. we have a system of
three PDEs (\ref{eq.PDE3.1}) - (\ref{eq.PDE3.3}) and only one unknown function, the potential
$V(x,y)$.

In the following sections, we solve these constraints and find the admitted
potentials which --as a rule-- are integrable. Subsequently, we apply Theorem
\ref{The first integrals of an autonomous holonomic dynamical system} to
each of these potentials in order to compute the corresponding LFIs/QFIs. From these results, we determine
which of these potentials are integrable and, especially, superintegrable.

\section{The constraint $L_{a}V^{,a} = s$}

\label{sec.const1}

The constraint $L_{a}V^{,a}=s$ gives the PDE (\ref{eq.PDE1}) which can be solved using the method of the characteristic equation.

To cover all possible occurrences, we have to consider the following cases:
a) $b_{3}=0$ and $b_{1}\neq 0$ (KVs $\partial _{x}$ and $\partial
_{x},\partial _{y})$; b) $b_{3}=b_{1}=0$ and $b_{2}\neq 0$ (KV $\partial
_{y})$; and c) $b_{3}\neq 0$ (KVs $y\partial _{x}-x\partial _{y}$; $%
\partial _{x},y\partial _{x}-x\partial _{y}$; and $\partial _{y},y\partial
_{x}-x\partial _{y})$. For each case the solution is shown in Table \ref{Table.LV.pots}.

\begin{longtable}{|c|c|c|}
\hline
Case & KV & $V(x,y)$ \\ \hline
a & $b_{3}=0,b_{1}\neq 0$ & $\frac{s}{b_{1}}x+F(b_{1}y-b_{2}x)$ \\
b & $b_{3}=b_{1}=0$, $b_{2}\neq 0$ & $\frac{s}{b_{2}}y+F(x)$ \\
c & $b_{3}\neq 0$ & $\frac{s}{b_{3}}\tan ^{-1}\left( \frac{y+\frac{b_{1}}{%
b_{3}}}{-x+\frac{b_{2}}{b_{3}}}\right) +F(b_{1}y+\frac{b_{3}}{2}y^{2}-b_{2}x+%
\frac{b_{3}}{2}x^{2})$ \\ \hline
\caption{\label{Table.LV.pots} Potentials that satisfy the constraint $L_{a}V^{,a}=s$.}
\end{longtable}

We shall refer to the above solutions of Table \ref{Table.LV.pots} as \textbf{Class I}\index{Potential! Class I} potentials. In order
to determine if these potentials admit QFIs, we apply Theorem \ref{The first
integrals of an autonomous holonomic dynamical system} to the following types of potentials resulting from Table \ref{Table.LV.pots}:
\[
V_{1}= cx+F(y-bx), \enskip V_{2}= cy+F(x), \enskip V_{3}= c\tan^{-1}\left( \frac{y+b_{1}}{-x+b_{2}}\right) +F\left( \frac{x^{2}+y^{2}}{2}+b_{1}y-b_{2}x\right)
\]
where $c$ and $b$ are arbitrary constants.

\subsection{The potential $V_{1}=cx+F(y-bx)$}

\label{subsec.V1}

\textbf{Case a.} $b=0$ and $F=\lambda y$.

The potential reduces to $V_{1a}=cx + \lambda y$.

The irreducible FIs are%
\begin{equation*}
L_{11}=\dot{x}+ct,\enskip L_{12}=\dot{y}+\lambda t,\enskip Q_{11}=\frac{1}{2}%
\dot{x}^{2}+cx,\enskip Q_{12}=\frac{1}{2}\dot{y}^{2}+\lambda y.
\end{equation*}%
We note that $Q_{11}+Q_{12}=\frac{1}{2}(\dot{x}^{2} +\dot{y}^{2}) +V =H$ is the
Hamiltonian. We compute $\{Q_{11},Q_{12}\}=0$ and $\{L_{11},Q_{11}\}=-c$.

The FIs $I_{1}=Q_{11}+Q_{12}$, $I_{2}=\lambda L_{11}-cL_{12}=\lambda \dot{x}%
-c\dot{y} $ and $I_{3}=Q_{11}$ are functionally independent and satisfy the relations $\{I_{1},I_{2}\}=\{I_{1},I_{3}\}=0$ and $\{I_{2},I_{3}\}= -c\lambda$. Therefore, the potential $V_{1a}$ is superintegrable.

We note that the FIs $I_{2}$ and $Q_{11}$ are, respectively, the FIs (3.1.4) and (3.2.20) of \cite{Hietarinta 1987}.
\bigskip

\textbf{Case b.} $\frac{d^{2}F}{dw^{2}}\neq 0$ and $w\equiv y-bx$.

The irreducible FIs are the following:
\begin{equation*}
L_{21}=\dot{x}+b\dot{y}+ct,\enskip L_{22}=t(\dot{x}+b\dot{y})-(x+by)+\frac{c%
}{2}t^{2},\enskip Q_{21}=(\dot{x}+b\dot{y})^{2}+2c(x+by).
\end{equation*}

For $F(y-bx)=-\frac{1}{2}\lambda ^{2}y^{2}$ and $b=0$, we have the potential $V_{1b}=cx-\frac{1}{2}\lambda ^{2}y^{2}$ where $\lambda \neq 0$, which admits the
additional time-dependent FI $L_{23}=e^{\lambda t}(\dot{y}-\lambda y)$. Observe that in this case thw QFI $Q_{21}$ reduces to $Q_{e1}=\frac{1}{2}\dot{x}%
^{2}+cx$ which using the Hamiltonian generates the QFI $Q_{e2}\equiv$ $H-Q_{e1}=\frac{1}{2}\dot{y}^{2}-$ $\frac{1}{2}\lambda ^{2}y^{2}$.

The LFI $L_{21}(c=0)$ is the (3.1.4) of \cite{Hietarinta 1987}.

We compute $\{H,L_{21}\}=\frac{\partial L_{21}}{\partial t}=c$ because $%
L_{21}$ is a time-dependent FI.

The potential of the case b is integrable because $\{H,Q_{21}\}=0$.

Moreover, $\{H,L_{22}\} = L_{21} = \frac{\partial L_{22}}{\partial t}$, $\{L_{21},L_{22}\}= 1 + b^{2}$,
$\{Q_{21},L_{21}\} = 2c(1+b^{2}) = 2c\{L_{21},L_{22}\}$ and
$\{Q_{21},L_{22}\}= 2(1+b^{2})L_{21} =2\{L_{21},L_{22}\} L_{21}$.

For the special case $V=cx-\frac{1}{2}\lambda ^{2}y^{2}$, we have: $\{H,L_{23}\}=\{Q_{e2},L_{23}\} =\lambda L_{23} =\frac{\partial L_{23}}{\partial t}$ and $\{Q_{e1},L_{23}\}=0$.
The triplet $Q_{e1}, Q_{e2}, L_{23}$ shows that this potential is superintegrable.

We note that in \cite{Hietarinta 1987} only the \textbf{Class II} potentials (see sec. \ref{sec.const2}) are examined for superintegrability (see p.108, eqs. (3.2.34) - (3.2.36) in \cite{Hietarinta 1987}).
\bigskip

For $c\neq 0$, the potential $V_{1}=cx+F(y-bx)$ is not included in \cite{Hietarinta 1987} because the author seeks for autonomous LFIs of the form (3.1.1); in that case $s=0$.

\subsection{The potential $V_{2}=cy+F(x)$}

\label{subsec.V2}

We consider the case $F^{\prime \prime }= \frac{d^{2}F}{dx^{2}}\neq 0$; otherwise, we
retrieve the potential $V_{1a}$ discussed above.

The irreducible FIs are
\begin{equation*}
L_{31}=\dot{y}+ct,\enskip Q_{31}=\frac{1}{2}\dot{x}^{2}+F(x),\enskip Q_{32}=%
\frac{1}{2}\dot{y}^{2}+cy.
\end{equation*}%
Therefore, the potential $V_{2}$ is integrable. We note that $V_{2}$ is of the separable form $V=F_{1}(x)+F_{2}(y)$ (see eq. (3.2.20) of \cite{Hietarinta 1987}).

For $F(x)=- \frac{1}{2} \lambda^{2}x^{2}$, we obtain the potential $V_{2a}=cy-\frac{1}{2}\lambda^{2}x^{2}$ with $\lambda\neq0$, which admits the additional LFI $L_{32}= e^{\lambda t}(\dot{x}-\lambda x)$. This potential is
superintegrable due to the functionally independent triplet $Q_{31}$, $Q_{32}$ and $L_{32}$.

\subsection{The potential $V_{3} = c \tan^{-1}\left( \frac{y+b_{1}}{-x +b_{2}} \right) + F\left( \frac{x^{2}+ y^{2}}{2} + b_{1}y - b_{2}x \right)$}

\label{subsec.V3}

We find the time-dependent LFI $L_{51}=y\dot{x}-x\dot{y}+b_{1}\dot{x}+b_{2}\dot{y}+ct$.

For $c=0$ the considered potential is integrable, while for $c\neq 0$ we do not know (needs further investigation).

- For $c=0$ and $F=\lambda \left(\frac{x^{2}+y^{2}}{2}+ b_{1}y-b_{2}x\right)$ where $\lambda \neq 0$, the irreducible FIs are: $L_{41}=y\dot{x}-x\dot{y} +b_{1}\dot{x} +b_{2}\dot{y}$, $Q_{41}=\frac{1}{2}
\dot{x}^{2}+\frac{1}{2}\lambda x^{2}-\lambda b_{2}x$, $Q_{42}=\frac{1}{2}\dot{y}^{2}+\frac{1}{2}\lambda y^{2} +\lambda b_{1}y$ and $Q_{43}=\dot{x}\dot{y}+\lambda (xy+b_{1}x-b_{2}y)$.

Observe that $Q_{41}+Q_{42}=H$ is the energy of the system. The LFI $L_{41}$ is the (3.1.6) of \cite{Hietarinta 1987}. The functionally independent triplet $H$, $L_{41}$, $Q_{41}$ proves that the considered potential is superintegrable. We compute: $\{H,L_{41}\}=\{H,Q_{41}\}=0$ and $\{L_{41},Q_{41}\}=-Q_{43} +\lambda b_{1}b_{2}$.

If $b_{1}=b_{2}=0$ and $\lambda = -k^{2}\neq0$, we obtain the superintegrable\footnote{
A subcase of the above superintegrable potential is the potential $V_{3a}=\lambda \left(
\frac{x^{2}+y^{2}}{2}+b_{1}y-b_{2}x\right)$.} potential $V_{3b} =-\frac{1}{2}%
k^{2} (x^{2}+ y^{2})$ which admits the additional time-dependent LFIs
\begin{equation*}
L_{42\pm}=e^{\pm k t} (\dot{x} \mp k x) \enskip \text{and} \enskip L_{43\pm}= e^{\pm k t}(\dot{y} \mp k y).
\end{equation*}
We also compute:  $\{L_{41},Q_{42}\}=Q_{43}-\lambda b_{1}b_{2}$, $\{L_{41},Q_{43}\}=2Q_{41}-2Q_{42}+\lambda (b_{2}^{2}-b_{1}^{2})$, $\{Q_{41},Q_{42}\}=0$ and $\{Q_{41},Q_{43}\}=\{Q_{43},Q_{42}\}= -\lambda L_{41}$.
\bigskip

In sec. 4 of \cite{Adlam 2007}, the author has found the superintegrable \textbf{Class I} potentials $V_{1a}$ and $V_{3a}$.

We note that in the review \cite{Hietarinta 1987} the time-dependent LFIs of the potentials $V_{1a}$ and $V_{2}$ are not discussed. In general, in \cite{Hietarinta 1987}, all the time-dependent FIs are ignored, although they can be used to decide the superintegrability of the system.

\subsection{Summary}

\label{sec.class1}

We collect the results for the \textbf{Class I} potentials in the Tables \ref{Table.Class1.1} and \ref{Table.Class1.2}.

\begin{longtable}{|l|l|l|}
\hline
{\large Potential} & {\large Ref. \cite{Hietarinta 1987}} & {\large LFIs and QFIs} \\ \hline
$V_{3}(c\neq 0)=c\tan ^{-1}\left( \frac{y+b_{1}}{-x+b_{2}}\right) +F\left(
\frac{x^{2}+y^{2}}{2}+b_{1}y-b_{2}x\right) $ & - & $L_{51}=y\dot{x}-x\dot{y}+b_{1}\dot{x}+b_{2}\dot{y}+ct$ \\ \hline
\multicolumn{3}{|c|}{\large Integrable potentials} \\ \hline
$V_{1}=cx+F(y-bx)$, $\frac{d^{2}F}{dw^{2}}\neq 0$, $w\equiv y-bx$ & - & %
\makecell[l]{$L_{21}=\dot{x}+b\dot{y}+ct$, \\
$L_{22}=t(\dot{x}+b\dot{y})-(x+by)+\frac{c}{2}t^{2}$, \\
$Q_{21}=(\dot{x}+b\dot{y})^{2}+2c(x+by)$} \\ \hline
$V_{2}=cy+F(x)$, $F^{\prime \prime }\neq 0$ & (3.2.20) & %
\makecell[l]{$L_{31}=\dot{y}+ct$, $Q_{31}=\frac{1}{2}\dot{x}^{2}+F(x)$, \\
$Q_{32}=\frac{1}{2}\dot{y}^{2}+cy$} \\ \hline
$V_{3}(c=0)$ & (3.1.6) & $L_{51}(c=0)$ \\ \hline
\caption{\label{Table.Class1.1} Integrable Class I potentials and the special non-integrable potential $V_{3}(c\neq0)$.}
\end{longtable}

\begin{longtable}{|l|l|l|}
\hline
\multicolumn{3}{|c|}{\large Superintegrable potentials} \\ \hline
{\large Potential} & {\large Ref. \cite{Hietarinta 1987} } & {\large LFIs and
QFIs} \\ \hline
$V_{1a}=cx+\lambda y$ & \makecell[l]{(3.1.4), \\ (3.2.20)} & %
\makecell[l]{$L_{11}=\dot{x}+ct$, $L_{12}= \dot{y} + \lambda t$, \\ $Q_{11}=
\frac{1}{2}\dot{x}^{2} + cx$, $Q_{12}= \frac{1}{2}\dot{y}^{2} + \lambda y$}
\\ \hline
$V_{1b}=cx-\frac{1}{2}\lambda ^{2}y^{2}$, $\lambda \neq 0$ & (3.2.20) & %
\makecell[l]{$L_{11}$, $L_{22}(b=0)=t\dot{x}-x + \frac{c}{2}t^{2}$, \\
$L_{23}=e^{\lambda t}(\dot{y}-\lambda y)$, $Q_{2e1}=\frac{1}{2}\dot{x}^{2}+cx$, \\
$Q_{2e2}=\frac{1}{2}\dot{y}^{2}-\frac{1}{2}\lambda ^{2}y^{2}$} \\ \hline
$V_{2a}=cy -\frac{1}{2}\lambda ^{2}x^{2}$, $\lambda \neq 0$ & (3.2.20) & \makecell[l]{$L_{31}$, $Q_{31a}=\frac{1}{2}\dot{x}^{2} -\frac{1}{2}\lambda ^{2}x^{2}$, $Q_{32}$, \\ $L_{32}=e^{\lambda t}(\dot{x}-\lambda x)$} \\ \hline
\makecell[l]{$V_{3a}=\lambda \left( \frac{x^{2}+y^{2}}{2} +b_{1}y-b_{2}x\right)$ \\ where $\lambda \neq 0$} & (3.1.6) & \makecell[l]{$L_{41}=y\dot{x}-x\dot{y}+b_{1}
\dot{x}+b_{2}\dot{y}$, \\ $Q_{41}= \frac{1}{2}\dot{x}^{2}+ \frac{1}{2}\lambda x^{2}-\lambda b_{2}x$, \\ $Q_{42}=\frac{1}{2}\dot{y}^{2}+\frac{1}{2}\lambda
y^{2}+\lambda b_{1}y$, \\ $Q_{43}=\dot{x}\dot{y}+\lambda (xy+b_{1}x-b_{2}y)$}
\\ \hline
$V_{3b}=-\frac{1}{2}k^{2}(x^{2}+y^{2})$, $k\neq 0$ & (3.1.5) & %
\makecell[l]{$L_{41b}=y\dot{x}-x\dot{y}$, $Q_{41b}= \frac{1}{2}\dot{x}^{2}-
\frac{1}{2}k^{2} x^{2}$, \\ $Q_{42b}= \frac{1}{2}\dot{y}^{2} - \frac{1}{2}
k^{2} y^{2}$, $Q_{43b}=\dot{x}\dot{y} - k^{2}xy$, \\
$L_{42\pm}=e^{\pm kt}(\dot{x}\mp kx)$, $L_{43\pm}=e^{\pm kt}(\dot{y} \mp ky)$} \\ \hline
\caption{\label{Table.Class1.2} Superintegrable Class I potentials.}
\end{longtable}

We note that the potentials $V_{1}$ and $V_{2}$ in Table \ref{Table.Class1.1} are also superintegrable due to additional time-dependent LFIs.

\section{The constraint $G_{,a}=2C_{ab}V^{,b}$}

\label{sec.const2}

In this case, we have the PDE (\ref{eq.PDE2})
\begin{eqnarray}
0 &=&(\gamma xy+\alpha x+\beta y-C)(V_{,xx}-V_{,yy})+\left[ \gamma(y^{2}-x^{2})-2\beta x+2\alpha y+A-B\right] V_{,xy}-  \notag \\
&&-3(\gamma x+\beta)V_{,y}+3(\gamma y+\alpha)V_{,x}.  \label{eq.Hie2}
\end{eqnarray}%
The potentials which follow from this PDE we call \textbf{Class II}\index{Potential! Class II} potentials. This PDE cannot be solved in full generality (see also \cite{Hietarinta 1987}); therefore, we consider various cases which produce the
known QFIs using Theorem \ref{The first integrals of an autonomous holonomic
dynamical system}. We emphasize that the potentials we find in this section
are only a subset of the possible potentials which will follow from the
general solution of (\ref{eq.Hie2}). However, the important point here is
that we recover the known results with a direct and unified approach which
can be used in the future by other authors to discover new integrable and
superintegrable potentials in $E^{2}$ and in other spaces.
\bigskip

1) $\gamma \neq 0$, $A=B$ and $\alpha=\beta =C=0$. Then, $C_{ab}=\left(
\begin{array}{cc}
\gamma y^{2}+A & -\gamma xy \\
-\gamma xy & \gamma x^{2}+A%
\end{array}%
\right) $ and equation (\ref{eq.Hie2}) becomes
\begin{equation}
xy(V_{,xx}-V_{,yy})+(y^{2}-x^{2})V_{,xy}-3xV_{,y}+3yV_{,x}=0.
\label{eq.Hie3a}
\end{equation}%
The solution of (\ref{eq.Hie3a}) is the potential
\begin{equation}
V_{21}= \frac{F_{1}\left( \frac{y}{x}\right) }{d_{1}x^{2}+d_{2}y^{2}}%
+F_{2}(x^{2}+y^{2})  \label{eq.Hie3b}
\end{equation}
where $d_{1}$ and $d_{2}$ are arbitrary constants.

- For the subcase $d_{1}=d_{2}=1$ with $A=0$, we find the QFI
\begin{equation}
I_{11}= (y\dot{x}-x\dot{y})^{2}+ 2F_{1}\left( \frac{y}{x}\right) =(r^{2}\dot{\theta})^{2}-\Phi (\theta)  \label{eq.Hie3bb}
\end{equation}%
where $r^{2}=x^{2}+y^{2}$ and $\theta =\tan ^{-1}\left( \frac{y}{x}\right)$. This is the well-known \textbf{Ermakov-Lewis invariant}\index{Invariant! Ermakov-Lewis} (see eq. (3.2.11) of \cite{Hietarinta 1987}).

- For $d_{1}\neq0$, the potential (\ref{eq.Hie3b}) is written equivalently as $V_{21} = \frac{F_{1}\left(\frac{y}{x}\right)}{x^{2}+cy^{2}} +
F_{2}(x^{2}+y^{2})$ where $c$ is an arbitrary constant.

This potential admits QFIs for $F_{1}= \frac{cky^{2}+ k x^{2}}{%
x^{2}+(2-c)y^{2}}$. Therefore,
\begin{equation*}
V_{21a}= \frac{k}{x^{2}+(2-c)y^{2}} + F_{2}(x^{2}+y^{2})= \frac{k}{x^{2}+\ell y^{2}} + F_{2}(x^{2}+y^{2})
\end{equation*}
and the associated QFI is
\begin{equation}
I_{11a} = (y\dot{x}-x\dot{y})^{2} + \frac{2k(c-1) y^{2}}{x^{2}+(2-c)y^{2}}=
(y\dot{x}-x\dot{y})^{2} + \frac{2k(1-\ell) y^{2}}{x^{2}+\ell y^{2}}
\label{eq.Hie1a}
\end{equation}
where $\ell \equiv 2-c$.

- For $d_{1}=0$ and $d_{2}\neq0$, the potential (\ref{eq.Hie3b}) becomes $V_{21} = \frac{F_{1}\left(\frac{y}{x}\right)}{y^{2}} + F_{2}(x^{2}+y^{2})$.

This potential admits QFIs for $F_{1}= \frac{ky^{2}}{2x^{2}+y^{2}}$. Then, $V_{21b} = \frac{k}{2x^{2}+y^{2}} + F(x^{2}+y^{2})$ and the associated QFI is $I_{11b} = (y\dot{x}-x\dot{y})^{2} + \frac{ky^{2}}{2x^{2}+ y^{2}}$. Observe that $V_{21b}$ is of the form $V_{21a}(c=3/2)$ or $V_{21a}(\ell=1/2)$ with $\bar{k}\equiv 2k$. Therefore, $V_{21b}$ is included in case $V_{21a}$.
\bigskip

2) $\gamma =1$ and $\alpha= \beta =B=C=0$. Then, $C_{ab}=\left(
\begin{array}{cc}
y^{2}+A & -xy \\
-xy & x^{2}%
\end{array}%
\right) $ and equation (\ref{eq.Hie2}) becomes
\begin{equation}
xy(V_{,xx}-V_{,yy})+(y^{2}-x^{2}+A)V_{,xy}-3xV_{,y}+3yV_{,x}=0.
\label{eq.Hie3c}
\end{equation}

- For $A=0$ equation (\ref{eq.Hie3c}) reduces to (\ref{eq.Hie3a}). \bigskip

- For $A\neq 0$ the PDE (\ref{eq.Hie3c}) gives the \textbf{Darboux solution}\index{Solution! Darboux}
\begin{equation}
V_{22}= \frac{F_{1}(u)-F_{2}(v)}{u^{2}-v^{2}}  \label{eq.Hie3d}
\end{equation}%
where $r^{2}=x^{2}+y^{2}$, $u^{2}=r^{2}+A+\left[ (r^{2}+A)^{2}-4Ax^{2}\right]
^{1/2}$ and $v^{2}=r^{2}+A-\left[ (r^{2}+A)^{2}-4Ax^{2}\right] ^{1/2}$.

We find the QFI (see eq. (3.2.9) of \cite{Hietarinta 1987}).
\begin{equation}
I_{21}=(y\dot{x}-x\dot{y})^{2} +A\dot{x}^{2} +\frac{v^{2}F_{1}(u)-u^{2}F_{2}(v)}{u^{2}-v^{2}}.  \label{eq.Hie3dd}
\end{equation}
\bigskip

3) $\gamma =1$, $B=-A$, $C=\pm iA\neq 0$ and $\alpha=\beta =0$. Then,
\begin{equation*}
C_{ab}=\left(
\begin{array}{cc}
y^{2}+A & -xy\pm iA \\
-xy\pm iA & x^{2}-A%
\end{array}%
\right)
\end{equation*}%
and equation (\ref{eq.Hie2}) gives again a potential of the form (\ref%
{eq.Hie3d}), but with $u^{2}=r^{2}+\left[ r^{4}-4A(x\pm iy)^{2}\right]
^{1/2} $ and $v^{2}=r^{2}-\left[ r^{4}-4A(x\pm iy)^{2}\right] ^{1/2}$.

We find the QFI (see eq. (3.2.13) of \cite{Hietarinta 1987})
\begin{equation}
I_{31}=(y\dot{x}-x\dot{y})^{2}+A(\dot{x}\pm i\dot{y})^{2}+\frac{%
v^{2}F_{1}(u)-u^{2}F_{2}(v)}{u^{2}-v^{2}}.  \label{eq.Hie3db}
\end{equation}%
\bigskip

4a) $\alpha=1$ and $\beta =\gamma =A=B=C=0$. Then,
\begin{equation*}
C_{ab}=\left(
\begin{array}{cc}
2y & -x \\
-x & 0%
\end{array}%
\right)
\end{equation*}%
and equation (\ref{eq.Hie2}) becomes
\begin{equation}
x(V_{,xx}-V_{,yy})+2yV_{,xy}+3V_{,x}=0  \label{eq.Hie4a}
\end{equation}%
which gives the potential
\begin{equation}
V_{24}=\frac{F_{1}(r+y)+F_{2}(r-y)}{r}  \label{eq.Hie4b}
\end{equation}%
where $r^{2}=x^{2}+y^{2}$.

We find the QFI (see eq. (3.2.15) of \cite{Hietarinta 1987})
\begin{equation}
I_{41}=\dot{x}(y\dot{x}-x\dot{y})+\frac{(r+y)F_{2}(r-y)- (r-y)F_{1}(r+y)}{r}.
\label{eq.Hie4c}
\end{equation}
\bigskip

4b) $\beta=1$ and $\alpha =\gamma =A=B=C=0$. Then,
\begin{equation*}
C_{ab}=\left(
\begin{array}{cc}
0 & -y \\
-y & 2x%
\end{array}%
\right)
\end{equation*}%
and equation (\ref{eq.Hie2}) becomes
\begin{equation}
y(V_{,xx}-V_{,yy})-2xV_{,xy}-3V_{,y}=0  \label{eq.Hie4a.1}
\end{equation}%
which gives the potential
\begin{equation}
V_{24b}=\frac{F_{1}(r+x)+F_{2}(r-x)}{r}  \label{eq.Hie4b.2}
\end{equation}%
where $r^{2}=x^{2}+y^{2}$.

We find the QFI
\begin{equation}
I_{41b}=\dot{y}(x\dot{y}-y\dot{x})+\frac{(r+x)F_{2}(r-x)-(r-x) F_{1}(r+x)}{r}.
\label{eq.Hie4c.3}
\end{equation}

\begin{remark} \label{remark.new.8}
Observe that under the rotation $x \leftrightarrow y$ the potential (\ref{eq.Hie4b}) becomes the potential (\ref{eq.Hie4b.2}) and all the results of the case 4b) follow. As a result, the case 4b) can be overlooked when we assess integrability. This is not the case for establishing superintegrability where the PDE (\ref{eq.Hie4a.1}) plays a crucial role (see the superintegrable potential (\ref{eq.Hie12d}) in sec. \ref{sec.super}).
\end{remark}

5) $\alpha=1$, $\beta =-i$, $A=-B=\frac{i}{4}$, $C=\frac{1}{4}$ and $\gamma =0$. Then,
\begin{equation*}
C_{ab}=\left(
\begin{array}{cc}
2y+\frac{i}{4} & -x+iy+\frac{1}{4} \\
-x+iy+\frac{1}{4} & -2ix-\frac{i}{4}%
\end{array}%
\right)
\end{equation*}%
and equation (\ref{eq.Hie2}) becomes
\begin{equation}
(x-iy-\frac{1}{4})\left( V_{,xx}-V_{,yy}\right) +2\left( y+ix+\frac{i}{4}%
\right) V_{,xy}+3iV_{,y}+3V_{,x}=0.  \label{eq.Hie5a}
\end{equation}%
This PDE is written equivalently as
\begin{equation}
(x-iy)\left( \partial _{x}+i\partial _{y}\right) ^{2}V-\frac{1}{4}(\partial
_{x}-i\partial _{y})^{2}V+3(\partial _{x}+i\partial _{y})V=0
\label{eq.Hie5aa}
\end{equation}%
which gives the potential
\begin{equation}
V_{25} =w^{-1/2}\left[ F_{1}(z+\sqrt{w})+F_{2}(z-\sqrt{w})\right]
\label{eq.Hie5b}
\end{equation}%
where $z=x+iy$ and $w=x-iy$.

We find the QFI (see eq. (3.2.17) of \cite{Hietarinta 1987})
\begin{equation}
I_{51}=(y\dot{x}-x\dot{y})(\dot{x}+i\dot{y}) +\frac{i}{8}(\dot{x}-i\dot{y}%
)^{2}+i\left( 1-\frac{z}{\sqrt{w}}\right) F_{1}(z+\sqrt{w})+ i\left( -1-\frac{z}{\sqrt{w}}\right) F_{2}(z-\sqrt{w}). \label{eq.Hie5c}
\end{equation}
\bigskip

6) $\alpha=1$, $\beta =\mp i$ and $\gamma =A=B=C=0$. Then,
\begin{equation*}
C_{ab}=\left(
\begin{array}{cc}
2y & -x\pm iy \\
-x\pm iy & \mp 2ix%
\end{array}%
\right)
\end{equation*}%
and equation (\ref{eq.Hie2}) becomes
\begin{equation}
(x\mp iy)\left( V_{,xx}-V_{,yy}\right) +2\left( y\pm ix\right) V_{,xy}\pm
3iV_{,y}+3V_{,x}=0  \label{eq.Hie6a}
\end{equation}%
from which follows the potential
\begin{equation}
V_{26}=\frac{F_{1}(z)}{r}+F_{2}^{\prime }(z)  \label{eq.Hie6b}
\end{equation}%
where $F_{2}^{\prime }=\frac{dF_{2}}{dz}$ and $z=x\pm iy$.

We find the QFI (see eq. (3.2.18) of \cite{Hietarinta 1987})
\begin{equation}
I_{61}=(y\dot{x}-x\dot{y})(\dot{x}\pm i\dot{y})-izV+iF_{2}(z).
\label{eq.Hie6c}
\end{equation}%
\bigskip

7) $AB\neq 0$, $A\neq B$ and $\alpha=\beta =\gamma =C=0$. Then, $C_{ab}=\left(
\begin{array}{cc}
A & 0 \\
0 & B%
\end{array}%
\right) $ and equation (\ref{eq.Hie2}) becomes
\begin{equation}  \label{eq.Hie7a}
(A-B)V_{,xy}=0 \implies V_{,xy}=0.
\end{equation}

Solving the PDE (\ref{eq.Hie7a}), we find the separable potential\index{Potential! separable}
\begin{equation}  \label{eq.Hie7b}
V_{27}= F_{1}(x) + F_{2}(y)
\end{equation}
where $F_{1}$ and $F_{2}$ are smooth functions of their arguments.

We find the irreducible QFIs (see eq. (3.2.20) of \cite{Hietarinta 1987}):
\[
I_{71a}=\dot{x}^{2}+2F_{1}(x) \enskip \text{and} \enskip I_{71b}=\dot{y}^{2}+2F_{2}(y).
\]

It can be proved that there are four special potentials of the form (\ref{eq.Hie7b}) which admit additional time-dependent QFIs and, hence, are superintegrable. These are the following:

7a. The potential $V_{271} =\frac{k_{1}}{\left( x+c_{1}\right) ^{2}}+\frac{k_{2}}{\left(y+c_{2}\right) ^{2}}$ which admits the independent QFIs:
\begin{eqnarray*}
I_{72a} &=& -\frac{t^{2}}{2} \dot{y}^{2} + t (y+c_{2})\dot{y} - t^{2}\frac{%
k_{2}}{(y+ c_{2})^{2}} - \frac{1}{2}y^{2} - c_{2}y \\
I_{72b} &=& -\frac{t^{2}}{2} \dot{x}^{2} + t (x+c_{1})\dot{x} - t^{2}\frac{%
k_{1}}{(x+ c_{1})^{2}} - \frac{1}{2}x^{2} - c_{1}x.
\end{eqnarray*}

7b. The potential $V_{272}=F_{1}(x)+\frac{k_{2}}{\left( y+c_{2}\right) ^{2}}$ which admits the QFI $I_{72a}$.

7c. The potential $V_{273}=F_{2}(y)+ \frac{k_{1}}{\left( x+c_{1}\right) ^{2}}$ which admits the QFI $I_{72b}$.

7d. The potential (see \cite{Fris}) $V_{274}=-\frac{\lambda ^{2}}{8}(x^{2}+y^{2})-\frac{\lambda ^{2}}{4}\left(
c_{1}x+ c_{2}y\right) -\frac{k_{1}}{(x+c_{1})^{2}} -\frac{k_{2}}{(y+c_{2})^{2}}$ which admits the independent QFIs:
\begin{eqnarray*}
I_{73a} &=& e^{\lambda t}\left[ -\dot{x}^{2}+\lambda (x+c_{1})\dot{x}-\frac{%
\lambda^{2}}{4}(x+c_{1})^{2}+ \frac{2k_{1}}{(x+c_{1})^{2}}\right] \\
I_{73b} &=& e^{\lambda t}\left[ -\dot{y}^{2}+\lambda (y+c_{2})\dot{y}-\frac{%
\lambda^{2}}{4}(y+c_{2})^{2}+\frac{2k_{2}}{(y+c_{2})^{2}}\right].
\end{eqnarray*}

The parameters $\lambda, c_{1}, c_{2}, k_{1}$, and $k_{2}$ are arbitrary constants.
\bigskip

8) $C\neq0$ and $\alpha=\beta=\gamma=0$.

Then, $C_{ab}=\left(
\begin{array}{cc}
A & C \\
C & B%
\end{array}%
\right) $ and equation (\ref{eq.Hie2}) becomes
\begin{equation}
C(V_{,yy}-V_{,xx})+(A-B)V_{,xy}=0.  \label{eq.Hie8a}
\end{equation}

Solving equation (\ref{eq.Hie8a}), we find the potential
\begin{equation}  \label{eq.Hie8b}
V_{28} = F_{1}\left(y + b_{0}x + \sqrt{b_{0}^{2}+1}x\right) + F_{2}\left(y +
b_{0}x - \sqrt{b_{0}^{2}+1}x\right)
\end{equation}
where $b_{0} \equiv \frac{A-B}{2C}$.

This potential admits the QFI
\begin{equation}  \label{eq.Hie8bb}
I_{81}= A\dot{x}^{2} + B\dot{y}^{2} + 2C\dot{x}\dot{y} + (A+B)V + 2C \sqrt{b_{0}^{2}+1} (F_{1}- F_{2}).
\end{equation}
We note that $b_{0}(A,B,C)$; therefore, $A, B, C$ are defining parameters of the potential and, as a result, they do not generate independent QFIs.

For $b_{0}=0$, we have: $A=B$, $V_{,yy} - V_{,xx}=0$ and the potential reduces to
\begin{equation}  \label{eq.Hie8c}
V_{28}(b_{0}=0)= F_{1}(y+x) + F_{2}(y-x)
\end{equation}
which is the solution of the 1d wave equation.\index{Equation! wave} The QFI (\ref{eq.Hie8bb}) gives (besides the Hamiltonian) the independent QFI
\begin{equation*}
I_{82}= \dot{x}\dot{y} + F_{1}(y+x) - F_{2}(y-x).
\end{equation*}
\bigskip

9) $A=2$, $C=\pm i$ and $\alpha=\beta =\gamma =B=0$. Then,
\begin{equation*}
C_{ab}=\left(
\begin{array}{cc}
2 & \pm i \\
\pm i & 0%
\end{array}%
\right)
\end{equation*}%
and equation (\ref{eq.Hie2}) becomes
\begin{equation}
\mp i(V_{,xx}-V_{,yy})+2V_{,xy}=0.  \label{eq.Hie9a}
\end{equation}

Solving equation (\ref{eq.Hie9a}), we find the potential
\begin{equation}  \label{eq.Hie9b}
V_{29} = r^{2} F_{1}^{\prime \prime }(z) + F_{2}(z)
\end{equation}
where $F_{1}^{\prime \prime }= \frac{d^{2}F_{1}}{dz^{2}}$ and $z=x\pm iy$.

The associated QFI is (see eq. (3.2.21) of \cite{Hietarinta 1987})
\begin{equation}
I_{91}=\dot{x}(\dot{x}\pm i\dot{y})+V_{29}+ 2zF_{1}^{\prime}(z) -2F_{1}(z). \label{eq.Hie9c}
\end{equation}
\bigskip

10) $A=1$, $B=-1$, $C=i$ and $\alpha=\beta=\gamma=0$.

Then, $C_{ab}=
\left(
  \begin{array}{cc}
    1 & i \\
    i & -1 \\
  \end{array}
\right)$ and equation (\ref{eq.Hie2}) becomes $i \left( V_{,yy} -V_{,xx} \right) +2V_{,xy}=0$. This PDE admits the solution
\begin{equation}
V_{211}= F_{1}(z) +\bar{z}F_{2}(z) \label{eq.Hien.1}
\end{equation}
where $z=x+iy$ and $F_{1}, F_{2}$ are arbitrary complex functions.

The associated QFI is $I_{211}= \dot{z}^{2} +4\int F_{2}(z)dz$.
\bigskip

11) $A=1$, $B=-1$, $C=-i$ and $\alpha=\beta=\gamma=0$.

As in the previous case, we find the potential
\begin{equation}
V_{212}= F_{1}(\bar{z}) +zF_{2}(\bar{z}) \label{eq.Hien.2}
\end{equation}
and the associated QFI $I_{212}= \dot{\bar{z}}^{2} +4\int F_{2}(\bar{z})d\bar{z}$.

\begin{remark} \label{remark.new.4}
For the trivial KT $C_{ab}= A\delta_{ab}$ in $E^{2}$, where $A$ is an arbitrary constant, the condition $G_{,a}=2C_{ab}V^{,b}$ implies that $G=2AV$ for all potentials $V(x,y)$. Therefore, all 2d Newtonian potentials $V(x,y)$ admit the QFI
\begin{equation*}
I=A(\dot{x}^{2}+\dot{y}^{2}+2V)= 2AH
\end{equation*}
where $H$ is the Hamiltonian.
\end{remark}

Comparing with previous works, we see that the potentials $V_{21a}$ and $V_{28}$ are new, while the potential $V_{274}(c_{1}=c_{2}=0)$ is mentioned in \cite{Fris}.

\subsection{The potential given in eq. (21) of \cite{Sen 1985}}

\label{sec.Sen.pot}

The Bertrand-Darboux PDE (\ref{eq.PDE2}) for $\gamma=0$ has been solved in \cite{Sen 1985} and the following potential found:
\begin{equation}
V(x,y)= \frac{F_{1}\left( \beta x +\alpha y -\sqrt{D^{2} +E^{2}} \right) +F_{2}\left( \beta x +\alpha y +\sqrt{D^{2} +E^{2}} \right)}{\sqrt{D^{2} +E^{2}}} \label{eq.sen1}
\end{equation}
where $D(x,y)= \alpha x +\beta y -C$, $E(x,y)= -\beta x +\alpha y +\frac{A-B}{2}$ and $D^{2} +E^{2}\neq0$. We note that for $A=B$, $C=\beta =0$ and $\alpha=1$ the potential (\ref{eq.sen1}) reduces to the integrable potential (\ref{eq.Hie4b}).

The potential (\ref{eq.sen1}) is integrable if there exists a function $G(x,y)$ such that $G_{,a}= 2C_{ab}V^{,b}$. Then, the associated QFI is
\begin{eqnarray}
I_{1}&=& (2\alpha y +A)\dot{x}^{2} +(2\beta x +B)\dot{y}^{2} -2(\alpha x +\beta y -C)\dot{x}\dot{y} +G \notag \\
&=& 2(\beta \dot{y} -\alpha\dot{x})L +A\dot{x}^{2} +2C\dot{x}\dot{y} +B\dot{y}^{2} +G \label{eq.sen2}
\end{eqnarray}
where $L\equiv x\dot{y} -y\dot{x}$ is the angular momentum. We note that in eq. (22) of \cite{Sen 1985} there is an unnecessary square over the constant $c_{2}$.

\subsection{The superintegrable potentials}

\label{sec.super}

When a potential belongs to two of the above nine \textbf{Class II} cases simultaneously, it is superintegrable (e.g. the potentials given in eqs. (3.2.34) - (3.2.36) of \cite{Hietarinta 1987}) because in that case it admits two additional autonomous QFIs besides the Hamiltonian. From the results of the section \ref{sec.const2}, we find the following \textbf{Class II} superintegrable potentials \cite{Ranada 1997, Kalnins 2001}:
\bigskip

S1) The potential (see eq. (3.2.34) of \cite{Hietarinta 1987}, case (b) in \cite{Ranada 1997} and \cite{Fris})
\begin{equation}
V_{s1}=\frac{k}{2}(x^{2}+y^{2})+\frac{b}{x^{2}}+\frac{c}{y^{2}}
\label{eq.Hie10a}
\end{equation}
where $k,b,c$ are arbitrary constants.

This potential is of the form (\ref{eq.Hie3b}) for $d_{1}=d_{2}=1$, $F_{1}\left( \frac{y}{x}\right) =b\left( \frac{y}{x}\right) ^{2}+c\left(
\frac{x}{y}\right) ^{2}$ and $F_{2}(x^{2}+y^{2})= \frac{k}{2}(x^{2}+y^{2}) +\frac{b+c}{x^{2}+y^{2}}$;
and also of the separable form (\ref{eq.Hie7b}). Therefore, $V_{s1}$ admits the additional QFIs:
\begin{eqnarray}
I_{s1a} &=&(y\dot{x}-x\dot{y})^{2}+2b\frac{y^{2}}{x^{2}} +2c\frac{x^{2}}{y^{2}%
}  \label{eq.Hie10b} \\
I_{s1b} &=&\frac{1}{2}\dot{x}^{2}+\frac{k}{2}x^{2}+\frac{b}{x^{2}}
\label{eq.Hie10c} \\
I_{s1c} &=&\frac{1}{2}\dot{y}^{2}+\frac{k}{2}y^{2}+\frac{c}{y^{2}}.
\label{eq.Hie10d}
\end{eqnarray}%
We note that $V_{s1}\left( k=-\frac{\lambda^{2}}{4}, b=-k_{1}, c=-k_{2}\right)$, where $\lambda \neq 0$, coincides with the potential $V_{274}$ for $c_{1}=c_{2}=0$; therefore, it admits also the time-dependent FIs $I_{73a}$ and $I_{73b}$.
\bigskip

S2) Potentials which are both of the form (\ref{eq.Hie4b}) and (\ref{eq.Hie7b}). Then, we have to solve the system of the PDEs (\ref{eq.Hie4a}) and $V_{,xy}=0$. We find 
\begin{equation}
V_{s2}=\frac{k_{1}}{2}(x^{2}+4y^{2})+\frac{k_{2}}{x^{2}}+k_{3}y
\label{eq.Hie11a}
\end{equation}%
and the QFIs:
\begin{eqnarray}
I_{s2a} &=&\dot{x}(y\dot{x}-x\dot{y})-k_{1}yx^{2}+\frac{2k_{2}y}{x^{2}}-%
\frac{k_{3}}{2}x^{2}  \label{eq.Hie11b} \\
I_{s2b} &=&\frac{1}{2}\dot{x}^{2}+\frac{k_{1}}{2}x^{2}+\frac{k_{2}}{x^{2}}
\label{eq.Hie11c} \\
I_{s2c} &=&\frac{1}{2}\dot{y}^{2}+2k_{1}y^{2}+k_{3}y \label{eq.Hie11d}
\end{eqnarray}
where $k_{1}, k_{2}, k_{3}$ are arbitrary constants.

The potential (\ref{eq.Hie11a}) is the superintegrable potential given in the case (a) of \cite{Ranada 1997}. We note that the QFI $I_{3}^{a}$ in \cite{Ranada 1997} is not correct. The correct is the QFI $I_{s2a}$ given in equation (\ref{eq.Hie11b}).

Moreover, the potential (3.2.35) of \cite{Hietarinta 1987} is superintegrable only for $b=4a$. In this case, the resulting potential coincides with $V_{s2}$ for $k_{1}=2a$, $k_{2}=c$ and $k_{3}=0$.
\bigskip

S3) Potentials which are both of the form (\ref{eq.Hie3b}) and (\ref{eq.Hie4b}). Then, we have to solve system of the PDEs (\ref{eq.Hie3a}) and (\ref{eq.Hie4a}). We find
\begin{equation}
V_{s3} = \frac{k_{1}}{x^{2}} + \frac{k_{2}}{r} + \frac{k_{3}y}{rx^{2}}
\label{eq.Hie12a}
\end{equation}
and the QFIs:
\begin{eqnarray}
I_{s3a} &=& (y\dot{x} - x\dot{y})^{2} + 2k_{1}\frac{y^{2}}{x^{2}} + 2k_{3}%
\frac{ry}{x^{2}}  \label{eq.Hie12b} \\
I_{s3b} &=& \dot{x}(y\dot{x} - x\dot{y}) + 2k_{1} \frac{y}{x^{2}} + k_{2}%
\frac{y}{r} + k_{3}\frac{x^{2}+2y^{2}}{rx^{2}}  \label{eq.Hie12c}
\end{eqnarray}
where $r^{2}=x^{2}+y^{2}$.

The superintegrable potential (\ref{eq.Hie12a}) is symmetric ($x
\leftrightarrow y$) to the superintegrable potential of case (c) in \cite{Ranada 1997}. In order to obtain directly the superintegrable potential of \cite{Ranada 1997}, we should consider the form (\ref{eq.Hie4b.2}) instead of the form (\ref{eq.Hie4b}).

We note that if we rename the constants in (\ref{eq.Hie12a}) as $k_{1}=b+c$, $k_{2}=a$ and $k_{3}=c-b$, we recover the superintegrable potential (3.2.36) of \cite{Hietarinta 1987}. Indeed, we have $V_{s3}=\frac{a}{r}+ \frac{\frac{b}{r+y} +\frac{c}{r-y}}{r}$.
\bigskip

S4) If we substitute the solution (\ref{eq.Hie4b}) of the PDE (\ref{eq.Hie4a}) in the PDE (\ref{eq.Hie4a.1}), we find that both the PDEs (\ref{eq.Hie4a}) and (\ref{eq.Hie4a.1}) are satisfied for $F_{1}(r+y)= k_{1}+k_{2}\sqrt{r+y}$ and $F_{2}(r-y)= k_{3}\sqrt{r-y}$. Therefore, the potential (see case (d) in \cite{Ranada 1997})
\begin{equation}
V_{s4} = \frac{k_{1}}{r} + k_{2} \frac{\sqrt{r+y}}{r} + k_{3}\frac{\sqrt{r-y}}{r}  \label{eq.Hie12d}
\end{equation}
is superintegrable with additional QFIs:
\begin{eqnarray*}
I_{s4a} &=& \dot{x}(y\dot{x}-x\dot{y}) + \frac{k_{1}y}{r} + \frac{k_{3}(r+y)%
\sqrt{r-y}-k_{2}(r-y)\sqrt{r+y}}{r}  \label{eq.Hie12e} \\
I_{s4b} &=& \dot{y}(x\dot{y}-y\dot{x}) + G(x,y)  \label{eq.Hie12f}
\end{eqnarray*}
provided that there exists a function $G(x,y)$ such that $G_{,x} + yV_{s4,y}=0$ and $G_{,y} + yV_{s4,x} -2xV_{s4,y}=0$.

We note that in the case (d) in \cite{Ranada 1997} the corresponding QFIs $I_{2}^{d}$ and $I_{3}^{d}$ are not correct, because $\{H,I_{2}^{d}\}\neq0$ and $\{H,I_{3}^{d}\}\neq0$. Moreover, the superintegrable potential (\ref{eq.Hie12d}) coincides with the potential of the case (E20) in \cite{Kalnins 2001}; and it is not mentioned in the review \cite{Hietarinta 1987}.
\bigskip

S5) If we use the relations $z\bar{z}=r^{2}$ and $z=re^{i\theta}$ where $\tan\theta=\frac{y}{x}$ and $z=x+iy$, the potentials $V_{21}$ and (\ref{eq.Hien.1}) are equal iff
\[
V_{21}=V_{211} \implies \frac{A_{1}(\theta)}{r^{2}} +A_{2}(r) = A_{3}(z) +A_{4}(z)r^{2} \implies
\]
\[
A_{1}= k_{1}e^{-4i\theta} +k_{2}e^{-2i\theta}, \enskip A_{2}=k_{3}r^{2}, \enskip A_{3}=\frac{k_{2}}{z^{2}}, \enskip A_{4}= \frac{k_{1}}{z^{4}} +k_{3}
\]
where $k_{1}, k_{2}, k_{3}$ are arbitrary constants. Therefore, we find the superintegrable potential
\begin{equation}
V_{s5}= \frac{k_{1}r^{2}}{z^{4}} +\frac{k_{2}}{z^{2}} +k_{3}r^{2} \label{eq.Hie13a}
\end{equation}
with associated QFIs:
\begin{eqnarray}
I_{s5a}&=& \frac{1}{2}M^{2} +k_{1}e^{-4i\theta} +k_{2}e^{-2i\theta} \label{eq.Hie13b} \\
I_{s5b}&=& \frac{1}{2} \dot{z}^{2} -\frac{k_{1}}{z^{2}} +k_{3}z^{2}. \label{eq.Hie13c}
\end{eqnarray}
\bigskip

In Tables \ref{Table.Class2.1} and \ref{Table.Class2.2}, we collect the results on \textbf{Class II} potentials together with the corresponding reference to the review paper \cite{Hietarinta 1987}. Concerning the notation: $r^{2}= x^{2} +y^{2}$, $z=x+iy=re^{i\theta}$ and $M_{3}= x\dot{y} -y\dot{x}$ is the angular momentum.

\begin{longtable}{|l|c|l|}
\hline
\multicolumn{3}{|c|}{Integrable potentials} \\ \hline
{\large Potential} & {\large Ref. \cite{Hietarinta 1987}} & {\large LFIs and
QFIs} \\ \hline
$V_{21}=\frac{F_{1}\left( \frac{y}{x}\right) }{x^{2}+y^{2}}%
+F_{2}(x^{2}+y^{2})$ & (3.2.10) & $I_{11}=M_{3}^{2}+2F_{1}\left( \frac{y}{x}\right) $ \\ \hline
$V_{21a}= \frac{k}{x^{2}+\ell y^{2}} + F_{2}(x^{2}+y^{2})$ & - & $I_{11a} =M_{3}^{2} + \frac{2k(1-\ell) y^{2}}{x^{2}+\ell y^{2}}$ \\
\hline
\makecell[l]{$V_{22}=\frac{F_{1}(u)-F_{2}(v)}{u^{2}-v^{2}}$, \\
$u^{2}=r^{2}+A+\left[ (r^{2}+A)^{2}-4Ax^{2}\right] ^{1/2}$, \\
$v^{2}=r^{2}+A-\left[ (r^{2}+A)^{2}-4Ax^{2}\right] ^{1/2}$} & (3.2.7,8) & $%
I_{21}=M_{3}^{2}+A\dot{x}^{2}+\frac{v^{2}F_{1}(u)-u^{2}F_{2}(v)%
}{u^{2}-v^{2}}$ \\ \hline
\makecell[l]{$V_{23}=\frac{F_{1}(u)-F_{2}(v)}{u^{2}-v^{2}}$, \\
$u^{2}=r^{2}+\left[ r^{4}-4A(x\pm iy)^{2}\right] ^{1/2}$, \\
$v^{2}=r^{2}-\left[ r^{4}-4A(x\pm iy)^{2}\right] ^{1/2}$} & (3.2.7,12) & $%
I_{31}=M_{3}^{2}+A(\dot{x}\pm i\dot{y})^{2}+\frac{%
v^{2}F_{1}(u)-u^{2}F_{2}(v)}{u^{2}-v^{2}}$ \\ \hline
$V_{24}=\frac{F_{1}(r+y)+F_{2}(r-y)}{r}$ & (3.2.15) & $I_{41}=M_{3}\dot{x} +\frac{(r-y)F_{1}(r+y) -(r+y)F_{2}(r-y)}{r}$ \\ \hline
\makecell[l]{$V_{25}=w^{-1/2}\left[
F_{1}(z+\sqrt{w})+F_{2}(z-\sqrt{w})\right] $, \\ $z=x+iy$, $w=x-iy$} & (3.2.17) & \makecell[l]{$I_{51}=-M_{3}(\dot{x}+i\dot{y})+
\frac{i}{8}(\dot{x}-i\dot{y})^{2}+$ \\ \qquad \enskip $+i\left(
1-\frac{z}{\sqrt{w}}\right) F_{1}(z+\sqrt{w})+$ \\ \qquad \enskip $+i\left(
-1-\frac{z}{\sqrt{w}}\right) F_{2}(z-\sqrt{w})$} \\ \hline
\makecell[l]{$V_{26}=\frac{F_{1}(z)}{r}+F_{2}^{\prime }(z)$, \\
$F_{2}^{\prime }=\frac{dF_{2}}{dz}$, $z=x\pm iy$} & (3.2.18) & $I_{61}=-M_{3}(\dot{x}\pm i\dot{y})-izV+iF_{2}(z)$ \\ \hline
$V_{27}=F_{1}(x)+F_{2}(y)$ & (3.2.20) & $I_{71a}=\frac{1}{2}\dot{x}%
^{2}+F_{1}(x)$, $I_{71b}=\frac{1}{2}\dot{y}^{2}+F_{2}(y)$ \\ \hline
\makecell[l]{$V_{28}=F_{1}\left( y+b_{0}x+\sqrt{b_{0}^{2}+1}x\right) +$ \\
\qquad \enskip $+ F_{2}\left(y+b_{0}x-\sqrt{b_{0}^{2}+1}x\right) $,
$b_{0}\equiv \frac{A-B}{2C}$} & - & \makecell[l]{$I_{81}=A\dot{x}^{2}+B\dot{y}^{2}+ 2C\dot{x}\dot{y}+(A+B)V+$ \\ \qquad \enskip $+
2C\sqrt{b_{0}^{2}+1}(F_{1}-F_{2})$} \\ \hline
$V_{28}(b_{0}=0)=F_{1}(y+x)+F_{2}(y-x)$ & - & $I_{82}=\dot{x}\dot{y}%
+F_{1}(y+x)-F_{2}(y-x)$ \\ \hline
\makecell[l]{$V_{29}=r^{2}F_{1}^{\prime \prime }(z)+F_{2}(z)$,
$F_{1}^{\prime \prime }=\frac{d^{2}F_{1}}{dz^{2}}$, \\ $z=x\pm iy$} & (3.2.21) & $I_{91}=\dot{x}(\dot{x}\pm i\dot{y})+V_{29} +2zF_{1}^{\prime
}(z)- 2F_{1}(z)$ \\ \hline
\makecell[l]{$V_{210}= \frac{F_{1}\left( \beta x +\alpha y -\sqrt{D^{2} +E^{2}} \right)}{\sqrt{D^{2} +E^{2}}}+$ \\ \qquad $+\frac{F_{2}\left( \beta x +\alpha y +\sqrt{D^{2} +E^{2}} \right)}{\sqrt{D^{2} +E^{2}}}$ \\ where $D(x,y)= \alpha x +\beta y -C$ \\ and $E(x,y)= -\beta x +\alpha y +\frac{A-B}{2}$} & - & \makecell[l]{$I_{210}= 2M_{3}(\beta \dot{y} -\alpha\dot{x}) +A\dot{x}^{2} +2C\dot{x}\dot{y}+$ \\ \qquad  \quad $+B\dot{y}^{2} +G(x,y)$ \\ $\frac{G_{,x}}{2}= \left( 2\alpha y +A \right)V_{,x} -\left( \alpha x +\beta y -C \right)V_{,y}$ \\ $\frac{G_{,y}}{2}= \left( 2\beta x+ B \right)V_{,y} -\left( \alpha x +\beta y -C \right) V_{,x}$} \\ \hline
$V_{211}= F_{1}(z) +\bar{z}F_{2}(z)$ & - & $I_{211}= \dot{z}^{2} +4\int F_{2}(z)dz$ \\ \hline
$V_{212}= F_{1}(\bar{z}) +zF_{2}(\bar{z})$ & - & $I_{211}= \dot{\bar{z}}^{2} +4\int F_{2}(\bar{z})d\bar{z}$ \\ \hline
\caption{\label{Table.Class2.1} Integrable Class II potentials.}
\end{longtable}

\begin{longtable}{|l|c|l|}
\hline
\multicolumn{3}{|c|}{Superintegrable potentials} \\ \hline
{\large Potential} & {\large Ref. \cite{Hietarinta 1987}} & {\large LFIs and
QFIs} \\ \hline
$V_{s1}= \frac{k}{2}(x^{2}+y^{2}) + \frac{b}{x^{2}} + \frac{c}{y^{2}}$ &
(3.2.34) & \makecell[l]{$I_{s1a}= M_{3}^{2} + 2b
\frac{y^{2}}{x^{2}} + 2c \frac{x^{2}}{y^{2}}$, \\ $I_{s1b} =
\frac{1}{2}\dot{x}^{2} + \frac{k}{2}x^{2} + \frac{b}{x^{2}}$, \\ $I_{s1c} =
\frac{1}{2}\dot{y}^{2} + \frac{k}{2}y^{2} + \frac{c}{y^{2}}$ \\ - For $k=0$:
$I_{72a}$, $I_{72b}$ \\ where $c_{1}=c_{2}=0, k_{1}=b, k_{2}=c$ \\ - For $k=-\frac{\lambda^{2}}{4}\neq0$: $I_{73a}$, $I_{73b}$ \\ where $c_{1}=c_{2}=0, k_{1}=-b, k_{2}=-c$} \\ \hline
$V_{s2}= \frac{k_{1}}{2}(x^{2}+4y^{2}) + \frac{k_{2}}{x^{2}} + k_{3}y$ & %
\makecell[l]{(3.2.35) \\ $k_{3}=0$} & \makecell[l]{$I_{s2a} =
M_{3}\dot{x} +k_{1}yx^{2} -\frac{2k_{2}y}{x^{2}} +\frac{k_{3}}{2} x^{2}$, \\ $I_{s2b}= \frac{1}{2}\dot{x}^{2} +
\frac{k_{1}}{2}x^{2} + \frac{k_{2}}{x^{2}}$, \\ $I_{s2c}=
\frac{1}{2}\dot{y}^{2} + 2k_{1}y^{2}+ k_{3}y$} \\ \hline
$V_{s3} = \frac{k_{1}}{x^{2}} + \frac{k_{2}}{r} + \frac{k_{3}y}{rx^{2}}$ & (3.2.36) & \makecell[l]{$I_{s3a}= M_{3}^{2} +2k_{1}\frac{y^{2}}{x^{2}} + 2k_{3}\frac{ry}{x^{2}}$, \\ $I_{s3b}=-M_{3}\dot{x} + 2k_{1} \frac{y}{x^{2}} + k_{2}\frac{y}{r} +k_{3}\frac{x^{2}+2y^{2}}{rx^{2}}$} \\ \hline
$V_{s4} = \frac{k_{1}}{r} + k_{2} \frac{\sqrt{r+y}}{r} + k_{3}\frac{\sqrt{r-y%
}}{r}$ & - & \makecell[l]{$I_{s4a} = -M_{3}\dot{x} +\frac{k_{1}y}{r} +$ \\ \qquad \quad $+ \frac{k_{3}(r+y)\sqrt{r-y} -k_{2}(r-y)\sqrt{r+y}}{r}$, \\ $I_{s4b}= M_{3}\dot{y} + G(x,y)$ \\ where $G_{,x} + yV_{s4,y}=0$ and \\ $G_{,y} + yV_{s4,x} -2xV_{s4,y}=0$} \\ \hline
$V_{s5}= \frac{k_{1}r^{2}}{z^{4}} +\frac{k_{2}}{z^{2}} +k_{3}r^{2}$ & - & \makecell[l]{ $I_{s5a}= \frac{1}{2}M^{2} +k_{1}e^{-4i\theta} +k_{2}e^{-2i\theta}$ \\ $I_{s5b}= \frac{1}{2} \dot{z}^{2} -\frac{k_{1}}{z^{2}} +k_{3}z^{2}$} \\ \hline
$V_{271}=\frac{k_{1}}{\left( x+c_{1} \right)^{2}}+ \frac{k_{2}}{\left(
y+c_{2} \right)^{2}}$ & (3.2.20) & \makecell[l]{$I_{71a}$, $I_{71b}$, \\
$I_{72a}=-\frac{t^{2}}{2}\dot{y}^{2}+t(y+c_{2})\dot{y}-
t^{2}\frac{k_{2}}{(y+c_{2})^{2}}- $ \\ \qquad \quad $-\frac{1}{2}y^{2}-c_{2}y$, \\
$I_{72b}=-\frac{t^{2}}{2}\dot{x}^{2}+t(x+c_{1})\dot{x}-
t^{2}\frac{k_{1}}{(x+c_{1})^{2}}-$ \\ \qquad \quad $-\frac{1}{2}x^{2}-c_{1}x$} \\ \hline
$V_{272}=F_{1}(x)+\frac{k_{2}}{\left( y+c_{2}\right) ^{2}}$ & (3.2.20) & $%
I_{71a}$, $I_{71b}$, $I_{72a}$ \\ \hline
$V_{273}=F_{2}(y)+\frac{k_{1}}{\left( x+c_{1}\right) ^{2}}$ & (3.2.20) & $%
I_{71a}$, $I_{71b}$, $I_{72b}$ \\ \hline
\makecell[l]{$V_{274}=-\frac{\lambda ^{2}}{8}(x^{2}+y^{2})-\frac{\lambda
^{2}}{4}\left( c_{1}x+ c_{2}y\right)-$ \\ \qquad \quad
$-\frac{k_{1}}{(x+c_{1})^{2}}-\frac{k_{2}}{(y+c_{2})^{2}}$, $\lambda \neq0$}
& (3.2.20) &  \makecell[l]{$I_{71a}$, $I_{71b}$, \\ $I_{73a}=e^{\lambda t}\left[ -\dot{x}^{2}+\lambda (x+c_{1})\dot{x} - \right.$ \\ \qquad \quad $\left. -\frac{\lambda^{2}}{4}(x+c_{1})^{2} +\frac{2k_{1}}{(x +c_{1})^{2}}\right]$, \\
$I_{73b}=e^{\lambda t}\left[ -\dot{y}^{2}+\lambda
(y+c_{2})\dot{y} - \right.$ \\ \qquad \quad $\left. -\frac{\lambda^{2}}{4}(y+c_{2})^{2}+\frac{2k_{2}}{(y+c_{2})^{2}} \right]$} \\ \hline
\caption{\label{Table.Class2.2} Superintegrable Class II potentials.}
\end{longtable}

\section{The constraint $\left(L_{b}V^{,b}\right)_{,a} = -2 L_{(a;b)} V^{,b}- \protect\lambda^{2}L_{a}$}

\label{sec.const3}

The integrability condition of the constraint $\left(
L_{b}V^{,b}\right)_{,a}=-2L_{(a;b)}V^{,b}-\lambda ^{2}L_{a}$ gives the PDE (\ref{eq.PDE3.3}). As mentioned in section \ref{sec.find.Pots}, in order to find new potentials from the PDE (\ref{eq.PDE3.3}) one (or both) of the conditions $\alpha= \beta =0$ and $a_{1}=C$ must be relaxed. However, even if we do find a new potential, this solution should satisfy also the remaining PDEs (\ref{eq.PDE3.1}) and (\ref{eq.PDE3.2}) in order to admit the time-dependent QFI $I_{3}$ given in case \textbf{Integral 3} of theorem \ref{The first integrals of an autonomous holonomic dynamical system}. New potentials which admit the QFI $I_{3}$ shall be referred to as \textbf{Class III} potentials.\index{Potential! Class III}

We note that the PB $\{H,I_{3}\}= \frac{\partial I_{3}}{\partial t} \neq 0$. Therefore, to find a new integrable potential, we should find a \textbf{Class III} potential admitting two independent FIs of the form $I_{3}$, say $I_{3a} $ and $I_{3b}$, such that $\{I_{3a}, I_{3b}\} =0$.

After relaxing one, or both, of the conditions $\alpha=\beta =0$ and $a_{1}=C$, we found that the only non-trivial \textbf{Class III} potential is the superintegrable potential $V_{3b}= -\frac{\lambda ^{2}}{2}r^{2}$ (see sec. \ref{subsec.V3}), which is derived for $\alpha\neq 0$ or $\beta\neq 0$. Therefore, there are no new \textbf{Class III} potentials.

\section{Using FIs to find the solution of 2d integrable dynamical systems}

\label{sec.example.2d}

In this section, we consider examples which show how one uses the 2d (super-)integrable potentials to find the solution of the dynamical equations.
\bigskip

1) The superintegrable potential $V_{3b}= -\frac{1}{2}k^{2}(x^{2}+y^{2})$ where $k\neq 0$.

We find the solution by using the time-dependent LFIs $L_{42\pm}= e^{\pm kt}(\dot{x}\mp kx)$ and $L_{43\pm}=e^{\pm kt}(\dot{y} \mp ky)$. Specifically, we have
\[
\begin{cases}
e^{kt}(\dot{x} -kx) = c_{1+} \\
e^{-kt}(\dot{x} +kx) = c_{1-}
\end{cases}
\implies
\begin{cases}
\dot{x} -kx = c_{1+}e^{-kt} \\
\dot{x} +kx = c_{1-}e^{kt}
\end{cases}
\implies
x(t) = \frac{c_{1-}}{2k}e^{kt} - \frac{c_{1+}}{2k}e^{-kt}.
\]
Similarly from the LFIs $L_{43\pm}$, we find
\[
y(t) = \frac{c_{2-}}{2k}e^{kt} - \frac{c_{2+}}{2k}e^{-kt}
\]
where $c_{1\pm}$ and $c_{2\pm}$ are arbitrary constants
\bigskip

2) The integrable potential $V_{2}=cy+F(x)$ where $F''\neq0$.

Using the LFI $L_{31}=\dot{y}+ct=c_{1}$, we find directly $y(t)= -\frac{c}{2}t^{2} +c_{1}t + c_{2}$ where $c,c_{1},c_{2}$ are arbitrary constants.

Using the QFI $2Q_{31} =\dot{x}^{2}+2F(x)=const=c_{3}$, we get
\[
\frac{dx}{dt} = \pm \left[-2F(x)+c_{3}\right]^{1/2} \implies dt= \pm \left[-2F(x)+c_{3}\right]^{-1/2}dx \implies t= \pm \int\left[-2F(x)+c_{3}\right]^{-1/2}dx + c_{0}
\]
where $c_{0}$ is an integration constant. The inverse function of $t=t(x)$ is the solution of the system. If the function $F(x)$ is given, the solution can be explicitly determined.
\bigskip

3) For the integrable potential $V_{27}= F_{1}(x) + F_{2}(y)$, by using the QFIs
\[
I_{71a}= \frac{1}{2}\dot{x}^{2} + F_{1}(x) \enskip \text{and} \enskip I_{71b}=\frac{1}{2}\dot{y}^{2} + F_{2}(y)
\]
we find:
\[
t= \int\left[c_{1}-2F_{1}(x)\right]^{-1/2}dx + c_{0}, \enskip t= \int\left[c_{2}-2F_{2}(y)\right]^{-1/2}dy + c_{3}
\]
where $c_{0}, c_{1}= 2I_{71a}, c_{2}=2I_{71b}$ and $c_{3}$ are arbitrary constants.

\section{Conclusions}

\label{sec.conclusions}

By using Theorem \ref{The first integrals of an autonomous holonomic dynamical
system}, we have reproduced in a systematic way most integrable
and superintegrable 2d potentials of autonomous conservative Newtonian dynamical systems. The method used, being covariant, it is directly applicable to spaces
of higher dimensions and to metrics with any signature and curvature.

We have found two classes of potentials and, in each class, we have determined
the integrable and the superintegrable potentials together with their QFIs. As
the general solution of the PDE (\ref{eq.PDE2}) is not possible, we have
found the potentials due to certain solutions only. New solutions of this
PDE will lead to new integrable and, possibly, superintegrable 2d
potentials.

It appears that the most difficult part in the application of Theorem \ref{The first integrals of an autonomous holonomic dynamical system} to higher
dimensions and curved configuration spaces is the determination of the KTs. The use
of algebraic computing is limited once one considers higher dimensions, since
then the number of the components of the KT increases dramatically.
Fortunately, today new techniques in differential geometry have been
developed (see e.g. \cite{Kalnins 1980, Barnes 2003, Garfinkle 2010, Coll 2006, Crampin 2008}), especially in the case of spaces of constant curvature and decomposable spaces, which can
help to deal with this problem.

%% file: QFIs_damping.tex
\chapter{Quadratic first Integrals of autonomous holonomic dynamical systems with a linear damping term}

\label{ch.QFIs.damping}

\section{Introduction}

In this chapter, we consider autonomous holonomic dynamical systems of the form
\begin{equation}
\ddot{q}^{a}= \omega ^{a}(q,\dot{q})  \label{FLII.0}
\end{equation}%
where $\omega ^{a}=-\Gamma_{bc}^{a}(q)\dot{q}^{b}\dot{q}^{c} -V^{,a}(q)+F^{a}(q,\dot{q})$, $\Gamma_{bc}^{a}(q)$ are the Riemannian connection coefficients of the kinetic metric   $\gamma_{ab}(q)$, $V(q)$ is the potential for all conservative forces, and $F^{a}$ are all non-conservative generalized forces. By assuming, then, that the generalized forces $F^{a}$ are linear in the velocities, we determine all LFIs/QFIs for the resulting systems. We use the direct method by making more general assumptions, for the tensors $K_{a}(t,q)$ and $K_{ab}(t,q)$, from those made in chapter \ref{ch1.QFIs.conservative}, and we state our results as Theorem \ref{Theorem2}. Moreover, we show that a general $m$th-order FI for dynamical systems of the form (\ref{FLII.0}) is associated to a gauged weak Noether symmetry, which we compute. Finally, we apply Theorem \ref{Theorem2} in the 2d problem of two autonomous linearly coupled damped harmonic oscillators and we find a plethora of new FIs.

\section{Determination of the weak Noether symmetry associated to polynomial FIs}

\label{sec.weak.Noether}

In this section, we apply the Inverse Noether Theorem \ref{Inverse Noether Theorem} to a general QFI of the form
\begin{equation}
\Lambda= K_{ab}(t,q) \dot{q}^{a} \dot{q}^{b} +K_{a}(t,q) \dot{q}^{a} +K(t,q) \label{eq.inv15}
\end{equation}
where $K_{ab}$ is a symmetric tensor, $K_{a}$ is a vector and $K$ is a scalar.

From conditions (\ref{eq.inv14a}) - (\ref{eq.inv14c}), we deduce that the general QFI (\ref{eq.inv15}) is associated to the gauged weak Noether symmetry:
\begin{eqnarray}
\eta^{a} &=& -\gamma^{ab} \left( 2K_{bc}\dot{q}^{c} +K_{b} \right) \label{eq.InvNoe.g2a} \\
\phi^{a} \frac{\partial L}{\partial \dot{q}^{a}} &=& -F^{a} \left( 2K_{ab}\dot{q}^{b} +K_{a} \right) \label{eq.InvNoe.g2b} \\
f &=& K_{ab}\dot{q}^{a}\dot{q}^{b} + K_{a}\dot{q}^{a} + K -\gamma^{ab} \left( 2K_{bc}\dot{q}^{c} +K_{b} \right) \frac{\partial L}{\partial \dot{q}^{a}} \label{eq.InvNoe.g2c}
\end{eqnarray}
where $\frac{\partial \Lambda}{\partial \dot{q}^{a}}= 2K_{ab}\dot{q}^{b} +K_{a}$, $L=L(t,q,\dot{q})$ is the Lagrangian of the system and $F^{a}(t,q,\dot{q})$ are the generalized non-conservative forces.

In the case that $L= \frac{1}{2} \gamma_{ab}(q) \dot{q}^{a} \dot{q}^{b} - V(q)$, conditions (\ref{eq.InvNoe.g2a}) - (\ref{eq.InvNoe.g2c}) become:
\begin{eqnarray}
\eta^{a} &=& -\gamma^{ab} \left( 2K_{bc}\dot{q}^{c} +K_{b} \right) \label{eq.InvNoe.g2na} \\
0&=& \left(\phi_{a} +2K_{ab}F^{b} \right)\dot{q}^{a} + K_{a}F^{a} \label{eq.InvNoe.g2nb} \\
f &=& -K_{ab}\dot{q}^{a}\dot{q}^{b} + K. \label{eq.InvNoe.g2nc}
\end{eqnarray}

Therefore, a QFI of the general form (\ref{eq.inv15}) is associated with the gauged weak Noether symmetry
\begin{equation}
\left( \xi=0, \enskip \eta_{a}= -2K_{ab}\dot{q}^{b} -K_{a}; \enskip \phi_{a}, \enskip f= -K_{ab}\dot{q}^{a}\dot{q}^{b} +K \right) \enskip \text{such that} \enskip  \left( \phi_{a} +2K_{ab}F^{b} \right)\dot{q}^{a} +K_{a}F^{a}=0. \label{eq.inv17a}
\end{equation}
It is easy to check that the gauged weak Noether symmetry (\ref{eq.inv17a}) does produce the Noether FI (\ref{eq.inv15}).

We note that since (\ref{eq.inv15}) is a FI the associated function $M(\Lambda)$ given by (\ref{eq.inv13}) must vanish identically. This implies that
\[
M(\Lambda)=0 \implies \frac{\partial \Lambda}{\partial t} +\frac{\partial \Lambda}{\partial q^{a}}\dot{q}^{a} +\frac{\partial \Lambda}{\partial \dot{q}^{a}}\gamma^{ab} \left( F_{b} +\frac{\partial L}{\partial q^{b}} - \frac{\partial^{2} L}{\partial \dot{q}^{b}\partial t} - \frac{\partial^{2} L}{\partial \dot{q}^{b} \partial q^{c}} \dot{q}^{c} \right) =0\implies
\]
\begin{eqnarray}
0&=& K_{(ab;c)} \dot{q}^{a} \dot{q}^{b} \dot{q}^{c} + \left( K_{ab,t} +K_{(a;b)}\right) \dot{q}^{a} \dot{q}^{b} +2 K_{ab} \dot{q}^{(b}(F^{a)} -V^{,a)}) + \left( K_{a,t} +K_{,a}\right) \dot{q}^{a}+ \notag \\
&& +K_{a}(F^{a} -V^{,a}) +K_{,t} \label{eq.inv16}
\end{eqnarray}
which (as expected) coincides with the QFI conditions found in section \ref{sec.ch1.conditions.QFIs.1}.

As a final application of the Inverse Noether Theorem, we consider $m$th-order FIs of the form
\begin{equation}
I^{(m)} = \sum_{r=0}^{m} M_{i_{1}i_{2}...i_{r}} \dot{q}^{i_{1}} \dot{q}%
^{i_{2}} ...\dot{q}^{i_{r}} =M + M_{i_{1}}\dot{q}^{i_{1}} + M_{i_{1}i_{2}}%
\dot{q}^{i_{1}}\dot{q}^{i_{2}}+ ... + M_{i_{1}i_{2}...i_{m}} \dot{q}^{i_{1}}
\dot{q}^{i_{2}} ...\dot{q}^{i_{m}}  \label{eq.inv18}
\end{equation}
where $M_{i_{1}...i_{r}}(t,q)$ with $r=0,1,...,m$ are totally symmetric $r$-rank tensors and the index $(m)$ denotes the order of the FI. Then, we find that the polynomial FI (\ref{eq.inv18}) is associated with the gauged weak Noether symmetry
\begin{equation}
\left( \xi=0, \enskip \eta_{i_{1}}= -\frac{\partial I^{(m)}}{\partial \dot{q}^{i_{1}}}; \enskip \phi_{a}, \enskip f =I^{(m)} -\frac{\partial I^{(m)}}{\partial \dot{q}^{i_{1}}} \dot{q}^{i_{1}} \right) \enskip \text{such that} \enskip
\phi_{a} \dot{q}^{a} +F^{a} \frac{\partial I^{(m)}}{\partial \dot{q}^{a}}= 0  \label{eq.inv19}
\end{equation}
where
\begin{eqnarray*}
\frac{\partial I^{(m)}}{\partial \dot{q}^{i_{1}}} &=& M_{i_{1}} +2M_{i_{1}i_{2}}\dot{q}^{i_{2}} + 3M_{i_{1}i_{2}i_{3}} \dot{q}^{i_{2}} \dot{q}^{i_{3}} + ... + m M_{i_{1}i_{2}...i_{m}} \dot{q}^{i_{2}}...\dot{q}^{i_{m}} \\
&=& \sum^{m-1}_{r=0} (r+1) M_{i_{1}i_{2}...i_{r+1}} \dot{q}^{i_{2}}...\dot{q}^{i_{r+1}}.
\end{eqnarray*}

\section{The conditions for a QFI}

\label{section.1}

As it has been mentioned in the Introduction, the purpose of the present chapter is to generalize the results of chapter \ref{ch1.QFIs.conservative} to the case of autonomous holonomic dynamical systems of the form (\ref{FLII.0}), which move in a Riemannian configuration space under the action of generalized forces of the form $F^{a}=-P^{a}(q) +A_{b}^{a}(q)\dot{q}^{b}$. The dynamical equations for these systems are
\begin{equation}
\ddot{q}^{a}=-\Gamma_{bc}^{a}(q)\dot{q}^{b}\dot{q}^{c} -Q^{a}(q)+A_{b}^{a}(q)\dot{q}^{b}  \label{FL.5.1}
\end{equation}%
where now the generalized forces $Q^{a}\equiv V^{,a}+P^{a}$ contain all the forces, conservative and non-conservative.

We assume again the general QFI (\ref{FL.5}) and, by using the dynamical equations (\ref{FL.5.1}) to replace the accelerations $\ddot{q}^{a}$, we write the condition $\frac{dI}{dt}=0$ as
\begin{align*}
0& =K_{(ab;c)}\dot{q}^{a}\dot{q}^{b}\dot{q}^{c}+\left(
K_{ab,t}+K_{a;b}+2K_{c(a}A_{b)}^{c}\right) \dot{q}^{a}\dot{q}^{b}+\left(
K_{a,t}+K_{,a}-2K_{ab}Q^{b}+\right. \\
& \quad \left. +K_{b}A_{a}^{b}\right) \dot{q}^{a}+K_{,t}-K_{a}Q^{a}.
\end{align*}
From the last equation, we obtain the following system of PDEs:
\begin{eqnarray}
K_{(ab;c)} &=&0  \label{eq.veldep4.1} \\
K_{ab,t}+K_{(a;b)}+2K_{c(a}A_{b)}^{c} &=&0  \label{eq.veldep4.2} \\
-2K_{ab}Q^{b}+K_{a,t}+K_{,a}+K_{b}A_{a}^{b} &=&0  \label{eq.veldep4.3} \\
K_{,t}-K_{a}Q^{a} &=&0.  \label{eq.veldep4.4}
\end{eqnarray}
Condition (\ref{eq.veldep4.1}) implies that $K_{ab}$ is a KT of order two (possibly zero) of the kinetic metric $\gamma_{ab}$.

The most general choice\footnote{We recall that in section \ref{sec.ch1.conditions.QFIs} we considered two simpler assumptions, i.e. the choices made in equations (\ref{choice1}) and (\ref{choice2}).} for the KT $K_{ab}$ in the case of an autonomous system is\footnote{%
Equivalently, we may assume $K_{ab}= \sum^{n}_{N=1} f_{N}(t) D_{(N)ab}(q)$,
where $f_{N}(t)$ is a sequence of analytic functions and $D_{(N)ab}$ is a
sequence of KTs of $\gamma_{ab}$. This expression is equivalent to (\ref{eq.aspm1}) because if we set
\begin{equation*}
f_{N}(t)= \sum^{n}_{M=0} d_{(N)M} t^{M} = d_{(N)0} + d_{(N)1}t + ... +d_{(N)n} t^{n}
\end{equation*}
then
\begin{equation*}
K_{ab}= \sum^{n}_{N=1} \sum^{n}_{M=0} d_{(N)M} t^{M} D_{(N)ab}(q) =
\sum^{n}_{M=0} \underbrace{\left( \sum^{n}_{N=1} d_{(N)M} D_{(N)ab}(q)
\right)}_{\equiv \bar{D}_{(M)ab}(q)} t^{M}= \sum^{n}_{M=0} \bar{D}%
_{(M)ab}(q) t^{M}.
\end{equation*}%
}
\begin{equation}
K_{ab}(t,q)=C_{(0)ab}(q) + \sum_{N=1}^{n}C_{(N)ab}(q)\frac{t^{N}}{N}
\label{eq.aspm1}
\end{equation}%
where $C_{(N)ab}(q)$, $N=0,1,...,n$, is a sequence of arbitrary KTs of order two of the kinetic metric $\gamma_{ab}$. This choice of $K_{ab}$ and equation (\ref{eq.veldep4.2}) indicate that we
set
\begin{equation}
K_{a}(t,q)=\sum_{M=0}^{m}L_{(M)a}(q)t^{M}  \label{eq.aspm2}
\end{equation}%
where $L_{(M)a}(q)$, $M=0,1,...,m$, are arbitrary vectors.

We note that both powers $n$ and $m$ in the above polynomial expressions may be infinite.

Substituting (\ref{eq.aspm1}) and (\ref{eq.aspm2}) in the system of equations (\ref{eq.veldep4.1}) -(\ref{eq.veldep4.4}), we obtain the following equations\footnote{Equation (\ref{eq.veldep4.1}) is identically zero because the quantities $C_{(N)ab}$ are assumed to be KTs.}:
\begin{eqnarray}
0&=& C_{(1)ab}+C_{(2)ab}t+...+C_{(n)ab}t^{n-1}+L_{(0)(a;b)} +L_{(1)(a;b)}t+...+L_{(m)(a;b)}t^{m}+
\notag \\
&& +2C_{(0)c(a}A_{b)}^{c}+2C_{(1)c(a}A_{b)}^{c}t+... +2C_{(n)c(a}A_{b)}^{c} \frac{t^{n}}{n}
\label{eq.veldep6} \\
0 &=&-2C_{(0)ab}Q^{b}-2C_{(1)ab}Q^{b}t-... -2C_{(n)ab}Q^{b} \frac{t^{n}}{n} +L_{(1)a}+2L_{(2)a}t +...+mL_{(m)a}t^{m-1} +  \notag \\
&& +K_{,a} +L_{(0)b}A_{a}^{b} +L_{(1)b}A_{a}^{b}t +...+L_{(m)b}A_{a}^{b}t^{m}  \label{eq.veldep7} \\
0&=&K_{,t}-L_{(0)a}Q^{a}-L_{(1)a}Q^{a}t-...-L_{(m)a}Q^{a}t^{m}.
\label{eq.veldep8}
\end{eqnarray}

Conditions (\ref{eq.veldep6}) - (\ref{eq.veldep8}) must be supplemented with the integrability conditions $K_{,at}=K_{,ta}$ and $K_{,[ab]}=0$ for the scalar $K$. The integrability condition $K_{,at}=K_{,ta}$ gives --if we make use of (\ref{eq.veldep7}) and (\ref{eq.veldep8})-- the equation
\begin{eqnarray}
0 &=& -2C_{(1)ab}Q^{b}-2C_{(2)ab}Q^{b}t... -2C_{(n)ab}Q^{b}t^{n-1}+2L_{(2)a}+6L_{(3)a}t+ ... + \notag \\
&& +m(m-1)L_{(m)a}t^{m-2} +\left( L_{(0)b}Q^{b}\right)_{,a} +\left( L_{(1)b}Q^{b}\right)_{,a}t +...+\left( L_{(m)b}Q^{b}\right)_{,a}t^{m}+ \notag \\
&& + L_{(1)b}A_{a}^{b} +2L_{(2)b}A_{a}^{b}t +...+mL_{(m)b}A_{a}^{b}t^{m-1}.  \label{eq.veldep9}
\end{eqnarray}%
The condition $K_{,[ab]}=0$ gives the equation
\begin{eqnarray}
0 &=&2\left( C_{(0)[a\left\vert c\right\vert }Q^{c}\right) _{;b]}+2\left(
C_{(1)[a\left\vert c\right\vert }Q^{c}\right) _{;b]}t+
...+2\left(C_{(n)[a\left\vert c\right\vert }Q^{c}\right) _{;b]}\frac{t^{n}}{n} - L_{(1)\left[a;b\right] }- \notag \\
&& -2L_{(2)\left[ a;b\right]}t -... -mL_{(m)\left[a;b\right] }t^{m-1} -L_{(0)c;[b}A_{a]}^{c}-L_{(1)c;[b}A_{a]}^{c}t-...
-L_{(m)c;[b}A_{a]}^{c}t^{m} -\notag \\
&& - L_{(0)c}A_{[a;b]}^{c}
-L_{(1)c}A_{[a;b]}^{c}t-... -L_{(m)c}A_{[a;b]}^{c}t^{m}  \label{eq.veldep10}
\end{eqnarray}%
which for 2d systems with $F^{a}=0$ reduces to the second order Bertrand-Darboux equation \cite{Darboux} (see section \ref{sec.find.Pots}).\index{Equation! Bertrand-Darboux}

Equations (\ref{eq.veldep6}) - (\ref{eq.veldep10}) constitute the system of equations we have to solve.

\section{The Theorem}

\label{section.2}

The solution of the system of equations (\ref{eq.veldep6}) - (\ref{eq.veldep10}) is stated\footnote{The proof of Theorem \ref{Theorem2} is given in Appendix \ref{app.proof.QFIs.damping}.} in Theorem \ref{Theorem2} below.

\begin{theorem}
\label{Theorem2} The independent QFIs of a dynamical system of the form (\ref{FL.5.1}) are the following\footnote{%
We note that the QFI $J_{1}$ is for $n$ finite, whereas $J_{2}$ is for $n$ infinite (hence the term $e^{\lambda t}$).}:
\bigskip

\textbf{Integral 1.}
\begin{eqnarray*}
J_{1} &=&\left( \frac{t^{n}}{n}C_{(n)ab}+...+\frac{t^{2}}{2}%
C_{(2)ab}+tC_{(1)ab}+C_{(0)ab}\right) \dot{q}^{a} \dot{q}^{b}+t^{n}L_{(n)a}\dot{q}^{a}+ ... +t^{2}L_{(2)a} \dot{q}^{a}+tL_{(1)a}\dot{q}^{a}+ \\
&&+L_{(0)a}\dot{q}^{a} +\frac{t^{n+1}}{n+1}L_{(n)a}Q^{a} +...+\frac{t^{2}}{2}L_{(1)a}Q^{a}+tL_{(0)a}Q^{a}+G(q)
\end{eqnarray*}%
where\footnote{
We note that for $n=0$, the conditions for the QFI $J_{1}(n=0)$ can be
derived if we set equal to zero the quantities $C_{(N)ab}$ and $L_{(N)a}$
for $N\neq 0$.} $C_{(N)ab}$ for $N=0,1,...,n$ are KTs, $%
C_{(1)ab}=-L_{(0)(a;b)}-2C_{(0)c(a}A_{b)}^{c}$, $C_{(k+1)ab}=-L_{(k)(a;b)}-%
\frac{2}{k}C_{(k)c(a}A_{b)}^{c}$ for $k=1,...,n-1$, $L_{(n)(a;b)}=-\frac{2}{n%
}C_{(n)c(a}A_{b)}^{c}$, $\left( L_{(k-1)b}Q^{b}\right)
_{,a}=2C_{(k)ab}Q^{b}-k(k+1)L_{(k+1)a}-kL_{(k)b}A_{a}^{b}$ for $k=1,...,n-1$%
, $\left( L_{(n-1)b}Q^{b}\right) _{,a}=2C_{(n)ab}Q^{b}-nL_{(n)b}A_{a}^{b}$, $%
L_{(n)a}Q^{a}=s$ and $G_{,a}=2C_{(0)ab}Q^{b}-L_{(1)a}-L_{(0)b}A_{a}^{b}$.

\textbf{Integral 2.}
\begin{equation*}
J_{2}= e^{\lambda t} \left( \lambda C_{ab} \dot{q}^{a} \dot{q}^{b} + \lambda
L_{a}\dot{q}^{a} + L_{a}Q^{a} \right)
\end{equation*}
where $\lambda \neq 0$, $C_{ab}$ is a KT, $\lambda C_{ab} = - L_{(a;b)} -
2C_{c(a} A^{c}_{b)}$ and $\left(L_{b}Q^{b}\right)_{,a} = 2\lambda C_{ab}
Q^{b} - \lambda^{2}L_{a} - \lambda L_{b}A^{b}_{a}$.
\end{theorem}

We note that in all cases $C_{(N)ab}$ are KTs of order two, while in many special cases the vector $K^{a}$ is a KV. This emphasizes the already known result from previous studies (see e.g. \cite{StephaniB, Katzin 1981, kalotas}) of the important role played by the KTs and the KVs of the kinetic metric
in the determination of the FIs of the dynamical system (\ref{FL.5.1}).

In the case that\footnote{%
If in addition $F^{a}=0$, then $Q^{a}=V^{,a}$ and the resulting dynamical equations describe an autonomous conservative system.} $A_{b}^{a}(q)=0$, Theorem \ref{Theorem2} takes the following form.

\begin{theorem}
\label{theorem3} The independent QFIs of the dynamical system (\ref{FL.5.1}) for $A^{a}_{b}=0$ are the following:
\bigskip

\textbf{Integral 1.}
\begin{eqnarray*}
I_{(1)} &=& \left( - \frac{t^{2\ell}}{2\ell} L_{(2\ell-1)(a;b)} - ... - \frac{%
t^{4}}{4} L_{(3)(a;b)} - \frac{t^{2}}{2} L_{(1)(a;b)} + C_{ab} \right) \dot{q%
}^{a} \dot{q}^{b} + t^{2\ell-1} L_{(2\ell-1)a}\dot{q}^{a} + ... +
t^{3}L_{(3)a}\dot{q}^{a} + \\
&& + t L_{(1)a}\dot{q}^{a} + \frac{t^{2\ell}}{2\ell} L_{(2\ell-1)a}Q^{a} +
... + \frac{t^{4}}{4} L_{(3)a}Q^{a} + \frac{t^{2}}{2} L_{(1)a}Q^{a} + G(q)
\end{eqnarray*}
where\footnote{%
We note that for $\ell=0$, the conditions for the QFI $I_{(1)}(\ell=0)$ are
given by nullifying all the vectors $L_{(M)a}$.} $C_{ab}$ and $L_{(M)(a;b)}$
for $M=1,3,...,2\ell-1$ are KTs, $\left( L_{(2\ell-1)b} Q^{b} \right)_{,a} =
-2L_{(2\ell-1)(a;b)}Q^{b}$, $\left( L_{(k-1)b} Q^{b} \right)_{,a} =
-2L_{(k-1)(a;b)}Q^{b} - k(k+1)L_{(k+1)a}$ for $k=2,4,...,2\ell-2$, and $G_{,a}= 2C_{ab}Q^{b} -L_{(1)a}$.

\textbf{Integral 2.}
\begin{eqnarray*}
I_{(2)} &=& \left( - \frac{t^{2\ell+1}}{2\ell+1} L_{(2\ell)(a;b)} - ... -
\frac{t^{3}}{3} L_{(2)(a;b)} - t L_{(0)(a;b)} \right) \dot{q}^{a} \dot{q}%
^{b} + t^{2\ell} L_{(2\ell)a}\dot{q}^{a} + ... + t^{2}L_{(2)a}\dot{q}^{a} +
\\
&& + L_{(0)a}\dot{q}^{a}+ \frac{t^{2\ell+1}}{2\ell+1} L_{(2\ell)a}Q^{a} +
... + \frac{t^{3}}{3} L_{(2)a}Q^{a} +t L_{(0)a}Q^{a}
\end{eqnarray*}
where $L_{M(a;b)}$ for $M=0,2,...,2\ell$ are KTs, $\left( L_{(2\ell)b} Q^{b}
\right)_{,a} = -2L_{(2\ell)(a;b)}Q^{b}$ and $\left( L_{(k-1)b}
Q^{b}\right)_{,a} = -2L_{(k-1)(a;b)}Q^{b} - k(k+1)L_{(k+1)a}$ for $%
k=1,3,...,2\ell-1$.

\textbf{Integral 3.}
\begin{equation*}
I_{(3)} = e^{\lambda t} \left(-L_{(a;b)}\dot{q}^{a}\dot{q}^{b} + \lambda L_{a} \dot{q}^{a} + L_{a}Q^{a} \right)
\end{equation*}
where the vector $L_{a}$ is such that $L_{(a;b)}$ is a KT and $\left(L_{b}Q^{b}\right)_{,a} = -2L_{(a;b)} Q^{b} - \lambda^{2} L_{a}$.
\end{theorem}

We observe that for $A_{b}^{a}=0$ the QFI $J_{1}$ breaks into two independent QFIs, the $I_{(1)}$ and $I_{(2)}$, corresponding to even and odd powers of $t$, respectively. Theorem \ref{theorem3} is a generalized version of Theorem \ref{The first integrals of an autonomous holonomic dynamical system} --even for $Q^{a}=V^{,a}$-- because the assumptions (\ref{eq.aspm1}) and (\ref{eq.aspm2}) are more general than those made in section \ref{sec.ch1.conditions.QFIs.1}.

It is apparent that before one attempts to compute the QFIs of a given
dynamical system of the form (\ref{FL.5.1}) using Theorem \ref{Theorem2}, one has to know the collineations of the kinetic metric including the second order KTs. This is not a trivial requirement for a general curved configuration space (see chapter \ref{ch.collineations}). For such spaces, one has to use special methods to compute the KTs (see e.g. \cite{Kalnins 1980, Barnes 2003, Garfinkle 2010, Coll 2006, Crampin 2008, Sommers 1973}).

\section{Computing the QFI $J_{1}\equiv I_{n}$ in terms of the fundamental QFI $I_{0}$}

\label{sec.In}

We prove that all QFIs $I_{N}$, where $N=1,2,...,n$, of the case \textbf{Integral 1} of Theorem \ref{Theorem2} can be constructed from the fundamental QFI\index{First integral! fundamental} $I_{0}$ by using the following systematic algorithm: \newline
1) Write the QFI $I_{0}$.\newline
2) Introduce a new KT $C_{(1)ab}$ and a new vector $L_{(1)a}$. \newline
3) Construct the QFI $I_{1}$ by adding to the expression of $I_{0}$ the time-dependent terms $tC_{(1)ab}\dot{q}^{a}\dot{q}^{b}$, $tL_{(1)a}\dot{q}^{a}$ and $\frac{t^{2}}{2}L_{(1)a}Q^{a}$. \newline
4) Expand the conditions for $I_{0}$ so as to satisfy the requirement $\frac{dI_{1}}{dt}=0$ along the dynamical equations. \newline
5) Continue in a similar manner with the construction of the QFI $I_{2}$ by using $I_{1}$. \newline
6) After some steps, use the QFI $I_{n-1}$ to construct the QFI $I_{n}$ by adding the terms $\frac{t^{n}}{n} C_{(n)ab}\dot{q}^{a} \dot{q}^{b}$, $t^{n}L_{(n)a}\dot{%
q}^{a}$ and $\frac{t^{n+1}}{n+1} L_{(n)a}Q^{a}$.
\bigskip

We illustrate the above procedure for the small values of $n$.
\bigskip

- For $n=0$:

We have the QFI
\begin{equation*}
I_{0}=C_{(0)ab}\dot{q}^{a}\dot{q}^{b}+L_{(0)a}\dot{q}^{a}+ st +G(q)
\end{equation*}%
where $C_{(0)ab}$ is a KT and the quantities $L_{(0)a}$ and $G$ are computed from the expressions:
\begin{equation*}
L_{(0)(a;b)}=-2C_{(0)c(a}A_{b)}^{c},\enskip L_{(0)b}Q^{b}=s,\enskip %
G_{,a}=2C_{(0)ab}Q^{b}-L_{(0)b}A_{a}^{b}.
\end{equation*}

- For $n=1$.

We have the QFI
\begin{equation*}
I_{1}=\left( tC_{(1)ab}+C_{(0)ab}\right) \dot{q}^{a}\dot{q}%
^{b}+tL_{(1)a}\dot{q}^{a}+L_{(0)a}\dot{q}^{a}%
+\frac{t^{2}}{2}s+tL_{(0)a}Q^{a}+G(q)
\end{equation*}%
where $C_{(1)ab}$ is a KT computed from the relation
\begin{equation*}
C_{(1)ab}=-L_{(0)(a;b)}-2C_{(0)c(a}A_{b)}^{c}
\end{equation*}%
while the vector $L_{(1)a}$ and the quantity $G$ are computed from the relations:
\begin{equation*}
L_{(1)(a;b)}=-2C_{(1)c(a}A_{b)}^{c},\enskip L_{(1)a}Q^{a}=s,\enskip\left(
L_{(0)b}Q^{b}\right) _{,a}=2C_{(1)ab}Q^{b}-L_{(1)b}A_{a}^{b},
\end{equation*}%
\begin{equation*}
L_{(1)a}=2C_{(0)ab}Q^{b}-L_{(0)b}A_{a}^{b}-G_{,a}.
\end{equation*}

- For $n=2$.

We have the QFI
\begin{eqnarray*}
I_{2} &=&\left( \frac{t^{2}}{2}C_{(2)ab}%
+tC_{(1)ab}+C_{(0)ab}\right) \dot{q}^{a}\dot{q}^{b}%
+t^{2}L_{(2)a}\dot{q}^{a}+tL_{(1)a}\dot{q}^{a}+L_{(0)a}\dot{%
q}^{a}+\frac{t^{3}}{3}s+\frac{t^{2}}{2}L_{(1)a}Q^{a}+ \\
&&+tL_{(0)a}Q^{a}+G(q)
\end{eqnarray*}%
where $C_{(2)ab}$ is a KT computed from the relations:
\begin{equation*}
C_{(1)ab}=-L_{(0)(a;b)}-2C_{(0)c(a}A_{b)}^{c},\enskip %
C_{(2)ab}=-L_{(1)(a;b)}-2C_{(1)c(a}A_{b)}^{c}
\end{equation*}%
while the vector $L_{(2)a}$ and the quantity $G$ are computed from the relations:
\begin{equation*}
L_{(2)(a;b)}=-C_{(2)c(a}A_{b)}^{c},\enskip L_{(2)a}Q^{a}=s,\enskip\left(
L_{(1)b}Q^{b}\right) _{,a}=2C_{(2)ab}Q^{b}-2L_{(2)b}A_{a}^{b}
\end{equation*}%
\begin{equation*}
L_{(1)a}=2C_{(0)ab}Q^{b}-L_{(0)b}A_{a}^{b}-G_{,a},\enskip %
L_{(2)a}=C_{(1)ab}Q^{b}-\frac{1}{2}L_{(1)b}A_{a}^{b}-\frac{1}{2}\left(
L_{(0)b}Q^{b}\right) _{,a}.
\end{equation*}

- For $n=3$.

We have the QFI
\begin{eqnarray*}
I_{3} &=&\left( \frac{t^{3}}{3}C_{(3)ab}+\frac{t^{2}}{2}%
C_{(2)ab}+tC_{(1)ab}+C_{(0)ab}\right) \dot{q}^{a}\dot{q}^{b}%
+t^{3}L_{(3)a}\dot{q}^{a}+t^{2}L_{(2)a}\dot{q}^{a}+tL_{(1)a}%
\dot{q}^{a}+L_{(0)a}\dot{q}^{a}+ \\
&&+\frac{t^{4}}{4}s+\frac{t^{3}}{3}L_{(2)a}Q^{a}+\frac{t^{2}%
}{2}L_{(1)a}Q^{a}+tL_{(0)a}Q^{a}+G(q)
\end{eqnarray*}%
where $C_{(3)ab}$ is a KT computed from the relations:
\begin{equation*}
C_{(1)ab}=-L_{(0)(a;b)}-2C_{(0)c(a}A_{b)}^{c},\enskip %
C_{(2)ab}=-L_{(1)(a;b)}-2C_{(1)c(a}A_{b)}^{c},\enskip %
C_{(3)ab}=-L_{(2)(a;b)}-C_{(2)c(a}A_{b)}^{c}
\end{equation*}%
while the vector $L_{(3)a}$ and the quantity $G$ are computed
from the
relations:
\begin{equation*}
L_{(3)a}=\frac{1}{3}C_{(2)ab}Q^{b} -\frac{1}{3}L_{(2)b}A_{a}^{b}-\frac{1}{6}%
\left( L_{(1)b}Q^{b}\right) _{,a}
\end{equation*}%
\begin{equation*}
L_{(3)(a;b)}=-\frac{2}{3}C_{(3)c(a}A_{b)}^{c},\enskip
L_{(3)a}Q^{a}=s,\enskip%
\left( L_{(2)b}Q^{b}\right)
_{,a}=2C_{(3)ab}Q^{b}-3L_{(3)b}A_{a}^{b}
\end{equation*}%
\begin{equation*}
L_{(1)a}=2C_{(0)ab}Q^{b}-L_{(0)b}A_{a}^{b}-G_{,a},\enskip %
L_{(2)a}=C_{(1)ab}Q^{b}-\frac{1}{2}L_{(1)b}A_{a}^{b} -\frac{1}{2}\left(L_{(0)b}Q^{b}\right) _{,a}.
\end{equation*}

In a similar manner, we continue for higher values of $n.$ We observe that for all values of $n$, the KTs $C_{(N)ab}$, the vectors $L_{(N)a}$ and, hence, the conditions for $I_{n}$
can be written in terms of the triplet $\{G(q), L_{(0)a}, C_{(0)ab}=KT\}$.

\section{Applications}

\label{applications}

In this section, we discuss various applications of Theorem \ref{Theorem2}.

\subsection{The problem of geodesics}

\label{sec.geodesics}

We apply Theorem \ref{Theorem2} to the geodesic equations\index{Equations! geodesic} in order to recover the results of \cite{Katzin 1981} in a simple and straightforward manner. In that case, $Q^{a}=0$ and $A_{b}^{a}=0$, and the conditions of the QFI \textbf{Integral 1} imply that $I_{n>2}=0$. Therefore, the only QFI which survives is the
\begin{equation*}
I_{2}=\left( \frac{t^{2}}{2}G_{;ab}-tL_{(0)(a;b)}+C_{(0)ab}\right) \dot{q}%
^{a}\dot{q}^{b}-tG_{,a}\dot{q}^{a}+L_{(0)a}\dot{q}^{a}+G(q)
\end{equation*}%
where $C_{(0)ab}$, $G_{;ab}$ and $L_{(0)(a;b)}$ are KTs.

The QFI $I_{2}$ consists of the three independent QFIs\footnote{We ignore the index $(0)$ in order to simplify the notation.} (see also Table\footnote{In Table \ref{Table.QFIs.geodesics}, we found the same results by applying Theorem \ref{The first integrals of an autonomous holonomic dynamical system}.} \ref{Table.QFIs.geodesics}):
\begin{equation*}
I_{2a}=C_{ab}\dot{q}^{a}\dot{q}^{b},\enskip I_{2b}=\frac{t^{2}}{2}G_{;ab}%
\dot{q}^{a}\dot{q}^{b}-tG_{,a}\dot{q}^{a}+G(q),\enskip I_{2c}=-tL_{(a;b)}%
\dot{q}^{a}\dot{q}^{b}+L_{a}\dot{q}^{a}.
\end{equation*}
The time-dependent QFIs $I_{2b}$ and $I_{2c}$ are the ones found in \cite{Katzin 1981}.  The QFI $I_{2a}$ is not found because the authors were looking only for time-dependent FIs.
\bigskip

As an application of the above general results, let us compute the QFIs of the geodesic equations of the 3d metric\footnote{
If we set $z=it$,  the line element (\ref{eq.exa4.1}) takes the form
$ds^{2}= -dt^{2} -t^{2} \left( dx^{2} + dy^{2} \right)$,
which is a conformally flat spacetime.}
\begin{equation}
ds^{2}=z^{2}\left( dx^{2}+dy^{2}\right) +dz^{2}. \label{eq.exa4.1}
\end{equation}

In this case, the kinetic metric is  $g_{ab}=diag(z^{2},z^{2},1)$ and the Ricci Scalar $R=-\frac{2}{z^{2}}$. Therefore, this metric is not of constant curvature, and consequently, the number of KTs is less than twenty (see Proposition \ref{pro.KT.1}).

The geodesic equations are:
\begin{eqnarray}
\ddot{x} &=& -\frac{2}{z} \dot{x}\dot{z} \label{eq.exa4.2a} \\
\ddot{y} &=& -\frac{2}{z} \dot{y}\dot{z} \label{eq.exa4.2b} \\
\ddot{z} &=& z(\dot{x}^{2}+\dot{y}^{2}). \label{eq.exa4.2c}
\end{eqnarray}

Solving the condition $C_{(ab;c)}=0$, we find that the metric (\ref{eq.exa4.1}) admits the following KTs:
\begin{equation*}
C_{ab}=
\left(
  \begin{array}{ccc}
    \left( \frac{c_{1}}{z^{2}} +\frac{c_{2}}{2}y^{2} +c_{3}y +c_{4} \right) z^{4} & -\frac{1}{2}\left( c_{2}xy +c_{3}x +c_{5}y -2c_{7} \right)z^{4} & 0 \\
    -\frac{1}{2}\left( c_{2}xy +c_{3}x +c_{5}y -2c_{7} \right)z^{4} & \left( \frac{c_{1}}{z^{2}} +\frac{c_{2}}{2}x^{2} +c_{5}x +c_{6} \right)z^{4} & 0 \\
    0 & 0 & c_{1} \\
  \end{array}
\right)
\end{equation*}
where $c_{\kappa}$, $\kappa=1,2,...,7$, are arbitrary constants. Therefore, there exist seven linearly independent KTs as many as the free parameters involved.

In order to find the reducible KTs, we solve the constraint $C_{ab}=L_{(a;b)}$ for a vector $L_{a}(x,y,z)$. We have the following system of equations:
\begin{eqnarray*}
L_{1,1}+zL_{3} &=& c_{1}z^{2} +\frac{c_{2}}{2}y^{2}z^{4} +c_{3}yz^{4} +c_{4}z^{4} \\
L_{1,2} + L_{2,1} &=& -c_{2}xyz^{4} -c_{3}xz^{4} -c_{5}yz^{4} +2c_{7}z^{4} \\
zL_{1,3} + zL_{3,1} - 2L_{1} &=& 0 \\
L_{2,2}+zL_{3} &=& c_{1}z^{2} +\frac{c_{2}}{2}x^{2}z^{4} +c_{5}xz^{4} +c_{6}z^{4} \\
zL_{2,3} + zL_{3,2} -2L_{2} &=& 0 \\
L_{3,3} &=& c_{1}.
\end{eqnarray*}

The solution of the above system is the vector $L_{a} =
\left(
  \begin{array}{c}
    z^{2}(b_{1}y+b_{2}) \\
    -z^{2}(b_{1}x+b_{3}) \\
    c_{1}z \\
  \end{array}
\right)$. We compute $L_{(a;b)}= c_{1}g_{ab}$, that is, $L_a$ is a HV with homothetic factor $c_{1}$.

In the case that the generating vector $L_{a}=G_{,a}$ (i.e. gradient), we find that $b_{1}=b_{2}=b_{3}=0$ and $G= \frac{c_{1}}{2}z^{2}$. Then, we have $G_{,a}=
\left(
  \begin{array}{c}
    0 \\
    0 \\
    c_{1}z \\
  \end{array}
\right)$ and $G_{;ab}= c_{1}g_{ab}$.

In order to compute the QFIs for the geodesic equations of the metric (\ref{eq.exa4.1}), we apply the results of Table \ref{Table.QFIs.geodesics}. We have the following:
\bigskip

1) The QFI $I_{2a}$:
\begin{eqnarray*}
I_{2a} &=& C_{ab}\dot{q}^{a}\dot{q}^{b} \\
&=& \left( \frac{c_{1}}{z^{2}} + \frac{c_{2}}{2}y^{2} +c_{3}y +c_{4} \right)z^{4}\dot{x}^{2} - \left( c_{2}xy +c_{3}x +c_{5}y -2c_{7} \right)z^{4}\dot{x}\dot{y} + \left( \frac{c_{1}}{z^{2}} +\frac{c_{2}}{2}x^{2} +c_{5}x +c_{6} \right)z^{4} \dot{y}^{2} +c_{1}\dot{z}^{2} \\
&=& 2c_{1} \underbrace{\frac{1}{2} \left(z^{2}\dot{x}^{2} + z^{2}\dot{y}^{2} + \dot{z}^{2} \right)}_{=\text{kinetic energy}} - \frac{c_{2}}{2} z^{4} \left( x\dot{y} - y\dot{x} \right)^{2} + c_{3}z^{4}\dot{x} \left( x\dot{y} - y\dot{x} \right) + c_{4}z^{4}\dot{x}^{2} -c_{5}z^{4}\dot{y} \left( x\dot{y} - y\dot{x} \right) + \\
&& +c_{6}z^{4}\dot{y}^{2} + 2c_{7}z^{4}\dot{x}\dot{y}.
\end{eqnarray*}
This expression contains the independent FIs:
\[
T= \frac{1}{2} \left(z^{2}\dot{x}^{2} + z^{2}\dot{y}^{2} + \dot{z}^{2} \right), \enskip I_{2a1}=z^{2}\dot{x}, \enskip I_{2a2}=z^{2}\dot{y}, \enskip I_{2a3}= z^{2}\left( x\dot{y} - y\dot{x} \right).
\]
We note that $T = \frac{1}{2} \left( \dot{x}I_{2a1} + \dot{y}I_{2a2} + \dot{z}^{2} \right)$ and $I_{2a3}= xI_{2a2} - yI_{2a1}$.

2) The QFI $I_{2b}$:
\begin{equation*}
I_{2b}= \frac{t^{2}}{2}G_{;ab}\dot{q}^{a}\dot{q}^{b} -tG_{,a}\dot{q}^{a} + G(q)= c_{1}\frac{t^{2}}{2}\left( z^{2}\dot{x}^{2} + z^{2}\dot{y}^{2} + \dot{z}^{2} \right) -c_{1}tz\dot{z} + \frac{c_{1}}{2}z^{2}
\end{equation*}
which gives the QFI $I_{2b}= -t^{2}T +tz\dot{z} -\frac{z^{2}}{2}$.

3) The QFI $I_{2c}$:
\begin{equation*}
I_{2c} = -tL_{(a;b)}\dot{q}^{a}\dot{q}^{b} + L_{a}\dot{q}^{a} = -c_{1}t\left( z^{2}\dot{x}^{2} + z^{2}\dot{y}^{2} + \dot{z}^{2} \right) + z^{2}(b_{1}y+b_{2})\dot{x} -z^{2}(b_{1}x+b_{3})\dot{y} +c_{1}z\dot{z}.
\end{equation*}
This QFI contains the new irreducible QFI $I_{2c1}=-tT + \frac{z\dot{z}}{2} = \frac{1}{2} \frac{d}{dt} \left(-t^{2}T + \frac{z^{2}}{2} \right)$. We observe that $I_{2b}= tI_{2c1} + \frac{tz\dot{z}}{2} - \frac{z^{2}}{2}$.
\bigskip

We collect the above results in Table \ref{Table.geod.exa1}.

\begin{longtable}{|l|}
\hline
$T= \frac{1}{2} \left(z^{2}\dot{x}^{2} + z^{2}\dot{y}^{2} + \dot{z}^{2} \right)$, \enskip $I_{2a1}=z^{2}\dot{x}$, \enskip $I_{2a2}=z^{2}\dot{y}$, \enskip $I_{2a3}= xI_{2a2} -yI_{2a1}$ \\
$I_{2c}=-tT + \frac{z\dot{z}}{2}$, \enskip $I_{2b}= -t^{2}T +tz\dot{z} -\frac{z^{2}}{2}$ \\ \hline
\caption{\label{Table.geod.exa1} The LFIs/QFIs of geodesics of (\ref{eq.exa4.1}).}
\end{longtable}

Since the metric $g_{ab}$ is not flat, the conjugate momenta $p_{a}$ of the Hamiltonian formalism are not equal to the velocities $\dot{q}^{a}$. Hence, to compute the PBs of the FIs, we have to make the required transformation.

The conjugate momenta\index{Momentum! conjugate} are $p_{a} \equiv \frac{\partial T}{\partial \dot{q}^{a}} = g_{ab} \dot{q}^{b} =
\left(
  \begin{array}{c}
    z^{2}\dot{x} \\
    z^{2}\dot{y} \\
    \dot{z} \\
  \end{array}
\right)$. Then, the FIs take the form shown in Table \ref{Table.geod.exa2}.

\begin{longtable}{|l|}
\hline
$T= \frac{1}{2} \left( \frac{p_{1}^{2}}{z^{2}} + \frac{p_{2}^{2}}{z^{2}} + p_{3}^{2} \right)$, \enskip $I_{2a1}=p_{1}$, \enskip $I_{2a2}=p_{2}$, \enskip $I_{2a3}= xp_{2} -yp_{1}$ \\
$I_{2c}=-tT + \frac{zp_{3}}{2}$, \enskip $I_{2b}= -t^{2}T +tzp_{3} -\frac{z^{2}}{2}$ \\
\hline
\caption{\label{Table.geod.exa2} The LFIs/QFIs of geodesics of (\ref{eq.exa4.1}) in the phase space $(q^{a}, p_{a})$.}
\end{longtable}

We compute the PBs: $\{T, I_{2a1}\}= \{T, I_{2a2}\}= \{T, I_{2a3}\}= 0$, $\{T, I_{2b}\}= \frac{\partial I_{2b}}{\partial t}$, $\{T, I_{2c}\}= \frac{\partial I_{2c}}{\partial t}$ and $\{I_{2a1}, I_{2a2}\}=0$.

The system is (Liouville) integrable because the three FIs $T, I_{2a1}, I_{2a2}$ are linearly independent and in involution. Therefore, we can find the solution of the system by quadrature using these FIs. However, it is simpler to use instead of $T$ the time-dependent FI $I_{2c}$. Indeed, we have:
\[
\begin{cases}
z^{2}\dot{x} = k_{1} \\
z^{2}\dot{y} = k_{2} \\
2z\dot{z} = 2k_{4}t + k_{3}
\end{cases}
\implies
\frac{dz^{2}}{dt}= 2k_{4}t + k_{3}  \implies z(t)= \pm \left( k_{4}t^{2} + k_{3}t + k_{0} \right)^{1/2}
\]
where $k_{0}, k_{1}\equiv I_{2a1}, k_{2}\equiv I_{2a2}, k_{3}\equiv 4I_{2c}$ and $k_{4}\equiv 2T$ are arbitrary constants.

Substituting the solution $z(t)$ in the two remaining FIs, we find:
\[
\dot{x}=\frac{k_{1}}{z^{2}} \implies x(t)= \frac{2k_{1}}{(4k_{0}k_{4} - k_{3}^{2})^{1/2}} \tan^{-1} \left[ \frac{2k_{4}t + k_{3}}{(4k_{0}k_{4} - k_{3}^{2})^{1/2}} \right] +c
\]
and
\[
\dot{y}=\frac{k_{2}}{z^{2}} \implies y(t)= \frac{2k_{2}}{(4k_{0}k_{4} - k_{3}^{2})^{1/2}} \tan^{-1} \left[ \frac{2k_{4}t + k_{3}}{(4k_{0}k_{4} - k_{3}^{2})^{1/2}} \right] +c'
\]
where $c, c'$ are arbitrary constants.

The geodesic curves for the metric (\ref{eq.exa4.1}) are
\begin{equation}
q^{a}(t)=
\left(
  \begin{array}{c}
    x(t) \\
    y(t) \\
    z(t) \\
  \end{array}
\right) =
\left(
  \begin{array}{c}
    \frac{2k_{1}}{(4k_{0}k_{4} - k_{3}^{2})^{1/2}} \tan^{-1} \left[ \frac{2k_{4}t + k_{3}}{(4k_{0}k_{4} - k_{3}^{2})^{1/2}} \right] +c \\
    \frac{2k_{2}}{(4k_{0}k_{4} - k_{3}^{2})^{1/2}} \tan^{-1} \left[ \frac{2k_{4}t + k_{3}}{(4k_{0}k_{4} - k_{3}^{2})^{1/2}} \right] +c' \\
    \pm \left( k_{4}t^{2} + k_{3}t + k_{0} \right)^{1/2} \\
  \end{array}
\right). \label{eq.sol.geod}
\end{equation}

\subsection{The Whittaker dynamical system}

\label{sec.Whittaker}

The Whittaker dynamical system is a 2d Newtonian system with dynamical equations:\index{Dynamical system! Whittaker}
\begin{equation*}
\ddot{x}=x, \enskip \ddot{y}=\dot{x}.
\end{equation*}
For that system the kinetic metric is the Euclidean metric $\delta_{ab}$ of $E^{2}$.

In the notation of Theorem \ref{Theorem2}, we have $A_{b}^{a}=\delta
_{2}^{a}\delta _{b}^{1}=\left(
\begin{array}{cc}
0 & 0 \\
1 & 0%
\end{array}%
\right)$ and $Q^{a}=V^{,a}= -x\delta _{1}^{a}=\left(
\begin{array}{c}
-x \\
0%
\end{array}%
\right)$, where $V=-\frac{1}{2}x^{2}$.

We apply Theorem \ref{Theorem2} to determine the QFIs\footnote{We use the geometric quantities of section \ref{sec.KTE2}.}.
\bigskip

\textbf{Integral 1.}
\begin{eqnarray*}
I_{n} &=& \left( \frac{t^{n}}{n} C_{(n)ab} + ... + \frac{t^{2}}{2} C_{(2)ab}
+ t C_{(1)ab} + C_{(0)ab} \right) \dot{q}^{a} \dot{q}^{b} + t^{n} L_{(n)a}\dot{q}^{a} + ... + t^{2}L_{(2)a}\dot{q}^{a} + t L_{(1)a}\dot{q}^{a} + \\
&& + L_{(0)a}\dot{q}^{a} + \frac{t^{n+1}}{n+1} L_{(n)a}Q^{a} + ... + \frac{t^{2}}{2} L_{(1)a}Q^{a}
+t L_{(0)a}Q^{a} + G(q)
\end{eqnarray*}
where $C_{(N)ab}$ are KTs. Taking into consideration the quantities mentioned above, we find:
\begin{equation*}
C_{(1)11} = -L_{(0)(1;1)} - 2C_{(0)12}, \enskip C_{(1)12} = -L_{(0)(1;2)} -
C_{(0)22}, \enskip C_{(1)22} = -L_{(0)(2;2)}
\end{equation*}
\begin{equation*}
C_{(k+1)11} = -L_{(k)(1;1)} - 2\frac{1}{k}C_{(k)12}, \enskip C_{(k+1)12} =
-L_{(k)(1;2)} - \frac{1}{k}C_{(k)22}, \enskip C_{(k+1)22} = -L_{(k)(2;2)}
\end{equation*}
\begin{equation*}
\left(-xL_{(k-1)1}\right)_{,1} = -2xC_{(k)11} - k(k+1)L_{(k+1)1} -kL_{(k)2}, %
\enskip \left(-xL_{(k-1)1}\right)_{,2} = -2xC_{(k)12} - k(k+1)L_{(k+1)2}
\end{equation*}
\begin{equation*}
L_{(n)(1;1)} =- \frac{2}{n}C_{(n)12}, \enskip L_{(n)(1;2)} = -\frac{1}{n}
C_{(n)22}, \enskip L_{(n)(2;2)} =0
\end{equation*}
\begin{equation*}
-xL_{(n)1}=s, \enskip \left(-xL_{(n-1)1}\right)_{,1} = -2xC_{(n)11} -
nL_{(n)2}, \enskip \left(-xL_{(n-1)1}\right)_{,2} = -2xC_{(n)12}
\end{equation*}
\begin{equation*}
G_{,1}= -2xC_{(0)11} - L_{(1)1} -L_{(0)2}, \enskip G_{,2}= -2xC_{(0)12} -L_{(1)2}
\end{equation*}
where $k=1,...,n-1$. We note that all the QFIs $I_{n}(n>1)$ reduce to the QFI $I_{1}$. Therefore, we continue only with the case $n=1$. We have the QFI
\begin{equation*}
I_{1}=\left( tD_{ab}+C_{ab}\right) \dot{q}^{a}\dot{q}^{b}+tL_{a}\dot{q}%
^{a}+B_{a}\dot{q}^{a}+\frac{t^{2}}{2}s+tB_{a}Q^{a}+G(q)
\end{equation*}%
where $C_{ab}, D_{ab}$ are KTs and the following conditions are satisfied:
\begin{equation}
D_{11}=-B_{(1;1)}-2C_{12},\enskip D_{12}=-B_{(1;2)}-C_{22},\enskip %
D_{22}=-B_{(2;2)}  \label{eq.wh1}
\end{equation}%
\begin{equation}
L_{(1;1)}=-2D_{12},\enskip L_{(1;2)}=-D_{22},\enskip L_{(2;2)}=0
\label{eq.wh2}
\end{equation}%
\begin{equation}
-xL_{1}=s  \label{eq.wh3}
\end{equation}%
\begin{equation}
\left( -xB_{1}\right) _{,1}=-2xD_{11}-L_{2},\enskip\left( -xB_{1}\right)
_{,2}=-2xD_{12}  \label{eq.wh4}
\end{equation}%
\begin{equation}
G_{,1}=-2xC_{11}-L_{1}-B_{2},\enskip G_{,2}=-2xC_{12}-L_{2}.  \label{eq.wh5}
\end{equation}

The KTs $C_{ab}$ and $D_{ab}$ are of the form (see section \ref{sec.KTE2.1}):
\begin{equation*}
C_{ab}=\left(
\begin{array}{cc}
\gamma _{0}y^{2}+2a_{0}y+A_{0} & -\gamma _{0}xy-a_{0}x-\beta _{0}y+C_{0} \\
-\gamma _{0}xy-a_{0}x-\beta _{0}y+C_{0} & \gamma _{0}x^{2}+2\beta _{0}x+E_{0}%
\end{array}%
\right)
\end{equation*}%
and
\begin{equation*}
D_{ab}=\left(
\begin{array}{cc}
\gamma _{1}y^{2}+2a_{1}y+A_{1} & -\gamma _{1}xy-a_{1}x-\beta _{1}y+C_{1} \\
-\gamma _{1}xy-a_{1}x-\beta _{1}y+C_{1} & \gamma _{1}x^{2}+2\beta _{1}x+E_{1}%
\end{array}%
\right) .
\end{equation*}

Solving conditions (\ref{eq.wh2}), we find that
\begin{equation*}
L_{a}= \left(
\begin{array}{c}
\gamma_{1}x^{2}y+ a_{1}x^{2} + 2\beta_{1}xy - 2C_{1}x + k_{1} \\
-\gamma_{1}x^{3} - 3\beta_{1}x^{2} - 2E_{1}x + k_{2} \\
\end{array}
\right).
\end{equation*}
Substituting the last vector in (\ref{eq.wh3}), we get $a_{1}=\beta_{1}= \gamma_{1}= C_{}=
k_{1}=0$ and $s=0$. Therefore,
\begin{equation*}
L_{a}= \left(
\begin{array}{c}
0 \\
-2E_{1}x + k_{2} \\
\end{array}
\right) \enskip \text{and} \enskip D_{ab}= \left(
\begin{array}{cc}
A_{1} & 0 \\
0 & E_{1}%
\end{array}%
\right).
\end{equation*}

Solving conditions (\ref{eq.wh1}), we find
\begin{equation*}
B_{a}= \left(
\begin{array}{c}
\gamma_{0}x^{2}y + 2\beta_{0}xy + a_{0}x^{2} - (A_{1}+2C_{0})x + k_{3} \\
-\gamma_{0}x^{3} - 3\beta_{0}x^{2} - 2E_{0}x - E_{1}y + k_{4} \\
\end{array}
\right)
\end{equation*}
which when replaced in (\ref{eq.wh4}) gives $a_{0}=\beta_{0}=\gamma_{0}=0$, $k_{2}=k_{3}$
and $E_{1}=2(A_{1}+C_{0})$. It follows that
\begin{equation*}
B_{a}= \left(
\begin{array}{c}
- (A_{1}+2C_{0})x + k_{2} \\
- 2E_{0}x - 2(C_{0}+A_{1})y + k_{4} \\
\end{array}
\right) \enskip \text{and} \enskip C_{ab} = \left(
\begin{array}{cc}
A_{0} & C_{0} \\
C_{0} & E_{0}%
\end{array}%
\right).
\end{equation*}

Substituting the above results in the integrability condition of (\ref{eq.wh5}), we find $A_{1}=0$ $\implies E_{1}=2C_{0}$. Therefore,
\begin{equation*}
L_{a}= \left(
\begin{array}{c}
0 \\
-4C_{0}x + k_{2} \\
\end{array}
\right), \enskip D_{ab}= \left(
\begin{array}{cc}
0 & 0 \\
0 & 2C_{0}%
\end{array}%
\right), \enskip B_{a}= \left(
\begin{array}{c}
- 2C_{0}x + k_{2} \\
- 2E_{0}x - 2C_{0}y + k_{4} \\
\end{array}
\right), \enskip C_{ab} = \left(
\begin{array}{cc}
A_{0} & C_{0} \\
C_{0} & E_{0}%
\end{array}%
\right).
\end{equation*}
Finally, integrating conditions (\ref{eq.wh5}), we find $G(x,y)= (E_{0}-A_{0})x^{2} + 2C_{0}xy - k_{4}x - k_{2}y$.

The QFI is
\begin{eqnarray*}
J_{1} &=& 2tC_{0}\dot{y}^{2} + A_{0}\dot{x}^{2} + 2C_{0}\dot{x} \dot{y} +
E_{0}\dot{y}^{2} - 4tC_{0}x\dot{y} + tk_{2}\dot{y} - 2C_{0}x\dot{x} + k_{2}%
\dot{x} - 2E_{0}x\dot{y} - 2C_{0}y\dot{y} + k_{4}\dot{y} + \\
&& + 2tC_{0}x^{2} - tk_{2}x + E_{0}x^{2}-A_{0}x^{2} + 2C_{0}xy - k_{4}x -k_{2}y
\end{eqnarray*}
which consists of the FIs: $J_{1a}= (\dot{y} - x) \left[ t(\dot{y}-x) + \dot{x}-y \right]$, $J_{1b}= \dot{x}^{2} -x^{2}$, $J_{1c}= (\dot{y} - x)^{2}$, $J_{1d}= t(\dot{y}-x) + \dot{x} -y$ and $J_{1e}= \dot{y}-x$.

The independent FIs are the following:
\begin{equation*}
J_{11}= \dot{x}^{2}-x^{2}, \enskip J_{12}=\dot{y}-x, \enskip J_{13}= t(\dot{y%
}-x) + \dot{x} - y.
\end{equation*}
\bigskip

\textbf{Integral 2.}
\begin{equation*}
J_{2}= e^{\lambda t} \left( \lambda C_{ab} \dot{q}^{a} \dot{q}^{b} + \lambda
L_{a}\dot{q}^{a} + L_{a}Q^{a} \right)
\end{equation*}
where $\lambda \neq 0$, $C_{ab}$ is a KT, $\lambda C_{11} = - L_{(1;1)} -
2C_{12}$, $\lambda C_{12}= - L_{(1;2)} - C_{22}$, $\lambda C_{22}= -
L_{(2;2)}$ and $\left(-xL_{1}\right)_{,a} = -2\lambda xC_{1a} -
\lambda^{2}L_{a} - \lambda L_{2} \delta^{1}_{a}$.

We have the following conditions:
\begin{eqnarray}
L_{1,1} &=& - \lambda C_{11} - 2C_{12} \label{Wh9} \\
L_{1,2} + L_{2,1} &=& - 2\lambda C_{12} - 2C_{22} \label{Wh10} \\
L_{2,2} &=& - \lambda C_{22}  \label{Wh11} \\
\left(-xL_{1}\right)_{,a} &=& -2\lambda xC_{1a} - \lambda^{2} L_{a} -\lambda L_{2} \delta^{1}_{a}.  \label{Wh12}
\end{eqnarray}

We recall that the KT $C_{ab}=\left(
\begin{array}{cc}
2ay+A & -ax-\beta y+C \\
-ax-\beta y+C & 2\beta x+B%
\end{array}%
\right)$,

Solving the system of PDEs (\ref{Wh9}) - (\ref{Wh11}), we find that
\begin{equation*}
L_{a}=\left(
\begin{array}{c}
ax^{2}+2\lambda \beta y^{2}+2(\beta -\lambda a)xy+k_{1}y-(\lambda
A+2C)x+k_{2} \\
(2\lambda a-3\beta )x^{2}-2\lambda \beta xy-\lambda By-(2\lambda
C+2B+k_{1})x+k_{3} \\
\end{array}%
\right).
\end{equation*}

Substituting the above quantities in the remaining condition (\ref{Wh12}), we get $a=\beta =B=k_{3}=0.$ We consider the following subcases:
\bigskip

i) Case $\lambda =\pm 1$.

We find $A=k_{1}=0$. Then, $L_{a}=\left(
\begin{array}{c}
-2Cx+k_{2} \\
\mp 2Cx%
\end{array}%
\right)$ and $C_{ab}=C\left(
\begin{array}{cc}
0 & 1 \\
1 & 0%
\end{array}%
\right)$.

The QFI is
\begin{equation*}
J_{2a}= e^{\pm t} \left[ \pm 2C\dot{x}\dot{y} \pm (-2Cx + k_{2})\dot{x} - 2Cx%
\dot{y} + 2Cx^{2} - k_{2}x \right]
\end{equation*}
which contains the FIs: $J_{21} = e^{\pm t} (\dot{x} \mp x)$ and $J_{21b}= e^{\pm t} (\dot{y}-x)(\dot{x} \mp x) = J_{21} J_{12}$.

We note that the QFI $J_{11}$ can be expressed in terms of the LFIs $J_{21\pm}$ as follows:
\begin{equation*}
J_{21+} J_{21-}= e^{t}(\dot{x} - x) e^{-t}(\dot{x}+x) = \dot{x}^{2} - x^{2}
= J_{11}
\end{equation*}
that is, $J_{11}$ is not an independent FI.
\bigskip

ii) Case $\lambda =\pm 2$.

We find $C=k_{1}=k_{2}=0$. Then,
$L_{a}=\left(
\begin{array}{c}
\mp 2Ax \\
0%
\end{array}%
\right)$ and $C_{ab}=A\left(
\begin{array}{cc}
1 & 0 \\
0 & 0%
\end{array}%
\right)$. The resulting QFI gives already known FIs.
\bigskip

We collect the above results in Table \ref{Table.Whittaker}.

\begin{longtable}{|c|}
\hline
\multicolumn{1}{|l|}{$J_{12}=\dot{y}-x$} \\
\multicolumn{1}{|l|}{$J_{13}=t(\dot{y}-x)+\dot{x}-y=tJ_{12} +\dot{x}-y$} \\
\multicolumn{1}{|l|}{$J_{21\pm}=e^{\pm t}(\dot{x}\mp x)$} \\ \hline
\caption{\label{Table.Whittaker} The LFIs/QFIs of the Whittaker system.}
\end{longtable}

In order to study the integrability of the Whittaker system, we compute the PBs of the independent FIs. We have:
\[
\{J_{12},J_{13}\}=0, \enskip \{J_{12},J_{21\pm}\}= -e^{\pm t}, \enskip \{J_{13},J_{21\pm}\}= -e^{\pm t}(t\mp1), \enskip \{J_{21+},J_{21-}\}=-2.
\]
Therefore, the 2d Whittaker system is integrable because the FIs $J_{12}$ and $J_{13}$ are (functionally) independent and in involution. However, the solution of the system can be found immediately by using $J_{12}$ and, instead of $J_{13}$, the time-dependent FIs $J_{21\pm}$. It follows that:
\begin{equation}
x(t) = \frac{1}{2}(c_{-}e^{t} - c_{+}e^{-t}), \enskip y(t)= c_{0}t + \frac{1%
}{2}(c_{-}e^{t} + c_{+}e^{-t}) + c_{1}
\end{equation}
where $c_{\pm}$, $c_{0}$ and $c_{1}$ are arbitrary constants.

\subsection{Two autonomous linearly coupled damped harmonic oscillators}

\label{sec.damped.oscillator}

This is the 2d dynamical system with
equations of motion:\index{Oscillator! coupled harmonic}
\begin{eqnarray}
\ddot{x} + kx &=& py - 2m\dot{x}  \label{eq.Dj.1} \\
\ddot{y} + ky &=& - px - 2m\dot{y}  \label{eq.Dj.2}
\end{eqnarray}
where $m$, $p$, $k$ are (real or imaginary, non-zero) constants and $q^1=x$, $q^2=y$. The determination of the QFIs of this system have been discussed before (see example 6.5 in \cite{Djukic 1975}), where it has been found one new time-dependent QFI by giving arbitrary values to the quantities involved in the weak Noether condition (equivalently the NBH equation). Using Theorem \ref{Theorem2}, we shall recover this QFI plus a plethora of new QFIs not found before.

A Lagrangian that describes this system is the Lagrangian of the 2d simple harmonic oscillator
\begin{equation}
L=T-V=\frac{1}{2}\left( \dot{x}^{2}+\dot{y}^{2}\right) -\frac{1}{2}k\left(x^{2}+y^{2}\right)  \label{eq.Dj.3}
\end{equation}%
with external generalized forces
\begin{equation}
F^{a}=-P^{a}+A_{b}^{a}\dot{q}^{b}  \label{eq.Dj.4}
\end{equation}%
where $P^{a}=\left(
\begin{array}{c}
-py \\
px%
\end{array}%
\right)$ and $A_{b}^{a}=-2m\delta _{b}^{a}$.

We observe that the dynamical equations (\ref{eq.Dj.1}) and (\ref{eq.Dj.2}) are of the general form
\begin{equation}
\ddot{q}^{a}=-Q^{a}+A_{b}^{a}\dot{q}^{b}  \label{eq.Dj.5}
\end{equation}%
where $Q^{a}=V^{,a}+P^{a}=\left(
\begin{array}{c}
kx-py \\
ky+px%
\end{array}%
\right)$ and the kinetic metric is the Euclidean metric $\delta_{ab}$ of $E^{2}$.

We apply Theorem \ref{Theorem2} to determine the QFIs of that system\footnote{We use the geometric quantities of section \ref{sec.KTE2}.}.
\bigskip

\textbf{Integral 1.}

The conditions of the QFI $I_{n}$ become:
\begin{eqnarray}
C_{(1)ab} &=& -L_{(0)(a;b)} +4mC_{(0)ab}  \label{eq.dam1} \\
C_{(k+1)ab} &=& -L_{(k)(a;b)} + \frac{4m}{k}C_{(k)ab}, \enskip k=1,...,n-1
\label{eq.dam2} \\
\left(L_{(k-1)b} Q^{b} \right)_{,a} &=& 2C_{(k)ab}Q^{b} - k(k+1)L_{(k+1)a}
+2mkL_{(k)a}, \enskip k=1,...,n-1  \label{eq.dam3} \\
L_{(n)(a;b)} &=& \frac{4m}{n} C_{(n)ab}  \label{eq.dam4} \\
L_{(n)a}Q^{a}&=&s  \label{eq.dam5} \\
\left(L_{(n-1)b} Q^{b} \right)_{,a} &=& 2C_{(n)ab}Q^{b} +2mnL_{(n)a}
\label{eq.dam6} \\
G_{,a}&=& 2C_{(0)ab}Q^{b} - L_{(1)a} + 2mL_{(0)a}.  \label{eq.dam7}
\end{eqnarray}

Since $C_{(N)ab}=0$, $L_{(N)a}=0$ for $N=2,3,...,n$, it follows that only the QFI $I_{1}$ survives. Therefore\footnote{
For simplicity, we set $C_{(0)ab}\equiv C_{ab}$, $L_{(0)a}\equiv B_{a},$ $%
C_{(1)ab}\equiv D_{ab}$ and $L_{(1)a}\equiv L_{a}$.},
\begin{equation*}
I_{1}=\left( tD_{ab}+C_{ab}\right) \dot{q}^{a}\dot{q}^{b}+tL_{a}\dot{q}%
^{a}+B_{a}\dot{q}^{a}+\frac{t^{2}}{2}s+tB_{a}Q^{a}+G(q)
\end{equation*}%
where $C_{ab}=\frac{1}{4m}B_{(a;b)}+\frac{1}{16m^{2}}L_{(a;b)}$ and $D_{ab}=\frac{1}{4m}L_{(a;b)}$ are KTs such that:
\begin{eqnarray}
L_{a}Q^{a} &=&s  \label{eq.DH1} \\
\left( B_{b}Q^{b}\right) _{,a} &=&2D_{ab}Q^{b}+2mL_{a}  \label{eq.DH2} \\
G_{,a} &=&2C_{ab}Q^{b}+2mB_{a}-L_{a}.  \label{eq.DH3}
\end{eqnarray}

Since $C_{ab}$ and $D_{ab}$ are KTs, the quantities $B_{(a;b)}$ and $L_{(a;b)}$ are reducible KTs. Therefore, from section \ref{sec.KTE2.1}, we write the following:
\begin{equation*}
B_{a}=\left(
\begin{array}{c}
-2\beta_{1} y^{2}+2a_{1}xy+A_{1}x+n_{8}y+n_{11} \\
-2a_{1}x^{2}+2\beta_{1} xy+n_{10}x+B_{1}y+n_{9}%
\end{array}
\right), \enskip B_{(a;b)}= \left(
\begin{array}{cc}
2a_{1}y+A_{1} & -a_{1}x-\beta_{1} y+C_{1} \\
-a_{1}x-\beta_{1} y+C_{1} & 2\beta_{1} x+B_{1}%
\end{array}
\right)
\end{equation*}
\begin{equation*}
L_{a}=\left(
\begin{array}{c}
-2\beta_{2} y^{2}+2a_{2}xy+A_{2}x+w_{8}y+w_{11} \\
-2a_{2}x^{2}+2\beta_{2} xy+w_{10}x+B_{2}y+w_{9}%
\end{array}%
\right), \enskip L_{(a;b)}= \left(
\begin{array}{cc}
2a_{2}y+A_{2} & -a_{2}x-\beta_{2} y+C_{2} \\
-a_{2}x-\beta_{2} y+C_{2} & 2\beta_{2} x+B_{2}%
\end{array}
\right)
\end{equation*}
where $2C_{1}=n_{8}+n_{10}$ and $2C_{2}=w_{8}+w_{10}$.

Substituting $L_{a}$ in (\ref{eq.DH1}), we obtain \underline{$a_{2}=\beta _{2}=s=0$}, and there remain the following two cases: \newline
1) $k=\pm ip$ with $w_{9}=\mp i w_{11}$, $w_{8}=\pm iB_{2},$ $w_{10}=\mp
iA_{2}$; and \newline
2) $w_{9}=w_{11}=0$, $A_{2}=B_{2},$ $w_{8}= -w_{10}=\frac{k}{p}A_{2}$.

We continue the consideration of the remaining conditions for these two cases.
\bigskip

1) Case $k=\pm ip$ with $w_{9}=\mp i w_{11}$, $w_{8}=\pm i B_{2}$ and $%
w_{10}=\mp i A_{2}$.
\bigskip

1.1. The subcase $k=ip$.

Then, $L_{a}=\left(
\begin{array}{c}
A_{2}x+iB_{2}y+w_{11} \\
-iA_{2}x+B_{2}y-iw_{11}%
\end{array}%
\right)$, $L_{(a;b)}=\left(
\begin{array}{cc}
A_{2} & \frac{i}{2}(B_{2}-A_{2}) \\
\frac{i}{2}(B_{2}-A_{2}) & B_{2}%
\end{array}%
\right)$ and condition (\ref{eq.DH2}) gives
\begin{equation*}
a_{1}=\beta _{1}=0,\enskip(p-4im^{2})A_{2}=0,\enskip A_{2}=B_{2},\enskip %
w_{11}=0,\enskip n_{8}=iB_{1},\enskip n_{10}=-iA_{1},\enskip n_{9}=-in_{11}.
\end{equation*}
Therefore, $B_{a}=\left(
\begin{array}{c}
A_{1}x+iB_{1}y+n_{11} \\
-iA_{1}x+B_{1}y-in_{11}%
\end{array}%
\right)$ and $B_{(a;b)}=\left(
\begin{array}{cc}
A_{1} & \frac{i}{2}(B_{1}-A_{1}) \\
\frac{i}{2}(B_{1}-A_{1}) & B_{1}%
\end{array}%
\right)$

The condition $(p-4im^{2})A_{2}=0$ implies the following subcases:

1.1.1. Subcase $A_{2}=B_{2}=0$ $\implies L_{a}=0$, $D_{ab}=0$ and $C_{ab}=
\frac{1}{4m} B_{(a;b)}$.

From the integrability condition of (\ref{eq.DH3}), we find $(p-
4im^{2})(A_{1}+B_{1})=0$ which gives the following:

1.1.1.A. $A_{1}=-B_{1}$.

We have
\begin{equation*}
B_{a}=\left(
\begin{array}{c}
-B_{1}x+iB_{1}y+n_{11} \\
iB_{1}x+B_{1}y-in_{11}%
\end{array}%
\right) ,\enskip B_{(a;b)}=B_{1}\left(
\begin{array}{cc}
-1 & i \\
i & 1%
\end{array}%
\right) ,\enskip C_{ab}=\frac{B_{1}}{4m}\left(
\begin{array}{cc}
-1 & i \\
i & 1%
\end{array}%
\right).
\end{equation*}

Integrating (\ref{eq.DH3}), we find $G(x,y)= mB_{1}(-x^{2}+y^{2})+2imB_{1}xy+2mn_{11}(x-iy)$.

The QFI is
\begin{eqnarray*}
J_{1}(11a) &=& -\frac{B_{1}}{4m} (\dot{x}^{2} - 2i\dot{x}\dot{y} -\dot{y}%
^{2}) + (-B_{1}x+iB_{1}y+n_{11})\dot{x} + (iB_{1}x+B_{1}y-in_{11})\dot{y} + \\
&& +mB_{1}(-x^{2}+y^{2}) + 2imB_{1}xy + 2mn_{11}(x-iy)
\end{eqnarray*}
which consists of the independent FIs:
\begin{eqnarray*}
J_{1a}(11a) &=& - \frac{1}{4m}(\dot{x} -i\dot{y})^{2} + (-x+iy) \dot{x} +
(ix +y)\dot{y} -m (x-iy)^{2} \equiv J_{11+} \\
J_{1b}(11a) &=& i\dot{x} +\dot{y} + 2m(ix+y) \equiv J_{12+}.
\end{eqnarray*}

1.1.1.B. For $p=4im^{2}$.

Integrating condition (\ref{eq.DH3}), we find $G(x,y)=-miA_{1}xy +miB_{1}xy -\frac{m}{2}(B_{1}-A_{1})(x^{2}-y^{2}) +2mn_{11}(x-iy)$.

The QFI is
\begin{eqnarray*}
J_{1}(11b) &=& \frac{A_{1}}{4m}\dot{x}^{2}-\frac{i}{4m}(A_{1}-B_{1})\dot{x}%
\dot{y} +\frac{B_{1}}{4m}\dot{y}^{2} +(A_{1}x+iB_{1}y)\dot{x}+in_{9}\dot{x}%
+(-iA_{1}x+B_{1}y)\dot{y}+n_{9}\dot{y}+ \\
&&+\frac{1}{2} m(A_{1}-B_{1})(x^{2}-y^{2})-im(A_{1}-B_{1})xy+2imn_{9}(x-iy)
\end{eqnarray*}%
which consists of the irreducible FIs:
\begin{eqnarray*}
J_{1a}(11b) &=&\frac{1}{4m}\dot{x}^{2}-\frac{i}{4m}\dot{x}\dot{y}+ x\dot{x}%
-ix\dot{y}+\frac{1}{2}m(x^{2}-y^{2})-imxy \\
J_{1b}(11b) &=&\frac{1}{4m}\dot{y}^{2}+\frac{i}{4m}\dot{x}\dot{y}+ iy\dot{x}%
+y\dot{y}-\frac{1}{2}m(x^{2}-y^{2})+imxy \\
J_{1c}(11b) &=&i\dot{x}+\dot{y}+2imx+2my.
\end{eqnarray*}

1.1.2. Subcase $p=4im^{2}$.

We have
\begin{equation*}
L_{a} = A_{2}\left(
\begin{array}{c}
x+iy \\
-ix+y%
\end{array}
\right), \enskip L_{(a;b)}= A_{2}\delta_{ab}, \enskip C_{ab}= \frac{1}{4m}
\left(
\begin{array}{cc}
A_{1}+\frac{A_{2}}{4m} & \frac{i}{2}(B_{1}-A_{1}) \\
\frac{i}{2}(B_{1}-A_{1}) & B_{1}+\frac{A_{2}}{4m}%
\end{array}
\right).
\end{equation*}

From the integrability condition of (\ref{eq.DH3}), we find $A_{2}=0$. Therefore, $L_{a}=0$ and $C_{ab}=\frac{1}{4m} B_{(a;b)}$.

We retrieve the FI $J_{1}(11b)$.
\bigskip

1.2. The subcase $k=-ip$.

Working similarly to the case $k=ip$, we find the following FIs:

1.2.1. $A_{1}=-B_{1}$.
\begin{eqnarray*}
J_{1}(k=-ip) &=&-\frac{B_{1}}{4m}\dot{x}^{2}-\frac{iB_{1}}{2m}\dot{x}\dot{y}+%
\frac{B_{1}}{4m}\dot{y}^{2}-B_{1}(x+iy)\dot{x}-in_{9}\dot{x}+ B_{1}(-ix+y)%
\dot{y}+n_{9}\dot{y}- \\
&&-mB_{1}(x^{2}-y^{2})-2imB_{1}xy-2imn_{9}x+2mn_{9}y
\end{eqnarray*}%
which gives the irreducible FIs:
\begin{eqnarray*}
J_{11-} &=&-\frac{1}{4m}\dot{x}^{2}-\frac{i}{2m}\dot{x}\dot{y}+\frac{1}{4m}%
\dot{y}^{2}-(x+iy)\dot{x}+(-ix+y)\dot{y}-m(x^{2}-y^{2})-2imxy \\
J_{12-} &=&-i\dot{x}+\dot{y}-2imx+2my.
\end{eqnarray*}

1.2.2. $p=-4im^{2}$.
\begin{eqnarray*}
J_{1}(k=-ip=-4m^{2}) &=&\frac{A_{1}}{4m}\dot{x}^{2}+\frac{i}{4m}(A_{1}-B_{1}) \dot{x}\dot{y} +\frac{B_{1}}{4m}\dot{y}^{2}+ (A_{1}x-iB_{1}y)\dot{x}-in_{9}\dot{x}+(iA_{1}x+B_{1}y)\dot{y}+ \\
&&+n_{9}\dot{y}+\frac{1}{2}m(A_{1}-B_{1})(x^{2}-y^{2}) +im(A_{1}-B_{1})xy-2imn_{9}x+2mn_{9}y
\end{eqnarray*}%
which consists of the irreducible FIs:
\begin{eqnarray*}
J_{1a} &=&\frac{1}{4m}\dot{x}^{2}+\frac{i}{4m}\dot{x}\dot{y}+x\dot{x}+ix\dot{%
y}+\frac{1}{2}m(x^{2}-y^{2})+imxy \\
J_{1b} &=&\frac{1}{4m}\dot{y}^{2}-\frac{i}{4m}\dot{x}\dot{y}-iy\dot{x}+y\dot{%
y}-\frac{1}{2}m(x^{2}-y^{2})-imxy \\
J_{1c} &=&-i\dot{x}+\dot{y}-2imx+2my.
\end{eqnarray*}

We note that the FIs $J_{11\pm}$ and $J_{12\pm}$ are written collectively as follows:
\begin{eqnarray*}
J_{11\pm} &=& -\frac{1}{4m} (\dot{x} \mp i\dot{y})^{2} + (-x\pm iy)\dot{x}%
+(\pm ix+y)\dot{y} -m(x^{2}-y^{2})\pm 2imxy \\
J_{12\pm} &=& \dot{x} \mp i\dot{y} + 2m(x \mp iy).
\end{eqnarray*}

We observe that $J_{11\pm} = - \frac{1}{4m} (J_{12\pm})^{2}$; therefore, the FIs $J_{11\pm}$ are not irreducible.
\bigskip

2) Case $w_{9}=w_{11}=0$, $A_{2}=B_{2}$ and $w_{8}=-w_{10}= \frac{k}{p}B_{2}$ $\implies C_{2}=0$.

We have $L_{a}= B_{2} \left(
\begin{array}{c}
x+ \frac{k}{p}y \\
-\frac{k}{p}x +y \\
\end{array}
\right)$ and $L_{(a;b)} =B_{2} \delta_{ab}$. Then,
\begin{equation*}
4mC_{ab} = B_{(a;b)} + \frac{1}{4m}L_{(a;b)}= \left(
\begin{array}{cc}
2a_{1}y+A_{1}+\frac{B_{2}}{4m} & -a_{1}x-\beta_{1} y+ C_{1} \\
-a_{1}x-\beta_{1} y+C_{1} & 2\beta_{1} x+B_{1}+\frac{B_{2}}{4m}%
\end{array}
\right)
\end{equation*}
which when replaced in (\ref{eq.DH2}) gives $a_{1}=\beta_{1}=0$ and produces the following subcases:

2.1. $k=\pm ip$, $n_{11}=\pm in_{9}$, $n_{8}=\pm iB_{1}$, $n_{10}= \mp iA_{1} $ and $B_{2}(p \mp 4im^{2})=0$.

This subcase gives again the FIs found in case 1).

2.2. $n_{9}=n_{11}=0$, $n_{8}=-n_{10}= \frac{k}{p}B_{1} - \frac{B_{2}}{4mp}%
\left( k+4m^{2} \right)$, $A_{1}=B_{1}$ and $B_{2}(p^{2}-4m^{2}k)=0$.

We have $B_{a}=\left(
\begin{array}{c}
B_{1}x+\frac{k}{p}B_{1}y - \frac{B_{2}}{4mp}\left( k+4m^{2} \right)y \\
-\frac{k}{p}B_{1}x + \frac{B_{2}}{4mp}\left( k+4m^{2} \right)x+B_{1}y%
\end{array}
\right)$ and $B_{(a;b)}= B_{1}\delta_{ab}, \enskip 4mC_{ab}= \left(B_{1}+ \frac{B_{2}}{4m} \right) \delta_{ab}$.

2.2.A. $B_{2}=0$.

We have $L_{a}=0$, $B_{a}=\left(
\begin{array}{c}
B_{1}x+\frac{k}{p}B_{1}y \\
-\frac{k}{p}B_{1}x+B_{1}y%
\end{array}%
\right)$, $B_{(a;b)}=B_{1}\delta _{ab}$ and $4mC_{ab}= B_{1}\delta_{ab}$.

The integrability condition of (\ref{eq.DH3}) implies that $p^{2}=4m^{2}k$ for non-trivial FIs, and we compute \newline
$G(x,y)= B_{1} \left( \frac{k}{4m} + m \right)(x^{2} + y^{2})$.

The QFI is
\begin{equation*}
J_{1}=\frac{1}{4m}(\dot{x}^{2}+\dot{y}^{2})+\left( x+\frac{k}{p}y\right)
\dot{x}+\left( y-\frac{k}{p}x\right) \dot{y}+\left( \frac{k}{4m}+m\right)
(x^{2}+y^{2})\implies
\end{equation*}%
\begin{equation*}
\bar{J}_{1}=\frac{p}{k}J_{1}=\frac{1}{4m} \frac{p}{k}(\dot{x}^{2}+\dot{y}%
^{2})+\left( \frac{p}{k}x+y\right) \dot{x}+\left( \frac{p}{k}y-x\right) \dot{%
y}+p\left( \frac{1}{4m}+\frac{m}{k}\right) (x^{2}+y^{2}).
\end{equation*}

2.2.B. $p^{2}=4m^{2}k$.

The integrability condition of (\ref{eq.DH3}) implies that $p^{2}=-4m^{4}$ $\implies k=-m^{2}$ and $p=\pm 2im^{2}$. Integrating, we find $G(x,y)=\frac{3m}{4}B_{1}(x^{2}+y^{2}) -\frac{9}{16}B_{2}(x^{2}+y^{2})$.

The QFI is
\begin{eqnarray*}
J_{1}(2.2) &=& t \frac{B_{2}}{4m} (\dot{x}^{2} + \dot{y}^{2}) + \frac{B_{1}}{%
4m}(\dot{x}^{2} + \dot{y}^{2}) + \frac{B_{2}}{16m^{2}}(\dot{x}^{2} + \dot{y}%
^{2}) + tB_{2} \left( x \pm \frac{i}{2}y \right) \dot{x} + \\
&& + tB_{2} \left( \mp \frac{i}{2}x + y\right) \dot{y} + B_{1} \left( x \pm
\frac{i}{2}y \right) \dot{x} \pm B_{2}\frac{3i}{8m}y\dot{x} + B_{1} \left( y
\mp \frac{i}{2}x \right) \dot{y} \mp B_{2}\frac{3i}{8m}x\dot{y} + \\
&& + t \frac{3m}{4} B_{2}(x^{2} + y^{2})+ \frac{3m}{4}B_{1}(x^{2}+y^{2}) -
\frac{9}{16} B_{2} (x^{2}+y^{2})
\end{eqnarray*}
which consists of the irreducible FIs:
\begin{eqnarray*}
J_{1a}(2.2) &=& t \frac{1}{4m} (\dot{x}^{2} + \dot{y}^{2}) + \frac{1}{16m^{2}%
}(\dot{x}^{2} + \dot{y}^{2}) + t \left( x \pm \frac{i}{2}y \right) \dot{x} +
t \left( \mp \frac{i}{2}x + y\right) \dot{y}\pm \\
&& \pm \frac{3i}{8m}(y\dot{x} - x\dot{y}) + t \frac{3m}{4} (x^{2} + y^{2})-
\frac{9}{16} (x^{2}+y^{2}) \\
J_{1b}(2.2) &=& \frac{1}{4m}(\dot{x}^{2} + \dot{y}^{2}) + \left( x \pm \frac{%
i}{2}y \right) \dot{x} + \left( y \mp \frac{i}{2}x \right) \dot{y} + \frac{3m%
}{4}(x^{2}+y^{2}).
\end{eqnarray*}
\bigskip

\textbf{Integral 2.}
\begin{equation*}
J_{2}= e^{\lambda t} \left( \lambda C_{ab} \dot{q}^{a} \dot{q}^{b} + \lambda
L_{a}\dot{q}^{a} + L_{a}Q^{a} \right)
\end{equation*}
where $\lambda \neq 0$ and the following conditions are satisfied:
\begin{eqnarray}
L_{(a;b)} &=& (4m - \lambda) C_{ab}  \label{cond8a} \\
\left(L_{b}Q^{b}\right)_{,a} &=& 2\lambda C_{ab} Q^{b} + \lambda(2m -\lambda)L_{a}.  \label{cond8b}
\end{eqnarray}

Since $C_{ab}$ is a KT, condition (\ref{cond8a}) implies that $L_{(a;b)}$ is a reducible KT.

We consider the following cases:
\bigskip

1) Case $\lambda =4m$.

From (\ref{cond8a}), we find that $L_{a}$ is a KV, i.e. $L_{a}=(b_{1}+b_{3}y)%
\partial _{x}+(b_{2}-b_{3}x)\partial _{y}$.

Then, condition (\ref{cond8b}) becomes
\begin{equation}
\left( L_{b}Q^{b}\right) _{,a}=8mC_{ab}Q^{b}-8m^{2}L_{a}.  \label{cond8bb}
\end{equation}

Substituting the KV $L_{a}$ and the KT (\ref{FL.14b}) in (\ref{cond8bb}), we
have the following six subcases:

1.1. $k=\frac{p^{2}}{4m^{2}}$, $C=0$, $A=B$, $a=\beta=\gamma=0$, $b_{1}=
b_{2}= 0$ and $b_{3}=-\frac{p}{m}A$.

The QFI is
\begin{equation*}
J_{2}\left( k=\frac{p^{2}}{4m^{2}} \right) = e^{4mt} \left[ \dot{x}^{2} +
\dot{y}^{2} - \frac{p}{m} (y\dot{x} - x\dot{y}) + \frac{p^{2}}{4m^{2}}
(x^{2}+y^{2}) \right].
\end{equation*}

1.2. $k=\frac{p^{2}}{4m^{2}}$, $C=0$, $A=B$, $a=\beta=\gamma=0$, $b_{3}=-%
\frac{p}{m}A$, $p=\pm i(k+8m^{2})$ and $b_{1}=\mp ib_{2}$.

Substituting $k=\frac{p^{2}}{4m^{2}}$ in $p=\pm i(k+8m^{2})$, we get a second degree algebraic equation in $p$ with solutions
\begin{equation*}
p_{1} = \mp 8im^{2} \implies k_{1}=-16m^{2}, \enskip p_{2}= \pm 4im^{2}
\implies k_{2}=-4m^{2}.
\end{equation*}

We have the following QFI\footnote{All the derived QFIs satisfy the condition $k=\frac{p^{2}}{4m^{2}}$.}:
\begin{eqnarray*}
J_{2} &=& e^{4mt} \left[ 4mA(\dot{x}^{2} + \dot{y}^{2}) + 4m \left( \mp i
b_{2} - \frac{p}{m}Ay \right)\dot{x} + 4m \left( b_{2} + \frac{p}{m}Ax
\right) \dot{y} + \right. \\
&& + \left. \left( \mp i b_{2} - \frac{p}{m}Ay \right)(kx-py) + \left( b_{2}
+ \frac{p}{m}Ax \right)(ky +px) \right] \\
&=& e^{4mt} \left[ 4mA(\dot{x}^{2} + \dot{y}^{2}) + 4m \left( \mp i b_{2} -
\frac{p}{m}Ay \right)\dot{x} + 4m \left( b_{2} + \frac{p}{m}Ax \right) \dot{y%
} + \right. \\
&& + \left. A\frac{p^{2}}{m} (x^{2}+y^{2}) + b_{2} (px + ky \mp ikx \pm ipy) %
\right]
\end{eqnarray*}
which consists of the FIs $J_{2}\left( k=\frac{p^{2}}{4m^{2}} \right)$ found earlier in the subcase 1.1 and the LFI
\begin{equation*}
J_{21}= e^{4mt} \left( \mp 4mi\dot{x} + 4m\dot{y} + px + ky \mp ikx \pm ipy
\right).
\end{equation*}
Specifically, we have
\begin{equation*}
J_{21}(k=-16m^{2}, p=\mp 8im^{2}) = J_{21}(k=-4m^{2}, p=\pm 4im^{2}) =
e^{4mt} \left( \mp i\dot{x} + \dot{y} \pm 2im x -2m y \right).
\end{equation*}

1.3. $C=0$, $A=B=0$, $a=\beta=\gamma=0$, $b_{3}=0$, $b_{1}=\mp b_{2}$ and $%
p=\pm i (k+8m^{2})$.

We find again the FI $J_{21}$ of the case 1.2., that is,
\begin{equation*}
J_{2}(p=\pm i (k+8m^{2}))= e^{4mt} \left( \mp 4mi\dot{x} + 4m\dot{y} + px +
ky \mp ikx \pm ipy \right).
\end{equation*}

1.4. $k=\pm ip$: $L_{a}=0$ and $C_{ab}=B\left(
\begin{array}{cc}
-1 & \pm i \\
\pm i & 1%
\end{array}%
\right)$.

We have the QFI $J_{2}(k=\pm ip)= e^{4mt} ( \dot{x}^{2} - \dot{y}^{2} \mp 2i \dot{x}\dot{y})$.

1.5. $k=\pm ip$: $L_{a}=\mp ib_{2}\partial _{x}+b_{2}\partial _{y}$, $%
C_{ab}=B\left(
\begin{array}{cc}
-1 & \pm i \\
\pm i & 1%
\end{array}%
\right) $ and $p=\pm i(k+8m^{2})=\pm 4im^{2}$, which implies $k=-4m^{2}$.

We have the irreducible FIs $J_{2}(k=\pm ip)$ found in the subcase 1.4 and
the FI (already found)
\begin{equation*}
J_{2}(1.5)= e^{4mt} \left( \mp 4mi\dot{x} + 4m\dot{y} + ky + px \mp ikx \pm
ipy \right).
\end{equation*}

1.6. $k=\pm ip$: $A=B \pm 2iC$, $b_{3}= 4m (C \mp iB)$, $b_{1} = \mp ib_{2}$
and $p= \pm i (k+8m^{2})=\pm 4im^{2}$, which implies $k=-4m^{2}$. Then, we
write the QFI
\begin{eqnarray*}
J_{2}(1.6) &=& e^{4mt} \left[ (B \pm 2iC)\dot{x}^{2} + B \dot{y}^{2} + 2C%
\dot{x}\dot{y} \mp ib_{2} \dot{x} + 4m (C \mp iB)y \dot{x} + \right. \\
&& + \left. b_{2}\dot{y} - 4m(C \mp iB)x\dot{y} \pm 2m i b_{2} x - 2m b_{2}y
\mp 4m^{2}i (C \mp iB) (x^{2} + y^{2}) \right]
\end{eqnarray*}
which consists of the FIs:
\begin{eqnarray*}
J_{21}(1.6) &=& e^{4mt} \left[ \dot{x}^{2} + \dot{y}^{2} \mp 4mi (y \dot{x}
- x\dot{y}) - 4m^{2} (x^{2} + y^{2}) \right] \\
J_{22}(1.6) &=& e^{4mt} \left[ \pm i \dot{x}^{2} + \dot{x}\dot{y} + 2m (y
\dot{x} - x\dot{y}) \mp 2m^{2}i (x^{2} + y^{2}) \right] \\
J_{23}(1.6) &=& e^{4mt} \left( \mp i \dot{x} + \dot{y} \pm 2m i x - 2m y
\right).
\end{eqnarray*}
\bigskip

2) Case $\lambda \neq 4m$.

Condition (\ref{cond8a}) gives the KT
\begin{equation}
C_{ab}=\frac{1}{4m-\lambda }L_{(a;b)}  \label{cond8c}
\end{equation}%
where the vector $L_{a}=\left(
\begin{array}{c}
-2\beta y^{2}+2axy+Ax+a_{8}y+a_{11} \\
-2ax^{2}+2\beta xy+a_{10}x+By+a_{9}%
\end{array}%
\right)$ generates the reducible KT
\begin{equation*}
L_{(a;b)}=\left(
\begin{array}{cc}
2ay+A & -ax-\beta y+C \\
-ax-\beta y+C & 2\beta x+B%
\end{array}%
\right)
\end{equation*}%
where $2C=a_{8}+a_{10}$.

Substituting (\ref{cond8c}) in (\ref{cond8b}), we get the condition
\begin{equation}
\left(L_{b}Q^{b}\right)_{,a} = \frac{2\lambda}{4m-\lambda} L_{(a;b)} Q^{b} +
\lambda(2m- \lambda)L_{a}  \label{cond8d}
\end{equation}
which implies the following conditions:
\begin{eqnarray}
0 &=& a(\lambda -3m) \label{intc1} \\
0 &=& \beta (\lambda -3m) \label{intc2} \\
0 &=&  pa - \frac{3}{5}(m^{2}-k)\beta \label{intc3} \\
0 &=& \frac{3}{5}(m^{2}-k)a + p\beta \label{intc4} \\
0 &=& A + B - \frac{\lambda^{2}}{2p} (a_{8} - a_{10}) \label{intc5} \\
0 &=& 4p(\lambda - 2m) B + \frac{\lambda^{3}}{2} (3a_{10} - a_{8}) -2m\lambda^{2}(a_{10} + 2a_{8}) + 2k\lambda(a_{8} + a_{10})+ \notag \\
&& + 8m^{2}\lambda a_{8} - 4km(a_{8}+a_{10}) \label{intc6} \\
0 &=& \left[ \lambda^{3} - 6m\lambda^{2} + 4(2m^{2}+k)\lambda - 8km \right] A + p\lambda a_{8} + p (3\lambda - 8m)a_{10}  \label{intc7} \\
0 &=& \left[ \lambda^{3} - 6m\lambda^{2} + 4(2m^{2}+k)\lambda - 8km \right] B - p(3\lambda - 8m)a_{8} - p\lambda a_{10}  \label{intc8} \\
0 &=& pa_{9} + \frac{\lambda^{3} - 6m\lambda^{2} + (8m^{2}+k)\lambda -4km}{\lambda-4m} a_{11} \label{intc9} \\
0 &=& \frac{\lambda^{3} - 6m\lambda^{2} + (8m^{2}+k)\lambda -4km}{\lambda -4m}a_{9} - p a_{11}. \label{intc10}
\end{eqnarray}

The set of the above conditions leads to three distinct QFIs because
conditions (\ref{intc1}) - (\ref{intc4}) concern only the parameters $a$ and $\beta$; conditions (\ref{intc5}) - (\ref{intc8}) the parameters $A$, $B$, $a_{8}$ and $a_{10}$; and conditions (\ref{intc9}) - (\ref{intc10}) the
parameters $a_{9}$ and $a_{11}$. Therefore, when we write the final form of the QFI, this will consist of three independent FIs one for each set of parameters.

The crucial parameter is the $\lambda$. We consider two cases $\lambda \neq2m$ and $\lambda =2m$ (where in both cases it is assumed that $\lambda\neq 4m$).

2.1. The subcase $\lambda \neq 2m$.

When $\lambda \neq 2m$, we have the following subcases for each set of
parameters considered above.

2.1.1. Non-vanishing parameters $a$ and $\beta$ (i.e. $A=B=a_{8}=a_{10}=0$, $a_{9}=a_{11}=0)$. Then, the conditions (\ref{intc1}) - (\ref{intc4}) survive and\footnote{This gives a non-trivial FI because then (\ref{intc1}) and (\ref{intc2}) vanish identically.} $\lambda= 3m$.

In that case, the linear system (\ref{intc3}) - (\ref{intc4}) has the non-zero solution $\beta =\pm ia$ when $p=\pm i\frac{3}{5}(m^{2}-k)$, and we have for this set of parameters the QFI
\begin{eqnarray*}
J_{2}(2.1.1) &=&e^{3mt}\left[ 3y\dot{x}^{2}\pm 3ix\dot{y}^{2}-3(x\pm iy)\dot{%
x}\dot{y}+3m(\mp iy^{2}+xy)\dot{x}+3m(-x^{2}\pm ixy)\dot{y}+\right. \\
&&\left. +(\mp iy^{2}+xy)(kx-py)+(-x^{2}\pm ixy)(ky+px)\right].
\end{eqnarray*}

2.1.2. Non-vanishing parameters $A$, $B$, $a_{8}$ and $a_{10}$.

In this case, we have the conditions (\ref{intc5}) - (\ref{intc8}).

From (\ref{intc5}) and (\ref{intc6}), the parameters $A$ and $B$ are expressed as linear combinations of $a_{8}$ and $a_{10}$ since $\lambda \neq 2m$. These expressions, say $A(a_{8},a_{10})$ and $B(a_{8},a_{10})$, when replaced in (\ref%
{intc7}) and (\ref{intc8}), respectively, give a homogeneous linear system.
This system has non-vanishing solution of the form $a_{8}=Da_{10}$ only when
$D=1$. In that case, $a_{8}=a_{10}$ with $p=\pm \frac{i}{4}(\lambda
^{2}-4m\lambda +4k)$, which implies that $A=-B=\mp ia_{8}$.

Then, $C_{ab}=\frac{a_{8}}{4m-\lambda }\left(
\begin{array}{cc}
\mp i & 1 \\
1 & \pm i \\
\end{array}%
\right)$, $L_{a}=a_{8}\left(
\begin{array}{c}
\mp ix+y \\
x\pm iy \\
\end{array}%
\right)$ and the QFI for this set of parameters is
\begin{eqnarray*}
J_{2}(2.1.2) &=&e^{\lambda t}\left[ \frac{\lambda }{4m-\lambda }(\mp i\dot{x}%
^{2}\pm i\dot{y}^{2}+2\dot{x}\dot{y})+\lambda (\mp ix+y)\dot{x}+\lambda
(x\pm iy)\dot{y}+\right. \\
&&\left. +(p\mp ik)(x^{2}-y^{2})+2(k\pm ip)xy\right] .
\end{eqnarray*}

2.1.3. Non-vanishing parameters $a_{9}$ and $a_{11}$.

In this case, we have the homogeneous linear system (\ref{intc9}) - (\ref%
{intc10}) which has the non-zero solution $a_{11}=\mp ia_{9}$ for $p=\pm i%
\frac{\lambda ^{3}-6m\lambda ^{2}+(8m^{2}+k)\lambda -4km}{\lambda -4m}$.
Therefore, $C_{ab}=0$ and $L_{a}= a_{9} (\mp i \partial_{x} + \partial_{y})$.

Since
\begin{equation*}
p=\pm i\frac{\lambda ^{3}-6m\lambda ^{2}+(8m^{2}+k)\lambda -4km}{\lambda -4m}%
=\pm i\frac{(\lambda -4m)(\lambda ^{2}-2\lambda m+k)}{\lambda -4m}=\pm
i(\lambda ^{2}-2\lambda m+k)
\end{equation*}%
we end up with the FI which we find in case 3) below.

2.2. The subcase $\lambda =2m$.

In this case, condition (\ref{cond8d}) becomes (since $\lambda \neq 4m$)
\begin{equation}
\left( L_{b}Q^{b}\right) _{,a}=2L_{(a;b)}Q^{b}.  \label{cond8e}
\end{equation}

From (\ref{cond8e}), we find that $a=\beta =0$, $A=-B$ and $a_{8}=a_{10}\implies$ $C=a_{8}$, and we end up with the system
\newline $\begin{cases}
pa_{9}+ka_{11}=0 \\
ka_{9}-pa_{11}=0%
\end{cases}$ which leads to two subcases: 2.2.1) $a_{9}=a_{11}=0$, and 2.2.2) $k=\pm ip$
and $a_{9}=\mp ia_{11}$.

2.2.1. Subcase $a_{9}=a_{11}=0$.

We have $C_{ab}=\frac{1}{2m}\left(
\begin{array}{cc}
A & a_{8} \\
a_{8} & -A%
\end{array}%
\right)$ and $L_{a}=\left(
\begin{array}{c}
Ax+a_{8}y \\
a_{8}x-Ay%
\end{array}%
\right)$.

The QFI is
\begin{eqnarray*}
J_{2}(\lambda =2m) &=&e^{2mt}\left[ A\dot{x}^{2}-A\dot{y}^{2}+2a_{8}\dot{x}%
\dot{y}+2mAx\dot{x}+2ma_{8}y\dot{x}+2ma_{8}x\dot{y} -2mAy\dot{y}+\right. \\
&&\left. +Ak(x^{2}-y^{2})-Apxy+a_{8}kxy-a_{8}py^{2}+a_{8}kxy+a_{8}px^{2}
-Apxy \right]
\end{eqnarray*}%
which consists of the irreducible FIs:
\begin{eqnarray*}
J_{2a}(2.2.1) &=&e^{2mt}\left[ \dot{x}^{2}-\dot{y}^{2}+2m(x\dot{x}-y\dot{y}%
)+k(x^{2}-y^{2})-2pxy\right] \\
J_{2b}(2.2.1) &=&e^{2mt}\left[ \dot{x}\dot{y}+m(y\dot{x}+x\dot{y})+\frac{p}{2%
}(x^{2}-y^{2})+kxy\right] .
\end{eqnarray*}%
The FI $J_{2b}(2.2.1)$ is the one found in eq. (38) of \cite{Djukic 1975}.

2.2.2. Subcase $a_{9}=\mp ia_{11}$.

Then, $k=\pm ip$, $C_{ab}=\frac{1}{2m}\left(
\begin{array}{cc}
A & a_{8} \\
a_{8} & -A%
\end{array}%
\right)$ and $L_{a}=\left(
\begin{array}{c}
Ax+a_{8}y+a_{11} \\
a_{8}x-Ay\mp ia_{11}%
\end{array}%
\right)$.

We find again the two FIs of the case 3.1 and the additional LFI
$J_{2}(2.2.2)= e^{2mt} (\dot{x} \mp i\dot{y})$. We note that the FI of the case 1.4 can be derived from the above FI as
follows:
\begin{equation*}
\left[ J_{2}(2.2.2) \right]^{2}= e^{4mt} (\dot{x} \mp i\dot{y})^{2} =
e^{4mt} ( \dot{x}^{2} - \dot{y}^{2} \mp 2i \dot{x}\dot{y})= J_{2}(1.4).
\end{equation*}
\bigskip

3) Case $C_{ab}=0$.

Condition (\ref{cond8a}) implies that $L_{a}=(b_{1}+b_{3}y)\partial
_{x}+(b_{2}-b_{3}x)\partial _{y}$ is a KV, and the remaining condition (\ref%
{cond8b}) becomes
\begin{equation}
\left( L_{b}Q^{b}\right) _{,a}=\lambda (2m-\lambda )L_{a}.  \label{cond8f}
\end{equation}

Substituting the KV $L_{a}$ in (\ref{cond8f}), we get $b_{3}=0$ and non-vanishing values for $b_{1}, b_{2}$ only when
\begin{equation*}
p=\pm i(\lambda^{2} - 2\lambda m + k) \implies
\begin{cases}
\lambda_{1}= m + \sqrt{m^{2} - k \mp ip} \\
\lambda_{2}= m - \sqrt{m^{2} - k \mp ip}.%
\end{cases}%
\end{equation*}
Then, $b_{1} = \mp ib_{2}$.

Observe that since $p\neq 0$, also $\lambda^{2} - 2\lambda m + k \neq 0$.

The FI is $J_{2(3)}=e^{\lambda t}(\mp i\lambda \dot{x}+\lambda \dot{y}\mp ikx\pm
ipy+px+ky)$ where $p=\pm i(\lambda ^{2}-2\lambda m+k)$.
\bigskip

We collect the above results in Table \ref{Table.FIs.damped.SHO}.

\begin{longtable}{|l|l|}
\hline
Condition & LFIs/QFIs \\ \hline
- & $J_{2a}=e^{2mt}\left[ \dot{x}^{2}-\dot{y}^{2}+2m(x\dot{x}-y\dot{y}%
)+k(x^{2}-y^{2})-2pxy\right] $ \\
-, FI (38) in \cite{Djukic 1975} & $J_{2b}=e^{2mt}\left[ \dot{x}\dot{y}+m(y%
\dot{x}+x\dot{y})+\frac{p}{2}(x^{2}-y^{2})+kxy\right] $ \\ \hline
$k=\pm ip$ & $J_{12}=\dot{x}\mp i\dot{y}+2m(x\mp iy)$ \\
& $J_{29}=e^{2mt}(\dot{x}\mp i\dot{y})$ \\
& $J_{2a}, J_{2b}$ \\ \hline
$p=\pm 4im^{2}$, $k=-4m^{2}$ & $J_{13}=\frac{1}{4m}\dot{x}^{2}\mp \frac{i}{4m%
}\dot{x}\dot{y}+x\dot{x}\mp ix\dot{y}+\frac{1}{2}m(x^{2}-y^{2})\mp imxy$ \\
& $J_{14}=\frac{1}{4m}\dot{y}^{2}\pm \frac{i}{4m}\dot{x}\dot{y}\pm iy\dot{x}%
+y\dot{y}-\frac{1}{2}m(x^{2}-y^{2})\pm imxy$ \\
& $J_{26}=e^{4mt}\left[ \pm i\dot{x}^{2}+\dot{x}\dot{y}+2m(y\dot{x}-x\dot{y}%
)\mp 2m^{2}i(x^{2}+y^{2})\right] $ \\
& $J_{2a}, J_{2b}, J_{12}, J_{29}, J_{11}, J_{21}, J_{23}$
\\ \hline
$p=\pm 2im^{2}$, $k=-m^{2}$ & \makecell[l]{$J_{15} = t \frac{1}{4m}
(\dot{x}^{2} + \dot{y}^{2}) + \frac{1}{16m^{2}}(\dot{x}^{2} + \dot{y}^{2}) +
t \left( x \pm \frac{i}{2}y \right) \dot{x} + t \left( \mp \frac{i}{2}x +
y\right) \dot{y}\pm $ \\ \qquad \enskip $\pm \frac{3i}{8m}(y\dot{x} -
x\dot{y}) + t \frac{3m}{4} (x^{2} + y^{2})- \frac{9}{16} (x^{2}+y^{2})$} \\
& $J_{16}=\frac{1}{4m}(\dot{x}^{2}+\dot{y}^{2})+\left( x\pm \frac{i}{2}%
y\right) \dot{x}+\left( y\mp \frac{i}{2}x\right) \dot{y}+\frac{3m}{4}%
(x^{2}+y^{2})$ \\
& $J_{2a}, J_{2b}$, $J_{11}$, $J_{21}$ \\ \hline
$k=\frac{p^{2}}{4m^{2}}$ & $J_{11}=\frac{1}{4m}(\dot{x}^{2}+\dot{y}%
^{2})+\left( x+\frac{k}{p}y\right) \dot{x}+\left( y-\frac{k}{p}x\right) \dot{%
y}+\left( \frac{k}{4m}+m\right) (x^{2}+y^{2})$ \\
& $J_{21}=e^{4mt}\left[ \dot{x}^{2}+\dot{y}^{2}-\frac{p}{m}(y\dot{x}-x\dot{y}%
)+\frac{p^{2}}{4m^{2}}(x^{2}+y^{2})\right] $ \\
& $J_{2a}, J_{2b}$ \\ \hline
$p=\pm i(k+8m^{2})$ & $J_{23}=e^{4mt}\left( \mp 4mi\dot{x}+4m\dot{y}%
+px+ky\mp ikx\pm ipy\right) $ \\
& $J_{2a}, J_{2b}$ \\ \hline
\makecell[l]{$p = \pm \frac{i}{4} (\lambda^{2} - 4 m \lambda + 4k)$, \\
$\lambda \neq 2m,4m$} & \makecell[l]{$J_{27}= e^{\lambda t} \left[
\frac{\lambda}{4m - \lambda} ( \mp i\dot{x}^{2} \pm i\dot{y}^{2} +
2\dot{x}\dot{y} ) + \lambda (\mp i x + y)\dot{x} + \lambda(x \pm iy) \dot{y}
+ \right.$ \\ \qquad \enskip $\left. + (p \mp ik)(x^{2} - y^{2})+ 2(k \pm
ip)xy \right]$} \\
& $J_{2a}, J_{2b}$ \\ \hline
$p=\pm i(\lambda ^{2}-2\lambda m+k)$ & $J_{28}=e^{\lambda t}(\mp i\lambda
\dot{x}+\lambda \dot{y}\mp ikx\pm ipy+px+ky)$ \\
& $J_{2a}, J_{2b}$ \\ \hline
$p=\pm i\frac{3}{5}(m^{2}-k)$ & \makecell[l]{$J_{24}= e^{3mt} \left[ 3y
\dot{x}^{2} \pm 3ix\dot{y}^{2} - 3(x \pm iy)\dot{x}\dot{y} + 3m (\mp i y^{2}
+ xy) \dot{x} + 3m (-x^{2} \pm ixy)\dot{y} + \right.$ \\ \qquad \enskip
$\left. + (\mp i y^{2} + xy)(kx-py) + (-x^{2} \pm ixy)(ky+px) \right]$} \\
& $J_{2a}, J_{2b}$ \\ \hline
\caption{\label{Table.FIs.damped.SHO} The LFIs/QFIs of two linearly coupled harmonic oscillators with linear damping.}
\end{longtable}

\begin{remark} \label{remark.damp.1}
In Table \ref{Table.FIs.damped.SHO}, some sets of conditions are a subset of other
more general conditions. In that case, the FIs corresponding to that more
general conditions are also FIs for the special subset of these conditions;
however, the opposite does not hold. For example: \newline
- The expressions $J_{2a}$ and $J_{2b}$ are FIs for all values of $k$,$p$,$m$. \newline
- The set of conditions ($k=-4m^{2}$, $p=\pm 4im^{2}$) gives $p=\mp ik$ $\implies k=\pm ip$, which means that the $J_{12}$ and $J_{29}$ are FIs of that set in addition to $J_{13}$, $J_{14}$ and $J_{26}$. Observe that for that special set of conditions $J_{13}-J_{14}= \frac{1}{4m} (J_{12})^{2}$ and $J_{11}=J_{13}+J_{14}$.
\end{remark}

\begin{remark} \label{remark.2.note1}
The set of conditions $k=\pm ip$ with $p_{+}=i(k+8m^{2})$ and $p_{-}=-i(k+8m^{2})$ implies that $k=-4m^{2}$ and $p=\pm 4im^{2}$.
\end{remark}

\begin{remark} \label{remark.3.note2}
For $k=\frac{p^{2}}{4m^{2}}$ and $p=\pm i(k+8m^{2})$, we find that $p_{+}= 4im^{2}, -8im^{2}$ and $p_{-}= -4im^{2}, 8im^{2}$. In that case, the corresponding FI $J_{23}$ reduces to
\begin{equation*}
J_{23}(k=p^{2}/4m^{2})=e^{4mt}\left( \mp i\dot{x}+\dot{y}\pm 2imx-2my\right).
\end{equation*}
\end{remark}

\subsubsection{Discussing integrability}

\label{sec.damped.oscillator.integrability}

- For arbitrary values of $k, p, m$.

For arbitrary values of $k, p, m$, the dynamical system admits two time-dependent QFIs the $J_{2a}$ and $J_{2b}$. These
FIs are not in involution, that is, their PB does not vanish. Therefore, an arbitrary 2d harmonic oscillator with arbitrary external forces is not in
general integrable; in order to achieve  integrability, we have to look for special values
of $k, p, m$ where more FIs are admitted.
\bigskip

- $k=\pm ip$.

The FIs $J_{12}$ and $J_{29}$ are functionally independent and in involution
since $\{J_{12\pm },J_{29\pm }\}=0$. Therefore, the system in that special
case is Liouville integrable and can be integrated by quadratures. Indeed, we have:
\begin{equation}
\begin{cases}
\dot{x}\mp i\dot{y}+2m(x\mp iy)=c_{1\pm } \\
e^{2mt}(\dot{x}\mp i\dot{y})=c_{2\pm }%
\end{cases}%
\implies z(t)=-\frac{c_{2\pm }}{2m}e^{-2mt}+\frac{c_{1\pm }}{2m}
\label{eq.intdj.1}
\end{equation}%
where $z(t)\equiv x(t)\mp iy(t)$ and $c_{1\pm}$, $c_{2\pm}$ are arbitrary complex constants.

- $p= \pm 4im^{2}$, $k=-4m^{2}$.

These values satisfy the conditions $k=\pm ip$, $k=\frac{p^{2}}{4m^{2}}$ and
$p= \pm i(k + 8m^{2})$. Since $k=\pm ip$, it is straightforward (see previous case) that the system is integrable.

%% file: Ermakov.tex
\chapter{The integrability of the generalized Ermakov conservative system}

\label{ch.Ermakov}

The 2d generalized Ermakov system has attained attention in 90's (see e.g. \cite{Ermakov, Ray 1979A, Athorne 1990, Leach 1991, Leach 1994, Goedert 1998}), where most of its properties have been revealed. A review of these studies can be found in \cite{Leach Andriopoulos 2008}. However, the interest in the topic is still alive, and a recent article has appeared \cite{Naz 2020} presenting new results. The purpose of the present chapter is to show that these latter results can be obtained as special cases of older and more recent results on the integrability of 2d dynamical systems (see chapter \ref{ch.2d.pots}).

\section{The 2d generalized Ermakov system}

\label{sec.erma.automous}

The 2d generalized Ermakov system is defined by the equations:\index{Dynamical system! generalized Ermakov}
\begin{eqnarray}
\ddot{x} &=& -\omega^{2}(t)x + \frac{1}{x^{2}y} f\left(\frac{y}{x}\right)
\label{eq.erm1a} \\
\ddot{y} &=& -\omega^{2}(t)y + \frac{1}{xy^{2}} g\left(\frac{x}{y}\right)
\label{eq.erm1b}
\end{eqnarray}
where $f$ and $g$ are arbitrary functions. This system admits the \textbf{Ermakov FI}\index{First integral! Ermakov}
\begin{equation}
I_{0}= \frac{1}{2} (x\dot{y}-y\dot{x})^{2} + \int^{y/x} f(u)du + \int^{x/y}
g(v)dv  \label{eq.erm2}
\end{equation}
where $u= v^{-1}= \frac{y}{x}$. It is well-known that the 2d generalized Ermakov system generalizes the 1d
time-dependent oscillator.\index{Oscillator! 1d time-dependent}

Introducing the functions $F$ and $G$ by the relations $f\left(\frac{y}{x}\right)= \frac{y}{x} F\left(\frac{y}{x}\right)$ and $g\left(\frac{x}{y}\right)= \frac{x}{y} G\left(\frac{y}{x}\right)$, respectively,
equations (\ref{eq.erm1a}) - (\ref{eq.erm1b}) take the
equivalent form:
\begin{eqnarray}
\ddot{x} &=& -\omega^{2}(t)x + \frac{1}{x^{3}} F\left(\frac{y}{x}\right)
\label{eq.erm3a} \\
\ddot{y} &=& -\omega^{2}(t)y + \frac{1}{y^{3}} G\left(\frac{y}{x}\right)
\label{eq.erm3b}
\end{eqnarray}
while the Ermakov FI (\ref{eq.erm2}) becomes
\begin{equation}
I_{0}= \frac{1}{2} (x\dot{y}-y\dot{x})^{2} + \int^{y/x} \left[ uF(u) - u^{-3}G(u)\right]du.  \label{eq.erm4}
\end{equation}

If one introduces the variables \cite{Leach 1991}
\begin{equation}
T= \int \rho^{-2}dt, \enskip X=\rho^{-1}x, \enskip Y=\rho^{-1}y
\label{eq.erm5}
\end{equation}
where $\rho(t)$ is a solution of the 1d time-dependent oscillator
\begin{equation}
\ddot{\rho} + \omega^{2}(t)\rho =0  \label{eq.erm6}
\end{equation}
then equations (\ref{eq.erm3a}) - (\ref{eq.erm3b}) become
the autonomous system:
\begin{eqnarray}
X^{\prime \prime }&=& \frac{1}{X^{3}} F\left(\frac{Y}{X}\right)
\label{eq.erm7a} \\
Y^{\prime \prime }&=& \frac{1}{Y^{3}} G\left(\frac{Y}{X}\right)
\label{eq.erm7b}
\end{eqnarray}
and the Ermakov FI
\begin{equation}
I_{0}= \frac{1}{2} \left( XY^{\prime} -YX^{\prime} \right)^{2} + \int^{Y/X} \left[ uF(u) -u^{-3}G(u)\right]du. \label{eq.erm8}
\end{equation}
Concerning the notation: $f^{\prime }\equiv\frac{df(T)}{dT}$ and $\dot{f}\equiv \frac{df(t)}{dt}$.

For general functions $F$ and $G$ the autonomous dynamical system (\ref{eq.erm7a}) - (\ref{eq.erm7b}) is not conservative. In the following sections, we determine the family of conservative Ermakov systems together with their FIs using collineations of the metric defined by these dynamical equations (i.e. the kinetic metric).

\section{Integrability of the 2d generalized Ermakov system}

\label{sec.erma.integrability}

Since the system (\ref{eq.erm7a}) - (\ref{eq.erm7b}) is autonomous, the second FI will be the Hamiltonian $H$. To find $H$, we do not have to do any new calculations because in chapter \ref{ch.2d.pots} all the integrable and superintegrable 2d autonomous conservative systems have been determined. From these results, we find (see section \ref{sec.const2}, case 1) that the 2d integrable potential
\begin{equation}
V_{21}= \frac{F_{1}(u)}{X^{2}+Y^{2}} +F_{2}(X^{2}+Y^{2})  \label{eq.ermint3a}
\end{equation}
where $u=\frac{Y}{X}$ and $F_{1}, F_{2}$ are arbitrary functions of their arguments, admits the QFI
\begin{equation}
I_{11}= \frac{1}{2} \left( XY^{\prime }-YX^{\prime} \right)^{2} +F_{1}(u).
\label{eq.ermint3b}
\end{equation}

If we consider $F_{1}(u)= (u^{2}+1)N(u)$ and $F_{2}=0$, we find that $V_{21}=\frac{N(u)}{X^{2}}$ while the resulting equations are:
\begin{eqnarray}
X^{\prime \prime }&=& \frac{2N +u \frac{dN}{du}}{X^{3}}  \label{eq.ermint8a}
\\
Y^{\prime \prime }&=& -\frac{1}{X^{3}} \frac{dN}{du}.  \label{eq.ermint8b}
\end{eqnarray}

Therefore, if we define the functions
\begin{equation}
F(u)= 2N +u\frac{dN}{du} \enskip \text{and} \enskip G(u)= -u^{3}\frac{dN}{du} \label{eq.ermint4b}
\end{equation}
then equations (\ref{eq.ermint8a}) - (\ref{eq.ermint8b}) become the Ermakov
equations (\ref{eq.erm7a}) - (\ref{eq.erm7b}) while $I_{0}=I_{11}$.

We conclude that the family of the conservative 2d Ermakov systems is defined by the potential $V=\frac{N(u)}{X^{2}}$, where $N(u)$ is an arbitrary function while the Hamiltonian is given by the expression
\begin{equation}
H=\frac{1}{2}\left( X^{\prime 2}+Y^{\prime 2} \right) +\frac{N(u)}{X^{2}}.
\label{eq.ermint4c}
\end{equation}

In the original coordinates, the system (\ref{eq.ermint8a}) - (\ref{eq.ermint8b}) becomes:
\begin{eqnarray}
\ddot{x} &=& -\omega^{2}(t)x +\frac{2N(u) +u\frac{dN}{du}}{x^{3}}
\label{eq.ermint7a} \\
\ddot{y} &=&-\omega^{2}(t)y -\frac{1}{x^{3}}\frac{dN}{du}
\label{eq.ermint7b}
\end{eqnarray}
where $u = \frac{y}{x} =\frac{Y}{X}$.

For $N(u)=\frac{u^{-2}}{2}$ we find, respectively,
\begin{eqnarray}
\ddot{x} &=&-\omega ^{2}(t)x  \label{eq.ermint9a} \\
\ddot{y} &=&-\omega ^{2}(t)y+\frac{1}{y^{3}}  \label{eq.ermint9b}
\end{eqnarray}%
while the Ermakov FI becomes the well-known Lewis invariant \cite{Lewis 1968} \index{Invariant! Lewis}
\begin{equation}
I_{0}=\frac{1}{2}(x\dot{y}-y\dot{x})^{2}+\frac{1}{2}\left( \frac{x}{y}%
\right) ^{2}.  \label{eq.ermint9c}
\end{equation}
Equation (\ref{eq.ermint9a}) is the 1d time-dependent harmonic oscillator,\index{Oscillator! 1d time-dependent} and (\ref{eq.ermint9b}) is the auxiliary equation\index{Equation! auxiliary} with which one determines the frequency $\omega(t)$ for a given function $y(t)$. These justify the characterization of the Ermakov system as a generalization of the harmonic oscillator.

\section{The FIs of the conservative Ermakov system}

\label{sec.FIs.conservative.Ermakov}

There are two ways to find the FIs of the Ermakov system. One way is to use the results of chapter \ref{ch.2d.pots}, and read the FIs for the potential $V_{21}$ given in (\ref{eq.ermint3a}) for $F_{1}(u)=(u^{2}+1)N(u)$ and $F_{2}=0$. In this section, we shall follow another way which can be useful in many similar problems. We shall use Theorem 2 of \cite{TsaPal 2011}, where it is stated that \emph{the generators of the
point Noether symmetries\index{Symmetry! point Noether} of autonomous conservative systems are the elements
of the homothetic algebra of the metric defined by the kinetic energy
(kinetic metric)}. In the Ermakov case, this metric is the Euclidean 2d metric $\gamma_{ab}=diag(1,1)$. For the convenience of the reader, we state Theorem 2 of \cite{TsaPal 2011}.

\begin{theorem}
\label{thm.point.Noether.autonomous.systems} Autonomous conservative
dynamical systems of the form
\begin{equation}
\ddot{q}^{a} = - \Gamma^{a}_{bc}(q) \dot{q}^{b} \dot{q}^{c} - V^{,a}(q)
\label{eq.PN1}
\end{equation}
where $\Gamma_{bc}^{a}$ are the Riemannian connection coefficients
determined of the kinetic metric $\gamma_{ab}(q)$ (kinetic energy) and $V(q)$
the potential of the system, admit the following point Noether symmetries:\index{Symmetry! point Noether}
\bigskip

Case 1. The point Noether symmetry
\begin{equation}
\mathbf{A}_{1}=\partial _{t}, \enskip f_{1}=const\equiv0  \label{eq.PN4a}
\end{equation}%
which produces the Noether FI (Hamiltonian)
\begin{equation}
H=\frac{1}{2}\gamma_{ab}\dot{q}^{a}\dot{q}^{b} + V(q).  \label{eq.PN4b}
\end{equation}

Case 2. The point Noether symmetry
\begin{equation}
\mathbf{A}_{2}=2\psi_{B}t\partial _{t}+B^{a}\partial_{q^{a}}, \enskip %
f_{2}=c_{1}t  \label{eq.PN5a}
\end{equation}
where $c_{1}$ is an arbitrary constant and $B^{a}$ is a KV ($\psi_{B}=0$) or
the HV ($\psi_{B}=1$) such that
\begin{equation}
B_{a}V^{,a}+2\psi_{B}V+c_{1}=0.  \label{eq.PN5b}
\end{equation}
The associated Noether FI is
\begin{equation}
I_{2} = 2\psi_{B}tH -B_{a}\dot{q}^{a} +c_{1}t.  \label{eq.PN5c}
\end{equation}

Case 3. The point Noether symmetry
\begin{equation}
\mathbf{A}_{3}=2\psi \int C(t)dt \partial_{t}
+C(t)\Phi^{,a}\partial_{q^{a}}, \enskip f_{3} =C_{,t}\Phi(q)+ D(t)
\label{eq.PN6a}
\end{equation}
where $\Phi^{,a}(q)$ is a gradient KV ($\psi=0$) or a gradient HV ($\psi=1$) such that ($c_{2}$ and $c_{3}$ are arbitrary constants)
\begin{equation}
\Phi_{,a}V^{,a}+2\psi V=c_{2}\Phi+c_{3}  \label{eq.PN6b}
\end{equation}
and the functions $C(t)$, $D(t)$ are determined by the relations ($C_{,t}\neq 0$)
\begin{equation}
C_{,tt}=-c_{2}C, \enskip D_{,t}= -c_{3} C.  \label{eq.PN6c}
\end{equation}
The associated Noether FI is
\begin{equation}
I_{3}=2\psi H\int C(t) dt- C(t) \Phi_{,a}\dot{q}^{a} +C_{,t}\Phi -c_{3}\int
C(t)dt.  \label{eq.PN6d}
\end{equation}
\end{theorem}

We apply Theorem \ref{thm.point.Noether.autonomous.systems} in
the case of the autonomous integrable Ermakov system (\ref{eq.ermint8a}) - (\ref{eq.ermint8b}) which has potential $V=\frac{N(u)}{X^{2}}$, where $u=Y/X$,
and kinetic metric $\gamma_{ab}=diag(1,1)$.

The homothetic algebra\index{Algebra! homothetic} of $\gamma_{ab}$ consists of two gradient KVs $\partial_{X}$ and $\partial_{Y}$, one non-gradient KV (rotation) $%
Y\partial_{X} -X\partial_{Y}$, and the gradient HV $%
X\partial_{X} +Y\partial_{Y}$.

For each case of Theorem \ref{thm.point.Noether.autonomous.systems} we have
the following.

\subsection{The vector $\partial_{T}$}

Case 1. In this case the point Noether symmetry $\mathbf{A}_{1}=\partial_{T}$, $f_{1}=0$ produces the Hamiltonian (as expected)
\begin{equation}
H= \frac{1}{2} \left( X^{\prime 2} + Y^{\prime 2} \right) + V= \frac{1}{2}
\left( X^{\prime 2} + Y^{\prime 2} \right) + \frac{N(u)}{X^{2}}.
\label{eq.PNerm1}
\end{equation}

\subsection{The gradient HV $X\partial_{X} +Y\partial_{Y}$}

\label{sec.erm.HV}

Case 2. Consider the gradient HV $B^{a}=(X,Y)$ with homothetic factor $\psi_{B}=1$.

Substituting in condition (\ref{eq.PN5b}), we find that
\begin{equation*}
XV_{,X}+ YV_{,Y} + 2V + c_{1}=0 \implies -\frac{2N + u\frac{dN}{du}}{X^{2}}
+ \frac{u}{X^{2}}\frac{dN}{du} +2\frac{N}{X^{2}}+c_{1} =0 \implies c_{1}=0.
\end{equation*}
Therefore, the point Noether symmetry is
\begin{equation}
\mathbf{A}_{2}= 2T\partial_{T} + X\partial_{X} + Y\partial_{Y}, \enskip %
f_{2}=0  \label{eq.PNerm2a}
\end{equation}
and the associated Noether FI
\begin{equation}
I_{2}= 2TH -(XX^{\prime }+YY^{\prime}).  \label{eq.PNerm2b}
\end{equation}

It can be shown that the three FIs $I_{0}, H, I_{2}$ are independent; therefore, the conservative generalized Ermakov system is superintegrable.

Although the remaining FIs will be expressible in terms of the $I_{0}, H, I_{2}$, we continue in order to show that we recover the results of \cite{Leach 1994} which were obtained using Lie symmetries.

Case 3. The point Noether symmetry is
\begin{equation}
\mathbf{A}_{3} = T^{2}\partial_{T} + TX\partial_{X} +TY\partial_{Y}, \enskip %
f_{3}= \frac{X^{2}+Y^{2}}{2}  \label{eq.PNerm3b}
\end{equation}
with associated Noether FI
\begin{equation}
I_{3}= T^{2}H - T(XX^{\prime }+YY^{\prime }) + \frac{X^{2}+Y^{2}}{2}= \frac{%
I_{2}^{2}+2I_{0}}{4H}.  \label{eq.PNerm3c}
\end{equation}
We observe that the Lie
symmetries (2.9a), (2.9b), (2.9c) found in \cite{Leach 1994} are the point
Noether symmetries $\mathbf{A}_{1}$, (\ref{eq.PNerm2a}), (\ref{eq.PNerm3b}). Concerning the remaining FIs of \cite{Leach 1994}, we have: (4.11) $I^{\prime }=2I_{0}$,
(4.12) $J_{1}^{\prime}=2H$, (4.13) $J^{\prime}_{2}=I_{2}$ and (4.14) $J^{\prime}_{3}=2I_{3}$. Using these relations, eq. (4.18) is equivalent to the expression (\ref{eq.PNerm3c}).

\subsection{The gradient KV $b_{1}\partial_{X} +b_{2}\partial_{Y}$}

\label{sec.erm.grKV}

The potential becomes\footnote{%
This is a superintegrable potential of the form $F(b_{1}Y-b_{2}X)$ (see chapter \ref{ch.2d.pots}).} $V_{1}= \frac{k}{(b_{1}Y - b_{2}X)^{2}}$ where $k,
b_{1}, b_{2}$ are arbitrary constants.

Case 2. The Noether generator, the Noether function and the FI are $\mathbf{A}_{21}= b_{1}\partial_{X} +b_{2}\partial_{Y}$, $f_{21}= 0$ and $I_{21}= b_{1}X^{\prime }+b_{2}Y^{\prime }$, respectively.

Case 3. The Noether generator, the Noether function and the FI are $\mathbf{A}_{31}= Tb_{1}\partial_{X} +Tb_{2}\partial_{Y}$, $f_{31}=b_{1}X +b_{2}Y$ and $I_{31}= b_{1} (-TX^{\prime }+X) +b_{2}(-TY^{\prime }+Y)$, respectively.

In order to compare these results with the ones of \cite{Naz 2020}, we use polar coordinates\index{Coordinates! polar} $X=r\cos\theta $ and $Y=r\sin \theta$. Then, we find:
\begin{equation*}
V_{1}= \frac{k}{r^{2}(b_{1}\sin\theta -b_{2}\cos\theta)^{2}}
\end{equation*}%
\begin{equation*}
\mathbf{A}_{21}= \left(b_{1}\cos\theta +b_{2}\sin\theta\right) \partial_{r}
+ \frac{1}{r} \left(b_{2}\cos\theta -b_{1}\sin\theta \right)
\partial_{\theta}, \enskip f_{21}= 0
\end{equation*}
\begin{equation*}
\mathbf{A}_{31}= \left(b_{1}T\cos\theta +b_{2}T\sin\theta
\right)\partial_{r} + \frac{1}{r}\left(b_{2}T\cos\theta -b_{1}T\sin\theta
\right) \partial_{\theta}, \enskip f_{31}= r(b_{1}\cos\theta +
b_{2}\sin\theta)
\end{equation*}
\begin{eqnarray*}
I_{21}&=& b_{1} \left( r^{\prime }\cos\theta - r\theta^{\prime }\sin\theta
\right) + b_{2} \left( r^{\prime }\sin\theta + r\theta^{\prime }\cos\theta
\right) \\
&=& b_{1} \left( \bar{p}_{1}\cos\theta -\frac{\bar{p}_{2}\sin\theta}{r}
\right) +b_{2} \left( \bar{p}_{1}\sin\theta +\frac{\bar{p}_{2}\cos\theta}{r}
\right)
\end{eqnarray*}
and
\begin{eqnarray*}
I_{31}&=& b_{1} \left(-Tr^{\prime }\cos\theta +Tr\theta^{\prime }\sin\theta
+r\cos\theta\right) +b_{2} \left(-Tr^{\prime }\sin\theta -Tr\theta^{\prime
}\cos\theta +r\sin\theta \right) \\
&=& b_{1} \left( -T\bar{p}_{1}\cos\theta + \frac{T\bar{p}_{2} \sin\theta}{r}
+ r\cos\theta \right) +b_{2}\left( -T\bar{p}_{1}\sin\theta - \frac{T\bar{p}%
_{2}\cos\theta}{r} + r\sin\theta\right)
\end{eqnarray*}
where $\bar{p}_{a}= \bar{\gamma}_{ab}\bar{q}^{b\prime}$ are the generalized momenta.\index{Momentum! generalized} Replacing with $\bar{q}_{a}= (r,\theta)$ and $\bar{\gamma}%
_{ab}=diag(1,r^{2})$, we find that $\bar{p}_{1}=r^{\prime }$ and $\bar{p}%
_{2}= r^{2}\theta^{\prime }$.

It is straightforward to show that the point Noether symmetries $\mathbf{A}_{21}, \mathbf{A}_{31}$, the Noether
functions $f_{21}, f_{31}$ and the FIs $I_{21}, I_{31}$ are the symmetries $%
\mathbf{X}_{9}, \mathbf{X}_{10}$, the functions $B_{9}, B_{10}$ and the FIs $I_{9}, I_{10}$, respectively, of \cite{Naz 2020}; while \\
- for $b_{1}=0, b_{2}=1$, they reduce to
the symmetries $\mathbf{X}_{5}, \mathbf{X}_{6}$, the functions $B_{5}, B_{6}$
and the FIs $I_{5}, I_{6}$, respectively, of \cite{Naz 2020} and \\
- for $b_{1}=1$, $b_{2}=0$, they reduce to
the symmetries $\mathbf{X}_{7}, \mathbf{X}_{8}$, the functions $B_{7}, B_{8}$
and the FIs $I_{7}, I_{8}$, respectively, of \cite{Naz 2020}.

As expected, the non-gradient KV (rotation) $Y\partial _{X}-X\partial _{Y}$ leads to the  LFI of angular momentum.\index{Momentum! angular}
\bigskip

Finally, using the three FIs $H, I_{0}, I_{2}$, we integrate the system (\ref{eq.ermint8a}) - (\ref{eq.ermint8b}) and find that in polar coordinates the solution is
\begin{eqnarray}
r^{2}(T)&=& \frac{1}{2H}\left( 2HT-I_{2}\right) ^{2}+\frac{I_{0}}{H}  \label{eq.int2} \\
\int \frac{d\theta }{\sqrt{I_{0}-\bar{F}(\theta )}} &=& \pm \int \frac{\sqrt{2}}{%
r^{2}(T)}dT=\pm \frac{1}{\sqrt{I_{0}}}\tan ^{-1}\left[ \frac{1}{\sqrt{2I_{0}}%
}(2HT-I_{2})\right]  \label{eq.int3}
\end{eqnarray}%
where $\bar{F}(\theta )=(\tan ^{2}\theta +1)N(\tan \theta)$. In Table 3 of \cite{Naz 2020}, the corresponding formula of (\ref{eq.int3}) gives $\tanh^{-1}$ instead of $\tan ^{-1}=\arctan$ which is the correct result.

\section{Conclusions}

\label{sec.Ermakov.conclusions}

Using recent results on the integrability of 2d conservative dynamical systems, we proved that the generalized Ermakov system is superintegrable and determined all the QFIs. We showed that the recent results of \cite{Naz 2020} can be obtained from the more general method outlined in chapter \ref{ch.2d.pots} by using Theorem \ref{thm.point.Noether.autonomous.systems}. Obviously, the methods discussed in the present chapter can be used by other authors in the study of similar dynamical systems.

%% file: higher_order_FIs.tex
\chapter{Higher order first integrals of autonomous dynamical systems}

\label{ch.Higher.order.FIs}

\section{Introduction}

\label{sec.intro.higher}

In the review paper \cite{Hietarinta 1987}, as we have seen in chapter \ref{ch.2d.pots}, the author considers the integrability of autonomous conservative dynamical systems with two degrees of freedom by means of mainly autonomous QFIs. The time-dependent FIs are totally absent, whereas there are occasional references to cubic FIs (CFIs) and, to a lesser extent, to quartic FIs (QUFIs). However, as it has been indicated in sections \ref{sec.int.FI.4} and \ref{subsec.Kep.3}, the time-dependent FIs are equally appropriate for establishing integrability \cite{Kozlov 1983, Vozmishcheva 2005}; the same applies to a greater degree for the higher order FIs. These two types of FIs are not usually considered because their determination is difficult, especially, when algebraic methods are employed. Fortunately, this does not apply to the geometric method where one uses the general results of differential geometry, concerning the collineations (symmetries) of the kinetic metric, to compute the FIs. An early example, in this direction, determines the time-dependent FIs of higher order of the geodesic equations in a general Riemannian manifold \cite{Katzin 1981}.

In this chapter, we apply the direct method --in the form established in chapter \ref{ch.QFIs.damping}-- and we determine the time-dependent and autonomous higher order polynomial FIs of autonomous dynamical systems. The results are stated in Theorem \ref{thm.mFIs}, which we apply in order to find new third order integrable/superintegrable systems (i.e. systems which allow CFIs that cannot be reduced to LFIs or QFIs).

\section{The conditions for an $m$th-order FI of an autonomous dynamical system}

\label{sec.higher.order.FIs}

We consider the autonomous holonomic dynamical system
\begin{equation}
\ddot{q}^{a}=-\Gamma _{bc}^{a}(q)\dot{q}^{b}\dot{q}^{c}-Q^{a}(q)
\label{eq.hfi1}
\end{equation}%
where $\Gamma _{bc}^{a}$ are the coefficients of the Riemannian connection of the kinetic metric $\gamma _{ab}(q)$ of the
system and $-Q^{a}(q)$ are the generalized forces.

We look for $m$th-order FIs of the form\index{First integral! $m$th-order}
\begin{equation}
I^{(m)}=\sum_{r=0}^{m}M_{i_{1}i_{2}...i_{r}}\dot{q}^{i_{1}} \dot{q}^{i_{2}}...\dot{q}^{i_{r}}= M+M_{i_{1}} \dot{q}^{i_{1}}+M_{i_{1}i_{2}}\dot{q}^{i_{1}} \dot{q}^{i_{2}}+ ...+M_{i_{1}i_{2}...i_{m}}\dot{q}^{i_{1}} \dot{q}^{i_{2}}...\dot{q}^{i_{m}}  \label{eq.hfi2}
\end{equation}%
where $M_{i_{1}...i_{r}}(t,q)$, with $r=0,1,...,m$, are totally symmetric $r$-rank tensors and the index $(m)$ denotes the order of the FI.

The condition $\frac{dI^{(m)}}{dt}=0$ along the dynamical equations (\ref{eq.hfi1}) leads to the following system of PDEs:
\begin{eqnarray}
M_{(i_{1}i_{2}...i_{m};i_{m+1})} &=&0  \label{eq.hfi4a} \\
M_{i_{1}i_{2}...i_{m},t}+M_{(i_{1}i_{2}...i_{m-1};i_{m})} &=&0
\label{eq.hfi4b} \\
M_{i_{1}i_{2}...i_{r},t}+M_{(i_{1}i_{2}...i_{r-1};i_{r})} -(r+1)M_{i_{1}i_{2}...i_{r}i_{r+1}}Q^{i_{r+1}} &=&0,\enskip r=1,2,...,m-1  \label{eq.hfi4c} \\
M_{,t}-M_{i_{1}}Q^{i_{1}} &=&0.  \label{eq.hfi4d}
\end{eqnarray}%
Equation (\ref{eq.hfi4a}) implies that $M_{i_{1}i_{2}...i_{m}}$ is an $m$th-order KT of the kinetic metric $\gamma_{ab}$.

Equations (\ref{eq.hfi4a}) - (\ref{eq.hfi4d})
must be supplemented with the integrability conditions $M_{,i_{1}t}=M_{,ti_{1}}$ and $M_{,[i_{1}i_{2}]}=0$ of the scalar $M$:
\begin{eqnarray}
M_{i_{1},tt}-2M_{i_{1}i_{2},t}Q^{i_{2}}+\left( M_{c}Q^{c}\right) _{,i_{1}}
&=&0  \label{eq.hfi5.1} \\
2\left( M_{[i_{1}|c|}Q^{c}\right) _{;i_{2}]}-M_{[i_{1};i_{2}],t} &=&0.
\label{eq.hfi5.2}
\end{eqnarray}

Equations (\ref{eq.hfi4a}) - (\ref{eq.hfi5.2}) constitute the system of equations which has to be solved.

\section{Determination of the $m$th-order FIs}

\label{sec.direct.meth}

In order to solve the system of equations (\ref{eq.hfi4a}) - (\ref{eq.hfi5.2}), we assume a polynomial form in $t$ for both the $m$th-order KT $%
M_{i_{1}...i_{m}}(t,q)$ and the $r$-rank totally symmetric tensors $%
M_{i_{1}...i_{r}}(t,q)$, where $r=1,2,...,m-1$, with coefficients depending
only on $q^{a}$. In particular, we assume that:\newline
a. The $\mathbf{m}${\textbf{th-order KT}} $M_{i_{1}...i_{m}}(t,q)$ has the form
\begin{equation}
M_{i_{1}...i_{m}}(t,q)=C_{(0)i_{1}...i_{m}}(q)+%
\sum_{N=1}^{n}C_{(N)i_{1}...i_{m}}(q)\frac{t^{N}}{N}  \label{eq.hfi5.3}
\end{equation}%
where $C_{(N)i_{1}...i_{m}}$, $N=0,1,...,n$, is a sequence of arbitrary $\mathbf{m}${\textbf{th-order KTs}} of the kinetic metric $\gamma_{ab}$ and $n$ is the
degree of the considered polynomial. \newline
b. The $\mathbf{r}${\textbf{-rank totally symmetric tensors}} (not in general KTs!) $M_{i_{1}...i_{r}}(t,q)$, where $r=1,2,...,m-1$, have the form
\begin{equation}
M_{i_{1}...i_{r}}(t,q)=%
\sum_{N_{r}=0}^{n_{r}}L_{(N_{r})i_{1}...i_{r}}(q)t^{N_{r}},\enskip r=1,2,...,m-1  \label{eq.hfi5.4}
\end{equation}%
where $L_{(N_{r})i_{1}...i_{r}}(q)$, $N_{r}=0,1,...,n_{r}$, are arbitrary $\mathbf{r}${\textbf{-rank totally symmetric tensors}} and $n_{r}$ is the degree of the considered polynomial.

The degrees $n$ and $n_{r}$ of the above polynomial expressions of $t$ may be infinite.

Substituting (\ref{eq.hfi5.3}) and (\ref{eq.hfi5.4}) in the system of PDEs (\ref{eq.hfi4b}) - (\ref{eq.hfi5.2}) (eq. (\ref{eq.hfi4a}) is
identically satisfied since $C_{(N)i_{1}...i_{m}}$ are assumed to be $m$th-order KTs), we find the
solution given in Theorem\footnote{The proof of Theorem \ref{thm.mFIs} is given in appendix \ref{app.proof.theorem.higher}.} \ref{thm.mFIs}.

\section{The Theorem}

\label{sec.theorem}

\begin{theorem}
\label{thm.mFIs} The independent $m$th-order FIs of the dynamical system (\ref{eq.hfi1}) are the following\footnote{The notation $I^{(m)}_{n}$ refers to an $m$th-order FI with time-dependence fixed by $n$.}: \bigskip

\textbf{Integral 1.}
\begin{eqnarray}
I_{n}^{(m)} &=&\left( -\frac{t^{n}}{n}L_{(n-1)(i_{1}...i_{m-1};i_{m})}-...-%
\frac{t^{2}}{2}%
L_{(1)(i_{1}...i_{m-1};i_{m})}-tL_{(0)(i_{1}...i_{m-1};i_{m})}+C_{(0)i_{1}...i_{m}}\right)
\dot{q}^{i_{1}}...\dot{q}^{i_{m}}+ \notag \\
&&+\sum_{r=1}^{m-1}\left(t^{n}L_{(n)i_{1}...i_{r}} +...+tL_{(1)i_{1}...i_{r}} +L_{(0)i_{1}...i_{r}}%
\right) \dot{q}^{i_{1}}...\dot{q}^{i_{r}} +s\frac{t^{n+1}}{n+1}+ \notag \\
&&+L_{(n-1)c}Q^{c}\frac{t^{n}}{n}+...+L_{(1)c}Q^{c}\frac{t^{2}}{2}%
+L_{(0)c}Q^{c}t+G(q) \label{integral.1}
\end{eqnarray}
where $C_{(0)i_{1}...i_{m}}$ and $L_{(N)(i_{1}...i_{m-1};i_{m})}$ for $%
N=0,1,...,n-1$ are $\mathbf{m}${\textbf{th-order KTs}}, $L_{(n)i_{1}...i_{m-1}}$ is an $\mathbf{(m-1)}${\textbf{th-order KT}}, $s$ is an arbitrary constant defined by the
condition
\begin{equation}
L_{(n)i_{1}}Q^{i_{1}}=s  \label{eq.FI1f}
\end{equation}%
while the vectors $L_{(N)i_{1}}$ and the {\textbf{totally symmetric tensors}} $%
L_{(A)i_{1}...i_{r}}$, $A=0,1,...,n$, $r=2,3,...,m-2$ satisfy the conditions:
\begin{eqnarray}
L_{(n)(i_{1}...i_{m-2};i_{m-1})} &=&-\frac{m}{n}%
L_{(n-1)(i_{1}...i_{m-1};i_{m})}Q^{i_{m}}  \label{eq.FI1a} \\
L_{(k-1)(i_{1}...i_{m-2};i_{m-1})} &=&-\frac{m}{k-1}%
L_{(k-2)(i_{1}...i_{m-1};i_{m})}Q^{i_{m}}-kL_{(k)i_{1}...i_{m-1}},\enskip %
k=2,3,...,n  \label{eq.FI1b} \\
L_{(0)(i_{1}...i_{m-2};i_{m-1})}
&=&mC_{(0)i_{1}...i_{m-1}i_{m}}Q^{i_{m}}-L_{(1)i_{1}...i_{m-1}}
\label{eq.FI1c} \\
L_{(n)(i_{1}...i_{r-1};i_{r})}
&=&(r+1)L_{(n)i_{1}...i_{r}i_{r+1}}Q^{i_{r+1}},\enskip r=2,3,...,m-2
\label{eq.FI1d} \\
L_{(k-1)(i_{1}...i_{r-1};i_{r})}
&=&(r+1)L_{(k-1)i_{1}...i_{r}i_{r+1}}Q^{i_{r+1}}-kL_{(k)i_{1}...i_{r}},%
\enskip k=1,2,...,n,\enskip r=2,3,...,m-2  \label{eq.FI1e} \\
\left( L_{(n-1)c}Q^{c}\right) _{,i_{1}} &=&2nL_{(n)i_{1}i_{2}}Q^{i_{2}}
\label{eq.FI1g} \\
\left( L_{(k-2)c}Q^{c}\right) _{,i_{1}}
&=&2(k-1)L_{(k-1)i_{1}i_{2}}Q^{i_{2}}-k(k-1)L_{(k)i_{1}},\enskip k=2,3,...,n
\label{eq.FI1h} \\
G_{,i_{1}} &=&2L_{(0)i_{1}i_{2}}Q^{i_{2}}-L_{(1)i_{1}}.  \label{eq.FI1i}
\end{eqnarray}

\textbf{Integral 2.}
\begin{equation}
I^{(m)}_{e}= \frac{e^{\lambda t}}{\lambda} \left(
-L_{(i_{1}...i_{m-1};i_{m})} \dot{q}^{i_{1}} ... \dot{q}^{i_{m}} + \lambda
\sum_{r=1}^{m-1} L_{i_{1}...i_{r}} \dot{q}^{i_{1}} ... \dot{q}^{i_{r}} +
L_{i_{1}}Q^{i_{1}} \right) \label{integral.2}
\end{equation}
where $\lambda\neq0$, $L_{(i_{1}...i_{m-1};i_{m})}$ is an $m$th-order KT and
the remaining totally symmetric tensors satisfy the conditions:
\begin{eqnarray}
L_{(i_{1}...i_{m-2};i_{m-1})}&=& -\frac{m}{\lambda}
L_{(i_{1}...i_{m-1};i_{m})} Q^{i_{m}} -\lambda L_{i_{1}...i_{m-1}}
\label{eq.FI2a} \\
L_{(i_{1}...i_{r-1};i_{r})}&=&(r+1) L_{i_{1}...i_{r}i_{r+1}} Q^{i_{r+1}}
-\lambda L_{i_{1}...i_{r}}, \enskip r=2,3,...,m-2  \label{eq.FI2b} \\
\left(L_{c}Q^{c}\right)_{,i_{1}}&=& 2\lambda L_{i_{1}i_{2}} Q^{i_{2}}
-\lambda^{2}L_{i_{1}}.  \label{eq.FI2c}
\end{eqnarray}
\end{theorem}

Theorem \ref{thm.mFIs} for $m=2$ reduces to Theorem \ref{theorem3} for the QFIs of autonomous dynamical systems.

Using mathematical induction, one also proves the following recursion formulae concerning the independent $m$th-order FIs (\ref{integral.1}) and (\ref{integral.2}).

\begin{proposition}
\label{pro.recursion.formulae}
For the independent $m$th-order FIs $I_{n}^{(m)}$ and $I^{(m)}_{e}$, the following recursion formulae hold:\newline
a. $I_{n}^{(k)}<I_{n}^{(k+1)}$, that is, each $k$th-order FI $I_{n}^{(k)}$ is a subcase of the next $(k+1)$th-order FI $I_{n}^{(k+1)}$ with the same degree $n$ of time-dependence for all $k\in \mathbb{N}$. \newline
b. $I_{\ell}^{(m)}<I_{\ell+1}^{(m)}$, that is, the $m$th-order FI $I_{\ell}^{(m)}$ with time-dependence fixed by $\ell$ is a subcase of the $m$th-order FI $I_{\ell+1}^{(m)}$ with time-dependence $\ell+1$ for all $\ell\in \mathbb{N}$. \newline
c. $I_{e}^{(k)}<I_{e}^{(k+1)}$, that is, each $k$th-order FI $I_{e}^{(k)}$ is a subcase of the next $(k+1)$th-order FI $I_{e}^{(k+1)}$ for all $k\in \mathbb{N}$.
\end{proposition}

\section{The independent FIs contained in the FI $I^{(m)}_{n}$}

\label{sec.independent.FIs}

In section \ref{section.2}, for $A^{a}_{b}(q)=0$, it is proved (see also appendix \ref{app.proof.QFIs.damping} and Theorem \ref{theorem3}) that the QFI $I^{(2)}_{n}$ consists of the following two independent QFIs: \newline
a. The QFI $J^{(2,1)}_{\ell}$ involving the odd vectors $L_{(2k+1)i_{1}}$, the KT $C_{(0)i_{1}i_{2}}\equiv C_{i_{1}i_{2}}$ and the function $G(q)$.\newline
b. The QFI $J^{(2,2)}_{\ell}$ involving the even vectors $L_{(2k)i_{1}}$, where $\ell\in \mathbb{N}$.

For the convenience of the reader we restate these FIs below.

a.
\begin{eqnarray}
J^{(2,1)}_{\ell} &=& \left( - \frac{t^{2\ell}}{2\ell} L_{(2\ell-1)(a;b)} - ... - \frac{%
t^{4}}{4} L_{(3)(a;b)} - \frac{t^{2}}{2} L_{(1)(a;b)} + C_{ab} \right) \dot{q%
}^{a} \dot{q}^{b} + t^{2\ell-1} L_{(2\ell-1)a}\dot{q}^{a} + ... +
t^{3}L_{(3)a}\dot{q}^{a} + \notag \\
&& + t L_{(1)a}\dot{q}^{a} + \frac{t^{2\ell}}{2\ell} L_{(2\ell-1)a}Q^{a} +
... + \frac{t^{4}}{4} L_{(3)a}Q^{a} + \frac{t^{2}}{2} L_{(1)a}Q^{a} + G(q) \label{eq.FI3a}
\end{eqnarray}
where $C_{ab}$ and $L_{(N)(a;b)}$ for $N=1,3,...,2\ell-1$ are KTs and the vectors $L_{(N)a}$ satisfy the conditions:
\begin{eqnarray*}
\left( L_{(2\ell-1)b} Q^{b} \right)_{,a} &=&
-2L_{(2\ell-1)(a;b)}Q^{b} \\
\left( L_{(k-1)b} Q^{b} \right)_{,a} &=&
-2L_{(k-1)(a;b)}Q^{b} - k(k+1)L_{(k+1)a}, \enskip k=2,4,...,2\ell-2 \\
G_{,a}&=& 2C_{ab}Q^{b} - L_{(1)a}.
\end{eqnarray*}

b.
\begin{eqnarray}
J^{(2,2)}_{\ell} &=& \left( - \frac{t^{2\ell+1}}{2\ell+1} L_{(2\ell)(a;b)} - ... -
\frac{t^{3}}{3} L_{(2)(a;b)} - t L_{(0)(a;b)} \right) \dot{q}^{a} \dot{q}%
^{b} + t^{2\ell} L_{(2\ell)a}\dot{q}^{a} + ... + t^{2}L_{(2)a}\dot{q}^{a} + \notag \\
&& + L_{(0)a}\dot{q}^{a}+ \frac{t^{2\ell+1}}{2\ell+1} L_{(2\ell)a}Q^{a} + ... + \frac{t^{3}}{3} L_{(2)a}Q^{a} +t L_{(0)a}Q^{a} \label{eq.FI3b}
\end{eqnarray}
where $L_{N(a;b)}$ for $N=0,2,...,2\ell$ are KTs and the involved vectors satisfy the conditions:
\begin{eqnarray*}
\left( L_{(2\ell)b} Q^{b}\right)_{,a} &=& -2L_{(2\ell)(a;b)}Q^{b} \\
\left( L_{(k-1)b}Q^{b}\right)_{,a} &=& -2L_{(k-1)(a;b)}Q^{b} - k(k+1)L_{(k+1)a}, \enskip k=1,3,...,2\ell-1.
\end{eqnarray*}

The notation $J^{(2,\kappa)}_{\ell}$, where $\ell=0,1,2,...$ and $\kappa=1,2$, denotes the two independent QFIs with time-dependence fixed by $\ell$. The index $\kappa$ counts the independent QFIs.

We note that:
\begin{eqnarray}
I^{(2)}_{2k}&=& J^{(2,1)}_{k} + J^{(2,2)}_{k}(L_{(2k)a}=KV) \label{eq.FI3c} \\
I^{(2)}_{2k+1}&=& J^{(2,1)}_{k+1}(L_{(2k+1)a}=KV) + J^{(2,2)}_{k} \label{eq.FI3d}
\end{eqnarray}
where\footnote{The notation $J^{(2,2)}_{k}(L_{(2k)a}=KV)$ means that the FI $J^{(2,2)}_{k}$ is computed over the constraint $L_{(2k)(a;b)}=0$, i.e. $L_{(2k)a}$ is a KV. Similarly for $J^{(2,1)}_{k+1}(L_{(2k+1)a}=KV$).} $k\in \mathbb{N}$. We note that \emph{the independent FIs $J_{\ell }^{(2,1)}$ and $J_{\ell }^{(2,2)}$, which generate the FI $I_{n}^{(2)}$, are constrained
only by second order KTs, while the FI $I_{n}^{(2)}$ is constrained by both
second order KTs and KVs. Therefore, $I_{n}^{(2)}$ depends not only to second
order symmetries (KTs) but also to lower order symmetries (KVs). In this
sense, $I_{n}^{(2)}$ is of an incomplete form because the lower symmetry can
be easily removed by adding in it an extra term of maximal order (here this is $m=2$) in velocities.}

It can be proved by induction (see also the results in section \ref{sec.practice}) that the above result for $m=2$ is generalized to $m$th-order FIs. Specifically, it holds the following proposition.

\begin{proposition}
\label{pro.independent.FIs}
The $m$th-order FI $I^{(m)}_{n}$ consists of the following two independent $m$th-order FIs:\newline
a. The FI $J^{(m,1)}_{\ell}$ whose coefficients are polynomials of $t$ containing even powers of $t$ for even products of velocities and odd powers of $t$ for odd products of velocities. \newline
b. The FI $J^{(m,2)}_{\ell}$ whose coefficients are polynomials of $t$ containing even powers of $t$ for odd products of velocities and odd powers of $t$ for even products of velocities.
\end{proposition}

If we choose\footnote{This choice is without loss of generality, because the odd order FIs can be derived as subcases from the FIs of even order, and vice versa.} an even order $m=2\nu$ ($\nu\in\mathbb{N}$), then the two independent FIs of the Proposition \ref{pro.independent.FIs} are given by the following formulae ($\ell\in\mathbb{N}$):

a.
\begin{eqnarray}
J^{(m=2\nu,1)}_{\ell}&=& \left( -\frac{t^{2\ell}}{2\ell} L_{(2\ell-1)(i_{1}...i_{m-1};i_{m})} - ... - \frac{t^{2}}{2} L_{(1)(i_{1}...i_{m-1};i_{m})} +C_{(0)i_{1}...i_{m}} \right) \dot{q}^{i_{1}} ... \dot{q}^{i_{m}} + \notag \\
&& + \sum_{1\leq r \leq m-1}^{odd} \left( t^{2\ell-1} L_{(2\ell-1)i_{1}...i_{r}} + ... + t^{3}L_{(3)i_{1}...i_{r}} +tL_{(1)i_{1}...i_{r}} \right) \dot{q}^{i_{1}} ... \dot{q}^{i_{r}} + \notag \\
&& + \sum_{1\leq r \leq m-1}^{even} \left( t^{2\ell} L_{(2\ell)i_{1}...i_{r}} + ... + t^{2}L_{(2)i_{1}...i_{r}} +L_{(0)i_{1}...i_{r}} \right) \dot{q}^{i_{1}} ... \dot{q}^{i_{r}} + \notag \\
&& + \frac{t^{2\ell}}{2\ell}L_{(2\ell-1)c}Q^{c} + ... + \frac{t^{2}}{2}L_{(1)c}Q^{c} + G(q) \label{eq.FI4a}
\end{eqnarray}
where $C_{(0)i_{1}...i_{m}}$ and $L_{(N)(i_{1}...i_{m-1};i_{m})}$ for $N=1,3,...,2\ell-1$ are $m$th-order KTs and the following conditions are satisfied:
\begin{eqnarray}
L_{(2\ell)(i_{1}...i_{m-2};i_{m-1})}&=& -\frac{m}{2\ell} L_{(2\ell-1)(i_{1}...i_{m-1};i_{m})} Q^{i_{m}} \label{eq.FI4.1} \\
L_{(k-1)(i_{1}...i_{m-2};i_{m-1})}&=&-\frac{m}{k-1} L_{(k-2)(i_{1}...i_{m-1};i_{m})} Q^{i_{m}} -kL_{(k)i_{1}...i_{m-1}}, \enskip k=3,5,...,2\ell-1 \label{eq.FI4.2} \\
L_{(0)(i_{1}...i_{m-2};i_{m-1})}&=& mC_{(0)i_{1}...i_{m-1}i_{m}} Q^{i_{m}} -L_{(1)i_{1}...i_{m-1}} \label{eq.FI4.3} \\
L_{(2\ell)(i_{1}...i_{r-1};i_{r})}&=& (r+1) L_{(2\ell)i_{1}...i_{r}i_{r+1}} Q^{i_{r+1}}, \enskip r=3,5,...,m-3 \label{eq.FI4.4} \\ L_{(k-1)(i_{1}...i_{r-1};i_{r})}&=& (r+1) L_{(k-1)i_{1}...i_{r}i_{r+1}} Q^{i_{r+1}} -kL_{(k)i_{1}...i_{r}}, \enskip k=1,3,...,2\ell-1, r=3,5,...,m-3 \notag \\
\label{eq.FI4.5} \\
L_{(k-1)(i_{1}...i_{r-1};i_{r})}&=& (r+1) L_{(k-1)i_{1}...i_{r}i_{r+1}} Q^{i_{r+1}} -kL_{(k)i_{1}...i_{r}}, \enskip k=2,4,...,2\ell, \enskip r=2,4,...,m-2 \notag \\
\label{eq.FI4.6} \\
\left( L_{(2\ell-1)c}Q^{c} \right)_{,i_{1}} &=& 4\ell L_{(2\ell)i_{1}i_{2}}Q^{i_{2}} \label{eq.FI4.7} \\
\left( L_{(k-2)c}Q^{c} \right)_{,i_{1}} &=& 2(k-1)L_{(k-1)i_{1}i_{2}}Q^{i_{2}} -k(k-1)L_{(k)i_{1}}, \enskip k=3,5,...,2\ell-1 \label{eq.FI4.8} \\
G_{,i_{1}}&=& 2L_{(0)i_{1}i_{2}}Q^{i_{2}} -L_{(1)i_{1}}. \label{eq.FI4.9}
\end{eqnarray}

b.
\begin{eqnarray}
J^{(m=2\nu,2)}_{\ell}&=& \left( -\frac{t^{2\ell+1}}{2\ell+1} L_{(2\ell)(i_{1}...i_{m-1};i_{m})} - ... - \frac{t^{3}}{3} L_{(2)(i_{1}...i_{m-1};i_{m})} - tL_{(0)(i_{1}...i_{m-1};i_{m})} \right) \dot{q}^{i_{1}} ... \dot{q}^{i_{m}} + \notag \\
&& + \sum_{1\leq r \leq m-1}^{odd} \left( t^{2\ell} L_{(2\ell)i_{1}...i_{r}} + ... + t^{2}L_{(2)i_{1}...i_{r}} +L_{(0)i_{1}...i_{r}} \right) \dot{q}^{i_{1}} ... \dot{q}^{i_{r}}+ \notag \\
&& + \sum_{1\leq r \leq m-1}^{even} \left( t^{2\ell+1} L_{(2\ell+1)i_{1}...i_{r}} + ... + t^{3}L_{(3)i_{1}...i_{r}} +tL_{(1)i_{1}...i_{r}} \right) \dot{q}^{i_{1}} ... \dot{q}^{i_{r}} + \notag \\
&& + \frac{t^{2\ell+1}}{2\ell+1}L_{(2\ell)c}Q^{c} + ... + \frac{t^{3}}{3}L_{(2)c}Q^{c} + tL_{(0)c}Q^{c} \label{eq.FI4b}
\end{eqnarray}
where $L_{(N)(i_{1}...i_{m-1};i_{m})}$ for $N=0,2,...,2\ell$ are $m$th-order KTs and the following conditions are satisfied:
\begin{eqnarray}
L_{(2\ell+1)(i_{1}...i_{m-2};i_{m-1})}&=& -\frac{m}{2\ell+1} L_{(2\ell)(i_{1}...i_{m-1};i_{m})} Q^{i_{m}} \label{eq.FI5.1} \\
L_{(k-1)(i_{1}...i_{m-2};i_{m-1})}&=& -\frac{m}{k-1} L_{(k-2)(i_{1}...i_{m-1};i_{m})} Q^{i_{m}} -kL_{(k)i_{1}...i_{m-1}}, \enskip k=2,4,...,2\ell \label{eq.FI5.2} \\
L_{(2\ell+1)(i_{1}...i_{r-1};i_{r})}&=& (r+1) L_{(2\ell+1)i_{1}...i_{r}i_{r+1}} Q^{i_{r+1}}, \enskip r=3,5,...,m-3 \label{eq.FI5.3} \\
L_{(k-1)(i_{1}...i_{r-1};i_{r})}&=& (r+1) L_{(k-1)i_{1}...i_{r}i_{r+1}} Q^{i_{r+1}} -kL_{(k)i_{1}...i_{r}}, \enskip k=1,3,...,2\ell+1, r=2,4,...,m-2 \notag \\
\label{eq.FI5.4} \\
L_{(k-1)(i_{1}...i_{r-1};i_{r})}&=& (r+1) L_{(k-1)i_{1}...i_{r}i_{r+1}} Q^{i_{r+1}} -kL_{(k)i_{1}...i_{r}}, \enskip k=2,4,...,2\ell, \enskip r=3,5,...,m-3 \notag \\
\label{eq.FI5.5} \\
\left( L_{(2\ell)c}Q^{c} \right)_{,i_{1}} &=& 2(2\ell+1)L_{(2\ell+1)i_{1}i_{2}}Q^{i_{2}} \label{eq.FI5.6} \\
\left( L_{(k-2)c}Q^{c} \right)_{,i_{1}} &=& 2(k-1)L_{(k-1)i_{1}i_{2}}Q^{i_{2}} -k(k-1)L_{(k)i_{1}}, \enskip k=2,4,...,2\ell. \label{eq.FI5.7}
\end{eqnarray}

For $\nu=1 \implies m=2$, the QFIs (\ref{eq.FI4a}) and (\ref{eq.FI4b}) reduce to the QFIs (\ref{eq.FI3a}) and (\ref{eq.FI3b}), respectively.

Finally, for the even order FIs we note that:
\begin{eqnarray*}
I^{(2\nu)}_{2k}&=& J^{(2\nu,1)}_{k} + J^{(2\nu,2)}_{k} \left( L_{(2k)(i_{1}...i_{m-1};i_{m})}=0; L_{(2k+1)(i_{1}...i_{r})}=0, 1\leq r \leq m-1, r=even \right) \\
I^{(2\nu)}_{2k+1}&=& J^{(2\nu,1)}_{k+1}\left( L_{(2k+1)(i_{1}...i_{m-1};i_{m})}=0; L_{(2k+2)(i_{1}...i_{r})}=0, 1\leq r \leq m-1, r=even \right) + J^{(2\nu,2)}_{k}
\end{eqnarray*}
where $m=2\nu$, while for the odd order FIs:
\begin{eqnarray*}
I^{(2\nu+1)}_{2k}&=& J^{(2\nu+2,1)}_{k} \left(M_{i_{1}...i_{m}}=0\right) + J^{(2\nu+2,2)}_{k} \left( M_{i_{1}...i_{m}}=0; L_{(2k+1)(i_{1}...i_{r})}=0, 1\leq r \leq m-1, r=even \right) \\
I^{(2\nu+1)}_{2k+1}&=& J^{(2\nu+2,1)}_{k+1}\left( M_{i_{1}...i_{m}}=0; L_{(2k+2)(i_{1}...i_{r})}=0, 1\leq r \leq m-1, r=even \right) + J^{(2\nu+2,2)}_{k}\left(M_{i_{1}...i_{m}}=0\right)
\end{eqnarray*}
where $m=2\nu+2$ and $M_{i_{1}...i_{m}}$ is the coefficient containing the $m$th-order KTs. For an odd order $m=2\nu+1$, the corresponding independent FIs $J^{(m,1)}_{\ell}$ and $J^{(m,2)}_{\ell}$ of the FI $I^{(m)}_{n}$ are given by the relations:
\[
J^{(2\nu+1,1)}_{\ell} = J^{(2\nu+2,1)}_{\ell} \left(M_{i_{1}...i_{m}}=0\right) \enskip \text{and} \enskip J^{(2\nu+1,2)}_{\ell} = J^{(2\nu+2,2)}_{\ell} \left(M_{i_{1}...i_{m}}=0\right).
\]

\section{The use of $I_{n}^{(m)}$ and $I_{e}^{(m)}$ in practice}

\label{sec.practice}

Undoubtedly, the results stated in Theorem \ref{thm.mFIs} are complicated and it would be rather hard to be of practical value as they are stated. Therefore, in order to show how they are used in practice, we write the FIs $I_{n}^{(m)}$ and $I_{e}^{(m)}$ given by (\ref{integral.1}) and (\ref{integral.2}), respectively, explicitly for the case of QFIs, CFIs and QUFIs which are the cases most likely to be used in practice.

\subsection{The FI $I^{(m)}_{n}$}

\label{sec.FImn}

For the values $m=2,3,4$, the FI $I^{(m)}_{n}$ given by (\ref{integral.1}) gives the following\footnote{For $m=0$ we have the trivial FI $I^{(0)}_{n}=const$.}.

\subsubsection{For $m=2$ (QFIs)}

\begin{eqnarray*}
I^{(2)}_{n}&=& \left( -\frac{t^{n}}{n} L_{(n-1)(i_{1};i_{2})} - ... - \frac{t^{2}}{2} L_{(1)(i_{1};i_{2})} - tL_{(0)(i_{1};i_{2})} +C_{(0)i_{1}i_{2}} \right) \dot{q}^{i_{1}} \dot{q}^{i_{2}} + \\
&& + \left( t^{n} L_{(n)i_{1}} + ... + tL_{(1)i_{1}} +L_{(0)i_{1}} \right) \dot{q}^{i_{1}} + s\frac{t^{n+1}}{n+1} + L_{(n-1)c}Q^{c}\frac{t^{n}}{n} + ... + \\
&& +L_{(1)c}Q^{c}\frac{t^{2}}{2} + L_{(0)c}Q^{c}t + G(q)
\end{eqnarray*}
where $C_{(0)i_{1}i_{2}}$ and $L_{(N)(i_{1};i_{2})}$ for $N=0,1,...,n-1$ are second order KTs, $L_{(n)i_{1}}$ is a KV, $s$ is an arbitrary constant defined by the requirement $L_{(n)i_{1}}Q^{i_{1}}= s$ and the vectors $L_{(A)i_{1}}$, $A=0,1,...,n$, satisfy the conditions:
\begin{eqnarray*}
\left( L_{(n-1)c}Q^{c} \right)_{,i_{1}} &=& -2L_{(n-1)(i_{1};i_{2})}Q^{i_{2}} \\
\left( L_{(k-1)c}Q^{c} \right)_{,i_{1}} &=& -2L_{(k-1)(i_{1};i_{2})}Q^{i_{2}} -k(k+1)L_{(k+1)i_{1}}, \enskip k=1,2,...,n-1 \\
G_{,i_{1}}&=& 2C_{(0)i_{1}i_{2}}Q^{i_{2}} -L_{(1)i_{1}}.
\end{eqnarray*}
As we have shown in section \ref{sec.independent.FIs}, the QFI $I^{(2)}_{n}$ consists of the independent QFIs $J^{(2,1)}_{\ell}$ and $J^{(2,2)}_{\ell}$ given by (\ref{eq.FI3a}) and (\ref{eq.FI3b}), respectively.

\subsubsection{For $m=3$ (CFIs)}

\begin{eqnarray*}
I^{(3)}_{n}&=& \left( -\frac{t^{n}}{n} L_{(n-1)(i_{1}i_{2};i_{3})} - ... - \frac{t^{2}}{2} L_{(1)(i_{1}i_{2};i_{3})} - tL_{(0)(i_{1}i_{2};i_{3})} +C_{(0)i_{1}i_{2}i_{3}} \right) \dot{q}^{i_{1}} \dot{q}^{i_{2}} \dot{q}^{i_{3}} + \\
&& + \left( t^{n} L_{(n)i_{1}i_{2}} + ... + tL_{(1)i_{1}i_{2}} +L_{(0)i_{1}i_{2}} \right) \dot{q}^{i_{1}} \dot{q}^{i_{2}} + \left( t^{n} L_{(n)i_{1}} + ... + tL_{(1)i_{1}} +L_{(0)i_{1}} \right) \dot{q}^{i_{1}} + \\
&& + s\frac{t^{n+1}}{n+1} + L_{(n-1)c}Q^{c}\frac{t^{n}}{n} + ... + L_{(1)c}Q^{c}\frac{t^{2}}{2} + L_{(0)c}Q^{c}t + G(q)
\end{eqnarray*}
where $C_{(0)i_{1}i_{2}i_{3}}$ and $L_{(N)(i_{1}i_{2};i_{3})}$ for $N=0,...,n-1$ are third order KTs, $L_{(n)i_{1}i_{2}}$ is a second order KT, $s$ is an arbitrary constant defined by the condition $L_{(n)i_{1}}Q^{i_{1}}=s$, while the vectors $L_{(A)i_{1}}$ and the symmetric tensors $L_{(A)i_{1}i_{2}}$, $A=1,2,..,n$, satisfy the conditions:
\begin{eqnarray*}
L_{(0)(i_{1};i_{2})}&=& 3 C_{(0)i_{1}i_{2}i_{3}} Q^{i_{3}} -L_{(1)i_{1}i_{2}} \\
L_{(k)(i_{1};i_{2})}&=& -\frac{3}{k} L_{(k-1)(i_{1}i_{2};i_{3})} Q^{i_{3}} -(k+1)L_{(k+1)i_{1}i_{2}}, \enskip k=1,2,...,n-1 \\ L_{(n)(i_{1};i_{2})}&=& -\frac{3}{n} L_{(n-1)(i_{1}i_{2};i_{3})} Q^{i_{3}} \\
\left( L_{(n-1)c}Q^{c} \right)_{,i_{1}} &=& 2nL_{(n)i_{1}i_{2}}Q^{i_{2}} \\
\left( L_{(k-1)c}Q^{c} \right)_{,i_{1}} &=& 2kL_{(k)i_{1}i_{2}}Q^{i_{2}} -k(k+1)L_{(k+1)i_{1}}, \enskip k=1,2,...,n-1 \\
G_{,i_{1}}&=& 2L_{(0)i_{1}i_{2}}Q^{i_{2}} -L_{(1)i_{1}}.
\end{eqnarray*}

For various values of $n$ (i.e. the degree of the time-dependence), we have the following CFIs:

- For $n=0$.

\begin{equation*}
I^{(3)}_{0}= C_{(0)i_{1}i_{2}i_{3}} \dot{q}^{i_{1}} \dot{q}^{i_{2}} \dot{q}^{i_{3}} +L_{(0)i_{1}i_{2}} \dot{q}^{i_{1}} \dot{q}^{i_{2}} +L_{(0)i_{1}} \dot{q}^{i_{1}} + st + G(q)
\end{equation*}
where $C_{(0)i_{1}i_{2}i_{3}}$ is a third order KT, $L_{(0)i_{1}i_{2}}$ is a second order KT, $L_{(0)i_{1}}Q^{i_{1}}= s$, $L_{(0)(i_{1};i_{2})}= 3C_{(0)i_{1}i_{2}i_{3}} Q^{i_{3}}$ and $G_{,i_{1}}= 2L_{(0)i_{1}i_{2}}Q^{i_{2}}$.

This CFI consists of the two independent FIs
\begin{eqnarray*}
I^{(3,1)}_{0} &=& L_{(0)i_{1}i_{2}} \dot{q}^{i_{1}} \dot{q}^{i_{2}} + G(q) = J^{(3,1)}_{0} \\
I^{(3,2)}_{0} &=& C_{(0)i_{1}i_{2}i_{3}} \dot{q}^{i_{1}} \dot{q}^{i_{2}} \dot{q}^{i_{3}} +L_{(0)i_{1}} \dot{q}^{i_{1}} + st = J^{(3,2)}_{0}\left(L_{(1)i_{1}i_{2}}=0\right).
\end{eqnarray*}

- For $n=1$.

\begin{eqnarray*}
I^{(3)}_{1}&=& \left( - tL_{(0)(i_{1}i_{2};i_{3})} +C_{(0)i_{1}i_{2}i_{3}} \right) \dot{q}^{i_{1}} \dot{q}^{i_{2}} \dot{q}^{i_{3}} + \left( tL_{(1)i_{1}i_{2}} +L_{(0)i_{1}i_{2}} \right) \dot{q}^{i_{1}} \dot{q}^{i_{2}} + \left( tL_{(1)i_{1}} +L_{(0)i_{1}} \right) \dot{q}^{i_{1}}+ \\
&& + s\frac{t^{2}}{2} + L_{(0)c}Q^{c}t + G(q)
\end{eqnarray*}
where $C_{(0)i_{1}i_{2}i_{3}}$ and $L_{(0)(i_{1}i_{2};i_{3})}$ are third order KTs, $L_{(1)i_{1}i_{2}}$ is a second order KT and the following conditions are satisfied:
\begin{eqnarray*}
L_{(0)(i_{1};i_{2})}&=& 3 C_{(0)i_{1}i_{2}i_{3}} Q^{i_{3}} -L_{(1)i_{1}i_{2}} \\
L_{(1)(i_{1};i_{2})}&=& -3 L_{(0)(i_{1}i_{2};i_{3})} Q^{i_{3}} \\
L_{(1)i_{1}}Q^{i_{1}}&=& s \\
\left( L_{(0)c}Q^{c} \right)_{,i_{1}} &=& 2L_{(1)i_{1}i_{2}}Q^{i_{2}} \\
G_{,i_{1}}&=& 2L_{(0)i_{1}i_{2}}Q^{i_{2}} -L_{(1)i_{1}}.
\end{eqnarray*}

This CFI consists of the two independent CFIs:
\begin{eqnarray*}
I^{(3,1)}_{1}&=& - tL_{(0)(i_{1}i_{2};i_{3})} \dot{q}^{i_{1}} \dot{q}^{i_{2}} \dot{q}^{i_{3}} + L_{(0)i_{1}i_{2}} \dot{q}^{i_{1}} \dot{q}^{i_{2}} + tL_{(1)i_{1}}\dot{q}^{i_{1}}+ s\frac{t^{2}}{2} + G(q) =J^{(3,1)}_{1}\left( L_{(2)i_{1}i_{2}}=0 \right) \\
I^{(3,2)}_{1}&=& C_{(0)i_{1}i_{2}i_{3}} \dot{q}^{i_{1}} \dot{q}^{i_{2}} \dot{q}^{i_{3}} + tL_{(1)i_{1}i_{2}} \dot{q}^{i_{1}} \dot{q}^{i_{2}} + L_{(0)i_{1}} \dot{q}^{i_{1}}+ tL_{(0)c}Q^{c}= J^{(3,2)}_{0}.
\end{eqnarray*}

- For $n=2$.

\begin{eqnarray*}
I^{(3)}_{2}&=& \left( - \frac{t^{2}}{2} L_{(1)(i_{1}i_{2};i_{3})} - tL_{(0)(i_{1}i_{2};i_{3})} +C_{(0)i_{1}i_{2}i_{3}} \right) \dot{q}^{i_{1}} \dot{q}^{i_{2}} \dot{q}^{i_{3}} + \left( t^{2} L_{(2)i_{1}i_{2}} +tL_{(1)i_{1}i_{2}} +L_{(0)i_{1}i_{2}} \right) \dot{q}^{i_{1}} \dot{q}^{i_{2}} + \\
&& +\left( t^{2} L_{(2)i_{1}} + tL_{(1)i_{1}} +L_{(0)i_{1}} \right) \dot{q}^{i_{1}} + s\frac{t^{3}}{3} + L_{(1)c}Q^{c}\frac{t^{2}}{2} + L_{(0)c}Q^{c}t + G(q)
\end{eqnarray*}
where $C_{(0)i_{1}i_{2}i_{3}}$ and $L_{(N)(i_{1}i_{2};i_{3})}$ for $N=0,1$ are third order KTs, $L_{(2)i_{1}i_{2}}$ is a second order KT and the following conditions are satisfied:
\begin{eqnarray*}
L_{(0)(i_{1};i_{2})}&=& 3 C_{(0)i_{1}i_{2}i_{3}} Q^{i_{3}} -L_{(1)i_{1}i_{2}} \\
L_{(1)(i_{1};i_{2})}&=& -3 L_{(0)(i_{1}i_{2};i_{3})} Q^{i_{3}} -2L_{(2)i_{1}i_{2}} \\
L_{(2)(i_{1};i_{2})}&=& -\frac{3}{2} L_{(1)(i_{1}i_{2};i_{3})} Q^{i_{3}} \\
L_{(2)i_{1}}Q^{i_{1}}&=& s \\
\left( L_{(1)c}Q^{c} \right)_{,i_{1}} &=& 4L_{(2)i_{1}i_{2}}Q^{i_{2}} \\
\left( L_{(0)c}Q^{c} \right)_{,i_{1}} &=& 2L_{(1)i_{1}i_{2}}Q^{i_{2}} -2L_{(2)i_{1}} \\
G_{,i_{1}}&=& 2L_{(0)i_{1}i_{2}}Q^{i_{2}} -L_{(1)i_{1}}.
\end{eqnarray*}

This CFI consists of the two independent CFIs:
\begin{eqnarray*}
I^{(3,1)}_{2}&=& - tL_{(0)(i_{1}i_{2};i_{3})} \dot{q}^{i_{1}} \dot{q}^{i_{2}} \dot{q}^{i_{3}} + \left( t^{2} L_{(2)i_{1}i_{2}} +L_{(0)i_{1}i_{2}} \right) \dot{q}^{i_{1}} \dot{q}^{i_{2}}  + tL_{(1)i_{1}} \dot{q}^{i_{1}} + L_{(1)c}Q^{c}\frac{t^{2}}{2} + G(q)\\
&=& J^{(3,1)}_{1} \\
I^{(3,2)}_{2}&=& \left( - \frac{t^{2}}{2} L_{(1)(i_{1}i_{2};i_{3})} +C_{(0)i_{1}i_{2}i_{3}} \right) \dot{q}^{i_{1}} \dot{q}^{i_{2}} \dot{q}^{i_{3}} +tL_{(1)i_{1}i_{2}} \dot{q}^{i_{1}} \dot{q}^{i_{2}} +\left( t^{2} L_{(2)i_{1}} +L_{(0)i_{1}} \right) \dot{q}^{i_{1}} + s\frac{t^{3}}{3} + L_{(0)c}Q^{c}t \\
&=& J^{(3,2)}_{1}\left( L_{(3)i_{1}i_{2}}=0\right).
\end{eqnarray*}

\subsubsection{For $m=4$ (QUFIs)}

\begin{eqnarray*}
I^{(4)}_{n}&=& \left( -\frac{t^{n}}{n} L_{(n-1)(i_{1}i_{2}i_{3};i_{4})} - ... - \frac{t^{2}}{2} L_{(1)(i_{1}i_{2}i_{3};i_{4})} - tL_{(0)(i_{1}i_{2}i_{3};i_{4})} +C_{(0)i_{1}i_{2}i_{3}i_{4}} \right) \dot{q}^{i_{1}} \dot{q}^{i_{3}} \dot{q}^{i_{3}} \dot{q}^{i_{4}} + \\
&& + \left( t^{n} L_{(n)i_{1}i_{2}i_{3}} + ... + tL_{(1)i_{1}i_{2}i_{3}} +L_{(0)i_{1}i_{2}i_{3}} \right) \dot{q}^{i_{1}}\dot{q}^{i_{2}} \dot{q}^{i_{3}} + \left( t^{n} L_{(n)i_{1}i_{2}} + ... + tL_{(1)i_{1}i_{2}} +L_{(0)i_{1}i_{2}} \right) \dot{q}^{i_{1}} \dot{q}^{i_{2}} + \\
&&+ \left( t^{n} L_{(n)i_{1}} + ... + tL_{(1)i_{1}} +L_{(0)i_{1}} \right) \dot{q}^{i_{1}} + s\frac{t^{n+1}}{n+1} + L_{(n-1)c}Q^{c}\frac{t^{n}}{n} + ... + L_{(1)c}Q^{c}\frac{t^{2}}{2} + L_{(0)c}Q^{c}t + G(q)
\end{eqnarray*}
where $C_{(0)i_{1}i_{2}i_{3}i_{4}}$ and $L_{(N)(i_{1}i_{2}i_{3};i_{4})}$ for $N=0,...,n-1$ are fourth order KTs, $L_{(n)i_{1}i_{2}i_{3}}$ is a third order KT, $s$ is an arbitrary constant defined by the condition $L_{(n)i_{1}}Q^{i_{1}}= s$, while the vectors $L_{(A)i_{1}}$ and the totally symmetric tensors $L_{(A)i_{1}i_{2}}, L_{(A)i_{1}i_{2}i_{3}}$ satisfy the conditions:
\begin{eqnarray*}
L_{(n)(i_{1}i_{2};i_{3})}&=& -\frac{4}{n} L_{(n-1)(i_{1}i_{2}i_{3};i_{4})} Q^{i_{4}} \\ L_{(k-1)(i_{1}i_{2};i_{3})}&=& -\frac{4}{k-1} L_{(k-2)(i_{1}i_{2}i_{3};i_{4})} Q^{i_{4}} -kL_{(k)i_{1}i_{2}i_{3}}, \enskip k=2,3,...,n \\
L_{(0)(i_{1}i_{2};i_{3})}&=& 4 C_{(0)i_{1}i_{2}i_{3}i_{4}} Q^{i_{4}} -L_{(1)i_{1}i_{2}i_{3}} \\
L_{(n)(i_{1};i_{2})}&=& 3 L_{(n)i_{1}i_{2}i_{3}} Q^{i_{3}} \\
L_{(k-1)(i_{1};i_{2})}&=& 3L_{(k-1)i_{1}i_{2}i_{3}} Q^{i_{3}} -kL_{(k)i_{1}i_{2}}, \enskip k=1,2,...,n \\
\left( L_{(n-1)c}Q^{c} \right)_{,i_{1}} &=& 2nL_{(n)i_{1}i_{2}}Q^{i_{2}} \\
\left( L_{(k-2)c}Q^{c} \right)_{,i_{1}} &=& 2(k-1)L_{(k-1)i_{1}i_{2}}Q^{i_{2}} -k(k-1)L_{(k)i_{1}}, \enskip k=2,3,...,n \\
G_{,i_{1}}&=& 2L_{(0)i_{1}i_{2}}Q^{i_{2}} -L_{(1)i_{1}}.
\end{eqnarray*}

For $n=0$ we have:
\[
I^{(4)}_{0}= C_{(0)i_{1}i_{2}i_{3}i_{4}} \dot{q}^{i_{1}} \dot{q}^{i_{3}} \dot{q}^{i_{3}} \dot{q}^{i_{4}} + L_{(0)i_{1}i_{2}i_{3}} \dot{q}^{i_{1}}\dot{q}^{i_{2}} \dot{q}^{i_{3}} + L_{(0)i_{1}i_{2}} \dot{q}^{i_{1}} \dot{q}^{i_{2}} +L_{(0)i_{1}} \dot{q}^{i_{1}} + st + G(q)
\]
where $C_{(0)i_{1}i_{2}i_{3}i_{4}}$ is a fourth order KT, $L_{(0)i_{1}i_{2}i_{3}}$ is a third order KT and the following conditions are satisfied:
\begin{eqnarray*}
L_{(0)i_{1}}Q^{i_{1}}&=& s \\
L_{(0)(i_{1}i_{2};i_{3})}&=& 4 C_{(0)i_{1}i_{2}i_{3}i_{4}} Q^{i_{4}} \\
L_{(0)(i_{1};i_{2})}&=& 3L_{(0)i_{1}i_{2}i_{3}} Q^{i_{3}} \\
G_{,i_{1}}&=& 2L_{(0)i_{1}i_{2}}Q^{i_{2}}.
\end{eqnarray*}

This FI consists of the two independent FIs:
\begin{eqnarray*}
I^{(4,1)}_{0}&=& C_{(0)i_{1}i_{2}i_{3}i_{4}} \dot{q}^{i_{1}} \dot{q}^{i_{3}} \dot{q}^{i_{3}} \dot{q}^{i_{4}} + L_{(0)i_{1}i_{2}} \dot{q}^{i_{1}} \dot{q}^{i_{2}} + G(q) =J^{(4,1)}_{0} \\
I^{(4,2)}_{0}&=& L_{(0)i_{1}i_{2}i_{3}} \dot{q}^{i_{1}}\dot{q}^{i_{2}} \dot{q}^{i_{3}} +L_{(0)i_{1}} \dot{q}^{i_{1}} + st =J^{(4,2)}_{0}\left( L_{(0)i_{1}i_{2}i_{3}}=KT; L_{(1)i_{1}i_{2}}=0 \right).
\end{eqnarray*}

\subsection{The FI $I^{(m)}_{e}$}

\label{sec.exponentialFI}

For the values $m=2,3,4$ the FI $I^{(m)}_{e}$ given by (\ref{integral.2}) gives the following\footnote{The final expressions of the FIs have been multiplied with $\lambda$ in order to be simplified. This does not change the associated conditions because $\lambda$ is a non-zero constant. However, to avoid mistakes, the derivation of the conditions for the case $m=2$ should be done by using the original formula (\ref{integral.2}).}.

\subsubsection{For $m=2$ (QFIs)}

\begin{equation*}
I^{(2)}_{e}= e^{\lambda t} \left( -L_{(i_{1};i_{2})} \dot{q}^{i_{1}} \dot{q}^{i_{2}} + \lambda L_{i_{1}} \dot{q}^{i_{1}} + L_{i_{1}}Q^{i_{1}} \right)
\end{equation*}
where $\lambda\neq0$, $L_{(i_{1};i_{2})}$ is a second order KT and\footnote{This condition is obtained directly from equation (\ref{eq.FI2a}). The scalar $L$ that appears in the right-hand side (since for $m=2$ all indices vanish) of (\ref{eq.FI2a}) is equal to $\frac{1}{\lambda} L_{i_{1}}Q^{i_{1}}$.} $\left(L_{c}Q^{c}\right)_{,i_{1}}= -2 L_{(i_{1};i_{2})} Q^{i_{2}} -\lambda^{2} L_{i_{1}}$.

This is the Integral 3 of Theorem \ref{theorem3}.

\subsubsection{For $m=3$ (CFIs)}

\begin{equation*}
I^{(3)}_{e}= e^{\lambda t} \left( -L_{(i_{1}i_{2};i_{3})} \dot{q}^{i_{1}} \dot{q}^{i_{2}} \dot{q}^{i_{3}} + \lambda L_{i_{1}i_{2}} \dot{q}^{i_{1}} \dot{q}^{i_{2}} + \lambda L_{i_{1}} \dot{q}^{i_{1}} + L_{i_{1}}Q^{i_{1}} \right)
\end{equation*}
where $\lambda\neq0$, $L_{(i_{1}i_{2};i_{3})}$ is a third order KT, $L_{(i_{1};i_{2})}= -\frac{3}{\lambda} L_{(i_{1}i_{2};i_{3})} Q^{i_{3}} -\lambda L_{i_{1}i_{2}}$ and $\left(L_{c}Q^{c}\right)_{,i_{1}}=$ $2\lambda L_{i_{1}i_{2}} Q^{i_{2}} -\lambda^{2}L_{i_{1}}$.

\subsubsection{For $m=4$ (QUFIs)}

\begin{equation*}
I^{(4)}_{e}= e^{\lambda t} \left( -L_{(i_{1}i_{2}i_{3};i_{4})} \dot{q}^{i_{1}} \dot{q}^{i_{2}} \dot{q}^{i_{3}} \dot{q}^{i_{4}} + \lambda L_{i_{1}i_{2}i_{3}} \dot{q}^{i_{1}} \dot{q}^{i_{2}} \dot{q}^{i_{3}} + \lambda L_{i_{1}i_{2}} \dot{q}^{i_{1}} \dot{q}^{i_{2}} + \lambda L_{i_{1}} \dot{q}^{i_{1}} + L_{i_{1}}Q^{i_{1}} \right)
\end{equation*}
where $\lambda\neq0$, $L_{(i_{1}i_{2}i_{3};i_{4})}$ is a fourth order KT, $L_{(i_{1}i_{2};i_{3})}= -\frac{4}{\lambda} L_{(i_{1}i_{2}i_{3};i_{4})} Q^{i_{4}} -\lambda L_{i_{1}i_{2}i_{3}}$, $L_{(i_{1};i_{2})}=3 L_{i_{1}i_{2}i_{3}} Q^{i_{3}} -\lambda L_{i_{1}i_{2}}$ and $\left(L_{c}Q^{c}\right)_{,i_{1}}= 2\lambda L_{i_{1}i_{2}} Q^{i_{2}} -\lambda^{2}L_{i_{1}}$.

\section{The independent autonomous polynomial $m$th-order FIs}

\label{sec.autFIs}

From expressions (\ref{eq.FI4a}) and (\ref{eq.FI4b}) for zero time-dependence, we find the following two independent autonomous polynomial FIs:

a. The \textbf{even order FI}
\begin{equation}
J_{0}^{(2\nu,1)}= \sum_{1\leq r \leq 2\nu}^{even} L_{i_{1}...i_{r}} \dot{q}^{i_{1}} ... \dot{q}^{i_{r}} +G(q) \label{eq.aut.1}
\end{equation}
where $\nu \in \mathbb{N}$, $L_{i_{1}...i_{2\nu}}(q)$ is a KT of order $2\nu$, $G(q)$ is an invariant, and the even rank totally symmetric tensors\footnote{These tensors exist only for $\nu>1$.} $L_{i_{1}...i_{r}}(q)$, $r=2,4,...,2\nu-2$, are such that:
\begin{eqnarray}
L_{(i_{1}...i_{r-1};i_{r})}&=& (r+1) L_{i_{1}...i_{r}i_{r+1}} Q^{i_{r+1}}, \enskip r=3,5,...,2\nu-1 \label{eq.aut.1.1} \\
G_{,i_{1}}&=& 2L_{i_{1}i_{2}}Q^{i_{2}}. \label{eq.aut.1.2}
\end{eqnarray}

For example, the autonomous sixth order FI ($\nu=3$) is
\begin{equation}
J_{0}^{(6,1)}= L_{i_{1}...i_{6}} \dot{q}^{i_{1}} ... \dot{q}^{i_{6}} + L_{i_{1}...i_{4}} \dot{q}^{i_{1}} ... \dot{q}^{i_{4}} +L_{i_{1}i_{2}}\dot{q}^{i_{1}} \dot{q}^{i_{2}} +G(q) \label{eq.aut.2}
\end{equation}
where $L_{i_{1}...i_{6}}(q)$ is a sixth order KT, $G(q)$ is an invariant, and the even rank totally symmetric tensors $L_{i_{1}...i_{r}}(q)$, $r=2,4$, are such that:
\begin{eqnarray}
L_{(i_{1}...i_{4};i_{5})}&=& 6 L_{i_{1}...i_{5}i_{6}} Q^{i_{6}} \label{eq.aut.2.1} \\
L_{(i_{1}i_{2};i_{3})}&=& 4L_{i_{1}i_{2}i_{3}i_{4}} Q^{i_{4}} \label{eq.aut.2.2} \\
G_{,i_{1}}&=& 2L_{i_{1}i_{2}}Q^{i_{2}}. \label{eq.aut.2.3}
\end{eqnarray}

b. The \textbf{odd order FI}
\begin{equation}
J_{0}^{(2\nu,2)}= \sum_{1\leq r \leq 2\nu-1}^{odd} L_{i_{1}...i_{r}} \dot{q}^{i_{1}} ... \dot{q}^{i_{r}} \label{eq.aut.3}
\end{equation}
where $\nu \in \mathbb{N}$, $L_{i_{1}...i_{2\nu-1}}(q)$ is a KT of order $2\nu-1$, and the odd rank totally symmetric tensors\footnote{These tensors exist only for $\nu>1$.} $L_{i_{1}...i_{r}}(q)$, $r=1,3,...,2\nu-3$, are such that:
\begin{eqnarray}
L_{(i_{1}...i_{r-1};i_{r})}&=& (r+1) L_{i_{1}...i_{r}i_{r+1}} Q^{i_{r+1}}, \enskip r=2,4,...,2\nu-2 \label{eq.aut.3.1} \\
L_{a}Q^{a}&=&0. \label{eq.aut.3.2}
\end{eqnarray}

For example, the autonomous fifth order FI ($\nu=3$) is
\begin{equation}
J_{0}^{(6,2)}= L_{i_{1}...i_{5}} \dot{q}^{i_{1}} ... \dot{q}^{i_{5}} + L_{i_{1}i_{2}i_{3}} \dot{q}^{i_{1}} \dot{q}^{i_{2}} \dot{q}^{i_{3}} +L_{i_{1}}\dot{q}^{i_{1}} \label{eq.aut.4}
\end{equation}
where $L_{i_{1}...i_{5}}(q)$ is a fifth order KT, and the odd rank totally symmetric tensors $L_{i_{1}...i_{r}}(q)$, $r=1,3$, are such that:
\begin{eqnarray}
L_{(i_{1}i_{2}i_{3};i_{4})}&=& 5 L_{i_{1}i_{2}i_{3}i_{4}i_{5}} Q^{i_{5}} \label{eq.aut.4.1} \\
L_{(i_{1};i_{2})}&=& 3 L_{i_{1}i_{2}i_{3}} Q^{i_{3}} \label{eq.aut.4.2} \\
L_{a}Q^{a}&=&0. \label{eq.aut.4.3}
\end{eqnarray}

\section{The $m$th-order FIs of geodesic equations using Theorem \ref{thm.mFIs}}

\label{sec.geodesics}

We have the following well-known result \cite{Katzin 1981}.

\begin{theorem}
\label{thm.mFIs.1}
The $m$th-order FI (\ref{eq.hfi2}) for the geodesic equations
\begin{equation}
\ddot{q}^{i}= -\Gamma^{i}_{jk}\dot{q}^{j}\dot{q}^{k} \label{eq.hfi8}
\end{equation}
is (see eq. (2.10) in \cite{Katzin 1981})
\begin{equation}
I^{(m)} = \sum_{r=0}^{m} \sum_{b=0}^{r} \frac{(-t)^{r-b}}{(r-b)!} C_{(i_{1}...i_{b};i_{b+1}..i_{r})} \dot{q}^{i_{1}} \dot{q}^{i_{2}} ...\dot{q}^{i_{r}} = \sum_{r=0}^{m} \sum_{b=0}^{r} \frac{(-t)^{r-b}}{(r-b)!} C_{i_{1}...i_{b};i_{b+1}..i_{r}} \dot{q}^{i_{1}} \dot{q}^{i_{2}} ...\dot{q}^{i_{r}} \label{eq.hfi9}
\end{equation}
where the $b$-rank totally symmetric tensors $C_{i_{1}...i_{b}}$ satisfy the relation\footnote{Relation (\ref{eq.hfi9}) does not vanish, because condition (\ref{eq.hfi7}) holds only for $r=m+1$. In equation (\ref{eq.hfi9}) $r$ runs from 0 to $m$.} (see eq. (2.9) in \cite{Katzin 1981})
\begin{equation}
C_{(i_{1}...i_{b};i_{b+1}..i_{m+1})} =0, \enskip b=0,1,2...,m
\label{eq.hfi7}
\end{equation}
that is, the totally symmetric tensors $C_{;i_{1}...i_{m}}$, $C_{(i_{1};i_{2}...i_{m})}$, $C_{(i_{1}i_{2};i_{3}...i_{m})}$, ..., $C_{i_{1}i_{2}...i_{m}}$ are $m$th-order KTs of the kinetic metric $\gamma_{ij}$.
\end{theorem}

As a first application of Theorem \ref{thm.mFIs} and a good working example, we reproduce the $m$th-order FI (\ref{eq.hfi9}) by assuming $Q^{a}=0$ in the independent $m$th-order FIs $I^{(m)}_{n}$ and $I^{(m)}_{e}$ of Theorem \ref{thm.mFIs}. We have the following:
\bigskip

1) The FI $I^{(m)}_{n}$ for $Q^{a}=0$ becomes
\begin{eqnarray}
I^{(m)}_{n}&=& \left( -\frac{t^{n}}{n} L_{(n-1)(i_{1}...i_{m-1};i_{m})} - ... - \frac{t^{2}}{2} L_{(1)(i_{1}...i_{m-1};i_{m})} - tL_{(0)(i_{1}...i_{m-1};i_{m})} +C_{(0)i_{1}...i_{m}} \right) \dot{q}^{i_{1}} ... \dot{q}^{i_{m}} + \notag \\
&& + \sum_{r=1}^{m-1} \left( t^{n} L_{(n)i_{1}...i_{r}} + ... + tL_{(1)i_{1}...i_{r}} +L_{(0)i_{1}...i_{r}} \right) \dot{q}^{i_{1}} ... \dot{q}^{i_{r}} + G(q) \label{eq.FIgeo1}
\end{eqnarray}
where $C_{(0)i_{1}...i_{m}}$ and $L_{(N)(i_{1}...i_{m-1};i_{m})}$ for $N=0,1,...,n-1$ are $m$th-order KTs, $L_{(n)i_{1}...i_{m-1}}$ is an $(m-1)$th-order KT and the following conditions are satisfied:
\begin{eqnarray}
L_{(n)(i_{1}...i_{m-2};i_{m-1})}&=& 0 \label{eq.FIgeo2a} \\
L_{(k)i_{1}...i_{m-1}}&=& -\frac{1}{k} L_{(k-1)(i_{1}...i_{m-2};i_{m-1})}, \enskip k=2,3,...,n \label{eq.FIgeo2b} \\
L_{(1)i_{1}...i_{m-1}} &=&-L_{(0)(i_{1}...i_{m-2};i_{m-1})} \label{eq.FIgeo2c} \\
L_{(n)(i_{1}...i_{r-1};i_{r})}&=& 0, \enskip r=2,3,...,m-2 \label{eq.FIgeo2d} \\
L_{(k)i_{1}...i_{r}} &=& -\frac{1}{k} L_{(k-1)(i_{1}...i_{r-1};i_{r})}, \enskip k=1,2,...,n, \enskip r=2,3,...,m-2 \label{eq.FIgeo2e} \\
L_{(k)i_{1}}&=& 0, \enskip k=2,3,...,n \label{eq.FIgeo2f} \\
L_{(1)i_{1}}&=& -G_{,i_{1}}. \label{eq.FIgeo2g}
\end{eqnarray}

From conditions (\ref{eq.FIgeo2f}) and (\ref{eq.FIgeo2g}), we find that the only surviving vectors are the $L_{(0)i_{1}}$ and $L_{(1)i_{1}}=-G_{,i_{1}}$.

From condition (\ref{eq.FIgeo2e}), we find that the only surviving $r$-rank totally symmetric tensors $L_{(k)i_{1}...i_{r}}$, where $r=2,3,...,m-2$, are those with $k=0,1,2,...,r$. For example, for $r=2$, the only surviving second order symmetric tensors are the
\[
L_{(0)i_{1}i_{2}}, \enskip L_{(1)i_{1}i_{2}}= -L_{(0)(i_{1};i_{2})}, \enskip L_{(2)i_{1}i_{2}}= \frac{1}{2}G_{;i_{1}i_{2}}
\]
and, for $r=3$, the only surviving third order totally symmetric tensors are the
\[
L_{(0)i_{1}i_{2}i_{3}}, \enskip L_{(1)i_{1}i_{2}i_{3}}= -L_{(0)(i_{1}i_{2};i_{3})}, \enskip L_{(2)i_{1}i_{2}i_{3}}= \frac{1}{2}L_{(0)(i_{1};i_{2}i_{3})}, \enskip L_{(3)i_{1}i_{2}i_{3}}= -\frac{1}{3!} G_{;i_{1}i_{2}i_{3}}.
\]

Taking into account the previous general result for $r=m-2$, conditions (\ref{eq.FIgeo2b}) and (\ref{eq.FIgeo2c}) imply that the only surviving $(m-1)$-rank totally symmetric tensors $L_{(k)i_{1}...i_{m-1}}$ are those with $k=0,1,2,...,m-1$. Since the $m$th-order KTs $L_{(N)(i_{1}...i_{m-1};i_{m})}$, where $N=0,1,2,..,n-1$, depend on $L_{(k)i_{1}...i_{m-1}}$ we deduce that $N=k$; therefore, the degree of the time-dependence $n=m$ and condition (\ref{eq.hfi7}) is reproduced.

The remaining conditions (\ref{eq.FIgeo2a}) and (\ref{eq.FIgeo2d}) are satisfied identically.

By substituting the surviving totally symmetric tensors in the associated $m$th-order FI (\ref{eq.FIgeo1}), we reproduce the general formula (\ref{eq.hfi9}). We note that in the $m$th-order FI (\ref{eq.hfi9}) the degree of each time polynomial coefficient is equal to the order of its associated velocity term.
\bigskip

2) The FI $I^{(m)}_{e}$ for $Q^{a}=0$ becomes
\begin{equation}
I^{(m)}_{e}= \frac{e^{\lambda t}}{\lambda} \left( -L_{(i_{1}...i_{m-1};i_{m})} \dot{q}^{i_{1}} ... \dot{q}^{i_{m}} + \lambda \sum_{r=1}^{m-1} L_{i_{1}...i_{r}} \dot{q}^{i_{1}} ... \dot{q}^{i_{r}} \right) \label{eq.FIgeo3}
\end{equation}
where $\lambda\neq0$, $L_{(i_{1}...i_{m-1};i_{m})}$ is an $m$th-order KT and the following conditions are satisfied:
\begin{eqnarray}
L_{(i_{1}...i_{m-2};i_{m-1})}&=& -\lambda L_{i_{1}...i_{m-1}} \label{eq.FIgeo3a} \\
L_{(i_{1}...i_{r-1};i_{r})}&=& -\lambda L_{i_{1}...i_{r}}, \enskip r=2,3,...,m-2 \label{eq.FIgeo3b} \\
L_{i_{1}}&=& 0. \label{eq.FIgeo3c}
\end{eqnarray}

Conditions (\ref{eq.FIgeo3a}) - (\ref{eq.FIgeo3c}) imply that all the totally symmetric tensors $L_{i_{1}...i_{k}}$, where $k=1,2,...,m-1$, vanish. Therefore, the FI $I^{(m)}_{e}=0$.

\section{Applications}

Theorem \ref{thm.mFIs} is covariant, independent of the dimension, and
applies to a curved Riemannian space provided its geometric elements can be
determined. In that respect, it can be used to determine the higher order (time-dependent and autonomous) FIs of autonomous holonomic dynamical
systems. In the following, we demonstrate the application of Theorem \ref{thm.mFIs} to the rather simple case of Newtonian autonomous conservative dynamical systems with two degrees of
freedom, which has been a research topic for many years. For these systems the kinetic metric\footnote{The geometric quantities of this metric have been computed in section \ref{sec.KTE2}.} $\gamma_{ab} =\delta_{ab}=diag(1,1)$ and $Q^{a}=-V^{,a}$, where $V(x,y)$ indicates the potential of the dynamical system.

The known integrable and superintegrable systems of that type that admit QFIs are reviewed in chapter \ref{ch.2d.pots}. Using Theorem \ref{thm.mFIs}, we shall show that: \newline
a. CFIs, which have been determined by other methods, follow as subcases directly from Theorem \ref{thm.mFIs}. \newline
b. New integrable potentials, which admit only CFIs, are found. \newline
c. Dynamical systems, which were considered to be integrable, admit an additional time-dependent CFI; therefore, are, in fact, superintegrable.

\subsection{Known CFIs}

In \cite{Fokas 1980}, the authors determined all potentials of the form $%
V=F(x^{2}+\nu y^{2})$, where $\nu $ is an arbitrary constant and $F$ an
arbitrary smooth function, that admit autonomous CFIs. They found the following three potentials\footnote{%
There is a misprint in the FI (3.15b) of \cite{Fokas 1980}, where the $p_{1}=\dot{x}$. In the last term, it must be $p_{2}=\dot{y}$.} (see eqs. (3.15a), (3.15b) and (3.19) of \cite{Fokas 1980}):
\begin{equation}
V_{(1a)}=\frac{1}{2}x^{2}+\frac{9}{2}y^{2},\enskip V_{(1b)}=\frac{1}{2}x^{2}+\frac{1}{18}y^{2},\enskip V_{(1c)}=(x^{2}-y^{2})^{-2/3}.  \label{Fok1}
\end{equation}%
Using Theorem \ref{thm.mFIs}, we found the new superintegrable\footnote{It is superintegrable because it is of the separable form $V(x,y)= F_{1}(x) +F_{2}(y)$, where $F_{1}, F_{2}$ are arbitrary smooth functions. It is well-known (see chapter \ref{ch.2d.pots}) that such potentials admit also the QFIs $I_{1}= \frac{1}{2}\dot{x}^{2} +F_{1}(x)$ and $I_{2}= \frac{1}{2}\dot{y}^{2} +F_{2}(y)$.} potential
\begin{equation}
V_{1}= c_{0}(x^{2} +9y^{2}) +c_{1}y \label{Fok2}
\end{equation}
where $c_{0}$ and $c_{1}$ are arbitrary constants, which admits the associated CFI
\begin{equation}
J_{1}= (x\dot{y} -y\dot{x})\dot{x}^{2} -\frac{c_{1}}{18c_{0}} \dot{x}^{3} +\frac{c_{1}}{3}x^{2}\dot{x} +6c_{0}x^{2}y\dot{x} - \frac{2c_{0}}{3}x^{3}\dot{y} \label{Fok3}
\end{equation}
and the integrable potential
\begin{equation}
V_{2}= k(x^{2}-y^{2})^{-2/3} \label{Fok4}
\end{equation}
where $k$ is an arbitrary constant, which admits the CFI
\begin{equation}
J_{2}= \left( x\dot{y} -y\dot{x} \right) \left( \dot{y}^{2} -\dot{x}^{2} \right) +4V_{2}(y\dot{x} +x\dot{y}). \label{Fok6}
\end{equation}

We note that the potentials (\ref{Fok1}) are special cases of $V_{1}$ and $V_{2}$ as follows:
\[
V_{(1a)}= V_{1}\left( c_{1}=0, c_{0}=\frac{1}{2} \right), \enskip V_{(1b)}= V_{1}\left( x \leftrightarrow y; c_{1}=0, c_{0}=\frac{1}{18} \right), \enskip V_{(1c)} =V_{2}(k=1).
\]

Working in the same manner, one recovers all known potentials which are integrable or superintegrable and admit higher order FIs \cite{Tsiganov 2000, Tsiganov 2008, Karlovini 2000, Karlovini 2002}.

\subsection{New superintegrable potentials}

Using Theorem \ref{thm.mFIs}, we found the integrable potential (see eq. (4.8) in \cite{Tsiganov 2000})
\begin{equation}
V_{3}= \frac{k_{1}}{(a_{2}y-a_{5}x)^{2}} +\frac{k_{2}}{r} + \frac{k_{3}(a_{2}x+a_{5}y)}{r(a_{2}y-a_{5}x)^{2}} \label{new1}
\end{equation}
where $k_{1}, k_{2}, k_{3}, a_{2}, a_{5}$ are arbitrary constants and $r=\sqrt{x^{2}+y^{2}}$, which admits the CFI
\begin{eqnarray}
J_{3} &=& M_{3}^{2}(a_{2}\dot{x}+a_{5}\dot{y}) +\frac{2k_{1}r^{2}}{(a_{2}y-a_{5}x)^{2}}(a_{2}\dot{x} +a_{5}\dot{y}) -\frac{k_{2}(a_{2}y-a_{5}x)}{r}M_{3} -\frac{k_{3}(a_{2}x+a_{5}y)}{r(a_{2}y-a_{5}x)}M_{3}+ \notag \\
&& +\frac{k_{3}r}{a_{2}y-a_{5}x} (a_{2}\dot{y}-a_{5}\dot{x}) +\frac{2k_{3} (a_{2}x+a_{5}y)r}{(a_{2}y-a_{5}x)^{2}}(a_{2}\dot{x} +a_{5}\dot{y}) \label{new2}
\end{eqnarray}
where $M_{3}= x\dot{y} -y\dot{x}$ is the angular momentum of the system.

Furthermore, for $a_{5}=0$, this potential becomes the superintegrable potential (see Table \ref{Table.Class2.2})
\begin{equation}
V_{4}= V_{3}(a_{5}=0) =\frac{c_{1}}{y^{2}} +\frac{c_{2}}{r} +\frac{c_{3}x}{ry^{2}} \label{new3}
\end{equation}
where $c_{1}=\frac{k_{1}}{a_{2}^{2}}$, $c_{2}=k_{2}$ and $c_{3}=\frac{k_{3}}{a_{2}}$ are arbitrary constants, which admits the CFI
\begin{equation}
J_{4}=J_{3}(a_{5}=0) = M_{3}^{2} \dot{x} -\frac{c_{2}y}{r}M_{3} +\frac{2c_{1}r^{2}}{y^{2}}\dot{x}+ \frac{c_{3} x(2x^{2} +3y^{2})}{ry^{2}}\dot{x} +\frac{c_{3}y}{r} \dot{y}. \label{new4}
\end{equation}
We note that under the transformation $c_{1}=B+C, c_{2}=A, c_{3}=C-B$, where $A, B, C$ are the new constants, the potential $V_{4}$ becomes $V_{4}=\frac{A}{r} + \frac{B}{r(r+x)} +\frac{C}{r(r-x)}$ which coincides with the potential (3.2.36) of \cite{Hietarinta 1987}.

For $k_{2}=0$, we have the special potential
\begin{equation}
V_{5} =V_{3}(k_{2}=0)= \frac{k_{1}}{(a_{2}y-a_{5}x)^{2}} + \frac{k_{3}(a_{2}x+a_{5}y)}{r(a_{2}y-a_{5}x)^{2}} \label{eq.su1}
\end{equation}
which admits the additional time-dependent CFI
\begin{equation}
J_{5}= -tJ_{3}(k_{2}=0) +(a_{2}x +a_{5}y)M_{3}^{2} + \frac{2k_{1}r^{2}(a_{2}x +a_{5}y)}{(a_{2}y -a_{5}x)^{2}} +\frac{2k_{3}r(a_{2}x +a_{5}y)^{2}}{(a_{2}y -a_{5}x)^{2}} +k_{3}r. \label{eq.su2}
\end{equation}
We conclude that $V_{5}$ is not just an integrable but a new superintegrable potential. This result illustrates the importance of the time-dependent FIs in the establishment of the integrability/superintegrability.

\section{Conclusions}

Theorem  \ref{thm.mFIs} provides a general method for determining higher
order FIs of autonomous holonomic dynamical systems in a general Riemannian space provided one
knows, or is able to calculate, the KTs of all orders --up to the order of the
FI-- of the kinetic metric. It is shown that an autonomous  dynamical system is possible to admit two families
of independent FIs of a given order. The results of
Theorem \ref{thm.mFIs} are covariant and do not depend on the number of degrees of
freedom of the dynamical system.

Theorem \ref{thm.mFIs} generalizes the results of \cite{Katzin 1981} in the case of autonomous holonomic dynamical systems. The geodesic equations are obtained as a special case for $Q^{a}=0$. In the latter case, the system of PDEs resulting from the condition $\frac{dI}{dt}=0$ is integrated directly without the need of additional assumptions.

From the application of Theorem \ref{thm.mFIs} in the rather simple --but widely studied-- case of autonomous
conservative dynamical systems with two degrees of freedom, we have achieved the following goals: \newline
a. We have shown that one is possible to obtain the known integrable potentials, which have been computed using other methods but the direct method, in a simple, direct, and concrete geometrical approach. \newline
b. We have found a new superintegrable potential whose integrability is established only by means of autonomous and time-dependent CFIs.

%% file: QFIs_timedependent.tex
\chapter{Quadratic first integrals of time-dependent dynamical systems of the form $\ddot{q}^{a}= -\Gamma^{a}_{bc}(q) \dot{q}^{b} \dot{q}^{c} -\omega(t)Q^{a}(q)$}

\label{ch.QFIs.timedependent}

\section{Introduction}

\label{sec.timedep.into}

As we have seen in the previous chapters, the standard way to determine the FIs of a differential equation is the use of Lie/Noether symmetries, which applies to the point as well as to the generalized Lie/Noether symmetries. The relation of the Lie/Noether symmetries with the symmetries of the kinetic metric has been considered, mostly, in the case of point symmetries for autonomous conservative dynamical systems moving in a Riemannian space. In particular, it has been shown (see e.g. \cite{Paliathanasis 2012, Katzin 1974, TsaPal 2011, TsamparlisHV 2012}) that the Lie point symmetries are generated by the special projective algebra of the kinetic metric, whereas the Noether point symmetries are generated by the homothetic algebra of the kinetic metric; the latter being a subalgebra of the projective algebra\footnote{A recent clear statement of these results is discussed in \cite{Tsamparlis 2015}.}.

In addition to the autonomous conservative systems, this method has been applied to the time-dependent potentials\index{Potential! time-dependent} $W(t,q)= \omega(t)V(q)$, that is, for dynamical equations of the form $\ddot{q}^{a}= -\Gamma_{bc}^{a}(q) \dot{q}^{b} \dot{q}^{c} -\omega(t)V^{,a}(q)$ (see e.g. \cite{Katzin 1976, Katzin II1977, Katzin 1977, Ray 1979B, Prince 1980, Ray 1980, LeoTsampAndro 2017}). In
this case, it has been shown that the Lie point symmetries, the Noether point symmetries and the
associated FIs are computed in terms of the collineations of the kinetic
metric, plus a set of constraints involving the time-dependent
potential and the collineation vectors. These time-dependent potentials are
important because (among others) they contain the time-dependent oscillator (see e.g. \cite{Ray 1979A, Lewis 1968, Katzin 1977, Prince 1980, Gunther 1977}) and the time-dependent Kepler potential (see e.g. \cite{Leach 1985, Katzin 1982, LeoTsampAndro 2017, Prince 1981}). A further development in the same line is the extension of this method to time-dependent potentials $W(t,q)$ with linear damping terms \cite{LeoTsampAndro 2017}. It has been shown that under a suitable time transformation the damping term can be removed and the problem reduces to a time-dependent potential of the form $W(t,q)=\bar{\omega}(t)V(q)$, where $\bar{\omega}(t)$ is a `frequency' different from the original. Finally, the method of the Lie/Noether symmetries has been applied to the study of PDEs (see e.g. \cite{Paliathanasis 2012, Bozhkov 2010, RoseKatizn 1994, Tsamp 2015}).

In this chapter, we shall use again the direct method (see section \ref{sec.methods.determine.FIs}), which has been employed in the literature (see e.g. \cite{Katzin 1973, Katzin 1981, Katzin 1982, Katzin 1983}) both for autonomous (see Part \ref{part3}) and time-dependent dynamical systems, in order to compute the QFIs of time-dependent dynamical systems of the form  $\ddot{q}^{a} =-\Gamma_{bc}^{a}(q) \dot{q}^{b}\dot{q}^{c} -\omega(t)Q^{a}(q)$. Because many well-known dynamical systems fall in this category, we intend to recover in a direct single approach all the known results, which are scattered in a large number of papers and have been derived mainly from the method of the Lie/Noether symmetries.

The application of the direct method implies that the symmetric tensor $K_{ab}$ of the considered QFI (\ref{FL.5}) is a KT of the kinetic metric. In general, the computation of the KTs of a metric is a major task. However, for spaces of constant curvature this problem has been solved (see sections \ref{sub.KT.1} and \ref{sec.detour.KTs.3}). Therefore, in this chapter, we restrict our discussion to Euclidean spaces only. Since the KT $K_{ab}$ is a function of $t$ and $q^a$, we suggest two procedures of work: a. The polynomial method, and b. The basis method.

In the \textbf{polynomial method},\index{Method! polynomial} one assumes a general polynomial form in the variable $t$ both for the KT $K_{ab}$ and the vector $K_{a}$, and replaces these expressions in the system of PDEs resulting from the condition $\frac{dI}{dt}=0$. On the other hand, in the \textbf{basis method},\index{Method! basis} one computes first a basis of the KTs of order two of the kinetic metric and, then, expresses in this basis the KT $K_{ab}$ with the coefficients to be functions of $t$. The vector $K_{a}$ and the FIs follow from the solution of the system of PDEs, which is the same for both methods. We note that \emph{both methods are suitable for autonomous dynamical systems, but for time-dependent systems it appears that the basis method is preferable.}

Concerning the quantities $\omega(t)$ and $Q^{a}(q)$, again, there are two ways to proceed: \newline
a) Consider a general form for the function $\omega(t)$ and let the quantities $Q^{a}$ unspecified. In this case, the quantities $Q^{a}$ act as constraints.\newline
b) Specify the quantities $Q^{a}$ and determine for which functions $\omega(t)$ the resulting dynamical system admits QFIs.

In the following, we shall consider both the polynomial method and the basis method, starting from the former. As a first application, we assume the KT $K_{ab}=N(t)\gamma_{ab}$, where $N(t)$ is an arbitrary function, and show that we recover all the point Noether FIs found in \cite{LeoTsampAndro 2017}. As a second application, we assume that $\omega(t)= b_{0}+ b_{1}t+...+b_{\ell }t^{\ell }$ with $b_{\ell }\neq 0$ and $\ell \geq 1$, while the quantities $Q^{a}$ are unspecified. We find that, in this case, the system admits two families of independent QFIs as stated in Theorem \ref{thm.polynomial.omega}.

Subsequently, we consider the basis method. This is carried out in two steps: In the first step, we assume that we know a basis $\{C_{(N)ab}(q)\}$ of the vector space of KTs of the kinetic metric, and require that $K_{ab}$ has the form $K_{ab}(t,q) =\sum_{N=1}^{m}\alpha _{N}(t)C_{(N)ab}(q)$. In the second step, we specify the generalized forces to be conservative with the time-dependent Newtonian generalized Kepler potential $V=-\frac{\omega (t)}{r^{\nu}}$, where $\nu$ is a non-zero real constant and $r=\sqrt{x^{2}+y^{2}+z^{2}}$. This potential for $\nu= -2, 1$ includes, respectively, the 3d time-dependent oscillator and the time-dependent Kepler potential. For other values of $\nu$, it reduces to other important dynamical systems; for example, for $\nu=2$, one obtains the Newton-Cotes potential \cite{Ibragimov 1998}. We determine the QFIs of the time-dependent generalized Kepler potential and recover in a systematic way the known results concerning the QFIs of the 3d time-dependent oscillator, the time-dependent Kepler potential  and the Newton-Cotes potential.

Using the well-known result that by a reparameterization the linear damping term $\phi(t)\dot{q}^{a}$ of a dynamical system is absorbed to a time-dependent force of the form $\omega(t)Q^{a}(q)$, we also study the integrability of the non-linear differential equation $\ddot{x}=-\omega(t)x^{\mu }+\phi (t)\dot{x}$ $(\mu \neq -1)$. Specifically, we compute the relation between the coefficients $\omega(t)$ and $\phi(t)$ for which QFIs are admitted. It is found that a family of `frequencies' $\bar{\omega}(s)$ is admitted, which for $\mu =0, 1, 2$ is parameterized with functions, whereas for $\mu \neq -1,0,1,2$ is parameterized with constants. As a further application, we study the integrability of the well-known generalized Lane-Emden equation.

\section{The system of PDEs}

\label{sec.conditions}

We consider the time-dependent dynamical system
\begin{equation}
\ddot{q}^{a} = - \Gamma^{a}_{bc}(q) \dot{q}^{b} \dot{q}^{c} - \omega(t)Q^{a}(q) \label{eq.red1}
\end{equation}
where $\Gamma^{a}_{bc}$ are the Riemannian connection coefficients determined by the kinetic metric $\gamma_{ab}$ of the system and $-\omega(t)Q^{a}(q)$ are the time-dependent generalized forces.

Next, we consider a function $I(t,q^{a},\dot{q}^{a})$ of the form
\begin{equation}
I=K_{ab}(t,q)\dot{q}^{a}\dot{q}^{b}+K_{a}(t,q)\dot{q}^{a}+K(t,q)
\label{FI.5}
\end{equation}%
where $K_{ab}$ is a symmetric tensor, $K_{a}$ is a vector and $K$ is an invariant.

We demand $(\ref{FI.5})$ to be a QFI of (\ref{eq.red1}) by imposing the condition $\frac{dI}{dt}=0$ along trajectories of the system. Using the dynamical equations (\ref{eq.red1}) to replace $\ddot{q}^{a}$ whenever it appears, we find the following system of PDEs:
\begin{eqnarray}
K_{(ab;c)} &=&0  \label{eq.TKN1} \\
K_{ab,t}+K_{(a;b)} &=&0  \label{eq.TKN2} \\
-2\omega K_{ab}Q^{b}+K_{a,t}+K_{,a} &=&0  \label{eq.TKN3} \\
K_{,t}-\omega K_{a}Q^{a} &=&0  \label{eq.TKN4} \\
K_{a,tt}+\omega \left( K_{b}Q^{b}\right) _{,a}-2\omega
_{,t}K_{ab}Q^{b}-2\omega K_{ab,t}Q^{b} &=&0  \label{eq.TKN5} \\
K_{[a;b],t}-2\omega \left( K_{[a|c|}Q^{c}\right)_{;b]} &=&0
\label{eq.TKN6}
\end{eqnarray}
where equations (\ref{eq.TKN5}) and (\ref{eq.TKN6}) express the integrability conditions $K_{,[at]}=0$ and $K_{,[ab]}=0$, respectively, for the scalar $K$.

Equation (\ref{eq.TKN1}) implies that $K_{ab}$ is a KT of order two (possibly zero) of the kinetic metric $\gamma_{ab}$.

The solution of the system (\ref{eq.TKN1}) - (\ref{eq.TKN6}) requires the function $\omega(t)$ and the quantities $Q^{a}(q)$; both being quantities which are characteristic of the given dynamical system. There are two ways to proceed: \newline
a) Consider a general form for the function $\omega(t)$ and let the quantities $Q^{a}(q)$ unspecified. In this case, the
quantities $Q^{a}(q)$ act as constraints.\newline
b) Specify the quantities $Q^{a}(q)$ and determine for which functions $\omega(t)$ the resulting dynamical system admits QFIs.

However, before continuing with this kind of considerations, we first proceed with the simple geometric choice $K_{ab}= N(t)\gamma_{ab}$, where $N(t)$ is an arbitrary smooth function. By specifying the KT $K_{ab}$ like this, both the function $\omega(t)$ and the quantities $Q^{a}(q)$ stay unspecified and can act as constraints.

\section{The point Noether FIs of the time-dependent dynamical system (\ref{eq.red1})}

\label{sec.point.integrals}

We consider the simplest choice
\begin{equation}
K_{ab}=N(t)\gamma_{ab} \label{eq.KTtr}
\end{equation}
where $N(t)$ is an arbitrary smooth function. This choice is purely geometric; therefore, the function $\omega(t)$ and the quantities $Q^{a}(q)$ are unspecified and act as constraints, whereas the vector $K_{a}$ is identified with one collineation of the kinetic metric. With this $K_{ab}$, the system of PDEs (\ref{eq.TKN1}) - (\ref{eq.TKN6}) becomes (eq. (\ref{eq.TKN1}) vanishes trivially):
\begin{eqnarray}
N_{,t}\gamma_{ab} +K_{(a;b)} &=&0  \label{eq.Tpoint.16a} \\
-2\omega N Q_{a}+K_{a,t}+K_{,a} &=&0  \label{eq.Tpoint.16b} \\
K_{,t}-\omega K_{a}Q^{a} &=&0  \label{eq.Tpoint.16c} \\
K_{a,tt} +\omega\left(K_{b}Q^{b}\right)_{,a} -2\omega_{,t}NQ_{a} -2\omega N_{,t}Q_{a} &=& 0 \label{eq.Tpoint.16d} \\
K_{[a;b],t} -2 \omega NQ_{[a;b]} &=&0.  \label{eq.Tpoint.16e}
\end{eqnarray}

We consider the following cases.

\subsection{Case $K_{a}=K_{a}(q)$ is the HV of $\gamma_{ab}$ with homothetic factor $\psi$}

\label{sec.p1}

In this case, $K_{a,t}=0$ and $K_{(a;b)}= \psi\gamma_{ab}$, where $\psi$ is an arbitrary constant.

Equation (\ref{eq.Tpoint.16a}) gives $N_{,t}=-\psi \implies N=-\psi t + c$, where $c$ is an arbitrary constant.

Equation (\ref{eq.Tpoint.16e}) implies that (take $\omega \neq 0$) $Q_{[a;b]}=0 \implies Q_{a}=V_{,a}$, where $V=V(q)$ is an arbitrary potential.

Replacing in (\ref{eq.Tpoint.16b}), we find that
\[
K_{,a} =2\omega (-\psi t +c)V_{,a} \implies K= 2\omega (-\psi t +c)V + M(t)
\]
where $M(t)$ is an arbitrary function.

Substituting the function $K(t,q)$ in (\ref{eq.Tpoint.16c}), we get
\begin{equation}
\omega K_{a}V^{,a} - 2\omega_{,t} (-\psi t +c)V + 2\omega \psi V -M_{,t} =0. \label{eq.Tpoint.17}
\end{equation}
The remaining condition (\ref{eq.Tpoint.16d}) is just the partial derivative of (\ref{eq.Tpoint.17}) and, hence, is satisfied trivially.

Moreover, since $\omega \neq 0$, equation (\ref{eq.Tpoint.17}) can be written in the form
\begin{equation}
K_{a}V^{,a} - 2(\ln\omega)_{,t} (-\psi t +c)V + 2\psi V -\frac{M_{,t}}{\omega} =0 \label{eq.Tpoint.18}
\end{equation}
which implies the conditions:
\begin{eqnarray}
2(\ln\omega)_{,t} (-\psi t +c) &=& c_{1} \label{eq.Tpoint.18a} \\
M_{,t} &=& c_{2}\omega \label{eq.Tpoint.18b}
\end{eqnarray}
where $c_{1}$ and $c_{2}$ are arbitrary constants.

Therefore, equation (\ref{eq.Tpoint.18}) becomes
\begin{equation}
K_{a}V^{,a} +(2\psi-c_{1}) V -c_{2} =0. \label{eq.Tpoint.18c}
\end{equation}

The QFI is
\begin{equation}
I_{1} = (-\psi t+c)\gamma_{ab}\dot{q}^{a}\dot{q}^{b} + K_{a}(q)\dot{q}^{a} + 2\omega (-\psi t +c)V + M(t) \label{eq.Tpoint.18d}
\end{equation}
where $Q_{a}=V_{,a}$ and the quantities $\omega(t), M(t), V(q), K_{a}(q)$ satisfy the conditions (\ref{eq.Tpoint.18a}) - (\ref{eq.Tpoint.18c}).

\subsection{Case $K_{a}= -M(t)S_{,a}(q)$ where $S_{,a}$ is the gradient HV of $\gamma_{ab}$}

\label{sec.p2}

In this case, $S_{;ab}=\psi \gamma_{ab}$ and $M(t)\neq0$ is an arbitrary function.

Equation (\ref{eq.Tpoint.16a}) implies $N_{,t}= \psi M$.

From equation (\ref{eq.Tpoint.16e}), we find that there exists a potential function $V(q)$ such that $Q_{a}=V_{,a}$.

Replacing the above results in (\ref{eq.Tpoint.16b}), we obtain
\[
K_{,a}= 2\omega NV_{,a} +M_{,t}S_{,a} \implies K= 2\omega NV + M_{,t}S + C(t)
\]
where $C(t)$ is an arbitrary function.

Substituting in (\ref{eq.Tpoint.16c}), we get (take $\omega M\neq 0$)
\[
\omega M S_{,a}V^{,a} +2\omega_{,t} NV + 2\omega \psi MV + M_{,tt}S + C_{,t} =0 \implies
\]
\[
S_{,a}V^{,a} + 2\psi V + \frac{2(\ln\omega)_{,t} N}{M} V + \frac{M_{,tt}}{\omega M}S + \frac{C_{,t}}{\omega M} =0
\]
which implies that:
\begin{eqnarray}
\frac{2(\ln\omega)_{,t} N}{M} &=& d_{1} \label{eq.Tpoint.19a} \\
\frac{M_{,tt}}{\omega M} &=& m \label{eq.Tpoint.19b} \\
\frac{C_{,t}}{\omega M} &=& k \label{eq.Tpoint.19c} \\
S_{,a}V^{,a} + (2\psi + d_{1})V + mS + k &=& 0 \label{eq.Tpoint.19d}
\end{eqnarray}
where $d_{1}, m, k$ are arbitrary constants. The remaining condition (\ref{eq.Tpoint.16d}) is satisfied identically.

The QFI is
\begin{equation}
I_{2} = N\gamma_{ab} \dot{q}^{a}\dot{q}^{b} - MS_{,a}\dot{q}^{a} +2\omega NV + M_{,t}S + C(t) \label{eq.Tpoint.19e}
\end{equation}
where $Q_{a}=V_{,a}$, $N_{,t}= \psi M$ and the conditions (\ref{eq.Tpoint.19a}) - (\ref{eq.Tpoint.19d}) must be satisfied.

\subsection{Case $Q_{a}=V_{,a}$ and $K_{a}= -M(t)V_{,a}(q)$ where $V_{,a}$ is the gradient HV of $\gamma_{ab}$}

\label{sec.p3}

Equation (\ref{eq.Tpoint.16a}) implies $N_{,t}= \psi M$, where $\psi$ is the homothetic factor of $V_{,a}$.

From equation (\ref{eq.Tpoint.16b}), we obtain
\[
K_{,a}= 2\omega NV_{,a} +M_{,t}V_{,a} \implies K= 2\omega NV + M_{,t}V + C(t)
\]
where $C(t)$ is an arbitrary function.

Substituting in (\ref{eq.Tpoint.16c}), we get (take $\omega M\neq 0$)
\[
\omega M V_{,a}V^{,a} +2\omega_{,t} NV + 2\omega \psi MV + M_{,tt}V + C_{,t} =0 \implies
\]
\[
V_{,a}V^{,a} + 2\psi V + \frac{2(\ln\omega)_{,t} N}{M} V + \frac{M_{,tt}}{\omega M}V + \frac{C_{,t}}{\omega M} =0
\]
which implies that:
\begin{eqnarray}
\frac{M_{,tt}}{\omega M} + \frac{2(\ln\omega)_{,t} N}{M} &=& d_{2} \label{eq.Tpoint.20a} \\
\frac{C_{,t}}{\omega M} &=& k \label{eq.Tpoint.20b} \\
V_{,a}V^{,a} + (2\psi + d_{2})V + k &=& 0 \label{eq.Tpoint.20c}
\end{eqnarray}
where $d_{2}$ and $k$ are arbitrary constants. The remaining conditions are satisfied identically.

The QFI is
\begin{equation}
I_{3} = N\gamma_{ab} \dot{q}^{a}\dot{q}^{b} - MV_{,a}\dot{q}^{a} + \left( 2\omega N + M_{,t} \right)V + C \label{eq.Tpoint.20d}
\end{equation}
where $Q_{a}=V_{,a}$, $N_{,t}= \psi M$ and the conditions (\ref{eq.Tpoint.20a}) - (\ref{eq.Tpoint.20c}) must be satisfied.
\bigskip

The above results reproduce Theorem 2 of \cite{LeoTsampAndro 2017} which states that \emph{the point Noether symmetries
of the time-dependent potentials of the form $\omega(t)V(q)$ are generated by the homothetic algebra of the kinetic metric (provided the Lagrangian is regular)}.

It is interesting to observe that \emph{the QFIs (\ref{eq.Tpoint.18d}), (\ref{eq.Tpoint.19e}) and (\ref{eq.Tpoint.20d}) produced by point Noether symmetries can be also produced by generalized (gauged) Noether symmetries using the Inverse Noether Theorem \ref{Inverse Noether Theorem} (see last comments in section \ref{sec.tables.theorem}). This proves that a Noether FI is not associated with a unique Noether symmetry.}

\section{The polynomial method for computing the QFIs}

\label{sec.direct.solution}

In the polynomial method,\index{Method! polynomial} one assumes a polynomial form in $t$ for the KT $K_{ab}(t,q)$ and the vector $K_{a}(t,q)$ and, then, solves the resulting system for given $\omega(t)$ and $Q^{a}(q)$. For example, this method is applied in chapter \ref{ch.QFIs.damping}, where a
general theorem is given (see Theorem \ref{Theorem2}) which allows the finding of the QFIs of an
autonomous holonomic dynamical system with a linear damping term. In the present chapter, we follow the assumptions
made in section \ref{section.1}, and assume that the KT $K_{ab}(t,q)$ and the vector $K_{a}(t,q)$ are given by equations (\ref{eq.aspm1}) and (\ref{eq.aspm2}), respectively.

Substituting (\ref{eq.aspm1}) and (\ref{eq.aspm2}) in the system of PDEs (\ref{eq.TKN1}) - (\ref{eq.TKN6}) (eq. (\ref{eq.TKN1}) is identically zero since $C_{(N)ab}$ are KTs), we obtain the system of equations:
\begin{eqnarray}
0&=& C_{(1)ab} + C_{(2)ab}t + ... + C_{(n)ab} t^{n-1} + L_{(0)(a;b)} + L_{(1)(a;b)}t +... + L_{(m)(a;b)}t^{m} \label{eq.red2a} \\
0 &=& -2\omega C_{(0)ab}Q^{b} -2\omega C_{(1)ab}Q^{b} t - ... - 2\omega C_{(n)ab} Q^{b} \frac{t^{n}}{n} + L_{(1)a} + 2L_{(2)a}t + ... + \notag \\
&& +mL_{(m)a}t^{m-1} + K_{,a} \label{eq.red2b} \\
0&=& K_{,t}- \omega L_{(0)a}Q^{a} - \omega L_{(1)a}Q^{a}t - ... - \omega L_{(m)a}Q^{a}t^{m} \label{eq.red2c} \\
0 &=& \left(-2C_{(0)ab}Q^{b} -2C_{(1)ab}Q^{b} t - ... - 2C_{(n)ab} Q^{b} \frac{t^{n}}{n}\right) \omega_{,t} -2\omega C_{(1)ab}Q^{b} -2\omega C_{(2)ab}Q^{b}t -... - \notag \\
&& -2 \omega C_{(n)ab}Q^{b}t^{n-1} +2L_{(2)a} +6L_{(3)a}t +...+m(m-1)L_{(m)a}t^{m-2}+ \omega\left( L_{(0)b}Q^{b}\right) _{,a} +\notag \\
&& + \omega \left( L_{(1)b}Q^{b}\right)_{,a}t +...+ \omega \left( L_{(m)b}Q^{b}\right)_{,a}t^{m} \label{eq.red2d} \\
0 &=&2\omega\left( C_{(0)[a\left\vert c\right\vert }Q^{c}\right) _{;b]}+ 2\omega \left(C_{(1)[a\left\vert c\right\vert }Q^{c}\right) _{;b]}t + ...+ 2\omega \left(C_{(n)[a\left\vert c\right\vert }Q^{c}\right) _{;b]}\frac{t^{n}}{n} - L_{(1)\left[a;b\right] } - \notag \\
&&- 2L_{(2)\left[ a;b\right] }t-...-mL_{(m)\left[ a;b\right] }t^{m-1}. \label{eq.red2e}
\end{eqnarray}

In this system of PDEs, the pairs $\omega(t), Q^{a}(q)$ are not specified. As we explained in section \ref{sec.timedep.into}, we shall fix a general form of $\omega$ and find the admitted QFIs in terms of the (unspecified) $Q^a$. In the following section, we choose $\omega(t)$ to be a general polynomial in $t$; however, any other choice is possible.

\section{The case $\mathbf{\omega(t)=b_{0} +b_{1}t + ... + b_{\ell}t^{\ell}}$ with $\mathbf{b_{\ell}\neq0}$ and $\mathbf{\ell\geq1}$}

\label{sec.polynomial.omega}

We assume that
\begin{equation}
\omega(t)= b_{0}+b_{1}t+ ... + b_{\ell}t^{\ell}, \enskip b_{\ell}\neq0, \enskip \ell \geq 1 \label{pol}
\end{equation}
where $\ell$ is the degree of the polynomial. Substituting the function (\ref{pol}) in the system of equations (\ref{eq.red2a}) - (\ref{eq.red2e}), we find\footnote{The proof of Theorem \ref{thm.polynomial.omega} is given in appendix \ref{app.proof.QFIs.time}.} that there are two independent QFIs as given in Theorem \ref{thm.polynomial.omega}.

\begin{theorem}
\label{thm.polynomial.omega}
The independent QFIs of the time-dependent dynamical system (\ref{eq.red1}), where $\omega(t)= b_{0} +b_{1}t + ... +b_{\ell}t^{\ell }$ with $b_{\ell }\neq 0$ and $\ell \geq 1$, are the following:
\bigskip

\textbf{Integral 1.}

\begin{equation*}
I_{n}= \left( C_{(0)ab} +\sum^{n}_{k=1} \frac{t^{k}}{k} C_{(k)ab} \right) \dot{q}^{a} \dot{q}^{b} + \sum^{n}_{k=0} t^{k} L_{(k)a} \dot{q}^{a} +\sum^{n}_{k=0} \sum^{\ell}_{r=0} \left( L_{(k)a}Q^{a} b_{r}\frac{t^{k+r+1}}{k+r+1} \right) +G(q)
\end{equation*}
where $n=0, 1, 2, ...$, $C_{(0)ab}$ is a KT, the KTs $C_{(N)ab} = -L_{(N-1)(a;b)}$ for $N=1,...,n$, $L_{(n)a}$ is a KV, $G(q)$ is an arbitrary function defined by the condition
\begin{equation}
G_{,a}= 2b_{0}C_{(0)ab}Q^{b} -L_{(1)a} \label{thm0}
\end{equation}
$c$ is an arbitrary constant defined by the condition
\begin{equation}
L_{(n)a}Q^{a}=c \label{thm1}
\end{equation}
and the following conditions are satisfied:
\begin{align}
0&= \sum_{s=0}^{\ell-1}\left[  -\frac{2(r+s) b_{(r+s\leq\ell)}}{n-s} C_{(n-s\geq0)ab}Q^{b} -2b_{(r+s\leq\ell)} C_{(n-s>0)ab}Q^{b} + \right. \notag \\
& \quad \left. +b_{(r+s\leq\ell)} \left( L_{(n-s-1\geq0)b}Q^{b}\right)_{,a} \right], \enskip r=1,2,...,\ell \label{thm2} \\
0&= -\sum_{s=1}^{\ell}\left[ \frac{2sb_{s}}{n-s} C_{(n-s\geq0)ab}Q^{b} \right] + \sum_{s=0}^{\ell} \left[ -2b_{s} C_{(n-s>0)ab}Q^{b} + b_{s} \left(L_{(n-s-1\geq0)b}Q^{b} \right)_{,a} \right] \label{thm3} \\
0&= k(k-1)L_{(k)a} - \sum_{s=1}^{\ell} \left[ \frac{2sb_{s}}{k-s-1} C_{(k-s-1\geq0)ab}Q^{b} \right] + \notag \\
& \quad +\sum_{s=0}^{\ell} \left[ -2b_{s}C_{(k-s-1>0)ab}Q^{b} +b_{s}\left( L_{(k-s-2\geq0)b}Q^{b} \right)_{,a} \right], \enskip k=2,3,...n. \label{thm4}
\end{align}

\textbf{Integral 2.}

\begin{equation*}
I_{e} =I_{e}(\ell=1) = - e^{\lambda t} L_{(a;b)} \dot{q}^{a}\dot{q}^{b} + \lambda e^{\lambda t} L_{a}\dot{q}^{a} + \left( b_{0} - \frac{b_{1}}{\lambda} \right) e^{\lambda t}L_{a}Q^{a} + b_{1}te^{\lambda t} L_{a}Q^{a}
\end{equation*}%
where $L_{(a;b)}$ is a KT, $\left( L_b Q^{b}\right)_{,a}= \frac{\lambda^{3}}{b_{1}}L_{a}$ and $\lambda^3 L_a = -2b_{1} L_{(a;b)} Q^{b}$.

We note that the FI $I_{e}$ exists only when $\omega(t)=b_{0}+b_{1}t$, that is, for $\ell=1$.
\end{theorem}

\section{Special cases of the QFI $I_{n}$}

\label{sec.In}

The parameter $n$ in the case Integral 1 of Theorem \ref{thm.polynomial.omega} runs over all positive integers, i.e. $n=0,1,2,...$. This results in a sequence of QFIs $I_{0}, I_{1}, I_{2}, ...$, that is, one QFI $I_{n}$ for each value $n$. A significant characteristic of this sequence is that $I_{k} < I_{k+1}$, that is, each QFI $I_{k}$, where $k=0,1,2,...$, can be derived from the next QFI $I_{k+1}$ as a subcase.

In the following, we consider some special cases of the QFI $I_{n}$ for small values of $n$.

\subsection{The QFI $I_{0}$}

For $n=0$, the QFI is
\begin{equation*}
I_{0}= C_{(0)ab} \dot{q}^{a} \dot{q}^{b} +L_{(0)a}\dot{q}^{a} + b_{\ell}s \frac{t^{\ell+1}}{\ell+1} + ... + b_{1}s \frac{t^{2}}{2} + b_{0}st
\end{equation*}
where $C_{(0)ab}$ is a KT, $L_{(0)a}$ is a KV, $L_{(0)a}Q^{a}=s$ and $C_{(0)ab}Q^{b}=0$.

This QFI consists of the independent FIs:
\[
I_{0a}= C_{(0)ab} \dot{q}^{a} \dot{q}^{b} \enskip \text{and} \enskip I_{0b}= L_{(0)a}\dot{q}^{a} + b_{\ell}s \frac{t^{\ell+1}}{\ell+1} + ... + b_{1}s \frac{t^{2}}{2} + b_{0}st.
\]

\subsection{The QFI $I_{1}$}

For $n=1$, the conditions (\ref{thm1}) - (\ref{thm4}) become:
\begin{eqnarray}
L_{(1)a}Q^{a}&=& s \label{eq.polc0.2} \\
\left( L_{(0)b} Q^{b} \right)_{,a} &=& -2(\ell+1)L_{(0)(a;b)} Q^{b} \label{eq.polc0.3} \\
kb_{k} C_{(0)ab}Q^{b} &=& - (\ell-k+1)b_{k-1}L_{(0)(a;b)}Q^{b}, \enskip k=1,...,\ell. \label{eq.polc0.4}
\end{eqnarray}

Since $b_{\ell}\neq 0$, the last condition for $k=\ell$ gives $C_{(0)ab}Q^{b} =$ $- \frac{b_{\ell-1}}{\ell b_{\ell}} L_{(0)(a;b)} Q^{b}$ and the remaining equations become
\[
\left[(\ell-k+1)b_{k-1} -\frac{kb_{k}b_{\ell-1}}{\ell b_{\ell}} \right] L_{(0)(a;b)} Q^{b} =  0, \enskip k=1,...,\ell-1.
\]
The last set of equations exist only for $\ell\geq2$. By using mathematical induction and after successive substitutions, we find
\[
\left( b_{0} - \frac{b^{\ell}_{\ell-1}}{\ell^{\ell} b^{\ell-1}_{\ell}} \right) L_{(0)(a;b)} Q^{b}=0.
\]

The QFI is ($I_{0}<I_{1}$)
\begin{eqnarray*}
I_{1} &=& \left( -tL_{(0)(a;b)} + C_{(0)ab} \right) \dot{q}^{a} \dot{q}^{b} + tL_{(1)a}\dot{q}^{a} +L_{(0)a}\dot{q}^{a} +sb_{\ell}\frac{t^{\ell+2}}{\ell+2} + \left( sb_{\ell-1} + b_{\ell}L_{(0)a}Q^{a} \right) \frac{t^{\ell+1}}{\ell+1} + ... + \\
&& + \left( sb_{0} + b_{1}L_{(0)a}Q^{a} \right) \frac{t^{2}}{2} + b_{0}L_{(0)a}Q^{a}t + G(q)
\end{eqnarray*}
where $C_{(0)ab}$ and $L_{(0)(a;b)}$ are KTs, $L_{(1)a}$ is a KV, $L_{(1)a}Q^{a}=s$, $\left( L_{(0)b} Q^{b} \right)_{,a} = -2(\ell+1)L_{(0)(a;b)}Q^{b}$, $C_{(0)ab}Q^{b} =$ $- \frac{b_{\ell-1}}{\ell b_{\ell}} L_{(0)(a;b)} Q^{b}$, $\left[(\ell-k+1)b_{k-1} -\frac{kb_{k}b_{\ell-1}}{\ell b_{\ell}} \right] L_{(0)(a;b)} Q^{b} =0$ with  $k=1,...,\ell-1$, and $G_{,a}= 2b_{0}C_{(0)ab}Q^{b} - L_{(1)a}$.
\bigskip

For some values of the degree $\ell$ of the polynomial $\omega(t)$, we have the following:

1) For $\ell=1$.

We have $\omega=b_{0}+b_{1}t$ and the QFI is
\begin{equation*}
I_{1}= \left( -tL_{(0)(a;b)} + C_{(0)ab} \right) \dot{q}^{a} \dot{q}^{b} + tL_{(1)a}\dot{q}^{a} +L_{(0)a}\dot{q}^{a} +sb_{1}\frac{t^{3}}{3} + \left( sb_{0} + b_{1}L_{(0)a}Q^{a} \right) \frac{t^{2}}{2} + b_{0}L_{(0)a}Q^{a}t + G(q)
\end{equation*}
where $C_{(0)ab}$ and $L_{(0)(a;b)}$ are KTs, $L_{(1)a}$ is a KV, $L_{(1)a}Q^{a}=s$, $\left( L_{(0)b} Q^{b} \right)_{,a} = -4L_{(0)(a;b)}Q^{b}$, $C_{(0)ab}Q^{b} = - \frac{b_{0}}{b_{1}} L_{(0)(a;b)} Q^{b}$, and $G_{,a}= 2b_{0}C_{(0)ab}Q^{b} - L_{(1)a}$.

2) For $\ell=2$.

We have $\omega=b_{0}+b_{1}t+b_{2}t^{2}$ and the QFI is
\begin{eqnarray*}
I_{1} &=& \left( -tL_{(0)(a;b)} + C_{(0)ab} \right) \dot{q}^{a} \dot{q}^{b} + tL_{(1)a}\dot{q}^{a} +L_{(0)a}\dot{q}^{a} +sb_{2}\frac{t^{4}}{4} + \left( sb_{1} + b_{2}L_{(0)a}Q^{a} \right) \frac{t^{3}}{3}+ \\
&& + \left( sb_{0} + b_{1}L_{(0)a}Q^{a} \right) \frac{t^{2}}{2} + b_{0}L_{(0)a}Q^{a}t + G(q)
\end{eqnarray*}
where $C_{(0)ab}$ and $L_{(0)(a;b)}$ are KTs, $L_{(1)a}$ is a KV, $L_{(1)a}Q^{a}=s$, $\left( L_{(0)b} Q^{b} \right)_{,a} = -6L_{(0)(a;b)}Q^{b}$, $C_{(0)ab}Q^{b} = - \frac{b_{1}}{2b_{2}} L_{(0)(a;b)} Q^{b}$, $\left( b_{0} -\frac{b_{1}^{2}}{4b_{2}} \right) L_{(0)(a;b)} Q^{b} =0$, and $G_{,a}= 2b_{0}C_{(0)ab}Q^{b} - L_{(1)a}$.

3) For $\ell=3$.

We have $\omega=b_{0} + b_{1}t + b_{2}t^{2}+ b_{3}t^{3}$ and the QFI is
\begin{eqnarray*}
I_{1} &=& \left( -tL_{(0)(a;b)} + C_{(0)ab} \right) \dot{q}^{a} \dot{q}^{b} + tL_{(1)a}\dot{q}^{a} +L_{(0)a}\dot{q}^{a} +sb_{3}\frac{t^{5}}{5} + \left( sb_{2} + b_{3}L_{(0)a}Q^{a} \right) \frac{t^{4}}{4} + \\
&& + \left( sb_{1} + b_{2}L_{(0)a}Q^{a} \right) \frac{t^{3}}{3}+ \left( sb_{0} + b_{1}L_{(0)a}Q^{a} \right) \frac{t^{2}}{2} + b_{0}L_{(0)a}Q^{a}t + G(q)
\end{eqnarray*}
where $C_{(0)ab}$ and $L_{(0)(a;b)}$ are KTs, $L_{(1)a}$ is a KV, $L_{(1)a}Q^{a}=s$, $\left( L_{(0)b} Q^{b} \right)_{,a} = -8L_{(0)(a;b)}Q^{b}$, $C_{(0)ab}Q^{b} =$ $-\frac{b_{2}}{3b_{3}} L_{(0)(a;b)} Q^{b}$, $\left( b_{0} - \frac{b_{1}b_{2}}{9b_{3}} \right) L_{(0)(a;b)} Q^{b} =0$, $\left( b_{1} - \frac{b_{2}^{2}}{3b_{3}} \right) L_{(0)(a;b)} Q^{b} =0$, and $G_{,a}= 2b_{0}C_{(0)ab}Q^{b} - L_{(1)a}$.

\section{The basis method for computing QFIs}

\label{sec.general.approach}

As it has been explained in section \ref{sec.timedep.into}, in the basis method,\index{Method! basis} instead of
considering the KT $K_{ab}$ to be given as a polynomial in $t$
with coefficients arbitrary KTs (see eq. (\ref{eq.aspm1}) ), one defines the KT $K_{ab}(t,q)$ by the requirement
\begin{equation}
K_{ab}(t,q)=\sum_{N=1}^{m}\alpha_{N}(t)C_{(N)ab}(q)  \label{eq.secapr.1}
\end{equation}%
where $\alpha_{N}(t)$ are arbitrary smooth functions and the $m$ linearly independent KTs $C_{(N)ab}(q)$ constitute a basis of the vector space of KTs of the kinetic metric $\gamma_{ab}(q)$. In this case, one does not assume a form for the vector
$K_{a}(t,q)$, which is determined from the resulting system of equations (\ref{eq.TKN1}) - (\ref{eq.TKN6}).

The basis method has been used previously by Katzin and
Levine in \cite{Katzin 1982}, in order to determine the QFIs for the time-dependent Kepler potential. As we shall apply the basis method to 3d Newtonian systems, we need a basis of KTs (and other collineations) of the Euclidean space $E^{3}$. This basis and, in general, the geometric quantities of $E^{3}$ have been already computed in section \ref{sec.KTE3}.

\section{The time-dependent Newtonian generalized Kepler potential}

\label{sec.TimeGKepler}

The time-dependent Newtonian generalized Kepler potential \index{Potential! time-dependent generalized Kepler} is $V=-\frac{\omega(t)}{r^{\nu}}$, where $\nu$ is a non-zero real constant and $r=(x^{2}+y^{2}+z^{2})^{\frac{1}{2}}$. This potential contains (among others) the 3d time-dependent oscillator \cite{Ray 1979A, Lewis 1968, Katzin 1977, Prince 1980, Gunther 1977} for $\nu =-2$, the time-dependent Kepler potential \cite{Leach 1985, LeoTsampAndro 2017, Prince 1981, Katzin 1982} for $\nu =1$, and the Newton-Cotes potential for $\nu=2$ \cite{Ibragimov 1998}. The integrability of these systems has been studied in numerous works over the years using various methods; mainly the Noether symmetries. Our purpose is to recover the results of these works --and also new ones-- using the basis method.

The Lagrangian of the system is
\begin{equation}
L=\frac{1}{2}(\dot{x}^{2}+\dot{y}^{2}+\dot{z}^{2})+\frac{\omega(t) }{r^{\nu}}
\label{eq.TGKep.1}
\end{equation}%
and the corresponding E-L equations are:
\begin{equation}
\ddot{x}=-\frac{\nu \omega(t)}{r^{\nu+2}}x, \enskip \ddot{y}=-\frac{\nu \omega(t)}{r^{\nu +2}}y, \enskip \ddot{z}=-\frac{\nu \omega(t)}{r^{\nu +2}}z.
\label{eq.TGKep.1a}
\end{equation}
For this dynamical system, the $Q^{a}= \frac{\nu q^{a}}{r^{\nu+2}}$ where $q^{a}= (x,y,z)$, whereas the $\omega(t)$ is unspecified. We shall determine the `frequencies' $\omega(t)$ for which the resulting LFIs/QFIs are not combinations of the angular momentum.

The LFIs/QFIs of the autonomous generalized Kepler potential,\index{Potential! generalized Kepler} that is, $\omega(t)=k=const$, have been determined in chapter \ref{ch1.QFIs.conservative} using the direct method and are listed in Table \ref{T1} (see also Table \ref{Table.QFIs.Kepler.gen}).

In Table \ref{T1}, $H_{\nu}$ is the Hamiltonian of the system, $L_{i}$ are the components of the angular momentum, $R_{i}$ are the components of the Runge-Lenz vector,\index{Vector! Runge-Lenz} and $B_{ij}$ are the components of the Jauch-Hill-Fradkin tensor.\index{Tensor! Jauch-Hill-Fradkin}

\begin{longtable}{|l|l|}
\hline
$V=-\frac{k}{r^{\nu}}$ & LFIs and QFIs \\ \hline
$\forall$ $\nu$ & $L_{1} = y\dot{z} - z\dot{y}$, $L_{2}= z\dot{x} - x\dot{z}$, $L_{3}= x\dot{y} - y\dot{x}$, $H_{\nu}= \frac{1}{2}(\dot{x}^{2} + \dot{y}^{2} + \dot{z}^{2}) - \frac{k}{r^{\nu}}$ \\
$\nu=-2$ & $B_{ij} = \dot{q}_{i} \dot{q}_{j} - 2k q_{i}q_{j}$ \\
$\nu=-2$, $k>0$ & $I_{3a\pm}= e^{\pm \sqrt{2k} t}(\dot{q}_{a} \mp \sqrt{2k} q_{a})$ \\
$\nu=-2$, $k<0$ & $I_{3a\pm}= e^{\pm i \sqrt{-2k} t}(\dot{q}_{a} \mp i \sqrt{-2k} q_{a})$ \\
$\nu=1$ & $R_{i}= (\dot{q}^{j} \dot{q}_{j}) q_{i} - (\dot{q}^{j}q_{j})\dot{q}_{i}- \frac{k}{r}q_{i}$ \\
$\nu=2$ & $I_{1}= - H_{2}t^{2} + t(\dot{q}^{i}q_{i}) - \frac{r^{2}}{2}$, $I_{2}= - H_{2}t + \frac{1}{2} (\dot{q}^{i}q_{i})$ \\ \hline
\caption{\label{T1} The LFIs/QFIs of the autonomous generalized Kepler potential for $\omega(t)=k=const$.}
\end{longtable}

Replacing $Q^{a}=\frac{\nu q^{a}}{r^{\nu+2}}$ in the conditions (\ref{eq.TKN1}) - (\ref{eq.TKN6}), we obtain the following system of PDEs \cite{Katzin 1982}:
\begin{eqnarray}
K_{(ab;c)} &=&0  \label{eq.TKNq1} \\
K_{(a;b)} + K_{ab,t} &=& 0 \label{eq.TKNq2} \\
K_{,a} -\frac{2\nu\omega}{r^{\nu+2}}K_{ab}q^{b} +K_{a,t} &=&0 \label{eq.TKNq3} \\
K_{,t} -\frac{\nu \omega}{r^{\nu+2}}K_{a}q^{a} &=&0 \label{eq.TKNq4} \\
K_{a,tt} + \nu\omega\left( \frac{K_{b}q^{b}}{r^{\nu+2}} \right)_{,a} -\frac{2\nu\omega_{,t}}{r^{\nu+2}}K_{ab}q^{b} -\frac{2\nu\omega}{r^{\nu+2}}K_{ab,t}q^{b} &=& 0 \label{eq.TKNq5} \\
K_{[a;b],t} -2 \nu\omega\left( \frac{K_{[a|c|}q^{c}}{r^{\nu+2}} \right)_{;b]} &=&0.  \label{eq.TKNq6}
\end{eqnarray}

From the Lagrangian (\ref{eq.TGKep.1}), we infer that the kinetic metric is $\delta_{ij}=diag(1,1,1)$.

According to the basis approach, the KT $K_{ab}(t,q)$ of (\ref{eq.TKNq1}) is given by (\ref{FL.E3}) provided that the twenty arbitrary constants $a_{I}$ are assumed to be time-dependent functions $a_{I}(t)$.

Condition (\ref{eq.TKNq2}) gives $K_{a,b} + K_{b,a} = -2K_{ab,t}$ which implies:
\begin{eqnarray}
K_{1,1} &=& -K_{11,t} \label{eq.TKNq7a} \\
K_{2,2} &=& -K_{22,t} \label{eq.TKNq7b} \\
K_{3,3} &=& -K_{33,t} \label{eq.TKNq7c} \\
K_{1,2} + K_{2,1} &=& -2K_{12,t} \label{eq.TKNq7d} \\
K_{1,3} + K_{3,1} &=& -2K_{13,t} \label{eq.TKNq7e} \\
K_{2,3} + K_{3,2} &=& -2K_{23,t}. \label{eq.TKNq7f}
\end{eqnarray}

From the first three conditions (\ref{eq.TKNq7a}) - (\ref{eq.TKNq7c}), we find:
\begin{eqnarray*}
K_{1} &=& -\frac{\dot{a}_{6}}{2}xy^{2} -\frac{\dot{a}_{1}}{2}
xz^{2} -\dot{a}_{4}xyz -\dot{a}_{5}xy -\dot{a}_{2}xz -\dot{a}_{3}x +A(y,z,t) \\
K_{2} &=& -\frac{\dot{a}_{6}}{2}yx^{2} -\frac{\dot{a}_{7}}{2} yz^{2} -\dot{a}_{14}xyz -\dot{a}_{15}xy -\dot{a}_{12}yz -\dot{a}_{13}y + B(x,z,t) \\
K_{3} &=& -\frac{\dot{a}_{1}}{2}zx^{2} -\frac{\dot{a}_{7}}{2}zy^{2} -\dot{a}_{10}xyz -\dot{a}_{11}xz -\dot{a}_{8}yz -\dot{a}_{9}z + C(x,y,t)
\end{eqnarray*}
where $A, B, C$ are arbitrary functions.

Substituting these results in (\ref{eq.TKNq7d}) - (\ref{eq.TKNq7f}), we obtain:
\begin{align}
0&= \dot{a}_{10}z^{2} -3\dot{a}_{6}xy -2\dot{a}_{4}xz -2\dot{a}_{14}yz -2\dot{a}_{5}x  -2\dot{a}_{15}y +2\dot{a}_{16}z +2\dot{a}_{17} + A_{,2} + B_{,1} \label{eq.TKNq8a} \\
0&= \dot{a}_{14}y^{2} -2\dot{a}_{4}xy -3\dot{a}_{1}xz -2\dot{a}_{10}yz -2\dot{a}_{2}x +2\dot{a}_{18}y -2\dot{a}_{11}z +2\dot{a}_{19} + A_{,3} + C_{,1} \label{eq.TKNq8b}
\\
0&= \dot{a}_{4}x^{2} -2\dot{a}_{14}xy -2\dot{a}_{10}xz -3\dot{a}_{7}yz -2(\dot{a}_{16}+\dot{a}_{18})x -2\dot{a}_{12}y -2\dot{a}_{8}z +2\dot{a}_{20} + \notag \\
& \quad +B_{,3} + C_{,2}. \label{eq.TKNq8c}
\end{align}

By taking the second partial derivatives of (\ref{eq.TKNq8a}) wrt $x, y$, of (\ref{eq.TKNq8b}) wrt $x, z$, and of (\ref{eq.TKNq8c}) wrt $y,z$, we find that: $a_{1}=c_{1}$, $a_{6}=c_{2}$ and $a_{7}=c_{3}$ are arbitrary constants.

Then, equations (\ref{eq.TKNq8a}) - (\ref{eq.TKNq8c}) become:
\begin{eqnarray}
0&=& \dot{a}_{10}z^{2} -2\dot{a}_{4}xz -2\dot{a}_{14}yz -2\dot{a}_{5}x  -2\dot{a}_{15}y +2\dot{a}_{16}z +2\dot{a}_{17} + A_{,2} + B_{,1} \label{eq.TKNq9a} \\
0&=& \dot{a}_{14}y^{2} -2\dot{a}_{4}xy -2\dot{a}_{10}yz -2\dot{a}_{2}x +2\dot{a}_{18}y -2\dot{a}_{11}z +2\dot{a}_{19} + A_{,3} + C_{,1} \label{eq.TKNq9b} \\
0&=& \dot{a}_{4}x^{2} -2\dot{a}_{14}xy -2\dot{a}_{10}xz -2(\dot{a}_{16}+\dot{a}_{18})x -2\dot{a}_{12}y -2\dot{a}_{8}z +2\dot{a}_{20} + B_{,3} + C_{,2}. \label{eq.TKNq9c}
\end{eqnarray}

By suitable differentiations of the above equations, we obtain:
$A_{,22}= 2\dot{a}_{14}z +2\dot{a}_{15}$, $A_{,33}= 2\dot{a}_{10}y + 2\dot{a}_{11}$, $B_{,11}= 2\dot{a}_{4}z + 2\dot{a}_{5}$, $B_{,33}= 2\dot{a}_{10}x +2\dot{a}_{8}$, $C_{,11}= 2\dot{a}_{4}y + 2\dot{a}_{2}$ and $C_{,22}= 2\dot{a}_{14}x +2\dot{a}_{12}$. Then,
\begin{eqnarray*}
A&=& \dot{a}_{14}zy^{2} +\dot{a}_{10}yz^{2} +\dot{a}_{15}y^{2} + \dot{a}_{11}z^{2} + \sigma_{1}(t)yz + \sigma_{2}(t)y + \sigma_{3}(t)z + \sigma_{4}(t) \\
B&=& \dot{a}_{4}zx^{2} + \dot{a}_{10}xz^{2} + \dot{a}_{5}x^{2} + \dot{a}_{8}z^{2} + \tau_{1}(t)xz + \tau_{2}(t)x + \tau_{3}(t)z + \tau_{4}(t) \\
C&=& \dot{a}_{4}yx^{2} + \dot{a}_{14}xy^{2} + \dot{a}_{2}x^{2} + \dot{a}_{12}y^{2} + \eta_{1}(t)xy + \eta_{2}(t)x + \eta_{3}(t)y + \eta_{4}(t)
\end{eqnarray*}
where $\sigma_{k}(t), \tau_{k}(t), \eta_{k}(t)$ for $k=1,2,3,4$ are arbitrary functions.

Substituting the above results in (\ref{eq.TKNq9a}) - (\ref{eq.TKNq9c}), we find:
\begin{eqnarray*}
(\ref{eq.TKNq9a}) &\implies& a_{10}=c_{4}, \enskip \sigma_{1}= -\tau_{1} -2\dot{a}_{16}, \enskip \sigma_{2}= - \tau_{2} -2\dot{a}_{17} \\
(\ref{eq.TKNq9b}) &\implies& a_{14}=c_{5}, \enskip \eta_{1}= -\sigma_{1} -2\dot{a}_{18}, \enskip \eta_{2}= - \sigma_{3} -2\dot{a}_{19} \\
(\ref{eq.TKNq9c}) &\implies& a_{4}=c_{6}, \enskip \tau_{1}= -\eta_{1} +2(\dot{a}_{16} + \dot{a}_{18}), \enskip \tau_{3}= -\eta_{3} -2\dot{a}_{20}
\end{eqnarray*}
from which, we have finally: $a_{10}=c_{4}$, $a_{14}=c_{5}$, $a_{4}=c_{6}$, $\tau_{1}=2\dot{a}_{18}$, $\eta_{1}=2\dot{a}_{16}$, $\sigma_{1}= -2(\dot{a}_{16} + \dot{a}_{18})$, $\tau_{2}= - \sigma_{2} -2\dot{a}_{17}$, $\eta_{2}= - \sigma_{3} -2\dot{a}_{19}$ and $\eta_{3}= -\tau_{3} -2\dot{a}_{20}$, where $c_{4}, c_{5}, c_{6}$ are arbitrary constants.

Therefore, the KT $K_{ab}$ is
\begin{eqnarray}
K_{11} &=&\frac{c_{2}}{2}y^{2}+\frac{c_{1}}{2}%
z^{2}+c_{6}yz+a_{5}y+a_{2}z+a_{3} \notag \\
K_{12} &=&\frac{c_{4}}{2}z^{2}-\frac{c_{2}}{2}xy-\frac{c_{6}}{2}xz-\frac{%
c_{5}}{2}yz-\frac{a_{5}}{2}x-\frac{a_{15}}{2}y+a_{16}z+a_{17}  \notag \\
K_{13} &=&\frac{c_{5}}{2}y^{2}-\frac{c_{6}}{2}xy-\frac{c_{1}}{2}xz-\frac{%
c_{4}}{2}yz-\frac{a_{2}}{2}x+a_{18}y-\frac{a_{11}}{2}z+a_{19} \label{eq.KT}
\\
K_{22} &=&\frac{c_{2}}{2}x^{2}+\frac{c_{3}}{2}%
z^{2} +c_{5}xz+a_{15}x+a_{12}z+a_{13} \notag \\
K_{23} &=&\frac{c_{6}}{2}x^{2}-\frac{c_{5}}{2}xy -\frac{c_{4}}{2}xz-\frac{c_{3}}{2}yz -(a_{16}+a_{18})x-\frac{a_{12}}{2}y -\frac{a_{8}}{2}z+a_{20} \notag \\
K_{33} &=&\frac{c_{1}}{2}x^{2}+\frac{c_{3}}{2}%
y^{2}+c_{4}xy+a_{11}x+a_{8}y+a_{9} \notag
\end{eqnarray}
and the vector $K_{a}$ is
\begin{eqnarray}
K_{1} &=& \dot{a}_{15}y^{2} + \dot{a}_{11}z^{2} -\dot{a}_{5}xy -\dot{a}_{2}xz -2(\dot{a}_{16}+\dot{a}_{18})yz -\dot{a}_{3}x + \sigma_{2}y + \sigma_{3}z + \sigma_{4} \notag \\
K_{2} &=& \dot{a}_{5}x^{2} + \dot{a}_{8}z^{2} -\dot{a}_{15}xy + 2\dot{a}_{18}xz -\dot{a}_{12}yz -(\sigma_{2}+2\dot{a}_{17})x -\dot{a}_{13}y + \tau_{3}z + \tau_{4} \label{eq.K} \\
K_{3} &=& \dot{a}_{2}x^{2} + \dot{a}_{12}y^{2} +2\dot{a}_{16}xy -\dot{a}_{11}xz -\dot{a}_{8}yz -(\sigma_{3}+2\dot{a}_{19})x -(\tau_{3}+2\dot{a}_{20})y -\dot{a}_{9}z + \eta_{4}. \notag
\end{eqnarray}

Replacing the above quantities in the constraint (\ref{eq.TKNq6}), we find the following set of conditions:
\begin{equation}
a_{2}=a_{12}, \enskip a_{5}=a_{8}, \enskip a_{11}=a_{15}, \enskip a_{16}=a_{18}=0 \label{eq.conTK1a}
\end{equation}
\begin{equation}
(\nu-1)a_{2}=0, \enskip (\nu-1)a_{5}=0, \enskip (\nu-1)a_{11}=0 \label{eq.conTK1b}
\end{equation}
\begin{equation}
(\nu+2)a_{17}=0, \enskip (\nu+2)a_{19}=0, \enskip (\nu+2)a_{20}=0, \enskip (\nu+2)(a_{3}-a_{9})=0, \enskip (\nu+2)(a_{3}-a_{13})=0 \label{eq.conTK1c}
\end{equation}
\begin{equation}
\ddot{a}_{2}=\ddot{a}_{5}=\ddot{a}_{11}=0, \enskip \dot{\sigma}_{2}= -\ddot{a}_{17}, \enskip \dot{\sigma}_{3}= -\ddot{a}_{19}, \enskip \dot{\tau}_{3}= -\ddot{a}_{20}. \label{eq.conTK1d}
\end{equation}

We consider three cases depending on the value of $\nu$: \newline
- $\forall \nu$. The general case. \newline
- $\nu =1$. Time-dependent Kepler potential. \newline
- $\nu =-2$. Time-dependent 3d oscillator.

The Newton-Cotes potential ($\nu=2$) is contained as a subcase of the general case.

\section{The general case}

\label{3.2}

Because this case holds for any value of $\nu$, conditions (\ref{eq.conTK1a}) - (\ref{eq.conTK1d}) give: \newline
$a_{2}=a_{5}=a_{8}=a_{11}=a_{12}=a_{15}=a_{16}=a_{17} =a_{18}=a_{19}=a_{20}=0$, $a_{3}=a_{9}=a_{13}$, $\sigma_{2}=c_{7}$, $\sigma_{3}= c_{8}$ and $\tau_{3}= c_{9}$,
where $c_{7}, c_{8}, c_{9}$ are arbitrary constants.

Substituting these results in the constraint (\ref{eq.TKNq5}), we find that
\begin{equation}
\dddot{a}_{3}=0, \enskip (\nu-2)\omega\dot{a}_{3} - 2\dot{\omega}a_{3}=0 \label{eq.gen1.1}
\end{equation}
and
\[
\ddot{\sigma}_{4}= \ddot{\tau}_{4}= \ddot{\eta}_{4}=0, \enskip \omega\sigma_{4}=\omega\tau_{4}=\omega\eta_{4}=0 \implies \sigma_{4}=\tau_{4}=\eta_{4}=0.
\]
Therefore, the KT $K_{ab}$ becomes
\begin{equation}
K_{ab}=
\left(
  \begin{array}{ccc}
    \frac{c_{2}}{2}y^{2}+\frac{c_{1}}{2}z^{2}+c_{6}yz+a_{3} & \frac{c_{4}}{2}z^{2}-\frac{c_{2}}{2}xy-\frac{c_{6}}{2}xz-\frac{%
c_{5}}{2}yz & \frac{c_{5}}{2}y^{2}-\frac{c_{6}}{2}xy-\frac{c_{1}}{2}xz-\frac{%
c_{4}}{2}yz \\
    \frac{c_{4}}{2}z^{2}-\frac{c_{2}}{2}xy-\frac{c_{6}}{2}xz-\frac{%
c_{5}}{2}yz & \frac{c_{2}}{2}x^{2}+\frac{c_{3}}{2}%
z^{2} +c_{5}xz +a_{3} & \frac{c_{6}}{2}x^{2}-\frac{c_{5}}{2}xy -\frac{c_{4}}{2}xz-\frac{c_{3}}{2}yz \\
    \frac{c_{5}}{2}y^{2}-\frac{c_{6}}{2}xy-\frac{c_{1}}{2}xz-\frac{%
c_{4}}{2}yz & \frac{c_{6}}{2}x^{2}-\frac{c_{5}}{2}xy -\frac{c_{4}}{2}xz-\frac{c_{3}}{2}yz & \frac{c_{1}}{2}x^{2} +\frac{c_{3}}{2} y^{2}+c_{4}xy +a_{3} \\
  \end{array}
\right) \label{eq.gencas.1}
\end{equation}
and the vector
\begin{equation}
K_{a}=
\left(
  \begin{array}{c}
    -\dot{a}_{3}x + c_{7}y + c_{8}z \\
    -c_{7}x -\dot{a}_{3}y + c_{9}z \\
    -c_{8}x -c_{9}y -\dot{a}_{3}z \\
  \end{array}
\right). \label{eq.gencas.2}
\end{equation}

Since the ten parameters $a_{3}(t)$ and $c_{A}$, where $A=1, 2, ..., 9$, are independent (i.e. they generate different FIs), we consider the following two cases.

\subsection{$a_{3}(t)=0$}

\label{sec.gen1}

In this case, conditions (\ref{eq.gen1.1}) are satisfied identically leaving the function $\omega(t)$ free to be any function.

The KT (\ref{eq.gencas.1}) becomes
\begin{equation*}
K_{ab}=
\left(
\begin{array}{ccc}
\frac{c_{2}}{2}y^{2}+\frac{c_{1}}{2}z^{2}+c_{6}yz & \frac{c_{4}}{2}z^{2}-\frac{c_{2}}{2}xy-\frac{c_{6}}{2}xz-\frac{%
c_{5}}{2}yz & \frac{c_{5}}{2}y^{2}-\frac{c_{6}}{2}xy-\frac{c_{1}}{2}xz-\frac{%
c_{4}}{2}yz \\
\frac{c_{4}}{2}z^{2}-\frac{c_{2}}{2}xy-\frac{c_{6}}{2}xz-\frac{%
c_{5}}{2}yz & \frac{c_{2}}{2}x^{2}+\frac{c_{3}}{2}%
z^{2} +c_{5}xz & \frac{c_{6}}{2}x^{2}-\frac{c_{5}}{2}xy -\frac{c_{4}}{2}xz-\frac{c_{3}}{2}yz \\
    \frac{c_{5}}{2}y^{2}-\frac{c_{6}}{2}xy-\frac{c_{1}}{2}xz-\frac{%
c_{4}}{2}yz & \frac{c_{6}}{2}x^{2}-\frac{c_{5}}{2}xy -\frac{c_{4}}{2}xz-\frac{c_{3}}{2}yz & \frac{c_{1}}{2}x^{2} +\frac{c_{3}}{2} y^{2}+c_{4}xy \\
  \end{array}
\right)
\end{equation*}
and the vector (\ref{eq.gencas.2}) becomes the general non-gradient KV $K_{a}=
\left(
  \begin{array}{c}
   c_{7}y + c_{8}z \\
   -c_{7}x +c_{9}z \\
   -c_{8}x -c_{9}y \\
  \end{array}
\right)$.

Then, the constraint (\ref{eq.TKNq4}) implies that (since $K_{a}q^{a}=0$) $K=G(x,y,z)$, which when replaced in (\ref{eq.TKNq3}) gives (since $K_{ab}q^{b}=0$) $G_{,a}=0$. Hence, $K=const \equiv0$.

The QFI $I=K_{ab}\dot{q}^{a}\dot{q}^{b} + K_{a}\dot{q}^{a}$ leads only to the three components $L_{i}$ of the angular momentum.\index{Momentum! angular} In particular, $I$ contains nine independent parameters each of them defining a FI: a) $c_{7}$, $c_{8}$, $c_{9}$ lead to the components $L_{1}= y\dot{z} - z\dot{y}$, $L_{2}= z\dot{x} - x\dot{z}$, $L_{3}= x\dot{y} - y\dot{x}$ of the angular momentum (LFIs); and b) $c_{1}$, $c_{2}$, $c_{3}$, $c_{4}$, $c_{5}$, $c_{6}$ lead to the products (QFIs depending on $L_{i}$) $L_{1}^{2}$, $L_{2}^{2}$, $L_{3}^{2}$, $L_{1}L_{2}$, $L_{1}L_{3}$ and $L_{2}L_{3}$.

We have the following proposition.

\begin{proposition} \label{pro.lfis}
The time-dependent generalized Kepler potential $V(t,q)= -\frac{\omega(t)}{r^{\nu}}$, for a general smooth function $\omega(t)$, admits only the LFIs of the angular momentum $L_{i}$. Independent QFIs in general do not exist; they are all quadratic combinations of $L_{i}$.
\end{proposition}

\subsection{$c_{A}=0$ where $A=1, 2, ..., 9$}

\label{sec.gen2}

In this case, conditions (\ref{eq.gen1.1}) imply that $a_{3}(t)= b_{0} + b_{1}t + b_{2}t^{2}$ and
\begin{equation}
\omega_{(\nu)}(t)= k\left(b_{0} + b_{1}t + b_{2}t^{2} \right)^{\frac{\nu-2}{2}} \label{eq.TKFI8}
\end{equation}
where $k, b_{0}, b_{1}, b_{2}$ are arbitrary constants and the index $(\nu)$ denotes the dependence of $\omega(t)$ on the value of $\nu$.

Since $c_{A}=0$, the quantities (\ref{eq.gencas.1}) and (\ref{eq.gencas.2}) become, respectively, $K_{ab}= a_{3}\delta_{ab}$ and $K_{a}= -\dot{a}_{3}q_{a}$. Substituting in the remaining constraints (\ref{eq.TKNq3}) and (\ref{eq.TKNq4}), we find $K= b_{2}r^{2} - \frac{2k(b_{0}+b_{1}t +b_{2}t^{2})^{\nu/2}}{r^{\nu}}$.

The QFI is
\begin{equation}
J_{\nu}= (b_{0} + b_{1}t + b_{2}t^{2}) \left[ \frac{\dot{q}^{i}\dot{q}_{i}}{2} - \frac{k(b_{0} +b_{1}t + b_{2}t^{2})^{\frac{\nu-2}{2}}}{r^{\nu}} \right]-\frac{b_{1} + 2b_{2}t}{2} q^{i}\dot{q}_{i} +\frac{b_{2} r^{2}}{2}. \label{eq.TKFI7}
\end{equation}

We note that the resulting time-dependent generalized Kepler potential\index{Potential! time-dependent generalized Kepler}
\begin{equation}
V= -\frac{\omega_{\nu}(t)}{r^{\nu}}, \enskip \omega_{\nu}= k\left(b_{0} + b_{1}t + b_{2}t^{2} \right)^{\frac{\nu-2}{2}} \label{eq.gen1}
\end{equation}
is a subcase of the Case III potential of \cite{Leach 1985} if we set  $U\left( \frac{r}{\phi} \right)=$ $k_{1}\frac{r^{2}}{\phi^{2}} -\frac{k\phi^{\nu}}{r^{\nu}}$
with $\phi= \sqrt{b_{0}+ b_{1}t +b_{2}t^{2}}$ and $k_{1}= \frac{b_{0}b_{2}}{2} -\frac{b_{1}^{2}}{8}$. Then, the associated QFI (3.13) of \cite{Leach 1985} (for $K_{1}=K_{2}=0$) reduces to the QFI $J_{\nu}$.

For some values of $\nu$, we have the following results:

- $\nu=1$ (time-dependent Kepler potential).\index{Potential! time-dependent Kepler}

The $\omega_{(1)}(t)= k\left(b_{0} + b_{1}t + b_{2}t^{2} \right)^{-1/2}$ and the QFI $J_{1}=E_{3}$ (see section \ref{sec.omega.3} below).

- $\nu=2$ (Newton-Cotes potential \cite{Ibragimov 1998}).\index{Potential! Newton-Cotes}

The $\omega_{(2)}=k=const$ and the QFI is
\begin{equation*}
J_{2}= (b_{0} + b_{1}t + b_{2}t^{2})\left( \frac{\dot{q}^{i} \dot{q}_{i}}{2} - \frac{k}{r^{2}} \right) -\frac{b_{1} + 2b_{2} t}{2} q^{i}\dot{q}_{i} +\frac{b_{2}}{2} r^{2} =b_{0}H_{2} -b_{1} I_{2} -b_{2}I_{1}.
\end{equation*}
This expression contains the independent QFIs:
\[
H_{2}= \frac{\dot{q}^{i}\dot{q}_{i}}{2} - \frac{k}{r^{2}}, \enskip I_{1}= -t^{2}H_{2} + tq^{i}\dot{q}_{i} -\frac{r^{2}}{2}, \enskip I_{2}= -tH_{2} +\frac{q^{i}\dot{q}_{i}}{2}
\]
where $H_{2}$ is the Hamiltonian of the system. These are the
FIs found in chapter \ref{ch1.QFIs.conservative} (see also Table \ref{T1}) in the case of the autonomous generalized Kepler potential for $\nu=2$.

- $\nu=-2$ (time-dependent oscillator).\index{Oscillator! time-dependent}

The $\omega_{(-2)}= k\left(b_{0} + b_{1}t + b_{2}t^{2} \right)^{-2}$ and the QFI is
\[
J_{-2}= (b_{0} + b_{1}t + b_{2}t^{2}) \left[ \frac{\dot{q}^{i}\dot{q}_{i}}{2} - \frac{k}{(b_{0} +b_{1}t + b_{2}t^{2})^{2}}r^{2} \right]-\frac{b_{1} + 2b_{2}t}{2} q^{i}\dot{q}_{i} +\frac{b_{2} r^{2}}{2}.
\]
This is the trace of the QFIs (\ref{eq.osc.FI3}) found below for $a_{3}(t)= b_{0} + b_{1}t + b_{2}t^{2}$. Substituting this $a_{3}(t)$ in (\ref{eq.osc.FI2}) and (\ref{eq.osc.FI3}), we find, respectively, that the $\omega= \omega_{(-2)}$ with constant $k=-\frac{1}{8}(b_{1}^{2} -4b_{2}b_{0} +2c_{0})$ and the QFIs are
\begin{equation}
I_{ij}= \Lambda_{ij}(a_{3}=b_{0}+b_{1}t+b_{2}t^{2}) = (b_{0}+b_{1}t+b_{2}t^{2}) \left( \dot{q}_{i}\dot{q}_{j} -2\omega q_{i}q_{j} \right) -(b_{1} +2b_{2}t)q_{(i}\dot{q}_{j)} +b_{2}q_{i}q_{j}. \label{eq.TKFI9}
\end{equation}
Therefore, the trace $Tr[I_{ij}]= I_{11}+I_{22}+I_{33} =2J_{-2}$. We note that $r^{2}=q^{i}q_{i}$.

We infer the following new general result, which includes the time-dependent Kepler potential and the time-dependent oscillator as subcases.

\begin{proposition}[3d time-dependent generalized Kepler potentials which admit FIs] \label{oscillator} For all \\ functions $\omega(t)$ the time-dependent generalized Kepler potential\index{Potential! time-dependent generalized Kepler} $V(t,q)= -\frac{\omega(t)}{r^{\nu}}$ admits the LFIs of the angular momentum and QFIs, which are products of the components of the angular momentum. However, for the function $\omega(t)=\omega_{(\nu)}(t) =k\left(b_{0} + b_{1}t + b_{2}t^{2} \right)^{\frac{\nu-2}{2}}$ the resulting time-dependent generalized Kepler potential admits the additional QFI $J_{\nu}$ given by (\ref{eq.TKFI7}).
\end{proposition}

\section{The time-dependent Kepler potential}

\label{sec.timedep.Kepler}

In this case, $\nu=1$ and conditions (\ref{eq.conTK1a}) - (\ref{eq.conTK1d}) give: $a_{16}=a_{17}=a_{18}=a_{19}=a_{20}=0$, $a_{5}=a_{8}$, $a_{2}=a_{12}$, $a_{3}=a_{9}=a_{13}$,$ a_{11}=a_{15}$, $\ddot{a}_{2}= \ddot{a}_{5}= \ddot{a}_{11}=0$, $\sigma_{2}=c_{7}$, $\sigma_{3}= c_{8}$ and $\tau_{3}= c_{9}$.

Then, the constraint (\ref{eq.TKNq5}) gives $\dddot{a}_{3}=0$, $\sigma_{4}= \tau_{4}= \eta_{4}=0$, $a_{3}\omega^{2}=c_{10}$, $a_{2}\omega=c_{11}$, $a_{5}\omega=c_{12}$ and $a_{11}\omega=c_{13}$, where $c_{10}, c_{11}, c_{12}, c_{13}$ are arbitrary constants. Finally, we have:
\begin{eqnarray*}
K_{11} &=&\frac{c_{2}}{2}y^{2}+\frac{c_{1}}{2}%
z^{2}+c_{6}yz+a_{5}y+a_{2}z+a_{3} \\
K_{12} &=&\frac{c_{4}}{2}z^{2}-\frac{c_{2}}{2}xy-\frac{c_{6}}{2}xz-\frac{%
c_{5}}{2}yz-\frac{a_{5}}{2}x-\frac{a_{11}}{2}y  \\
K_{13} &=&\frac{c_{5}}{2}y^{2}-\frac{c_{6}}{2}xy-\frac{c_{1}}{2}xz-\frac{%
c_{4}}{2}yz-\frac{a_{2}}{2}x-\frac{a_{11}}{2}z
\\
K_{22} &=&\frac{c_{2}}{2}x^{2}+\frac{c_{3}}{2}%
z^{2} +c_{5}xz+a_{11}x+a_{2}z+a_{3}  \\
K_{23} &=&\frac{c_{6}}{2}x^{2}-\frac{c_{5}}{2}xy -\frac{c_{4}}{2}xz-\frac{c_{3}}{2}yz -\frac{a_{2}}{2}y -\frac{a_{5}}{2}z \\
K_{33} &=&\frac{c_{1}}{2}x^{2}+\frac{c_{3}}{2}%
y^{2}+c_{4}xy+a_{11}x+a_{5}y+a_{3}
\end{eqnarray*}
and
\begin{eqnarray*}
K_{1} &=& \dot{a}_{11}y^{2} + \dot{a}_{11}z^{2} -\dot{a}_{5}xy -\dot{a}_{2}xz -\dot{a}_{3}x + c_{7}y + c_{8}z \\
K_{2} &=& \dot{a}_{5}x^{2} + \dot{a}_{5}z^{2} -\dot{a}_{11}xy -\dot{a}_{2}yz -c_{7}x -\dot{a}_{3}y + c_{9}z \\
K_{3} &=& \dot{a}_{2}x^{2} + \dot{a}_{2}y^{2} -\dot{a}_{11}xz -\dot{a}_{5}yz -c_{8}x -c_{9}y -\dot{a}_{3}z
\end{eqnarray*}
where
\begin{equation}
\ddot{a}_{2}= \ddot{a}_{5}= \ddot{a}_{11}=0, \enskip \dddot{a}_{3}=0, \enskip a_{3}\omega^{2}=c_{10}, \enskip a_{2}\omega=c_{11}, \enskip a_{5}\omega=c_{12}, \enskip a_{11}\omega=c_{13}. \label{Kepler.1}
\end{equation}

From the last conditions, it follows that in order QFIs to be allowed the function $\omega(t)$ can have only three possible forms: \newline
- $\omega(t)$ a general function;\newline
- $\omega(t)= \omega_{2K}(t)= \frac{c_{11}}{b_{0} +b_{1}t}$, where $c_{11}b_{1}\neq0$; and\newline
- $\omega(t)= \omega_{3K}(t)=\frac{k}{(b_{0}+b_{1}t +b_{2}t^{2})^{1/2}}$, where $k\neq0$ and $b_{1}^{2} -4b_{2}b_{0} \neq0$.

This result confirms the results found previously in \cite{Leach 1985, Katzin 1982, LeoTsampAndro 2017}. We note that the time-dependent Kepler potential $V= -\frac{\omega_{2K}(t)}{r}$ is a subcase of the Case II potential of \cite{Leach 1985} for $\mu_{0}=c_{11}$ and $\phi=b_{0}+b_{1}t$, whereas the potential $V= -\frac{\omega_{3K}(t)}{r}$ is a subcase of the Case III potential of \cite{Leach 1985} (see section \ref{sec.gen2}).

In the following, we discuss the cases for the special functions $\omega_{2K}(t)$ and $\omega_{3K}(t)$ because the case for a general function $\omega(t)$ reproduces the results of the subsection \ref{sec.gen1}.

\subsection{$\omega(t)= \omega_{2K}(t)= \frac{c_{11}}{b_{0} +b_{1}t}$, $c_{11}b_{1}\neq0$}

\label{sec.omega.2}

In this case, conditions (\ref{Kepler.1}) give: $a_{2}=b_{0}+b_{1}t$, $a_{3}= \frac{c_{10}}{c_{11}^{2}} (b_{0}+b_{1}t)^{2}$, $a_{5}= \frac{c_{12}}{c_{11}} (b_{0}+b_{1}t)$ and $a_{11}= \frac{c_{13}}{c_{11}} (b_{0}+b_{1}t)$.

Substituting the resulting vector $K_{a}$ and the KT $K_{ab}$ in (\ref{eq.TKNq4}), we find the scalar $K = -\frac{2c_{10}b_{1}t}{c_{11}r} + G(q)$. Replacing this function in the remaining constraint (\ref{eq.TKNq3}), we find
\[
G(x,y,z)= - \frac{2c_{10}b_{0}}{c_{11}r} - \frac{c_{13}x + c_{12}y + c_{11}z}{r} + \frac{c_{10}b_{1}^{2}}{c_{11}^{2}} r^{2}.
\]
Therefore,
\[
K(x,y,z,t)= \frac{c_{10}b_{1}^{2} r^{2}}{c_{11}^{2}} -\frac{2c_{10}(b_{0}+b_{1}t)}{c_{11}r} -\frac{c_{13}x +c_{12}y + c_{11}z}{r}.
\]

The QFI is
\begin{eqnarray*}
I&=& \frac{c_{3}}{2} L_{1}^{2} +\frac{c_{1}}{2} L_{2}^{2} +\frac{c_{2}}{2} L_{3}^{2} -c_{4}L_{1}L_{2} -c_{5}L_{1}L_{3} -c_{6}L_{2}L_{3} -c_{9}L_{1} + c_{8}L_{2} - c_{7}L_{3} + \\
&&+ \frac{2c_{10}}{c_{11}^{2}}E_{2} + \frac{c_{13}}{c_{11}} A_{1}+ \frac{c_{12}}{c_{11}} A_{2} + A_{3}
\end{eqnarray*}
where $\omega_{2K}(t)=\frac{c_{11}}{b_{0}+b_{1}t}$ and
\begin{eqnarray}
L_{i} &\equiv& q_{i+1} \dot{q}_{i+2} - q_{i+2} \dot{q}_{i+1} \label{eq.TKFI1} \\
E_{2} &\equiv& (b_{0}+b_{1}t)^{2} \left[ \frac{\dot{q}^{i} \dot{q}_{i}}{2} -\frac{c_{11}}{r(b_{0}+b_{1}t)} \right] -b_{1}(b_{0}+b_{1}t) q^{i}\dot{q}_{i} + \frac{b_{1}^{2}r^{2}}{2} \label{eq.TKFI1a} \\
\tilde{R}_{i} &\equiv& (\dot{q}^{j}\dot{q}_{j})q_{i} - (\dot{q}^{j}q_{j})\dot{q}_{i} - \frac{c_{11}}{r(b_{0}+b_{1}t)} q_{i} \label{eq.TKFI1b} \\
A_{i} &\equiv& (b_{0}+b_{1}t) \tilde{R}_{i} + b_{1}\left( q_{i+2}L_{i+1} - q_{i+1}L_{i+2} \right). \label{eq.TKFI1c}
\end{eqnarray}
We note that $i=1,2,3$, $q_{i}=(x,y,z)$ and $q_{i}\equiv q_{i+3k}$ for all $k \in \mathbb{N}$, that is, $x=q_{1}=q_{4}=$ $q_{7}=...,$ $y=q_{2}=$ $q_{5}=q_{8}=...,$ and $z=q_{3}=q_{6}$ $=q_{9}=...~$.

The QFI $I$ contains the already found LFIs $L_{i}$ of the angular momentum; the QFI $E_{2}$, which for $b_{1}=0$ reduces to the Hamiltonian of the Kepler potential $V= -\frac{c_{11}}{b_{0}r}$; and the QFIs $A_{i}$, which may be considered as a generalization of the Runge-Lenz vector\index{Vector! Runge-Lenz} $R_{i}\left( k=\frac{c_{11}}{b_{0}} \right)$ for time-dependence $\omega_{2K}(t)= \frac{c_{11}}{b_{0}+b_{1}t}$. Indeed, we have $A_{i}(b_{1}=0)= b_{0}R_{i}\left( k=\frac{c_{11}}{b_{0}} \right)$.

The expressions (\ref{eq.TKFI1a}) - (\ref{eq.TKFI1c}) are written compactly as follows:
\begin{eqnarray}
E_{2} &\equiv& c_{11}^{2} \left[ \frac{1}{\omega_{2K}^{2}} \left( \frac{\dot{q}^{i}\dot{q}_{i}}{2} -\frac{\omega_{2K}}{r} \right) -\frac{1}{2} \frac{d}{dt}\left(\frac{1}{\omega_{2K}}\right)^{2} q^{i}\dot{q}_{i} + \frac{d^{2}}{dt^{2}} \left( \frac{1}{\omega_{2K}} \right)^{2} \frac{r^{2}}{4} \right] \label{eq.TKFI3a} \\
\tilde{R}_{i} &\equiv& (\dot{q}^{j}\dot{q}_{j})q_{i} - (\dot{q}^{j}q_{j})\dot{q}_{i} - \frac{\omega_{2K}}{r}q_{i} \label{eq.TKFI3b} \\
A_{i} &\equiv& c_{11} \left[ \frac{1}{\omega_{2K}} \tilde{R}_{i} - \frac{(\ln \omega_{2K})^{\cdot}}{\omega_{2K}}\left( q_{i+2}L_{i+1} - q_{i+1}L_{i+2} \right) \right] \label{eq.TKFI3c}
\end{eqnarray}
where $\omega_{2K}(t)= \frac{c_{11}}{b_{0}+b_{1}t}$.

We remark that only five of the seven FIs $E_{2}$, $L_{i}$, $A_{i}$ are (functionally) independent because they are related as follows:
\begin{equation}
\mathbf{A} \cdot \mathbf{L} = 0, \enskip 2E_{2}\mathbf{L}^{2} + c_{11}^{2} = \mathbf{A}^{2}. \label{eq.TKFI4}
\end{equation}
For $b_{1}=0$ and $b_{0}\neq0$, we have $\omega_{2K}=\frac{c_{11}}{b_{0}}\equiv k=const$, $E_{2}=b_{0}^{2}H$, $\tilde{R}_{i}=R_{i}$ and $A_{i}= b_{0}R_{i}$, where $H$ is the Hamiltonian and $R_{i}$ the Runge-Lenz vector for the Kepler potential $V=-\frac{k}{r}$. Then, as expected, equation (\ref{eq.TKFI4}) reduces to the well-known relation
\begin{equation}
2H\mathbf{L}^{2} + k^{2} = \mathbf{R}^{2}. \label{eq.kepler.identity}
\end{equation}

\subsection{$\omega(t)= \omega_{3K}(t)= \frac{k}{(b_{0}+b_{1}t +b_{2}t^{2})^{1/2}}$, $k\neq0$, $b_{1}^{2} -4b_{2}b_{0}\neq0$}

\label{sec.omega.3}

In this case, conditions (\ref{Kepler.1}) give\footnote{Observe that if $b_{1}^{2}-4b_{2}b_{0}=0$, this case reduces to the case of the section \ref{sec.omega.2} because equation $b_{0}+b_{1}t+b_{2}t^{2}=0$ has a double root $t_{0}$ and can be factored in the form $b_{2}(t-t_{0})^{2}$.}: $a_{2}=a_{5}=a_{11}=0$, $c_{11}= c_{12}= c_{13}=0$ and $a_{3}= \frac{c_{10}}{k^{2}}(b_{0}+b_{1}t+b_{2}t^{2})$.

Substituting the resulting quantities $K_{a}$ and $K_{ab}$ in (\ref{eq.TKNq4}), we find the scalar $K= -\frac{2c_{10}}{r\omega_{3K}} + G(q)$. When this scalar is replaced in the remaining constraint (\ref{eq.TKNq3}), it gives $G(x,y,z)= \frac{b_{2}c_{10}}{k^{2}}r^{2}$. Therefore,
\[
K(x,y,z,t) = \frac{b_{2}c_{10}}{k^{2}}r^{2} -\frac{2c_{10}}{r\omega_{3K}}.
\]

The QFI is
\begin{equation*}
I = \frac{c_{3}}{2} L_{1}^{2} +\frac{c_{1}}{2} L_{2}^{2} +\frac{c_{2}}{2} L_{3}^{2} -c_{4}L_{1}L_{2} -c_{5}L_{1}L_{3} -c_{6}L_{2}L_{3} -c_{9}L_{1} + c_{8}L_{2} - c_{7}L_{3} + \frac{2c_{10}}{k^{2}} E_{3}
\end{equation*}
where
\begin{equation}
E_{3} \equiv (b_{0}+b_{1}t+b_{2}t^{2}) \left[ \frac{\dot{q}^{i}\dot{q}_{i}}{2} -\frac{k}{r(b_{0}+b_{1}t +b_{2}t^{2})^{1/2}} \right] - \frac{b_{1}+2b_{2}t}{2} q^{i}\dot{q}_{i} + \frac{b_{2}r^{2}}{2} \label{eq.TKFI5a}
\end{equation}
is the only new independent QFI. This QFI is written equivalently as
\begin{equation}
E_{3} = k^{2} \left[ \frac{1}{\omega_{3K}^{2}} \left( \frac{\dot{q}^{i}\dot{q}_{i}}{2} -\frac{\omega_{3K}}{r} \right) -\frac{1}{2} \frac{d}{dt}\left(\frac{1}{\omega_{3K}}\right)^{2} q^{i}\dot{q}_{i} + \frac{d^{2}}{dt^{2}} \left( \frac{1}{\omega_{3K}} \right)^{2} \frac{r^{2}}{4} \right]. \label{eq.TKFI5b}
\end{equation}
For $b_{1}=b_{2}=0$, $E_{2}$ reduces to the well-known Hamiltonian of the time-independent Kepler potential.

We note also that the QFIs (\ref{eq.TKFI1a}) and (\ref{eq.TKFI5a}) can be written compactly as (see eq. (2.86) in \cite{Katzin 1982})
\begin{equation}
E_{\mu} = k^{2} \left[ \frac{1}{\omega_{\mu K}^{2}} \left( \frac{\dot{q}^{i}\dot{q}_{i}}{2} -\frac{\omega_{\mu K}}{r} \right) -\frac{1}{2} \frac{d}{dt}\left(\frac{1}{\omega_{\mu K}}\right)^{2} q^{i}\dot{q}_{i} + \frac{d^{2}}{dt^{2}} \left( \frac{1}{\omega_{\mu K}} \right)^{2} \frac{r^{2}}{4} \right] \label{eq.TKFI6com}
\end{equation}
where $\mu=2,3$, $\omega_{2K}(t)= \frac{k}{b_{0}+b_{1}t}$ and $\omega_{3K}(t)= \frac{k}{(b_{0}+b_{1}t+b_{2}t^{2})^{1/2}}$.

\begin{proposition}[Time-dependent Kepler potentials which admit additional FIs \cite{Katzin 1982}] \label{kepler} The \\ time-dependent Kepler\index{Potential! time-dependent} potential $V(t,q)=-\frac{\omega (t)}{r}$, for the function $\omega _{2K}(t)=\frac{c_{11}}{b_{0}+b_{1}t%
}$, where $c_{11}b_{1}\neq 0$, and the function $\omega_{3K}(t) =\frac{k}{(b_{0}+b_{1}t+b_{2}t^{2})^{1/2}}$, where $k\neq 0$ and $b_{1}^{2}-4b_{2}b_{0}\neq 0$, admits additional QFIs given by (\ref{eq.TKFI1a}), (\ref{eq.TKFI1c}) and (\ref{eq.TKFI5a}), respectively.
\end{proposition}

\section{The 3d time-dependent oscillator}

In this case, we have $\nu =-2$ and conditions (\ref{eq.conTK1a}) - (\ref{eq.conTK1d}) give:\index{Oscillator! 3d time-dependent} \newline $a_{2}=a_{5}=a_{8}=a_{11}=a_{12}=a_{15}=a_{16}=a_{18}=0$ and
\begin{equation}
\enskip \dot{\sigma}_{2}= -\ddot{a}_{17}, \enskip \dot{\sigma}_{3}= -\ddot{a}_{19}, \enskip \dot{\tau}_{3}= -\ddot{a}_{20}. \label{eq.osc.pde1}
\end{equation}
Then, the constraint (\ref{eq.TKNq5}) implies that:
\begin{equation}
\ddot{\sigma}_{4} - 2\omega\sigma_{4}=0, \enskip \ddot{\tau}_{4} - 2\omega\tau_{4}=0, \enskip \ddot{\eta}_{4} - 2\omega\eta_{4}=0, \label{eq.osc.pde1a}
\end{equation}
\begin{equation}
\dddot{a}_{3} - 8\omega\dot{a}_{3} -4\dot{\omega}a_{3}=0, \enskip \dddot{a}_{9} - 8\omega\dot{a}_{9} -4\dot{\omega}a_{9}=0, \enskip \dddot{a}_{13} - 8\omega\dot{a}_{13} -4\dot{\omega}a_{13}=0, \label{eq.osc.pde1b}
\end{equation}
\begin{equation}
\dddot{a}_{17} - 8\omega\dot{a}_{17} -4\dot{\omega}a_{17}=0,
\enskip \dddot{a}_{19} - 8\omega\dot{a}_{19} -4\dot{\omega}a_{19}=0, \enskip \dddot{a}_{20} - 8\omega\dot{a}_{20} -4\dot{\omega}a_{20}=0. \label{eq.osc.pde1c}
\end{equation}

Therefore,
\begin{eqnarray}
K_{11} &=&\frac{c_{2}}{2}y^{2}+\frac{c_{1}}{2}%
z^{2}+c_{6}yz+a_{3} \notag \\
K_{12} &=&\frac{c_{4}}{2}z^{2}-\frac{c_{2}}{2}xy-\frac{c_{6}}{2}xz-\frac{%
c_{5}}{2}yz+a_{17} \notag \\
K_{13} &=&\frac{c_{5}}{2}y^{2}-\frac{c_{6}}{2}xy-\frac{c_{1}}{2}xz-\frac{%
c_{4}}{2}yz+a_{19} \notag
\\
K_{22} &=&\frac{c_{2}}{2}x^{2}+\frac{c_{3}}{2}%
z^{2} +c_{5}xz+a_{13} \label{eq.osc.pde2} \\
K_{23} &=&\frac{c_{6}}{2}x^{2}-\frac{c_{5}}{2}xy -\frac{c_{4}}{2}xz-\frac{c_{3}}{2}yz +a_{20} \notag \\
K_{33} &=&\frac{c_{1}}{2}x^{2}+\frac{c_{3}}{2}
y^{2}+c_{4}xy +a_{9} \notag
\end{eqnarray}
and
\begin{eqnarray}
K_{1} &=& -\dot{a}_{3}x + \sigma_{2}y + \sigma_{3}z + \sigma_{4} \notag \\
K_{2} &=& -(\sigma_{2}+2\dot{a}_{17})x -\dot{a}_{13}y + \tau_{3}z + \tau_{4} \label{eq.osc.pde3} \\
K_{3} &=& -(\sigma_{3}+2\dot{a}_{19})x -(\tau_{3}+2\dot{a}_{20})y -\dot{a}_{9}z + \eta_{4}. \notag
\end{eqnarray}

Before we proceed with considering various subcases, it is important that we discuss the ODEs (\ref{eq.osc.pde1b}) and (\ref{eq.osc.pde1c}).

\subsection{The Lewis invariant}

\label{Note 1}

Equations of the form
\begin{equation}
\dddot{a} - 8\omega\dot{a} -4\dot{\omega}a=0 \label{eq.TKNq10}
\end{equation}
where $a=a(t)$, can be written as follows:
\begin{equation}
a\ddot{a} - \frac{1}{2}\dot{a}^{2} -4\omega a^{2} = c_{0}=const. \label{eq.TKNq11}
\end{equation}
By putting $a=-\rho^{2}$, where $\rho=\rho(t)$, equation (\ref{eq.TKNq11}) becomes
\begin{equation}
\ddot{\rho} -2\omega \rho - \frac{c_{0}}{2\rho^{3}}=0. \label{eq.TKNq12}
\end{equation}
For $2\omega(t)=-\psi^{2}(t)$, equation (\ref{eq.TKNq12}) is written as
\begin{equation}
\ddot{\rho} + \psi^{2}\rho - \frac{c_{0}}{2\rho^{3}}=0. \label{eq.TKNq13}
\end{equation}
Equation (\ref{eq.TKNq13}) is the \textbf{auxiliary equation}\index{Equation! auxiliary} (see \cite{Tsamparlis 2012, Leach 1991, Katzin 1977}) that should be introduced in order to derive the Lewis invariant for the 1d time-dependent oscillator
\begin{equation}
\ddot{x} + \psi^{2}x =0. \label{eq.TKNq14}
\end{equation}
By eliminating the $\psi^{2}$, using (\ref{eq.TKNq14}), and by multiplying with the factor $x\dot{\rho}-\rho\dot{x}$, equation (\ref{eq.TKNq13}) gives
\[
\ddot{\rho} - \frac{\rho}{x}\ddot{x} - \frac{c_{0}}{2\rho^{3}}=0 \implies
\left[ \frac{1}{2}\left( x\dot{\rho} - \rho\dot{x} \right)^{2} + \frac{c_{0}}{4}\left(\frac{x}{\rho}\right)^{2} \right]^{\cdot} =0 \implies
\]
\begin{equation}
I \equiv \frac{1}{2}\left( x\dot{\rho} - \rho\dot{x} \right)^{2} +\frac{c_{0}}{4}\left(\frac{x}{\rho}\right)^{2} =const \label{eq.TKNq15}
\end{equation}
which is the well-known \textbf{Lewis invariant}\index{Invariant! Lewis} for the 1d time-dependent harmonic oscillator or, equivalently, a FI for the 2d time-dependent system with equations of motion (\ref{eq.TKNq13}) and (\ref{eq.TKNq14}).

\subsection{The system of equations (\ref{eq.osc.pde1}) - (\ref{eq.osc.pde1c})}

Conditions (\ref{eq.osc.pde1a}) are not involved into conditions (\ref{eq.osc.pde1}), (\ref{eq.osc.pde1b}) and (\ref{eq.osc.pde1c}). This means that the parameters $\sigma_{4}, \tau_{4}, \eta_{4}$ give different independent FIs from the remaining parameters $a_{3}, a_{9}, a_{13}, a_{17}, a_{19}, a_{20}$. Therefore, without loss of generality, they can be treated separately. This leads to the following two cases.

\subsubsection{Case: $a_{3}\neq0$ and $\sigma_{4} =\tau_{4} =\eta_{4} =0$}

\label{2.3}

Because the ODEs (\ref{eq.osc.pde1b}) and (\ref{eq.osc.pde1c}) are independent (i.e. each one leads to a different FI) and of the same form, without loss of generality, we assume: $a_{9}=k_{1}a_{3}$, $a_{13}=k_{2}a_{3}$, $a_{17}=k_{3}a_{3}$, $a_{19}= k_{4}a_{3}$ and $a_{20}= k_{5}a_{3}$, where $k_{1}, k_{2}, k_{3}, k_{4}, k_{5}$ are arbitrary constants.

From the discussion of section \ref{Note 1} and the assumption $a_{3}\neq0$, condition (\ref{eq.osc.pde1b}) concerning $a_{3}(t)$ becomes (see eq. (9.2) in \cite{Katzin 1977})
\begin{equation}
\dddot{a}_{3} - 8\omega\dot{a}_{3} -4\dot{\omega}a_{3}=0 \implies a_{3}\ddot{a}_{3} - \frac{1}{2}\dot{a}_{3}^{2} -4\omega a_{3}^{2} = c_{0} \implies \omega(t)= \frac{\ddot{a}_{3}}{4a_{3}} - \frac{1}{8} \left(\frac{\dot{a}_{3}}{a_{3}}\right)^{2} -\frac{c_{0}}{4a_{3}^{2}} \label{eq.osc.FI2}
\end{equation}
where $c_{0}$ is an arbitrary constant and $a_{3}(t)$ is an arbitrary non-zero function.

Moreover, conditions (\ref{eq.osc.pde1}) become $\sigma_{2}=-\dot{a}_{17}$, $\sigma_{3}= -\dot{a}_{19}$ and $\tau_{3}= -\dot{a}_{20}$ because any additional constant (in general $\sigma_{2}= -\dot{a}_{17} +m_{1}$, where $m_{1}$ is a constant) leads to the usual LFIs of the angular momentum.

Then, the KT (\ref{eq.osc.pde2}) and the vector (\ref{eq.osc.pde3}) become\footnote{We set $c_{1}= ... = c_{6} =0$ because they generate the already found FIs of the angular momentum.}:
\begin{equation*}
K_{ab}=a_{3}\left(
\begin{array}{ccc}
1 & k_{3} & k_{4} \\
k_{3} & k_{2} & k_{5} \\
k_{4} & k_{5} & k_{1}%
\end{array}%
\right) ,\enskip K_{a}=-\dot{a}_{3}\left(
\begin{array}{c}
x+k_{3}y+k_{4}z \\
k_{3}x+k_{2}y+k_{5}z \\
k_{4}x+k_{5}y+k_{1}z%
\end{array}%
\right).
\end{equation*}

Substituting in the constraints (\ref{eq.TKNq3}) and (\ref{eq.TKNq4}), we find
\begin{equation*}
K=\frac{\dot{a}_{3}^{2}+2c_{0}}{4a_{3}}\left( x^{2}+k_{2}y^{2} +k_{1}z^{2}+2k_{3}xy+2k_{4}xz+2k_{5}yz\right).
\end{equation*}
Using equation (\ref{eq.osc.FI2}), we can write $\frac{\dot{a}_{3}^{2}+2c_{0}%
}{4a_{3}}=\frac{\ddot{a}_{3}}{2}-2\omega a_{3}$.

The QFI is
\begin{eqnarray*}
I &=& a_{3}\left( \dot{x}^{2} + k_{2}\dot{y}^{2} +k_{1}\dot{z}^{2} +
2k_{3}\dot{x}\dot{y} + 2k_{4}\dot{x}\dot{z} + 2k_{5}\dot{y}\dot{z} \right) -%
\dot{a}_{3}(x +k_{3}y +k_{4}z)\dot{x} - \\
&& -\dot{a}_{3}(k_{3}x +k_{2}y+ k_{5}z)\dot{y} -\dot{a}_{3}(k_{4}x + k_{5}y
+k_{1}z)\dot{z}+ \\
&& + \left( \frac{\ddot{a}_{3}}{2} -2\omega a_{3} \right) \left( x^{2} +
k_{2}y^{2} +k_{1}z^{2} +2k_{3}xy +2k_{4}xz +2k_{5}yz \right).
\end{eqnarray*}
This expression contains six QFIs, which are the components of the symmetric tensor (see eqs. (1.4) and (6.24) in \cite{Katzin 1977})
\begin{equation}
\Lambda_{ij}= a_{3} \left( \dot{q}_{i}\dot{q}_{j} -2\omega q_{i} q_{j} \right) -\dot{a}_{3} q_{(i}\dot{q}_{j)} +\frac{\ddot{a}_{3}}{2} q_{i}q_{j}. \label{eq.osc.FI3}
\end{equation}
This tensor for $a_{3}=const\neq0$ reduces to the Jauch-Hill-Fradkin tensor $B_{ij}$ with $\omega= -\frac{c_{0}}{4a_{3}^{2}} =const$.\index{Tensor! Jauch-Hill-Fradkin}

If we make the transformation (see section \ref{Note 1}) $a_{3}(t)=-\rho^{2}(t)$ and $2\omega(t)=-\psi^{2}(t)$, equation (\ref{eq.TGKep.1a}) becomes
\begin{equation}
\ddot{q}^{a} -2\omega q^{a} =0 \implies \ddot{q}^{a} + \psi^{2}q^{a} =0 \label{eq.osc.FI3a}
\end{equation}
and the QFIs (\ref{eq.osc.FI3}) give
\begin{equation}
\Lambda_{ij}= -\left(\rho\dot{q}_{i} - \dot{\rho}q_{i}\right) \left(\rho\dot{q}_{j} - \dot{\rho}q_{j}\right) -\frac{c_{0}}{2} \rho^{-2}q_{i}q_{j} \label{eq.osc.FI3b}
\end{equation}
where condition (\ref{eq.osc.FI2}) takes the form (\ref{eq.TKNq13}).

The symmetric tensor (\ref{eq.osc.FI3b}) may be thought of as a 3d generalization of the 1d Lewis invariant (\ref{eq.TKNq15}). Moreover, equation (\ref{eq.osc.FI3b}) coincides with eq. (8) in \cite{Gunther 1977} and eq. (1.4) in \cite{Katzin 1977} when $c_{0}=2$.

\subsubsection{Case: $a_{3}=a_{9}=a_{13} =a_{17}=a_{19} =a_{20}=0$ and $\sigma_{4}\neq0$}

\label{2.4}

In this case, conditions (\ref{eq.osc.pde1b}) and (\ref{eq.osc.pde1c}) vanish identically, and conditions (\ref{eq.osc.pde1}) imply that $\sigma_{2}=c_{7}$, $\sigma_{3}=c_{8}$ and $\tau_{3}=c_{9}$.

Since the remaining ODEs (\ref{eq.osc.pde1a}) are all independent (i.e. each one generates an independent FI) and of the same form, without loss of generality, we assume $\tau_{4}=k_{1}\sigma_{4}$ and $\eta_{4}=k_{2}\sigma_{4}$, where $k_{1}$ and $k_{2}$ are arbitrary constants.

From (\ref{eq.osc.pde1a}) for $\sigma_{4}\neq0$, we get
\begin{equation}
\omega(t)= \frac{\ddot{\sigma}_{4}}{2\sigma_{4}}. \label{eq.osc.FI4}
\end{equation}

The parameters $c_{A}$, where $A=1,2,...,9$, produce the FIs of the angular momentum and we fix them to zero. Therefore,
$K_{ab}=0$ and $K_{a}=\sigma_{4} \left( 1, k_{1}, k_{2} \right)$. Substituting in the remaining constraints (\ref{eq.TKNq3}) and (\ref{eq.TKNq4}), we find $K= -\dot{\sigma}_{4} \left(x + k_{1}y +k_{2}z \right)$.

The QFI is $I= \sigma_{4}\dot{x} -\dot{\sigma}_{4}x + k_{1} \left( \sigma_{4}\dot{y} -\dot{\sigma}_{4} y \right) + k_{2} \left( \sigma_{4}\dot{z} -\dot{\sigma}_{4}z \right)$, which contains the irreducible LFIs (see eq. (6.25) in \cite{Katzin 1977})
\begin{equation}
I_{4i}= f\dot{q}_{i} -\dot{f}q_{i} \label{eq.osc.FI5}
\end{equation}
where $f(t)$ is an arbitrary non-zero function satisfying (\ref{eq.osc.FI4}) for $\sigma_{4}=f$. We note that the LFIs (\ref{eq.osc.FI5}) can be derived directly from the equations of motion for $\omega(t)= \frac{\ddot{f}}{2f}$.
\bigskip

From the above two cases, we arrive at the following conclusion.

\begin{proposition}[3d time-dependent oscillators which admit additional FIs]
\label{oscillator} For the function \\ $\omega(t)= \frac{\ddot{a}_{3}}{4a_{3}} - \frac{1}{8} \left(\frac{\dot{a}_{3}}{a_{3}}\right)^{2} -\frac{c_{0}}{4a^{2}_{3}}$, where $a_{3}(t)\neq0$ and $c_{0}$ is an arbitrary constant, and the function $\omega(t)= \frac{\ddot{f}}{2f}$, where $f(t)\neq0$, the resulting 3d time-dependent oscillator\index{Oscillator! time-dependent} $V(t,q)= -\omega(t)r^{2}$ admits the QFIs (\ref{eq.osc.FI3}) and the LFIs (\ref{eq.osc.FI5}), respectively.
\end{proposition}

\section{A special class of time-dependent oscillators}

\label{sec.discussion.GK}

In Proposition \ref{oscillator}, it has been shown that the time-dependent oscillator ($\nu=-2$) for the frequency\index{Oscillator! time-dependent}
\begin{equation}
\omega_{1O}(t)=\frac{\ddot{f}}{4f(t)}-\frac{1}{8}\left( \frac{\dot{f}}{f}%
\right) ^{2}-\frac{c_{0}}{4f^{2}}  \label{eq.disosc0.a}
\end{equation}%
where $f(t)$ is an arbitrary non-zero function, admits the six QFIs
\begin{equation}
\Lambda _{ij}=f(t)\left( \dot{q}_{i}\dot{q}_{j}-2\omega q_{i}q_{j}\right) -\dot{f}q_{(i}\dot{q}_{j)} +\frac{\ddot{f}}{2}q_{i}q_{j}  \label{eq.disosc0.a1}
\end{equation}%
and for the frequency
\begin{equation}
\omega_{2O}(t)=\frac{\ddot{g}}{2g(t)}  \label{eq.disosc0.b}
\end{equation}
where $g(t)$ is an arbitrary non-zero function, admits the three
LFIs
\begin{equation}
I_{4i}=g(t)\dot{q}_{i}-\dot{g}q_{i}.  \label{eq.disosc0.b1}
\end{equation}

We consider the class of the 3d time-dependent oscillators for which $\omega_{1O}(t)= \omega_{2O}(t)$. These oscillators admit both the six QFIs $\Lambda _{ij}$ and the three LFIs $I_{4i}$.

The condition $\omega _{1O}(t) =\omega _{2O}(t)$ relates the functions $f(t)$ and $g(t)$ as follows:
\begin{equation}
\omega_{3O}(t)=\frac{\ddot{f}}{4f(t)}-\frac{1}{8}\left( \frac{\dot{f}}{f}%
\right) ^{2}-\frac{c_{0}}{4f^{2}}=\frac{\ddot{g}}{2g(t)}.  \label{eq.disosc1}
\end{equation}

It can be easily proved that both the choices
\begin{equation}
g=f^{1/2} \cos\theta, \enskip \dot{\theta}= \left( \frac{c_{0}}{2} \right)^{1/2} f^{-1} \implies  \theta(t)= \left( \frac{c_{0}}{2} \right)^{1/2} \int \frac{dt}{f(t)} \label{eq.disosc2a}
\end{equation}
and
\begin{equation}
g=f^{1/2} \sin\theta, \enskip \dot{\theta}= \left( \frac{c_{0}}{2} \right)^{1/2} f^{-1} \implies  \theta(t)= \left( \frac{c_{0}}{2} \right)^{1/2} \int \frac{dt}{f(t)} \label{eq.disosc2b}
\end{equation}
satisfy the requirement (\ref{eq.disosc1}) for any non-zero function $f(t)$. In other words, all the time-dependent oscillators with frequency
\begin{equation}
\omega_{3O}(t)=\frac{\ddot{f}}{4f(t)} - \frac{1}{8} \left(\frac{\dot{f}}{f}\right)^{2} -\frac{c_{0}}{4f^{2}} \label{eq.disosc3}
\end{equation}
admit the six QFIs
\begin{equation}
\Lambda_{ij}= f(t) \left( \dot{q}_{i}\dot{q}_{j} -2\omega q_{i} q_{j} \right) -\dot{f} q_{(i}\dot{q}_{j)} +\frac{\ddot{f}}{2} q_{i}q_{j} \label{eq.disosc4a}
\end{equation}
and the six LFIs
\begin{eqnarray}
I_{41i}&=& \left( \frac{c_{0}}{2} \right)^{1/2} f^{-1/2}q_{i} \sin\theta + \left( f^{1/2} \dot{q}_{i} - \frac{\dot{f}}{2} f^{-1/2}q_{i} \right) \cos\theta \label{eq.disosc4b} \\
I_{42i}&=& -\left( \frac{c_{0}}{2} \right)^{1/2} f^{-1/2}q_{i} \cos\theta + \left( f^{1/2} \dot{q}_{i} - \frac{\dot{f}}{2} f^{-1/2}q_{i} \right) \sin\theta. \label{eq.disosc4c}
\end{eqnarray}
These are the LFIs $J^{k}_{3}$ and $J^{k}_{4}$ derived in eqs. (44) and (45) in \cite{Prince 1980} using point Noether symmetries and Noether's Theorem.

We note that $\frac{dI_{42i}}{d\theta }=I_{41i}$ and
\begin{equation}
\Lambda_{ij} = I_{41i}I_{41j} + I_{42i}I_{42j}. \label{eq.disosc5}
\end{equation}

Next, we consider the LFIs of the angular momentum\index{Momentum! angular} $L_{i}= q_{i+1} \dot{q}_{i+2} - q_{i+2} \dot{q}_{i+1}$. These LFIs can be expressed equivalently as components of the totally antisymmetric tensor
\begin{equation}
L_{ij}= q_{i}\dot{q}_{j} - q_{j}\dot{q}_{i} =\varepsilon_{ijk} L^{k} \label{eq.disosc6}
\end{equation}
where $\varepsilon_{ijk}$ is the 3d Levi-Civita symbol and $L^{i}= L_{i}$ since the kinetic metric $\gamma_{ij} =\delta_{ij}$. Then, (see eq. (51) in \cite{Prince 1980})
\begin{equation}
L_{ij}= \left( \frac{2}{c_{0}} \right)^{1/2} \left( I_{41i}I_{42j} - I_{41j}I_{42i} \right). \label{eq.disosc7}
\end{equation}

\begin{proposition} \label{pro.indLFIs}
For the class of 3d time-dependent oscillators\index{Oscillator! time-dependent} with potential $V(t,q)=-\omega(t)r^{2}$, where $\omega(t)$ is defined in terms of an arbitrary non-zero (smooth) function $f(t)$ as in (\ref{eq.disosc3}), the only independent FIs are the LFIs $I_{41i}$ and $I_{42i}$.
\end{proposition}

In order to recover the results of \cite{Prince 1980}, we assume a time-dependent oscillator with $\omega_{3O}(t)$ given by (\ref{eq.disosc3}), and we write the non-zero function $f(t)$ in the form $f(t)= \rho^{2}(t)$. Then, equation (\ref{eq.disosc3}) becomes
\begin{equation}
\omega_{3O}(t)= \frac{\ddot{\rho}}{2\rho} - \frac{c_{0}}{4\rho^{4}}. \label{eq.disosc8}
\end{equation}
The relations (\ref{eq.disosc2a}) and (\ref{eq.disosc2b}) become:
\begin{equation}
g=\rho \cos\theta, \enskip \dot{\theta}= \left( \frac{c_{0}}{2} \right)^{1/2} \rho^{-2} \implies  \theta(t)= \left( \frac{c_{0}}{2} \right)^{1/2} \int \frac{dt}{\rho^{2}} \label{eq.disosc8a}
\end{equation}
\begin{equation}
g=\rho \sin\theta, \enskip \dot{\theta}= \left( \frac{c_{0}}{2} \right)^{1/2} \rho^{-2} \implies  \theta(t)= \left( \frac{c_{0}}{2} \right)^{1/2} \int \frac{dt}{\rho^{2}} \label{eq.disosc8b}
\end{equation}
and the LFIs (\ref{eq.disosc4b}) and (\ref{eq.disosc4c}) take the form:
\begin{eqnarray}
I_{41i} &=& \left( \frac{c_{0}}{2} \right)^{1/2} \rho^{-1}q_{i} \sin\theta + \left( \rho \dot{q}_{i} - \dot{\rho}q_{i} \right) \cos\theta \label{eq.disosc9a} \\
I_{42i}&=& -\left( \frac{c_{0}}{2} \right)^{1/2} \rho^{-1}q_{i} \cos\theta + \left( \rho \dot{q}_{i} - \dot{\rho}q_{i} \right) \sin\theta. \label{eq.disosc9b}
\end{eqnarray}
These latter expressions for $c_{0}=2$ coincide with the independent LFIs (44) and (45) found in \cite{Prince 1980}.

Finally, we note that if we consider in this special class of oscillators the simple case $f=1$, we find $\omega_{3O}(t)= const= -\frac{c_{0}}{4} \equiv k$, which is the 3d autonomous oscillator (for $k<0$). Then, it can be shown that the exponential LFIs $I_{3i\pm}$ (see Table \ref{T1}) found in chapter \ref{ch1.QFIs.conservative} can be written in terms of $I_{41i}$ and $I_{42j}$. Indeed, we have $I_{3i\pm}(k>0) = I_{41i} \mp iI_{42i}$ and $I_{3i\pm}(k<0)= I_{41i} \pm iI_{42i}$.

\section{Collection of results}

\label{Table 2}

We collect the results concerning the time-dependent generalized Kepler potential for all values of $\nu$ in Table \ref{T2}. We note that for $\nu=-2,1,2$ the dynamical system is the time-dependent 3d oscillator, the time-dependent Kepler potential, and the Newton-Cotes potential, respectively. Concerning notation, we have: $q^{i}=(x,y,z)$, $q_{i}\equiv q_{i+3k}$ for all $k \in \mathbb{N}$ and $\tilde{R}_{i}= (\dot{q}^{j}\dot{q}_{j})q_{i} - (\dot{q}^{j} q_{j})\dot{q}_{i} - \frac{k}{r(b_{0}+b_{1}t)}q_{i}$.
\newpage

\begin{longtable}{|c|c|l|}
\hline
$\nu$ & $\omega(t)$ & LFIs and QFIs \\ \hline
\multirow{3}{*}{$\forall$ $\nu$} & $\forall$ $\omega$ & $L_{i} = q_{i+1}\dot{q}_{i+2} - q_{i+2}\dot{q}_{i+1}$, $L_{ij}= q_{i}\dot{q}_{j} - q_{j}\dot{q}_{i} =\varepsilon_{ijk} L^{k}$ \\
& $k$ & $H_{\nu}= \frac{1}{2}\dot{q}^{i}\dot{q}_{i} - \frac{k}{r^{\nu}}$ \\
& $\omega_{\nu}= k\left(b_{0} + b_{1}t + b_{2}t^{2} \right)^{\frac{\nu-2}{2}}$ & $J_{\nu}=  (b_{0} + b_{1}t + b_{2}t^{2}) \left( \frac{\dot{q}^{i}\dot{q}_{i}}{2} - \frac{\omega_{\nu}}{r^{\nu}} \right) -\frac{b_{1} + 2b_{2}t}{2} q^{i}\dot{q}_{i} +\frac{b_{2} r^{2}}{2}$ \\ \hline
\multirow{10}{*}{$-2$} & $k$ & $B_{ij} = \dot{q}_{i} \dot{q}_{j} - 2k q_{i}q_{j}$ \\
& $k>0$ & $I_{3a\pm}= e^{\pm \sqrt{2k} t}(\dot{q}_{a} \mp \sqrt{2k} q_{a})$ \\
& $k<0$ & $I_{3a\pm}= e^{\pm i \sqrt{-2k} t}(\dot{q}_{a} \mp i \sqrt{-2k} q_{a})$ \\
& $\frac{k}{(b_{0} +b_{1}t +b_{2}t^{2})^{2}}$ & $I_{ij}= (b_{0}+b_{1}t+b_{2}t^{2}) \left( \dot{q}_{i}\dot{q}_{j} -2\omega q_{i}q_{j} \right) -(b_{1} +2b_{2}t)q_{(i}\dot{q}_{j)} +b_{2}q_{i}q_{j}$ \\ \cline{2-3}
& $\frac{\ddot{f}}{4f(t)} - \frac{1}{8} \left(\frac{\dot{f}}{f}\right)^{2} -\frac{c_{0}}{4f^{2}}$ & \makecell[l]{$L_{ij}= \left( \frac{2}{c_{0}} \right)^{1/2} \left( I_{41i}I_{42j} - I_{41j}I_{42i} \right)$, \\ $\Lambda_{ij}= f(t) \left( \dot{q}_{i}\dot{q}_{j} -2\omega q_{i} q_{j} \right) -\dot{f} q_{(i}\dot{q}_{j)} +\frac{\ddot{f}}{2} q_{i}q_{j} = I_{41i}I_{41j} + I_{42i}I_{42j}$, \\ $I_{41i} =\left( \frac{c_{0}}{2} \right)^{1/2} f^{-1/2}q_{i} \sin\theta + \left( f^{1/2} \dot{q}_{i} - \frac{\dot{f}}{2} f^{-1/2}q_{i} \right) \cos\theta$, \\ $I_{42i}= -\left( \frac{c_{0}}{2} \right)^{1/2} f^{-1/2}q_{i} \cos\theta + \left( f^{1/2} \dot{q}_{i} - \frac{\dot{f}}{2} f^{-1/2}q_{i} \right) \sin\theta$ \\ where $\theta=  \left( \frac{c_{0}}{2} \right)^{1/2} \int f^{-1}dt$} \\ \cline{2-3}
& $\frac{\ddot{g}}{2g(t)}$ & $I_{4i}= g(t)\dot{q}_{i} -\dot{g}q_{i}$ \\ \hline
\multirow{5}{*}{$1$} & $k$ & $R_{i}= (\dot{q}^{j} \dot{q}_{j}) q_{i} - (\dot{q}^{j}q_{j})\dot{q}_{i}- \frac{k}{r}q_{i}$ \\ \cline{2-3}
& $\frac{k}{b_{0}+b_{1}t}$ & \makecell[l]{$E_{2}= (b_{0}+b_{1}t)^{2} \left[ \frac{\dot{q}^{i} \dot{q}_{i}}{2} -\frac{k}{r(b_{0}+b_{1}t)} \right] -b_{1}(b_{0}+b_{1}t) q^{i}\dot{q}_{i} + \frac{b_{1}^{2}r^{2}}{2}$, \\ $A_{i}= (b_{0}+b_{1}t) \tilde{R}_{i} + b_{1}\left( q_{i+2}L_{i+1} - q_{i+1}L_{i+2} \right)$ \\ where $\tilde{R}_{i}= (\dot{q}^{j} \dot{q}_{j})q_{i} - (\dot{q}^{j}q_{j})\dot{q}_{i} - \frac{k}{r(b_{0}+b_{1}t)} q_{i}$} \\ \cline{2-3}
& $\frac{k}{(b_{0}+b_{1}t+b_{2}t^{2})^{1/2}}$ & $E_{3} = (b_{0}+b_{1}t+b_{2}t^{2}) \left[ \frac{\dot{q}^{i} \dot{q}_{i}}{2} -\frac{k}{r ( b_{0}+b_{1}t +b_{2}t^{2} )^{1/2}} \right] - \frac{b_{1}+2b_{2}t}{2} q^{i}\dot{q}_{i} + \frac{b_{2}r^{2}}{2}$ \\
\hline
$2$ & $k$ & $I_{1}= - H_{2}t^{2} + t(\dot{q}^{i}q_{i}) - \frac{r^{2}}{2}$, $I_{2}= - H_{2}t + \frac{1}{2} (\dot{q}^{i}q_{i})$ \\ \hline
\caption{\label{T2} The LFIs/QFIs of the time-dependent generalized Kepler potential $V=-\frac{\omega(t)}{r^{\nu}}$.}
\end{longtable}

\section{Integrating the dynamical equations}

\label{sec.applications.GK}

In this section, we use the independent LFIs $I_{41i}$ and $I_{42i}$ to integrate the dynamical equations of the special class of 3d time-dependent oscillators ($\nu= -2$) defined in section \ref{sec.discussion.GK} with $\omega(t)$ given by (\ref{eq.disosc3}). We also use the FIs $L_{i}$, $E_{2}$, $A_{i}$ to integrate the time-dependent Kepler potential ($\nu=1$) with $\omega(t)= \frac{k}{b_{0}+b_{1}t}$, where $kb_{1}\neq0$ (see section \ref{sec.omega.2}).

\subsection{The 3d time-dependent oscillator with $\omega(t)$ given by (\ref{eq.disosc3})}

\label{sec.exa.oscillator}

Using the LFIs (\ref{eq.disosc4b}) and (\ref{eq.disosc4c}), we find
\begin{equation}
q_{i}(t) = \left( \frac{2}{c_{0}} \right)^{1/2} f^{1/2} \Big( I_{41i}\sin\theta - I_{42i}\cos\theta \Big) \label{eq.exaGK.1}
\end{equation}
where $I_{41i}$ and $I_{42i}$, $i=1,2,3$, are arbitrary constants (real or imaginary), and $\theta(t)=  \left( \frac{c_{0}}{2} \right)^{1/2} \int f^{-1}dt$.

The solution (\ref{eq.exaGK.1}) coincides with the solution (52) of \cite{Prince 1980}.

In the case of the 1d time-dependent oscillator, if we set $2\omega(t)=-\psi^{2}(t)$, $c_{0}=2$ and $f(t)=\rho^{2}(t)$, equation (\ref{eq.TGKep.1a}) and the defining relation (\ref{eq.disosc3}) for $\omega(t)$ become:
\begin{eqnarray}
\ddot{x} &=& -\psi^{2}x \label{eq.ermexa1} \\
\ddot{\rho} &=& -\psi^{2}\rho + \rho^{-3}. \label{eq.ermexa2}
\end{eqnarray}
The LFIs (\ref{eq.disosc9a}) and (\ref{eq.disosc9b}) become:
\begin{eqnarray}
I_{41} &=& \rho^{-1}x \sin\theta + \left( \rho \dot{x} - x\dot{\rho} \right) \cos\theta \label{eq.ermexa3} \\
I_{42} &=& -\rho^{-1}x \cos\theta + \left( \rho \dot{x} - x\dot{\rho} \right) \sin\theta. \label{eq.ermexa4}
\end{eqnarray}
The general solution (\ref{eq.exaGK.1}) is
\begin{equation}
x(t) = \rho(t) \Big( I_{41}\sin\theta - I_{42}\cos\theta \Big) \label{eq.ermexa5}
\end{equation}
where $\dot{\theta}= \rho^{-2}$ and $\rho(t)$ is a given non-zero function which defines $\psi(t)$ through (\ref{eq.ermexa2}). This is the 1d solution (9) in \cite{Prince 1980}.

\subsection{The solution of the time-dependent Kepler potential with $\omega_{2K}(t)= \frac{k}{b_{0}+b_{1}t}$, where $kb_{1}\neq0$}

In section \ref{sec.omega.2}, it is shown that this system admits the following FIs:
\[
L_{1}=y\dot{z} -z\dot{y}, \enskip L_{2}= z\dot{x} -x\dot{z}, \enskip L_{3}= x\dot{y} -y\dot{x}
\]
\[
E_{2}= (b_{0}+b_{1}t)^{2} \left[ \frac{\dot{q}^{i} \dot{q}_{i}}{2} -\frac{k}{r(b_{0}+b_{1}t)} \right] -b_{1}(b_{0}+b_{1}t) q^{i}\dot{q}_{i} + \frac{b_{1}^{2}r^{2}}{2}
\]
\[
A_{i}= (b_{0}+b_{1}t) \tilde{R}_{i} + b_{1}\left( q_{i+2}L_{i+1} - q_{i+1}L_{i+2} \right)
\]
where $\tilde{R}_{i}= (\dot{q}^{j}\dot{q}_{j})q_{i} - (\dot{q}^{j}q_{j})\dot{q}_{i} - \frac{k}{r(b_{0}+b_{1}t)} q_{i}$. The components of the generalized Runge-Lenz vector are written as:
\begin{eqnarray*}
A_{1}&=& (b_{0}+b_{1}t)(\dot{y}L_{3} -\dot{z}L_{2}) + b_{1}\left( zL_{2} - yL_{3} \right) - \frac{k}{r}x \\
A_{2}&=& (b_{0}+b_{1}t)(\dot{z}L_{1} -\dot{x}L_{3}) + b_{1}\left( xL_{3} - zL_{1} \right) - \frac{k}{r}y \\
A_{3}&=& (b_{0}+b_{1}t)(\dot{x}L_{2} -\dot{y}L_{1}) + b_{1}\left( yL_{1} - xL_{2} \right) -\frac{k}{r}z.
\end{eqnarray*}

Since the angular momentum is a FI, the motion is on a plane. We choose, without loss of generality, the plane $z=0$ and on that the polar coordinates $x=r\cos\theta$ and $y=r\sin\theta$. Then,
\[
L_{1}=L_{2}=0, \enskip L_{3}=r^{2}\dot{\theta}, \enskip E_{2}= (b_{0}+b_{1}t)^{2} \left[ \frac{\dot{r}^{2} +r^{2} \dot{\theta}^{2}}{2} -\frac{k}{r(b_{0}+b_{1}t)} \right] -b_{1}(b_{0}+b_{1}t) r\dot{r} + \frac{b_{1}^{2}r^{2}}{2}
\]
\[
A_{1}= L_{3} \Big[ (b_{0}+b_{1}t)\dot{r} -b_{1}r \Big] \sin\theta + \Big[ (b_{0}+b_{1}t)L_{3}r\dot{\theta} - k \Big] \cos\theta
\]
\[
A_{2}= -L_{3} \Big[ (b_{0}+b_{1}t)\dot{r} -b_{1}r \Big] \cos\theta + \Big[ (b_{0}+b_{1}t) L_{3} r\dot{\theta} -k \Big]\sin\theta, \enskip  A_{3}=0.
\]

Using the relation $\dot{\theta}=\frac{L_{3}}{r^{2}}$ to replace $\dot{\theta}$, the above relations are written as:
\begin{eqnarray}
E_{2}&=& (b_{0}+b_{1}t)^{2} \left[ \frac{\dot{r}^{2}}{2} + \frac{L_{3}^{2}}{2r^{2}} -\frac{k}{r(b_{0}+b_{1}t)} \right] -b_{1}(b_{0}+b_{1}t) r\dot{r} + \frac{b_{1}^{2}r^{2}}{2} \label{eq.exaGK.2a} \\
A_{1}&=& L_{3} \Big[ (b_{0}+b_{1}t)\dot{r} -b_{1}r \Big] \sin\theta + \Big[ (b_{0}+b_{1}t)\frac{L_{3}^{2}}{r} - k \Big] \cos\theta \label{eq.exaGK.2b} \\
A_{2}&=& -L_{3} \Big[ (b_{0}+b_{1}t)\dot{r} -b_{1}r \Big] \cos\theta + \Big[ (b_{0}+b_{1}t) \frac{L_{3}^{2}}{r} -k \Big] \sin\theta. \label{eq.exaGK.2c}
\end{eqnarray}

By multiplying equation (\ref{eq.exaGK.2b}) with $\cos\theta$ and (\ref{eq.exaGK.2c}) with $\sin\theta$, we find that
\begin{equation}
\frac{1}{r} = \frac{k}{L_{3}^{2}(b_{0}+b_{1}t)} \left( 1 + k_{1}\cos\theta + k_{2} \sin\theta \right) \implies r= \frac{L_{3}^{2}(b_{0}+b_{1}t)}{k\left( 1 + k_{1}\cos\theta + k_{2} \sin\theta \right)} \label{eq.exaGK.3}
\end{equation}
where $k_{1}\equiv \frac{A_{1}}{k}$ and $k_{2}\equiv \frac{A_{2}}{k}$.

Applying the transformation $k_{1}= \alpha \cos\beta$ and $k_{2}= \alpha \sin\beta$, equation (\ref{eq.exaGK.3}) is written (see also section 5 in \cite{Katzin 1982})
\begin{equation}
\frac{1}{r} = \frac{\omega_{2K}}{L_{3}^{2}} \Big[ 1 + \alpha \cos \left( \theta -\beta \right) \Big] \implies r = \frac{L_{3}^{2}\omega_{2K}^{-1}}{1 + \alpha \cos \left(\theta - \beta\right)} \label{eq.exaGK.4}
\end{equation}
which for $\omega_{2K}(t)=const$ (standard Kepler problem) reduces to the analytical equation of a conic section in polar coordinates. In that case, $\alpha$ is the eccentricity.\index{Eccentricity}

It is also worthwhile to mention that the relation (\ref{eq.TKFI4}) becomes $2E_{2}L_{3}^{2} = k^{2}(\alpha^{2}-1)$.

Moreover, equation (\ref{eq.exaGK.2a}) gives
\[
\left[ \frac{d}{dt} \left( \frac{r}{b_{0}+b_{1}t} \right) \right]^{2} = -2(b_{0}+b_{1}t)^{-2} \left[ \frac{L_{3}^{2}}{2r^{2}} -\frac{k}{r(b_{0}+b_{1}t)} -\frac{E_{2}}{(b_{0}+b_{1}t)^{2}} \right].
\]

Finally, in the polar plane, the equations of motion (\ref{eq.TGKep.1a}) for $\nu=1$ become:
\begin{eqnarray}
\ddot{r} - r\dot{\theta}^{2} +\frac{\omega_{2K}}{r^{2}} &=& 0 \label{eq.exaGK.5a} \\
r\ddot{\theta} + 2\dot{r}\dot{\theta}&=&0. \label{eq.exaGK.5b}
\end{eqnarray}
Equation (\ref{eq.exaGK.5b}) implies the FI of the angular momentum $L_{3} =r^{2}\dot{\theta}$. It can be easily checked that the solution (\ref{eq.exaGK.3}) satisfies equation (\ref{eq.exaGK.5a}) by replacing $\ddot{\theta}$ from (\ref{eq.exaGK.5b}) and $\dot{\theta}$ with $\frac{L_{3}}{r^{2}}$. The solution (\ref{eq.exaGK.3}) into the LFI $L_{3}$ gives
\begin{equation}
\int \frac{k^{2}dt}{L_{3}^{3}(b_{0}+b_{1}t)^{2}}= \int \frac{d\theta}{\left( 1 + k_{1}\cos\theta + k_{2} \sin\theta \right)^{2}} \implies \frac{k}{L_{3}^{2} (b_{0}+b_{1}t)} = - \frac{b_{1} L_{3}}{k}\int \frac{d\theta}{\left( 1 + k_{1}\cos\theta + k_{2} \sin\theta \right)^{2}}. \label{eq.exaGK.6}
\end{equation}
Substituting (\ref{eq.exaGK.6}) in (\ref{eq.exaGK.3}), we obtain the trajectory\index{Trajectory}
\begin{equation}
\frac{1}{r} = - \frac{b_{1}L_{3}}{k} \left( 1 + k_{1}\cos\theta + k_{2} \sin\theta \right) \int \frac{d\theta}{\left( 1 + k_{1}\cos\theta + k_{2} \sin\theta \right)^{2}} \label{eq.exaGK.7}
\end{equation}
which coincides with eq. (5.17) in \cite{Katzin 1982}.

\section{A class of 1d non-linear time-dependent equations}

\label{sec.nonlin}

In this section, we use the well-known result \cite{LeoTsampAndro 2017} that the non-linear dynamical system
\begin{equation}
\ddot{q}^{a}= -\Gamma^{a}_{bc}\dot{q}^{b}\dot{q}^{c} -\omega(t)Q^{a}(q) +\phi(t)\dot{q}^{a} \label{eq.damp0a}
\end{equation}
is equivalent to the dynamical system (without damping term)
\begin{equation}
\frac{d^{2}q^{a}}{ds^{2}}= -\Gamma^{a}_{bc}\frac{dq^{b}}{ds} \frac{dq^{c}}{ds} -\bar{\omega}(s)Q^{a}(q) \label{eq.damp0b}
\end{equation}
where $\phi(t)$ is an arbitrary function such that
\begin{equation}
s(t)= \int e^{\int\phi(t)dt} dt, \enskip \bar{\omega}(s)= \omega(t(s)) \left(\frac{dt}{ds}\right)^{2} \iff \omega(t)= \bar{\omega}(s(t)) e^{2\int\phi(t)dt}. \label{eq.damp0c}
\end{equation}

We apply this result to the following problem:

\emph{Consider the second order ODE}
\begin{equation}
\ddot{x}=-\omega(t)x^{\mu }+\phi(t)\dot{x}  \label{eq.nonl1}
\end{equation}%
\emph{where the constant $\mu\neq-1$, and determine the relation
between the functions $\omega(t)$ and $\phi(t)$ for which the ODE (\ref{eq.nonl1}) admits a QFI; therefore, it is integrable.}

This problem has been considered previously in \cite{Da Silva 1974, Sarlet 1980} (see eq. (28a) in \cite{Da Silva 1974} and eq. (17) in \cite{Sarlet 1980}) and has been answered partially using different methods. In \cite{Da Silva 1974}, the author used the Hamiltonian formalism, where one looks for a canonical transformation to bring the Hamiltonian in a time-separable form; whereas in \cite{Sarlet 1980} the author used a direct method for constructing FIs by multiplying the equation with an integrating factor. In \cite{Sarlet 1980}, it is shown that both methods are equivalent and that the results of \cite{Sarlet 1980} generalize those of \cite{Da Silva 1974}. In the following, we shall generalize the results of \cite{Sarlet 1980}; in addition, we discuss a number of applications.

Equation (\ref{eq.nonl1}) is equivalent to equation
\begin{equation}
\frac{d^{2}x}{ds^{2}}= -\bar{\omega}(s)x^{\mu}, \enskip \mu\neq-1 \label{eq.nonl2}
\end{equation}
where the function $\bar{\omega}(s)$ is given by (\ref{eq.damp0c}).

Replacing with $Q^{1}=x^{\mu}$ in the system of equations (\ref{eq.TKN1}) - (\ref{eq.TKN6}), we find that\footnote{
In 1d Euclidean space, the KT condition (\ref{eq.TKN1}) gives $K_{11,1}=0$ $\implies K_{11}=K_{11}(s)$, that is, the KT is an arbitrary function of $s$.
} $K_{11}= K_{11}(s)$ and the following conditions:
\begin{align}
K_{1}(s,x) &= -\frac{dK_{11}}{ds}x + b_{1}(s) \label{eq.nonl3a} \\
K(s,x) &= 2\bar{\omega} K_{11} \frac{x^{\mu+1}}{\mu+1} + \frac{d^{2}K_{11}}{ds^{2}} \frac{x^{2}}{2} -\frac{db_{1}}{ds}x +b_{2}(s) \label{eq.nonl3b} \\
0 &= \left( \frac{2\frac{d\bar{\omega}}{ds} K_{11}}{\mu+1} +\frac{2\bar{\omega} \frac{dK_{11}}{ds}}{\mu+1} +\bar{\omega}\frac{dK_{11}}{ds} \right) x^{\mu+1} -\bar{\omega} b_{1} x^{\mu} + \frac{d^{3}K_{11}}{ds^{3}}\frac{x^{2}}{2} - \frac{d^{2}b_{1}}{ds^{2}}x + \frac{db_{2}}{ds} \label{eq.nonl3c}
\end{align}
where $b_{1}(s)$ and $b_{2}(s)$ are arbitrary functions. Then, the general QFI (\ref{FI.5}) becomes
\begin{equation}
I= K_{11}(s) \left( \frac{dx}{ds} \right)^{2} +K_{1}(s,x)\frac{dx}{ds} + K(s,x). \label{eq.nonl4}
\end{equation}

We consider the solution of the system (\ref{eq.nonl3a}) - (\ref{eq.nonl3c}) for various values of $\mu$.

As it will be shown, for $\mu \neq -1$, there results a family of `frequencies' $\bar{\omega}(s)$ parameterized with constants; whereas, for the specific values $\mu =0,1,2$, there results a family of `frequencies' $\bar{\omega}(s)$ parameterized with functions.
\bigskip

1) Case $\mu=0$.

We find the QFI
\begin{equation}
I= K_{11} \left( \frac{dx}{ds} \right)^{2} - \frac{dK_{11}}{ds} x\frac{dx}{ds} +b_{1}(s)\frac{dx}{ds} + c_{3}x^{2} +2\bar{\omega}(s)K_{11}x -\frac{db_{1}}{ds}x + \int b_{1}(s) \bar{\omega}(s) ds \label{eq.nonl4.1}
\end{equation}
where $K_{11}= c_{1} +c_{2}s + c_{3}s^{2}$, $c_{1}, c_{2}, c_{3}$ are arbitrary constants, and the functions $b_{1}(s), \bar{\omega}(s)$ satisfy the condition
\begin{equation}
\frac{d^{2}b_{1}}{ds^{2}}= 2\frac{d\bar{\omega}}{ds}K_{11} +3\bar{\omega}\frac{dK_{11}}{ds}. \label{eq.nonl4.2}
\end{equation}

Using the transformation (\ref{eq.damp0c}), equations (\ref{eq.nonl4.1}) and (\ref{eq.nonl4.2}) become, respectively, as follows:
\begin{eqnarray}
I&=& \left[ c_{1} +c_{2}\int e^{\int\phi(t)dt} dt +c_{3}\left(\int e^{\int\phi(t)dt} dt\right)^{2} \right] e^{-2\int \phi(t)dt} \dot{x}^{2}  -\left[ c_{2} +2c_{3}\int e^{\int\phi(t)dt} dt \right] e^{-\int \phi(t)dt} x\dot{x} + \notag\\
&& +b_{1}(s(t)) e^{-\int \phi(t)dt} \dot{x} + c_{3}x^{2} + 2\omega(t) \left[ c_{1} +c_{2}\int e^{\int\phi(t)dt} dt +c_{3}\left(\int e^{\int\phi(t)dt} dt\right)^{2} \right] e^{-2\int \phi(t)dt} x - \notag \\
&& -\dot{b}_{1} e^{-\int \phi(t)dt}x + \int b_{1}(s(t)) \omega(t) e^{-\int \phi(t)dt} dt \label{eq.nonl4.2.1}
\end{eqnarray}
and
\begin{eqnarray}
\ddot{b}_{1} -\phi\dot{b}_{1} &=& 2e^{-\int \phi(t)dt} \left( \dot{\omega} -2\phi \omega \right) \left[ c_{1} +c_{2}\int e^{\int\phi(t)dt} dt +c_{3}\left(\int e^{\int\phi(t)dt} dt\right)^{2} \right] + \notag \\
&& + 3\omega \left[ c_{2} +2c_{3}\int e^{\int\phi(t)dt} dt \right]. \label{eq.nonl4.2.2}
\end{eqnarray}

2) Case $\mu=1$.

We derive again the results of the time-dependent oscillator\index{Oscillator! time-dependent} (see Table \ref{T2} for $\nu=-2$) in one dimension. Using the transformation (\ref{eq.damp0c}), we deduce that the original equation
\begin{equation}
\ddot{x}= -\omega(t)x +\phi(t)\dot{x} \label{eq.nonl4.2.5}
\end{equation}
for the frequency
\begin{equation}
\omega(t)= -\rho^{-1}\ddot{\rho} +\phi (\ln \rho)^{\cdot} +\rho^{-4} e^{2\int\phi(t)dt} \label{eq.nonl4.2.6}
\end{equation}
admits the general solution
\begin{equation}
x(t)= \rho(t) \left( A \sin\theta + B\cos\theta \right) \label{eq.nonl4.2.7}
\end{equation}
where $\rho(t)\equiv \rho(s(t))$ and $\theta(s(t))= \int \rho^{-2}(t) e^{\int\phi(t)dt}dt$.

3) Case $\mu=2$.

We find the function $\bar{\omega}= K_{11}^{-5/2}$ and the QFI
\begin{equation}
I= K_{11}(s) \left( \frac{dx}{ds} \right)^{2} -\frac{dK_{11}}{ds} x\frac{dx}{ds} +(c_{4}+c_{5}s) \frac{dx}{ds} + \frac{2}{3}K_{11}^{-3/2} x^{3} + \frac{d^{2}K_{11}}{ds^{2}} \frac{x^{2}}{2} -c_{5}x \label{eq.nonl4.3}
\end{equation}
where $c_{4}, c_{5}$ are arbitrary constants and the function $K_{11}(s)$ is given by
\begin{equation}
\frac{d^{3}K_{11}}{ds^{3}}= 2(c_{4}+c_{5}s)K_{11}^{-5/2}. \label{eq.nonl4.4}
\end{equation}

Using the transformation (\ref{eq.damp0c}), the above results become:
\begin{equation}
\omega(t)= K_{11}^{-5/2} e^{2\int\phi(t)dt} \label{eq.nonl4.4.1}
\end{equation}
\begin{eqnarray}
I&=& K_{11} e^{-2\int \phi(t)dt} \dot{x}^{2} -\dot{K}_{11} e^{-2\int \phi(t)dt} x\dot{x} +\left[ c_{4} +c_{5}\int e^{\int\phi(t)dt} dt \right] e^{-\int \phi(t)dt}\dot{x} + \frac{2}{3}K_{11}^{-3/2} x^{3} + \notag \\
&&+ \left( \ddot{K}_{11} -\phi\dot{K}_{11} \right)e^{-2\int \phi(t)dt} \frac{x^{2}}{2} -c_{5}x \label{eq.nonl4.4.2}
\end{eqnarray}
and
\begin{equation}
\dddot{K}_{11} -3\phi \ddot{K}_{11} -\dot{\phi}\dot{K}_{11} +2\phi^{2}\dot{K}_{11}= 2\left[ c_{4} +c_{5}\int e^{\int\phi(t)dt} dt \right] e^{3\int\phi(t)dt} K_{11}^{-5/2} \label{eq.nonl4.4.3}
\end{equation}
where the function $K_{11}=K_{11}(s(t))$.

We note that for $\mu=2$ equation (\ref{eq.nonl1}), or to be more specific its equivalent (\ref{eq.nonl2}), arises in the solution of Einstein field equations when the gravitational field is spherically symmetric and the matter source is a shear-free perfect fluid\index{Fluid! shear-free perfect} (see e.g. \cite{Stephani, Stephani 1983, Srivastana 1987, Leach 1992, LeachMaartens 1992, Maharaj 1996}).

4) Case $\mu \neq -1$.

In this case $b_{1}=b_{2}=0$, $K_{11}= c_{1} +c_{2}s +c_{3}s^{2}$ and $\bar{\omega}(s)= (c_{1} +c_{2}s +c_{3}s^{2})^{-\frac{\mu+3}{2}}$, where $c_{1}, c_{2}, c_{3}$ are arbitrary constants.

The QFI (\ref{eq.nonl4}) becomes
\begin{equation}
I= (c_{1} +c_{2}s +c_{3}s^{2})\left( \frac{dx}{ds} \right)^{2}  -(c_{2} +2c_{3}s)x\frac{dx}{ds} +\frac{2}{\mu+1} (c_{1} +c_{2}s +c_{3}s^{2})^{-\frac{\mu +1}{2}} x^{\mu+1} + c_{3}x^{2} \label{eq.nonl5}
\end{equation}
and the function
\begin{equation}
\bar{\omega}(s)= (c_{1} +c_{2}s +c_{3}s^{2})^{-\frac{\mu +3}{2}}. \label{eq.nonl6}
\end{equation}
It can be checked that (\ref{eq.nonl5}) and (\ref{eq.nonl6}) for $\mu=0, 1, 2$ give results compatible with the ones we found for these values of $\mu$.

Using the transformation (\ref{eq.damp0c}), we deduce that the original system (\ref{eq.nonl1}) is integrable iff the functions $\omega(t), \phi(t)$ are related as follows:
\begin{equation}
\omega(t)= \left[ c_{1} +c_{2}\int e^{\int\phi(t)dt} dt +c_{3}\left(\int e^{\int\phi(t)dt} dt\right)^{2} \right]^{-\frac{\mu+3}{2}} e^{2\int \phi(t)dt}. \label{eq.nonl6.1}
\end{equation}
In this case, the associated QFI (\ref{eq.nonl5}) is
\begin{eqnarray}
I&=& \left[ c_{1} +c_{2}\int e^{\int\phi(t)dt} dt +c_{3}\left(\int e^{\int\phi(t)dt} dt\right)^{2} \right] e^{-2\int \phi(t)dt} \dot{x}^{2} -\left[ c_{2} +2c_{3}\int e^{\int\phi(t)dt} dt \right] e^{-\int \phi(t)dt} x\dot{x} + \notag \\
&& +\frac{2}{\mu+1} \left[ c_{1} +c_{2}\int e^{\int\phi(t)dt} dt +c_{3}\left(\int e^{\int\phi(t)dt} dt\right)^{2} \right]^{-\frac{\mu +1}{2}} x^{\mu+1} + c_{3}x^{2}. \label{eq.nonl6.2}
\end{eqnarray}

These expressions generalize the ones given in \cite{Sarlet 1980}. Indeed, if we introduce the notation $\omega(t)\equiv \alpha(t)$ and $\phi(t)\equiv -\beta(t)$, then equations (\ref{eq.nonl6.1}) and (\ref{eq.nonl6.2}) for $c_{3}=0$ become eqs. (25) and (26) of \cite{Sarlet 1980}.

\subsection{The generalized Lane-Emden equation}

\label{sec.emden}

Consider the 1d \textbf{generalized Lane-Emden equation}\index{Equation! generalized Lane-Emden} (see eq. (6) in \cite{Muatje 2011})
\begin{equation}
\ddot{x}= -\omega(t)x^{\mu} -\frac{k}{t}\dot{x} \label{eq.emd1}
\end{equation}
where $k$ is an arbitrary constant. This equation is well-known in the literature because of its \emph{many applications in astrophysical problems} (see citations in \cite{Muatje 2011}). In general, to find explicit analytic solutions of equation (\ref{eq.emd1}) is a major task. For example, such solutions have been found only for the special values $\mu=0,1,5$, in the case that the function $\omega(t)=1$ and the constant $k=2$. New exact solutions, or at least the Liouville integrability, of equation (\ref{eq.emd1}) are guaranteed, if we find a way to determine its FIs. We see that equation (\ref{eq.emd1}) is a subcase of the original equation (\ref{eq.nonl1}) for $\phi(t)= -\frac{k}{t}$; therefore, we can apply the results found earlier in section \ref{sec.nonlin}.

In what follows, we discuss only the fourth case where $\mu\neq-1$ in order to compare our results with those found in\footnote{In \cite{Muatje 2011}, the authors used point Noether symmetries and Noether's Theorem.} Table 1 of \cite{Muatje 2011}. In particular, for $\phi(t)= -\frac{k}{t}$ the function (\ref{eq.nonl6.1}) and the associated QFI (\ref{eq.nonl6.2}) become:
\begin{equation}
\omega(t)= t^{-2k} \left( c_{1} +c_{2}M +c_{3}M^{2} \right)^{-\frac{\mu+3}{2}} \label{eq.emd2}
\end{equation}
and
\begin{equation}
I= t^{2k} \left( c_{1} +c_{2}M +c_{3}M^{2} \right) \dot{x}^{2} -t^{k} \left( c_{2} +2c_{3}M \right) x\dot{x} +\frac{2}{\mu+1} \left( c_{1} +c_{2}M +c_{3}M^{2} \right)^{-\frac{\mu +1}{2}} x^{\mu+1} + c_{3}x^{2} \label{eq.emd3}
\end{equation}
where the function $M(t)= \int t^{-k}dt$.

Concerning the form of the function $M(t)$, there are two cases to be considered: a) $k=1$, and b) $k\neq1$.
\bigskip

a) Case $k=1$.

We have $M=\ln t$. Then, equations (\ref{eq.emd2}) and (\ref{eq.emd3}) become:
\begin{equation}
\omega(t)= t^{-2} \left[ c_{1} +c_{2}\ln t +c_{3}(\ln t)^{2} \right]^{-\frac{\mu+3}{2}} \label{eq.emd4.1}
\end{equation}
and
\begin{equation}
I= t^{2} \left[ c_{1} +c_{2}\ln t +c_{3}(\ln t)^{2} \right] \dot{x}^{2} -t \left( c_{2} +2c_{3}\ln t \right) x\dot{x} +\frac{2}{\mu+1} \left[ c_{1} +c_{2}\ln t +c_{3}(\ln t)^{2} \right]^{-\frac{\mu +1}{2}} x^{\mu+1} + c_{3}x^{2}. \label{eq.emd4.2}
\end{equation}

We consider the following subcases:

-$c_{2}=c_{3}=0$ and $c_{1}\neq0$.

Equations (\ref{eq.emd4.1}) and (\ref{eq.emd4.2}) give, respectively, the function $\omega(t)= At^{-2}$ and the QFI (divide $I$ with $2c_{1}$) $I= \frac{t^{2}}{2}\dot{x}^{2} + \frac{A}{\mu+1}x^{\mu+1}$ where the constant $A=c_{1}^{-\frac{\mu+3}{2}}$. This is the Case 5 in Table 1 of \cite{Muatje 2011}.

- $c_{1}=c_{3}=0$ and $c_{2}\neq0$.

Equations (\ref{eq.emd4.1}) and (\ref{eq.emd4.2}) give, respectively, the function $\omega(t)= At^{-2}(\ln t)^{-\frac{\mu+3}{2}}$ and the QFI (divide $I$ with $2c_{2}$)
$I= \frac{1}{2}t^{2}(\ln t)\dot{x}^{2} -\frac{t}{2} x\dot{x} +\frac{A}{\mu+1} (\ln t)^{-\frac{\mu+1}{2}} x^{\mu+1}$ where the constant $A= c_{2}^{-\frac{\mu+3}{2}}$. This is the Case 6 in Table 1 of \cite{Muatje 2011}.

- $c_{1}=c_{2}=0$ and $c_{3}\neq0$.

Equations (\ref{eq.emd4.1}) and (\ref{eq.emd4.2}) give, respectively, the function $\omega(t)= At^{-2}(\ln t)^{-\mu-3}$ and the QFI (divide $I$ with $2c_{3}$) $I= \frac{1}{2} (t\ln t)^{2} \dot{x}^{2} -t(\ln t)x\dot{x} +\frac{A}{\mu+1} (\ln t)^{-\mu-1}x^{\mu+1} +\frac{x^{2}}{2}$ where the constant $A= c_{3}^{-\frac{\mu+3}{2}}$. This is the Case 7 in Table 1 of \cite{Muatje 2011}.
\bigskip

b) Case $k\neq1$.

We have $M= \frac{t^{1-k}}{1-k}$. Then, equations (\ref{eq.emd2}) and (\ref{eq.emd3}) become, respectively, as follows:
\begin{equation}
\omega(t)= t^{-2k} \left[ c_{1} +\frac{c_{2}}{1-k}t^{1-k} +\frac{c_{3}}{(1-k)^{2}}t^{2(1-k)} \right]^{-\frac{\mu+3}{2}} \label{eq.emd5.1}
\end{equation}
and
\begin{eqnarray}
I&=& t^{2k} \left[ c_{1} +\frac{c_{2}}{1-k}t^{1-k} +\frac{c_{3}}{(1-k)^{2}}t^{2(1-k)} \right] \dot{x}^{2} -t^{k} \left( c_{2} +\frac{2c_{3}}{1-k}t^{1-k} \right) x\dot{x} +\notag \\
&& +\frac{2}{\mu+1} \left[ c_{1} +\frac{c_{2}}{1-k}t^{1-k} +\frac{c_{3}}{(1-k)^{2}}t^{2(1-k)} \right]^{-\frac{\mu +1}{2}} x^{\mu+1} + c_{3}x^{2}. \label{eq.emd5.2}
\end{eqnarray}

We consider the following subcases:

- $c_{2}=c_{3}=0$ and $c_{1}\neq0$.

Equations (\ref{eq.emd5.1}) and (\ref{eq.emd5.2}) give, respectively, the function $\omega(t)= At^{-2k}$ and the QFI (divide $I$ with $2c_{1}$) $I= \frac{t^{2k}}{2}\dot{x}^{2} + \frac{A}{\mu+1}x^{\mu+1}$ where the constant $A=c_{1}^{-\frac{\mu+3}{2}}$. This is the Case 2 in Table 1 of \cite{Muatje 2011}.

- $c_{1}=c_{3}=0$ and $c_{2}\neq0$.

Equations (\ref{eq.emd5.1}) and (\ref{eq.emd5.2}) give, respectively, the function $\omega(t)= At^{\frac{1}{2}(k\mu -k -\mu -3)}$ and the QFI (multiply $I$ with $\frac{1-k}{c_{2}}$)
$I= t^{k+1}\dot{x}^{2} +(k-1)t^{k}x\dot{x} +\frac{2A}{\mu+1} t^{\frac{1}{2}(\mu+1)(k-1)} x^{\mu+1}$ where the constant $A= \left( \frac{c_{2}}{1-k} \right)^{-\frac{\mu+3}{2}}$. This is the Case 3 in Table 1 of \cite{Muatje 2011}.

We note also that for $k=\frac{\mu+3}{\mu-1}$ where $\mu\neq1$  the function $\omega(t)=A=const$. This reproduces the first subcase of Case 1 in Table 1 of \cite{Muatje 2011}, which is the Case 5.1 of \cite{Khalique 2008}.

- $c_{1}=c_{2}=0$ and $c_{3}\neq0$.

Equations (\ref{eq.emd5.1}) and (\ref{eq.emd5.2}) give, respectively, the function $\omega(t)= At^{k\mu +k -\mu -3}$ and the QFI (multiply $I$ with $\frac{(1-k)^{2}}{2c_{3}}$) $I= \frac{t^{2}}{2}\dot{x}^{2} +(k-1)tx\dot{x} +\frac{A}{\mu+1} t^{(\mu+1)(k-1)}x^{\mu+1} +\frac{1}{2}(k-1)^{2}x^{2}$ where the constant $A= \left( \frac{1-k}{\sqrt{c_{3}}} \right)^{\mu+3}$. This is the Case 4 in Table 1 of \cite{Muatje 2011}.

We note also that for $k=\frac{\mu+3}{\mu+1}$ the function $\omega(t)=A=const$. This recovers the second subcase of Case 1 in Table 1 of \cite{Muatje 2011}, which is the Case 5.2 of \cite{Khalique 2008}.
\bigskip

We conclude that the seven cases 1-7 found in Table 1 of \cite{Muatje 2011} are just subcases of the above two general cases a) and b). To compare with these results, one may adopt the notation: $\omega=f$, $k=n$ and $\mu=p$.

%% file: Brans_Dicke.tex
\chapter{New conservation laws and exact cosmological solutions in Brans-Dicke cosmology with an extra scalar field}

\label{ch.BransDicke}

\section{Introduction}

\label{sec.Brans.intro}

The detailed analysis of the recent cosmological observations indicates that the universe has been through two accelerating phases \cite{Riess 1998, Perlmutter 1999, Riess 2007, Suzuki 2012}. The current acceleration era\index{Acceleration era} is assumed to be driven by an unknown source known as dark energy\index{Dark energy} whose main characteristic is the negative pressure, which provides an anti-gravity effect \cite{DiValentino 2021}. On the other hand, the early-universe acceleration era, known as inflation,\index{Inflation} is described by a scalar field, the inflaton,\index{Inflaton} which is used to explain the homogeneity and isotropy of the present universe. In particular, this scalar field dominates the dynamics and explains the expansion era \cite{Starobinsky 1980, Guth 1981}.\index{Expansion era} Nevertheless, the scalar
field inflationary models are mainly defined on homogeneous spacetimes, or on background spaces with small inhomogeneities \cite{Muller 1988, Kofman 1985}. In \cite{Wald 1983}, it was found that the presence of a positive cosmological constant in Bianchi cosmologies leads to expanding Bianchi spacetimes, evolving towards the de Sitter universe. That was the first result to support the cosmic `no-hair' conjecture \cite{Gibbons 1977, Hawking 1982}. This latter conjecture states that all expanding universes with a positive cosmological constant admit as asymptotic solution the de Sitter universe. The necessity of the de Sitter expansion is that it provides a rapid expansion for the size of the universe such that the latter, effectively, loses its memory on the initial conditions, which implies that the de Sitter expansion solves the `flatness', `horizon' and monopole problem \cite{Sato 1981, Barrow 1983}.

In the literature, scalar fields\index{Scalar field} have been introduced in the gravitational theory in various ways. The simplest scalar field model is the quintessence model, which consists of a scalar field minimally coupled to gravity \cite{Ratra, Barrow}. Another family of scalar fields are those which belong to
the scalar-tensor theory. In this theory, the scalar field is non-minimally coupled
to gravity, which makes it essential for the physical state of the theory.
Another important characteristic of the scalar-tensor theories is that they
are in consistence with Mach's principle. The most common scalar-tensor theory
is the Brans-Dicke theory \cite{Brans} which is considered in this study. For other scalar-tensor theories and generalizations, we refer the reader to \cite{faraonibook, Horndeski 1974, OHanlon 1972, Nicolis 2009, Deffayet 2009, Belinchon 2017, Fomin 2018, Fomin 2019} and references therein.

The Einstein field equations of general relativity are a set of ten nonlinear second-order PDEs with independent variables the spacetime coordinates and dependent variables the components of the metric tensor. However, by assuming
specific forms for the metric tensor and the existence of collineations, the field equations
are simplified by reducing the number of the independent variables (see e.g. \cite{Saridakis Tsamparlis 1991, Shapovalov 1979, Osetrin 2017, Tsamparlis 2019GR, Obukhov 2021} and references therein). According to the cosmological principle, in large scale the universe is
assumed to be homogeneous, isotropic, and spatially flat. This implies that the
background space is described by the Friedmann - Lema\^{\i}tre - Robertson - Walker (FLRW) spacetime.\index{Spacetime! FLRW} This spacetime is characterized by the scale factor,\index{Scale factor} which defines the radius of the 3d Euclidean space. Since
General Relativity is a second order theory, the
field equations involve second order derivatives of the scale factor. For simple
cosmological fluids, such are the ideal gas or the cosmological constant, the field
equations can be solved explicitly \cite{amendolaB}. However, when additional
degrees of freedom are introduced, such as a scalar field, the field equations
cannot be solved with the use of closed-form functions and techniques of
analytic mechanics, and one looks for FIs which establish their (Liouville)
integrability  \cite{sym1, sym2, sym3, sym4}. In the following, we shall determine FIs (i.e. conservation laws) of the field equations, by applying the direct method of chapter \ref{ch.QFIs.timedependent}.

In this chapter, we consider a cosmological model in which the gravitational action integral is that of Brans-Dicke theory with an additional scalar field minimally coupled to gravity \cite{Mukherjee2019, Giacomini 2020}. This two-scalar field model belongs to the
family of multi-scalar field models, which have been used  as unified dark
energy models \cite{sf6, sf7, sf8} or as alternative models for the description of the acceleration phases of the universe \cite{sf9, sf10, sf11, sf12}. Indeed, multifield inflationary models provide an alternative mechanism for the description of the early acceleration phase of the universe. The mechanism for the end of the inflation is much more simple. Specifically, the scalar fields at the beginning, and at the end, of the inflation are not necessary the same. Thus, this can lead to different number of e-folds and affect the curvature perturbations \cite{Lyth, Choi}. The latter, in the non-adiabatic perturbations, can provide detectable non-Gaussianities in the power spectrum \cite{Langlois}. As far as the late-time acceleration phase, multifield cosmological models have been introduced to describe dark energy models with varying equation of state parameter, which can cross the phantom
divide line without the appearance of ghosts \cite{Langlois}. Such models can solve the Hubble-tension problem \cite{sf10}. Furthermore, multi-scalar field models can attribute the additional degrees of freedom provided by the alternative theories of gravity \cite{lan1, lan2, lan3}.

\section{Cosmological model}

\label{sec2}

For the gravitational action integral, we consider that of Brans-Dicke scalar field theory with an additional matter source. We have the following expression \cite{Brans, faraonibook}:
\begin{equation}
S= \int d^{4}x \sqrt{-g} \left[  \frac{1}{2}\phi R -\frac{1}{2} \frac{\omega_{BD}}{\phi}g^{\mu\nu}\phi_{;\mu}\phi_{;\nu} +L_{\psi} \left( \psi, \psi_{;\mu} \right)  \right]  +S_{m} \label{bd.01}%
\end{equation}
where $x^{k}$ are the spacetime coordinates, $g_{\mu\nu}$ is the metric, $\phi\left( x^{\kappa}\right)$ denotes the Brans-Dicke scalar field\index{Scalar field! Brans-Dicke}
and $\omega_{BD}$ is the Brans-Dicke parameter.\index{Parameter! Brans-Dicke} The action $S_{m}$ is assumed
to describe an ideal gas\index{Gas! ideal} with constant equation of state parameter and the Lagrangian function $L_{\psi}\left( \psi, \psi_{;\mu}\right)$ corresponds to the second scalar field $\psi\left(  x^{\kappa}\right)$, which is assumed to
be that of quintessence\index{Scalar field! quintessence} and minimally coupled to the Brans-Dicke scalar field. With these assumptions the action integral (\ref{bd.01}) takes the following form:
\begin{equation}
S=\int d^{4}x\sqrt{-g}\left[  \frac{1}{2}\phi R -\frac{1}{2}\frac{\omega_{BD}}{\phi} g^{\mu\nu}\phi_{;\mu} \phi_{;\nu} -\frac{1}{2}g^{\mu\nu}\psi_{;\mu}\psi_{;\nu} -V\left(\psi\right)  \right]  +S_{m}. \label{bd.01a}
\end{equation}

The gravitational field equations follow from the variation (see $\delta_{0}$-variation in chapter \ref{ch.VarCalSym}) of the action integral (\ref{bd.01a}) wrt the metric tensor $g_{\mu\nu}$. They are
\begin{equation}
G_{\mu\nu}=\frac{\omega_{BD}}{\phi^{2}}\left(  \phi_{;\mu}\phi_{;\nu}-\frac
{1}{2}g_{\mu\nu}g^{\kappa\lambda}\phi_{;\kappa}\phi_{;\lambda}\right)
+\frac{1}{\phi}\left(  \phi_{;\mu\nu}-g_{\mu\nu}g^{\kappa\lambda}\phi
_{;\kappa\lambda}\right)  +\frac{1}{\phi}T_{\mu\nu}\label{bd.02}%
\end{equation}
where $G_{\mu\nu}=R_{\mu\nu}-\frac{1}{2}Rg_{\mu\nu}$ is the Einstein tensor.\index{Tensor! Einstein}
The total energy momentum tensor\index{Tensor! energy momentum} $T_{\mu\nu}= {}^{\psi}T_{\mu\nu}+{}^{m}T_{\mu\nu}$,
where ${}^{m}T_{\mu\nu}$ corresponds to the ideal gas and ${}^{\psi}T_{\mu\nu}$ provides the contribution of the field $\psi\left(  x^{k}\right)$ in the
field equations.

Concerning the equations of motion for the matter source and the two scalar fields, we have ${}^{m}T_{\mu\nu;\sigma}g^{\mu\sigma}=0,$ while variation wrt the fields $\phi\left(  x^{\kappa}\right)$ and $\psi\left(
x^{\kappa}\right)$ provides the second order PDEs:
\begin{equation}
g^{\mu\nu}\phi_{;\mu\nu} -\frac{1}{2\phi}g^{\mu\nu}\phi_{;\mu} \phi_{;\nu}
+\frac{\phi}{2\omega_{BD}}R=0 \label{bd.02a}%
\end{equation}%
\begin{equation}
g^{\mu\nu}\psi_{;\mu\nu} -\frac{dV}{d\psi} =0. \label{bd.02b}%
\end{equation}

We assume the background space to be the \textbf{FLRW spacetime}\index{Spacetime! FLRW} with line element
\begin{equation}
ds^{2}=-dt^{2}+a^{2}(t)\left(  dx^{2}+dy^{2}+dz^{2}\right)  \label{bd.05}%
\end{equation}
where $a(t)$ is the scale factor\index{Scale factor} of the universe and $H\left(  t\right)
=\frac{\dot{a}}{a}$ is the Hubble function.\index{Function! Hubble} We note that a dot indicates
derivative wrt the cosmic time $t$.\index{Time! cosmic}

From the line element (\ref{bd.05}), it follows that the Ricci scalar is $R=6\left[ \frac{\ddot{a}}{a}+\left(  \frac{\dot{a}}{a}\right)^{2}\right]$. Replacing in the gravitational field equations (\ref{bd.02}), we obtain:
\begin{equation}
3\left(  \frac{\dot{a}}{a}\right)  ^{2}=\frac{\omega_{BD}}{2} \left(
\frac{\dot{\phi}}{\phi} \right)  ^{2} -3 \frac{\dot{a}}{a} \frac{\dot{\phi}%
}{\phi} +\frac{\rho_{m}+\rho_{\psi}}{\phi} \label{bd.10}%
\end{equation}%
\begin{equation}
2\frac{\ddot{a}}{a} +\left(  \frac{\dot{a}}{a}\right)  ^{2} = -\frac
{\omega_{BD}}{2}\left(  \frac{\dot{\phi}}{\phi}\right)  ^{2}-2 \frac{\dot{a}%
}{a} \frac{\dot{\phi}}{\phi} -\frac{\ddot{\phi}}{\phi} -\frac{p_{m}+p_{\psi}%
}{\phi} \label{bd.11}%
\end{equation}
where $\rho_{m}$ and $p_{m}$ are the mass density and the isotropic pressure, respectively, of the
ideal gas. For the quintessence field, we have:
\begin{equation}
\rho_{\psi}=\frac{1}{2}\dot{\psi}^{2}+V\left(\psi\right), \enskip p_{\psi}= \frac{1}{2}\dot{\psi}^{2}-V\left( \psi\right). \label{bd.12}%
\end{equation}

For the equations of motion of the scalar fields, we find:
\begin{equation}
\ddot{\phi}+3 \frac{\dot{a}}{a} \dot{\phi} = \frac{\left(  \rho_{m}%
-3p_{m}\right)  +\left(  \rho_{\psi}-3p_{\psi}\right)  }{2\omega_{BD}+3}
\label{bd.13}%
\end{equation}
and%
\begin{equation}
\ddot{\psi}+3H\dot{\psi}+ \frac{dV}{d\psi}=0. \label{bd.14a}%
\end{equation}

Finally, for the matter source, the continuity equation ${}^{m}T_{\mu\nu;\sigma}g^{\mu\sigma}=0$ reads
\begin{equation}
\dot{\rho}_{m}+3\frac{\dot{a}}{a}\left(  \rho_{m}+p_{m}\right)  =0.
\label{bd.15}%
\end{equation}
For an ideal gas, the equation of state is\index{Equation! state} $p_{m}=w_{m}\rho_{m}$, where $w_{m}$ is an arbitrary
constant. Substituting in equation (\ref{bd.15}), we
find the solution
\begin{equation}
\rho_{m} =\rho_{m0} a^{-3\left(  1+w_{m}\right)  } \label{bd.16}%
\end{equation}
where $\rho_{m0}$ is an arbitrary constant.

The system of the ODEs that should be solved
consists of the differential equations (\ref{bd.10}), (\ref{bd.11}), (\ref{bd.13}) and (\ref{bd.14a}).

\section{Exact cosmological solutions}

\label{sec.exact.solution}

We can use the direct results\footnote{
To be consistent with the notation of this chapter, we change the notation of section \ref{sec.nonlin} as follows: the power $\mu$ changes into $n$, the new parameter $s$ into $\tau$, and the function $\phi(t)$ into $\Phi(t)$.
} of section \ref{sec.nonlin} as an alternative to the Euler-Duarte-Moreira method\index{Method! Euler-Duarte-Moreira} of integrability of the anharmonic oscillator \cite{Duarte1991} in order to find exact solutions in the modified Brans-Dicke (BD) theory.

Specifically, we consider the equation of motion for the quintessence scalar
field $\psi\left(  t\right)$ with potential function $V(\psi)=\frac
{\psi^{n+1}}{n+1}$, where $n\neq-1$. Then, equation (\ref{bd.14a})
becomes
\begin{equation}
\ddot{\psi}=-\psi^{n}-3\frac{\dot{a}}{a}\dot{\psi} \label{eq.nonl8}
\end{equation}
which is a subcase of (\ref{eq.nonl1}) for $\omega(t)=1$ and $\Phi(t)=-3(\ln a)^{\cdot}$. Replacing in the transformation (\ref{eq.damp0c}), we find that
\begin{equation}
\tau(t)=\int a^{-3}(t)dt,\enskip\bar{\omega}(\tau(t))=a^{6}(t).\label{new0}%
\end{equation}
In the new parameter $\tau$, equation (\ref{eq.nonl8}) reads
\begin{equation}
\psi^{\prime\prime}+a^{6}\psi^{n}=0\label{eq.01}%
\end{equation}
where $\psi^{\prime}=\frac{d\psi\left(  \tau\right)  }{d\tau}$.

The line element of the background space becomes
\begin{equation}
ds^{2}=-a^{6}\left(  \tau\right)  d\tau^{2}+a^{2}\left(  \tau\right)  \left(
dx^{2}+dy^{2}+dz^{2}\right)  \label{eq.02}%
\end{equation}
which means that the rest of the field equations read:
\begin{align}
6\phi\left(  \frac{a^{\prime}}{a}\right)  ^{2} +6 \frac{a^{\prime}}{a}
\phi^{\prime} -\omega_{BD}\frac{\phi^{\prime2}}{\phi} -(\psi^{\prime})^{2}
-\frac{2}{n+1}a^{6}\psi^{n+1}  &  = 2a^{6}\rho_{m}\label{bd.17a}\\
4\phi\frac{a^{\prime\prime}}{a} -10\phi\left(  \frac{a^{\prime}}{a}\right)
^{2} -2 \frac{a^{\prime}}{a} \phi^{\prime} +\omega_{BD}\frac{\left(
\phi^{\prime}\right)  ^{2}}{\phi} +2\phi^{\prime\prime} +(\psi^{\prime})^{2}
-\frac{2}{n+1}a^{6}\psi^{n+1}  &  = -2a^{6}p_{m}\label{bd.17b}\\
6\phi\frac{a^{\prime\prime}}{a} -\omega_{BD}\left[  2\phi^{\prime\prime}%
-\frac{(\phi^{\prime})^{2}}{\phi}\right]  -12\phi\left(  \frac{a^{\prime}}%
{a}\right)  ^{2}  &  =0. \label{bd.17c}%
\end{align}

Next, we apply the results of section \ref{sec.nonlin} for equation (\ref{eq.01}) and we determine for several cases of the parameter $n$ the corresponding QFIs. The resulting QFIs are expressed in terms of the scale factor $a(\tau)$ and the other arbitrary functions of $\tau$ (i.e. $K_{11}(\tau)$, $b_{1}(\tau)$, $b_{2}(\tau)$), which satisfy additional conditions. Solving these conditions, whenever it is possible, we find a scale factor $a(\tau)$ for which equation (\ref{eq.01}) is integrable. Replacing this scale factor in the original equation of motion (\ref{eq.nonl8}) of the quintessence field, we end up with a new integrable second-order ODE for $\psi$, which, most times, can be solved using standard methods (e.g. Lie symmetries) from the symmetries of differential equations. As a final step, for the computation of exact solutions, we replace the solutions $a(t)$ and $\psi(t)$ in the remaining equations (\ref{bd.10}) - (\ref{bd.13}), and we determine the BD scalar field $\phi(t)$ and the BD parameter $\omega_{BD}$.

\subsection{Case $n=0$}

For $n=0$, the associated QFI (\ref{eq.nonl4.1}) becomes
\begin{equation}
I=K_{11}\left(  \psi^{\prime}\right)  ^{2} -K_{11}^{\prime} \psi\psi^{\prime}
+b_{1}(\tau)\psi^{\prime} +c_{3}\psi^{2} +2a^{6}K_{11}\psi-b_{1}^{\prime}%
\psi+\int b_{1}(\tau) a^{6} d\tau\label{new1}%
\end{equation}
where $K_{11}=c_{1}+c_{2}\tau+c_{3}\tau^{2}$, the parameters $c_{1}, c_{2}, c_{3}$ are arbitrary constants and the functions $b_{1}(\tau), a(\tau)$
satisfy the condition
\begin{equation}
b_{1}^{\prime\prime} =12a^{5}a^{\prime} K_{11} +3a^{6}K_{11}^{\prime}. \label{new2}
\end{equation}
We note that for $b_{1}=0$, we find the results of section \ref{sec.case4} below when $n=0$.

\subsection{Case $n=1$}

Using transformation (\ref{new0}), equation $\psi^{\prime\prime}=-a^{6}\psi$
admits the solution
\begin{equation}
\psi(\tau)= \rho(\tau) \left(  A \sin\theta+B\cos\theta\right)  \label{new3}%
\end{equation}
where $\theta= \int\rho^{-2}d\tau$ and the functions $\rho(t(\tau)),
a(t(\tau))$ satisfy the condition
\begin{equation}
\rho^{\prime\prime} +\rho a^{6} -\rho^{-3}=0. \label{new4}%
\end{equation}

\subsection{Case $n=2$}

For $n=2$, we have $K_{11}= a^{-12/5}$ and the associated QFI (\ref{eq.nonl4.3}) becomes
\begin{equation}
I= a^{-12/5} \left(  \psi^{\prime}\right)  ^{2} +\frac{12}{5} a^{-17/5}a^{\prime} \psi\psi^{\prime} +(c_{4}+c_{5}\tau) \psi^{\prime} +\frac{2}{3} a^{18/5} \psi^{3} +\frac{6}{5} \left[  \frac{17}{5} a^{-22/5} (a^{\prime})^{2}-a^{-17/5} a^{\prime\prime}\right]\psi^{2} -c_{5}\psi \label{new5}
\end{equation}
where $c_{4}, c_{5}$ are arbitrary constants and the function $a(t(\tau))\equiv
a(\tau)$ is given by
\begin{equation}
a^{\prime\prime\prime}-\frac{51}{5} \frac{a^{\prime}}{a}a^{\prime\prime}%
+\frac{374}{25} \left(  \frac{a^{\prime}}{a} \right)  ^{2}a^{\prime} +\frac
{5}{6} (c_{4}+c_{5}\tau) a^{47/5}=0. \label{eq.nonl4.4}%
\end{equation}

Substituting the given functions $\omega(t)$ and $\Phi(t)$ in equations
(\ref{eq.nonl4.4.1}) - (\ref{eq.nonl4.4.3}), we find equivalently that
\begin{equation}
a(t)=K_{11}^{-\frac{5}{12}} \label{eq.nonl8.0.6}%
\end{equation}
and
\begin{align}
I  &  =K_{11}^{-3/2}\dot{\psi}^{2}-K_{11}^{-5/2}\dot{K}_{11}\psi\dot{\psi
}+\left(  c_{4}+c_{5}\int K_{11}^{5/4}dt\right)  K_{11}^{-5/4}\dot{\psi}%
+\frac{2}{3}K_{11}^{-3/2}\psi^{3} + \notag \\
& \quad +\left[  \ddot{K}_{11}-\frac{5}{4}\left(  \ln K_{11}\right)  ^{\cdot}%
\dot{K}_{11}\right]  K_{11}^{-5/2}\frac{\psi^{2}}{2}-c_{5}\psi.
\label{eq.nonl8.0.7}%
\end{align}
where the function $K_{11}=K_{11}(t)$ is given by the ODE
\begin{equation}
\dddot{K}_{11}-\frac{15}{4}\left(  \ln K_{11}\right)  ^{\cdot}\ddot{K}%
_{11}-\frac{5}{4}\left(  \ln K_{11}\right)  ^{\cdot\cdot}\dot{K}_{11}%
+\frac{25}{8}\frac{\dot{K}_{11}^{3}}{K_{11}^{2}}=2\left[  c_{4}+c_{5}\int
K_{11}^{5/4}dt\right]  K_{11}^{5/4}. \label{eq.nonl8.0.8}%
\end{equation}

Equation (\ref{eq.nonl8}) becomes $\ddot{\psi}=-\psi^{2}+\frac{5}{4}\left(\ln K_{11}\right)^{\cdot} \dot{\psi}$. We note that for $c_{4}=c_{5}=0$ we
retrieve the results of section \ref{sec.case4} below for $n=2$.

In the special case with $c_{5}=0$, we find for equation (\ref{eq.nonl8.0.8})
the special solution $K_{11}\left(  t\right)  =k_{0}t^{-12}$ with constraint
$c_{4}k_{0}^{1/4}=-192$, where $k_{0}$ is an arbitrary constant. Moreover, from
equation (\ref{eq.nonl8.0.6}), the scale factor is
\begin{equation}
a(t)=K_{11}^{-\frac{5}{12}}=k_{0}^{-5/12}t^{5}.
\end{equation}
Therefore, the Klein-Gordon equation\index{Equation! Klein-Gordon} (\ref{eq.nonl8}) becomes
\begin{equation}
\ddot{\psi}+\frac{15}{t}\dot{\psi}+\psi^{2}=0. \label{eq.KG1}%
\end{equation}
The latter equation can be solved by quadratures. In particular, it admits the Lie symmetries: $\Gamma^{1}= \psi\partial_{\psi}-\frac{t}{2}\partial_{t}$ and $\Gamma^{2}=\left(3\psi t^{2}-48\right)\partial_{\psi} -\frac{t^{3}}{2} \partial_{t}$.
Using the vector field $\Gamma^{1}$, we find the reduced equation $\frac{1}{2}\frac{d f^{2}}{d\lambda} +2\lambda\frac{df}{d\lambda} +12f+\lambda^{2}=0$, where\footnote{It can be easily checked that if we replace $\frac{df}{d\lambda}= \frac{df}{dt} \frac{dt}{d\lambda}$ in the reduced form, we obtain again equation (\ref{eq.KG1}).} $f\left(  \lambda\right)  =t^{3}\dot{\psi}$ and $\lambda=t^{2}\psi$.
The latter ODE is an Abel equation of second type. Moreover, if we assume that $\lambda$ is a constant, $\lambda=\lambda_{0}$, then we find $\psi= \lambda_{0}t^{-2}$. By replacing this $\psi(t)$ in (\ref{eq.KG1}), it follows that $\lambda_{0}=24$. Therefore, we end up with the solution $\psi= \frac{24}{t^{2}}$. Let us now find, for this particular exact solution, the complete solution for the gravitational field equations.

Replacing these results in the rest of the field equations for dust fluid source, that is, $p_{m}=0$ and $\rho_{m}=\rho_{0} a^{-3}$ where $\rho_{0}$ is a constant, the evolution equation for the BD field becomes
\[
\ddot{\phi} +\frac{15}{t}\dot{\phi} = \frac{1}{2\omega_{BD}+3} \left( \rho_{0} a^{-3} -\dot{\psi}^{2} +\frac{4}{3}\psi^{3} \right)
\]
which admits the general solution
\[
\phi(t)= -\frac{1}{2\omega_{BD}+3} \left(  \frac{2016}{5} t^{-4} +\frac{\rho
_{0}k_{0}^{5/4}}{13}t^{-13} \right)  +\frac{k_{1}}{14}t^{-14}
\]
where $k_{1}$ is an arbitrary constant. Finally, by replacing this solution in the constraint (\ref{bd.10}), it follows that (eq. (\ref{bd.11}) is satisfied identically) $\omega_{BD}=-\frac{45}{16}$ and $k_{1}=\rho_{0}=0$.

We conclude that the gravitational field equations for this model, with the use of the QFI for equation (\ref{eq.nonl8}), admit the following exact solution:
\begin{equation}
\omega=-\frac{45}{16}, \enskip a(t)= k_{0}^{-5/12} t^{5}, \enskip \psi(t)=24t^{-2}, \enskip \phi(t)= \frac{768}{5}t^{-4} \label{sol1}%
\end{equation}
with physical quantities $\rho_{m}=p_{m}=0$, $\rho_{\psi}= 5760 t^{-6}$ and $p_{\psi}= -3456 t^{-6}$.

For solution (\ref{sol1}), transformation (\ref{new0}) gives
\begin{equation}
\tau= -\frac{k_{0}^{5/4}}{14}t^{-14} \implies t= \left(  -14 k_{0}^{-5/4}
\right)  ^{-1/14} \tau^{-1/14}. \label{sol1.1}%
\end{equation}
Then, the transformed field equations (\ref{eq.01}) and (\ref{bd.17a}) - (\ref{bd.17c}) admit the solution:
\begin{equation}
\omega=-\frac{45}{16}, \enskip a= k_{0}^{-5/12} (-14 k_{0}^{-5/4})^{-5/14}\tau^{-5/14}, \enskip \psi=24(-14 k_{0}^{-5/4})^{1/7}\tau^{1/7}, \enskip \phi=
\frac{768}{5}(-14 k_{0}^{-5/4})^{2/7}\tau^{2/7}. \label{sol3}%
\end{equation}

\subsection{Case $n\neq-1$}

\label{sec.case4}

In this case, the associated QFI (\ref{eq.nonl5}) becomes
\begin{equation}
I=(c_{1}+c_{2}\tau+c_{3}\tau^{2})\left(  \psi^{\prime} \right)  ^{2}%
-(c_{2}+2c_{3}\tau) \psi\psi^{\prime} +\frac{2}{n+1}(c_{1} +c_{2}\tau
+c_{3}\tau^{2})^{-\frac{n+1}{2}}\psi^{n+1} +c_{3}\psi^{2} \label{new6}%
\end{equation}
and the function
\begin{equation}
a(\tau) =(c_{1}+c_{2}\tau+c_{3}\tau^{2})^{-\frac{n+3}{12}}. \label{new7}%
\end{equation}

Substituting the given functions $\omega(t)$ and $\Phi(t)$ in the relation (\ref{eq.nonl6.1}), we find equivalently that
\begin{equation}
a^{6}(t)= \left[  c_{1} +c_{2}\int a^{-3}(t) dt +c_{3}\left(  \int a^{-3}(t)
dt\right)  ^{2} \right]  ^{-\frac{n+3}{2}} \label{eq.nonl8.1}%
\end{equation}
and the associated QFI (\ref{eq.nonl6.2}) becomes
\begin{align}
I  &  = \left[  c_{1} +c_{2}\int a^{-3}(t) dt +c_{3}\left(  \int a^{-3}(t)dt\right)  ^{2} \right]  a^{6}(t) \dot{\psi}^{2}- \left[  c_{2} +2c_{3}\int
a^{-3}(t) dt \right]  a^{3}(t) \psi\dot{\psi}+ \notag \\
& \quad +\frac{2}{n+1} \left[  c_{1} +c_{2}\int a^{-3}(t) dt +c_{3}\left(  \int
a^{-3}(t) dt\right)  ^{2} \right]  ^{-\frac{n+1}{2}} \psi^{n+1} + c_{3}\psi^{2}. \label{eq.nonl8.2}
\end{align}

We consider the following special cases for which equation (\ref{eq.nonl8}) admits a closed-form solution for $n\neq-3,1$. In the case $n=-3$, the spacetime is that of Minkowski. Hence, we omit the analysis.

\subsubsection{Subcase $\left\vert \tau\right\vert <<1$}

For small values of $\left\vert \tau\right\vert $ (i.e. $c_{1}=c_{3}=0$) the scale
factor (\ref{new7}) is approximated as $a\left(  \tau\right)  \simeq
\tau^{-\frac{n+3}{12}}$; therefore, it follows
\begin{equation}
a(t)=B_{0} (t-t_{0})^{\frac{n+3}{3(n-1)}} \label{eq.nonl9}%
\end{equation}
where $B_{0}= \left[  -\frac{c_{2}(n-1)}{4} \right]^{\frac{n+3}{3(n-1)}}$
and $t_{0}$ is an arbitrary constant.

For this asymptotic solution, the equation of motion (\ref{eq.nonl8}) for the
second field $\psi$ becomes
\begin{equation}
\ddot{\psi}=-\psi^{n}-\frac{n+3}{n-1}\frac{1}{t-t_{0}} \dot{\psi}.
\label{eq.nonl9.1}%
\end{equation}
For the latter equation, the QFI (\ref{eq.nonl8.2}) is
\begin{equation}
I=\left[  -\frac{c_{2}(n-1)}{4}\right]  ^{\frac{2(n+1)}{n-1}}(t-t_{0}%
)^{\frac{2(n+1)}{n-1}}\left(  \dot{\psi}^{2}+\frac{2}{n+1}\psi^{n+1}\right)
-c_{2}\left[  -\frac{c_{2}(n-1)}{4}\right]  ^{\frac{n+3}{n-1}}(t-t_{0}%
)^{\frac{n+3}{n-1}}\psi\dot{\psi}. \label{eq.nonl10}%
\end{equation}
This QFI, which corresponds to the scale factor (\ref{eq.nonl9}), together with the results of
the cases $n=0,1,2$ produces new solutions $\psi(t)$ which  have not found before.

Furthermore, for the scale factor (\ref{eq.nonl9}) the closed-form solution of (\ref{eq.nonl9.1}) is found to be
\begin{equation}
\psi\left(  t\right)  =\psi_{0}(t-t_{0})^{-\frac{2}{n-1}}~,~\psi_{0}=\left(
\frac{2}{n-1}\right)  ^{\frac{2}{n-1}} \label{eq.nonl8.6.1}%
\end{equation}
whereas for the BD field $\phi\left(  t\right)$ it follows that $n=3,~\phi(t)=\frac{\phi_{0}}{(t-t_{0})^{2}}$ and $\omega_{BD} =-\frac{3}{2}$. However, this value for the BD parameter $\omega_{BD}$ is not physically accepted. Hence, we do not have any closed-form solution. In all discussion above, we have considered $\rho_{m}=0$.

\subsubsection{Subcase $\left\vert \tau\right\vert \gg1$}

For large values of $\tau$ (i.e. $c_{1}=c_{2}=0$), the scale factor
(\ref{new7}) is approximated as $a(\tau)\simeq\tau^{-\frac{n+3}{6}}$.
Therefore, the original equation (\ref{eq.nonl8.1}) becomes
\begin{equation}
a^{-\frac{6}{n+3}}=c_{3}^{\frac{1}{2}}\int a^{-3}dt \label{eq.nonl8.3}%
\end{equation}
which implies (see eq. (31) of \cite{Mukherjee2019})
\begin{equation}
a(t)=A_{0}(t-t_{0})^{\frac{n+3}{3(n+1)}} \label{eq.nonl8.4}%
\end{equation}
where $A_{0}=\left[ -\frac{\sqrt{c_{3}}(n+1)}{2} \right]  ^{\frac{n+3}{3(n+1)}}$ and $t_{0}$ is an arbitrary constant. The scale factor
(\ref{eq.nonl8.4}) describes a scaling solution, where the effective
cosmological fluid is that of an ideal gas with effective parameter\footnote{
It must hold that $a \propto (t -t_{0})^{\frac{2}{3\left( 1 +w_{eff} \right)}}$. Therefore, by equating with the power of (\ref{eq.nonl8.4}), we find that
\[
\frac{2}{3\left( 1 +w_{eff} \right)}= \frac{n+3}{3(n+1)} \implies w_{eff}=\frac{n-1}{n+3}.
\]
} for the equation of state $w_{eff}=\frac{n-1}{n+3}.$ Furthermore, for
$-3<n<-1$ and $-1<n<0$, the scale factor describes an accelerated universe. For
$-1<n<0$, we have $-1<w_{eff}<-\frac{1}{3}$; while for $-3<n<-1$, the effective parameter crosses the phantom divide line, that is, $w_{eff}<-1$.

For this asymptotic solution, the equation of motion (\ref{eq.nonl8}) for the scalar field $\psi$ becomes%
\begin{equation}
\ddot{\psi}= -\psi^{n} -\frac{n+3}{n+1} \frac{1}{t-t_{0}} \dot{\psi}
\label{eq.nonl8.3.1}%
\end{equation}
and the corresponding QFI (\ref{eq.nonl8.2}) is written as
\begin{equation}
I= c_{3} \left[  \frac{(n+1)(t-t_{0})}{2} \dot{\psi} +\psi\right]  ^{2} +
\frac{c_{3}(n+1)}{2} (t-t_{0})^{2} \psi^{n+1} \label{eq.nonl8.5}%
\end{equation}
where $t\neq t_{0}$.

However, the system admits the closed-form solution (see eq. (32) of \cite{Mukherjee2019})
\begin{equation}
\psi\left(  t\right)  =\psi_{0}\left(  t-t_{0}\right)  ^{-\frac{2}{n-1}%
}\label{eq.nonl8.5.1}%
\end{equation}
where $\psi_{0}$ is given by the expression $\psi_{0}=(-2)^{\frac{3}{n-1}}\left[  \left(  n+1\right)  \left(  n-1\right)^{2}\right]^{\frac{1}{1-n}}$. Replacing in the remaining equations (\ref{bd.10}) - (\ref{bd.13}) for the
BD field, we find
\begin{equation}
\phi\left(  t\right) =\phi_{0}(t-t_{0})^{-\frac{4}{n-1}} \label{eq.nonl8.5.2}
\end{equation}
where
\begin{align}
\phi_{0} &= \frac{(n-1)^{\frac{4}{1-n}}}{2(n+3)(2\omega_{BD}+3)}\left[
(-2)^{\frac{3(n+1)}{n-1}}(n+1)^{\frac{n+1}{1-n}} -(-2)^{\frac{6}{n-1}}(n+1)^{\frac{n-3}{n-1}}\right]  \label{eq.nonl8.5.3}\\
\omega_{BD} &= \frac{b_{1}-3b_{2}}{1+2b_{2}}. \label{eq.nonl8.5.4}
\end{align}
We note that we have assumed that there is not any other matter source, i.e.
$\rho_{m}=0$. The constants $b_{1}$ and $b_{2}$ are given by the relations:
\begin{align}
b_{1} &  =\frac{(n+3)(n-1)}{2(n+1)}\left[  \frac{(n+3)(n-1)}{12(n+1)}%
-1\right] \label{eq.nonl8.5.5}\\
b_{2} &  =\frac{n+3}{4}\cdot\frac{2(-2)^{\frac{6}{n-1}}(n+1)^{\frac{2}{1-n}} +(-2)^{\frac{3(n+1)}{n-1}}(n+1)^{\frac{2n}{1-n}}}{(-2)^{\frac{3(n +1)}{n-1}}(n+1)^{\frac{n+1}{1-n}} -(-2)^{\frac{6}{n-1}}(n +1)^{\frac{n-3}{n-1}}}. \label{eq.nonl8.5.6}%
\end{align}

In the following section, we perform a detailed study on the stability (see chapter \ref{ch.stability}) of the latter closed-form solutions.

\section{Stability of scaling solutions}

\label{sec.stab}

The analysis of the stability properties (see chapter \ref{ch.stability}) of the exact solutions provides us with important
information about the evolution of the background space on the asymptotic solutions.
In particular, we can infer if an exact solution is stable, which can be seen as a future
attractor for the original dynamical system. On the other hand, in the cases of unstable
solutions, the behavior of the asymptotic solution and its dynamics give us results for the
curvature and the dynamics of the metric space.

According to the methods in \cite{Ratra, Liddle 1998, Uzan 1999}, we consider the second order ODE
\begin{equation}
F(\ddot\psi, \dot\psi, \psi)=0 \label{bst.1}
\end{equation}
where $\psi(t)$ is a smooth function, which admits a singular power law solution
\begin{equation}
\psi_c(t)= \psi_0 t^\beta \label{bst.2}
\end{equation}
where $\phi_{0}$ and $\beta$ are arbitrary constants. To examine the stability of the solution $\psi_c$, we introduce the logarithmic time $T$ through $t= e^{T}$. Using this transformation, we compute
\begin{equation}
\label{eq.82}
\dot{\psi}= e^{-T}\psi', \enskip \ddot{\psi}= e^{-2 T} (\psi''-\psi'), \enskip \psi_{c}\left(t(T)\right)= \psi_{0}e^{\beta T}, \enskip \frac{\psi_c'}{\psi_c}=\beta
\end{equation}
where in this discussion $\psi'\equiv \frac{d \psi}{dT}$.

We introduce, also, the dimensionless function
\begin{equation}
u(T)= \frac{\psi(T)}{\psi_c(T)} \label{bst.3}
\end{equation}
and the stability analysis is translated into the analysis of the stability of the equilibrium point $u=1$ of a transformed dynamical system with $\psi(T)= \psi_{c}(T) u(T)$.

In this section, we use a similar procedure for analyzing the stability of the scaling solutions obtained in section \ref{sec.case4}.

\subsection{Case $\left\vert \tau\right\vert \gg 1$}

For the analysis of the solution \eqref{eq.nonl8.5.1} of equation \eqref{eq.nonl8.3.1}, we set $t_0=0$ by a time shift. Using \eqref{eq.82}, we have:
\begin{equation}
\label{eq.83}
\psi''= -\frac{2}{n+1}\psi' -e^{2T} \psi^n
\end{equation}
and
\begin{equation}
\psi_{c}\left(T\right)  =\psi_{0} e^{-\frac{2T}{n-1}} \label{eq.83.1}
\end{equation}
where $\psi=\psi(t(T))$.

Introducing the parameter $p= -\frac{2}{n-1}$, we have:
$u''(T
   )= \frac{p^2 e^{-p T } \psi (T
   )}{\psi_0}+\frac{e^{-p T } \psi ''(T )}{\psi_0}-\frac{2 p e^{-p T } \psi '(T )}{\psi_0}$, \newline $u'(T )= \frac{e^{-p T } \psi '(T )}{\psi_0}-\frac{p e^{-p
   T } \psi (T )}{\psi_0}$ and $u(T )=\frac{e^{-p T } \psi
   (T )}{\psi_0}$. Inverting these relations, we find:
\begin{equation*}
\psi ''(T )= \psi_0 e^{p T } \left[ p^2
   u(T )+2 p u'(T )+u''(T )\right], \enskip \psi '(T )= \psi_0 e^{p T } \left[ p u(T )+u'(T ) \right], \enskip \psi (T )= \psi_0 e^{p T } u(T ).
\end{equation*}

Then, equation \eqref{eq.83} becomes
\begin{equation}
u''= -2\left(p +\frac{1}{n+1}\right)u' -\psi_0^{n-1} e^{\left[ (n-1)p +2 \right]T} u^n -\frac{p (np+p+2)}{n+1}u. \label{eq.83.2}
\end{equation}
Substituting $p= -\frac{2}{n-1}$ and $\psi_{0}= (-2)^{\frac{3}{n-1}} \left[ \left(n+1\right)\left(n-1\right)^{2} \right]^{\frac{1}{1-n}}$ in (\ref{eq.83.2}), we obtain the second order ODE
\begin{equation}
u''= \frac{2(n+3)}{n^2-1}u' +\frac{8}{(n-1)^2(n+1)} \left( u^{n} -u \right). \label{eq.83.3}
\end{equation}
Introducing the variables $x= u(T)$ and $y=u'(T)$, we obtain the autonomous system:
\begin{eqnarray}
x'&=& y \label{eq.93}\\
y'&=&\frac{2(n+3)}{n^2-1}y +\frac{8}{(n-1)^2 (n+1)}\left( x^{n} -x \right). \label{eq.94}
\end{eqnarray}
The scaling solution \eqref{eq.nonl8.5.1} is transformed into the equilibrium point $P:= (x,y)=(1,0)$. The system \eqref{eq.93} - \eqref{eq.94} admits two additional equilibrium points: a) The trivial solution $O:= (x,y)=(0,0)$, and b) when $n$ is odd, the symmetrical point of $P$, that is, $\bar{P}:= (x,y)=(-1,0)$. The linearization matrix of the system \eqref{eq.93} - \eqref{eq.94} is $J(x,y)=\left(
\begin{array}{cc}
 0 & 1 \\
 \frac{8 \left(n x^{n-1}-1\right)}{(n-1)^2 (n+1)} & \frac{2(n+3)}{n^2-1}
\end{array}
\right)$.

For $n>1$, $J(0,0)$ is real-valued with eigenvalues $\left\{\frac{4}{n^2-1},\frac{2}{n-1}\right\}$. Then, the origin is an unstable equilibrium point.

The eigenvalues of $J(1,0)$ are $\left\{ -\frac{2}{n+1}, \frac{4}{n-1} \right\}$. Therefore, $P(1,0)$ is a sink for $-1<n<1$ and a saddle for $n<-1$ or $n>1$.

If $n$ is an odd number, say $n=2k+1$ with $k\in \mathbb{Z}$, the eigenvalues of $J(-1,0)$ are $\left\{ -\frac{1}{k+1}, \frac{2}{k} \right\}$ and, when they exist, $\bar{P}$ is a saddle.

\begin{figure}[H]
\begin{center}
\includegraphics[scale=0.7]{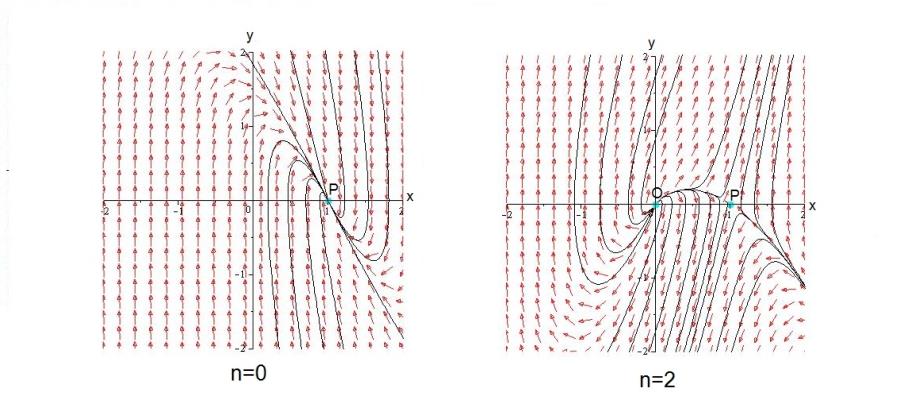}
\caption{\label{fig.n0} The phase portrait of the system \eqref{eq.93} - \eqref{eq.94} for $n=0$ and $n=2$. For $n=0$ the only equilibrium point is the sink $P(1,0)$, whereas for $n=2$ the point $O(0,0)$ is a source and the point $P(1,0)$ is a saddle.}
\end{center}
\end{figure}

\begin{figure}[H]
\begin{center}
\includegraphics[scale=0.6]{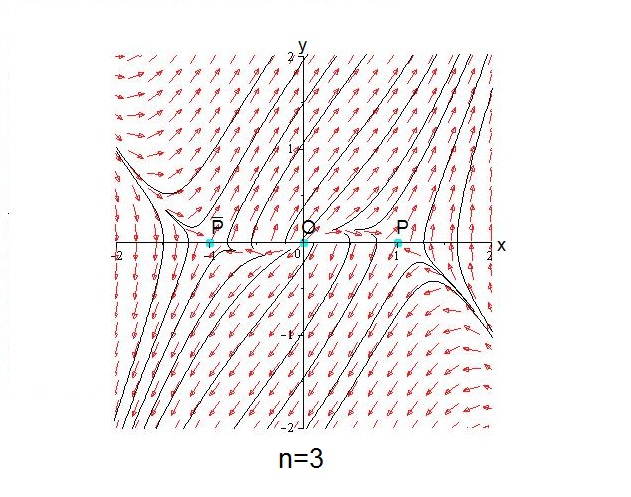}
\caption{\label{fig.n3} The phase portrait of the system \eqref{eq.93} - \eqref{eq.94} for $n=3$. In this case, the origin $O(0,0)$ is a source and the points $\bar{P}(-1,0)$, $P(1,0)$ are saddles.}
\end{center}
\end{figure}

In Figures \ref{fig.n0} and \ref{fig.n3}, we give the phase portraits of the system \eqref{eq.93} - \eqref{eq.94} for some characteristic values of $n$.

\section{Conclusions}

We considered a cosmological model consisted by a Brans-Dicke scalar field and a minimally coupled quintessence scalar field in a spatially flat FLRW background space. For this cosmological model, the gravitational field equations define a Hamiltonian system of six degrees of freedom, whose dynamical variables are the scale factor and the two scalar fields.

For a power law potential function of the quintessence scalar field, we found QFIs for the equation of motion of this field. Using these QFIs, we determined exact solutions for the field equations. In particular, we found scaling solutions for the scale factor which describe ideal gas solutions. Moreover, we were able to recover previous published results in the literature and also to find new QFIs.

Using well-known methods, we studied the stability of the scaling solution \eqref{eq.nonl8.5.1}. We showed that this solution is transformed to the equilibrium point $P:= (x,y)=(1,0)$, which is a sink for $-1<n<1$ and a saddle for $n<-1$ or $n>1$. Furthermore, we found that the associated dynamical system admits two additional equilibrium points: a) The trivial solution $O:= (x,y)=(0,0)$ which is a source for $n>1$, and b) the symmetrical point of $P$, $\bar{P}:= (x,y)=(-1,0)$, which is a saddle when $n$ is odd with $n \neq -1,1$.

Until now, the majority of this kind of studies, have been done mainly with the application of variational symmetries. Our approach is more general and does not required the existence of a point-like Lagrangian, that is, of a minisuperspace description. Therefore, that generic approach can be applied in other gravitational models without minisuperspace; for example, such models are the Class B Bianchi spacetimes.

%% file: td_central_pots.tex
\chapter{Integrable time-dependent central potentials}

\label{ch.td.central.pots}

\section{Introduction}

\label{sec.central.intro}

As we have seen in previous chapters, there are various methods for the determination of the FIs of the time-dependent conservative systems
\begin{equation}
\ddot{q}^{a}= -U^{,a}(t,q)  \label{NP.0}
\end{equation}
where $U(t,q)$ denotes the potential of the system and a general flat kinetic metric $\gamma_{ab}(q)$ is used for raising/lowering the indices. These methods include (see sections \ref{sec.methods.determine.FIs} and \ref{sec.timedep.into}) e.g.: Noether's theorem, the Lie theory of extended groups, Ermakov's method, theory of canonical transformations, Inverse Noether Theorem, and the direct method (see e.g. \cite{ Leach 1985, Djukic 1975, Katzin 1973, Fokas 1979, Katzin 1981, Horwood 2007, Karlovini 2000, Katzin 1982, Ermakov, Katzin 1977, Da Silva 1974, Leach 1981A}).

The case of QFIs of the general form\footnote{
The case of LFIs also included for $K_{ab}=0$.
} (\ref{FL.5}) is the one that has been considered mostly in the literature. There are two major methods which have been used in order to determine the QFIs (\ref{FL.5}): a. The method of the Inverse Noether Theorem (see Theorem \ref{Inverse Noether Theorem}), and b. The direct method. Both methods have been discussed in section \ref{sec.methods.determine.FIs}; however, it is necessary to recall some key points, essential for the current discussion.

Concerning method a., one assumes that the dynamical equations (\ref{NP.0}) possess a regular Lagrangian $L(t,q,\dot{q})$ and uses the Inverse Noether Theorem \ref{Inverse Noether Theorem} to associate to each QFI of the form (\ref{FL.5}) the generalized gauged Noether symmetry
\begin{equation}
\xi =0,\enskip\eta _{a}=-2K_{ab}\dot{q}^{b}-K_{a},\enskip f=-K_{ab}\dot{q}^{a}\dot{q}^{b}+K. \label{NP.2}
\end{equation}
From (\ref{NP.2}), it follows that the generator $\eta^{a}$ is a linear function of the velocities, that is, $\eta^{a} =A_{b}^{a}(t,q)\dot{q}^{b} +B^{a}(t,q)$ where $A^{a}_{b}(t,q)$ and $B^{a}(t,q)$ are tensor quantities. Subsequently, one substitutes $\eta^{a}$ and $L$ in the generalized Killing equations (see section \ref{con.mot.subsec.killing.2}) and finds a system of PDEs whose solution determines the time-dependent potentials that admit QFIs.

In the present chapter, we address the problem of finding all time-dependent central potentials\index{Potential! time-dependent central} of regular conservative Newtonian systems that admit QFIs of the form (\ref{FL.5}); therefore, all such potentials that are (Liouville) integrable. Previous work on this problem has been done in \cite{Leach 1985}, where the method of the Inverse Noether Theorem was used. It was concluded that there are three classes of time-dependent central potentials, which were classified as Cases I, II and III, given by the following expressions:
\[
V_{I}(t,r)= \frac{1}{2}\lambda (t)r^{2}, \enskip V_{II}(t,r)= -\frac{\ddot{\phi}}{2\phi } r^{2} -\frac{\mu_{0}}{\phi }r^{-1} \enskip \text{and} \enskip V_{III}(t,r)= -\frac{\ddot{\phi}}{2\phi }r^{2} +\phi ^{-2}F\left( \frac{r}{\phi }\right).
\]
Here $\phi(t)$ is an arbitrary smooth function, the frequency $\lambda(t)$ of the time-dependent oscillator may be either the function $\lambda_{1}(t)= \frac{K}{\phi^{4}} -\frac{\ddot{\phi}}{\phi}$ or the function $\lambda_{2}(t)= -\frac{\ddot{\phi}}{\phi}$, $\mu_{0}$ and $K$ are arbitrary constants, and $F$ is an arbitrary smooth function of its argument. For each potential, the corresponding QFI was provided. We note that, in fact, there is only one Case because Cases I and II are subcases of Case III for $F=0$, $F=\frac{K}{2}\phi^{-2}r^{2}$, and $F=-\mu_{0} \frac{\phi}{r}$. As it will be shown, this result is a partial answer to the problem. It is important also to note that in \cite{Leach 1985} the dynamical system was considered to be 2d with variables $r, \theta$.

In a different approach \cite{LewLea 1982}, the authors considered 1d Hamiltonian systems of the form $H=\frac{1}{2}p^{2} +U(t,q)$ and, using the direct method for QFIs of the form (\ref{FL.5}), determined all the time-dependent potentials $U(t,q)$ which admit QFIs\footnote{Obviously, in this case, central motion makes no sense.}. They concluded that these potentials are
\begin{equation}
U(t,r)= -\frac{\ddot{\rho}}{2\rho }q^{2}+\left( \alpha \frac{\ddot{\rho}}{\rho } -\ddot{\alpha}\right) q+\frac{1}{\rho ^{2}}G\left( \frac{q-\alpha }{\rho }\right)   \label{NP.3}
\end{equation}%
where $\rho (t)$, $\alpha (t)$ and $G$ are arbitrary smooth functions of their arguments. The associated QFI is
\begin{equation}
I=\frac{1}{2}\left[ \rho (\dot{\rho}-\dot{\alpha}) -\dot{\rho}(q-\alpha )\right] ^{2} +G\left( \frac{q-\alpha}{\rho }\right). \label{NP.11}
\end{equation}
Equivalently, the potential (\ref{NP.3}) may be written as
\begin{equation}
U(t,r)=\frac{1}{2}\Omega ^{2}(t)q^{2}-F_{1}(t)q+\frac{1}{\rho ^{2}}\widetilde{G}\left( \frac{q-\alpha }{\rho }\right) \label{NP.12}
\end{equation}
where $\widetilde{G}$ is an arbitrary smooth function of its argument and the functions $\Omega(t), F_{1}(t), \rho(t), \alpha(t)$ satisfy the conditions:
\begin{eqnarray}
\ddot{\rho} +\Omega^{2}(t)\rho -\frac{k}{\rho^{3}} &=& 0 \label{NP.13.1} \\
\ddot{\alpha} +\Omega^{2}(t)\alpha &=& F_{1}(t) \label{NP.13.2}
\end{eqnarray}
where $k$ is an arbitrary constant. In this notation, the associated QFI (\ref{NP.11}) becomes
\begin{equation}
I= \frac{1}{2} \left[ \rho (\dot{\rho} -\dot{\alpha}) -\dot{\rho} (q-\alpha) \right]^{2} +\frac{k}{2}\left( \frac{q-\alpha}{\rho} \right)^{2} +\widetilde{G}\left( \frac{q-\alpha}{\rho} \right). \label{NP.14}
\end{equation}

In the following sections, we consider the problem of central motion and determine the time-dependent LFIs/QFIs using the direct method of \cite{LewLea 1982}. In order to do this, we reduce the degrees of freedom to one by means of the LFI of angular momentum. Finally, we collect our results in Theorem \ref{theorem.central} and consider various applications.

\section{The integrable time-dependent central potentials $V(t,r)$}

\label{sec.cenpot}

The characteristic property of Newtonian central motion\index{Central motion} is that the angular momentum is conserved and the motion takes place on a plane normal to the angular momentum. On that plane, we assume polar coordinates $(r,\theta)$ so that the Lagrangian becomes
\begin{equation}
L= \frac{1}{2} \left( \dot{r}^{2} +r^{2}\dot{\theta}^{2} \right) -V(t,r). \label{eq.cp3}
\end{equation}

The E-L equations are:
\begin{eqnarray}
\ddot{r} &=& r\dot{\theta}^{2} -\frac{\partial V}{\partial r} \label{eq.cp4a} \\
L_{3}&=& r^{2}\dot{\theta} \label{eq.cp4b}
\end{eqnarray}
where $L_{3}= r^{2}\dot{\theta}$ is the LFI of the angular momentum. Replacing $\dot{\theta}$ in (\ref{eq.cp4a}), we find the time-dependent second order ODE in the variable $r(t)$:
\begin{equation}
\ddot{r}= \frac{L_{3}^{2}}{r^{3}} -\frac{\partial V(t,r)}{\partial r}. \label{eq.cp6}
\end{equation}
Solving equation (\ref{eq.cp6}), one finds a solution for $r(t)$ which when replaced in (\ref{eq.cp4b}), by integration, gives the $\theta(t)$.

In what follows, we study the integrability of the time-dependent system
\begin{equation}
\ddot{r}= -\frac{\partial U}{\partial r} \label{eq.nfx2}
\end{equation}
where the time-dependent potential
\begin{equation}
U(t,r)= \frac{L_{3}^{2}}{2r^{2}} +V(t,r). \label{eq.nfx3}
\end{equation}
so that
\begin{equation}
V(t,r)= U(t,r) - \frac{L_{3}^{2}}{2r^{2}}. \label{eq.nfx4}
\end{equation}
The problem of finding all the integrable potentials $U(t,r)$ has been solved in sections II and III of \cite{LewLea 1982} using the direct method. In the following, we update the approach of \cite{LewLea 1982}. Using (\ref{FL.5}) and the dynamical equations (\ref{NP.0}) to replace the term $\ddot{q}^{a}$ whenever it appears, condition $\frac{dI}{dt}=0$ leads to the following system of PDEs:
\begin{eqnarray}
K_{(ab,c)} &=&0  \label{eq.cp8a} \\
K_{ab,t} +K_{(a,b)} &=&0  \label{eq.cp8b} \\
K_{a,t} +K_{,a} -2K_{ab}U^{,b} &=&0  \label{eq.cp8c} \\
K_{,t} -K_{a}U^{,a} &=&0  \label{eq.cp8d} \\
K_{a,tt} -2\left(K_{ab}U^{,b}\right)_{,t} +\left( K_{b}U^{,b} \right)_{,a} &=& 0 \label{eq.cp8e} \\
2\left( K_{[a|c|}U^{,c} \right)_{,b]} -K_{[a,b],t} &=& 0. \label{eq.cp8f}
\end{eqnarray}

We note that the PDEs (\ref{eq.cp8a}) and (\ref{eq.cp8b}) are purely geometric because they do not involve the potential $U(t,q)$. The PDE (\ref{eq.cp8a}) implies that $K_{ab}$ is a KT of order two (possibly zero) of $\gamma_{ab}$. Moreover, the last two PDEs (\ref{eq.cp8e}) and (\ref{eq.cp8f}) express the integrability conditions for the scalar $K$.

Because the considered dynamical system (\ref{eq.nfx2}) is 1d, the variable $q^{1}=r$ and the KT $K_{ab}$ is of the form $K_{11}=g_{1}(t)$. Therefore, the system of PDEs (\ref{eq.cp8a}) - (\ref{eq.cp8f}) becomes:
\begin{eqnarray}
K_{11}&=& g_{1}(t) \label{eq.cp9.0} \\
K_{1}(t,r)&=& -\dot{g}_{1}r +g_{2}(t) \label{eq.cp9.1} \\
\frac{\partial K}{\partial r}&=& 2g_{1}\frac{\partial U}{\partial r} +\ddot{g}_{1}r -\dot{g}_{2} \label{eq.cp9.2} \\
\frac{\partial K}{\partial t}&=& \left(g_{2} -\dot{g}_{1} r \right) \frac{\partial U}{\partial r} \label{eq.cp9.3} \\
0&=& \left(\dot{g}_{1} r - g_{2}\right) \frac{\partial^{2} U}{\partial r^{2}} +2g_{1}\frac{\partial^{2}U}{\partial t\partial r} +3\dot{g}_{1}\frac{\partial U}{\partial r} +\dddot{g}_{1}r -\ddot{g}_{2} \label{eq.cp9.4}
\end{eqnarray}
where $g_{1}(t)$ and $g_{2}(t)$ are arbitrary smooth functions. The integrability condition (\ref{eq.cp9.4}) is ignored because it is satisfied identically due to the PDEs (\ref{eq.cp9.2}) and (\ref{eq.cp9.3}). Replacing $K_{1}$ and $K_{11}$ given by (\ref{eq.cp9.0}) and (\ref{eq.cp9.1}), respectively, in the expression (\ref{FL.5}) of the associated QFI, we find
\begin{equation}
I= g_{1}\dot{r}^{2} +\left( g_{2} -\dot{g}_{1}r \right)\dot{r} +K(t,r). \label{eq.cp9.5}
\end{equation}

There remains the scalar $K$ and the corresponding time-dependent potential $U$ which shall be determined from the PDEs (\ref{eq.cp9.2}) and (\ref{eq.cp9.3}). There are two cases to consider: a) $g_{1}(t)=0$ which provides the LFIs, and b) $g_{1}(t)\neq0$ which provides the QFIs.

\subsection{Case $g_{1}(t)=0$ (LFIs)}

\label{sec.fx1}

In this case, $K_{11}=0$, $K_{1}=g_{2}(t)\neq0$, and the PDEs (\ref{eq.cp9.2}) and (\ref{eq.cp9.3}) become:
\begin{eqnarray}
\frac{\partial K}{\partial r} &=& -\dot{g}_{2} \label{eq.nfx5.1} \\
\frac{\partial K}{\partial t}&=& g_{2} \frac{\partial U}{\partial r}. \label{eq.nfx5.2}
\end{eqnarray}
Integrating the PDE (\ref{eq.nfx5.1}), we find that $K(t,r)=-\dot{g}_{2}r +g(t)$ where $g(t)$ is an arbitrary function. Replacing this function in the PDE (\ref{eq.nfx5.2}), we find the potential (see eq. (2.9) in \cite{LewLea 1982})
\begin{equation}
U(t,r)= -\frac{\ddot{g}_{2}}{2g_{2}}r^{2} +\frac{\dot{g}}{g_{2}}r. \label{eq.nfx6.0}
\end{equation}
Substituting $U(t,r)$ in (\ref{eq.nfx4}), we find the integrable time-dependent central potential
\begin{equation}
V(t,r)= -\frac{\ddot{g}_{2}}{2g_{2}}r^{2} +\frac{\dot{g}}{g_{2}}r -\frac{L_{3}^{2}}{2r^{2}}. \label{eq.nfx6}
\end{equation}
This potential does not belong to any of the three classes found in \cite{Leach 1985} due to the additional term $\frac{\dot{g}}{g_{2}}r$. We observe that it is the sum of:  \newline
a. A repulsive time-dependent oscillator (term $-\frac{\ddot{g}_{2}}{2g_{2}}r^{2})$.\newline
b. An attractive Newton-Cotes potential (term $-\frac{L_{3}^{2}}{2r^{2}})$. \newline
c. A pure time-dependent central force (term $\frac{\dot{g}}{g_{2}}r$).

The associated LFI is
\begin{equation}
I= g_{2}\dot{r} -\dot{g}_{2}r +g \label{eq.nfx7}
\end{equation}
which coincides with the LFI (2.11) of \cite{LewLea 1982}.

\subsection{Case $g_{1}(t)\neq0$ (QFIs)}

\label{sec.fx2}

In this case, equation (\ref{eq.cp9.2}) can be integrated and gives the potential (see eq. (3.8) in \cite{LewLea 1982})
\begin{equation}
U= \frac{K}{2g_{1}} -\frac{\ddot{g}_{1}}{4g_{1}}r^{2} +\frac{\dot{g}_{2}}{2g_{1}}r +g(t) \label{eq.nfx8}
\end{equation}
where $g(t)$ is an arbitrary function. Replacing this potential in the PDE (\ref{eq.cp9.3}), we find the function (see eq. (3.14) in \cite{LewLea 1982}) $K= F\left( g_{1}^{-1/2}r +\frac{1}{2} \int g_{1}^{-3/2}g_{2}dt \right) +\frac{1}{4g_{1}} \left( \dot{g}_{1}r -g_{2} \right)^{2}$ where $F$ is an arbitrary smooth function of its argument.

Substituting the function $K$ in (\ref{eq.nfx8}), we find the potential\footnote{We choose the function $g= -\frac{g_{2}^{2}}{8 g_{1}^{2}}$ so as not to have an additive function of $t$ in the potential.}
\begin{equation}
U= \left[ \frac{1}{8} \left(\frac{\dot{g}_{1}}{g_{1}} \right)^{2} -\frac{\ddot{g}_{1}}{4g_{1}} \right] r^{2} +\frac{1}{2g_{1}} \left( \dot{g}_{2} -g_{2} \frac{\dot{g}_{1}}{2g_{1}} \right)r +\frac{1}{2g_{1}} F\left( g_{1}^{-1/2}r +\frac{1}{2} \int g_{1}^{-3/2}g_{2}dt \right).
\label{eq.fx8.1}
\end{equation}
Replacing $U$ in (\ref{eq.nfx4}), we obtain the integrable time-dependent central potential
\begin{equation}
V(t,r)= \left[ \frac{1}{8} \left(\frac{\dot{g}_{1}}{g_{1}} \right)^{2} -\frac{\ddot{g}_{1}}{4g_{1}} \right] r^{2} +\frac{1}{2g_{1}} \left( \dot{g}_{2} -g_{2} \frac{\dot{g}_{1}}{2g_{1}} \right)r +\frac{1}{2g_{1}} F\left( g_{1}^{-1/2}r +\frac{1}{2} \int g_{1}^{-3/2}g_{2}dt \right) -\frac{L_{3}^{2}}{2r^{2}}. \label{eq.nfx9}
\end{equation}

The associated QFI (\ref{eq.cp9.5}) is
\begin{equation}
I= g_{1}\dot{r}^{2} +(g_{2} -\dot{g}_{1}r)\dot{r} +F\left( g_{1}^{-1/2}r +\frac{1}{2} \int g_{1}^{-3/2}g_{2}dt \right) +\frac{1}{4g_{1}} \left( \dot{g}_{1}r -g_{2} \right)^{2}. \label{eq.nfx10}
\end{equation}

We note that for $g_{1}=\frac{\phi^{2}}{2}$, $g_{2}=0$ and $F= \bar{F} +\frac{L_{3}^{2}\phi^{2}}{2r^{2}}$, where $F$ and  $\bar{F}$ are functions of the same argument, we recover the Case III potentials of \cite{Leach 1985} as a special case of the potential (\ref{eq.nfx9}).

Using the Inverse Noether Theorem, we have that for the QFI (\ref{eq.nfx10}) the associated gauged Noether symmetry (\ref{NP.2}) is:
\begin{eqnarray}
\eta_{1}&=& -2g_{1}\dot{r} +\dot{g}_{1}r -g_{2} \label{eq.nfx10.1} \\
f&=& -g_{1}\dot{r}^{2} +F\left( g_{1}^{-1/2}r +\frac{1}{2} \int g_{1}^{-3/2}g_{2}dt \right) +\frac{1}{4g_{1}} \left( \dot{g}_{1}r -g_{2} \right)^{2}. \label{eq.nfx10.2}
\end{eqnarray}
These results coincide with the ones of \cite{Leach 1985} provided one replaces $\dot{\theta}$ from $L_{3}= r^{2}\dot{\theta}$.
\bigskip

We collect the results in Theorem \ref{theorem.central}.

\begin{theorem} \label{theorem.central}
The integrable time-dependent central Newtonian potentials\index{Potential! time-dependent central} $V(t,r)$ with angular momentum $L_{3}$ are the following: \newline
a. The potentials $V(t,r)= -\frac{\ddot{g}_{2}}{2g_{2}}r^{2} +\frac{\dot{g}}{g_{2}}r -\frac{L_{3}^{2}}{2r^{2}}$ which admit the LFIs $I=g_{2}\dot{r}-\dot{g}_{2}r +g$, where $g_{2}(t)\neq 0$ and $g(t)$ are arbitrary functions. \newline
b. The potentials $V(t,r)=\left[ \frac{1}{8}\left( \frac{\dot{g}_{1}}{g_{1}}%
\right) ^{2}-\frac{\ddot{g}_{1}}{4g_{1}}\right] r^{2}+\frac{1}{2g_{1}}\left(
\dot{g}_{2}-g_{2}\frac{\dot{g}_{1}}{2g_{1}}\right) r+\frac{1}{2g_{1}}F\left(
g_{1}^{-1/2}r+\frac{1}{2}\int g_{1}^{-3/2}g_{2}dt\right) -\frac{L_{3}^{2}}{%
2r^{2}}$ which admit the QFIs $I=g_{1}\dot{r}^{2}+(g_{2}-\dot{g}_{1}r)\dot{r}%
+F\left( g_{1}^{-1/2}r+\frac{1}{2}\int g_{1}^{-3/2}g_{2}dt\right) +\frac{1}{%
4g_{1}}\left( \dot{g}_{1}r-g_{2}\right)^{2}$, where $g_{1}(t)\neq0$ and $g_{2}(t)$ are arbitrary functions.
\end{theorem}

\section{Applications of Theorem \ref{theorem.central}}

\label{applications}

From Theorem \ref{theorem.central}, it follows that it is possible to classify all integrable time-dependent central Newtonian potentials in just two cases, according to if they admit a. LFIs or b. QFIs. In this section, we consider various applications of Theorem \ref{theorem.central}.

\subsection{The time-dependent oscillator}

\label{sec.app1}

In this case, the potential is of the form $V=-\omega(t)r^{2}$, where $\omega(t)$ is an arbitrary function. The LFIs and the QFIs are as follows:
\bigskip

a. For $g(t)=0$ and $L_{3}=0 \implies \theta(t)=const$, we have the time-dependent potential $V= -\frac{\ddot{g}_{2}}{2 g_{2}}r^{2}$ with the LFI $I= g_{2}\dot{r} -\dot{g}_{2}r$.

b. For $g_{2}=0$ and $F= \frac{c_{0}}{2g_{1}}r^{2} +L_{3}^{2} \frac{g_{1}}{r^{2}}$, where $c_{0}$ is an arbitrary constant, we obtain the time-dependent potential
\begin{equation}
V= -\left[ \frac{\ddot{g}_{1}}{4g_{1}} -\frac{1}{8}\left( \frac{\dot{g}_{1}}{g_{1}} \right)^{2} -\frac{c_{0}}{4g_{1}^{2}} \right] r^{2} \label{osc.3}
\end{equation}
with the QFI
\begin{equation}
I= g_{1}\left( \dot{r}^{2} + \frac{L_{3}^{2}}{r^{2}} \right) -\dot{g}_{1}r\dot{r} +\frac{\dot{g}_{1}^{2}}{4g_{1}} r^{2} +\frac{c_{0}}{2g_{1}} r^{2}. \label{osc.4}
\end{equation}
If we replace $L_{3}=r^{2}\dot{\theta}$, the QFI (\ref{osc.4}) is the sum of the diagonal components of the Jauch-Hill-Fradkin tensor.\index{Tensor! Jauch-Hill-Fradkin} If, in addition, $g_{1}=\frac{\phi^{2}}{2}$ and $c_{0}=\frac{K}{2}$, where $\phi(t)$ is an arbitrary function and $K$ is a constant, we derive the Case I QFI of \cite{Leach 1985}
\begin{equation}
I= \frac{1}{2} \left( \phi \dot{r} -\dot{\phi}r \right)^{2} +\frac{1}{2}r^{2}\phi^{2}\dot{\theta}^{2} +\frac{K}{2\phi^{2}} r^{2}. \label{osc.5}
\end{equation}

The above results for the time-dependent oscillator\index{Oscillator! time-dependent} coincide with the ones of Theorem 6.2 of \cite{Katzin 1977}.

\subsection{The time-dependent generalized Kepler potential}

\label{sec.app2}

In this case, $V= -\frac{\omega(t)}{r^{\nu}}$ where $\nu$ is an arbitrary non-zero constant.
\bigskip

a. No new integrable potentials.

b. For $g_{1}=\frac{\phi^{2}}{2}$, $g_{2}=0$, and $F= \bar{F} +\frac{L_{3}^{2}\phi^{2}}{2r^{2}}$, where $F, \bar{F}$ are functions of the same argument and $\phi(t)$ is an arbitrary smooth function, we find the Case III potential of \cite{Leach 1985}
\begin{equation}
V= -\frac{\ddot{\phi}}{2\phi} r^{2} +\phi^{-2} \bar{F} \left( \frac{r}{\phi} \right) \label{gen.1}
\end{equation}
with the QFI
\begin{equation}
I= \frac{1}{2} (\phi\dot{r} -r\dot{\phi})^{2} +\frac{L_{3}^{2} \phi^{2}}{2r^{2}} +\bar{F}\left( \frac{r}{\phi} \right) = \frac{1}{2} (\phi\dot{r} -r\dot{\phi})^{2} +\frac{1}{2} \phi^{2} r^{2} \dot{\theta}^{2} +\bar{F}\left( \frac{r}{\phi} \right) \label{gen.2}
\end{equation}
where we replaced $L_{3}=r^{2}\dot{\theta}$.

We choose $\bar{F}\left( \frac{r}{\phi} \right)= k_{1}\frac{r^{2}}{\phi^{2}} -\frac{k\phi^{\nu}}{r^{\nu}}$ with
$\phi= \sqrt{b_{0}+ b_{1}t +b_{2}t^{2}}$ and $k_{1}= \frac{b_{0}b_{2}}{2} -\frac{b_{1}^{2}}{8}$, where $\nu, k, b_{0}, b_{1}, b_{2}$ are arbitrary constants. Then, the potential (\ref{gen.1}) becomes the integrable time-dependent generalized Kepler potential\index{Potential! time-dependent generalized Kepler}
\begin{equation}
V= -\frac{\omega_{\nu}(t)}{r^{\nu}}, \enskip \omega_{\nu}= k\left(b_{0} + b_{1}t + b_{2}t^{2} \right)^{\frac{\nu-2}{2}} \label{gen.3}
\end{equation}
which admits the QFI
\begin{equation}
J_{\nu}=  (b_{0} + b_{1}t + b_{2}t^{2}) \left[ \frac{1}{2} \left( \dot{r}^{2} +r^{2}\dot{\theta}^{2} \right) - \frac{\omega_{\nu}}{r^{\nu}} \right] -\frac{b_{1} + 2b_{2}t}{2} r\dot{r} +\frac{b_{2} r^{2}}{2}. \label{gen.4}
\end{equation}
This result is in accordance with Propositions \ref{pro.lfis} and \ref{oscillator}.

We note that in the case of the standard Kepler potential\index{Potential! Kepler} (i.e. $\nu=1$) for $k_{1}=0 \implies b_{1}^{2} -4b_{0}b_{2}=0$, the `frequency' $\omega_{1}= \frac{k}{\sqrt{b_{0}+ b_{1}t+b_{2}t^{2}}}$ degenerates into the well-known result $\omega_{1}= \frac{k}{a_{0} +a_{1}t}$, where $a_{0}$ and $a_{1}$ are constants (see Theorem 2.1 of \cite{Katzin 1982} and Proposition \ref{kepler}).

\subsection{Integrating the equations of motion for the case a. potentials of Theorem \ref{theorem.central}}

\label{sec.app4}

In Theorem \ref{theorem.central}, we found the new class of time-dependent integrable central potentials (\ref{eq.nfx6}), which admit the LFIs (\ref{eq.nfx7}).

Using the transformation $R= \frac{g_{2}}{r}$, we compute $\dot{R}= \frac{\dot{g}_{2}r -g_{2}\dot{r}}{r^{2}}$. Replacing this in the LFI (\ref{eq.nfx7}), we find the solution
\begin{equation}
r(t)= g_{2} \int \frac{I -g}{g_{2}^{2}} dt +c \label{eq.cl3}
\end{equation}
where $c$ is an integration constant.

Substituting (\ref{eq.cl3}) in $L_{3}= r^{2}\dot{\theta}$, we find that
\begin{equation}
\theta(t)= \int \frac{L_{3}}{r^{2}(t)}dt +\theta_{0} \label{eq.cl4}
\end{equation}
where $\theta_{0}$ is another integration constant.

\subsection{Integrating the Scr\"{o}dinger equation for an integrated central potential}

\label{sec.app3}

We consider\index{Equation! Scr\"{o}dinger} a special class of integrable central potentials with fixed angular momentum $L_{3}$ generated from case b. of Theorem \ref{theorem.central} for $g_{1}= \frac{\phi^{2}}{2}$, $g_{2}=0$ and $F= -k\frac{\phi}{r}$, where $\phi(t)$ is an arbitrary non-zero function and $k\neq0$ is an arbitrary constant. Replacing these choices in the defining formula, we obtain the potential
\begin{equation}
V(t,r)= -\frac{\ddot{\phi}}{2\phi}r^{2} -\frac{k}{\phi}r^{-1} -\frac{L_{3}^{2}}{2r^{2}}. \label{eq.e1}
\end{equation}
The associated QFI is
\begin{equation}
I= \frac{1}{2} \left( \phi \dot{r} -\dot{\phi}r \right)^{2} -k\frac{\phi}{r}. \label{eq.e2}
\end{equation}

By introducing the transformation $R= \frac{\phi}{r}$, we compute $\dot{R}= \frac{\dot{\phi}r -\phi \dot{r}}{r^{2}}$ and, replacing in the QFI (\ref{eq.e2}), we find that $\frac{dR}{R^{2}\sqrt{2(I+kR)}}= \pm \frac{dt}{\phi^{2}}$. Using the LFI of the angular momentum, we find the following orbit:
\[
L_{3}= r^{2} \dot{\theta} \implies d\theta = L_{3} R^{2} \frac{dt}{\phi^{2}} \implies \int d\theta= \pm L_{3}\int \frac{dR}{\sqrt{2(I+kR)}} \implies
\]
\begin{equation}
\theta= \pm \frac{L_{3}}{k} \sqrt{2 \left( I +\frac{k\phi}{r} \right)} +\theta_{0} \label{eq.e4}
\end{equation}
where $\theta_{0}$ is an integration constant.

In \cite{Leach 1985}, it has been shown that the wavefunction\index{Wavefunction} for the potential $V_{III}(t,r)= -\frac{\ddot{\phi}}{2\phi }r^{2} +\phi ^{-2}F\left( \frac{r}{\phi }\right)$ is given by the relation
\begin{equation}
\psi(r,\theta,t)= |\phi|^{-1/2} e^{\frac{i}{2\hbar}\frac{\dot{\phi}}{\phi}r^{2}} \bar{\psi} \left( \phi^{-1}r, \theta, T(t) \right) \label{eq.schr12}
\end{equation}
where\footnote{We note that in eq. (6.14) of \cite{Leach 1985} the second term into the brackets should be multiplied by 2.}
$\bar{\psi}(R,\Theta,T)= e^{-i\lambda T/\hbar} e^{im\Theta} R^{-1/2} A(R)$, $\hbar$ is the Planck constant, $\lambda$ and $m$ are arbitrary constants, $R= \phi^{-1}r$, $\Theta= \theta$, $T(t)= \int \phi^{-2}dt \implies \dot{T}= \phi^{-2}$ and $A(R)$ is an arbitrary smooth function which satisfies the second order ODE (see eq. (6.6) of \cite{Leach 1985})
\begin{equation}
\frac{d^{2}A}{dR^{2}} +\left( \frac{2\lambda}{\hbar^{2}} -\frac{2}{\hbar^{2}}F -\frac{m^{2} -\frac{1}{4}}{R^{2}} \right) A =0. \label{eq.schr11}
\end{equation}

We observe that for $F\left( \frac{r}{\phi} \right)= -k\phi r^{-1} -\frac{L_{3}^{2} \phi^{2}}{2r^{2}} \implies F(R)= -\frac{k}{R} -\frac{L_{3}^{2}}{2R^{2}}$, the potential $V_{III}(t,r)$ reduces to the potential (\ref{eq.e1}) for which the orbit equation has been found. For this choice of $F$, the ODE (\ref{eq.schr11}) becomes
\begin{equation}
\frac{d^{2}A}{dR^{2}} +\left( \frac{2\lambda}{\hbar^{2}} +\frac{2k}{\hbar^{2}R} -\frac{m^{2} -\frac{1}{4} -\frac{L_{3}^{2}}{\hbar^{2}}}{R^{2}} \right) A =0. \label{eq.schr13}
\end{equation}
From Table 22.6 in p. 781 of \cite{Abramowitz}, we find that (\ref{eq.schr13}) admits the solution $A(R)= e^{-R/2} R^{\frac{a+1}{2}} L^{(a)}_{b}(R)$, where $L^{(a)}_{b}(R)$ is a generalized Laguerre polynomial\index{Laguerre polynomial} and the constants $\lambda, k, m$ are fixed as follows: $\lambda=-\frac{\hbar^{2}}{8}$, $k= \frac{\hbar^{2}}{4} \left( 2b +a +1 \right)$ and $m^{2}= \frac{a^{2}}{4} +\frac{L_{3}^{2}}{\hbar^{2}}$.

Substituting the above results in (\ref{eq.schr12}), we find the wavefunction
\begin{equation}
\psi(r,\theta,t)= |\phi|^{-1/2} e^{\frac{i}{2\hbar}\frac{\dot{\phi}}{\phi} r^{2}} e^{i\hbar T(t)/8} e^{i \sqrt{\frac{a^{2}}{4} +\frac{L_{3}^{2}}{\hbar^{2}}} \theta} \phi^{1/2} r^{-1/2} e^{-\frac{r}{2\phi}} \left( \frac{r}{\phi} \right)^{\frac{a+1}{2}} L^{(a)}_{b}(\phi^{-1}r) \label{eq.schr16}
\end{equation}
provided the defining constant of the potential (\ref{eq.e1}) is $k= \frac{\hbar^{2}}{4} \left( 2b +a +1 \right)$.

\subsection{The integrable two-body problem with variable mass}

\label{sec.app5}

In Newtonian Physics, the motion of a point of variable mass\index{Variable mass} $m(t)$ is described by the dynamical equation \cite{Hadjidemetriou 1963, Plastino 1992}
\begin{equation}
m(t)\ddot{\mathbf{R}}= \mathbf{F} + \sum_{i} \dot{m}_{i} \mathbf{u}_{i} \label{eq.bin1}
\end{equation}
where $\mathbf{R}$ is the position vector of $m(t)$ wrt a fixed frame of reference, $\mathbf{F}$ denotes the external forces, $\mathbf{u}_{i}$ is the relative velocity of the escaping mass from the $i$th-point of the surface surrounding $m(t)$, and $\dot{m}_{i}$ is the rate of loss of mass from the $i$th-point. If the loss of mass is continuous, the summation symbol may be replaced by a double integral over the whole surface around $m$.

It is said that \textbf{the loss of mass is isotropic} iff $\sum_{i} \dot{m}_{i} \mathbf{u}_{i}=0$. For example, this is the case when the stars lose mass by radiation.

The two-body problem with variable mass \cite{Hadjidemetriou 1963} (e.g. a binary system\index{Binary system} of stars) consists of two point masses $m_{1}(t)$ and $m_{2}(t)$ with only gravitational attraction. Assuming isotropic loss of mass, the equations of motion wrt a fixed inertial frame are:
\begin{equation}
m_{1}(t) \ddot{\mathbf{r}}_{1}= \frac{G m_{1}m_{2}}{r^{2}} \hat{\mathbf{r}}, \enskip m_{2}(t) \ddot{\mathbf{r}}_{2}= -\frac{G m_{1}m_{2}}{r^{2}} \hat{\mathbf{r}} \label{eq.bin2}
\end{equation}
where $G$ is the gravitational constant, $\mathbf{r}= \mathbf{r}_{2} -\mathbf{r}_{1}$ is the relative position of the star with mass $m_{2}$ wrt the other star with mass $m_{1}$, $r= \|\mathbf{r}\|$ and the unit vector $\hat{\mathbf{r}}= \frac{\mathbf{r}}{r}$. From equations (\ref{eq.bin2}), we find the dynamical equation
\begin{equation}
\ddot{\mathbf{r}}= -\frac{Gm(t)}{r^{2}} \hat{\mathbf{r}} \label{eq.bin3}
\end{equation}
where $m(t)\equiv m_{1}(t) +m_{2}(t)$ is the total mass of the binary system. If $m(t)=const$, the system is called closed and the orbit is a well-known conic section. The potential driving the system is the time-dependent Kepler potential\index{Potential! time-dependent Kepler}
\begin{equation}
V(t,r)= -\frac{\omega(t)}{r} \label{eq.bin4}
\end{equation}
where the `frequency' $\omega(t)= Gm(t)$. The problem which has been around for a long time was the determination of the mass $m(t)$ so that the potential (\ref{eq.bin4}) is integrable. It has been found  \cite{Gylden, Mestschersky 1893, Mestschersky 1902, Prieto 1997, Rahoma 2009} that this potential is integrable for the following $m(t)$:
\[
m_{I}(t)= \frac{1}{a_{0}+a_{1}t}, \enskip m_{II}(t)= \frac{1}{\sqrt{b_{0}+b_{1}t}}, \enskip m_{III}(t)= \frac{1}{\sqrt{b_{0}+b_{1}t+b_{2}t^{2}}}.
\]
Using Theorem \ref{theorem.central}, we derive these functions easily. They are subcases of the `frequency' $\omega_{\nu}$ of the integrable time-dependent generalized Kepler potential (\ref{gen.3}) for $\nu=1$ and $k=G$. Indeed, we have:
\[
\omega_{1}(t)= Gm(t) \implies \frac{G}{\sqrt{b_{0} +b_{1}t +b_{2}t^{2}}} =Gm(t) \implies m(t)= \frac{1}{\sqrt{b_{0}+b_{1}t+b_{2}t^{2}}}= m_{III}(t).
\]
For $b_{2}=0$, we obtain the mass $m_{II}(t)$, and for vanishing discriminant $b_{1}^{2} -4b_{0}b_{2}=0$ the mass $m_{I}(t)$.

\subsection{Time-dependent integrable Yukawa and interatomic potentials}

\label{sec.app6}

In plasma physics, solid-state physics and nuclear physics, the following types of central potentials are widely used:
\bigskip

1. The Yukawa type potentials\index{Potential! Yukawa}
\begin{equation}
V(r)= A\frac{e^{-Br}}{r} \label{eq.Y1}
\end{equation}
where $A$ and $B$ are arbitrary constants. This type of potentials describes the screened Coulomb potential \cite{Debye 1923} generated around a positive charged particle into a neutral fluid (e.g. a plasma of electrons in a background of heavy positive charged ions \cite{Bellan}), and also models successfully the neutron-proton interaction \cite{Yukawa 1935}.
\bigskip

2. The interatomic pair potentials \cite{Jones 1924A, Jones 1924B} \index{Potential! interatomic pair}
\begin{equation}
V(r)= \frac{A}{r^{m}} -\frac{B}{r^{n}} \label{eq.A1}
\end{equation}
where $A, B, m, n$ are arbitrary positive constants. These central potentials manifest between the atoms of diatomic molecules. From (\ref{eq.A1}), we observe that they consist of a repulsive term $\frac{A}{r^{m}}$ and an attractive term $-\frac{B}{r^{n}}$. The most well-known potential of this form is the Lennard-Jones\index{Potential! Lennard-Jones} potential \cite{LJ potential} in which $m=12$ and $n=6$.
\bigskip

Using Theorem \ref{theorem.central}, we shall answer to the following problem: \newline
\emph{Assume the parameters $A, B$ of the potentials (\ref{eq.Y1}) and (\ref{eq.A1}) to be time-dependent. Then, find for which functions $A(t)$ and $B(t)$ the resulting potentials are integrable.}

The case b. potentials of Theorem \ref{theorem.central} for
$g_{2}=0$ and $F= -\frac{c_{1}}{4g_{1}}r^{2} +\frac{L_{3}^{2}g_{1}}{r^{2}} +\bar{F} \left( g_{1}^{-1/2} r \right)$, where $c_{1}$ is an arbitrary constant and $\bar{F}$ a smooth function of its argument, reduce to the integrable potentials
\begin{equation}
V(t,r)= -\left[ \frac{\ddot{g}_{1}}{4g_{1}} -\frac{1}{8}\left( \frac{\dot{g}_{1}}{g_{1}} \right)^{2} +\frac{c_{1}}{8g_{1}^{2}} \right] r^{2} +\frac{1}{2g_{1}} \bar{F} \left( g_{1}^{-1/2} r \right) \label{eq.R1}
\end{equation}
which admit the QFIs
\begin{equation}
I= \left(\dot{r}^{2} +\frac{L_{3}^{2}}{r^{2}} \right) g_{1} -\dot{g}_{1}r\dot{r} -\frac{c_{1}}{4g_{1}} r^{2} +\frac{\dot{g}_{1}^{2}}{4g_{1}} r^{2} +\bar{F} \left( g_{1}^{-1/2} r \right). \label{eq.R2}
\end{equation}

From (\ref{eq.R1}), we see that integrable time-dependent central potentials of the form $V(t,r)= \frac{1}{2g_{1}} \bar{F} \left( g_{1}^{-1/2} r \right)$ exist only when $g_{1}(t)=b_{0}+b_{1}t+b_{2}t^{2}$ and $c_{1}=b_{1}^{2} -4b_{2}b_{0}$, where $b_{0}, b_{1}, b_{2}$ are arbitrary constants. For special choices of the function $\bar{F}$, we obtain the required integrable potentials as follows:
\bigskip

1. $\bar{F}= \frac{2k\exp{\left( -g_{1}^{-1/2}r \right)}}{g_{1}^{-1/2}r}$, where $k$ is an arbitrary constant.

We find the new integrable time-dependent Yukawa type potentials\index{Potential! time-dependent Yukawa}
\begin{equation}
V(t,r)=\frac{k}{\sqrt{b_{0}+b_{1}t+b_{2}t^{2}}} \frac{e^{-\frac{r}{\sqrt{b_{0}+b_{1}t+b_{2}t^{2}}}}}{r} \label{eq.R5}
\end{equation}
where $A(t)= \frac{k}{\sqrt{b_{0}+b_{1}t+b_{2}t^{2}}}$ and $B(t)= \frac{1}{\sqrt{b_{0}+b_{1}t+b_{2}t^{2}}}$.
\bigskip

2. $\bar{F}= \frac{2k_{1}g_{1}^{m/2}}{r^{m}} -\frac{2k_{2}g_{1}^{n/2}}{r^{n}}$, where $k_{1}, k_{2}$ are arbitrary constants.

We find the new integrable time-dependent interatomic pair potentials\index{Potential! time-dependent interatomic pair}
\begin{equation}
V(t,r)= \frac{k_{1}(b_{0}+b_{1}t+b_{2}t^{2})^{\frac{m-2}{2}}}{r^{m}} -\frac{k_{2}(b_{0}+b_{1}t+b_{2}t^{2})^{\frac{n-2}{2}}}{r^{n}} \label{eq.R6}
\end{equation}
where $A(t)= k_{1}(b_{0}+b_{1}t+b_{2}t^{2})^{\frac{m-2}{2}}$ and $B(t)= k_{2}(b_{0}+b_{1}t+b_{2}t^{2})^{\frac{n-2}{2}}$.

\section{Conclusions}

\label{conclusions.central.pots}

Using the LFI of angular momentum and the direct method, we have managed to compute the integrable time-dependent central potentials. These potentials are widely used in all branches of Physics. It is remarkable that so divergent potentials can be squeezed into two simple classes, i.e. the ones which admit LFIs and the ones which admit QFIs. One may ask: Why in \cite{Leach 1985} not all the integrable time-dependent central Newtonian potentials were found? The reason is that in \cite{Leach 1985} the generalized Killing equations (\ref{con.mot.eq.kil3}) and (\ref{con.mot.eq.kil4}), instead of the dynamical equations, were used. Indeed, the former result from the Noether condition, which is written in the form $A(t,q,\dot{q}) +B_{a}(t,q,\dot{q})\ddot{q}^{a}=0$, by requiring $B_{a}=0$ and
without using the dynamical equations to replace the $\ddot{q}^{a}$ (see section \ref{con.mot.subsec.killing.2}).

%% file: stability_append2.tex
\chapter{Properties of $2\times2$ real matrices with two equal eigenvalues $\lambda_{+}=\lambda_{-}$}

\label{append2}

Let $A=
\left(
  \begin{array}{cc}
    a & b \\
    c & d \\
  \end{array}
\right)$ $\in \mathbb{R}^{2\times2}$. We assume that its  eigenvalues $\lambda_{\pm}$ computed in (\ref{eq.fx19}) are equal, that is,
\begin{equation}
\Delta=0 \implies tr(A)^{2}= 4\det(A) \implies (a-d)^{2}= -4bc, \enskip \lambda_{\pm}\equiv \lambda = \frac{tr(A)}{2} = \frac{a+d}{2}. \label{app2.1}
\end{equation}
In order to find the corresponding eigenvectors $v_{\pm}$, we solve the linear system
\begin{equation}
Av=\lambda v \implies
\begin{cases}
av_{1} +bv_{2}= \frac{a+d}{2}v_{1} \\
cv_{1} +dv_{2}= \frac{a+d}{2}v_{2}.
\end{cases} \label{app2.2}
\end{equation}
We consider the following cases:
\bigskip

1) Case $b=0$. From equations (\ref{app2.1}), we find that $a=d$ and $\lambda=a$.

1.1. Subcase $c=0$. We have $A=
\left(
  \begin{array}{cc}
    a & 0 \\
    0 & a \\
  \end{array}
\right)= aI$ and $v=
\left(
  \begin{array}{c}
    v_{1} \\
    v_{2} \\
  \end{array}
\right)$.

1.2. Subcase $c\neq0$. We have $A=
\left(
  \begin{array}{cc}
    a & 0 \\
    c & a \\
  \end{array}
\right)$ and $v=
\left(
  \begin{array}{c}
    0 \\
    v_{2} \\
  \end{array}
\right)$.

2) Case $b\neq0$. We have $c= -\frac{(a-d)^{2}}{4b}$ and $\lambda= \frac{a+d}{2}$. The matrix $A=
\left(
  \begin{array}{cc}
    a & b \\
    -\frac{(a-d)^{2}}{4b} & d \\
  \end{array}
\right)$ and $v=
\left(
  \begin{array}{c}
    v_{1} \\
    \frac{d-a}{2b}v_{1} \\
  \end{array}
\right)$.

We observe that the matrix $A$ of the subcase 1.1 is the only type of such matrices which admits two distinct eigenvectors.

We collect the above results in the following proposition.

\begin{proposition} \label{pro.app2}
Consider an arbitrary square matrix $A \in \mathbb{R}^{2\times2}$ which admits one double eigenvalue $\lambda_{\pm}=\lambda$. Then, $\lambda=\frac{tr(A)}{2}$ and $tr(A)^{2}=4\det(A)$. There are three different types of such matrices with the following eigenvectors: \newline
i. $A=
\left(
  \begin{array}{cc}
    a & b \\
    -\frac{(a-d)^{2}}{4b} & d \\
  \end{array}
\right)$, $\lambda=\frac{a+d}{2}$ and $v=
\left(
  \begin{array}{c}
    v_{1} \\
    \frac{d-a}{2b}v_{1} \\
  \end{array}
\right)$, where $b\neq0$ and $v_{1}$ are arbitrary real constants. \newline
ii. $A=
\left(
  \begin{array}{cc}
    a & 0 \\
    0 & a \\
  \end{array}
\right)= aI$, $\lambda=a$ and $v=
\left(
  \begin{array}{c}
    v_{1} \\
    v_{2} \\
  \end{array}
\right)$, where $a, v_{1}$ and $v_{2}$ are arbitrary real constants. \newline
iii. $A=
\left(
  \begin{array}{cc}
    a & 0 \\
    c & a \\
  \end{array}
\right)$, $\lambda=a$ and $v=
\left(
  \begin{array}{c}
    0 \\
    v_{2} \\
  \end{array}
\right)$, where $c\neq0$, $a$ and $v_{2}$ are arbitrary real constants.
\end{proposition}

In the case that $tr(A)=0 \implies \det(A)=0$, $\lambda=0$, $d=-a$, and $a^{2}=-bc$. Then, proposition \ref{pro.app2} reduces to the following proposition.

\begin{proposition} \label{pro.app1}
An arbitrary square matrix $A \in \mathbb{R}^{2\times2}$ with $tr(A)= \det(A)=0$ has one eigenvalue $\lambda=0$ with multiplicity two. There are three different types of such matrices: \newline
i. $A=
\left(
  \begin{array}{cc}
    a & b \\
    -\frac{a^{2}}{b} & -a \\
  \end{array}
\right)$ with eigenvector $v=
\left(
  \begin{array}{c}
    v_{1} \\
    -\frac{a}{b}v_{1} \\
  \end{array}
\right)$, where $b\neq0$ and $v_{1}$ are arbitrary real constants.
\newline
ii. $A=0$ (zero matrix) with $v=
\left(
  \begin{array}{c}
    v_{1} \\
    v_{2} \\
  \end{array}
\right)$, where $v_{1}$ and $v_{2}$ are arbitrary real constants. \newline
iii. $A=
\left(
  \begin{array}{cc}
    0 & 0 \\
    c & 0 \\
  \end{array}
\right)$ and $v=
\left(
  \begin{array}{c}
    0 \\
    v_{2} \\
  \end{array}
\right)$, where $c\neq0$.
\end{proposition}

From the conditions (\ref{app2.1}), we deduce that \emph{the matrix $B= A -\lambda I$ has $tr(B)=\det(B)=0$. This implies that $B^{2}=0$; therefore, $B^{k}=0$ for any positive integer $k \geq 2$.} Indeed, we have $tr(B)= tr(A) -2\lambda = tr(A) -2\frac{tr(A)}{2} =0$ and $\det(B)= \det(A -\lambda I)= 0$.

Since $\lambda=\frac{tr(A)}{2}$, the matrix $B=
\left(
  \begin{array}{cc}
    \frac{a-d}{2} & b \\
    c & \frac{d-a}{2} \\
  \end{array}
\right)$ where $(a-d)^{2}= -4bc$. Therefore, for an arbitrary vector $u=
\left(
  \begin{array}{c}
    u_{1} \\
    u_{2} \\
  \end{array}
\right) \in \mathbb{R}^{2\times1}$ we find that
\begin{equation}
B^{2}u =0 \implies B(Bu)=0 \implies (A -\lambda I)(Bu)=0 \implies A(Bu) = \lambda (Bu) \implies Bu=v \label{app2.3}
\end{equation}
where $v$ is the eigenvector of $A$ corresponding to the double eigenvalue $\lambda$ (see proposition \ref{pro.app2}). \emph{We note that $Bu$ is an eigenvector of $A$.} If $u$ is linearly dependent with $v$, then $Bu=0$. Therefore, $Bu \neq 0$ only when $u, v$ are linearly independent. Replacing with the three different types of matrices found in proposition \ref{pro.app2}, we obtain the following relations between the vectors $u$ and $v$:

i.
\[
\left(
  \begin{array}{cc}
    \frac{a-d}{2} & b \\
    -\frac{(a-d)^{2}}{4b} & \frac{d-a}{2} \\
  \end{array}
\right)
\left(
  \begin{array}{c}
    u_{1} \\
    u_{2} \\
  \end{array}
\right) =
\left(
  \begin{array}{c}
    v_{1} \\
    \frac{d-a}{2b}v_{1} \\
  \end{array}
\right) \implies v_{1}= \frac{a-d}{2}u_{1} +bu_{2}.
\]

ii. In this case, $B=0$.

iii.
\[
\left(
  \begin{array}{cc}
    0 & 0 \\
    c & 0 \\
  \end{array}
\right)
\left(
  \begin{array}{c}
    u_{1} \\
    u_{2} \\
  \end{array}
\right) =
\left(
  \begin{array}{c}
    0 \\
    v_{2} \\
  \end{array}
\right) \implies
v_{2}=cu_{1}.
\]

We observe that $A$ admits two distinct eigenvectors only in the case ii. of proposition \ref{pro.app2}. In this case, $Bu=0$ for all vectors $u$. Moreover, the above relations determine which eigenvector $v$ of $A$ is associated with an arbitrary vector $u$ if its components $u_{1}$ and $u_{2}$ are given.

\begin{proposition} \label{pro.app3}
Let $A$ be a matrix of either type i. or type iii. of proposition \ref{pro.app2}. Then, $A$ admits only one distinct eigenvector $v$ and for an arbitrary vector $u \in \mathbb{R}^{2\times1}$ linearly independent with $v$, it holds that
\begin{equation}
A^{k}u =\lambda^{k}u +k\lambda^{k-1}v, \enskip \forall k \in \mathbb{N}. \label{app2.4}
\end{equation}
\end{proposition}

%% file: stability_append3.tex
\chapter{Complex solutions for the linear system $\dot{q}= Aq$}

\label{append3}

Consider the linear system $\dot{q}= Aq$, where $q=(q^{1}, ..., q^{n})$ and $A \in \mathbb{R}^{n\times n}$, and let
\begin{equation}
q(t)= x(t) +iy(t) \iff q^{a}(t) =x^{a}(t) +iy^{a}(t) \label{app3.1}
\end{equation}
where $x(t), y(t) \in \mathbb{R}^{n\times1}$, be a complex solution of the system. Then, $\dot{x}=Ax$ and $\dot{y}=Ay$. Therefore, the real vectors $x=Re(q)$ and $y=Im(q)$ are solutions of the system as well.

The complex conjugate $\bar{q}(t)= x(t) -iy(t)$ of the solution $q(t)$ is also a solution of the system. \emph{The solutions $q(t)$ and $\bar{q}(t)$ are linearly independent} because $\bar{q}(t)$ is generated by the complex conjugate eigenvalue $\bar{\lambda}$ of the eigenvalue $\lambda$, which produces the $q(t)$. We recall that $Av =\lambda v \implies A\bar{v}= \bar{\lambda}\bar{v}$.

Using $q(t)$ and $\bar{q}(t)$, the real solutions $x(t)$ and $y(t)$ are written as $x(t)= \frac{q(t) +\bar{q}(t)}{2}$ and $y(t)= \frac{q(t) -\bar{q}(t)}{2i}$. We shall prove now that the vectors $x(t)$ and $y(t)$ are also linearly independent.

Let $c_{1}, c_{2}$ be complex constants such that $c_{1}x(t) +c_{2}y(t)= 0$. Then, it is sufficient to show that $c_{1}=c_{2}=0$. Indeed, we have: $c_{1} \frac{q(t) +\bar{q}(t)}{2} +c_{2} \frac{q(t) -\bar{q}(t)}{2i} =0 \implies (c_{2} +ic_{1}) q(t) -(c_{2} -ic_{1})\bar{q}(t) = 0$. The last relation implies that $c_{2}=-ic_{1}$ and $c_{2}=ic_{1}$ due to the fact that $q(t)$ and $\bar{q}(t)$ are linearly independent. Therefore, $c_{1}=c_{2}=0$ which implies that $x(t)$ and $y(t)$ are linearly independent.

We collect the above results in the following proposition.

\begin{proposition} \label{pro.app4}
If $q(t)= x(t) +iy(t)$ is a complex solution of the linear system $\dot{q}=Aq(t)$, where $q(t) \in \mathbb{C}^{n\times1}$ and $A\in \mathbb{R}^{n\times n}$, then the real vectors $x(t)$ and $y(t)$ are linearly independent solutions of the system.
\end{proposition}

Let now $\lambda= \sigma +i\omega$, where $\sigma, \omega \in \mathbb{R}$, be a complex eigenvalue of $A$ with corresponding complex eigenvector (i.e. $Av =\lambda v$) $v= w+ iz$, where $w,z \in \mathbb{R}^{n\times1}$. Then, we have the special complex solution $q(t)= e^{\lambda t} v$ which is written as follows:
\begin{eqnarray*}
q(t)&=& e^{(\sigma +i\omega)t} (w +iz) = e^{\sigma t} \left[ \cos(\omega t) +i\sin(\omega t) \right] (w +iz) \\
&=& e^{\sigma t} \left\{ \left[ w\cos(\omega t) -z\sin(\omega t) \right] +i \left[ w\sin(\omega t) +z\cos(\omega t) \right] \right\} \\
&=& \phi(t) +i\psi(t)
\end{eqnarray*}
where we have introduced the real vectors
\begin{eqnarray}
\phi(t)&=& Re(q(t)) = e^{\sigma t} \left[ w\cos(\omega t) -z\sin(\omega t) \right] \label{app3.3} \\
\psi(t)&=& Im(q(t)) = e^{\sigma t} \left[ w\sin(\omega t) +z\cos(\omega t) \right]. \label{app3.4}
\end{eqnarray}
From proposition \ref{pro.app4}, the vector fields $\phi(t)$ and $\psi(t)$ are linearly independent real solutions of the system. The same result is obtained if we work with the complex conjugate eigenvalue $\bar{\lambda}$.

Using the above results, we end up with the following proposition.

\begin{proposition} \label{pro.app5}
The general solution of the real linear system (for $n=2$) $\dot{q}= Aq$, where $q(t) \in \mathbb{R}^{2\times 1}$ and $A \in \mathbb{R}^{2 \times 2}$, is of the form\footnote{Since $\phi(t)$ and $\psi(t)$ are linearly independent, the vector $q(0)= c_{1} \phi(0) +c_{2} \psi(0)$ represents any possible initial condition in the plane $q^{1}, q^{2}$.}
\begin{equation}
q(t)= c_{1}\phi(t) +c_{2}\psi(t) \label{app3.5}
\end{equation}
iff $A$ admits the complex eigenvalues $\lambda_{\pm}= \sigma \pm i\omega$ (see eq. (\ref{eq.fx19}) ) with corresponding eigenvectors $v_{\pm} =w \pm iz$. The parameters $c_{1}, c_{2}$ are arbitrary real constants computed by the initial conditions and $\phi(t), \psi(t)$ are given by (\ref{app3.3}) and (\ref{app3.4}), respectively.
\end{proposition}

Replacing (\ref{app3.3}) and (\ref{app3.4}) in the general solution (\ref{app3.5}), we find 
\begin{eqnarray}
q(t)&=& e^{\sigma t} \left\{ \left[ c_{1}\cos(\omega t) +c_{2}\sin(\omega t) \right] w -\left[ c_{1}\sin(\omega t) -c_{2}\cos(\omega t) \right] z \right\} \notag \\
&=& R e^{\sigma t}  \left[ \cos(\omega t +\theta) w - \sin(\omega t +\theta) z \right] \label{app3.6}
\end{eqnarray}
where $R= \sqrt{c_{1}^{2} +c_{2}^{2}}$, $\cos(-\theta)= \frac{c_{1}}{R}$ and $\sin(-\theta)= \frac{c_{2}}{R}$. We note that the vectors $w, z \in \mathbb{R}^{2\times1}$ are linearly independent due to the fact that the eigenvectors $v_{\pm}= w\pm iz$ produce the plane $q^{1}, q^{2}$. It can be easily checked that the vector (\ref{app3.6}) does satisfy equation $\dot{q}=Aq$ and, also, it is compatible with any initial condition $q(0)$. Indeed, we have $q(0)= R\cos\theta~w -R\sin\theta~z = c_{1}w +c_{2}z$ and from the eigenvalue problem of $A$ we get:
\[
Av_{\pm}= \lambda_{\pm} v_{\pm} \implies A(w \pm iz) = (\sigma \pm i\omega) (w \pm iz) \implies Aw \pm iAz = (\sigma w -\omega z) \pm i (\omega w + \sigma z) \implies
\]
\begin{equation}
Aw= \sigma w -\omega z, \enskip Az= \omega w + \sigma z. \label{app3.7}
\end{equation}
Using the last relations, it is straightforward that (\ref{app3.6}) is the general solution of the 2d linear system.

%% file: proof_thm_QFIs.tex
\chapter{Proof of Theorem \ref{The first integrals of an autonomous holonomic dynamical system}}

\label{app2.proof.QFIs}

We look for solutions in which $g(t)$ and $f(t)$ are analytic functions so that they can be represented by polynomial functions of $t$:
\begin{eqnarray*}
g(t)&=&\sum_{k=0}^{n}c_{k}t^{k}=c_{0}+c_{1}t+...+c_{n}t^{n} \\
f(t)&=&\sum_{k=0}^{m}d_{k}t^{k}=d_{0}+d_{1}t+...+d_{m}t^{m}
\end{eqnarray*}
where $n, m \in \mathbb{N}$ (or infinite) and $c_{k}, d_{k}\in \mathbb{R}$.

We recall that the quantities $C_{ab}(q)$ are KTs of order two. In what follows, we consider various cases\footnote{
Equation (\ref{FL.1.e}) is not necessary because the integrability condition $K_{,[ab]}=0$ does not intervene in the calculations. However, it has been checked that equation (\ref{FL.1.e}) is always satisfied identically from the solutions of the other equations of the system.
}.
\bigskip

\textbf{I. For both $\mathbf{n}$ and $\mathbf{m}$ finite.} \vspace{12pt}

\underline{\textbf{I.1. Case $\mathbf{n=m}$:}} \vspace{12pt}

\underline{\textbf{Subcase $\mathbf{(n=0, m=0)}$.}} $g=c_{0}$ and $f=d_{0}$.

\begin{equation*}
\begin{cases}
(\ref{FL.1.a}) \implies d_0 L_{(a;b)} + B_{(a;b)} = 0 \\
(\ref{FL.1.b}) \implies - 2 c_0 C_{ab} V^{,b} + K_{,a} = 0 \\
(\ref{FL.1.c}) \implies K_{,t} -d_0 L_b V^{,b} - B_b V^{,b}=0 \\
(\ref{FL.1.d}) \implies d_0 \left( L_b V^{,b} \right)_{;a} + \left( B_bV^{,b} \right)_{;a} = 0
\end{cases}%
\end{equation*}

We define the vector field $\tilde{L}_a \equiv d_0 L_a + B_a$. Then, equation (\ref{FL.1.a}) implies that $\tilde{L}_{(a;b)} = 0$, i.e. $\tilde{L}_a$ is a KV, and (\ref{FL.1.d}) gives $\tilde{L}_a V^{,a} = s_0 = const$.

Solving equation (\ref{FL.1.c}), we get $K = s_0 t + G(q)$, which when replaced into (\ref{FL.1.b}) gives $G_{,a} = 2 c_0 C_{ab} V^{,b}$.

The QFI is
\begin{equation*}
I_{00} = c_{0}C_{ab}\dot{q}^{a}\dot{q}^{b} + \tilde{L}_{a}\dot{q}^{a} + s_{0}t+G(q)
\end{equation*}%
where $c_0C_{ab}$ is a KT, $\tilde{L}_{a}$ is a KV such that $\tilde{L}_a V^{,a} = s_0$, and $G(q)=2c_{0} \int C_{ab} V^{,b} dq^{a}$.

The QFI $I_{00}$ consists of the independent FIs:
\begin{equation*}
Q_{1} = C_{ab}\dot{q}^{a}\dot{q}^{b} + G(q), \enskip Q_{2} = L_{a}\dot{q}^{a} + s_{1}t
\end{equation*}
where $C_{ab}$ is a KT, $L_{a}$ is a KV such that $L_{a}V^{,a} =s_{1}= const$, and $G_{,a}=2C_{ab}V^{,b}$.

\underline{\textbf{Subcase $\mathbf{(n=1, m=1)}$.}} $g = c_0 + c_1 t$ and $f = d_0 + d_1 t$ with $c_1d_1 \neq 0$.

\begin{equation*}
\begin{cases}
\eqref{FL.1.a} \implies c_1 C_{ab} + (d_0 + d_1t) L_{(a;b)} + B_{(a;b)} = 0
\\
\eqref{FL.1.b} \implies - 2 c_1 C_{ab} V^{,b} t - 2 c_0 C_{ab} V^{,b} + d_1
L_a + K_{,a} = 0 \\
\eqref{FL.1.c} \implies K_{,t} = \left( d_0 + d_1 t \right) L_a V^{,a} + B_a
V^{,a} \\
\eqref{FL.1.d} \implies \left( d_0 + d_1 t \right) \left( L_b V^{,b}
\right)_{;a} + \left( B_b V^{,b} \right)_{;a} - 2 c_1 C_{ab} V^{,b} = 0.%
\end{cases}%
\end{equation*}

From \eqref{FL.1.a}, $L_a$ is a KV and $c_1 C_{ab} + B_{(a;b)} = 0$.

From \eqref{FL.1.d}, we find that $L_aV^{,a}=s_{1}$ and $\left(
B_{b}V^{,b}\right) _{;a}=2c_{1}C_{ab}V^{,b}$. Then, \eqref{FL.1.c} gives $K=s_{1}\left( d_{0}t+\frac{d_{1}}{2}t^{2}\right) +B_{a}V^{,a}t+G(q)$, which when substituted into \eqref{FL.1.b} yields $G_{,a}=2c_{0}C_{ab}V^{,b}-d_{1}L_{a}$. Using the relation $\left( B_{b}V^{,b}\right)_{;a}=2c_{1}C_{ab}V^{,b}$, we
find that
\begin{equation*}
G_{,a}=\frac{c_{0}}{c_{1}}\left( B_{b}V^{,b}\right) _{;a}-d_{1}L_{a}
\implies L_{a} = \frac{c_0}{c_1d_1} \left(B_b V^{,b}\right)_{,a} - \frac{1}{d_1} G_{,a}.
\end{equation*}

The QFI is
\begin{equation*}
I_{11}= - \frac{1}{c_{1}} \left( c_{0}+c_{1}t\right) B_{(a;b)}\dot{q}^{a}\dot{q}^{b}+\left(
d_{0}+d_{1}t\right) L_{a}\dot{q}^{a}+B_{a}\dot{q}^{a}+s_{1}\left( d_{0}t+%
\frac{d_{1}}{2}t^{2}\right) +B_{a}V^{,a}t+G(q)
\end{equation*}%
where $B_{(a;b)}$ is a KT, $L_{a}=\frac{c_{0}}{%
c_{1}d_{1}}\left( B_{b}V^{,b}\right) _{,a}-\frac{1}{d_{1}}G_{,a} \equiv \Phi_{,a}$ is a gradient KV such that $L_{a}V^{,a}=s_{1}$ and the vector $B_{a}$ satisfies the condition $\left( B_{b}V^{,b}\right)_{;a}=-2B_{(a;b)}V^{,b}$.

We observe that
\begin{equation*}
L_{a} = \Phi_{,a} = \frac{c_{0}}{c_{1}d_{1}} \left(B_{b} V^{,b}\right)_{,a}
- \frac{1}{d_{1}} G_{,a} \implies G(q) = \frac{c_{0}}{c_{1}} B_{a}V^{,a} - d_{1}\Phi(q).
\end{equation*}
Therefore,
\begin{eqnarray*}
I_{11} &=& \frac{c_{0}}{c_{1}} Q_{3} + Q_{4} + d_{0}Q_{2} + d_{1} Q_{5}
\end{eqnarray*}
where
\begin{equation*}
Q_{3} = - B_{(a;b)} \dot{q}^{a} \dot{q}^{b} + B_{a}V^{,a}, \enskip Q_{4} = - t B_{(a;b)} \dot{q}^{a} \dot{q}^{b} + B_{a}\dot{q}^{a} + t
B_{a}V^{,a}, \enskip
Q_{5} = tL_{a}\dot{q}^{a} + \frac{s_{1}}{2}t^{2} - \Phi(q)
\end{equation*}
are independent FIs.

\underline{\textbf{Subcase $\mathbf{(n=2, m=2)}$.}} $g = c_0 + c_1 t + c_2t^2$ and $f = d_0 + d_1 t + d_2 t^2$ with $c_2d_2 \neq 0$.

\begin{equation*}
\begin{cases}
\eqref{FL.1.a} \implies (c_1 + 2c_2 t) C_{ab} + (d_0 + d_1t + d_2 t^2)
L_{(a;b)} + B_{(a;b)} = 0 \\
\eqref{FL.1.b} \implies - 2 (c_0 + c_1 t + c_2 t^2) C_{ab} V^{,b} + (d_1 +
2d_2 t) L_a + K_{,a} = 0 \\
\eqref{FL.1.c} \implies K_{,t} = \left( d_0 + d_1 t + d_2 t^2 \right) L_a
V^{,a} + B_a V^{,a} \\
\eqref{FL.1.d} \implies 2d_2 L_a + \left( d_0 + d_1 t + d_2 t^2 \right)
\left( L_b V^{,b} \right)_{;a} + \left( B_b V^{,b} \right)_{;a} - 2 (c_1 + 2
c_2 t) C_{ab} V^{,b} = 0.
\end{cases}
\end{equation*}

From \eqref{FL.1.a}, we get $C_{ab} = 0$ and the vectors $L_a$, $B_a$ are KVs.

From \eqref{FL.1.d}, we find that $L_aV^{,a}=s_{1}$ and $L_a = - \frac{1}{2d_2} \left(B_{b}V^{,b}\right)_{;a}$, that is, $L_a$ is a gradient KV.

The solution of \eqref{FL.1.c} is $K = s_{1} \left( d_{0}t + \frac{d_{1}}{2}t^{2} + \frac{d_2}{3} t^3 \right) +
B_{a}V^{,a}t + G(q)$ which when substituted into \eqref{FL.1.b} gives
\begin{equation*}
G_{,a} + d_1 L_a \underbrace{+ 2d_2 L_a t + \left( B_{b}V^{,b} \right)_{,a} t%
}_{=0} = 0 \implies G_{,a} = -d_1 L_a = \frac{d_1}{2d_2} \left(B_b
V^{,b}\right)_{,a} \implies G(q) = \frac{d_1}{2d_2} B_a V^{,a}.
\end{equation*}

The QFI is
\begin{equation*}
I_{22} = \left(d_{0}+d_{1}t+d_2t^2\right) L_{a}\dot{q}^{a} + B_{a}\dot{q}%
^{a} + s_{1} \left( d_{0}t + \frac{d_{1}}{2}t^{2} + \frac{d_2}{3} t^3
\right) + B_{a}V^{,a}t + \frac{d_1}{2d_2} B_a V^{,a}
\end{equation*}
where $L_{a} = - \frac{1}{2d_2} \left(B_{b}V^{,b}\right)_{;a}$ is a gradient KV such that $L_{a}V^{,a}=s_{1}$, and $B_a$ is a KV.

The QFI $I_{22}= d_{0}Q_{2} + d_{1}Q_{5} + F_{1}$. This expression contains the new independent LFI $F_{1} = t^{2} X_{a}\dot{q}^{a} + \frac{s}{3} t^{3} + B_{a}\dot{q}^{a} + B_{a}V^{,a}t$, where $X_{a}\equiv d_{2}L_{a}$ and $s\equiv d_{2}s_{1}$

\underline{\textbf{Subcase $\mathbf{(n=m>2)}$.}} $c_nd_n \neq 0$.

\begin{equation*}
\begin{cases}
\eqref{FL.1.a} \implies (c_1 + 2c_2 t + ... + n c_n t^{n-1}) C_{ab} + (d_0 +
d_1t + .. + d_n t^n) L_{(a;b)} + B_{(a;b)} = 0 \\
\eqref{FL.1.b} \implies - 2 (c_0 + c_1 t + ... + c_n t^n) C_{ab} V^{,b} +
(d_1 + 2d_2 t + ... + n d_n t^{n-1}) L_a + K_{,a} = 0 \\
\eqref{FL.1.c} \implies K_{,t} = \left( d_0 + d_1 t + ... + d_n t^n \right)
L_a V^{,a} + B_a V^{,a} \\
\eqref{FL.1.d} \implies \left[2d_2 + 3 \cdot 2 d_3 t + ... + n (n-1) d_n
t^{n-2}\right] L_a + \left( d_0 + d_1 t + ... + d_n t^n \right) \left( L_b
V^{,b} \right)_{;a} + \\
\qquad \qquad \quad +\left( B_b V^{,b} \right)_{;a} -2 (c_1 + 2c_2 t + ... + n c_n t^{n-1}) C_{ab} V^{,b} = 0.
\end{cases}
\end{equation*}

Equation \eqref{FL.1.a} implies that $C_{ab} = 0$ and $L_a$, $B_a$ are KVs.

From \eqref{FL.1.d}, we find that $L_a = 0$ and $B_a V^{,a} = s_2$.

The solution of \eqref{FL.1.c} is $K = s_2 t + G(q)$ which when substituted into \eqref{FL.1.b} gives $G = const$. Such a constant is ignored because any constant can be added to $I$ without changing the condition $\frac{dI}{dt}=0$.

The FI is (of the form $Q_{2}$) $I_{nn}(n>2)= B_{a}\dot{q}^{a}+s_{2}t$, where $B_{a}$ is a KV such that $B_{a}V^{,a}=s_{2}$.
\bigskip

We continue with the case $n>m$. This case is broken down equivalently into the cases $n=m+1$ and $n>m+1$. Both cases are analyzed below\footnote{It is much more convenient to follow this line of analysis because for $n > m+1$ equation \eqref{FL.1.a} implies directly that $C_{ab} =0$ and the derived FIs are linear. For $n=m+1$ all the derived FIs are quadratic.}.
\bigskip

\underline{\textbf{I.2. Case $\mathbf{n=m+1}$.}}
\bigskip

\underline{\textbf{Subcase $\mathbf{(n=1, m=0)}$.}} $g = c_0 + c_1 t$ and $f =d_0$ with $c_1 \neq 0$.

\begin{equation*}
\begin{cases}
\eqref{FL.1.a} \implies c_1 C_{ab} + \tilde{L}_{(a;b)} = 0 \\
\eqref{FL.1.b} \implies - 2 c_1 C_{ab} V^{,b} t - 2 c_0 C_{ab} V^{,b} +K_{,a} = 0 \\
\eqref{FL.1.c} \implies K_{,t} - \tilde{L}_a V^{,a} = 0 \\
\eqref{FL.1.d} \implies \left( \tilde{L}_b V^{,b} \right)_{;a} - 2 c_1 C_{ab}V^{,b} = 0.
\end{cases}
\end{equation*}

Solving \eqref{FL.1.c}, we get $K = \bar{L}_a V^{,a} t + G(q)$ which when substituted into \eqref{FL.1.b} gives $G_{,a} = 2 c_0 C_{ab} V^{,b}$. However, $2 C_{ab} V^{,b} = \frac{1}{c_1} \left( \tilde{L}_b V^{,b} \right)_{;a}$; therefore, $G_{,a} = \frac{c_0}{c_1} \left( \tilde{L}_b V^{,b} \right)_{,a} \implies G(q)= \frac{c_0}{c_1} \tilde{L}_a V^{,a}$.

The QFI is
\begin{equation*}
I_{10} = - \frac{1}{c_1}\left( c_{0} + c_{1}t \right) \tilde{L}_{(a;b)}\dot{q}^{a}\dot{q}^{b} + \tilde{L}_{a}\dot{q}^{a}+\left( t + \frac{c_{0}}{c_{1}}\right) \tilde{L}_{a}V^{,a}
\end{equation*}%
where $\tilde{L}_{a}$ is a vector such that $\tilde{L}_{(a;b)}$ is a reducible KT and $\left( \tilde{L}_b V^{,b}\right)_{;a} = - 2 \tilde{L}_{(a;b)} V^{,b}$.

We note that $I_{10} = \frac{c_{0}}{c_{1}} Q_{3}(\tilde{L}_{a}) + Q_{4}(\tilde{L}_{a})$.

\underline{\textbf{Subcase $\mathbf{(n=2, m=1)}$.}} $g = c_0 + c_1 t + c_2t^2$ and $f = d_0 + d_1 t$ with $c_2d_1 \neq 0$.

\begin{equation*}
\begin{cases}
\eqref{FL.1.a} \implies \left( c_1 + 2 c_2 t \right) C_{ab} + \left( d_0 +d_1 t \right) L_{(a;b)} + B_{(a;b)} = 0 \\
\eqref{FL.1.b} \implies - 2 \left( c_0 + c_1 t + c_2 t^2 \right) C_{ab}V^{,b} + d_1 L_a + K_{,a} = 0 \\
\eqref{FL.1.c} \implies K_{,t} = \left( d_0 + d_1 t \right) L_a V^{,a} + B_aV^{,a} \\
\eqref{FL.1.d} \implies \left( d_0 + d_1 t \right) \left( L_b V^{,b}\right)_{;a} + \left( B_b V^{,b} \right)_{;a} - 2 \left( c_1 + 2 c_2 t\right) C_{ab} V^{,b} = 0.
\end{cases}%
\end{equation*}

Equation \eqref{FL.1.a} gives $2c_2 C_{ab} + d_1 L_{(a;b)} = 0$ and $c_1 C_{ab} + d_0 L_{(a;b)} + B_{(a;b)} = 0$.

From \eqref{FL.1.d}, we have that $d_0 \left( L_b V^{,b} \right)_{;a} +
\left( B_b V^{,b} \right)_{;a} - 2 c_1 C_{ab} V^{,b} = 0$ and $d_1 \left(
L_b V^{,b} \right)_{;a} = 4 c_2 C_{ab} V^{,b}$.

Solving \eqref{FL.1.c}, we find that $K=\left( d_{0}t+\frac{d_{1}}{2}t^{2}\right) L_{a}V^{,a}+B_{a}V^{,a}t+G(q)$ which when substituted into \eqref{FL.1.b} gives
\begin{equation*}
G_{,a}=2c_{0}C_{ab}V^{,b}-d_{1}L_{a} \implies G_{,a}= \frac{c_0 d_1}{2c_2}
\left( L_b V^{,b} \right)_{,a} - d_1 L_a \implies G(q) = \frac{c_0 d_1}{2c_2}
L_a V^{,a} - d_1 \int L_a dq^a.
\end{equation*}

We note also that: $\begin{cases}
2c_2 C_{ab} + d_1 L_{(a;b)} = 0 \\
c_1 C_{ab} + d_0 L_{(a;b)} + B_{(a;b)} = 0%
\end{cases}
\implies B_{(a;b)} = \left( \frac{2 c_2 d_0}{d_1} - c_1 \right) C_{ab}$,
\begin{equation*}
\begin{cases}
d_0 \left( L_b V^{,b} \right)_{;a} + \left( B_b V^{,b} \right)_{;a} - 2 c_1
C_{ab} V^{,b} = 0 \\
d_1 \left(L_b V^{,b} \right)_{;a} = 4 c_2 C_{ab} V^{,b}%
\end{cases}
\implies \left( B_b V^{,b} \right)_{;a} = 2 c_1 C_{ab} V^{,b} - \frac{4c_2d_0%
}{d_1} C_{ab} V^{,b}
\end{equation*}
and $\frac{c_1 d_1}{2c_2} \left(L_b V^{,b} \right)_{;a} = d_0 \left(L_b
V^{,b} \right)_{;a} + \left(B_b V^{,b} \right)_{;a}$. Therefore
\begin{equation*}
\begin{cases}
\left( B_b V^{,b} \right)_{;a} = - 2 \left( \frac{2c_2d_0}{d_1} - c_1
\right) C_{ab} V^{,b} \\
B_{(a;b)} = \left( \frac{2 c_2 d_0}{d_1} - c_1 \right) C_{ab}%
\end{cases}
\implies \left( B_b V^{,b} \right)_{;a} = - 2 B_{(a;b)} V^{,b} \implies [
B^a, V^{,a} ] \equiv [ \mathbf{B}, \mathbf{\nabla}V ]^a \neq 0.
\end{equation*}

The QFI is
\begin{equation*}
I_{21}= -\frac{d_{1}}{2c_{2}} \left( c_{0}+c_{1}t+c_{2}t^{2}\right) L_{(a;b)} \dot{q}^{a}\dot{q}%
^{b}+\left( d_{0}+d_{1}t\right) L_{a}\dot{q}^{a}+B_{a}\dot{q}^{a}+\left(
d_{0}t+\frac{d_{1}}{2}t^{2}\right) L_{a}V^{,a}+B_{a}V^{,a}t+G(q)
\end{equation*}
where $L_a$ is a vector
such that $L_{(a;b)}$ is a KT and $\left( L_{b}V^{,b}\right)_{;a} = - 2 L_{(a;b)} V^{,b}$, $B_a$ is
a vector satisfying the relations $\left( B_b V^{,b} \right)_{;a} = - 2
B_{(a;b)} V^{,b}$ and $B_{(a;b)}= \frac{2c_2d_{0} - c_{1}d_{1}}{d_{1}}
C_{ab} $, and $G(q)= \frac{c_{0}d_{1}}{2c_{2}}L_{a} V^{,a} - d_{1} \int
L_{a}dq^{a}$.

We observe that $G_{,a} = \frac{c_{0}d_{1}}{2c_{2}} \left(L_{b} V^{,b}\right)_{,a} - d_{1} L_{a}$. Therefore, $L_{a}=\Phi_{,a}$, i.e. $L_{a}$ is a gradient, and the function $G(q)= \frac{c_{0}d_{1}}{2c_{2}}L_{a} V^{,a} - d_{1}\Phi(q)$.

Moreover, the condition $B_{(a;b)}= \left(\frac{ c_{1}d_{1}}{2c_{2}}- d_{0}\right)L_{(a;b)}$ implies that $-\frac{ c_{1}d_{1}}{2c_{2}} L_{(a;b)} = - B_{(a;b)} - d_{0}L_{(a;b)}$ and hence $B_{(a;b)}$ is a KT.

Substituting the above results in the QFI $I_{21}$, we find
\begin{equation*}
I_{21} = \frac{c_{0}d_{1}}{2c_{2}} Q_{3}(L_{a}) + Q_{4} + d_{0}Q_{4}(L_{a})+ d_{1}Q_{6}
\end{equation*}
where
\begin{equation*}
Q_{6} = - \frac{t^{2}}{2} L_{(a;b)} \dot{q}^{a} \dot{q}^{b} + t L_{a} \dot{q}^{a} + \frac{t^{2}}{2} L_{a}V^{,a} - \Phi(q)
\end{equation*}
is a new independent QFI.

We note that the expression
\[
Q_{1} + Q_{6} = - \frac{t^{2}}{2} L_{(a;b)} \dot{q}^{a} \dot{q}^{b} + C_{ab} \dot{q}^{a} \dot{q}^{b} + t L_{a} \dot{q}^{a} + \frac{t^{2}}{2} L_{a}V^{,a} - \Phi(q) + G(q)
\]
where $\Phi_{,a}=L_{a}$ and $G_{,a}=2C_{ab}V^{,b}$, leads to the new independent QFI
\[
Q_{16} = - \frac{t^{2}}{2} L_{(a;b)} \dot{q}^{a} \dot{q}^{b} + C_{ab} \dot{q}^{a} \dot{q}^{b} + t L_{a} \dot{q}^{a} + \frac{t^{2}}{2} L_{a}V^{,a} + H(q)
\]
where $H_{,a} = 2C_{ab}V^{,b} - L_{a}$. The FIs $Q_{1}$ and $Q_{6}$ are derived from $Q_{16}$ as follows: $Q_{16}(C_{ab}=0)=Q_{6}$ and $Q_{16}(L_{a}=0)=Q_{1}$.

\underline{\textbf{Subcase $\mathbf{(n=3, m=2)}$.}} $g = c_0 + c_1 t + c_2t^2 + c_3 t^3$ and $f = d_0 + d_1 t + d_2 t^2$ with $c_3d_2 \neq0 $.

\begin{equation*}
\begin{cases}
\eqref{FL.1.a} \implies \left( c_1 + 2 c_2 t + 3c_3t^2 \right) C_{ab} +
\left( d_0 + d_1 t + d_2 t^2\right) L_{(a;b)} + B_{(a;b)} = 0 \\
\eqref{FL.1.b} \implies - 2 \left( c_0 + c_1 t + c_2 t^2 + c_3 t^3 \right)
C_{ab} V^{,b} + (d_1 + 2d_2 t) L_a + K_{,a} = 0 \\
\eqref{FL.1.c} \implies K_{,t} = \left( d_0 + d_1 t + d_2 t^2 \right) L_a
V^{,a} + B_a V^{,a} \\
\eqref{FL.1.d} \implies 2d_2 L_a + \left( d_0 + d_1 t + d_2 t^2 \right)
\left( L_b V^{,b} \right)_{;a} + \left( B_b V^{,b} \right)_{;a} - 2 \left(
c_1 + 2 c_2 t + 3 c_3 t^2 \right) C_{ab} V^{,b} = 0.%
\end{cases}%
\end{equation*}

From \eqref{FL.1.a}, we have that $3c_3 C_{ab} + d_2 L_{(a;b)} = 0$, $2c_2C_{ab} + d_1 L_{(a;b)} = 0$ and $c_1 C_{ab} + d_0 L_{(a;b)} + B_{(a;b)} = 0$.

From \eqref{FL.1.d}, we find that $d_2 \left( L_b V^{,b} \right)_{;a} = 6 c_3C_{ab} V^{,b}$, $d_1 \left( L_b V^{,b} \right)_{;a} = 4 c_2 C_{ab} V^{,b}$ and $2d_2 L_a + d_0 \left( L_b V^{,b} \right)_{;a} + \left( B_b V^{,b}\right)_{;a} - 2 c_1 C_{ab} V^{,b} = 0$.

The solution of \eqref{FL.1.c} is $K = \left( d_{0}t + \frac{d_{1}}{2}t^{2} + \frac{d_2}{3}t^3 \right)
L_{a}V^{,a} + B_{a}V^{,a}t + G(q)$ which when substituted into \eqref{FL.1.b} and using the above derived conditions gives
\begin{equation*}
G_{,a}= 2c_{0}C_{ab}V^{,b}-d_{1}L_{a} \implies G_{,a}= \frac{c_0 d_2}{3c_3}
\left( L_b V^{,b} \right)_{,a} - d_1 L_a \implies G(q) = \frac{c_0 d_2}{3c_3}
L_a V^{,a} - d_1 \int L_a dq^a.
\end{equation*}

From the first set of conditions, we get:
\begin{equation*}
\begin{cases}
3c_3 C_{ab} + d_2 L_{(a;b)} = 0 \\
2c_2 C_{ab} + d_1 L_{(a;b)} = 0 \\
c_1 C_{ab} + d_0 L_{(a;b)} + B_{(a;b)} = 0%
\end{cases}
\implies
\begin{cases}
C_{ab} = - \frac{d_2}{3c_3} L_{(a;b)} \\
\left( d_1 - \frac{2c_2 d_2}{3c_3} \right) L_{(a;b)} = 0 \\
B_{(a;b)} = \left( \frac{3 c_3 d_0}{d_2} - c_1 \right) C_{ab}%
\end{cases}%
\end{equation*}
and, from the second set, we get:
\begin{equation*}
\begin{cases}
d_2 \left( L_b V^{,b} \right)_{;a} = 6 c_3 C_{ab} V^{,b} \\
d_1 \left(L_b V^{,b} \right)_{;a} = 4 c_2 C_{ab} V^{,b} \\
2d_2 L_a + d_0 \left( L_b V^{,b} \right)_{;a} + \left( B_b V^{,b}
\right)_{;a} - 2 c_1 C_{ab} V^{,b} = 0%
\end{cases}
\implies
\begin{cases}
\left( L_b V^{,b} \right)_{;a} = \frac{6 c_3}{d_2} C_{ab} V^{,b} \\
\left( \frac{6 c_3 d_1}{d_2} - 4c_2 \right) C_{ab} V^{,b} = 0 \\
L_a = \left( \frac{c_1}{6c_3} - \frac{d_0}{2d_2} \right) \left( L_b V^{,b}
\right)_{;a} - \frac{1}{2d_2} \left( B_b V^{,b} \right)_{;a}.%
\end{cases}%
\end{equation*}
Therefore, $L_a$ is a gradient vector and the function $G(q) = \left(\frac{c_0 d_2}{3c_3} - \frac{c_1 d_1}{6c_3} + \frac{d_0d_1}{2d_2} \right) L_a V^{,a} + \frac{d_1}{2d_2} B_a V^{,a}$.

The QFI is
\begin{eqnarray*}
I_{32} &=& - \frac{d_2}{3c_3} \left( c_{0}+c_{1}t+c_{2}t^{2} + c_3t^3 \right) L_{(a;b)} \dot{q}^{a}%
\dot{q}^{b}+ \left( d_{0}+d_{1}t + d_2t^2\right) L_{a}\dot{q}^{a} + B_{a}%
\dot{q}^{a} + \\
&& + \left( d_{0}t + \frac{d_{1}}{2}t^{2} + \frac{d_2}{3}t^3
\right) L_{a}V^{,a} + B_a V^{,a}t + \left(\frac{2c_0 d_2 - c_1 d_1}{6c_3} + \frac{d_0d_1}{2d_2%
} \right) L_a V^{,a} + \frac{d_1}{2d_2} B_a V^{,a}.
\end{eqnarray*}
where the vector $L_{a}=\left( \frac{c_{1}}{6c_{3}}-\frac{d_{0}}{2d_{2}}\right)
\left(L_{b}V^{,b}\right) _{;a}-\frac{1}{2d_{2}}\left( B_{b}V^{,b}\right)_{;a}$ is a gradient
such that $L_{(a;b)}$ is a KT, \\ $\left( \frac{2c_{2}d_{2}}{3c_{3}}- d_{1}\right) L_{(a;b)} =0$ and
$\left( L_{b}V^{,b}\right) _{;a}=-2L_{(a;b)}V^{,b}$, and $B_{a}$ is a vector satisfying the relation $B_{(a;b)}=\left( \frac{c_{1}d_{2}}{3c_{3}}-d_{0}\right) L_{(a;b)}$.

The vector $L_{a}=\Psi_{,a}$ where $\Psi(q) = \left( \frac{c_{1}}{6c_{3}}-\frac{d_{0}}{2d_{2}}\right)
L_{a}V^{,a} - \frac{1}{2d_{2}} B_{a}V^{,a}$.

We observe also that $\frac{c_{2}d_{2}}{3c_{3}} L_{(a;b)} = \frac{d_{1}}{2} L_{(a;b)}$ and $\frac{c_{1}d_{2}}{3c_{3}} L_{(a;b)} = B_{(a;b)} +d_{0} L_{(a;b)}$. The last condition implies that $B_{(a;b)}$ is a reducible KT.

Another useful relation is the following\footnote{This is the condition which must be satisfied in order $Q_{7}$ to be a FI.}:
\begin{equation*}
L_{a} = \Psi_{,a} \implies 2d_{2}L_{a} = \left( \frac{c_{1}d_{2}}{3c_{3}}%
-d_{0}\right) \left(L_{b}V^{,b}\right) _{,a}-\left( B_{b}V^{,b}\right)_{,a}
\implies
\end{equation*}
\begin{equation*}
\left( B_{b}V^{,b}\right)_{,a} = -2B_{(a;b)}V^{,b} -2d_{2}L_{a}.
\end{equation*}

Substituting the above relations in the QFI $I_{32}$, we find
\begin{equation*}
I_{32} = \frac{d_{2}c_{0}}{3c_{3}}Q_{3}(L_{a}) + Q_{7} + d_{0}Q_{4}(L_{a}) +
d_{1}Q_{6}(\Psi)
\end{equation*}
where
\begin{equation*}
Q_{7} = - \frac{t^{3}}{3} d_{2}L_{(a;b)}\dot{q}^{a}\dot{q}^{b} + t^{2}
d_{2}L_{a}\dot{q}^{a} + \frac{t^{3}}{3} d_{2}L_{a}V^{,a} - t B_{(a;b)} \dot{q%
}^{a}\dot{q}^{b} + B_{a}\dot{q}^{a} + t B_{a}V^{,a}
\end{equation*}
is a new independent QFI.

\underline{\textbf{Subcase $\mathbf{(n=m+1,m>2)}$.}} $c_{n}\neq
0$ and $d_{m}\neq 0$.

\begin{equation*}
\begin{cases}
\eqref{FL.1.a} \implies \left[ c_1 + 2 c_2 t + ... + (m+1) c_n t^{m} \right]
C_{ab} + \left( d_0 + d_1 t + ... + d_m t^m \right) L_{(a;b)} + B_{(a;b)} = 0
\\
\eqref{FL.1.b} \implies - 2 \left( c_0 + c_1 t + ... + c_n t^{m+1} \right)
C_{ab} V^{,b} + (d_1 + 2d_2 t + ... + m d_m t^{m-1}) L_a + K_{,a} = 0 \\
\eqref{FL.1.c} \implies K_{,t} = \left( d_0 + d_1 t + ... + d_m t^m \right)
L_a V^{,a} + B_a V^{,a} \\
\eqref{FL.1.d} \implies \left[ 2d_2 + 3 \cdot 2 d_3 t + ... + m (m-1) d_m
t^{m-2} \right] L_a + \left( d_0 + d_1 t + ... + d_m t^m \right) \left( L_b
V^{,b} \right)_{;a} + \\
\qquad \qquad \quad + \left( B_b V^{,b} \right)_{;a}- 2 \left[ c_1 + 2 c_2 t + ... + (m+1) c_n t^{m} \right]
C_{ab} V^{,b} = 0.
\end{cases}%
\end{equation*}

From \eqref{FL.1.a}, we find the conditions: $(k+1) c_{k+1} C_{ab} + d_k
L_{(a;b)} = 0$, where $k = 1,2,...,m$, and $c_1 C_{ab} + d_0 L_{(a;b)} +
B_{(a;b)} = 0$. For $k=m$, we get $C_{ab} = - \frac{d_m}{(m+1)c_{m+1}} L_{(a;b)}$ and the remaining conditions become:
\begin{equation*}
\left[ d_k - \frac{(k+1)c_{k+1}d_m}{(m+1)c_{m+1}} \right] L_{(a;b)} = 0, \enskip k=1,2,...,m-1
\end{equation*}
and
\begin{equation*}
B_{(a;b)} = \left[ \frac{c_1d_m}{(m+1)c_{m+1}} - d_0 \right] L_{(a;b)}.
\end{equation*}

From \eqref{FL.1.d}, we find that $2(k+1)c_{k+1} C_{ab} V^{,b} = d_k \left(
L_b V^{,b} \right)_{;a}$, where $k=m-1,m$, $(k+2)(k+1)d_{k+2} L_a + d_k
\left( L_b V^{,b} \right)_{;a} - 2 (k+1) c_{k+1} C_{ab} V^{,b} = 0$, where $%
k=1,...,m-2$, and $2d_2 L_a + d_0 \left(L_b V^{,b} \right)_{;a} + \left( B_b
V^{,b} \right)_{;a} - 2 c_1 C_{ab} V^{,b} = 0$, i.e. $L_a$ is a gradient
vector.

The first set of equations, for $k=m$, gives
\begin{equation*}
\left( L_b V^{,b} \right)_{,a} = \frac{2(m+1)c_{m+1}}{d_m} C_{ab} V^{,b}\implies \left( L_b V^{,b} \right)_{,a} = - 2 L_{(a;b)} V^{,b}
\end{equation*}
and, for $k=m-1$, $\left[ d_{m-1} - \frac{m c_m d_m}{(m+1)c_{m+1}}\right] \left( L_b V^{,b}\right)_{,a} = 0$.

The second set of equations, for $k=m-2$, gives
\begin{equation*}
L_a = \left[ \frac{c_{m-1}}{m(m+1)c_{m+1}} - \frac{d_{m-2}}{m (m-1) d_m} \right] \left( L_b V^{,b} \right)_{,a}
\end{equation*}
and for the remaining values of $k=1,2,...,m-3$ (exist only for $m>3$)
\begin{equation*}
\left\{ d_k + (k+2)(k+1)d_{k+2} \left[ \frac{c_{m-1}}{m(m+1)c_{m+1}} - \frac{%
d_{m-2}}{m (m-1) d_m} \right] - \frac{(k+1) c_{k+1} d_m}{(m+1)c_{m+1}}
\right\} \left( L_b V^{,b} \right)_{,a} = 0.
\end{equation*}

The third set of equations gives
\begin{equation*}
\left( B_b V^{,b} \right)_{,a} = \left[ \frac{c_1 d_m}{(m+1) c_{m+1}} -\frac{2 d_2 c_{m-1}}{m(m+1)c_{m+1}} + \frac{2d_2 d_{m-2}}{m (m-1) d_m} - d_0 \right] \left( L_b V^{,b} \right)_{,a}.
\end{equation*}

The solution of \eqref{FL.1.c} is $K = \left( d_{0}t + \frac{d_{1}}{2}t^{2} + ... + \frac{d_m}{m+1}t^{m+1}
\right) L_{a}V^{,a} + B_{a}V^{,a}t + G(q)$ which when substituted into \eqref{FL.1.b} gives\footnote{Use the conditions: $2d_2 L_a + d_0 \left(L_b
V^{,b} \right)_{;a} + \left( B_b V^{,b} \right)_{;a} - 2 c_1 C_{ab} V^{,b} =
0$, $2(k+1)c_{k+1} C_{ab} V^{,b} = d_k \left( L_b V^{,b} \right)_{;a}$, where
$k=m-1,m$, and $(k+2)(k+1)d_{k+2} L_a + d_k \left( L_b V^{,b} \right)_{;a} -
2 (k+1) c_{k+1} C_{ab} V^{,b} = 0$, where $k=1,...,m-2$.}
\begin{align*}
G_{,a} =& - \left( d_{0}t + \frac{d_{1}}{2}t^{2}+ ... + \frac{d_{m-1}}{m}%
t^{m} + \frac{d_m}{m+1}t^{m+1} \right) \left( L_{b}V^{,b} \right)_{,a} + 2
\left( c_0 + c_1 t + ...+ c_m t^m + c_{m+1} t^{m+1} \right) C_{ab} V^{,b} -
\\
& - \left( B_{b} V^{,b} \right)_{,a} t - (d_1 + 2d_2 t + ... + m d_m
t^{m-1}) L_a \\
=& 2 c_0 C_{ab} V^{,b} - d_1 L_a \\
=& \left[ \frac{c_0d_m}{(m+1)c_{m+1}} - \frac{c_{m-1} d_1}{m(m+1)c_{m+1}} +
\frac{d_1 d_{m-2}}{m (m-1) d_m} \right] \left( L_{b}V^{,b} \right)_{,a}
\implies
\end{align*}
\begin{equation*}
G(q)= \left[ \frac{c_0d_m}{(m+1)c_{m+1}} - \frac{c_{m-1} d_1}{%
m(m+1)c_{m+1}} + \frac{d_1 d_{m-2}}{m (m-1) d_m} \right] L_a V^{,a}.
\end{equation*}
Therefore, $K= \left( d_{0}t + \frac{d_{1}}{2}t^{2} + ... + \frac{d_m}{m+1}t^{m+1}
\right) L_{a}V^{,a} + B_{a}V^{,a}t +\left[ \frac{c_0d_m}{(m+1)c_{m+1}} -
\frac{c_{m-1} d_1}{m(m+1)c_{m+1}} + \frac{d_1 d_{m-2}}{m (m-1) d_m} \right]
L_a V^{,a}$.

The QFI is
\begin{eqnarray*}
I_{(m+1)m}(m>2) &=&-\frac{d_{m}}{(m+1)c_{m+1}}\left(
c_{0}+c_{1}t+...+c_{m+1}t^{m+1}\right)L_{(a;b)} \dot{q}^{a} \dot{q}^{b} + \\
&& +\left( d_{0}+d_{1}t+...+d_{m}t^{m}\right) L_{a} \dot{q}^{a} +B_{a}\dot{q}^{a}+B_{a}V^{,a}t+ \\
&& +\left( d_{0}t+\frac{d_{1}}{2}t^{2}+...+\frac{d_{m}}{m+1}%
t^{m+1}\right) L_{a}V^{,a}+G(q)
\end{eqnarray*}
where $c_{m+1}d_{m}\neq 0$ for a finite $m>2$. The vector
\begin{equation*}
L_{a}=\left[ \frac{c_{m-1}}{m(m+1)c_{m+1}}-\frac{d_{m-2}}{m(m-1)d_{m}}\right]
\left( L_{b}V^{,b}\right) _{,a}
\end{equation*}
is a gradient such that $L_{(a;b)}$ is a KT, $\left( L_{b}V^{,b}\right) _{,a}=-2L_{(a;b)}V^{,b}$, $\left[ d_{k}-
\frac{(k+1)c_{k+1}d_{m}}{(m+1)c_{m+1}}\right] L_{(a;b)}$ $=0$ with $k=1,2,...,m-1$, $\left[ d_{m-1}-\frac{mc_{m}d_{m}}{(m+1)c_{m+1}}\right] \left(
L_{b}V^{,b}\right) _{,a}$ $=0$ and
\begin{equation*}
\left\{ d_{k}+(k+2)(k+1)d_{k+2}\left[ \frac{c_{m-1}}{m(m+1)c_{m+1}}-\frac{%
d_{m-2}}{m(m-1)d_{m}}\right] -\frac{(k+1)c_{k+1}d_{m}}{(m+1)c_{m+1}}\right\}
\left( L_{b}V^{,b}\right) _{,a}=0
\end{equation*}%
with $k=1,2,...,m-3$. The vector $B_{a}$ satisfies the conditions
$B_{(a;b)}=\left[ \frac{c_{1}d_{m}}{(m+1)c_{m+1}}-d_{0}\right] L_{(a;b)}$ and
\begin{equation*}
\left( B_{b}V^{,b}\right) _{,a}=\left[ \frac{c_{1}d_{m}}{(m+1)c_{m+1}}%
-\right. \left. \frac{2d_{2}c_{m-1}}{m(m+1)c_{m+1}}+\frac{2d_{2}d_{m-2}}{%
m(m-1)d_{m}}-d_{0}\right] \left( L_{b}V^{,b}\right) _{,a}.
\end{equation*}%
The function $G(q)=\left[ \frac{c_{0}d_{m}}{(m+1)c_{m+1}}-\frac{c_{m-1}d_{1}}{m(m+1)c_{m+1}%
}+\frac{d_{1}d_{m-2}}{m(m-1)d_{m}}\right] L_{a}V^{,a}$.

However, the considered case $m>2$ implies that $m-1>1$ and $m-2>0$; therefore, there always exist $k$-values for $m-1$ and $m-2$.

For $k=m-2$, the condition $\left[ d_{k}-\frac{(k+1) c_{k+1}d_{m}}{(m+1)c_{m+1}}\right] L_{(a;b)}=0$ gives
\begin{equation*}
\left[ d_{m-2}-\frac{(m-1) c_{m-1}d_{m}}{(m+1)c_{m+1}}\right] L_{(a;b)}=0 \implies \left[ \frac{c_{m-1}}{m(m+1)c_{m+1}}-\frac{d_{m-2}}{m(m-1)d_{m}}\right]
\left( L_{b}V^{,b}\right) _{,a} = 0 \implies L_{a}=0
\end{equation*}
because $\left( L_{b}V^{,b}\right)_{,a}=-2L_{(a;b)}V^{,b}$.

Since $L_{a}=0$, we have $G=0$, $B_{(a;b)}=0$ (i.e. $B_{a}$ is a KV) and $B_{a}V^{,a}=s =const$. 

Therefore, $I_{(m+1)m}(m>2) = B_{a}\dot{q}^{a} + st = Q_{2}(B_{a})$.
\bigskip

\underline{\textbf{I.3. Case $\mathbf{n>m+1}$}.}
\bigskip

\underline{\textbf{Subcase $\mathbf{(n>1, m=0)}$.}} $c_n \neq 0$.
\begin{equation*}
\begin{cases}
\eqref{FL.1.a} \implies (c_1 + 2 c_2 t + ... + n c_n t^{n-1}) C_{ab} + \tilde{L}_{(a;b)} = 0 \\
\eqref{FL.1.b} \implies -2 (c_0 + c_1 t + ... + c_n t^n) C_{ab} V^{,b} +
K_{,a} = 0 \\
\eqref{FL.1.c} \implies K_{,t} = \tilde{L}_b V^{,b} \\
\eqref{FL.1.d} \implies \left( \tilde{L}_b V^{,b} \right)_{;a} - 2 (c_1 + 2c_2 t + ... + n c_n t^{n-1}) C_{ab} V^{,b} = 0.
\end{cases}
\end{equation*}

From \eqref{FL.1.a}, we find that $C_{ab} = 0$ and $\tilde{L}_{(a;b)} = 0$, i.e. $\tilde{L}_a$ is a KV. Then,  \eqref{FL.1.d} gives $\tilde{L}_aV^{,a} = s_0$, and \eqref{FL.1.c} yields $K = s_0 t + G(q)$. The last result when substituted into \eqref{FL.1.b} gives $G_{,a} = 0$ $\implies G= const \equiv 0$.

The FI is (of the form $Q_{2}$) $I_{n0}(n>1)= \tilde{L}_{a}\dot{q}^{a}+s_{0}t$, where $\tilde{L}_{a}\equiv d_{0}L_{a}+B_{a}$ is a KV such that $\tilde{L}_aV^{,a} = s_0$.

\underline{\textbf{Subcase $\mathbf{(n>2, m=1)}$.}} $c_nd_1\neq 0$.
\begin{equation*}
\begin{cases}
\eqref{FL.1.a} \implies \left( c_1 + 2 c_2 t + ... + n c_n t^{n-1} \right)
C_{ab} + \left( d_0 + d_1 t \right) L_{(a;b)} + B_{(a;b)} =0 \\
\eqref{FL.1.b} \implies - 2 \left( c_0 + c_1 t + ... + c_n t^n \right)
C_{ab} V^{,b} + d_1 L_a + K_{,a} = 0 \\
\eqref{FL.1.c} \implies K_{,t} = \left( d_0 + d_1 t \right) L_a V^{,a} + B_a
V^{,a} \\
\eqref{FL.1.d} \implies \left( d_0 + d_1 t \right) \left( L_b V^{,b}
\right)_{;a} + \left( B_b V^{,b} \right)_{;a} - 2 \left( c_1 + 2 c_2 t + ...
+ n c_n t^{n-1} \right) C_{ab} V^{,b} = 0.%
\end{cases}%
\end{equation*}

From \eqref{FL.1.a}, we find that $C_{ab} = 0$ and $L_a$, $B_a$ are KVs.

From \eqref{FL.1.d}, we have that $L_a V^{,a} = s_1$ and $B_a V^{,a} = s_2$, where $s_{1}$ and $s_{2}$ are arbitrary constants.

Then, equation \eqref{FL.1.c} gives $K = s_1 \left( d_0t + \frac{d_1}{2} t^2 \right) + s_2 t + G(q)$ which when substituted into \eqref{FL.1.b} yields $G_{,a} = - d_1 L_a$, that is, $L_a$ is a gradient KV.

The FI is (consists of FIs of the form $Q_{2}$ and $Q_{5}$)
\begin{equation*}
I_{n1}(n>2)=(d_{0}+d_{1}t)L_{a}\dot{q}^{a}+B_{a}\dot{q}^{a}+ \frac{s_{1}d_{1}%
}{2}t^{2}+(s_{1}d_{0}+s_{2})t -d_{1}\int L_{a}dq^{a}
\end{equation*}%
where $L_{a}$ and $B_{a}$ are KVs such that $L_{a}V^{,a}=s_{1}$ and $B_{a}V^{,a}=s_{2}$.

\underline{\textbf{Subcase $\mathbf{(n>3, m=2)}$.}} $c_n d_2
\neq 0$.

\begin{equation*}
\begin{cases}
\eqref{FL.1.a} \implies \left( c_1 + 2 c_2 t + ... + n c_n t^{n-1} \right)
C_{ab} + \left( d_0 + d_1 t + d_2 t^2 \right) L_{(a;b)} + B_{(a;b)} =0 \\
\eqref{FL.1.b} \implies - 2 \left( c_0 + c_1 t + ... + c_n t^n \right)
C_{ab} V^{,b} + (d_1 + 2d_2t) L_a + K_{,a} = 0 \\
\eqref{FL.1.c} \implies K_{,t} = \left( d_0 + d_1 t + d_2 t^2 \right) L_a
V^{,a} + B_a V^{,a} \\
\eqref{FL.1.d} \implies 2d_2 L_a + \left( d_0 + d_1 t + d_2t^2 \right)
\left( L_b V^{,b} \right)_{;a} + \left( B_b V^{,b} \right)_{;a} - \\
\qquad \qquad - 2 \left( c_1 + 2 c_2 t + ... + n c_n t^{n-1} \right) C_{ab}
V^{,b} = 0.%
\end{cases}%
\end{equation*}

From \eqref{FL.1.a}, we find that $C_{ab} = 0$ and $L_a$, $B_a$ are KVs.

From \eqref{FL.1.d}, we have that $L_a V^{,a} = s_1$ and $L_a = - \frac{1}{%
2d_2} \left( B_b V^{,b} \right)_{;a}$, that is, $L_a$ is a gradient KV.

Then, equation \eqref{FL.1.c} gives $K = s_1 \left( d_0t + \frac{d_1}{2} t^2 + \frac{d_2}{3} t^3 \right) + B_a
V^{,a} t + G(q)$ which when substituted into \eqref{FL.1.b} yields
\begin{equation*}
G_{,a} = - d_1 L_a = \frac{d_1}{2d_2} \left( B_b V^{,b} \right)_{,a}
\implies G(q) = \frac{d_1}{2d_2} B_a V^{,a}.
\end{equation*}

The FI is (consists of FIs of the form $Q_{2}$, $Q_{5}$ and $Q_{7}$)
\begin{equation*}
I_{n2}(n>3)=(d_{0}+d_{1}t+d_2t^2) L_{a}\dot{q}^{a} + B_{a} \dot{q}^{a} + s_1
\left( d_0t + \frac{d_1}{2} t^2 + \frac{d_2}{3} t^3 \right) + B_a V^{,a} t +
\frac{d_1}{2d_2} B_a V^{,a}
\end{equation*}
where $L_a = - \frac{1}{2d_2} \left( B_b V^{,b} \right)_{;a}$ is a gradient
KV such that $L_a V^{,a} = s_1$ and $B_{a}$ is a KV.

\underline{\textbf{Subcase $\mathbf{(n>m+1, m>2)}$.}} $c_nd_m
\neq 0$. Note that $n > n-1 > m > 2$.
\begin{equation*}
\begin{cases}
\eqref{FL.1.a} \implies \left( c_1 + 2 c_2 t + ... + n c_n t^{n-1} \right)
C_{ab} + \left( d_0 + d_1 t + ... + d_m t^m \right) L_{(a;b)} + B_{(a;b)} = 0
\\
\eqref{FL.1.b} \implies - 2 \left( c_0 + c_1 t + ... + c_n t^{n} \right)
C_{ab} V^{,b} + (d_1 + 2d_2 t + ... + m d_m t^{m-1}) L_a + K_{,a} = 0 \\
\eqref{FL.1.c} \implies K_{,t} = \left( d_0 + d_1 t + ... + d_m t^m \right)
L_a V^{,a} + B_a V^{,a} \\
\eqref{FL.1.d} \implies \left[ 2d_2 + 3 \cdot 2 d_3 t + ... + m (m-1) d_m
t^{m-2} \right] L_a + \left( d_0 + d_1 t + ... + d_m t^m \right) \left( L_b
V^{,b} \right)_{;a} + \\
\qquad \qquad \quad +\left( B_b V^{,b} \right)_{;a} - 2 \left( c_1 + 2 c_2 t + ... + n c_n t^{n-1} \right) C_{ab}
V^{,b} = 0.
\end{cases}%
\end{equation*}

From \eqref{FL.1.a}, we find that $C_{ab} = 0$ and $L_a$, $B_a$ are KVs.

From \eqref{FL.1.d}, we have that $L_a = 0$ and $B_a V^{,a} = s_2$.

Then, the solution of \eqref{FL.1.c} is $K = s_2 t + G(q)$ which when substituted into \eqref{FL.1.b} gives $G= const \equiv 0$.

The FI is (again of the form $Q_{2}$) $I_{nm}(n>m+1, m>2)= B_a \dot{q}^a + s_2 t$, where $B_{a}$ is a KV such that $B_a V^{,a} = s_2$.
\bigskip

\underline{\textbf{I.4. Case $\mathbf{n<m}$.}}
\bigskip

\underline{\textbf{Subcase $\mathbf{(n=0, m=1)}$.}} $g = c_0$ and $f = d_0 +d_1 t$ with $d_1 \neq 0$.
\begin{equation*}
\begin{cases}
\eqref{FL.1.a} \implies (d_0 + d_1 t) L_{(a;b)} + B_{(a;b)} = 0 \\
\eqref{FL.1.b} \implies -2 c_0 C_{ab} V^{,b} + d_1 L_a + K_{,a} = 0 \\
\eqref{FL.1.c} \implies K_{,t} = (d_0 + d_1t) L_b V^{,b} + B_b V^{,b} \\
\eqref{FL.1.d} \implies (d_0 + d_1t) \left( L_b V^{,b} \right)_{;a} + \left(
B_b V^{;b} \right)_{;a} = 0.%
\end{cases}%
\end{equation*}

Equation \eqref{FL.1.a} implies that $L_a$ and $B_a$ are KVs.

Equation \eqref{FL.1.d} gives $L_b V^{,b} = s_1 = const$ and $B_b V^{,b} = s_2 =const$.

Then, \eqref{FL.1.c} gives $K = s_1 \left(d_0 t + \frac{d_1}{2}t^2\right) + s_2 t + G(q)$ which when substituted into \eqref{FL.1.b} yields $G_{,a} = 2c_0C_{ab}V^{,b} - d_1 L_a$.

The QFI is ($c_{0}$ is absorbed by $C_{ab}$)
\begin{equation*}
I_{01} = C_{ab}\dot{q}^{a}\dot{q}^{b}+ \left( d_{0}+d_{1}t\right)
L_{a}\dot{q}^{a} + B_{a}\dot{q}^{a} + s_{1}\left( d_{0}t + \frac{d_{1}}{2}%
t^{2} \right)+ s_{2}t + G\left( q\right)
\end{equation*}
where $d_1 \neq 0$, $L_{a}$ and $B_{a}$ are KVs such that $L_{a}V^{,a}=s_{1}$ and $B_{a}V^{,a}=s_{2} $, and $C_{ab}$ is a KT such that $G_{,a}= 2 C_{ab} V^{,b} -d_1L_{a}$.

We see that $I_{01} = Q_{2}(B_{a}) + d_{0}Q_{2} + Q_{8}$, where
\begin{equation*}
Q_{8} = C_{ab}\dot{q}^{a}\dot{q}^{b} + d_{1}tL_{a}\dot{q}^{a} + d_{1} \frac{s_{1}}{2} t^{2} + G(q)
\end{equation*}
is just a subcase of $Q_{16}$ because $d_{1}L_{a}$ is a KV.

\underline{\textbf{Subcase $\mathbf{(n=0, m=2)}$.}} $g = c_0$ and $f = d_0 +d_1 t + d_2 t^2 $ with $d_2 \neq 0$.
\begin{equation*}
\begin{cases}
\eqref{FL.1.a} \implies \left( d_0 + d_1 t + d_2 t^2 \right) L_{(a;b)} +
B_{(a;b)} = 0 \\
\eqref{FL.1.b} \implies -2 c_0 C_{ab} V^{,b} + \left( d_1 + 2d_2 t \right)
L_a + K_{,a} = 0 \\
\eqref{FL.1.c} \implies K_{,t} = \left( d_0 + d_1t + d_2 t^2 \right) L_b
V^{,b} + B_b V^{,b} \\
\eqref{FL.1.d} \implies 2d_2 L_a + \left( d_0 + d_1t + d_2 t^2 \right)
\left( L_b V^{,b} \right)_{;a} + \left( B_b V^{,b} \right)_{;a}= 0.%
\end{cases}%
\end{equation*}

From \eqref{FL.1.a}, we have that $L_a$ and $B_a$ are KVs.

From \eqref{FL.1.d}, we get $L_{b}V^{,b}=s_{1}$ and $L_{a}= -\frac{1}{2d_{2}}
\left( B_{b}V^{,b}\right)_{,a}$, that is, $L_{a}$ is a gradient KV.

Equation \eqref{FL.1.c} yields $K = s_1 \left( d_0t + \frac{d_1}{2}t^2 + \frac{d_2}{3} t^3 \right) + B_b
V^{,b}t + G(q)$ and \eqref{FL.1.b} gives $G_{,a} = 2 c_0 C_{ab} V^{,b} - d_1L_a$. Using the
relation $L_{a}= -\frac{1}{2d_{2}} \left( B_{b}V^{,b} \right)_{,a}$, we find
that $G(q) = \frac{d_1}{2d_{2}} B_{a} V^{,a} + 2 c_0 \int C_{ab} V^{,b} dq^a$.

The QFI is ($c_{0}$ is absorbed by $C_{ab}$)
\begin{equation*}
I_{02} = C_{ab}\dot{q}^{a}\dot{q}^{b} + \left( d_{0} +
d_{1}t+d_{2}t^{2} \right) L_a \dot{q}^{a} + B_{a}\dot{q}^{a} + s_{1} \left(
d_0t + \frac{d_1}{2}t^{2} + \frac{d_2}{3} t^3 \right) + B_{a}V^{,a}t +
G\left( q\right)
\end{equation*}
where $d_{2}\neq 0$, $B_{a}$ is a KV, $L_{a} = - \frac{1}{2d_{2}} \left(
B_{b}V^{,b} \right)_{,a}$ is a gradient KV such that $L_{a}V^{,a} = s_{1}$,
and $C_{ab}$ is a KT satisfying the relation $G_{,a} - 2C_{ab}V^{,b} +
d_{1}L_{a} = 0$.

We note that $I_{02} = Q_{8} + d_{0}Q_{2} + Q_{7}(L_{a}=KV,B_{a}=KV)$.

\underline{\textbf{Subcase $\mathbf{(n=0, m>2)}$.}} $d_m \neq 0$.

\begin{equation*}
\begin{cases}
\eqref{FL.1.a} \implies \left( d_0 + d_1 t + ... + d_m t^m \right) L_{(a;b)}
+ B_{(a;b)} = 0 \\
\eqref{FL.1.b} \implies -2 c_0 C_{ab} V^{,b} + \left( d_1 + 2d_2 t + ... + m
d_m t^{m-1} \right) L_a + K_{,a} = 0 \\
\eqref{FL.1.c} \implies K_{,t} = \left( d_0 + d_1t + ... + d_m t^m \right)
L_b V^{;b} + B_b V^{;b} = 0 \\
\eqref{FL.1.d} \implies \left[ 2d_2 + 3 \cdot 2 d_3 t + ... + m (m-1) d_m
t^{m-2} \right] L_a + \\
\qquad \qquad \quad + \left( d_0 + d_1t + ... + d_m t^m \right) \left( L_b
V^{;b} \right)_{;a} + \left( B_b V^{;b} \right)_{;a}= 0.
\end{cases}%
\end{equation*}

From \eqref{FL.1.a}, $L_a$ and $B_a$ are KVs.

From \eqref{FL.1.d}, $L_a = 0$ and $B_a V^{,b} = s_2$.

Solving \eqref{FL.1.c}, we find $K = s_2t + G(q)$ which when substituted into \eqref{FL.1.b}
gives $G_{,a} = 2 c_0 C_{ab} V^{,b}$.

The QFI is (of the form $Q_{8}$)
\[
I_{0m}(m>2)=c_{0}C_{ab}\dot{q}^{a}\dot{q}^{b}+B_{a}\dot{q}^{a}+ s_{2}t +2c_{0}\int C_{ab}V^{,b}dq^{a}
\]
where $C_{ab}$ is a KT and $B_{a}$ is a KV such that $B_{a}V^{,a}=s_{2}$.

\underline{\textbf{Subcase $\mathbf{(n=1, m=2)}$.}} $g = c_0 + c_1 t$ and $f =d_0 + d_1 t + d_2 t^2$ with $c_1d_2 \neq 0$.
\begin{equation*}
\begin{cases}
\eqref{FL.1.a} \implies c_1 C_{ab} + (d_0 + d_1t + d_2t^2) L_{(a;b)} +
B_{(a;b)} = 0 \\
\eqref{FL.1.b} \implies - 2 \left( c_0 + c_1 t \right) C_{ab} V^{,b} +
\left( d_1 + 2 d_2 t \right) L_a + K_{,a} = 0 \\
\eqref{FL.1.c} \implies K_{,t} = \left( d_0 + d_1 t + d_2 t^2 \right) L_a
V^{,a} + B_a V^{,a} \\
\eqref{FL.1.d} \implies 2 d_2 L_a + \left( d_0 + d_1 t + d_2 t^2 \right)
\left( L_b V^{,b} \right)_{;a} + \left( B_b V^{,b} \right)_{;a} - 2 c_1C_{ab} V^{,b} = 0.
\end{cases}%
\end{equation*}

Equation \eqref{FL.1.a} implies that $L_a$ is a KV and $c_1 C_{ab} = -
B_{(a;b)}$.

From \eqref{FL.1.d}, we have $L_a V^{,a} = s_1$ and $2 d_2 L_a + \left( B_b V^{,b} \right)_{;a} - 2 c_1 C_{ab} V^{,b} = 0$.

The solution of \eqref{FL.1.c} is $K = s_1 \left( d_0t + \frac{d_1}{2} t^2 + \frac{d_2}{3} t^3 \right) + B_a
V^{,a} t + G(q)$ which when replaced into \eqref{FL.1.b} gives
\begin{equation*}
- 2 \left( c_0 + c_1 t \right) C_{ab} V^{,b} + \left( d_1 + 2 d_2 t \right)
L_a + G_{,a} + \left( B_b V^{,b} \right)_{,a} t = 0 \implies
\end{equation*}
\begin{equation*}
G_{,a} - 2 c_0 C_{ab} V^{,b} + d_1 L_a + \underbrace{\left[ -2 c_1 C_{ab}
V^{,b} + 2d_2 L_a + \left( B_b V^{,b} \right)_{,a} \right]}_{=0} t = 0
\implies
\end{equation*}
\begin{equation*}
G_{,a} = \underbrace{2 c_0 C_{ab} V^{,b}} - d_1 L_a = c_0 \underbrace{\frac{%
2d_2c_0}{c_1}L_a + \frac{c_0}{c_1} \left( B_b V^{,b} \right)_{;a}} - d_1 L_a.
\end{equation*}

The QFI is
\begin{eqnarray*}
I_{12} &=& -\frac{1}{c_{1}} \left( c_{0}+c_{1}t \right) B_{(a;b)}%
\dot{q}^{a} \dot{q}^{b} + \left( d_{0}+d_{1}t+d_{2}t^{2} \right) L_{a}\dot{q}%
^{a} + B_{a}\dot{q}^{a} + s_1 \left( d_{0}t + \frac{d_{1}}{2}t^{2} + \frac{%
d_{2}}{3}t^{3} \right) + \\
&& + B_{a} V^{,a}t + G\left(q\right)
\end{eqnarray*}
where $c_{1}d_{2}\neq 0$, $B_{(a;b)}$ is a KT and $L_{a} = -%
\frac{1}{2d_{2}} \left[ \left( B_{b}V^{,b} \right)_{,a} + 2B_{(a;b)} V^{,b} %
\right]$ is a KV such that $L_{a}V^{,a}=s_{1}$. The function $G\left(q\right)$ is defined by the condition $G_{,a} - \frac{c_0}{c_1} \left( B_{b}V^{,b}
\right)_{,a} + \left( d_{1} - \frac{2d_{2}c_{0}}{c_{1}} \right) L_{a}$ $=0$ which is analysed as follows:

1) For $d_{1} \neq \frac{2d_{2}c_{0}}{c_{1}}$.

Then, $L_{a}$ is a gradient KV, that is, $L_{a} = \Phi_{,a} \implies G(q) = \frac{c_0}{c_1} B_{a}V^{,a} - d_{1}
\Phi(q) + \frac{2d_{2}c_{0}}{c_{1}} \Phi(q)$.

The QFI becomes
\begin{equation*}
I_{12(1)} = \frac{c_{0}}{c_{1}} Q_{9} + d_{0}Q_{2} + d_{1} Q_{6}(L_{a}=KV) +
Q_{7}(L_{a}=KV)
\end{equation*}
where $Q_{9} = -B_{(a;b)}\dot{q}^{a}\dot{q}^{b} + B_{a}V^{,a} + 2d_{2} \Phi(q)$. We observe that $Q_{3}=Q_{9}(\Phi=0)$ and $\left( B_{b}V^{,b} \right)_{,a} = -
2B_{(a;b)} V^{,b} -2d_{2}\Phi_{,a}$.

2) For $d_{1} = \frac{2d_{2}c_{0}}{c_{1}}$.

We have $G(q) = \frac{c_0}{c_1} B_{a}V^{,a}$ and the QFI becomes
$I_{12(2)} = \frac{c_{0}}{c_{1}} Q_{3} + d_{0}Q_{2} + Q_{8}(C_{ab}= -B_{(a;b)}) + Q_{7}(L_{a}=KV)$.

\underline{\textbf{Subcase $\mathbf{(n=1,m>2)}$.}} $g=c_{0}+c_{1}t$ and $c_{1}d_{m}\neq 0$.

\begin{equation*}
\begin{cases}
\eqref{FL.1.a} \implies c_1 C_{ab} + (d_0 + d_1t + ... + d_mt^m) L_{(a;b)} +
B_{(a;b)} = 0 \\
\eqref{FL.1.b} \implies - 2 \left( c_0 + c_1 t \right) C_{ab} V^{,b} +
\left( d_1 + 2 d_2 t + ... + m d_m t^{m-1} \right) L_a + K_{,a} = 0 \\
\eqref{FL.1.c} \implies K_{,t} = \left( d_0 + d_1 t + ... + d_m t^m \right)
L_a V^{,a} + B_a V^{,a} \\
\eqref{FL.1.d} \implies \left[ 2d_2 + 2 \cdot 3 d_3 t + ... + (m-1) m d_m
t^{m-2} \right] L_a + \\
\qquad \qquad + \left( d_0 + d_1 t + ... + d_m t^m \right) \left( L_b V^{,b}
\right)_{;a} + \left( B_b V^{,b} \right)_{;a} - 2 c_1 C_{ab} V^{,b} = 0.%
\end{cases}%
\end{equation*}

Equation \eqref{FL.1.a} implies that $L_a$ is a KV and $c_1 C_{ab} =
-B_{(a;b)}$.

From \eqref{FL.1.d}, we have $L_a = 0$ and $\left( B_b V^{,b} \right)_{;a} =2 c_1 C_{ab} V^{,b}$.

The solution of \eqref{FL.1.c} is $K = B_a V^{,a} t + G(q)$ which when substituted into \eqref{FL.1.b} gives
\begin{equation*}
- 2 c_0 C_{ab} V^{,b} \underbrace{- 2 c_1 C_{ab} V^{,b} t + \left( B_bV^{,b} \right)_{;a} t}_{=0} + G_{,a} = 0 \implies G_{,a} = 2 c_0 C_{ab}V^{,b}.
\end{equation*}
But $\left( B_b V^{,b} \right)_{;a} = 2 c_1 C_{ab} V^{,b}$; therefore, $G_{,a} = \frac{c_0}{c_1} \left( B_b V^{,b} \right)_{;a} \implies G(q) =\frac{c_0}{c_1} B_a V^{,a}$.

The QFI is (consists of FIs of the form $Q_{1}$ and $Q_{4}$)
\begin{equation*}
I_{1m}(m>2)=\left( c_{0}+c_{1}t\right) C_{ab}\dot{q}^{a}\dot{q}^{b}+B_{a}%
\dot{q}^{a}+B_{a}V^{,a}t+\frac{c_{0}}{c_{1}}B_{a}V^{,a}
\end{equation*}%
where $C_{ab}=-\frac{1}{c_{1}}B_{(a;b)}$ is a KT and $B_{a}$ is a vector such that $\left( B_{b}V^{,b}\right) _{;a}+2B_{(a;b)}V^{,b}=0$.

\underline{\textbf{Subcase $\mathbf{(n>1, m>n)}$.}} $c_nd_m \neq 0$.

\begin{equation*}
\begin{cases}
\eqref{FL.1.a} \implies \left( c_1 + 2 c_2 t + ... + n c_n t^{n-1} \right)
C_{ab} + \left( d_0 + d_1 t + ... + d_m t^m \right) L_{(a;b)} + B_{(a;b)} = 0
\\
\eqref{FL.1.b} \implies - 2 \left( c_0 + c_1 t + ... + c_n t^{n} \right)
C_{ab} V^{,b} + (d_1 + 2d_2 t + ... + m d_m t^{m-1}) L_a + K_{,a} = 0 \\
\eqref{FL.1.c} \implies K_{,t} = \left( d_0 + d_1 t + ... + d_m t^m \right)
L_a V^{,a} + B_a V^{,a} \\
\eqref{FL.1.d} \implies \left[ 2d_2 + 3 \cdot 2 d_3 t + ... + m (m-1) d_m
t^{m-2} \right] L_a + \left( d_0 + d_1 t + ... + d_m t^m \right) \left( L_b
V^{,b} \right)_{;a}+ \\
\qquad \qquad \quad + \left( B_b V^{,b} \right)_{;a} -2 \left( c_1 + 2 c_2 t + ... + n c_n t^{n-1} \right) C_{ab}
V^{,b} = 0.
\end{cases}%
\end{equation*}

From \eqref{FL.1.a}, we find that $C_{ab} = 0$ and $L_a$, $B_a$ are KVs.

From \eqref{FL.1.d}, we have that $L_a = 0$ and $B_a V^{,a} = s_2$.

Then, the solution of \eqref{FL.1.c} is $K = s_2 t + G(q)$ which when substituted into \eqref{FL.1.b} gives $G=const \equiv 0$.

The FI is (of the form $Q_{2}$) $I_{nm}(n>1,m>n)= B_a \dot{q}^a + s_2 t$, where $B_{a}$ is a KV such that $B_a V^{,a} = s_2$.
\bigskip

\textbf{II. For $\mathbf{n \to \infty}$ and $\mathbf{m}$ finite.}
\bigskip

We find the equivalences: $(n=\infty, m=0) \equiv (n>1, m=0) \equiv (g=e^{\lambda t}, m=0)$, $(n=\infty, m=1) \equiv (n>2, m=1) \equiv (g=e^{\lambda t}, m=1)$, $(n=\infty, m=2) \equiv (n>3, m=2) \equiv (g=e^{\lambda t}, m=2)$, $(n=\infty, m>2) \equiv (n>m+1, m>2) \equiv (g=e^{\lambda t}, m>2)$.

Then, for each case, we have the following.
\bigskip

\underline{\textbf{II.1. Case $\mathbf{(g = e^{\lambda t}}$, $\mathbf{f = d_0)}$.}} $\lambda \neq 0$.

\begin{equation*}
\begin{cases}
\eqref{FL.1.a} \implies \lambda e^{\lambda t} C_{ab} + d_0 L_{(a;b)} +
B_{(a;b)} = 0 \\
\eqref{FL.1.b} \implies - 2 e^{\lambda t} C_{ab} V^{,b} + K_{,a} = 0 \\
\eqref{FL.1.c} \implies K_{,t} = d_0 L_a V^{,a} + B_a V^{,a} \\
\eqref{FL.1.d} \implies d_0 \left( L_b V^{,b} \right)_{;a} + \left( B_b
V^{,b} \right)_{;a} - 2 \lambda e^{\lambda t} C_{ab} V^{,b} = 0.%
\end{cases}%
\end{equation*}

From \eqref{FL.1.a}, we get $C_{ab}=0$ and $\tilde{L}_{a}\equiv
d_{0}L_{a}+B_{a} $ is a KV.

From \eqref{FL.1.d}, we have that $\tilde{L}_a V^{,a} = s_0$.

Equation \eqref{FL.1.c} gives $K = s_0t + G(q)$ which when substituted into \eqref{FL.1.b} yields $G = const \equiv 0$.

The FI is (of the form $Q_{2}$) $I_{e0}=\tilde{L}_{a}\dot{q}^{a} +s_{0}t$, where $\tilde{L}_{a}$ is a KV such that $\tilde{L}_{a}V^{,a}=s_{0}$.

\underline{\textbf{II.2. Case $\mathbf{(g=e^{\lambda t}}$, $\mathbf{f=d_{0}+d_{1}t)}$.}} $\lambda d_{1}\neq 0$.

\begin{equation*}
\begin{cases}
\eqref{FL.1.a} \implies \lambda e^{\lambda t} C_{ab} + \left( d_0 + d_1 t
\right) L_{(a;b)} + B_{(a;b)} = 0 \\
\eqref{FL.1.b} \implies - 2 e^{\lambda t} C_{ab} V^{,b} + d_1 L_a + K_{,a} =
0 \\
\eqref{FL.1.c} \implies K_{,t} = \left( d_0 + d_1 t \right) L_a V^{,a} + B_a
V^{,a} \\
\eqref{FL.1.d} \implies \left( d_0 + d_1 t \right) \left( L_b V^{,b}
\right)_{;a} + \left( B_b V^{,b} \right)_{;a} - 2 \lambda e^{\lambda t}
C_{ab} V^{,b} = 0.%
\end{cases}%
\end{equation*}

From \eqref{FL.1.a}, we have that $C_{ab} = 0$ and $L_a$, $B_a$ are KVs.

From \eqref{FL.1.d}, we get that $L_a V^{,a} = s_1$ and $B_a V^{,a} = s_2$.

Then, equation \eqref{FL.1.c} gives $K = s_1 \left( d_0t + \frac{d_1}{2} t^2 \right) + s_2 t + G(q)$ which when substituted into \eqref{FL.1.b} gives $G_{,a} = - d_1 L_a$.

The FI is (consists of $Q_{2}$, $Q_{5}$)
\begin{equation*}
I_{e1}=\left( d_{0}+d_{1}t\right) L_{a}\dot{q}^{a}+B_{a}\dot{q}%
^{a}+(s_{1}d_{0}+s_{2})t+\frac{s_{1}d_{1}}{2}t^{2} -d_{1}\int L_{a}dq^{a}
\end{equation*}%
where $L_{a}=-\frac{1}{d_{1}}G_{,a}$ is a gradient KV such that $L_{a}V^{,a}=s_{1}$ and $B_{a}$ is a KV such that $B_{a}V^{,a}=s_{2}$.

\underline{\textbf{II.3. Case $\mathbf{(g = e^{\lambda t}}$, $\mathbf{f = d_0 + d_1 t + d_2t^2)}$.}} $\lambda d_2 \neq 0$.

\begin{equation*}
\begin{cases}
\eqref{FL.1.a} \implies \lambda e^{\lambda t} C_{ab} + \left( d_0 + d_1 t +
d_2t^2 \right) L_{(a;b)} + B_{(a;b)} = 0 \\
\eqref{FL.1.b} \implies - 2 e^{\lambda t} C_{ab} V^{,b} + (d_1 + 2d_2t) L_a
+ K_{,a} = 0 \\
\eqref{FL.1.c} \implies K_{,t} = \left( d_0 + d_1 t + d_2t^2 \right) L_a
V^{,a} + B_a V^{,a} \\
\eqref{FL.1.d} \implies 2d_2 L_a + \left( d_0 + d_1 t + d_2 t^2 \right)
\left( L_b V^{,b} \right)_{;a} + \left( B_b V^{,b} \right)_{;a} - 2 \lambda
e^{\lambda t} C_{ab} V^{,b} = 0.%
\end{cases}%
\end{equation*}

From \eqref{FL.1.a}, we have that $C_{ab} = 0$ and $L_a$, $B_a$ are KVs.

From \eqref{FL.1.d}, we get that $L_a V^{,a} = s_1$ and $\left( B_b V^{,b}
\right)_{;a} = -2d_2 L_a$, that is, $L_a$ is a gradient KV.

Then, equation \eqref{FL.1.c} gives $K = s_1 \left( d_0t + \frac{d_1}{2} t^2 + \frac{d_2}{3} t^3 \right) + B_aV^{,a} t + G(q)$ which when substituted into \eqref{FL.1.b} yields $G_{,a} = - d_1 L_a$.

We observe that
\begin{equation*}
\begin{cases}
G_{,a} = - d_1 L_a \\
L_a = - \frac{1}{2 d_2} \left( B_b V^{,b} \right)_{;a}%
\end{cases}
\implies G_{,a} = \frac{d_1}{2 d_2} \left( B_b V^{,b} \right)_{,a} \implies
G = \frac{d_1}{2 d_2} B_b V^{,b}.
\end{equation*}

The FI is (consists of $Q_{2}$, $Q_{5}$ and $Q_{7}$)
\begin{equation*}
I_{e2}=\left( d_{0}+d_{1}t+d_{2}t^{2}\right) L_{a}\dot{q}^{a}+B_{a}\dot{q}%
^{a}+s_{1}\left( d_{0}t+\frac{d_{1}}{2}t^{2}+\frac{d_{2}}{3}t^{3}\right)
+B_{a}V^{,a}t+\frac{d_{1}}{2d_{2}}B_{b}V^{,b}
\end{equation*}%
where $L_{a}=-\frac{1}{2d_{2}}\left( B_{b}V^{,b}\right) _{;a}$ is a gradient KV such that $L_{a}V^{,a}=s_{1}$ and $B_{a}$ is a KV.

\underline{\textbf{II.4. Case $\mathbf{(g = e^{\lambda t}}$, $\mathbf{m>2)}$.}} $\lambda d_m \neq 0$.

\begin{equation*}
\begin{cases}
\eqref{FL.1.a} \implies \lambda e^{\lambda t} C_{ab} + \left( d_0 + d_1 t +
... + d_m t^m \right) L_{(a;b)} + B_{(a;b)} = 0 \\
\eqref{FL.1.b} \implies - 2 e^{\lambda t} C_{ab} V^{,b} + \left( d_1 + 2 d_2
t + ... + m d_m t^{m-1} \right) L_a + K_{,a} = 0 \\
\eqref{FL.1.c} \implies K_{,t} = \left( d_0 + d_1 t + ... + d_m t^m \right)
L_a V^{,a} + B_a V^{,a} \\
\eqref{FL.1.d} \implies \left[ 2 d_2 + 3 \cdot 2 t + ... + m (m-1) d_m
t^{m-2} \right] L_a + \left( d_0 + d_1 t + ... + d_m t^m \right) \left( L_b
V^{,b} \right)_{;a} + \\
\qquad \qquad + \left( B_b V^{,b} \right)_{;a} - 2 \lambda e^{\lambda t}
C_{ab} V^{,b} = 0.%
\end{cases}%
\end{equation*}

From \eqref{FL.1.a}, we have that $C_{ab} = 0$ and $L_a$, $B_a$ are KVs.

From \eqref{FL.1.d}, we get that $L_a = 0$ and $B_a V^{,a} = s_2$.

Then, equation \eqref{FL.1.c} gives $K = s_2 t + G(q)$ which when substituted into \eqref{FL.1.b} gives $G = const \equiv 0$.

Therefore, the FI is (of the form $Q_{2}$) $I_{em}(m>2)=B_{a}\dot{q}^{a}+s_{2}t$, where $B_{a}$ is a KV such that $B_{a}V^{,a}=s_{2}$. This is a time-dependent LFI.
\bigskip

\textbf{III. For $\mathbf{n}$ finite and $\mathbf{m \to \infty}$.}
\bigskip

We distinguish between two cases because in condition \eqref{FL.1.d} we have to compare polynomial coefficients of the infinite sums $f_{,tt}$ and $f$.
\bigskip

\textbf{III.1. Case with $\mathbf{f_{,tt} \neq \lambda^2 f}$.}

\begin{equation*}
(n=0, m=\infty) \equiv (n=0, m>2), \enskip (n=1, m=\infty) \equiv (n=1,
m>2), \enskip (n>1, m=\infty) \equiv (n>1, m>n).
\end{equation*}

\textbf{III.2. Case with $\mathbf{f_{,tt} = \lambda^2 f}$.}

\begin{equation*}
(n=0, m=\infty) \equiv (n=0, f=e^{\lambda t}), \enskip (n=1, m=\infty)
\equiv (n=1, f=e^{\lambda t}), \enskip (n>1, m=\infty) \equiv (n>1,
f=e^{\lambda t}).
\end{equation*}

For each subcase, we have the following.
\vspace{12pt}

\underline{\textbf{Subcase $\mathbf{(g=c_{0}}$, $\mathbf{f=e^{\lambda t})}$.}}  $\lambda \neq 0$.

\begin{equation*}
\begin{cases}
\eqref{FL.1.a} \implies e^{\lambda t} L_{(a;b)} + B_{(a;b)} = 0 \\
\eqref{FL.1.b} \implies -2 c_0 C_{ab} V^{,b} + \lambda e^{\lambda t} L_a +
K_{,a} = 0 \\
\eqref{FL.1.c} \implies K_{,t} = e^{\lambda t} L_b V^{,b} + B_b V^{,b} \\
\eqref{FL.1.d} \implies \lambda^2 e^{\lambda t} L_a + e^{\lambda t} \left(
L_b V^{,b} \right)_{;a} + \left( B_b V^{,b} \right)_{;a} = 0.%
\end{cases}%
\end{equation*}

Equation \eqref{FL.1.a} implies that $L_a$ and $B_a$ are KVs.

From \eqref{FL.1.d}, we find that $B_a V^{,b} = s_2$ and $L_a = - \frac{1}{\lambda^2} \left( L_b V^{,b} \right)_{,a}$, that is, $L_a$ is a gradient KV.

From \eqref{FL.1.c}, we find $K = \frac{1}{\lambda} e^{\lambda t} L_b V^{,b} + s_2 t + G(q)$ which when substituted into \eqref{FL.1.b} gives $G_{,a} = 2 c_0 C_{ab} V^{,b}$.

The QFI is ($c_{0}$ is absorbed by $C_{ab}$)
\begin{equation*}
I_{0e} = C_{ab}\dot{q}^{a}\dot{q}^{b} + e^{\lambda t} L_{a}\dot{q}%
^{a} +B_{a}\dot{q}^{a} + \frac{1}{\lambda } e^{\lambda t} L_{a}V^{,a} +
s_{2}t +G\left(q\right)
\end{equation*}%
where $\lambda \neq 0$, $L_{a} = - \frac{1}{\lambda^2} \left( L_bV^{,b}\right)_{,a}$ is a gradient KV, $B_{a}$ is a KV such that $B_{a}V^{;a}=s_{2}$, and $C_{ab}$ is a KT such that $G_{,a}-2C_{ab}V^{,b}=0$.

We note that $I_{0e} = Q_{1} + Q_{2}(B_{a}) + Q_{10}$ where
\begin{equation*}
Q_{10} = e^{\lambda t} \left( L_{a}\dot{q}^{a} + \frac{1}{\lambda}
L_{a}V^{,a} \right)
\end{equation*}
is a new independent LFI.

\underline{\textbf{Subcase $\mathbf{(g = c_0 + c_1 t}$, $\mathbf{f = e^{\lambda t})}$.}} $\lambda c_1 \neq 0$.

\begin{equation*}
\begin{cases}
\eqref{FL.1.a} \implies c_1 C_{ab} + e^{\lambda t} L_{(a;b)} + B_{(a;b)} = 0
\\
\eqref{FL.1.b} \implies - 2 \left( c_0 + c_1 t \right) C_{ab} V^{,b} +
\lambda e^{\lambda t} L_a + K_{,a} = 0 \\
\eqref{FL.1.c} \implies K_{,t} = e^{\lambda t} L_b V^{,b} + B_b V^{,b} \\
\eqref{FL.1.d} \implies \lambda^2 e^{\lambda t} L_a + e^{\lambda t} \left(
L_b V^{,b} \right)_{;a} + \left( B_b V^{,b} \right)_{;a} - 2 c_1 C_{ab}
V^{,b} = 0.%
\end{cases}%
\end{equation*}

From \eqref{FL.1.a}, we find that $L_a$ is a KV and $c_1 C_{ab} = - B_{(a;b)}$.

From \eqref{FL.1.d}, we get the conditions $L_{a}=-\frac{1}{\lambda ^{2}}%
\left( L_{b}V^{,b}\right) _{;a}$ and $\left( B_{b}V^{,b}\right)
_{;a}=2c_{1}C_{ab}V^{,b}$.

Equation \eqref{FL.1.c} gives $K=\frac{1}{\lambda }e^{\lambda t}L_{b}V^{,b}+B_{b}V^{,b}t+G(q)$. Substituting $K$ into \eqref{FL.1.b}, we obtain
\begin{equation*}
G_{,a} = 2c_0C_{ab}V^{,b} = \frac{c_0}{c_1} \left( B_b V^{,b} \right)_{;a} \implies G(q) = \frac{c_{0}}{c_{1}} B_{b} V^{,b}.
\end{equation*}

The QFI is
\begin{equation*}
I_{1e} = -\frac{1}{c_{1}}\left( c_{0}+c_{1}t \right) B_{(a;b)}\dot{%
q}^{a} \dot{q}^{b} + e^{\lambda t} L_{a}\dot{q}^{a} + B_{a}\dot{q}^{a} +
\frac{1}{\lambda} e^{\lambda t} L_{a}V^{,a} + B_{a}V^{,a} t + \frac{c_{0}}{%
c_{1}} B_{a}V^{,a}
\end{equation*}
where $\lambda c_{1} \neq 0$, $L_{a}=-\frac{1}{\lambda^{2}} \left(
L_{b}V^{,b}\right)_{,a}$ is a gradient KV, and $B_a$ is such that $B_{(a;b)}$ is a KT and $\left(B_{b}V^{,b}\right)_{,a} = -2B_{(a;b)}V^{,b}$.

We note that $I_{1e} = \frac{c_{0}}{c_{1}}Q_{3} + Q_{4} + Q_{10}$.

\underline{\textbf{Subcase $\mathbf{(n>1}$, $\mathbf{f = e^{\lambda t})}$.}} $\lambda c_n\neq 0$.

\begin{equation*}
\begin{cases}
\eqref{FL.1.a} \implies (c_1 + 2c_2t + ... + nc_n t^{n-1}) C_{ab} +
e^{\lambda t} L_{(a;b)} + B_{(a;b)} = 0 \\
\eqref{FL.1.b} \implies - 2 \left( c_0 + c_1 t + ... + c_n t^n \right)
C_{ab} V^{,b} + \lambda e^{\lambda t} L_a + K_{,a} = 0 \\
\eqref{FL.1.c} \implies K_{,t} = e^{\lambda t} L_b V^{,b} + B_b V^{,b} \\
\eqref{FL.1.d} \implies \lambda^2 e^{\lambda t} L_a + e^{\lambda t} \left(
L_b V^{,b} \right)_{;a} + \left( B_b V^{,b} \right)_{;a} - 2 (c_1 + 2c_2t +
... + nc_n t^{n-1}) C_{ab} V^{,b} = 0.%
\end{cases}%
\end{equation*}

From \eqref{FL.1.a}, we find that $C_{ab} = 0$ and $L_a$, $B_a$ are KVs. Then, \eqref{FL.1.d} implies that $B_a V^{,a} = s_2$ and $L_a = - \frac{1}{\lambda^2} \left( L_b V^{,b} \right)_{;a}$, i.e. $L_a$ is a gradient KV.

The solution of \eqref{FL.1.c} is $K = \frac{1}{\lambda} e^{\lambda t} L_b V^{,b} + s_2t + G(q)$ which when substituted into \eqref{FL.1.b} gives $G = const\equiv 0$.

The FI is (consists of $Q_{2}$ and $Q_{10}$)
\begin{equation*}
I_{ne}(n>1)=e^{\lambda t}L_{a}\dot{q}^{a}+B_{a}\dot{q}^{a}+\frac{1}{\lambda }%
e^{\lambda t}L_{a}V^{,a}+s_{2}t
\end{equation*}%
where $L_{a}=-\frac{1}{\lambda ^{2}}\left( L_{b}V^{,b}\right) _{;a}$ is a gradient KV and $B_{a}$ is a KV such that $B_{a}V^{,a}=s_{2}$.
\bigskip

\textbf{IV. Both $\mathbf{n}$ and $\mathbf{m}$ are infinite.} \bigskip

We consider three cases.
\bigskip

\textbf{IV.1. Case where $\mathbf{f_{,tt} = \lambda^2 f}$ and $\mathbf{g_{,t} \neq \lambda f}$.}

\begin{equation*}
(n=\infty, m=\infty) \equiv (g=e^{\mu t}, f=e^{\lambda t}, \lambda \neq \mu)
\equiv (n>1, f=e^{\lambda t}).
\end{equation*}

\underline{\textbf{Subcase $\mathbf{(g = e^{\lambda t}}$, $\mathbf{f = e^{\mu t})}$.}} $\lambda \mu \neq 0$.

\begin{equation*}
\begin{cases}
\eqref{FL.1.a} \implies \lambda e^{\lambda t} C_{ab} + e^{\mu t} L_{(a;b)} +
B_{(a;b)} = 0 \\
\eqref{FL.1.b} \implies - 2 e^{\lambda t} C_{ab} V^{,b} + \mu e^{\mu t} L_a
+ K_{,a} = 0 \\
\eqref{FL.1.c} \implies K_{,t} = e^{\mu t} L_a V^{,a} + B_a V^{,a} \\
\eqref{FL.1.d} \implies \mu^2 e^{\mu t} L_a + e^{\mu t} \left( L_b V^{,b}
\right)_{;a} + \left( B_b V^{,b} \right)_{;a} - 2 \lambda e^{\lambda t}
C_{ab} V^{,b} = 0.%
\end{cases}%
\end{equation*}

a) \underline{For $\lambda \neq \mu$:}

From \eqref{FL.1.a}, we have that $C_{ab} = 0$ and $L_a$, $B_a$ are KVs.

From \eqref{FL.1.d}, we find that $\mu^2 L_a + \left( L_b V^{,b} \right)_{;a}= 0$ and $B_b V^{,b} = s_2$.

The solution of \eqref{FL.1.c} is $K = \frac{1}{\mu} e^{\mu t} L_a V^{,a} + s_2 t + G(q)$ which when replaced into \eqref{FL.1.b} and using the relation $\mu^2 L_a + \left( L_b V^{,b}\right)_{;a} = 0$ gives $G = const \equiv 0$.

The FI is (consists of $Q_{2}$ and $Q_{10}$)
\begin{equation*}
I_{ee}(\lambda \neq \mu) =e^{\mu t}L_{a}\dot{q}^{a}+B_{a}\dot{q}^{a}+\frac{1%
}{\mu }e^{\mu t}L_{a}V^{a,}+s_{2}t
\end{equation*}%
where $L_{a}=-\frac{1}{\mu ^{2}}\left( L_{b}V^{,b}\right) _{;a}$ is a gradient KV and $B_{a}$ is a KV such that $B_{b}V^{,b}=s_{2}$. This is a time-dependent LFI.
\bigskip

b) \underline{For $\lambda = \mu$:}

From \eqref{FL.1.a}, we have that $\lambda C_{ab} + L_{(a;b)} = 0$ and $B_a$
is a KV.

From \eqref{FL.1.d}, we find that $\lambda^2 L_a + \left( L_b V^{,b}\right)_{;a} - 2 \lambda C_{ab} V^{,b} = 0$ and $B_b V^{,b} = s_2$.

The solution of \eqref{FL.1.c} is $K = \frac{1}{\lambda} e^{\lambda t} L_a V^{,a} + s_2 t + G(q)$ which when substituted into \eqref{FL.1.b} and using the relation $\lambda^2 L_a + \left( L_b V^{,b} \right)_{;a} - 2 \lambda C_{ab} V^{,b} = 0$ gives $G = const \equiv 0$.

The QFI is
\begin{equation*}
I_{ee}(\lambda = \mu) = -\frac{1}{\lambda } e^{\lambda t}
L_{(a;b)} \dot{q}^a \dot{q}^b + e^{\lambda t} L_a \dot{q}^a + B_a \dot{q}^a
+ \frac{1}{\lambda} e^{\lambda t} L_a V^{,a} + s_2 t
\end{equation*}
where $\lambda \neq 0$, $L_a$ is such that $L_{(a;b)}$ is a KT, $\lambda^{2}L_{a}+\left( L_{b}V^{,b}\right)_{,a} + 2L_{(a;b)}V^{,b}=0$, and $B_{a}$ is a KV such that $B_{a}V^{,a}=s_{2}$.

We note that $I_{ee}(\lambda=\mu) = Q_{2}(B_{a}) + Q_{11}$, where
\begin{equation*}
Q_{11} = e^{\lambda t} \left( -\frac{1}{\lambda} L_{(a;b)} \dot{q}^a \dot{q}%
^b + L_{a}\dot{q}^{a} + \frac{1}{\lambda} L_{a}V^{,a} \right)
\end{equation*}
is a new independent QFI. Then, $Q_{10}=Q_{11}(L_{a}=KV)$.
\bigskip

\textbf{IV.2. Case where $\mathbf{f_{,tt} = \lambda^2 f}$ and $\mathbf{g_{,t} = \lambda f}$.}

\begin{equation*}
(n=\infty, m=\infty) \equiv (g=e^{\lambda t}, f=e^{\lambda t}).
\end{equation*}
\bigskip

\textbf{IV.3. Case where $\mathbf{f_{,tt} \neq \lambda^2 f}$ and $\mathbf{g_{,t} \neq \lambda f}$ or $\mathbf{g_{,t} = \lambda f}$.}
\begin{equation*}
(n=\infty, m=\infty) \equiv (n>1, m>n) \equiv (n>m+1, m>2) \equiv (g=e^{\lambda t}, m>2) \equiv (n=m, m>2).
\end{equation*}
\bigskip

By collecting all the above LFIs/QFIs $Q_{A}$, the derivation of  Theorem \ref{The first integrals of an autonomous holonomic dynamical system} is straightforward. Specifically, we cover all the FIs mentioned in Theorem \ref{The first integrals of an autonomous holonomic dynamical system} as follows:
\[
I_{1}=Q_{16}, \enskip I_{2}=Q_{7}, \enskip I_{3}= Q_{11}.
\]
The FIs $Q_{1}$, $Q_{3}$, $Q_{5}$, $Q_{6}$, $Q_{8}$, $Q_{9}$ are subcases of $I_{1}$; the $Q_{2}$, $Q_{4}$ are subcases of $I_{2}$; and, finally, $Q_{10}$ is a subcase of $I_{3}$.

After a careful and extensive study, we have shown that the LFIs/QFIs of an autonomous conservative dynamical system can be produced by the three parameterized FIs listed in Theorem \ref{The first integrals of an autonomous holonomic dynamical system}.

%% file: proof_thm_damping.tex
\chapter{Proof of Theorem \ref{Theorem2}}

\label{app.proof.QFIs.damping}

Recall that
\begin{equation*}
K_{ab}(t,q)= C_{(0)ab}(q) + C_{(1)ab}(q) t + C_{(2)ab}(q) \frac{t^{2}}{2} +
... + C_{(n)ab}(q) \frac{t^{n}}{n}
\end{equation*}%
and
\begin{equation*}
K_{a}(t,q)= L_{(0)a}(q) + L_{(1)a}(q)t + L_{(2)a}(q)t^{2} + ... +
L_{(m)a}(q) t^{m}.
\end{equation*}
We consider various cases\footnote{Equation (\ref{eq.veldep10}) is not necessary because the integrability condition $K_{,[ab]}=0$ does not intervene in the calculations. However, it has been checked that equation (\ref{eq.veldep10}) is always
satisfied identically from the solutions of the other equations
of the system.}.
\bigskip

\underline{\textbf{I. Case $\mathbf{n=m}$}} (both $n$ and $m$ are finite)

Equation (\ref{eq.veldep6}) implies that: $C_{(1)ab} = -L_{(0)(a;b)} -2C_{(0)c(a}A_{b)}^{c}$, $C_{(k)ab} =
-L_{(k-1)(a;b)} - \frac{2}{k-1}C_{(k-1)c(a}A_{b)}^{c}$ with $k=2,...,n$, and $L_{(n)(a;b)} =-\frac{2}{n} C_{(n)c(a}A_{b)}^{c}$.

Equation (\ref{eq.veldep9}) gives: $L_{(n)a}Q^{a}=s$, $\left( L_{(n-1)b} Q^{b} \right)_{,a} =2C_{(n)ab}Q^{b} - nL_{(n)b}A^{b}_{a}$, and \newline
$\left( L_{(k-2)b} Q^{b} \right)_{,a} = 2C_{(k-1)ab}Q^{b} - k(k-1)L_{(k)a} -
(k-1)L_{(k-1)b}A^{b}_{a}$ with $k=2,...,n$.

The solution of (\ref{eq.veldep8}) is $K= L_{(0)a}Q^{a}t + L_{(1)a}Q^{a}\frac{t^{2}}{2} + ... + L_{(n)a}Q^{a} \frac{%
t^{n+1}}{n+1} + G(q)$ which when replaced into (\ref{eq.veldep7}) gives $G_{,a}= 2C_{(0)ab}Q^{b} - L_{(1)a} - L_{(0)b}A^{b}_{a}$.

The QFI is
\begin{eqnarray*}
I_{n} &=& \left( \frac{t^{n}}{n} C_{(n)ab} + ... + \frac{t^{2}}{2} C_{(2)ab} + t C_{(1)ab} + C_{(0)ab} \right) \dot{q}^{a} \dot{q}^{b} + t^{n} L_{(n)a}\dot{q}^{a} + ... + t^{2}L_{(2)a}\dot{q}^{a} + \\
&& + tL_{(1)a}\dot{q}^{a} +L_{(0)a}\dot{q}^{a} + \frac{t^{n+1}}{n+1} L_{(n)a}Q^{a} + ... + \frac{t^{2}}{2} L_{(1)a}Q^{a}
+t L_{(0)a}Q^{a} + G(q)
\end{eqnarray*}
where $C_{(N)ab}$ are KTs, $C_{(1)ab} = -L_{(0)(a;b)} -2C_{(0)c(a}A_{b)}^{c}$%
, $C_{(k+1)ab} = -L_{(k)(a;b)} - \frac{2}{k}C_{(k)c(a}A_{b)}^{c}$ for $%
k=1,...,n-1$, $L_{(n)(a;b)} =-\frac{2}{n} C_{(n)c(a}A_{b)}^{c}$, $%
L_{(n)a}Q^{a}=s$, $\left( L_{(n-1)b} Q^{b} \right)_{,a} = 2C_{(n)ab}Q^{b} -
nL_{(n)b}A^{b}_{a}$, $\left( L_{(k-1)b} Q^{b} \right)_{,a} = 2C_{(k)ab}Q^{b}
- k(k+1)L_{(k+1)a} - kL_{(k)b}A^{b}_{a}$ for $k=1,...,n-1$, and $G_{,a}=2C_{(0)ab}Q^{b} - L_{(1)a} - L_{(0)b}A^{b}_{a}$.

We note that $I_{0} < I_{1} < I_{2} < I_{3} < I_{4} < ...$, that is, each QFI
$I_{k}$ is a subcase of the next QFI $I_{k+1}$ for all $k \in \mathbb{N}$.
Therefore, we have only one independent QFI the $I_{n}$. The value of $n$ is
determined by the symmetries of the kinetic metric and the dynamics of each
specific system.

Observe that for $A^{a}_{b}=0$, the QFI $I_{n}$ reduces to
\begin{eqnarray*}
I_{ns} &=& \left( - \frac{t^{n}}{n} L_{(n-1)(a;b)} - ... - \frac{t^{2}}{2}
L_{(1)(a;b)} - t L_{(0)(a;b)} + C_{(0)ab} \right) \dot{q}^{a} \dot{q}^{b} +
t^{n} L_{(n)a}\dot{q}^{a} + ... + t^{2}L_{(2)a}\dot{q}^{a} + \\
&& + t L_{(1)a}\dot{q}^{a} + L_{(0)a}\dot{q}^{a}+ \frac{t^{n+1}}{n+1}
L_{(n)a}Q^{a} + ... + \frac{t^{2}}{2} L_{(1)a}Q^{a} +t L_{(0)a}Q^{a} + G(q)
\end{eqnarray*}
where $C_{(0)ab}$ and $L_{(N)(a;b)}$ are KTs, $L_{(n)a}$ is a KV, $L_{(n)a}Q^{a}=s$, $\left( L_{(n-1)b} Q^{b} \right)_{,a} =
-2L_{(n-1)(a;b)}Q^{b}$, \\ $\left( L_{(k-1)b} Q^{b} \right)_{,a} =-2L_{(k-1)(a;b)}Q^{b} - k(k+1)L_{(k+1)a}$ for $k=1,...,n-1$, and $G_{,a}=2C_{(0)ab}Q^{b} - L_{(1)a}$.

We shall prove that $I_{n}(A^{a}_{b}=0)$ consists of two independent FIs.

In the case that $A^{a}_{b}=0$, we have the following:

- For $n=0$.
\begin{equation*}
I_{0} = C_{(0)ab} \dot{q}^a \dot{q}^b + L_{(0)a} \dot{q}^a + st + G(q)
\end{equation*}
where $C_{(0)ab}$ is a KT, $L_{(0)a}$ is a KV, $L_{(0)a} Q^{a}=s$, and $G_{,a} = 2C_{(0)ab} Q^{b}$.

This QFI consists of the independent FIs: $I_{01}= C_{(0)ab} \dot{q}^a \dot{q}^b + G(q)$ and $I_{02}= L_{(0)a} \dot{q}^a + st$.

- For $n=1$.
\begin{eqnarray*}
I_{1}&=& \left( -tL_{(0)(a;b)} + C_{(0)ab} \right) \dot{q}^{a} \dot{q}^{b} +
tL_{(1)a}\dot{q}^{a} + L_{(0)a}\dot{q}^{a} + \frac{t^{2}}{2}s +
tL_{(0)a}Q^{a} + G(q)
\end{eqnarray*}
where $C_{(0)ab}$ and $L_{(0)(a;b)}$ are KTs, $L_{(1)a}$ is a KV, $L_{(1)a}Q^{a}=s$, $\left(L_{(0)b}Q^{b} \right)_{,a} = - 2L_{(0)(a;b)}Q^{b}$, and \\ $G_{,a} =2C_{(0)ab}Q^{b} - L_{(1)a}$.

This QFI consists of the independent FIs:
\begin{eqnarray*}
I_{11} &=& C_{(0)ab} \dot{q}^{a} \dot{q}^{b} + tL_{(1)a}\dot{q}^{a} + \frac{%
t^{2}}{2}s + G(q) \\
I_{12} &=& -tL_{(0)(a;b)} \dot{q}^{a} \dot{q}^{b} + L_{(0)a}\dot{q}^{a} +
tL_{(0)a}Q^{a}.
\end{eqnarray*}

- For $n=2$.
\begin{eqnarray*}
I_{2} &=& \left( - \frac{t^{2}}{2}L_{(1)(a;b)} -tL_{(0)(a;b)} + C_{(0)ab}
\right) \dot{q}^{a} \dot{q}^{b} + t^{2}L_{(2)a}\dot{q}^{a} + t L_{(1)a}\dot{q%
}^{a} + L_{(0)a}\dot{q}^{a} + \\
&& + \frac{t^{3}}{3}s + \frac{t^{2}}{2}
L_{(1)a}Q^{a} +t L_{(0)a}Q^{a} + G(q)
\end{eqnarray*}
where $C_{(0)ab}$ and $L_{(M)(a;b)}$ for $M=0,1$ are KTs, $L_{(2)a}$ is a KV, $L_{(2)a}Q^{a}=s$, $\left( L_{(1)b} Q^{b} \right)_{,a} = -2L_{(1)(a;b)}Q^{b}$, $\left( L_{(0)b} Q^{b} \right)_{,a} = -2L_{(0)(a;b)}Q^{b} -2L_{(2)a}$, and $G_{,a}= 2C_{(0)ab}Q^{b} - L_{(1)a}$.

This QFI consists of the independent FIs:
\begin{eqnarray*}
I_{21} &=& \left( - \frac{t^{2}}{2}L_{(1)(a;b)} + C_{(0)ab} \right) \dot{q}%
^{a} \dot{q}^{b} + t L_{(1)a}\dot{q}^{a} + \frac{t^{2}}{2} L_{(1)a}Q^{a} +
G(q) \\
I_{22} &=& -tL_{(0)(a;b)} \dot{q}^{a} \dot{q}^{b} + t^{2}L_{(2)a}\dot{q}^{a}
+ L_{(0)a}\dot{q}^{a} + \frac{t^{3}}{3}s + +t L_{(0)a}Q^{a}.
\end{eqnarray*}

- For $n=3$.
\begin{eqnarray*}
I_{3} &=& \left( - \frac{t^{3}}{3} L_{(2)(a;b)} - \frac{t^{2}}{2}
L_{(1)(a;b)} - t L_{(0)(a;b)} + C_{(0)ab} \right) \dot{q}^{a} \dot{q}^{b} +
t^{3} L_{(3)a}\dot{q}^{a} + t^{2}L_{(2)a}\dot{q}^{a} + t L_{(1)a}\dot{q}^{a}+ \\
&& + L_{(0)a}\dot{q}^{a} + \frac{t^{4}}{4} s + \frac{t^{3}}{3} L_{(2)a}Q^{a} + \frac{t^{2}}{2}
L_{(1)a}Q^{a} +t L_{(0)a}Q^{a} + G(q)
\end{eqnarray*}
where $C_{(0)ab}$ and $L_{(M)(a;b)}$ for $M=0,1,2$ are KTs, $L_{(3)a}$ is a KV, $L_{(3)a}Q^{a}=s$, $\left( L_{(2)b} Q^{b} \right)_{,a} = -2L_{(2)(a;b)}Q^{b}$, $\left( L_{(1)b} Q^{b} \right)_{,a} = -2L_{(1)(a;b)}Q^{b} - 6L_{(3)a}$, $\left( L_{(0)b} Q^{b} \right)_{,a} = -2L_{(0)(a;b)}Q^{b} - 2L_{(2)a}$, and $G_{,a}= 2C_{(0)ab}Q^{b} - L_{(1)a}$.

This QFI consists of the independent FIs:
\begin{eqnarray*}
I_{31} &=& \left( - \frac{t^{2}}{2} L_{(1)(a;b)} + C_{(0)ab} \right) \dot{q}%
^{a} \dot{q}^{b} + t^{3} L_{(3)a}\dot{q}^{a} + t L_{(1)a}\dot{q}^{a} + \frac{%
t^{4}}{4} s + \frac{t^{2}}{2} L_{(1)a}Q^{a} + G(q) \\
I_{32} &=& \left( - \frac{t^{3}}{3} L_{(2)(a;b)} - t L_{(0)(a;b)} \right)
\dot{q}^{a} \dot{q}^{b}+ t^{2}L_{(2)a}\dot{q}^{a} + L_{(0)a}\dot{q}^{a} +
\frac{t^{3}}{3} L_{(2)a}Q^{a} +t L_{(0)a}Q^{a}.
\end{eqnarray*}

- For $n=4$.
\begin{eqnarray*}
I_{4} &=& \left( -\frac{t^{4}}{4} L_{(3)(a;b)} - \frac{t^{3}}{3}
L_{(2)(a;b)} - \frac{t^{2}}{2}L_{(1)(a;b)} - t L_{(0)(a;b)} + C_{(0)ab}
\right) \dot{q}^{a} \dot{q}^{b} + t^{4} L_{(4)a}\dot{q}^{a} + \\
&& +t^{3} L_{(3)a}\dot{q}^{a} + t^{2}L_{(2)a}\dot{q}^{a} + t L_{(1)a}\dot{q}^{a} + L_{(0)a}\dot{q}^{a} + \frac{t^{5}}{5} s + \frac{%
t^{4}}{4} L_{(3)a}Q^{a} + \frac{t^{3}}{3} L_{(2)a}Q^{a} + \\
&& + \frac{t^{2}}{2}
L_{(1)a}Q^{a} +t L_{(0)a}Q^{a} + G(q)
\end{eqnarray*}
where $C_{(0)ab}$ and $L_{(M)(a;b)}$ for $M=0,...n-1$ are KTs, $L_{(4)a}$ ia a KV, $L_{(4)a}Q^{a}=s$, $\left( L_{(3)b} Q^{b} \right)_{,a} =
-2L_{(3)(a;b)}Q^{b}$, $\left( L_{(2)b} Q^{b} \right)_{,a} =
-2L_{(2)(a;b)}Q^{b} -12 L_{(4)a}$, $\left( L_{(1)b} Q^{b} \right)_{,a} =
-2L_{(1)(a;b)}Q^{b} - 6L_{(3)a}$, $\left( L_{(0)b} Q^{b} \right)_{,a} =
-2L_{(0)(a;b)}Q^{b} - 2L_{(2)a}$, and $G_{,a}= 2C_{(0)ab}Q^{b} - L_{(1)a}$.

This QFI consists of the independent FIs:
\begin{eqnarray*}
I_{41} &=& \left( -\frac{t^{4}}{4} L_{(3)(a;b)} - \frac{t^{2}}{2}%
L_{(1)(a;b)} + C_{(0)ab} \right) \dot{q}^{a} \dot{q}^{b} + t^{3} L_{(3)a}%
\dot{q}^{a} + t L_{(1)a}\dot{q}^{a} + \\
&& +\frac{t^{4}}{4} L_{(3)a}Q^{a} + \frac{%
t^{2}}{2} L_{(1)a}Q^{a}+ G(q) \\
I_{42} &=& \left( - \frac{t^{3}}{3} L_{(2)(a;b)} - t L_{(0)(a;b)} \right)
\dot{q}^{a} \dot{q}^{b} + t^{4} L_{(4)a}\dot{q}^{a} + t^{2}L_{(2)a}\dot{q}%
^{a} + L_{(0)a}\dot{q}^{a} + \\
&& +\frac{t^{5}}{5} s+ \frac{t^{3}}{3}
L_{(2)a}Q^{a} +t L_{(0)a}Q^{a}.
\end{eqnarray*}

-For $n=5$.
\begin{eqnarray*}
I_{5} &=& \left( -\frac{t^{5}}{5} L_{(4)(a;b)} - \frac{t^{4}}{4}
L_{(3)(a;b)} -\frac{t^{3}}{3}L_{(2)(a;b)} - \frac{t^{2}}{2} L_{(1)(a;b)} -t L_{(0)(a;b)} + C_{(0)ab} \right) \dot{q}^{a} \dot{q}^{b} + \\
&& + t^{5} L_{(5)a}%
\dot{q}^{a} + t^{4} L_{(4)a}\dot{q}^{a} +t^{3} L_{(3)a}\dot{q}^{a} + t^{2}L_{(2)a}\dot{q}^{a}+ t L_{(1)a}\dot{q}%
^{a} + L_{(0)a}\dot{q}^{a}+\frac{t^{6}}{6} s + \\
&&+ \frac{t^{5}}{5}L_{(4)a}Q^{a} + \frac{t^{4}}{4} L_{(3)a}Q^{a} + \frac{t^{3}}{3} L_{(2)a}Q^{a} + \frac{t^{2}}{2} L_{(1)a}Q^{a} +t L_{(0)a}Q^{a} + G(q)
\end{eqnarray*}
where $C_{(0)ab}$ and $L_{(M)(a;b)}$ for $M=0,...n-1$ are KTs, $L_{(5)a}$ is a
KV, $L_{(5)a}Q^{a}=s$, $\left( L_{(4)b} Q^{b} \right)_{,a} = -2L_{(4)(a;b)}
Q^{b}$, $\left( L_{(3)b} Q^{b} \right)_{,a} = -2L_{(3)(a;b)} Q^{b} -20
L_{(5)a}$, $\left( L_{(2)b} Q^{b} \right)_{,a} = -2L_{(2)(a;b)} Q^{b} -12
L_{(4)a}$, $\left( L_{(1)b} Q^{b} \right)_{,a} = -2L_{(1)(a;b)}Q^{b} -
6L_{(3)a}$, $\left( L_{(0)b} Q^{b} \right)_{,a} = -2L_{(0)(a;b)}Q^{b} -
2L_{(2)a}$, and $G_{,a}= 2C_{(0)ab}Q^{b} - L_{(1)a}$.

The QFI consists of the independent FIs:
\begin{eqnarray*}
I_{51} &=& \left( - \frac{t^{4}}{4} L_{(3)(a;b)} - \frac{t^{2}}{2}
L_{(1)(a;b)}+ C_{(0)ab} \right) \dot{q}^{a} \dot{q}^{b} + t^{5} L_{(5)a}\dot{%
q}^{a} + t^{3} L_{(3)a}\dot{q}^{a} + t L_{(1)a}\dot{q}^{a} + \\
&& + \frac{t^{6}}{6}
s+ \frac{t^{4}}{4} L_{(3)a}Q^{a} +\frac{t^{2}}{2} L_{(1)a}Q^{a} + G(q) \\
I_{52} &=& \left( -\frac{t^{5}}{5} L_{(4)(a;b)} - \frac{t^{3}}{3}%
L_{(2)(a;b)} - t L_{(0)(a;b)} \right) \dot{q}^{a} \dot{q}^{b} + t^{4}
L_{(4)a}\dot{q}^{a} + t^{2}L_{(2)a}\dot{q}^{a} + L_{(0)a}\dot{q}^{a} + \\
&& + \frac{t^{5}}{5} L_{(4)a}Q^{a}+ \frac{t^{3}}{3} L_{(2)a}Q^{a} +t L_{(0)a}Q^{a}.
\end{eqnarray*}

If we continue in the same way, we prove that for $A^{a}_{b}=0$ the QFI $I_{n}$ consists of the independent FIs:
\begin{eqnarray*}
I_{\ell1} &=& \left( - \frac{t^{2\ell}}{2\ell} L_{(2\ell-1)(a;b)} - ... -
\frac{t^{4}}{4} L_{(3)(a;b)} - \frac{t^{2}}{2} L_{(1)(a;b)} + C_{(0)ab}
\right) \dot{q}^{a} \dot{q}^{b} + t^{2\ell-1} L_{(2\ell-1)a}\dot{q}^{a} +
... + \\
&& + t^{3}L_{(3)a}\dot{q}^{a} +t L_{(1)a}\dot{q}^{a} + \frac{t^{2\ell}}{2\ell} L_{(2\ell-1)a}Q^{a} +
... + \frac{t^{4}}{4} L_{(3)a}Q^{a} + \frac{t^{2}}{2} L_{(1)a}Q^{a} + G(q)
\end{eqnarray*}
where $C_{(0)ab}$ and $L_{(M)(a;b)}$ for $M=1,3,...,2\ell-1$ are KTs, $\left(
L_{(2\ell-1)b} Q^{b} \right)_{,a} = -2L_{(2\ell-1)(a;b)}Q^{b}$, $\left(
L_{(k-1)b} Q^{b} \right)_{,a} = -2L_{(k-1)(a;b)}Q^{b} - k(k+1)L_{(k+1)a}$
for $k=2,4,...,2\ell-2$, and $G_{,a}= 2C_{(0)ab}Q^{b} - L_{(1)a}$; and
\begin{eqnarray*}
I_{\ell2} &=& \left( - \frac{t^{2\ell+1}}{2\ell+1} L_{(2\ell)(a;b)} - ... -
\frac{t^{3}}{3} L_{(2)(a;b)} - t L_{(0)(a;b)} \right) \dot{q}^{a} \dot{q}%
^{b} + t^{2\ell} L_{(2\ell)a}\dot{q}^{a} + ... + t^{2}L_{(2)a}\dot{q}^{a} +\\
&& + L_{(0)a}\dot{q}^{a}+ \frac{t^{2\ell+1}}{2\ell+1} L_{(2\ell)a}Q^{a} +
... + \frac{t^{3}}{3} L_{(2)a}Q^{a} +t L_{(0)a}Q^{a}
\end{eqnarray*}
where $L_{(M)(a;b)}$ for $M=0,2,...,2\ell$ are KTs, $\left( L_{(2\ell)b}
Q^{b} \right)_{,a} = -2L_{(2\ell)(a;b)}Q^{b}$, and \\ $\left( L_{(k-1)b} Q^{b}
\right)_{,a} = -2L_{(k-1)(a;b)}Q^{b} - k(k+1)L_{(k+1)a}$ for $%
k=1,3,...,2\ell-1$.

We note that the set of the constraints of the QFI $I_{n}(A^{a}_{b}=0)$ is divided into: a) One set involving the odd vectors $L_{(2k+1)a}$, the KT $C_{0ab}$ and the function $G(q)$; and b) A second set involving only the even vectors $L_{(2k)a}$. This explains why the QFI $I_{n}(A^{a}_{b}=0)$ breaks into two independent FIs.

\underline{\textbf{II. Case $\mathbf{n \neq m}$}.} ($n$ or $m$ may be infinite)

We find QFIs that are subcases of those found in \textbf{Case I} and \textbf{Case III} below.

\underline{\textbf{III. Both $\mathbf{n}$ and $\mathbf{m}$ are infinite.}}

In this case, we consider the solution to have the form\footnote{To find a solution, we consider $C_{(0)ab}=c_{0}C_{ab}$, $C_{(1)ab}=c_{1}C_{ab}$, ..., $C_{(n)ab}= nc_{n}C_{ab}$, and $L_{(0)a}= d_{0}L_{a}$, $L_{(1)a}=d_{1}L_{a}$, ...., $L_{(m)a}=d_{m}L_{a}$.}:
$K_{ab}(t,q) = g(t)C_{ab}(q)$ and $K_{a}(t,q)= f(t)L_{a}(q)$,
where the functions $g(t)$ and $f(t)$ are analytic so that they may be represented by polynomial functions as follows:
\[
g(t)= \sum^n_{k=0} c_k t^k = c_0 + c_1 t + ... + c_n t^n \enskip \text{and} \enskip f(t)= \sum^m_{k=0} d_k t^k = d_0 + d_1 t + ... + d_m t^m.
\]

Only the following subcase give a new (non-trivial) independent QFI\footnote{This is the QFI $J_{2}$ of Theorem \ref{Theorem2}.}.

\underline{\textbf{Subcase $\mathbf{(g = e^{\lambda t}}$, $\mathbf{f =e^{\mu t})}$.}} $\lambda \mu \neq 0$.

\begin{equation*}
\begin{cases}
\eqref{eq.veldep6} \implies \lambda e^{\lambda t} C_{ab} + e^{\mu t}
L_{(a;b)} + 2e^{\lambda t}C_{c(a}A^c_{b)} = 0 \\
\eqref{eq.veldep7} \implies - 2 e^{\lambda t} C_{ab} Q^{b} + \mu e^{\mu t}L_a + K_{,a} + e^{\mu t}L_bA^b_a = 0 \\
\eqref{eq.veldep8} \implies K_{,t} = e^{\mu t} L_a Q^{a} \\
\eqref{eq.veldep9} \implies \mu^2 e^{\mu t} L_a + \mu e^{\mu t} L_bA^b_a+e^{\mu t} \left( L_b Q^{b}\right)_{;a} - 2 \lambda e^{\lambda t} C_{ab}Q^{b} = 0.
\end{cases}%
\end{equation*}

We consider the following subcases:

a) \underline{For $\lambda \neq \mu$:}

From \eqref{eq.veldep6}, we have that $C_{ab} = -\frac{2}{\lambda}
C_{c(a}A^c_{b)}$ and $L_a$ is a KV.

From \eqref{eq.veldep9}, we find that $C_{ab}Q^{b}=0$ and $\mu^2 L_a + \mu
L_bA^b_a + \left( L_b Q^{b} \right)_{,a}= 0$.

The solution of \eqref{eq.veldep8} is $K=\frac{1}{\mu }e^{\mu t}L_{a}Q^{a} +G(q)$ which when replaced in \eqref{eq.veldep7} gives $G(q)=const \equiv0$.

The QFI is
\begin{equation*}
I_{e}(\lambda \neq \mu )= e^{\lambda t}C_{ab}\dot{q}^{a}\dot{q}^{b}+e^{\mu
t}L_{a}\dot{q}^{a}+\frac{1}{\mu }e^{\mu t}L_{a}Q^{a}
\end{equation*}
where $C_{ab}=-\frac{2}{\lambda }C_{c(a}A_{b)}^{c}$ is a KT such that $%
C_{ab}Q^{b}=0$, and $L_{a}=-\frac{1}{\mu ^{2}}\left( L_{b}Q^{b}\right) _{,a}-%
\frac{1}{\mu }L_{b}A_{a}^{b}$ is a KV.

We note that the QFI $I_{e}(\lambda \neq \mu)$ consists of the two independent FIs:
\begin{equation*}
J_{2a} = e^{\lambda t}C_{ab}\dot{q}^{a}\dot{q}^{b}, \enskip J_{2b}= e^{\mu
t}L_{a}\dot{q}^{a}+\frac{1}{\mu }e^{\mu t}L_{a}Q^{a}
\end{equation*}

b) \underline{For $\lambda = \mu$:}

From \eqref{eq.veldep6}, we have that $C_{ab} = - \frac{1}{\lambda} L_{(a;b)}
- \frac{2}{\lambda} C_{c(a}A^c_{b)}$.

From \eqref{eq.veldep9}, we find that $\lambda^2 L_a + \lambda L_bA^b_a +
\left( L_b Q^{b}\right)_{,a} - 2 \lambda C_{ab} Q^{b} = 0$.

The solution of \eqref{eq.veldep8} is $K=\frac{1}{\lambda }e^{\lambda t}L_{a}Q^{a} +G(q)$ which when replaced in \eqref{eq.veldep7} gives $G(q)=const \equiv 0$.

The QFI is
\begin{equation*}
I_{e}(\lambda =\mu )=e^{\lambda t}C_{ab}\dot{q}^{a}\dot{q}^{b}+e^{\lambda
t}L_{a}\dot{q}^{a}+ \frac{1}{\lambda }e^{\lambda t}L_{a}Q^{a} \equiv J_{2}
\end{equation*}%
where $C_{ab}=-\frac{1}{\lambda }L_{(a;b)}-\frac{2}{\lambda }%
C_{c(a}A_{b)}^{c}$ is a KT, and the vector $L_{a}=-\frac{1}{\lambda ^{2}}%
\left( L_{b}Q^{b}\right) _{,a}-\frac{1}{\lambda }L_{b}A_{a}^{b}+\frac{2}{%
\lambda }C_{ab}Q^{b}$.

We observe that the FIs $J_{2a}$ and $J_{2b}$ found previously are subcases of the new QFI $J_{2}$; indeed, \newline
$J_{2a}=$ $J_{2}(L_{a}=0)$ and $J_{2b}=$ $J_{2}(C_{ab}=0)$. Therefore, the \textbf{Case III} leads to only one independent QFI the $J_{2}$.
\bigskip

The above complete the proof of Theorem \ref{Theorem2}.

%% file: proof_thm_higher_FIs.tex
\chapter{Proof of Theorem \ref{thm.mFIs}}

\label{app.proof.theorem.higher}

Substituting (\ref{eq.hfi5.3}) and (\ref{eq.hfi5.4}) in the system of PDEs\footnote{Equation (\ref{eq.hfi4a}) is identically satisfied since the quantities $C_{(N)i_{1}...i_{m}}$ are assumed to be $m$th-order KTs.} (\ref{eq.hfi4b}) - (\ref{eq.hfi5.2}), we obtain the following system of equations\footnote{Equation (\ref{eq.hfi5.2}) is not necessary, because the integrability condition $M_{,[i_{1}i_{2}]}=0$ does not intervene in the calculations. In any case, it
has been checked  that equation (\ref{eq.hfi5.2}) is always
satisfied (as an identity $0=0$) from the solutions of the other equations of the system.}:
\begin{eqnarray}
0&=& -2 L_{(1)i_{1}i_{2}}Q^{i_{2}} -4 L_{(2)i_{1}i_{2}}Q^{i_{2}}t -... - 2n L_{(n)i_{1}i_{2}}Q^{i_{2}} t^{n-1} +2L_{(2)i_{1}}+6L_{(3)i_{1}}t+ ...+ \notag \\
&& +n(n-1)L_{(n)i_{1}}t^{n-2} + \left( L_{(0)c}Q^{c}\right)_{,i_{1}}+ \left( L_{(1)c}Q^{c}\right)_{,i_{1}}t +...+ \left( L_{(n)c}Q^{c}\right)_{,i_{1}}t^{n} \label{eq.hfi5.5a} \\
0&=& M_{,t} - L_{(0)i_{1}}Q^{i_{1}} - L_{(1)i_{1}}Q^{i_{1}}t - ... - L_{(n)i_{1}}Q^{i_{1}}t^{n} \label{eq.hfi5.5b} \\
0&=& -2 L_{(0)i_{1}i_{2}}Q^{i_{2}} -2 L_{(1)i_{1}i_{2}}Q^{i_{2}} t - ... - 2 L_{(n)i_{1}i_{2}} Q^{i_{2}} t^{n}
+ L_{(1)i_{1}} + 2L_{(2)i_{1}}t + ... + \notag \\
&& +nL_{(n)i_{1}}t^{n-1} + M_{,i_{1}} \label{eq.hfi5.5c} \\
0&=& -(r+1) L_{(0)i_{1}...i_{r}i_{r+1}}Q^{i_{r+1}} -(r+1) L_{(1)i_{1}...i_{r}i_{r+1}}Q^{i_{r+1}} t - ... -\notag \\
&&- (r+1) L_{(n)i_{1}...i_{r}i_{r+1}}Q^{i_{r+1}} t^{n}
+ L_{(1)i_{1}...i_{r}} + 2L_{(2)i_{1}...i_{r}}t + ... + nL_{(n)i_{1}...i_{r}}t^{n-1}+ \notag \\
&& + L_{(0)(i_{1}...i_{r-1};i_{r})} + L_{(1)(i_{1}...i_{r-1};i_{r})}t + ... + L_{(n)(i_{1}...i_{r-1};i_{r})} t^{n}, \enskip r=2,3,...,m-2
\label{eq.hfi5.5d} \\
0&=& -m C_{(0)i_{1}...i_{m-1}i_{m}}Q^{i_{m}} -m C_{(1)i_{1}...i_{m-1}i_{m}}Q^{i_{m}} t -m C_{(2)i_{1}...i_{m-1}i_{m}}Q^{i_{m}} \frac{t^{2}}{2} -... - \notag \\
&&- m C_{(n)i_{1}...i_{m-1}i_{m}}Q^{i_{m}} \frac{t^{n}}{n}
+L_{(1)i_{1}...i_{m-1}} + 2L_{(2)i_{1}...i_{m-1}}t + ... + nL_{(n)i_{1}...i_{m-1}}t^{n-1} + \notag \\
&& + L_{(0)(i_{1}...i_{m-2};i_{m-1})} + L_{(1)(i_{1}...i_{m-2};i_{m-1})}t + ... + L_{(n)(i_{1}...i_{m-2};i_{m-1})} t^{n}
\label{eq.hfi5.5e} \\
0&=& C_{(1)i_{1}...i_{m}} + C_{(2)i_{1}...i_{m}}t + ... + C_{(n)i_{1}...i_{m}} t^{n-1} + L_{(0)(i_{1}...i_{m-1};i_{m})} + L_{(1)(i_{1}...i_{m-1};i_{m})}t+ \notag \\
&& + ... + L_{(n)(i_{1}...i_{m-1};i_{m})} t^{n}
\label{eq.hfi5.5f}
\end{eqnarray}
where --without loss of generality-- the polynomial expressions (\ref{eq.hfi5.3}) and (\ref{eq.hfi5.4}) of $t$ are assumed to be of the same degree, that is, $n=n_{r}$ for all values of $r$. All the results with $n \neq n_{r}$ are derived as subcases from the case $n=n_{r}$.

We consider the following cases.
\bigskip

\underline{\textbf{I. Case with $n$ finite.}}

From (\ref{eq.hfi5.5a}), we obtain $L_{(n)i_{1}}Q^{i_{1}}= s= const$, $\left( L_{(n-1)c}Q^{c} \right)_{,i_{1}} =$ $2nL_{(n)i_{1}i_{2}}Q^{i_{2}}$ and
\[
\left( L_{(k-2)c}Q^{c} \right)_{,i_{1}} = 2(k-1)L_{(k-1)i_{1}i_{2}}Q^{i_{2}} -k(k-1)L_{(k)i_{1}}, \enskip k=2,3,...,n.
\]

The solution of (\ref{eq.hfi5.5b}) is $M= L_{(0)c}Q^{c}t + L_{(1)c}Q^{c}\frac{t^{2}}{2} + ... + L_{(n-1)c} Q^{c} \frac{t^{n}}{n} + s\frac{t^{n+1}}{n+1} + G(q)$, where $G(q)$ is an arbitrary function. Substituting the function $M$ in (\ref{eq.hfi5.5c}), we find that $G_{,i_{1}}= 2L_{(0)i_{1}i_{2}}Q^{i_{2}} -L_{(1)i_{1}}$.

Equation (\ref{eq.hfi5.5d}) implies that $L_{(n)(i_{1}...i_{r-1};i_{r})} = (r+1) L_{(n)i_{1}...i_{r}i_{r+1}} Q^{i_{r+1}}$ and 
\[
L_{(k-1)(i_{1}...i_{r-1};i_{r})}= (r+1) L_{(k-1)i_{1}...i_{r}i_{r+1}} Q^{i_{r+1}} -kL_{(k)i_{1}...i_{r}}, \enskip k=1,2,...,n
\]
where $r=2,3,...,m-2$.

Equation (\ref{eq.hfi5.5e}) gives $L_{(n)(i_{1}...i_{m-2};i_{m-1})}= \frac{m}{n} C_{(n)i_{1}...i_{m-1}i_{m}} Q^{i_{m}}$ and
\[
L_{(k-1)(i_{1}...i_{m-2};i_{m-1})}= \frac{m}{k-1} C_{(k-1)i_{1}...i_{m-1}i_{m}} Q^{i_{m}} -kL_{(k)i_{1}...i_{m-1}}, \enskip k=1,2,...,n.
\]
We note that \emph{when the undefined term $\frac{C_{(0)i_{1}...i_{m}}}{0}$ appears in the calculations, it must be replaced by $C_{(0)i_{1}...i_{m}}$ in order to have a consistent result.}

Finally, equation (\ref{eq.hfi5.5f}) implies that $L_{(n)(i_{1}...i_{m-1};i_{m})}= 0$ and
$L_{(k-1)(i_{1}...i_{m-1};i_{m})}= -C_{(k)i_{1}...i_{m}}$, where $k=1,2,...,n$.

The FI is
\begin{eqnarray}
I^{(m)}_{n}&=& \left( -\frac{t^{n}}{n} L_{(n-1)(i_{1}...i_{m-1};i_{m})} - ... - \frac{t^{2}}{2} L_{(1)(i_{1}...i_{m-1};i_{m})} - tL_{(0)(i_{1}...i_{m-1};i_{m})} +C_{(0)i_{1}...i_{m}} \right) \dot{q}^{i_{1}} ... \dot{q}^{i_{m}} + \notag \\
&& + \sum_{r=1}^{m-1} \left( t^{n} L_{(n)i_{1}...i_{r}} + ... + tL_{(1)i_{1}...i_{r}} +L_{(0)i_{1}...i_{r}} \right) \dot{q}^{i_{1}} ... \dot{q}^{i_{r}} + s\frac{t^{n+1}}{n+1} + \notag \\
&& + L_{(n-1)c}Q^{c}\frac{t^{n}}{n} + ... + L_{(1)c}Q^{c}\frac{t^{2}}{2} + L_{(0)c}Q^{c}t + G(q) \label{eq.FI1}
\end{eqnarray}
where $C_{(0)i_{1}...i_{m}}$ and $L_{(N)(i_{1}...i_{m-1};i_{m})}$ for $N=0,...,n-1$ are $m$th-order KTs, $L_{(n)i_{1}...i_{m-1}}$ is an $(m-1)$th-order KT, $s$ is an arbitrary constant and the following conditions are satisfied:
\begin{eqnarray*}
L_{(n)(i_{1}...i_{m-2};i_{m-1})}&=& -\frac{m}{n} L_{(n-1)(i_{1}...i_{m-1};i_{m})} Q^{i_{m}} \\
L_{(k-1)(i_{1}...i_{m-2};i_{m-1})}&=& -\frac{m}{k-1} L_{(k-2)(i_{1}...i_{m-1};i_{m})} Q^{i_{m}} -kL_{(k)i_{1}...i_{m-1}}, \enskip k=2,3,...,n \\ L_{(0)(i_{1}...i_{m-2};i_{m-1})}&=& m C_{(0)i_{1}...i_{m-1}i_{m}} Q^{i_{m}} -L_{(1)i_{1}...i_{m-1}} \\
L_{(n)(i_{1}...i_{r-1};i_{r})}&=& (r+1) L_{(n)i_{1}...i_{r}i_{r+1}} Q^{i_{r+1}}, \enskip r=2,3,...,m-2 \\
L_{(k-1)(i_{1}...i_{r-1};i_{r})}&=& (r+1) L_{(k-1)i_{1}...i_{r}i_{r+1}} Q^{i_{r+1}} -kL_{(k)i_{1}...i_{r}}, \enskip k=1,2,...,n, \enskip r=2,3,...,m-2 \\
L_{(n)i_{1}}Q^{i_{1}}&=& s \\
\left( L_{(n-1)c}Q^{c} \right)_{,i_{1}} &=& 2nL_{(n)i_{1}i_{2}}Q^{i_{2}} \\
\left( L_{(k-2)c}Q^{c} \right)_{,i_{1}} &=& 2(k-1)L_{(k-1)i_{1}i_{2}}Q^{i_{2}} -k(k-1)L_{(k)i_{1}}, \enskip k=2,3,...,n \\
G_{,i_{1}}&=& 2L_{(0)i_{1}i_{2}}Q^{i_{2}} -L_{(1)i_{1}}.
\end{eqnarray*}

The notation $I^{(m)}_{n}$ means the $m$th-order FI (upper index) with time-dependence $n$ (lower index). For example, in this notation the $I^{(2)}_{n}$ is a QFI whose coefficients are expressed as polynomials of $t$ of degree fixed by $n$.

We note that all totally symmetric tensors $C_{(0)i_{1}...i_{m}}$ and $L_{(N)(i_{1}...i_{m-1};i_{m})}$ for $N=0,...,n-1$ are $m$th-order KTs; therefore, in order to find the $m$th-order FIs, we need to calculate \emph{all} KTs (both reducible and irreducible) of the kinetic metric $\gamma_{ab}$ of all orders (i.e. $0,1,...,m$).

\underline{\textbf{II. Case with $n$ infinite.}}

In this case, we consider the $m$th-order KT $M_{i_{1}...i_{m}}(t,q)$ and the $r$-rank totally symmetric tensors $M_{i_{1}...i_{r}}(t,q)$, where $r=1,2,...,m-1$, to have the form:
\[
M_{i_{1}...i_{m}}(t,q)= f_{(m)}(t) C_{i_{1}...i_{m}}(q), \enskip M_{i_{1}...i_{r}}(t,q)= f_{(r)}(t) L_{i_{1}...i_{r}}(q), \enskip r=1,2,...,m-1
\]
where $L_{i_{1}...i_{r}}(q)$ are $r$-rank totally symmetric tensors, $C_{i_{1}...i_{m}}(q)$ is an $m$th-order KT of $\gamma_{ab}$ and the functions $f_{(s)}(t)$, $s=1,2,...,m$, are analytic so that they may be represented by polynomial functions of $t$ as follows:
\begin{equation*}
f_{(s)}(t) = \sum^n_{k=0} d_{(s)k} t^{k} = d_{(s)0} + d_{(s)1} t + ... + d_{(s)n} t^{n}.
\end{equation*}
The coefficients $d_{(s)0}, d_{(s)1}, ..., d_{(s)n}$ are arbitrary constants. 

It is found that all subcases give results already found in the previous case with $n$ finite except for the subcase considered below.

\textbf{\underline{Subcase $f_{(s)}= e^{\lambda_{s}t}$ with $\lambda_{s}\neq0$.}}

The system of equations (\ref{eq.hfi5.5a}) - (\ref{eq.hfi5.5f}) becomes:
\begin{eqnarray}
0&=& -2 \lambda_{2}e^{\lambda_{2}t}L_{i_{1}i_{2}}Q^{i_{2}} +\lambda_{1}^{2}e^{\lambda_{1}t}L_{i_{1}} + e^{\lambda_{1}t} \left( L_{c}Q^{c}\right)_{,i_{1}} \label{eq.hfi5.6a} \\
0&=& M_{,t} - e^{\lambda_{1}t}L_{i_{1}}Q^{i_{1}} \label{eq.hfi5.6b} \\
0&=& -2 e^{\lambda_{2}t}L_{i_{1}i_{2}}Q^{i_{2}} + \lambda_{1} e^{\lambda_{1}t}L_{i_{1}} + M_{,i_{1}} \label{eq.hfi5.6c} \\
0&=& -(r+1) e^{\lambda_{r+1}t} L_{i_{1}...i_{r}i_{r+1}}Q^{i_{r+1}}+ \lambda_{r} e^{\lambda_{r}t} L_{i_{1}...i_{r}} + e^{\lambda_{r-1}t} L_{(i_{1}...i_{r-1};i_{r})}, \enskip r=2,3,...,m-2
\label{eq.hfi5.6d} \\
0&=& -m e^{\lambda_{m}t} C_{i_{1}...i_{m-1}i_{m}}Q^{i_{m}}+ \lambda_{m-1}e^{\lambda_{m-1}t}L_{i_{1}...i_{m-1}} + e^{\lambda_{m-2}t} L_{(i_{1}...i_{m-2};i_{m-1})}
\label{eq.hfi5.6e} \\
0&=& \lambda_{m} e^{\lambda_{m}t}C_{i_{1}...i_{m}}+ e^{\lambda_{m-1}t} L_{(i_{1}...i_{m-1};i_{m})}.
\label{eq.hfi5.6f}
\end{eqnarray}
We note that in the calculations that follow, without loss of generality, all the constants $\lambda_{s}$ are fixed to the same non-zero constant $\lambda$. The FI produced from this assumption contains as subcases all the FIs arising from constants $\lambda_{s}$ which are not all the same.

Equation (\ref{eq.hfi5.6a}) implies that $\left(L_{c}Q^{c}\right)_{,i_{1}}= 2\lambda L_{i_{1}i_{2}} Q^{i_{2}} -\lambda^{2}L_{i_{1}}$.

The solution of (\ref{eq.hfi5.6b}) is $M= \frac{e^{\lambda t}}{\lambda} L_{i_{1}}Q^{i_{1}} +G(q)$ which when replaced in (\ref{eq.hfi5.6c}) gives $G(q)=const \equiv 0$.

From the remaining equations (\ref{eq.hfi5.6d}) - (\ref{eq.hfi5.6f}), we find the following conditions:
\[
C_{i_{1}...i_{m}}= -\frac{1}{\lambda} L_{(i_{1}...i_{m-1};i_{m})}
\]
\[
L_{(i_{1}...i_{r-1};i_{r})}=(r+1) L_{i_{1}...i_{r}i_{r+1}} Q^{i_{r+1}} -\lambda L_{i_{1}...i_{r}}, \enskip r=2,3,...,m-2
\]
and
\[
L_{(i_{1}...i_{m-2};i_{m-1})}= m C_{i_{1}...i_{m-1}i_{m}} Q^{i_{m}} -\lambda L_{i_{1}...i_{m-1}}.
\]

The FI is
\begin{equation}
I^{(m)}_{e}= \frac{e^{\lambda t}}{\lambda} \left( -L_{(i_{1}...i_{m-1};i_{m})} \dot{q}^{i_{1}} ... \dot{q}^{i_{m}} + \lambda \sum_{r=1}^{m-1} L_{i_{1}...i_{r}} \dot{q}^{i_{1}} ... \dot{q}^{i_{r}} + L_{i_{1}}Q^{i_{1}} \right) \label{eq.FI2}
\end{equation}
where $\lambda\neq0$, $L_{(i_{1}...i_{m-1};i_{m})}$ is an $m$th-order KT and the following conditions are satisfied:
\begin{eqnarray*}
L_{(i_{1}...i_{m-2};i_{m-1})} &=& -\frac{m}{\lambda} L_{(i_{1}...i_{m-1};i_{m})} Q^{i_{m}} -\lambda L_{i_{1}...i_{m-1}} \\
L_{(i_{1}...i_{r-1};i_{r})} &=&(r+1) L_{i_{1}...i_{r}i_{r+1}} Q^{i_{r+1}} -\lambda L_{i_{1}...i_{r}}, \enskip r=2,3,...,m-2\\
\left(L_{c}Q^{c}\right)_{,i_{1}} &=& 2\lambda L_{i_{1}i_{2}} Q^{i_{2}} -\lambda^{2}L_{i_{1}}.
\end{eqnarray*}
We note that the FI $I^{(m)}_{e}$ requires only the reducible KTs of $\gamma_{ab}$ of all orders (i.e. $0,1,...,m$).

%% file: proof_thm_timedependent.tex
\chapter{Proof of Theorem \ref{thm.polynomial.omega}}

\label{app.proof.QFIs.time}

Substituting the polynomial function (\ref{pol}) in the system of PDEs (\ref{eq.red2a}) - (\ref{eq.red2e}), we have the following cases.
\bigskip

\underline{\textbf{I. Case $\mathbf{n=m}$}} (both $n$ and $m$ are finite)

From (\ref{eq.red2a}), we find that $C_{(k)ab} = -L_{(k-1)(a;b)}$, $k=1,...,n$, and $L_{(n)a}$ is a KV of $\gamma_{ab}$.

Equation (\ref{eq.red2d}) gives
\begin{align*}
0 &= -2\left( b_{1} +2b_{2}t + ... +\ell b_{\ell}t^{\ell-1} \right) \left(C_{(0)ab}Q^{b} +C_{(1)ab}Q^{b} t + ... + C_{(n)ab} Q^{b} \frac{t^{n}}{n}\right) + 2L_{(2)a}+ \\
& \quad + 6L_{(3)a}t +... + n(n-1)L_{(n)a}t^{n-2} -2\left( b_{0}+b_{1}t+ ... + b_{\ell}t^{\ell} \right) \left( C_{(1)ab}Q^{b} + C_{(2)ab}Q^{b}t + \right. \\
& \quad \left. +... + C_{(n)ab}Q^{b} t^{n-1} \right)+ \left( b_{0}+b_{1}t+ ... + b_{\ell}t^{\ell} \right) \left[ \left( L_{(0)b}Q^{b}\right) _{,a} + \left( L_{(1)b}Q^{b} \right)_{,a}t +...+ \right. \\
& \quad \left.+ \left( L_{(n-1)b}Q^{b} \right)_{,a}t^{n-1} + \left( L_{(n)b}Q^{b} \right)_{,a}t^{n} \right].
\end{align*}
This is a polynomial equation of the general form $P_{(0)a}(q) + P_{(1)a}(q)t +... +P_{(n+\ell)a}(q)t^{n+\ell}=0$; therefore, from the vanishing of the coefficients $P_{(k)a}(q)$, we get the following conditions:
\begin{equation}
L_{(n)a}Q^{a}= c= const \label{eq.polc1}
\end{equation}
and
\begin{align}
0 &= \sum_{s=0}^{\ell-1}\left[  -\frac{2(k+s)b_{(k+s\leq\ell)}}{n-s} C_{(n-s\geq0)ab}Q^{b} -2b_{(k+s\leq\ell)} C_{(n-s>0)ab}Q^{b} + \right. \notag \\
& \quad \left.+ b_{(k+s\leq\ell)} \left( L_{(n-s-1\geq0)b}Q^{b}\right)_{,a} \right], \enskip k=1,2,...,\ell \label{eq.polc2} \\
0 &= -\sum_{s=1}^{\ell}\left[ \frac{2sb_{s}}{n-s} C_{(n-s\geq0)ab}Q^{b} \right] + \sum_{s=0}^{\ell} \left[ -2b_{s} C_{(n-s>0)ab}Q^{b} + b_{s} \left(L_{(n-s-1\geq0)b}Q^{b} \right)_{,a} \right] \label{eq.polc3} \\
0 &= \sum_{s=0}^{\ell} \left[ -2b_{s}C_{(k-s-1>0)ab}Q^{b} +b_{s}\left( L_{(k-s-2\geq0)b}Q^{b} \right)_{,a} \right] -\sum_{s=1}^{\ell} \left[ \frac{2sb_{s}}{k-s-1} C_{(k-s-1\geq0)ab}Q^{b} \right] + \notag \\
& \quad + k(k-1)L_{(k)a}, \enskip k=2,3,...n. \label{eq.polc4}
\end{align}

We note that in the $n+\ell+1$ equations (\ref{eq.polc2}) - (\ref{eq.polc4}), when the undefined quantity $\frac{C_{(0)ab}}{0}$ appears in the calculations, it must be replaced by $C_{(0)ab}$ in order to have a consistent result.

We continue with the remaining constraints (\ref{eq.red2b}) and (\ref{eq.red2c}) in order to determine the scalar coefficient $K(t,q)$.

The solution of (\ref{eq.red2c}) is
\begin{eqnarray*}
K_{,t}&=& L_{(0)a}Q^{a} \left( b_{0} +b_{1}t +... +b_{\ell}t^{\ell} \right) + L_{(1)a}Q^{a} \left( b_{0}t +b_{1}t^{2} +... +b_{\ell}t^{\ell+1} \right) + ... + \\
&& + L_{(n-1)a}Q^{a} \left( b_{0}t^{n-1} +b_{1}t^{n} +... +b_{\ell}t^{n+\ell-1} \right) + s \left( b_{0}t^{n} +b_{1}t^{n+1} +... +b_{\ell}t^{n+\ell} \right) \implies \\
K &=& L_{(0)a}Q^{a} \left( b_{0}t +b_{1}\frac{t^{2}}{2} +... +b_{\ell}\frac{t^{\ell+1}}{\ell+1} \right) + L_{(1)a}Q^{a} \left( b_{0}\frac{t^{2}}{2} +b_{1}\frac{t^{3}}{3} +... +b_{\ell}\frac{t^{\ell+2}}{\ell+2} \right) + ... + \\
&& + L_{(n-1)a}Q^{a} \left( b_{0}\frac{t^{n}}{n} +b_{1}\frac{t^{n+1}}{n+1} +... +b_{\ell}\frac{t^{n+\ell}}{n+\ell} \right) + s \left( b_{0}\frac{t^{n+1}}{n+1} +b_{1}\frac{t^{n+2}}{n+2} +... +b_{\ell}\frac{t^{n+\ell+1}}{n+\ell+1} \right) +G(q).
\end{eqnarray*}
Replacing $K$ in (\ref{eq.red2b}) and using the conditions (\ref{eq.polc1}) - (\ref{eq.polc4}), we find that $G_{,a}= 2b_{0}C_{(0)ab}Q^{b} -L_{(1)a}$.

Condition (\ref{eq.red2e}) is satisfied trivially from the above solutions.

The QFI is
\begin{eqnarray*}
I &=& \left( \frac{t^{n}}{n}C_{(n)ab} + ... + tC_{(1)ab} + C_{(0)ab} \right) \dot{q}^{a}\dot{q}^{b} + t^{n}L_{(n)a} \dot{q}^{a} + ... + tL_{(1)a}\dot{q}^{a} + L_{(0)a}\dot{q}^{a} + \\
&& + L_{(0)a}Q^{a} \left( b_{0}t +b_{1}\frac{t^{2}}{2} +... +b_{\ell}\frac{t^{\ell+1}}{\ell+1} \right) + L_{(1)a}Q^{a} \left( b_{0}\frac{t^{2}}{2} +b_{1}\frac{t^{3}}{3} +... +b_{\ell}\frac{t^{\ell+2}}{\ell+2} \right) + ... + \\
&& + L_{(n-1)a}Q^{a} \left( b_{0}\frac{t^{n}}{n} +b_{1}\frac{t^{n+1}}{n+1} +... +b_{\ell}\frac{t^{n+\ell}}{n+\ell} \right) + s \left( b_{0}\frac{t^{n+1}}{n+1} +b_{1}\frac{t^{n+2}}{n+2} +... +b_{\ell}\frac{t^{n+\ell+1}}{n+\ell+1} \right) +G(q)
\end{eqnarray*}
where $C_{(0)ab}$ is a KT, the KTs $C_{(k)ab} = -L_{(k-1)(a;b)}$ for $k=1,...,n$,  $L_{(n)a}$ is a KV such that $L_{(n)a}Q^{a}=s$, $G_{,a}= 2b_{0}C_{(0)ab}Q^{b} -L_{(1)a}$, and the conditions (\ref{eq.polc2}) - (\ref{eq.polc4}) are satisfied.

\underline{\textbf{II. Case $\mathbf{n \neq m}$}.} (either $n$ or $m$ may be infinite)

We find QFIs that are subcases of those found in \textbf{Case I} and \textbf{Case III} below.

\underline{\textbf{III. Both $\mathbf{n}$ and $\mathbf{m}$ are infinite.}}

In this case, we consider the solution to have the form $K_{ab}(t,q) = g(t)C_{ab}(q)$ and $K_{a}(t,q)= f(t)L_{a}(q)$, where the functions $g(t)$ and $f(t)$ are analytic so that
they may be represented by polynomial functions as follows:
\begin{equation*} 
g(t) = \sum^n_{k=0} c_k t^k = c_0 + c_1 t + ... + c_n t^n, \enskip f(t) = \sum^m_{k=0} d_k t^k = d_0 + d_1 t + ... + d_m t^m.
\end{equation*}
The coefficients $c_{0}, c_{1}, ..., c_{n}$ and $d_{0}, d_{1}, ..., d_{m}$ are arbitrary constants. We find that only the following subcase gives a new independent QFI. 

\textbf{\underline{Subcase $\mathbf{(g = e^{\lambda t}}$, $\mathbf{f = e^{\mu t})}$, $\mathbf{\lambda \mu \neq 0}$}}.

In this case, the system of PDEs (\ref{eq.red2a}) - (\ref{eq.red2d}) becomes\footnote{Equation (\ref{eq.red2e}) is
satisfied trivially from the solutions found below.}:
\begin{eqnarray}
0&=& \lambda e^{\lambda t} C_{ab} + e^{\mu t} L_{(a;b)} \label{eq.polc5.1} \\
0&=& -2\left( b_{0}+b_{1}t + ... + b_{\ell}t^{\ell} \right) e^{\lambda t} C_{ab} Q^{b} + \mu e^{\mu t} L_a + K_{,a}  \label{eq.polc5.2} \\
0&=& K_{,t} - (b_{0}+b_{1}t +... + b_{\ell}t^{\ell}) e^{\mu t} L_a Q^{a} \label{eq.polc5.3} \\
0&=& -2\left( b_{1} + 2b_{2}t + ... + \ell b_{\ell} t^{\ell-1} \right)e^{\lambda t} C_{ab}Q^{b} - 2 \lambda (b_{0}+b_{1}t+ ... + b_{\ell}t^{\ell})e^{\lambda t} C_{ab} Q^{b} + \notag \\
&& +\mu^2 e^{\mu t} L_a + (b_{0}+b_{1}t + ... + b_{\ell} t^{\ell})e^{\mu t} \left( L_b Q^{b}\right)_{,a} \label{eq.polc5.4}.
\end{eqnarray}

We consider the following subcases:

a. \underline{For $\lambda \neq \mu$:}

From (\ref{eq.polc5.1}), we have that $C_{ab} =0$ and $L_{a}$ is a KV. Equation (\ref{eq.polc5.4}) implies that $L_{a}=0$. Therefore, the QFI $I_{e}(\lambda \neq \mu) =const$ which is trivial.

b. \underline{For $\lambda = \mu$:}

From (\ref{eq.polc5.1}), we have that $C_{ab} = - \frac{1}{\lambda} L_{(a;b)}$; therefore, $L_{(a;b)}$ is a reducible KT.

We consider two cases according to the degree $\ell$ of the polynomial $\omega(t)$.

- Case $\ell=1$.

From (\ref{eq.polc5.4}), we get the conditions:
\begin{eqnarray}
\left( L_b Q^{b}\right)_{,a} &=& 2\lambda C_{ab}Q^{b} \label{eq.polc6.1} \\
\lambda^2 L_a + b_{0}\left( L_b Q^{b}\right)_{,a} - 2 (b_{1} + \lambda b_{0}) C_{ab} Q^{b} &=& 0. \label{eq.polc6.2}
\end{eqnarray}
Replacing with $C_{ab} = - \frac{1}{\lambda} L_{(a;b)}$ and by substituting (\ref{eq.polc6.1}) in (\ref{eq.polc6.2}), we obtain:
\begin{eqnarray}
\left( L_b Q^{b}\right)_{,a}&=& -2 L_{(a;b)} Q^{b} \label{eq.polc6.3} \\
\lambda^3 L_a + 2b_{1} L_{(a;b)} Q^{b} &=& 0. \label{eq.polc6.4}
\end{eqnarray}

The solution of (\ref{eq.polc5.3}) is $K=\left( \frac{b_{0}}{\lambda } - \frac{b_{1}}{\lambda^{2}} \right) e^{\lambda t}L_{a}Q^{a} + \frac{b_{1}}{\lambda} te^{\lambda t} L_{a}Q^{a} +G(q)$ which when replaced in (\ref{eq.polc5.2}) gives $G=const\equiv0$.

The QFI is
\begin{equation}
I_{e}(\ell=1) = - e^{\lambda t} L_{(a;b)} \dot{q}^{a}\dot{q}^{b} + \lambda e^{\lambda t} L_{a}\dot{q}^{a} + \left( b_{0} - \frac{b_{1}}{\lambda} \right) e^{\lambda t}L_{a}Q^{a} + b_{1}te^{\lambda t} L_{a}Q^{a} \label{eq.polc6.4a}
\end{equation}
where $L_{(a;b)}$ is a KT, $\left( L_b Q^{b}\right)_{,a}= \frac{\lambda^{3}}{b_{1}}L_{a}$ and $\lambda^3 L_a = -2b_{1} L_{(a;b)} Q^{b}$.

- Case $\ell>1$.

From (\ref{eq.polc5.4}), we find that $\left( L_b Q^{b}\right)_{,a}= 2\lambda C_{ab}Q^{b}$, $C_{ab}Q^{b}= 0$ and $\lambda^2 L_a = 2b_{1}C_{ab} Q^{b}$. Therefore, $L_{a}=0$ $\implies$ $C_{ab} = - \frac{1}{\lambda} L_{(a;b)}=0$ and we end up with the trivial FI $I_{e}=const$.